\input epsf.tex
\epsfclipon
 \documentclass[11pt,a4paper,amsart]{article}
 \usepackage{jheppub}
\usepackage[toc,page]{appendix}
\usepackage[compat=1.1.0]{tikz-feynman}

\usepackage{amsmath,amssymb,mathtools}
\usetikzlibrary{decorations.pathreplacing}
\usetikzlibrary{hobby}

\usepackage{hyperref}
\usepackage{multirow,tabularx}

 \usepackage{color}
\definecolor{myblue}{rgb}{.8, .8, 1}

\usepackage{amsmath}
\usepackage{empheq}
\usepackage{slashed}
\newlength\mytemplen
\newsavebox\mytempbox

\makeatletter
\newcommand\mybluebox{%
    \@ifnextchar[
       {\@mybluebox}%
       {\@mybluebox[0pt]}}

\def\@mybluebox[#1]{%
    \@ifnextchar[
       {\@@mybluebox[#1]}%
       {\@@mybluebox[#1][0pt]}}

\def\@@mybluebox[#1][#2]#3{
    \sbox\mytempbox{#3}%
    \mytemplen\ht\mytempbox
    \advance\mytemplen #1\relax
    \ht\mytempbox\mytemplen
    \mytemplen\dp\mytempbox
    \advance\mytemplen #2\relax
    \dp\mytempbox\mytemplen
    \colorbox{myblue}{\hspace{1em}\usebox{\mytempbox}\hspace{1em}}}

\makeatother

\usepackage{float}
\usepackage{mathtools}
\usepackage{color}

\usepackage{epsfig}
\usepackage{epstopdf}
\usepackage{epsf}
\input{epsf.sty}
\usepackage{graphicx,amsmath,amssymb}
\usepackage{graphicx}
\usepackage{epsfig,multicol}
\usepackage{multirow}
\allowdisplaybreaks

\usepackage{tikz} 

\usepackage[bottom,splitrule]{footmisc}

\usepackage{float}

\usepackage{bbm,bm,amsmath,amssymb}
 
\newcommand{\Su}{\color{blue}}

\usepackage[normalem]{ulem}

\usepackage{comment}
\def\be{\begin{equation}}
\def\ee{\end{equation}}
\def\figs/B{B}
\def\bea{\begin{eqnarray}}
\def\eea{\end{eqnarray}}
\def\bg{\begin{eqnarray}}
\def\nd{\end{eqnarray}}
\def\sin{{\rm sin}}
\def\cos{{\rm cos}}

\def\log{{\rm log}}
\def\ln{{\rm log}}

\renewcommand{\[}{\left[}
\renewcommand{\]}{\right]}
\renewcommand{\(}{\left(}
\renewcommand{\)}{\right)}

\def\be{\begin{equation}}
\def\ee{\end{equation}}
\def\tL{0}
\def\tR{10}
\def\yUp{0.9}
\def\yDn{-0.9}
\newcommand{\transpapers}{\cite{borel2,hetborel,dyson,dorigoni,unsal,grossper,gevrey,borelborel,ecalle,maximE,mirzakhanipaper}}

\newcommand{\gspapers}{\cite{joydeep,wdwpaper,coherbeta,coherbeta2,borel2,hetborel,dileep,desitter2}}

\def\doi{http://doi.org}

\newcommand{\ket}[1]{|#1\rangle}
\newcommand{\bra}[1]{\langle #1|}
\newcommand{\braket}[2]{\langle #1|#2\rangle}

\usepackage{tocloft}
\usepackage{etoolbox}

\makeatletter
\pretocmd{\tableofcontents}{\begingroup\small}{}{}
\apptocmd{\tableofcontents}{\endgroup}{}{}
\makeatother

\usepackage{physics}
\usepackage{tikz}
\usetikzlibrary{arrows.meta}
\usetikzlibrary{angles,quotes} 
\usetikzlibrary{decorations.markings,arrows.meta}
\tikzset{>=latex} 

\tikzset{
  midarr/.style={decoration={markings,mark=at position #1 with {\arrow{stealth}}},postaction={decorate}},
  midarr/.default=0.5
}
\colorlet{xcol}{blue!70!black}

\title{Coherent states versus Glauber-Sudarshan States: Bootstrapping, Schwinger-Keldysh Contours and Lefschetz Thimbles}
\author{{\Su Heliudson Bernardo}$^{1}$, {\Su Tatsuya Daniel}$^{2,3}$, {\Su Keshav Dasgupta}$^{2}$, {\Su Brayden Hull}$^1$, {\Su Yue Katherine Lei}$^2$ and
{\Su Yiya Selina Li}$^{1}$\\
	\vskip.03in
        ${}^1$ Department of Physics \& Astronomy, University of Lethbridge, Lethbridge, \\
        ~~Alberta, T1K 3M4, Canada \\   
        ${}^2$ Department of Physics, McGill University, Montr\'{e}al, Qu\'{e}bec, H3A 2T8, Canada \\  
        ${}^3$ Trottier Space Institute, McGill University, Montr\'{e}al, Qu\'{e}bec, H3A 2T8, Canada \\ 
        {\tt heliudson.bernardo@uleth.ca, tdaniel@physics.mcgill.ca, keshav@hep.physics.mcgill.ca, brayden.hull@uleth.ca, yue.lei@mail.mcgill.ca, selina.li@uleth.ca}}

\date{\today}

\abstract{ We investigate how, in a highly constrained system such as a four-dimensional diffeomorphism-invariant theory with vanishing bulk Hamiltonian and non-trivial interactions between the metric and additional degrees of freedom, transient excited states $-$ called the Glauber-Sudarshan states $-$ can be constructed over supersymmetric minima. These states are generically non-supersymmetric and, although they are not minima of any potential, they admit positive-energy metric configurations that effectively mimic four-dimensional quasi–de Sitter backgrounds. We analyze how such states differ from conventional coherent states by studying their time evolution both in the canonical formalism—via boundary Hamiltonians—and in the in–in path-integral framework through Schwinger–Keldysh contours. We also examine the consistency of the construction across three
complementary descriptions: the 1PI effective action, the Wilsonian
(or exact renormalization group) effective action, and the
Picard--Lefschetz (Lefschetz thimble) decomposition of the
Schwinger--Keldysh path integral, which provides the trans-series
organization of the theory. In the presence of gauge fixing and ghost sectors, we show that the dynamics of these transient configurations are governed by a nontrivial bootstrapping relation that simultaneously constrains their behavior near supersymmetric Minkowski vacua and along the quasi–de Sitter–like trajectories. Finally, we examine the Wheeler--DeWitt equation, the notion of time in the presence of the transient excited states, and the emergence of the bulk Schr\"odinger equation. We further show that these states admit a natural structural interpretation analogous to the vertex operators that arise in the two-dimensional worldsheet formulation of string theory.}

\begin{document}

\maketitle

\newpage

\hskip3.2in {\textcolor{blue}{\it ``Hey .... time to wake up!"}}

\hskip3.0in {\textcolor{blue}{David Lynch's Mulholland Drive (2001)} }

\section{Introduction and conceptual framework \label{sec001}}

Quantum gravity $-$ and more generically string and M-theories $-$ are non-trivial with subtle dynamics compared to standard QFTs on a given background \cite{adm, teitelboim,ghy,WdWorig,hh,hhlater,FPI,page,witten,Dirac-algebra,suvrat,tats}. Once dynamical gravity is inserted in, numerous subtleties show up that change the behavior of the theory drastically. In \cite{joydeep,wdwpaper,coherbeta,coherbeta2,borel2,hetborel,dileep,desitter2} we have discussed many such issues, and here we will
provide further elaborations on them before addressing the main topic of the temporal evolution of the Glauber-Sudarshan states. The latter basically hinges on the following two observations.

\vskip.1in

\noindent (1) If the spacetime slicing is fixed by external structure, as in QFT on a fixed gravitational background, then a quantum state defined on a Cauchy slice evolves naturally under a non-vanishing Hamiltonian.

\vskip.1in

\noindent (2) In dynamical gravity, where the metric itself is quantum, different Cauchy slices are physically equivalent unless a preferred time is fixed by boundary conditions, symmetries, or relational clocks. Only from such a time-fixing structure does genuine time evolution emerge. Alternatively, in dynamical gravity, a state involving both gravity and matter does not admit intrinsic time evolution unless additional structure fixes a preferred time. This structure may be supplied by boundary conditions (asymptotic time), symmetries (Killing vectors), or relational clocks. In the absence of such structure, different Cauchy slices are physically equivalent.

\vskip.1in

\noindent In fact the aforementioned two observations lie at the heart of the problems in quantum gravity or string and M-theories. It is one of the goals of this work to clarify some of these details as they are often misunderstood.

\noindent Our overreaching goal is to elucidate many aspects of how a curved spacetime would be described as an excited state within a class of states that we call Glauber-Sudarshan states. These are defined as specific excitations over a Minkowski state. Since we want to be as general as possible, we only write the explicit form of the theory action when we wish to describe how different sectors contribute to the calculations. So, the only condition on the underlying theory is that it possess a (possibly supersymmetric) Minkowski vacuum state. Broadly speaking, we focus on: {(i)} the ADM decomposition for diffeomorphism invariant theories and the definition of the GS states, {(ii)} how the expectation values of arbitrary operators would be computed in the path integral approach, and {(iii)} how the GS states should emerge consistently from the quantum gravity theory. In what follows, we first present a detailed summary of our main results. We then review several relevant developments in quantum gravity, with the aim of providing the necessary context for the discussion beginning from section \ref{sec2.0} onwards.

\subsection{Organization of the paper and short summaries of the sections}

This paper develops a framework for understanding how transient quasi--de Sitter configurations may arise in a diffeomorphism-invariant theory of gravity {\it without} being interpreted as genuine de Sitter vacua or as minima of an effective potential. The central idea is that such configurations can be realized as Glauber--Sudarshan states constructed over supersymmetric Minkowski sectors. These states are generically non-supersymmetric and time-dependent, but their semiclassical expectation values can temporarily mimic four-dimensional quasi--de Sitter geometries over a finite time interval.

The paper compares these GS states with ordinary coherent states, emphasizing that coherent states are naturally tied to perturbative fluctuations over fixed backgrounds, whereas GS states are designed to encode more general interacting semiclassical configurations. The construction is analyzed using canonical boundary Hamiltonians, Schwinger--Keldysh in-in path integrals, trans-series and Lefschetz-thimble decompositions, Schwinger--Dyson bootstrap equations, and Wheeler--DeWitt constraints. A recurring theme is that time evolution in gravity is not generated by an unconstrained bulk Hamiltonian, but emerges through boundary conditions, semiclassical WKB branches, or relational structures. The paper concludes by drawing a structural analogy between bulk GS insertions and world-sheet vertex operators, while emphasizing that this analogy is not meant to imply a dynamical equivalence between the two theories. In the following we provide a more detailed summary of each of the sections of the paper.

\subsubsection*{1.1.1~~~Summary of section \ref{sec001}}

Sections \ref{sec1.1} to \ref{sec1.5} set up the central conceptual problem of the paper: how time evolution should be understood in a diffeomorphism-invariant theory whose bulk Hamiltonian is a sum of constraints. The main point is that the absence of an intrinsic bulk time in quantum gravity does not preclude physical time evolution when a preferred time is fixed by boundary conditions, asymptotic symmetries, or a semiclassical background. This distinction is then used to motivate the construction of coherent and Glauber--Sudarshan states over supersymmetric Minkowski sectors. In the following we provide a brief summary of the remaining parts of section \ref{sec001} for those readers who would like to skip this section and go directly to section \ref{sec2.0}.

Subsection \ref{sec1.1} reviews the ADM decomposition of the Einstein--Hilbert action, including the role of the Gibbons--Hawking--York boundary term. The bulk Hamiltonian is shown to be a linear combination of the Hamiltonian and momentum constraints, while the nontrivial physical energy is carried by boundary terms. This provides the basic canonical explanation for why gravitational time evolution is tied to boundary structure rather than to an unconstrained bulk Hamiltonian.

Subsection \ref{sec1.2} extends the ADM discussion to a more general action containing higher-curvature corrections, fluxes, matter fields, gauge sectors, branes, and possible stringy contributions. The main claim is that diffeomorphism invariance preserves the universal form of the Hamiltonian as a sum of constraints plus boundary terms, even though the explicit constraint densities are modified. The subsection emphasizes that higher-derivative sectors may enlarge the phase space, but the lapse and shift still enforce generalized Hamiltonian and momentum constraints.

In subsection \ref{sec1.3} we organize the previous discussion into a systematic canonical procedure. One first isolates variational boundary terms, specifies the time and spatial boundaries, chooses the correct canonical variables, performs the Legendre transform, identifies the constraints as projections of the full Euler--Lagrange equations, and finally obtains the boundary Hamiltonian on the constraint surface. The purpose is to clarify that the bulk equations of motion remain nontrivial even though the bulk Hamiltonian vanishes weakly.

Subsection \ref{sec:Hphys_bulk_vs_boundary} distinguishes more carefully between the bulk constraints and the physical Hamiltonian after gauge fixing, ghost contributions, and higher-order sectors are included. It explains that the time-total-derivative terms in the action are associated with endpoint data and the variational principle, while spatial boundary terms contribute to the physical Hamiltonian. The result is that the boundary Hamiltonian should be understood as a properly completed ADM, Brown--York, or Iyer--Wald charge rather than as a naive leftover surface term.

In subsection \ref{sec1.5} we explain how ordinary quantum-mechanical time evolution reappears after choosing a fixed warped Minkowski background. Once a background with asymptotic time is selected, fluctuations around it may be quantized using an effective fluctuation Hamiltonian obtained from the boundary energy. Coherent graviton states then evolve in the usual Schr\"odinger sense, while the full background-independent theory remains governed by Wheeler--DeWitt-type constraints.

\subsubsection*{1.1.2~~~ Summary of section \ref{sec2.0}}

Section \ref{sec2.0} explains why one should not treat de Sitter as a fundamental vacuum or as a minimum of an effective potential. Instead, quasi--de Sitter geometries should be interpreted as transient excited configurations produced by GS states over supersymmetric Minkowski sectors. The section argues that this interpretation avoids several conceptual obstacles associated with de Sitter vacua in quantum gravity.

In subsection \ref{dubaitag1} we discuss the convexity properties of the 1PI effective action and explain why they constrain true vacuum configurations but not necessarily excited-state expectation values. A GS insertion may look algebraically similar to a source term, but it is not varied and no Legendre transform is performed with respect to it. Therefore GS-state expectation values are not interpreted as extrema of the 1PI effective action and are not subject to the same vacuum-minimization logic.

In subsection \ref{sec2.2} we argue that in an accelerating, time-dependent background the usual Wilsonian separation between light and heavy modes can become ill-defined. If the masses of towers of states vary on Hubble time scales, then integrating them out does not yield a local, time-independent Wilsonian effective action. The resulting description is instead closer to a nonlocal, time-dependent influence functional appropriate for an in-in or open-system framework.

In subsection \ref{sec:premiseP} we carefully formulate the obstruction to metastable de Sitter EFTs in terms of three possible failures: absence of quasi-stationarity, insufficient lifetime, and loss of scale separation due to towers of light states. If any of these failures occurs, a de Sitter configuration cannot be interpreted as a controlled metastable vacuum EFT. We conclude the section by showing that such configurations should instead be described as transient cosmological episodes, precisely the role assigned to GS states in the paper.

\subsubsection*{1.1.3~~~ Summary of section \ref{sec3.0}}

Section \ref{sec3.0} begins the main comparison between ordinary coherent states and Glauber--Sudarshan states in gravity. Coherent states are associated with controlled excitations over a background and usually rely on a free or perturbative vacuum, while GS states are defined over an interacting gravitational vacuum and are designed to describe more general semiclassical configurations. The section emphasizes their different behavior under Hamiltonian evolution, their different wavefunctional support, and their different implications for supersymmetry breaking.

Subsection \ref{sec3.1} explains how the notion of an interacting vacuum must be modified in gravity. Since the bulk Hamiltonian is a constraint, the interacting vacuum cannot be generated in the same way as in ordinary QFT by a bulk Hamiltonian. Instead, in asymptotically Minkowski sectors, the interacting vacuum is defined using the boundary Hamiltonian or, semiclassically, the effective Hamiltonian for fluctuations around a chosen background.

In subsection \ref{sec3.2} we compare the time evolution of coherent states and GS states. Ordinary coherent states remain simple under a free Hamiltonian but generally fail to remain coherent under interacting evolution. GS states, by contrast, are defined using the interacting vacuum and can encode interaction effects more naturally. This difference is crucial for using GS states to describe transient semiclassical geometries rather than merely free-field wave packets.

In subsection \ref{sec3.3} we compare the wavefunctional profiles of coherent and GS states. Coherent states are sharply peaked around classical configurations in the free or perturbative phase space, while GS states may have a more general spread over interacting configurations. This distinction allows GS states to describe semiclassical metric profiles that are not simply free coherent graviton excitations over a fixed background.

It is important to figure out how supersymmetry breaking happens for the coherent and the GS states. In subsection \ref{sec3.4} we analyze how supersymmetry breaking should be interpreted in the two constructions. A coherent state may break supersymmetry by exciting modes over a supersymmetric background, but its validity is tied to perturbative control. GS states break supersymmetry at the level of their expectation values while still being built over a supersymmetric reference minimum. Thus the non-supersymmetric quasi--de Sitter profile is interpreted as an excited-state effect, not as a new non-supersymmetric vacuum.

\subsubsection*{1.1.4~~~ Summary of section \ref{sec4.0}}

Section \ref{sec4.0} develops the Schwinger--Keldysh path-integral formulation for coherent-state expectation values. Since the physical configurations of interest are time-dependent expectation values rather than transition amplitudes, the in-in formalism is the natural framework. The section also introduces the role of gauge fixing, ghosts, trans-series structure, Euclidean continuation, and Lefschetz thimbles.

Subsection \ref{sec4.1} introduces the closed-time-path contour appropriate for computing expectation values in a time-dependent gravitational state. The forward and backward branches encode unitary evolution of the density matrix and allow one to compute real-time observables without imposing out-vacuum boundary conditions. This is the appropriate starting point for describing transient states.

In subsection \ref{sec4.2} we formulate expectation values of metric operators and other observables on the SK contour. The central object is not an in-out matrix element but an in-in expectation value evaluated with a density matrix or state prepared in the far past. This distinction is important because quasi--de Sitter profiles are understood as expectation values in a state, not as saddle points connecting fixed in- and out-vacua.

To quantify the SK contours we need to understand a more generic form of the total action.
In subsection \ref{sec4.3} we explain that the full gauge-fixed action should be understood as a trans-series including perturbative, nonperturbative, local, nonlocal, ghost, and gauge-fixing contributions. This trans-series organization is necessary because the backgrounds and states under study receive contributions from instantons, higher-order corrections, and nonperturbative sectors. The gauge-fixed action is therefore the appropriate object entering the exact path integral.

In subsection \ref{sec4.4} we address the crucial question as to why expectation values of Hermitian metric operators can be real in the SK formalism. We discuss how the forward and backward branches combine so that physical observables are computed from a normalized in-in expression rather than from a generally complex in-out amplitude. This provides the formal basis for interpreting $\langle g_{\mu\nu}\rangle_\sigma$ as a real semiclassical metric profile.

Subsection \ref{sec4.5} presents the Euclidean counterpart of the coherent-state construction. Euclidean methods are useful for defining vacuum wavefunctionals and for organizing saddle-point expansions, but they must be related carefully to Lorentzian time evolution. The subsection therefore serves as a bridge between canonical coherent-state intuition and Euclidean path-integral representations.

In subsection \ref{sec4.6} we give an alternative exact expression for the expectation value of the metric in the coherent-state framework. Our purpose here is to show that the same physical object can be represented in different but equivalent path-integral forms, provided the contours, insertions, and normalizations are treated correctly.

Subsection \ref{sec4.7} is more technical. 
We recast the path integral in terms of a Lefschetz-thimble decomposition. Our goal is to connect semiclassical saddles, complexified field configurations, and trans-series sectors. The thimble picture provides a geometric way of organizing different saddle contributions to coherent-state expectation values.

In subsection \ref{sec4.7.1} we explain how the Euclidean path integral may be decomposed into thimbles associated with complex saddles. Each thimble contributes a sector of the semiclassical expansion, and the total answer is obtained by summing over the relevant integration cycles with their intersection numbers.

Subsection \ref{sec4.7.2} then extends the thimble logic to the Lorentzian SK contour. We argue that the real-time path integral is more subtle because of oscillatory phases and doubled fields, but the same general idea applies: complex saddles and their associated thimbles organize the trans-series expansion of real-time expectation values.

\subsubsection*{1.1.5~~~ Summary of section \ref{sec5}}

Understanding the fate of the Minkowski minimum is important in the light of the trans-series form of the action itself. 
In section \ref{sec5} we study how the reference supersymmetric Minkowski minimum survives once the full trans-series structure of the theory is included. The main concern is whether nonperturbative effects, instanton sectors, and IR corrections shift or destabilize the chosen vacuum. In this section we argue that the UV Minkowski sectors can persist as reference branches even when semiclassical expectation values are shifted in the far IR.

In fact subsection \ref{sec4.7.3} decomposes the full action into perturbative and nonperturbative sectors. The goal is to make explicit that $S_{\rm tot}$ is not merely a finite local effective action but a trans-series object containing instanton contributions, nonlocal terms, and sector-dependent corrections. 
Since this structure is generically expected in any quantum gravity proposal, it is essential to describe both the reference vacuum and the GS-state deformation within it.

Subsection \ref{fitboushona} examines how localized vacua may mix through instanton effects. Rather than treating different minima as completely isolated, the trans-series perspective allows tunneling and mixing effects to be included systematically. The result is a richer picture in which semiclassical sectors remain meaningful but are embedded in a larger nonperturbative structure.

One of the concern is how solving the corrected equations of motion may shift the location of the chosen Minkowski minimum. In subsection \ref{sec5.3} we discuss this issue. The key point is that the reference minimum may receive controlled on-shell corrections without being eliminated as a valid semiclassical branch. This is important because the GS construction is built over such a corrected interacting background.

Continuing along the same logic
subsection \ref{sec5.4} argues that the UV Minkowski sectors can persist as meaningful reference sectors even after flowing to the far IR. Although the IR dynamics may generate time-dependent or quasi--de Sitter-like expectation values, these do not necessarily replace the underlying Minkowski reference branch. This helps justify the interpretation of quasi--de Sitter as a transient excited configuration rather than as a new vacuum.

\subsubsection*{1.1.6 ~~~ Summary of section \ref{sec6.0}}

In section \ref{sec6.0} we extend the SK and thimble constructions from coherent states to GS states. The GS state is represented through operator insertions on the contour, and the resulting expectation values are expressed using doubled fields and normalized path integrals. In this section we also compare in-in and in-out formulations and clarify when each representation is useful.

Comparing the operator structures underlying coherent and GS states is initiated in subsection \ref{sec6.1}. Coherent states are associated with the standard displacement algebra, while GS states involve insertions that act more generally on the interacting vacuum. In this subsection we prepare the ground for representing GS expectation values as path integrals with appropriate source-like but state-defining insertions.

Subsection \ref{sec6.2} then constructs the Lorentzian SK expression for GS-state expectation values. The forward and backward branches carry the appropriate GS insertions, and normalization by the corresponding vacuum-to-vacuum functional ensures that the result is an expectation value in a state rather than an unnormalized transition amplitude.

One of the offshoot of the analysis is the 
derivation of the first SK identity needed for the GS expectation value. This is done in subsection \ref{sec6.333}. It shows how the operator ordering and branch structure reproduce one of the two equations appearing in the GS representation. The purpose is to demonstrate that the GS insertions can be consistently distributed along the closed time contour.

Another offshoot is the derivation of the second SK identity required for the GS construction. This is performed in subsection \ref{6.444}. Together with the previous subsection, it completes the Lorentzian in-in representation and clarifies the role of forward and backward evolution in the presence of GS insertions.

Subsection \ref{sec6.3} then develops the Euclidean version of the GS-state path integral. The Euclidean representation is useful for analyzing convergence, vacuum preparation, and saddle structures, while remaining formally connected to the Lorentzian in-in expectation values. It also helps clarify how GS insertions differ from ordinary external sources.

The story is further elaborated in subsection \ref{biralador}. Here we show how the Euclidean GS expression may be reduced to a single-contour representation under appropriate assumptions. The proof is meant to demonstrate that the doubled or inserted structure of the GS expectation value can be reorganized into a compact form without changing the underlying physics.

We now get back to the similar technical side of our analysis as done for the coherent states. 
Subsection \ref{sec6.4} applies the Lefschetz-thimble perspective to the GS-state path integral. The GS insertion modifies the saddle structure and therefore the thimble decomposition. The resulting picture relates GS states to a trans-series sum over complexified saddle configurations.

Subsection \ref{sec6.4.1} then discusses the Euclidean thimble decomposition of the GS path integral. The Euclidean version provides the cleanest setting for identifying saddles and their associated descent cycles, thereby organizing the nonperturbative sectors contributing to GS expectation values.

In a similar vein, subsection \ref{sec6.4.2} extends the thimble analysis to the Lorentzian SK contour. The doubled real-time contour leads to a more complicated saddle structure, but it is the appropriate framework for transient time-dependent states. The result is a Lorentzian trans-series organization of GS expectation values.

One of the crucial question that arises in our framework is whether we can allow both in-in and in-out path integral analysis. This is addressed in subsection \ref{sec6.5}, where 
we compare the physical meaning of in-in and in-out quantities. The in-in formalism computes expectation values in a state and is therefore appropriate for time-dependent geometries, while the in-out formalism computes transition amplitudes between asymptotic states. The subsection clarifies when an in-out expression can be extracted from or related to an SK construction.

The story is continued in subsection \ref{dixiruporfatieachi} wherein 
we explain under which conditions the backward branch of the SK contour effectively decouples. Such decoupling can simplify the expression and help establish the reality of expectation values, but it is not automatic. It depends on the equality of sources, the placement of insertions, and the normalization of the state.

Furthermore subsection \ref{sec6.5.2} explains how an in-out-like quantity can be obtained from the SK formalism under special boundary and source choices. The main point is that the in-out object is not the fundamental observable for transient cosmology, but it can be useful as an auxiliary representation.

The story is solidified in subsections \ref{sec6.5.3} and \ref{sec6.5.4}. In subsection \ref{sec6.5.3} we revisit the single-contour in-out kernel and clarify its relation to the full SK contour. It helps identify precisely which pieces of the doubled contour are needed for expectation values and which pieces can be reorganized when one studies transition amplitudes.

In subsection \ref{sec6.5.4} we summarize the limitations of the in-out contour. The in-out formalism is useful for scattering amplitudes, vacuum persistence, and transition functions, but it is not the natural framework for real-time expectation values in transient cosmological states. For GS-induced quasi--de Sitter profiles, the in-in formalism is the correct approach.

\subsubsection*{1.1.7 ~~~ Summary of section \ref{sec7.0}}

Bootstrapping technique is a powerful way of finding solutions to complicated set of equations. In section \ref{sec7.0}
we develop the bootstrap equations obeyed by GS states. Starting from Schwinger--Dyson identities with GS insertions and external sources, we derive self-consistency conditions that determine whether a chosen GS profile is compatible with the full quantum equations of motion. In {that} section we {define a nonlinear map starting from these conditions and we} discuss nonuniqueness, fixed points, basins of attraction, and the interpretation of de Sitter-like configurations as bootstrap fixed points.

Subsection \ref{sec7.1} derives the Schwinger--Dyson identities in the presence of GS insertions. The GS deformation modifies the usual quantum equation of motion by producing additional terms from the variation of the insertion. These terms encode how the chosen state influences the expectation values of the fields.

The analysis continues in 
subsection \ref{sec7.2} where we define the generating functional appropriate to GS states on the SK contour. Our analysis includes both the usual doubled sources and the GS deformation. This functional is the central object from which correlation functions, expectation values, and bootstrap relations are derived.

Subsection \ref{sec7.3} introduces external SK sources $J^+_{\mu\nu}$ and $J^-_{\mu\nu}$ in addition to the GS deformation. Varying the generating functional yields Schwinger--Dyson equations on the two branches. These equations distinguish between genuine probing sources and the fixed GS data defining the state.

Our main computations appear from subsection \ref{sec7.4}.
This subsection formulates the bootstrap problem: the GS profile must be chosen so that its expectation values solve the quantum-corrected consistency equations. The bootstrap equation relates the variation of the full action to the variation of the GS insertion. A GS state is consistent only if the induced semiclassical configuration lies in the appropriate basin of solutions of this equation.

Sometimes naive computations can lead to wrong conclusions.
In subsection \ref{sec7.4.1} we warn against an incorrect interpretation in which the GS insertion is treated as a freely variable external source. Such an approach would incorrectly convert the state-defining data into a Legendre-transform source and would obscure the distinction between GS states and 1PI effective-action extrema. Moreover it would provide a wrong interpretation of the dark energy. 

Once we take the right direction, the analysis becomes straightforward. 
Subsection \ref{sec7.4.2} discusses the bootstrap equations for off-shell GS profiles. These configurations need not solve the classical equations pointwise, but they must satisfy the quantum Schwinger--Dyson consistency conditions appropriate to the GS-deformed state. This is the sector in which transient semiclassical profiles can be meaningfully discussed.

Similarly subsection \ref{sec7.4.3} specializes the bootstrap equations to configurations that are on-shell, or approximately on-shell, with respect to the quantum-corrected effective dynamics. In this sector the GS state reproduces semiclassical configurations that behave more like ordinary saddle-point geometries, while still being interpreted as excited states rather than vacua.

In subsection \ref{sec7.4.4} we discuss how one might solve the bootstrap equations in practice. The solution requires simultaneously choosing the GS data, the trans-series sector, and the semiclassical background so that the expectation values remain self-consistent. The emphasis is on consistency rather than on arbitrary engineering of a desired metric profile.

The issue of nonuniqueness of the bootstrap map, bootstrap fixed points and the different basins of attraction are now studied in subsection \ref{sec7.5}.
In fact, this subsection explains that the bootstrap map may have multiple fixed points and that the final semiclassical profile may depend on the basin of attraction associated with the initial GS data. This non-uniqueness is not necessarily a problem. Rather, it reflects the
existence of different semiclassical branches, transient configurations, or
basins of attraction within the same underlying quantum theory. Equivalently,
these branches may be interpreted as arising from the backreaction of different
sets of quantum modes on the reference Minkowski, or warped-Minkowski,
background.

Detailed confirmation of the aformentioned comment appears in subsections \ref{sec7.5.1}, \ref{sec7.5.2} and \ref{sec7.5.3}.
Subsection \ref{sec7.5.1} formulates the bootstrap procedure as a map on the space of possible GS profiles. Fixed points of this map correspond to self-consistent quantum states whose expectation values reproduce the assumed semiclassical configuration. Nonuniqueness means that different choices of GS data may flow to different fixed points.

Subsection \ref{sec7.5.2} discusses the basin structure of the bootstrap map. Nearby initial profiles may converge to the same semiclassical configuration, while profiles in different regions may converge to different fixed points. This provides a way to organize the space of possible GS-induced geometries.

In subsection \ref{sec7.5.3} applies the fixed-point picture to quasi--de Sitter configurations. The de Sitter-like state is interpreted as a local, possibly transient, bootstrap fixed point rather than as a global minimum of an effective potential. Stability is therefore understood in terms of the local behavior of the bootstrap map and the finite-time persistence of the state.

Finally in subsection \ref{sec7.5.44} we summarize how an unsplit Schwinger--Dyson equation governs the full space of admissible Glauber--Sudarshan states, whereas the split equations apply only to the special subset that admits a self-consistent semiclassical interpretation. The bootstrap procedure is therefore best understood as a way to isolate this distinguished semiclassical sector from the larger GS state space.

\subsubsection*{1.1.8~~~ Summary of section \ref{sec8.0}}

In section \ref{sec8.0} we connect the GS construction to the Wheeler--DeWitt equation and then to a structural analogy with world-sheet vertex operators. It explains how boundary time, WKB time, and the frozen WdW formalism coexist in different regimes. It also clarifies that the analogy between four-dimensional GS insertions and two-dimensional stringy vertex operators is structural, not a claim of dynamical equivalence.

Subsection \ref{sec8.1} is a detailed analysis of the  boundary time, WdW time, and their connection to the bulk Schr\"odinger evolution.
This subsection studies the relation between boundary Hamiltonian evolution and WdW evolution. Boundary time arises when an asymptotic or background structure fixes a preferred time, while WdW time emerges semiclassically through a WKB expansion of the wavefunctional. The subsection explains how both notions can be compatible when applied to the appropriate branches of the theory.

Subsection \ref{sec8.1.1} explains how the Schr\"odinger equation emerges from the Wheeler--DeWitt equation in a Born--Oppenheimer or WKB expansion. The WKB time $t_{\rm WKB}$ is not a fundamental external time but a parameter defined by the classical gravitational background. It governs the evolution of light degrees of freedom on a chosen semiclassical branch.

Subsection \ref{sec8.1.2} revisits the frozen formalism of the Wheeler--DeWitt equation that we initiated in section \ref{sec001}. The exact wavefunctional is annihilated by the Hamiltonian constraint, but this does not mean that all physical evolution disappears. Instead, time appears relationally, semiclassically, or through boundary structure, depending on the regime.

Subsection \ref{sec8.1.3} answers the question as to why the boundary and the bulk evolution match up.
In fact, this subsubsection explains why boundary Hamiltonian evolution and bulk constraint evolution can agree in the semiclassical regime. The boundary Hamiltonian generates physical time translations, while the bulk constraints enforce diffeomorphism invariance. Once a background branch is chosen, the two descriptions are compatible because the boundary time is tied to a particular slicing of the bulk solution.

Subsection \ref{sec8.1.4}  distinguishes on-shell and off-shell fluctuations around a reference Minkowski minimum. On-shell fluctuations correspond to physical excitations satisfying the relevant equations, while off-shell fluctuations appear inside the path integral and contribute to quantum expectation values. GS states provide a way to control or organize these off-shell configurations.

In subsection \ref{sec8.1.5} we explain that different supersymmetric Minkowski minima define different semiclassical branches. Each branch has its own background time, fluctuation spectrum, and effective Hamiltonian. GS states are constructed over such reference branches and can produce transient semiclassical profiles within them.

Subsection \ref{sec8.1.6} distinguishes the exact microscopic Wheeler--DeWitt equation from an emergent WdW-type equation acting on GS expectation-value variables. The exact wavefunctional depends on the full gauge-fixed configuration space, including ghosts and gauge-fixing data, while the emergent wavefunctional depends on reduced semiclassical variables such as $\langle \widehat\Xi\rangle_\sigma$.

In subsection \ref{hwlelo}, the GS states are naturally interpreted as physical superpositions within the constrained Hilbert space, provided that the displacement operator maps the chosen physical reference state back into that space. The bulk WdW and momentum constraints eliminate independent local bulk evolution, while the boundary Hamiltonian generates the genuine time dependence measured by the asymptotic clock. Although the GS state may formally be expanded in a complete basis of physical boundary-energy eigenstates, constructing that basis explicitly is generally intractable. This is precisely why the Schwinger--Keldysh in--in formalism is so useful: it allows physical expectation values and boundary-time evolution to be computed directly without first solving the full constrained spectral problem.

Finally in subsection \ref{sec8.1.7} we explain how the WdW equation inherits the trans-series structure of the underlying action. Since the full Hamiltonian constraint contains perturbative, nonperturbative, higher-derivative, and nonlocal contributions, the WdW equation should also be organized sector by sector. This gives a formal way of incorporating instanton and IR effects into the quantum constraint equation.

In subsection \ref{sec8.2} we change the gear a bit to study  the structural analogy between world-sheet vertex operators and bulk GS insertions. Both act as operator insertions that deform the path-integral weighting and select particular physical configurations. However, the comparison is explicitly structural and does not imply that the world-sheet and bulk theories have identical dynamics.

Subsection \ref{sec8.2.1} reviews the role of vertex operators in string theory. Physical vertex operators correspond to BRST-invariant world-sheet insertions satisfying conformal-dimension conditions. They create or absorb string states and encode spacetime fields through world-sheet operators.

However one shouldn't overstate the correspondence. 
In subsection \ref{sec8.2.2} we clarify the limits of the analogy between vertex operators and GS states. The analogy is meant to compare how operator insertions select states or backgrounds inside a path integral. It is not meant to identify the two theories dynamically, nor to claim that a four-dimensional GS insertion is literally a world-sheet vertex operator.

The issue of time that we discussed earlier, also appears on the worldsheet. Subsection \ref{sec8.2.3} extends the time discussion to the world-sheet theory. Since the world-sheet theory is constrained by diffeomorphism and Weyl invariance, there is no external target-space time built into the world-sheet path integral in the same way as in a fixed-background quantum mechanics. Time or propagation is recovered after choosing a target-space interpretation, a gauge such as light-cone gauge, or a semiclassical background in which one embedding coordinate plays the role of a clock.

In section \ref{sec9.0} we conclude by synthesizing the main message of the paper and by elucidating future directions. 
Throughout the draft, we use the mostly plus metric signature and work with natural units.

\subsection{Hamiltonian from the EH + GHY action with explicit boundary terms \label{sec1.1}}

\begin{figure}[h]
\centering
\begin{tabular}{c}
\includegraphics[width=5in]{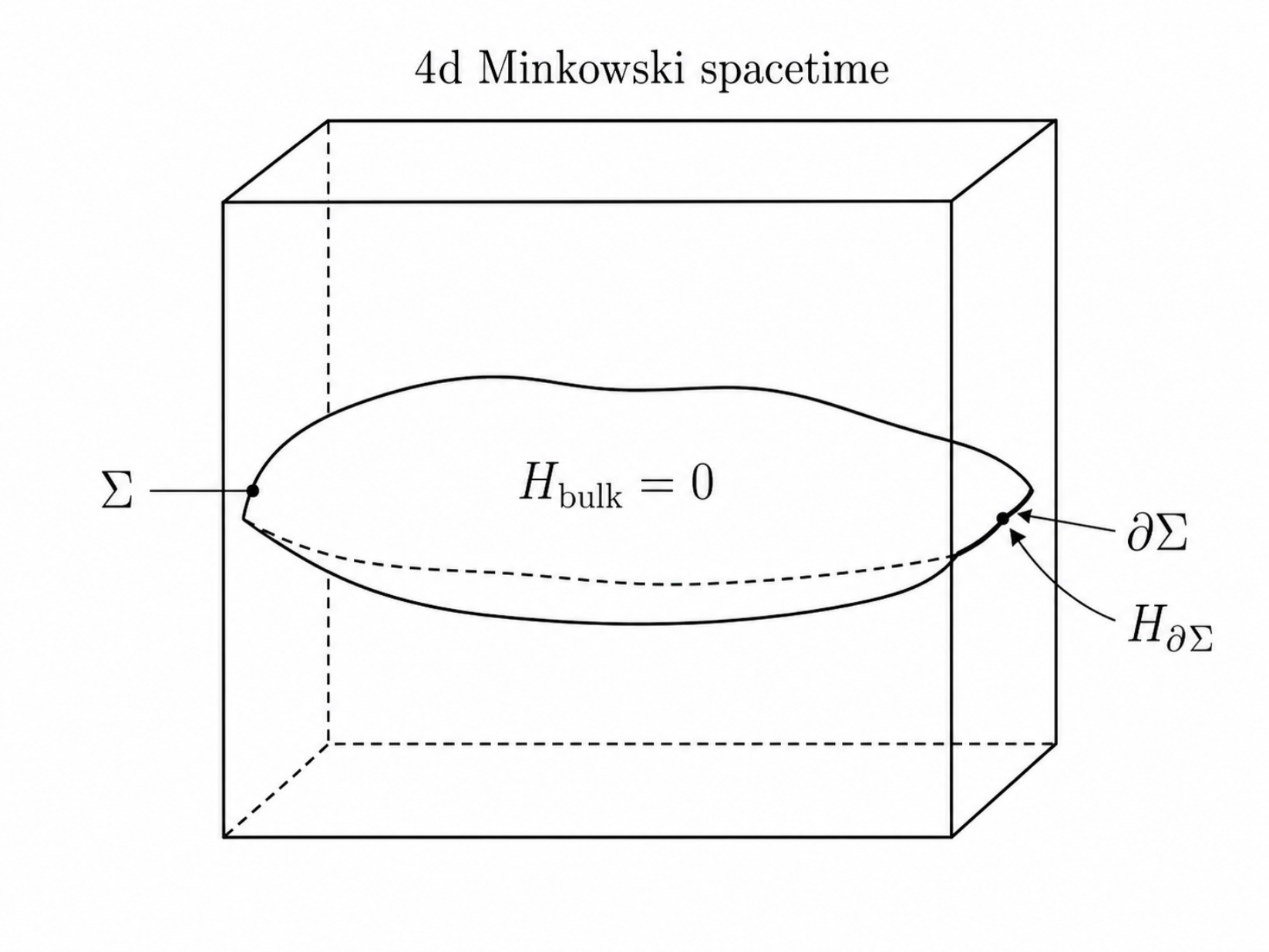}
\end{tabular}
\caption[]{A schematic representation of the the bulk and the boundary Hamiltonian. $\Sigma$ is the spatial Cauchy slice, while $\partial\Sigma$ is its asymptotic two-dimensional boundary.}
\label{bulkboundary}
\end{figure}

Our starting point is the vanishing Hamiltonian condition that is often stated as the reason for the {\it time problem} in 
QG \cite{WdWorig,hh,hhlater,FPI,page,witten}. This is a highly simplified version of the issue that hides numerous subtleties. The problem starts with the Gauss--Codazzi decomposition \cite{adm,teitelboim,ghy,WdWorig} of the Einstein--Hilbert Lagrangian density:
\bg\label{corichoshma}
\sqrt{-g}\,R
&= &
\sqrt{-g}\,g^{\mu\nu}\Big(\Gamma^\lambda_{\lambda\rho}\Gamma^\rho_{\mu\nu}
-\Gamma^\lambda_{\nu\rho}\Gamma^\rho_{\mu\lambda}\Big)
+\partial_\lambda\Big[\sqrt{-g}\,\big(g^{\mu\nu}\Gamma^\lambda_{\mu\nu}
-g^{\lambda\nu}\Gamma^\rho_{\rho\nu}\big)\Big]\nonumber\\
&=&
N\sqrt{h}\big({}^{(3)}R + K_{ij}K^{ij}-K^2\big)
+
2\,\partial_\mu\Big(\sqrt{-g}(n^\mu K-a^\mu)\Big),
\nd
where in the second line we used the ADM decomposition. The other parameters are: $n^\mu$ is the unit normal to the spatial slices, 
$a^\mu = n^\nu\nabla_\nu n^\mu$, 
$K_{ij}$ is the extrinsic curvature, 
and $\sqrt{-g}=N\sqrt{h}$. Note that the decomposition \eqref{corichoshma} is specifically for $3+1$ space-time dimensions, but could be generalized to any $d+1$ dimensions. For example if we have a $d+1$ dimensional spacetime with metric components $g_{00}, g_{0{\rm M}}$ and $g_{\rm MN}$, then the ADM decomposition \cite{adm} can be matched exactly with the way we define the vielbeins $e_0^a$ and $e_{\rm M}^a$, but the bulk and the boundary pieces would take the form given by the first line of \eqref{corichoshma}. What the Gibbons-Hawking-York (GHY) term \cite{ghy} does is to simply cancel the time--boundary part of the divergence. The action in $3+1$ dimensions then takes the following form \cite{ghy}:

{\footnotesize
\bg\label{corichosm2}
{\rm S}_{\rm EH} +{\rm S}_{\rm GHY}
=
\frac{1}{16\pi G}
\int dt\,d^3x
\Big[
N\sqrt{h}\big({}^{(3)}R + K_{ij}K^{ij}-K^2\big)
+
2\,\partial_i\Big(\sqrt{-g}(n^iK-a^i)\Big)
\Big],
\nd}
with the parameters as defined above (although we may not always display the ${\rm S}_{\rm GHY}$ piece explicitly). In fact this is all there is to it, at least if we are only considering the EH part of the total action. What happens for the most generic case will be discussed below. 
Using \eqref{corichosm2} we can define the canonical Lagrangian 
density as:
\bg\label{cori3}
L_a = L_{\rm bulk} + L_{\rm bdy}
=
\frac{1}{16\pi G}
\Big[
N\sqrt{h}\big({}^{(3)}R + K_{ij}K^{ij}-K^2\big)
+
2\,\partial_i\Big(\sqrt{-g}(n^iK-a^i)\Big)
\Big],
\nd
from where, using
$K_{ij}
=
\frac{1}{2N}\big(\dot h_{ij}-D_iN_j-D_jN_i\big)$, the momentum conjugate to $h_{ij}$ is
$\pi^{ij}
=
\frac{\partial L_a}{\partial \dot h_{ij}}
=
\frac{\sqrt{h}}{16\pi G}\big(K^{ij}-Kh^{ij}\big)$, such that 
$\pi = h_{ij}\pi^{ij}$. This leads to the following Legendre transformation, or generically the Hamiltonian density:
\bg\label{corimey}
\pi^{ij}\dot h_{ij} - L_a
=
\Big(2N\pi^{ij}K_{ij}-L_{\rm bulk}\Big)
+
2\pi^{ij}D_iN_j
-
L_{\rm bdy},
\nd
with $L_{\rm bulk}$ and $L_{\rm bdy}$ has the nice split given in \eqref{cori3}. Notice that the above way of expressing the Legendre transformation, and in particular the Hamiltonian density, is crucial because it lacks the $\pi^{00}$ and the $\pi^{0i}$ pieces. This is simply because after removing the temporal boundary piece using the GHY term, the remaining bulk and the boundary pieces do not have terms proportional to $\dot{g}_{00}$ or $\dot{g}_{0i}$. One may then wonder if such dependence could appear once we include the higher order corrections which are necessary from string and M-theories. The answer, as we will also briefly discuss later, is surprising: although the higher derivative terms allow the appearance of $\dot{g}_{00}$ and $\dot{g}_{0i}$, the corresponding canonical momenta are {\it constrained to vanish} due to the underlying diffeomorphism invariance! 
The lapse part from \eqref{corimey} becomes:
\bg\label{corimey3}
2N\pi^{ij}K_{ij}-L_{\rm bulk}
=
N\left[
\frac{16\pi G}{\sqrt{h}}
\Big(\pi^{ij}\pi_{ij}-\frac12\pi^2\Big)
-
\frac{\sqrt{h}}{16\pi G}\,{}^{(3)}R
\right]
\equiv N\,\mathcal H,
\nd
which basically yields one part of the bulk Hamiltonian density $\mathcal H$. 
For the remaining two pieces in \eqref{corimey}, we can first 
integrate the shift term by parts to get:
\begin{equation}
\int_\Sigma d^3x\,2\pi^{ij}D_iN_j
=
\int_\Sigma d^3x\,N^j\mathcal H_j
+
\oint_{\partial\Sigma} d^2x\,(2N_j\pi^{ij}s_i),
\end{equation}
where
$\mathcal H_j = -2D_i\pi^i{}_j$. Notice the appearance of the first boundary contribution to the Hamiltonian density. The second boundary contribution comes from the explicit divergence term $L_{\rm bdy}$ whose form can be easily read from \eqref{corimey}. This yields:
\begin{equation}
-\int_\Sigma d^3x\,L_{\rm bdy}
=
-\frac{1}{8\pi G}
\oint_{\partial\Sigma} d^2x\,
\sqrt{-g}(n^iK-a^i)s_i.
\end{equation}
This is all there is to it: for an action with higher derivative interactions (both metric and other DOFs), every piece of the action splits into a bulk and/or a boundary piece. The bulk pieces all contribute to either $\mathcal H$ or $\mathcal H_i$, whereas the boundary pieces all accumulate together. Together they produce the following four-dimensional Hamiltonian: 
\bg\label{homilbeta}
H
=
\int_\Sigma d^3x\,(N\mathcal H + N^i\mathcal H_i)
+
H_{\partial\Sigma},
\nd
where $H_{\partial \Sigma}$ is the generalized boundary term coming from accumulating all the boundary contributions from the usual $\sqrt{-g} R$ and the higher order interactions if any. For the simple case with pure Einstein gravity in four-dimensions, the surface term takes the form\footnote{\textcolor{blue}{Throughout this paper, we denote the boundary Hamiltonian by either
$H_{\partial\Sigma}$ or $H_{\partial}$, and the corresponding quantum
operator by either $\widehat{H}_{\partial\Sigma}$ or
$\widehat{H}_{\partial}$. These notations will be used interchangeably.
The additional subscript $\Sigma$ merely indicates the Cauchy
hypersurface whose boundary is under consideration.}}:
\bg\label{katkozz}
H_{\partial\Sigma}
 & = &
\oint_{\partial\Sigma} d^2x
\left[
2\,N_j\,\pi^{ij}s_i
-
\frac{1}{8\pi G}\sqrt{-g}(n^iK-a^i)s_i
\right] \nonumber\\
&  = &  \oint_{\partial\Sigma} d^2x
\left[
2\,N_j\,\pi^{ij}s_i
+
\frac{1}{8\pi G}\sqrt{h}\,(N^iK + D^iN)\,s_i
\right], \nd
where in the second line we have used the standard ADM decomposition:
$\sqrt{-g}=N\sqrt{h}$, $n^i=-N^i/N$, and $Na^i=D^iN$. On shell, the bulk constraints vanish and the Hamiltonian reduces entirely to this boundary term, which for asymptotically flat boundary conditions reproduces the ADM energy and momentum. See also {\bf figure \ref{bulkboundary}}.

Question is whether there is an alternative way to justify the vanishing of $\mathcal H$ and $\mathcal H_i$ on-shell. To see this, 
let us first define $n^\mu$ to be the unit normal to the hypersurfaces $\Sigma_t$, as
$n_\mu = (-N,0,0,0)$ such that
$n^\mu = \Big(\frac{1}{N},-\frac{N^i}{N}\Big)$
and $h^\mu{}_i$ to be the spatial projector. With these variables, one can easily show that the Hamiltonian and momentum constraints are given geometrically by the Einstein tensor $G_{\mu\nu}$ in the following way:
\bg\label{billicox2}
&& {\cal H}_i
=
-\frac{2\sqrt{h}}{16\pi G}\,G_{\mu\nu}n^\mu h^\nu{}_i = -{\sqrt{h}\over 8\pi G N}\Big(G_{0i}-N^jG_{ji}\Big)\nonumber\\
&& {\cal H}
=
-\frac{2\sqrt{h}}{16\pi G}\,G_{\mu\nu}n^\mu n^\nu = 
-{\sqrt{h}\over 8\pi G N^2} \Big(G_{00}-2N^iG_{0i}+N^iN^jG_{ij}\Big), 
\nd
and since on-shell $G_{00} = G_{0i} = G_{ij} = 0$ in the absence of any external energy momentum tensor, they clear suggest that $\mathcal H = \mathcal H_i = 0$. Thus the so-called {\it vanishing Hamiltonian} is just that: due to the on-shell constraints, the bulk contribution to the Hamiltonian vanishes. The only surviving piece is the {\it boundary} contribution \eqref{katkozz}\footnote{In this sense the Hamiltonian in gravity is {\it holographic}. This continues to be the case even in the presence of extra matter or interactions as we shall discuss briefly soon.}. A slightly weaker constraint, coming from the form of the total Hamiltonian \eqref{homilbeta} is:
\bg\label{homileta2}
G_{\mu\nu}n^\mu n^\nu = 0,
\qquad
G_{\mu\nu}n^\mu h^\nu{}_i = 0,
\nd
which may be argued from the fact that $\mathcal H$ and $\mathcal H_i$ attach themselves to Lagrange multiplies $N$ and $N^i$ respectively. Thus, as long as the boundary piece \eqref{katkozz} is non-vanishing, the system has a well-defined Hamiltonian which may be used for dynamical evolution of either classical or quantum states. 

\subsection{Total Hamiltonian when all corrections are packaged into $S_{\rm matter}$ \label{sec1.2}}

The vanishing of the bulk part of the Hamiltonian due to the on-shell EOMs now raises the following question: what happens when we have non-trivial interactions? More generically, what happens if we allow higher curvature as well as interactions from all other non-gravitational degrees of freedom?
Consider an effective action of the form
\begin{equation}
S
=
\frac{1}{16\pi G}\int d^4x\,\sqrt{-g}\,R
+
S_{\rm matter}[g,\Psi],
\end{equation}
where $\Psi$ denotes all additional degrees of freedom (scalars, $p$-forms, fermions, branes, \emph{etc.}) and where
$S_{\rm matter}$ also contains higher-derivative corrections (including higher curvature terms, curvature--flux couplings, and so on).
We can perform a $3+1$ split with ADM variables $(N,N^i,h_{ij})$. The full set of canonical variables consists of
$(h_{ij},\pi^{ij})$ {for the metric}, and 
$(\Psi^A,\Pi_A)$ {for all matter fields},
where $\Pi_A$ are the canonical momenta defined by the matter Lagrangian density.
Independently of the detailed form of $S_{\rm matter}$, diffeomorphism invariance implies that the \emph{Dirac total Hamiltonian} has the universal form:
\begin{equation}\label{hotat1}
H_{\rm tot}
=
\int_{\Sigma} d^3x
\Big(
N\,{\cal H}_{\rm tot}
+
N^i\,{\cal H}^{\rm tot}_i
\Big)
+
H_{\partial\Sigma}
+
H_{\rm prim},
\end{equation}
where $H_{\rm prim}$ enforces primary constraints associated with nondynamical variables (including the lapse and shift):
\begin{equation}
H_{\rm prim}
=
\int_{\Sigma} d^3x\,
\lambda_\mu\,\pi^{0\mu}
+
\sum_{\alpha}\int_{\Sigma} d^3x\,
u^\alpha\,\phi_\alpha .
\end{equation}
Here $\pi^{0\mu}\approx 0$ are the gravitational primary constraints and $\phi_\alpha\approx 0$ denotes any additional primary constraints from gauge symmetries of the matter sector (for example Gauss-law constraints for gauge fields, or additional constraints induced by higher-derivative terms). The total constraints split additively into gravitational and matter pieces:
\begin{equation}
{\cal H}_{\rm tot}
=
{\cal H}_{\rm grav}
+
{\cal H}_{\rm eff},
\qquad
{\cal H}^{\rm tot}_i
=
{\cal H}^{\rm grav}_i
+
{\cal H}^{\rm eff}_i.
\end{equation}
The Einstein--Hilbert contribution gives the standard ADM constraints
with ${\cal H}_{\rm grav}
=
\frac{16\pi G}{\sqrt{h}}
\Big(\pi^{ij}\pi_{ij}-\frac12\pi^2\Big)
-
\frac{\sqrt{h}}{16\pi G}\,{}^{(3)}R$ and 
${\cal H}^{\rm grav}_i
=
-2D_j\pi_i{}^{j}$ identified to \eqref{billicox2}. For the remaining, 
all corrections and extra fields contribute through effective constraint densities:
\begin{equation}
{\cal H}_{\rm eff} = {\cal H}_{\rm eff}[h_{ij},\pi^{ij};\Psi,\Pi],
\qquad
{\cal H}^{\rm eff}_i = {\cal H}^{\rm eff}_i[h_{ij},\pi^{ij};\Psi,\Pi],
\end{equation}
whose explicit forms depend on the detailed terms in $S_{\rm matter}$.
Geometrically, they can be expressed in terms of the effective stress tensor $T^{\rm eff}_{\mu\nu}
\equiv
-\frac{2}{\sqrt{-g}}\frac{\delta S_{\rm matter}}{\delta g^{\mu\nu}}$,
as ${\cal H}_{\rm eff} = \sqrt{h}\,\rho_{\rm eff}$
${\cal H}^{\rm eff}_i = \sqrt{h}\,j^{\rm eff}_i$,
with $(\rho_{\rm eff}, j_i^{\rm eff}) = (T^{\rm eff}_{\mu\nu}n^\mu n^\nu, -T^{\rm eff}_{\mu\nu}n^\mu h^\nu{}_i)$.
Equivalently, the total constraints are the normal and mixed projections of the full field equations:
\begin{equation}
G_{\mu\nu}
=
8\pi G\,T^{\rm eff}_{\mu\nu},
\end{equation}
namely $G_{\mu\nu}n^\mu n^\nu = 8\pi G\,\rho_{\rm eff}$ and 
$G_{\mu\nu}n^\mu h^\nu{}_i = 8\pi G\,j^{\rm eff}_i$.
The boundary term $H_{\partial\Sigma}$ is determined by the choice of boundary conditions (asymptotically flat, AdS, finite box, etc.) and generates physical time evolution when a notion of time is fixed at the boundary. The bulk Hamiltonian remains a linear combination of constraints:
\begin{equation}\label{hotat2}
H_{\rm bulk}
=
\int_{\Sigma} d^3x
\Big(
N\,{\cal H}_{\rm tot}
+
N^i\,{\cal H}^{\rm tot}_i
\Big).
\end{equation}
Therefore, on the full constraint surface,
${\cal H}_{\rm tot} \approx 0$ and 
${\cal H}^{\rm tot}_i \approx 0$
implying that
$H_{\rm bulk}\approx 0$
and $H_{\rm tot}\approx H_{\partial\Sigma} + H_{\rm prim}$.

If $S_{\rm matter}$ contains higher time derivatives (as in generic higher-curvature terms), one introduces additional canonical variables via an Ostrogradsky-type procedure (or uses an equivalent first-order formulation). This typically enlarges the phase space and can introduce extra primary/secondary constraints. Nevertheless, diffeomorphism invariance still implies the universal structure:
\begin{equation}\label{hotat3}
H_{\rm tot}
=
\int_{\Sigma} d^3x
\Big(
N\,{\cal H}_{\rm tot}
+
N^i\,{\cal H}^{\rm tot}_i
\Big)
+\text{(constraints)}+H_{\partial\Sigma},
\end{equation}
with lapse and shift acting as Lagrange multipliers enforcing the generalized Hamiltonian and momentum constraints\footnote{There is an interesting {\it non-gravitational} toy model with only a scalar field $\phi$ that can be used to explain the Hamiltonian constraint that we saw above in a {\it gravitational} theory.   
Consider a free scalar field in Minkowski spacetime with an action
$S[\phi]
=
-\frac12 \int d^4X \, \eta^{\mu\nu} \, \partial_\mu \phi(X) \, \partial_\nu \phi(X)$.
Introduce arbitrary coordinates $\sigma^a=(\tau,\sigma^i)$ and promote the spacetime
coordinates to embedding variables,
$X^\mu = X^\mu(\tau,\sigma),~
\phi(\tau,\sigma) \equiv \phi(X(\tau,\sigma))$.
The induced metric on the worldvolume is
$g_{ab}
=
\eta_{\mu\nu}\,\partial_a X^\mu\,\partial_b X^\nu$
with 
$g \equiv \det g_{ab}$.
The action becomes manifestly reparametrization invariant,
$S[X,\phi]
=
-\frac12 \int d^4\sigma \, \sqrt{-g} \, g^{ab} \, \partial_a \phi \, \partial_b \phi$.
Define the tangent vectors and unit normal,
$e_i^{\ \mu} \equiv \partial_i X^\mu,
n_\mu e_i^{\ \mu} = 0,
n_\mu n^\mu = -1$.
The induced spatial metric and volume element are
$h_{ij} = \eta_{\mu\nu} e_i^{\ \mu} e_j^{\ \nu},
\sqrt{-g} = N \sqrt{h}$.
The velocity of the embedding decomposes in lapse and shift as
$\dot X^\mu
=
N n^\mu + N^i e_i^{\ \mu}$.
The Lagrangian density becomes
\begin{equation}
\mathcal L_\phi
=
\frac12 N \sqrt{h}
\left[
\frac{1}{N^2}(\dot\phi - N^i \partial_i \phi)^2
-
h^{ij} \partial_i \phi \partial_j \phi
\right],\nonumber
\end{equation}
which is expressed in terms of ADM variables.
The momentum conjugate to $\phi$ is
$\pi
\equiv
\frac{\partial \mathcal L_\phi}{\partial \dot\phi}
=
\sqrt{h}\,\frac{1}{N}(\dot\phi - N^i \partial_i \phi)$
Solving for the velocity, we get
$\dot\phi = \frac{N}{\sqrt{h}} \pi + N^i \partial_i \phi$.
The scalar Hamiltonian density takes the ADM form
$\mathcal H_\phi
= \pi \dot\phi - \mathcal L_\phi 
= N \, \mathcal H_\perp^{(\phi)} + N^i \, \mathcal H_i^{(\phi)}$,
where:
\begin{equation}
\mathcal H_\perp^{(\phi)}
=
\frac{1}{2\sqrt{h}} \pi^2
+
\frac{\sqrt{h}}{2} h^{ij} \partial_i \phi \partial_j \phi,~~~
\mathcal H_i^{(\phi)}
=
\pi \, \partial_i \phi,\nonumber
\end{equation}
which looks surprisingly similar to the gravitational result from \eqref{billicox2}. We can also define the momenta conjugate to the embedding itself. This takes the form:
\begin{equation}
P_\mu(\sigma)
=
\frac{\partial \mathcal L_\phi}{\partial \dot X^\mu} 
=
-\sqrt{h}
\left(
\mathcal H_\perp^{(\phi)} n_\mu
+
\mathcal H_i^{(\phi)} e^i_{\ \mu}
\right), \nonumber
\end{equation}
where we used
$\frac{\partial N}{\partial \dot X^\mu} = - n_\mu$ and 
$\frac{\partial N^i}{\partial \dot X^\mu} = e^i_{\ \mu}$.
Projecting tangentially gives the momentum (diffeomorphism) constraint
$\mathcal H_i
\equiv
P_\mu \partial_i X^\mu + \pi \partial_i \phi
\approx 0$, whereas projecting along the normal gives the Hamiltonian constraint
$\mathcal H_\perp
\equiv
P_\mu n^\mu + \mathcal H_\perp^{(\phi)}
\approx 0$.
The full Legendre transform now reads as
$H_{\rm can}
=
\int d^3\sigma
\left(
\pi \dot\phi + P_\mu \dot X^\mu
\right)
-
L$. Using $\dot X^\mu = N n^\mu + N^i e_i^{\ \mu}$ and the definitions above, all bulk terms cancel, {\it i.e.}
$H_{\rm can}^{\rm bulk} = 0$. Following Dirac’s prescription, the total Hamiltonian is
$H
=
\int d^3\sigma
\left(
N \mathcal H_\perp
+
N^i \mathcal H_i
\right)
+ H_{\rm boundary}$
which is the same as the Hamiltonian that one would get from the original action using Noether's procedure.}.

\subsection{Canonical picture for $S_{\rm tot}=S_{\rm EH}+S_{\rm int}$: a corrected 6--step summary \label{sec1.3}}

Let us now summarize the general picture in terms of 6-steps. 
Consider a generally covariant theory in $D=d+1$ dimensions with fields
$\Phi \equiv \{g_{MN},\Psi\}$ with
$M,N=0,1,\dots,d$,
and action:
\begin{equation}
S_{\rm tot}[\Phi]
=
S_{\rm EH}[g]
+
S_{\rm int}[g,\Psi],
\qquad
S_{\rm EH}[g]=\frac{1}{16\pi G}\int_{\cal M} d^{D}x\,\sqrt{-g}\,R .
\end{equation}
The interaction sector $S_{\rm int}$ may contain higher--curvature terms and couplings to other degrees of freedom (fluxes, scalars, branes, Chern--Simons terms, etc.). We choose a foliation ${\cal M}\simeq \mathbb R\times \Sigma$ with ADM variables
\begin{equation}
ds^2=-N^2dt^2+h_{ij}(dx^i+N^i dt)(dx^j+N^j dt),
\qquad i,j=1,\dots,d,
\end{equation}
and $n^M$ the unit normal to $\Sigma_t$. The question is what is the generic procedure to extract the Hamiltonian of the system when all higher order corresctions are inserted in.

\subsubsection*{Step 1: Isolate the variational boundary term and choose $S_{\rm bnd}$}

For any local Lagrangian $D$--form $L(\Phi)$, the first variation has the universal structure
\begin{equation}\label{indigpaap1}
\delta S_{\rm tot}
=
\int_{\cal M} d^{D}x\,\sqrt{-g}\,
\Big(
E^{MN}\,\delta g_{MN}
+
E_A\,\delta\Psi^A
\Big)
+
\int_{\partial{\cal M}} \Theta(\Phi,\delta\Phi),
\end{equation}
where $E^{MN}=0$ and $E_A=0$ are the bulk Euler--Lagrange equations and $\Theta$ is the symplectic potential (a boundary term in the variation). In particular,
\begin{equation}
E^{MN}
\equiv
\frac{1}{\sqrt{-g}}\frac{\delta S_{\rm tot}}{\delta g_{MN}}
=
\frac{1}{16\pi G}G^{MN}
+
E^{MN}_{\rm int},
\end{equation}
and $E^{MN}_{\rm int}$ can, in general, be expressible as $-\frac12 T^{MN}$ even when $S_{\rm int}$ depends on curvature.
A well--posed variational principle requires adding a boundary action $S_{\rm bnd}$ such that
\begin{equation}
\delta(S_{\rm tot}+S_{\rm bnd})
=
\int_{\cal M} d^{D}x\,\sqrt{-g}\,
\Big(
E^{MN}\,\delta g_{MN}
+
E_A\,\delta\Psi^A
\Big)
+
\int_{\partial{\cal M}} \Big(\Theta+\delta B\Big),
\end{equation}
where $\Theta$ is defined in \eqref{indigpaap1}, 
and the chosen boundary conditions make the boundary term vanish:
\begin{equation}\label{indigpaap2}
\Big(\Theta+\delta B\Big)\Big|_{\rm b.c.}=0,
\qquad
S_{\rm bnd}\equiv \int_{\partial{\cal M}} B.
\end{equation}
For Einstein--Hilbert, $S_{\rm bnd}$ contains the usual GHY term on the relevant boundaries. For higher--curvature and curvature--matter couplings, $B$ is a generalized boundary functional (not necessarily a simple bulk total derivative).

\subsubsection*{Step 2: Handle time boundaries vs spatial boundaries}

The boundary contribution that appeared in \eqref{indigpaap2} is a bit more subtle. To see this,
decompose the boundary into initial/final time slices and spatial boundary:
\begin{equation}
\partial{\cal M} = \Sigma_{t_f}\cup \Sigma_{t_i}\cup (\mathbb R\times \partial\Sigma),
\end{equation}
where $\Sigma_t$ is the usual Cauchy slice at $t$.
One chooses boundary conditions on $\Sigma_{t_i}$ and $\Sigma_{t_f}$ (equivalently: specifies initial/final states, or fixes induced data) and includes the corresponding time--boundary terms in $S_{\rm bnd}$ so that the time--boundary variation cancels.

After this choice, the remaining boundary contributions relevant for the Hamiltonian are the spatial surface terms on $\partial\Sigma$. Schematically:
\begin{equation}
S_{\rm bnd}
=
S_{\rm time\;bnd}
+
\int dt \oint_{\partial\Sigma} d^{d-1}y\;{\cal B},
\end{equation}
where ${\cal B}$ depends on the theory and on the boundary data (and in AdS typically includes counterterms). In fact it is precisely one of the boundary part of the action that eliminates the troublesome $\partial_0(..)^0$ terms from the bulk.

\subsubsection*{Step 3: Choose correct canonical variables and fix the status of $\pi_N,\pi_i$}

The canonical variables on the Cauchy slice $\Sigma_t$ are
$(h_{ij},\pi^{ij}), 
(\Psi^A,\Pi_A)$ and 
$(N, N^i)$.
In pure EH (and many first--order formulations) the Lagrangian contains no $\dot N$ or $\dot N^i$, giving primary constraints:
\begin{equation}
\pi_N \equiv \frac{\partial L}{\partial \dot N} \approx 0,
\qquad
\pi_i \equiv \frac{\partial L}{\partial \dot N^i} \approx 0.
\end{equation}
However with generic higher--derivative $S_{\rm int}$, an ADM rewriting may introduce apparent $\dot N,\dot N^i$. The correct, general procedure then is one of the following two choices.
(1) Either rewrite the theory in an equivalent first--order/auxiliary--field form in which $\dot N,\dot N^i$ do not appear, so that $\pi_N\approx 0$, $\pi_i\approx 0$ remain primary constraints;
or (2) keep the higher--derivative form, enlarge phase space via Ostrogradsky/Dirac variables, and obtain additional primary constraints that eliminate any would--be propagating lapse/shift momenta as nonphysical. More importantly, in both cases, spacetime diffeomorphism invariance guarantees that lapse and shift do not become physical propagating degrees of freedom.

\subsubsection*{Step 4: Choose canonical bulk Lagrangian and perform the Legendre transform}

The next step is crucial and lies at the heart of the construction.
After adding $S_{\rm bnd}$ and choosing boundary conditions, one obtains a canonical Lagrangian of the form:
\begin{equation}
L_{\rm can}
=
\int_{\Sigma} d^{d}x\;
\Big(
\pi^{ij}\dot h_{ij}
+
\Pi_A \dot\Psi^A
-
N\,{\cal H}_{\rm tot}
-
N^i\,{\cal H}^{\rm tot}_i
\Big)
-
\int_{\Sigma} d^{d}x\;
\sum_\alpha u^\alpha \phi_\alpha
-
H_{\partial\Sigma},
\end{equation}
where $\phi_\alpha\approx 0$ denote any additional primary/secondary constraints from gauge symmetries (Gauss laws), auxiliary fields, Chern--Simons sectors, or higher--derivative degeneracies, and $H_{\partial\Sigma}$ is the spatial boundary Hamiltonian. Equivalently, the total Hamiltonian is obtained by the Legendre transform:
\bg\label{daitaylor}
H_{\rm tot}
& = &
\int_{\Sigma} d^{d}x\;
\Big(
\pi^{ij}\dot h_{ij}
+
\Pi_A \dot\Psi^A
\Big)
-
L_{\rm can} \nonumber\\
&= &
\int_{\Sigma} d^{d}x\;
\Big(
N\,{\cal H}_{\rm tot}
+
N^i\,{\cal H}^{\rm tot}_i
+
\sum_\alpha u^\alpha \phi_\alpha
\Big)
+
H_{\partial\Sigma}, \nd
together with the primary constraints enforcing the nondynamical multipliers. Notice the universal behavior of \eqref{daitaylor}: no matter what kind of action we have, the Hamiltonian always takes the form \eqref{daitaylor}. 

\subsubsection*{Step 5: Identify ${\cal H}_{\rm tot}$ and ${\cal H}^{\rm tot}_i$ as projections of the full bulk equations}

The {\it miracle} that happens at this point can be elucidated as follows.
Define the full bulk Euler-Lagrange tensor for the metric (including all interaction terms) as:
\begin{equation}
E_{MN}
\equiv
\frac{1}{\sqrt{-g}}\frac{\delta S_{\rm tot}}{\delta g^{MN}},
\end{equation}
where $S_{\rm tot}$ is the total action that includes all corrections.
Then the Hamiltonian and momentum constraints are the normal and mixed projections of $E_{MN}$:
\begin{equation}\label{taylorchase}
{\cal H}_{\rm tot}
\propto
\sqrt{h}\;E_{MN}\,n^M n^N,
\qquad
{\cal H}^{\rm tot}_i
\propto
\sqrt{h}\;E_{MN}\,n^M h^N{}_i,
\end{equation}
where $h^N{}_i$ is the spatial projector (equivalently $e^N{}_i$ a spatial frame on $\Sigma_t$). In the special case where $S_{\rm int}$ depends on the metric only algebraically (no curvature dependence),
\begin{equation}
E_{MN}
=
\frac{1}{16\pi G}G_{MN}
-\frac12 T^{({\rm matter})}_{MN},
\end{equation}
and the constraints become proportional to projections of $G_{MN}-8\pi G T^{({\rm matter})}_{MN}$. For genuine higher--curvature couplings, $E_{MN}$ is the correct replacement for $G_{MN}-8\pi G T^{({\rm matter})}_{MN}$. 

\subsubsection*{Step 6: On-shell vanishing of the bulk Hamiltonian and the generalized boundary Hamiltonian}

The {\it miracle} alluded above is the fact that we can always express $\mathcal H$ and $\mathcal H_i$ in terms of $E_{MN}$ as in \eqref{taylorchase}. 
The full equations of motion are
$E_{MN}=0,~
E_A=0,
\phi_\alpha \approx 0$. Therefore, on the full constraint surface, we expect
${\cal H}_{\rm tot}\approx 0$ and 
${\cal H}^{\rm tot}_i\approx 0$; 
and hence the bulk Hamiltonian vanishes on shell \cite{adm,teitelboim,ghy,WdWorig,hh,hhlater,FPI}:
\begin{equation}
H_{\rm bulk}
=
\int_{\Sigma} d^{d}x\,
\Big(
N\,{\cal H}_{\rm tot}
+
N^i\,{\cal H}^{\rm tot}_i
\Big)
\approx 0,
\end{equation}
which continues to be the case for {\it any} theory coupled to gravity. Thus all non-trivial {\it bulk interactions} are soaked in by gravity and nullified on-shell. Of-shell however nothing vanishes, and the theory continues to have a non-trivial bulk term. 
The physical Hamiltonian therefore is the spatial boundary term:
\begin{equation}
H_{\rm phys}=H_{\partial\Sigma},
\end{equation}
implying that any theory coupled to gravity, the Hamiltonian is always {\it holographic}! A fully general expression for $H_{\partial\Sigma}$ is provided by the covariant phase space (Iyer--Wald) formula. For a diffeomorphism generated by $\xi^M$ (time translation at infinity or at the boundary),
\begin{equation}
H_\xi
=
\oint_{\partial\Sigma}\Big(Q_\xi-\xi\cdot B\Big),
\end{equation}
where $Q_\xi$ is the Noether charge $(D-2)$--form computed from the full Lagrangian (including higher curvature and couplings to $\Psi$) and $B$ is the boundary $(D-1)$--form used in $S_{\rm bnd}=\int_{\partial{\cal M}}B$. For Lagrangians depending on the Riemann tensor, it is convenient to define a quantity
$P^{MNRS}\equiv \frac{\partial {\cal L}}{\partial R_{MNRS}}$,
so that the gravitational part of $Q_\xi$ contains the schematic structure:
\begin{equation}
Q_\xi
\sim
-\epsilon_{MN}\,P^{MNRS}\nabla_R\xi_S+\cdots,
\end{equation}
where the dots include contributions involving $\nabla P$ when ${\cal L}$ depends on $\nabla R$, as well as matter/gauge surface charges (electric/magnetic/Page/brane charges) when $\Psi$ includes gauge potentials and fluxes.


\subsubsection*{Summary: Equations of motion versus Hamiltonian constraints}

The six steps given above may be applied to any theory coupled to gravity. In particular, they explain why in string or M-theories the total Hamiltonian could become simple despite allowing a complicated action. The covariant field equations of the theory follow from the variational principle:
\begin{equation}
\frac{\delta S_{\rm tot}}{\delta g^{MN}} = 0,
\qquad
\frac{\delta S_{\rm tot}}{\delta \Phi} = 0,
\qquad
S_{\rm tot} = S_{\rm bulk} + S_{\rm bnd},
\end{equation}
where $\Phi$ collectively denotes all matter fields, fluxes and fermions.  Since
the boundary action $S_{\rm bnd}$ is supported only on $\partial M$, it does not
contribute to the bulk Euler--Lagrange equations, so that the full non--linear
dynamics arises entirely from
\begin{equation}
\frac{\delta S_{\rm bulk}}{\delta g^{MN}} = 0,
\qquad
\frac{\delta S_{\rm bulk}}{\delta \Phi} = 0.
\end{equation}
It is interesting to now note that the EOM is controlled by the bulk terms whereas the energy of the system, or any state, is controlled by the boundary term. Thus the pieces that contribute to the EOMs {\it do not} contribute to the energy. 
In the Hamiltonian formulation the total Hamiltonian takes the universal ADM form
\begin{equation}\label{chukka5}
H
=
\int d^3x
\left(
N \, \mathcal H_\perp
+
N^i \, \mathcal H_i
\right)
+ H_{\rm boundary},
\end{equation}
where $\mathcal H_\perp$ and $\mathcal H_i$ are the Hamiltonian and momentum
constraint densities.  As we saw earlier, physical configurations satisfy the constraints
$\mathcal H_\perp \approx 0,
\mathcal H_i \approx 0$,
so that on the constraint surface the canonical Hamiltonian reduces to the
boundary contribution,
$H_{\rm phys} = H_{\rm boundary}$.
Despite this apparent simplicity, the bulk equations of motion remain highly
non--trivial.  The Hamiltonian equations are
\begin{equation}
\dot q(x)
=
\{ q(x), H \}
=
\int d^3y
\left(
N(y) \{ q(x), \mathcal H_\perp(y) \}
+
N^i(y) \{ q(x), \mathcal H_i(y) \}
\right),
\end{equation}
and similarly for $\dot p(x)$.  Although the constraints vanish weakly,
$\mathcal H_\perp \approx 0$ and $\mathcal H_i \approx 0$, their functional
derivatives are generically non--vanishing.  Indeed:
\begin{equation}
\{ q(x), \mathcal H_\perp(y) \}
=
\frac{\delta \mathcal H_\perp(y)}{\delta p(x)},
\qquad
\{ p(x), \mathcal H_\perp(y) \}
=
- \frac{\delta \mathcal H_\perp(y)}{\delta q(x)} ,
\end{equation}
so that the equations of motion are determined by the functional form of the
constraint densities rather than by their on--shell values.
Since $\mathcal H_\perp$ and $\mathcal H_i$ contain the full non--linear dependence
on the metric, higher--curvature operators, fluxes, fermions and matter fields,
their functional derivatives reproduce exactly the covariant Euler--Lagrange
equations obtained from $\delta S_{\rm bulk}=0$.  The boundary term contributes
only to conserved charges and to the generator of asymptotic time translations,
while the entire bulk dynamics is encoded in the constraint algebra.

Thus the apparent reduction of the Hamiltonian to a boundary term reflects only
the universal constraint structure enforced by diffeomorphism invariance; the
highly non--trivial bulk equations of motion arise from the functional derivatives
of the Hamiltonian constraints and are completely equivalent to the covariant
field equations.

\subsection{Bulk constraints versus the physical Hamiltonian}
\label{sec:Hphys_bulk_vs_boundary}

The discussion in the previous section can be made a bit more quantitative once we include other degrees of freedom like ghosts and gauge fixing terms. Let:
\begin{equation}
S_{\rm tot}
=
S_{\rm EH}(g)
+
S_{\rm high}(g)
+
S_{\rm kin+int}(g,\phi)
+
S_{\rm ghost}(g,\phi,c,\bar c)
+
S_{\rm gf}(g,\phi,b),
\end{equation}
where $S_{\rm high}$ and $S_{\rm kin+int}$ may admit a trans-series expansion due
to nonperturbative (instanton) effects. (See details in section \ref{sec4.7.3}.) For each sector
$A\in\{{\rm EH},{\rm high},{\rm kin+int},{\rm ghost},{\rm gf}\}$ we write a formal
decomposition:
\begin{equation}
S_A
=
S_A^{\rm bulk}
+
\int d^4x\;\partial_0\!\left(S_A^{(1)}\right)^0
+
\int d^4x\;\partial_i\!\left(S_A^{(2)}\right)^i.
\label{eq:SA_decomp_again}
\end{equation}
The Hamiltonian and momentum constraint densities arise from the bulk dynamics
(via the lapse/shift variations) and may be written schematically as
\begin{equation}
\mathcal H_\perp^{\rm tot}
\propto
\frac{\delta S_{\rm tot}}{\delta g^{\mu\nu}}\,n^\mu n^\nu
=
\sum_A
\frac{\delta S_A^{\rm bulk}}{\delta g^{\mu\nu}}\,n^\mu n^\nu,
\end{equation}
with an analogous expression for $\mathcal H_i^{\rm tot}$.
The physical Hamiltonian is obtained after solving/imposing the constraints.
It is given by the boundary charge and is controlled by spatial boundary terms
(at fixed time), not by time total derivatives. Concretely,
\begin{equation}
H[N,N^i]
=
\int_{\Sigma_t} d^3x\,
\big(N\mathcal H_\perp^{\rm tot}+N^i\mathcal H_i^{\rm tot}\big)
+
H_{\partial\Sigma}^{\rm tot}[N,N^i],
\end{equation}
so that on the constraint surface (or on physical states)
$H_{\rm phys}
=
H_{\partial\Sigma}^{\rm tot}[1,0]$. This may be made precise in the following way.
In the decomposition \eqref{eq:SA_decomp_again}, the boundary contribution is
controlled by the spatial divergences and the gravitational boundary completion:
\begin{equation}\label{chukkamukha}
H_{\rm phys}
\ \propto\
-\oint_{\partial\Sigma_t} dS_i\,
\left[
\sum_A \left(S_A^{(2)}\right)^i
\right]
\;+\;
(\text{boundary completion from }S_{\rm EH}+S_{\rm high}).
\end{equation}
The terms $\partial_0(S_A^{(1)})^0$ contribute only to endpoint terms in the action
(and to the well-posed variational principle) and do not furnish an additional
generator of time evolution.

There are however two points that may require a bit more discussion: one, the boundary completion from $S_{\rm EH}+S_{\rm high}$ and two, why $\partial_0(S^{(1)})^0$ is not $H_{\rm phys}$. Assuming that each sector admits a formal decomposition \eqref{eq:SA_decomp_again}, 
the gravitational boundary \emph{completion} is the additional boundary functional
$S_{\rm bnd}[g]$ that must be added to $S_{\rm EH}^{\rm bulk}+S_{\rm high}^{\rm bulk}$
so that the variational principle is well posed under the chosen boundary conditions
(e.g.\ fixing the induced metric on $\partial\mathcal M$). In the present language,
the only terms that can contribute to $S_{\rm bnd}$ are those that integrate to
$\partial\mathcal M$, i.e. the total derivative pieces in \eqref{eq:SA_decomp_again}.
Accordingly, we may write schematically:

{\footnotesize
\begin{equation}
S_{\rm bnd}[g]
=
-\int_{\mathcal M} d^4x\;\partial_0\!\left[\left(S_{\rm EH}^{(1)}\right)^0+\left(S_{\rm high}^{(1)}\right)^0\right]
-\int_{\mathcal M} d^4x\;\partial_i\!\left[\left(S_{\rm EH}^{(2)}\right)^i+\left(S_{\rm high}^{(2)}\right)^i\right]
+
S_{\rm ct}[g] ,
\label{eq:Sbnd_in_S12}
\end{equation}}
where $S_{\rm ct}[g]$ denotes possible additional local counterterms on the boundary
needed to render the variational principle finite and/or to match the desired
asymptotic falloffs (for asymptotically flat spacetimes one typically chooses
a subtraction against a reference background). The key point is that the
\emph{completion} is a functional supported on $\partial\mathcal M$ and can always be
represented as an integral of total derivatives on $\mathcal M$, as in
\eqref{eq:Sbnd_in_S12}. Note that {\it absence} of scalar, matter, ghosts and gauge-fixing contributions: they generally do not require boundary completions and hence do not contribute to the boundary terms\footnote{Although \eqref{chukkamukha} is the correct formal decomposition, the
ghost and gauge-fixing contributions typically \emph{vanish} for standard asymptotically
flat boundary conditions. The reason is that one usually imposes falloffs such as
$c,\bar c,b \to 0$ {as }  $r\to\infty$ and 
$\delta c,\delta\bar c,\delta b \to 0$,
so that no nontrivial surface flux of ghost/gf degrees of freedom reaches $\partial\Sigma$.
In BRST language the ghost/gf sector is BRST-exact and does not change the value of the
physical charge on the BRST cohomology:
\begin{equation}
H_{\partial\Sigma}^{\rm ghost}[1,0]
+
H_{\partial\Sigma}^{\rm gf}[1,0]
\;\approx\;0
\qquad
\label{eq:ghost_gf_boundary_trivial}
\end{equation}
on physical states, where we have used the $[1, 0]$ formalism to fix the ADM data on the boundary.
Therefore, under standard boundary conditions one may set:
\begin{equation}
H_{\rm phys}
=
H_{\partial\Sigma}^{\rm grav}[1,0]
+
H_{\partial\Sigma}^{\rm kin+int}[1,0],\nonumber
\label{eq:Hphys_practical}
\end{equation}
with the further simplification that $H_{\partial\Sigma}^{\rm kin+int}[1,0]$ also vanishes
for sufficiently fast falloff of matter fields (so that the entire charge is the ADM/Iyer--Wald
gravitational charge evaluated on the full solution). See {\bf Table \ref{boundaricori}}.}.  In particular, the usual Einstein--Hilbert completion is the well-known
GHY term \cite{ghy,hh,hhlater}:
\begin{equation}
S_{\rm GHY}
=
\frac{1}{8\pi G}\int_{\partial\mathcal M} d^3y\,\sqrt{|h|}\,K,
\end{equation}
and higher-derivative terms in $S_{\rm high}$ generally require additional boundary
functionals (often expressible in Iyer--Wald/Noether-charge language) so that the
variational principle is well posed.

\begin{table}[h]
\centering
\begin{tabular}{|c|c|}
\hline
\textbf{Sector} & \textbf{Boundary completion needed?} \\
\hline
Einstein--Hilbert & Yes (GHY term) \\
\hline
Higher-curvature terms & Yes \\
\hline
Scalar / matter fields & Usually no \\
\hline
Gauge-fixing sector & Usually no \\
\hline
Ghost sector & Usually no \\
\hline
\end{tabular}
\caption{Boundary completions required for different sectors of the gauge-fixed gravitational action. 
The Einstein--Hilbert term requires the Gibbons--Hawking--York (GHY) boundary term because the action contains second derivatives of the metric. 
Higher-curvature terms typically require analogous boundary completions. 
For standard matter, gauge-fixing, and ghost sectors the boundary contributions vanish under appropriate boundary conditions.}
\label{boundaricori}
\end{table}

It is important to however note the following.
\eqref{eq:Sbnd_in_S12} is a \emph{bookkeeping} statement: it says the boundary
completion is built from total-derivative data of the gravitational sector. It does
\emph{not} assert that $\left(S_{\rm EH}^{(1)}\right)^0$ or $\left(S_{\rm high}^{(1)}\right)^0$
are uniquely defined or that the split into $(1)$ and $(2)$ pieces is canonical.
What is canonical is the existence of a boundary functional $S_{\rm bnd}$ that removes
unwanted boundary variations and yields a well-defined canonical generator. In other words, 
writing:
\begin{equation}
S_{\rm EH}+S_{\rm high}
=
(S_{\rm EH}+S_{\rm high})^{\rm bulk}
+
\int_{\mathcal M} d^4x\;\partial_\mu \Theta^\mu_{\rm grav},
\end{equation}
one should not choose the boundary completion $S_{\rm bnd}$ to cancel the total
derivative term $\partial_\mu\Theta^\mu_{\rm grav}$ itself. Rather, $S_{\rm bnd}$
is chosen so that the \emph{variation} of the combined action has a well-posed
Dirichlet form:
\begin{equation}
\delta\Big[(S_{\rm EH}+S_{\rm high})^{\rm bulk}+S_{\rm bnd}\Big]
=
\int_{\mathcal M} d^4x\;(\text{EOM})\cdot \delta g
+
\int_{\partial\mathcal M} d^3y\;\Pi^{ij}\,\delta h_{ij},
\label{eq:wellposed_variation}
\end{equation}
i.e.\ all terms involving normal derivatives of $\delta g_{\mu\nu}$ cancel.
The remaining boundary term in \eqref{eq:wellposed_variation} is \emph{not} set
to zero; it is precisely the term from which the canonical boundary charge
(ADM/Brown--York/Iyer--Wald, depending on the asymptotics) is obtained.
In particular, the spatial divergence pieces $\partial_i(S_{\rm EH}^{(2)})^i$
and $\partial_i(S_{\rm high}^{(2)})^i$ should be viewed as the raw boundary
data entering the canonical generator. The boundary completion reorganizes these
contributions so that the generator is functionally differentiable; it does not
eliminate the boundary charge.

Therefore, for asymptotically flat boundary conditions and $(N,N^i)\to(1,0)$ at infinity, the
gravitational boundary term reduces to the ADM (or, with higher-curvature terms, the
appropriate Iyer--Wald/Noether) energy evaluated on the asymptotic metric determined by
the full equations of motion (including trans-series corrections). Symbolically,
\begin{equation}
H_{\rm phys}
=
E_{\rm ADM/Wald}\big[g(\text{solution})\big],
\end{equation}
where $g(\text{solution})$ denotes the metric that solves the corrected mean-field
equations (possibly in the GS state, in which case $g_{\mu\nu}=\langle \hat g_{\mu\nu}\rangle_\sigma$).
The dependence on instanton/trans-series data enters through the asymptotic tail of this
solution and hence through the boundary charge itself.


\subsection{Fixed backgrounds, fluctuations, and the boundary Hamiltonian \label{sec1.5}}

Let us now come to a more streamlined discussion directly pertaining to the kind of background that we want to study. This means we will now adapt the general discussion of fixed backgrounds, quantum fluctuations, and boundary Hamiltonians to warped Minkowski compactifications arising in M--theory and heterotic string theory.
Such backgrounds play a central role in flux compactifications and admit multiple Minkowski minima associated with different choices of fluxes, bundles, and moduli.

Consider a fixed classical background geometry that is (possibly warped) Minkowski spacetime. (We will start with $3+1$ dimensions.)
The background metric is taken to be non-dynamical and admits a preferred notion of time, either due to asymptotic flatness or the presence of a timelike Killing vector. We decompose the full metric as:
\begin{equation}
g_{\mu\nu}(x) = g^{\rm bg}_{\mu\nu}(x) + h_{\mu\nu}(x),
\end{equation}
where $g^{\rm bg}_{\mu\nu}$ denotes the fixed background geometry and $h_{\mu\nu}$ represents small fluctuations satisfying $|h_{\mu\nu}| \ll 1$. Because the background is fixed, the lapse and shift are fixed asymptotically,
\begin{equation}
N \to 1,
\qquad
N^i \to 0,
\end{equation}
thereby selecting a preferred asymptotic Minkowski time. This is the formal advantage that we have with our choice of the Minkowski minimum. (Of course in the full M-theory set-up we will also require that such a minimum be supersymmetric, but we won't worry about it right away. We will come back to this issue a little later\footnote{The convexity of the supersymmetric 1PI effective action $-$ assuming that it exists $-$ clearly demands the presence of at least one Minkowski minimum. Multiple minima appears if we study the 1PI action sectorially. What this means will be discussed in the next section.}.)

To study the constraint structure for metric fluctuations around a fixed background,
we expand both the metric and the matter sector (if present) as
\begin{equation}
g_{\mu\nu}
=
g^{\rm bg}_{\mu\nu}
+
\epsilon\,h^{(1)}_{\mu\nu}
+
\epsilon^2 h^{(2)}_{\mu\nu}
+\cdots,
\qquad
\phi
=
\phi^{\rm bg}
+
\epsilon\,\varphi^{(1)}
+
\epsilon^2\varphi^{(2)}
+\cdots,
\end{equation}
where $\epsilon$ is a tunable bookkeeping parameter. Accordingly the Einstein tensor and
stress tensor admit the following expansions:
\bg\label{damac1}
&&G_{\mu\nu}(g)
=
G^{\rm bg}_{\mu\nu}
+
\epsilon\,G^{(1)}_{\mu\nu}[h^{(1)}]
+
\epsilon^2\Big(
G^{(1)}_{\mu\nu}[h^{(2)}]
+
G^{(2)}_{\mu\nu}[h^{(1)},h^{(1)}]
\Big)
+\cdots \nonumber\\
&& T_{\mu\nu}(g,\phi)
=
T^{\rm bg}_{\mu\nu}
+
\epsilon\,T^{(1)}_{\mu\nu}[h^{(1)},\varphi^{(1)}]
+
\epsilon^2\,T^{(2)}_{\mu\nu}[h^{(1)},h^{(2)},\varphi^{(1)},\varphi^{(2)}]
+\cdots,
\nd
where $\epsilon$ is the same tunable parameter.
We assume the background solves the full equations of motion
$G^{\rm bg}_{\mu\nu}
=
8\pi G\,T^{\rm bg}_{\mu\nu}$, but the question is what constraints do the diffeomorphism invariance play on the fluctuations. 
In ADM variables the lapse $N$ and shift $N^i$ appear without time derivatives and
therefore act as Lagrange multipliers. Their variations impose the Hamiltonian and
momentum constraints:

{\footnotesize
\begin{equation}
\mathcal H
\equiv
\frac{1}{\sqrt{h}}\frac{\delta S}{\delta N}
\;\propto\;
\big(G_{\mu\nu}-8\pi G\,T_{\mu\nu}\big)n^\mu n^\nu,
\qquad
\mathcal H_i
\equiv
\frac{1}{\sqrt{h}}\frac{\delta S}{\delta N^i}
\;\propto\;
\big(G_{\mu\nu}-8\pi G\,T_{\mu\nu}\big)n^\mu h^\nu{}_i,
\label{eq:constraints_covariant}
\end{equation}}
where $n^\mu$ is the unit normal to $\Sigma_t$ and $h^\nu{}_i$ is the spatial projector.
Because \eqref{eq:constraints_covariant} follows from diffeomorphism invariance and the
nondynamical nature of $N,N^i$, it holds as an operator statement (or as a statement on
physical states) independently of whether a given off--shell configuration satisfies the
dynamical equations. Expanding \eqref{eq:constraints_covariant} in powers of $\epsilon$ yields an infinite
tower of constraint equations. Defining:
$E_{\mu\nu}
\equiv
G_{\mu\nu}-8\pi G\,T_{\mu\nu}$ and 
$E_{\mu\nu}
=
E^{\rm bg}_{\mu\nu}
+
\epsilon\,E^{(1)}_{\mu\nu}
+
\epsilon^2\,E^{(2)}_{\mu\nu}
+\cdots$ 
with $E^{\rm bg}_{\mu\nu}=0$ by the background EOM, the constraints at order
$\epsilon^n$ take the universal form:
\bg\label{damac2}
&& \mathcal H^{(n)}
\;\propto\;
E^{(n)}_{\mu\nu}\,n^\mu n^\nu
=
0,
\qquad
\mathcal H^{(n)}_i
\;\propto\;
E^{(n)}_{\mu\nu}\,n^\mu h^\nu{}_i
=
0,
\qquad
n\ge 1\\
\implies && \Big(G^{(n)}_{\mu\nu}-8\pi G\,T^{(n)}_{\mu\nu}\Big)\,n^\mu n^\nu=0,
\qquad
\Big(G^{(n)}_{\mu\nu}-8\pi G\,T^{(n)}_{\mu\nu}\Big)\,n^\mu h^\nu{}_i=0,
\qquad
n\ge 1, \nonumber \nd
where $G^{(n)}_{\mu\nu}$ and $T^{(n)}_{\mu\nu}$ include, at each order, both the direct
$h^{(n)}$ (and $\varphi^{(n)}$) contribution and the nonlinear combinations built from
lower-order fluctuations. At $O(\epsilon)$,
$E^{(1)}_{\mu\nu}
=
G^{(1)}_{\mu\nu}[h^{(1)}]
-
8\pi G\,T^{(1)}_{\mu\nu}[h^{(1)},\varphi^{(1)}]$ 
so that
$\mathcal H^{(1)}\propto
\Big(G^{(1)}_{\mu\nu}[h^{(1)}]-8\pi G\,T^{(1)}_{\mu\nu}\Big)n^\mu n^\nu=0$ and 
$\mathcal H_i^{(1)}\propto
\Big(G^{(1)}_{\mu\nu}[h^{(1)}]-8\pi G\,T^{(1)}_{\mu\nu}\Big)n^\mu h^\nu{}_i=0$.
In ADM time gauge\footnote{To analyze the constraint structure around a background solution, it is important
to expand not only the spacetime metric but also the ADM variables themselves.
For a general metric written in ADM form,
we expand
$N = N_{\rm bg} + \epsilon\,\delta N^{(1)} + \epsilon^2 \delta N^{(2)} + \cdots,
N_i = N^{\rm bg}_i + \epsilon\,\delta N^{(1)}_i + \epsilon^2 \delta N^{(2)}_i + \cdots$, and 
$h_{ij}
=
h^{\rm bg}_{ij}
+\epsilon\,\gamma^{(1)}_{ij}
+\epsilon^2\gamma^{(2)}_{ij}
+\cdots$.
Here $(N_{\rm bg},N^{\rm bg}_i,h^{\rm bg}_{ij})$ describe the background geometry,
which is assumed to satisfy the full Einstein equations with background matter. Because the lapse and shift enter the action without time derivatives,
their variations impose the Hamiltonian and momentum constraints
$\frac{\delta S}{\delta N}=0$ and 
$\frac{\delta S}{\delta N^i}=0$. These relations hold independently of whether the metric fluctuations satisfy
the dynamical equations of motion.
Expanding in powers of $\epsilon$ yields a hierarchy of constraints.
These equations determine the nondynamical fluctuations $\delta N^{(1)}$
and $\delta N_i^{(1)}$ in terms of the spatial metric perturbations and
matter fluctuations. Only after expanding and writing the constraint equations may one impose a gauge
choice. In a background where
$N_{\rm bg}=1$
and $N^{\rm bg}_i=0$, 
one can choose the ADM time gauge
$\delta N^{(1)}=0$
and $\delta N^{(1)}_i=0$
which fixes the time slicing.} $N=1$, $N^i=0$, these become the familiar linearized constraint
equations:
\begin{equation}
G^{(1)}_{00}[h^{(1)}]
=
8\pi G\,T^{(1)}_{00},
\qquad
G^{(1)}_{0i}[h^{(1)}]
=
8\pi G\,T^{(1)}_{0i}.
\label{damacdamac}
\end{equation}
They determine the nondynamical components (lapse/shift perturbations) in terms of the
other fields and do not constitute the propagating equations for the graviton. At $O(\epsilon^2)$ one finds schematically
$E^{(2)}_{\mu\nu}
=
G^{(1)}_{\mu\nu}[h^{(2)}]
+
G^{(2)}_{\mu\nu}[h^{(1)},h^{(1)}]
-
8\pi G\,T^{(2)}_{\mu\nu}[h^{(1)},h^{(2)},\varphi^{(1)},\varphi^{(2)}]$
and therefore
$\mathcal H^{(2)}\propto
E^{(2)}_{\mu\nu}\,n^\mu n^\nu=0$
and $\mathcal H_i^{(2)}\propto
E^{(2)}_{\mu\nu}\,n^\mu h^\nu{}_i=0$.
In time gauge this reads:
\begin{equation}
G^{(1)}_{00}[h^{(2)}]
+
G^{(2)}_{00}[h^{(1)},h^{(1)}]
=
8\pi G\,T^{(2)}_{00},~~
G^{(1)}_{0i}[h^{(2)}]
+
G^{(2)}_{0i}[h^{(1)},h^{(1)}]
=
8\pi G\,T^{(2)}_{0i}.
\label{eq:constraints_quadratic_timegauge}
\end{equation}
These equations determine the nondynamical components of $h^{(2)}_{\mu\nu}$ in terms of
lower-order fluctuations (and matter perturbations), as required by diffeomorphism
invariance.

The relations \eqref{damacdamac} and \eqref{eq:constraints_quadratic_timegauge}
express the fact that $N$ and $N^i$ are Lagrange multipliers and encode the gauge
redundancy associated with spacetime diffeomorphisms. They do \emph{not} mean that a
generic off--shell field configuration integrated over in the path integral satisfies the
full Einstein equations component-by-component. Rather, after gauge fixing one integrates
over the remaining unconstrained variables (including ghosts and gauge-fixing auxiliaries),
and the constraints are imposed on physical states and on expectation values (or enforced
through the $N,N^i$ integrations in the canonical formulation). The dynamical equations for
classical perturbations or for the mean field in a saddle-point approximation correspond to
the remaining components of $E^{(n)}_{\mu\nu}$ (notably the $ij$ components), but they are
not satisfied by arbitrary off--shell configurations.
In summary, only the background geometry satisfies Einstein’s equations pointwise.
Metric fluctuations are integrated off shell. The Hamiltonian and momentum constraints
arise from lapse and shift as operator (or expectation--value) constraints, while the
equations $G^{(n)}_{ij}=0$ describe classical evolution or saddle--point dynamics and
should not be interpreted as restrictions on the quantum integration variables.


Expanding the Einstein--Hilbert action to quadratic order in $h_{\mu\nu}$ yields a theory of linearized gravitational fluctuations propagating on the fixed background $\bar g_{\mu\nu}$.  
After gauge fixing and solving the linearized constraints, the physical degrees of freedom reduce to the transverse--traceless graviton modes.
At this order, the Hamiltonian is no longer a constraint but a genuine generator of time evolution with respect to the background time\footnote{This may be justified in the following way. In the ADM formulation of general relativity, the total Hamiltonian takes the form \eqref{chukka5} and with the on-shell constraints $\mathcal H_{\perp} = \mathcal H_i = 0$, 
the physical Hamiltonian reduces entirely to the boundary contribution,
$H_{\text{phys}} = H_{\rm boundary}$.
For asymptotically flat (or warped Minkowski) spacetimes, the boundary Hamiltonian is
the ADM energy,
$H_{\rm boundary} = E_{\text{ADM}}$.
Choosing a background metric $g^{\rm bg}_{\mu\nu}$ that fixes a preferred asymptotic time, we can expand
$g_{\mu\nu} = g^{\rm bg}_{\mu\nu} + h_{\mu\nu}$ in the way discussed above such that
the ADM energy admits an expansion of the form:
\begin{equation}
E_{\text{ADM}}[g^{\rm bg}+h]
=
E_{\text{ADM}}[g^{\rm bg}]
+
E^{(1)}[h]
+
E^{(2)}[h]
+
\cdots ,
\end{equation}
where we have used Taylor expansion to get the higher order terms.
For a background satisfying the constraints, the linear term vanishes,
$E^{(1)}[h] = 0$,
and the leading nontrivial contribution comes from the quadratic term and higher order terms in the following way:
\begin{equation}
H_{\text{fluc}} \equiv E^{(2)}[h] + ....
\end{equation}
After gauge fixing and solving the constraints, $H_{\text{fluc}}$ generates the
dynamics of propagating graviton modes as we shall discuss below.
Thus the fluctuation Hamiltonian $H_{\text{fluc}}$ is precisely the quadratic expansion
of the ADM boundary Hamiltonian around a chosen background geometry. In general we can summarize this in the following way: Once a background geometry fixes a preferred asymptotic time, the ADM
boundary Hamiltonian generates genuine time evolution, and its quadratic (and higher order) expansions
define the standard Hamiltonian governing graviton fluctuations.},
\begin{equation}
H_{\text{fluc}} = \sum_k \omega_k\, a_k^\dagger a_k + \cdots .
\end{equation}
Thus, linearized gravity on a fixed background is equivalent, at the conceptual level, to quantum field theory on curved spacetime.
The metric fluctuations admit a mode expansion of the form
\begin{equation}
h_{\mu\nu}(x)
=
\sum_k \left(
a_k\, u^{(k)}_{\mu\nu}(x)
+
a_k^\dagger\, u^{(k)\,*}_{\mu\nu}(x)
\right),
\end{equation}
where the mode functions $u^{(k)}_{\mu\nu}(x)$ are defined with respect to the background geometry and its preferred time coordinate.
Canonical quantization leads to the commutation relations
$[a_k, a_{k'}^\dagger] = \delta_{k k'}$
and defines a Fock vacuum $|0\rangle$.
A coherent state of gravitational fluctuations is defined by
\begin{equation}
|\{\alpha_k\}\rangle
=
\exp\!\left(
\sum_k (\alpha_k a_k^\dagger - \alpha_k^* a_k)
\right)
|0\rangle,
\end{equation}
where $\alpha_k$ is in general a complex number. Note that we have chosen a free vacuum $\vert 0 \rangle$ to define our coherent state. 
Such states satisfy
\begin{equation}
\langle \{\alpha_k\} |\, h_{\mu\nu}(x)\, | \{\alpha_k\} \rangle
=
\sum_k \left(
\alpha_k\, u^{(k)}_{\mu\nu}(x)
+
\alpha_k^*\, u^{(k)\,*}_{\mu\nu}(x)
\right),
\end{equation}
which corresponds to a classical metric perturbation on top of the background geometry.  
Coherent graviton states therefore describe semiclassical geometries or weak gravitational waves. More interestingly, 
because the background geometry fixes a preferred notion of time, the coherent states evolve unitarily according to
\begin{equation}
|\{\alpha_k(t)\}\rangle
=
e^{-i H_{\text{fluc}} t}\,
|\{\alpha_k(0)\}\rangle ,
\end{equation}
with
$\alpha_k(t) = \alpha_k(0)\, e^{-i \omega_k t}$.
This evolution is genuine time evolution in the usual quantum-mechanical sense, rather than a gauge redundancy. The story continues practically unchanged once we incorporate the full warped geometry of the form:
\begin{equation}
ds^2
=
e^{2A(y)} \eta_{\mu\nu} dx^\mu dx^\nu
+
e^{2B(y)} g_{mn}(y) dy^m dy^n ,
\end{equation}
where $\eta_{\mu\nu}$ is the four-dimensional Minkowski metric, $y^m$ are coordinates on the internal manifold,
$g_{mn}(y)$ is the internal metric, and 
$A(y)$ and $B(y)$ are warp factors determined by fluxes and sources.
In heterotic theory, the internal space is typically a six-manifold with gauge bundles satisfying anomaly cancellation, while in M--theory the internal space is an eight-manifold in the presence of $G_4$ flux. All these internal manifold do not have to be Calabi-Yau manifolds: they can be non-K\"ahler manifolds preserving G-structures so as to preserve background supersymmetry. Another key assumption is that the backgrounds solve the higher-dimensional equations of motion and yield effective four-dimensional Minkowski vacua with vanishing cosmological constants. For example, choosing say one of the Heterotic theories, one may show that
the theory may admit multiple distinct warped Minkowski solutions labeled by an index $a$ as:
\begin{equation}
\bar g^{(a)}_{MN}, \qquad a = 1,2,\dots ,
\end{equation}
corresponding to different discrete flux choices, bundle data, or stabilized moduli. Each such solution defines a separate semiclassical branch of the theory with three characteristics: (1) vanishing four-dimensional cosmological constant, (2)
asymptotic four-dimensional Minkowski time, and (3)
identical local symmetries but different global data. Choosing one minimum corresponds to selecting a superselection sector of the full quantum theory. The question is how standard quantum time evolution emerges when gravity is treated semiclassically, and how this picture connects to the ADM Hamiltonian and its boundary term. To see this
fix one warped Minkowski background $\bar g^{(a)}_{MN}$ and expand the higher-dimensional fields as
\begin{equation}
g_{MN} = \bar g^{(a)}_{MN} + h_{MN},
\end{equation}
together with analogous fluctuations for fluxes, gauge fields, and moduli. We assume that all fluctuations are small,
{\it i.e.}
$|h_{MN}| \ll 1$,
so that the background geometry remains fixed to leading order.
Expanding the higher-dimensional action to quadratic order in fluctuations and performing Kaluza--Klein reduction yields an effective four-dimensional theory describing gravitons, moduli, and matter fields propagating on a warped Minkowski background.
After gauge fixing and solving the linearized constraints, the physical degrees of freedom admit a Hamiltonian description,
\begin{equation}
H^{(a)}_{\text{fluc}}
=
\sum_k \omega^{(a)}_k a_k^\dagger a_k
+
\cdots ,
\end{equation}
where the frequencies $\omega^{(a)}_k$ depend on the warp factors and internal geometry.
Crucially, this Hamiltonian is \emph{not} a constraint. It generates genuine time evolution with respect to the asymptotic four-dimensional Minkowski time.

The existence of a non-zero Hamiltonian also suggests the existence of a Wheeler--DeWitt equation. Recall that,
in background-independent quantum gravity, the quantum state is a wavefunctional $\Psi[h_{ij},\phi]$ satisfying the Wheeler--DeWitt equation \cite{WdWorig,hh,hhlater,FPI}:
\begin{equation}\label{shoottara}
\widehat{\mathcal H}(x)\,\Psi = 0,
\qquad
\widehat{\mathcal H}_i(x)\,\Psi = 0.
\end{equation}
This equation encodes refoliation invariance and leads to a frozen formalism with no preferred time. To recover semiclassical dynamics, one may adopt a Born--Oppenheimer or WKB ansatz,
\begin{equation}
\Psi[h_{ij},\phi]
\approx
\exp\!\left(\frac{i}{\hbar} S_0[h]\right)\,
\psi[h,\phi],
\end{equation}
where at leading order, $S_0[h]$ satisfies the Hamilton--Jacobi equation for classical gravity, selecting a background spacetime.  
At subleading order, $\psi$ satisfies an effective Schr\"odinger equation,
\begin{equation}
i\frac{\partial}{\partial t_{\text{WKB}}}\psi
=
\widehat{H}_{\text{fluc}}\,\psi,
\end{equation}
where $t_{\text{WKB}}$ is an emergent time parameter defined by the classical background. Each Minkowski minimum corresponds to a distinct semiclassical branch with its own notion of time and its own fluctuation Hamiltonian. We will come back to a more detailed discussion on this a little later. A summary of the situation is the following. (1) In the absence of boundaries, the bulk Hamiltonian generates only refoliations and leads to a frozen formalism.
(2) A fixed background geometry explicitly breaks refoliation invariance and provides a preferred notion of time.
(3) In asymptotically flat spacetimes, this preferred time arises from boundary conditions, and the corresponding generator is the boundary Hamiltonian.
(4) Quantizing small fluctuations around a fixed warped Minkowski background yields a standard quantum theory with unitary time evolution, governed semiclassically by the boundary Hamiltonian.

This framework explains how coherent states of gravitational fluctuations evolve naturally in background time while remaining consistent with the constrained Hamiltonian structure of full general relativity. Alternatively, we can express everything as: 
\vskip.1in

\noindent $\bullet$ Fixing a warped Minkowski background provides a preferred notion of time.

\vskip.1in

\noindent $\bullet$ Small gravitational fluctuations around such a background form a standard quantum field theory.

\vskip.1in

\noindent $\bullet$ Coherent states of metric fluctuations evolve unitarily with a non-vanishing Hamiltonian.

\vskip.1in

\noindent $\bullet$ Multiple Minkowski minima define distinct semiclassical sectors, each with its own time evolution.

\vskip.1in

\noindent $\bullet$ The vanishing bulk Hamiltonian and the Wheeler--DeWitt equation apply to the unreduced, background-independent theory.

\vskip.1in

\noindent $\bullet$ Ordinary Schr\"odinger evolution emerges semiclassically via boundary conditions or WKB time.

\vskip.1in 
\noindent This reconciles the existence of coherent-state evolution on fixed backgrounds with the constrained Hamiltonian structure of full quantum gravity.



\section{Why is de Sitter spacetime only realized as an excited state? \label{sec2.0}}

The discussion presented in section \ref{sec001} is necessarily schematic and conceals a number of important subtleties. As we will clarify below, it is essential to distinguish carefully between { coherent states} and {Glauber--Sudarshan (GS) states}, which play conceptually different roles in the construction. Before turning to a detailed analysis, however, it is useful to highlight several conceptual pitfalls that arise when attempting to describe a de Sitter minimum within any consistent quantum theory of gravity\footnote{This issue is also reflected in string theory through various no-go theorems \cite{GMN,KMMS}. However, since our objective here is to analyze the problem exclusively from a four-dimensional perspective, we shall refrain from invoking any intrinsically string-theoretic input.}.

\subsection{Possible constraints from the 1PI effective action \label{dubaitag1}}

The first conceptual issue comes from the 1PI effective action itself.
Assume that the full quantum theory admits a well--defined generating functional
for connected correlators,
\begin{equation}
Z[J]
=
\int \mathcal D\phi\;
\exp\!\left(
i S[\phi] + i \int d^dx\, J(x)\phi(x)
\right),
\end{equation}
with
$W[J] = - i \log Z[J]$.
We can define the mean field
$\bar\phi(x)
=
\frac{\delta W[J]}{\delta J(x)}$,
and assume that this relation can be inverted to express $J$ as a functional of
$\bar\phi$.  The one--particle--irreducible (1PI) effective action is then defined by
the Legendre transform
\begin{equation}
\Gamma[\bar\phi]
=
W[J]
-
\int d^dx\, J(x)\,\bar\phi(x).
\end{equation}
Standard argument then tells us that the second functional derivative of $W[J]$ is the connected two--point function,
\begin{equation}
\frac{\delta^2 W}{\delta J(x)\,\delta J(y)}
=
\langle \phi(x)\phi(y) \rangle_J
-
\langle \phi(x)\rangle_J \langle \phi(y)\rangle_J,
\end{equation}
which is positive semidefinite in Euclidean signature.  By properties of the
Legendre transform, the second functional derivative of $\Gamma$ satisfies
\begin{equation}
\frac{\delta^2 \Gamma}{\delta \bar\phi(x)\,\delta \bar\phi(y)}
=
\left(
\frac{\delta^2 W}{\delta J\,\delta J}
\right)^{-1}\!(x,y)
\;\ge\; 0.
\end{equation}
Therefore, $\Gamma$ is a convex functional of $\bar\phi$. From here conventional wisdom guarantees that 
a vacuum of the quantum theory is defined as a stationary configuration of the 1PI
effective action in the following sense:
\begin{equation}
\frac{\delta \Gamma[\bar\phi]}{\delta \bar\phi(x)} = 0,
\end{equation}
evaluated at vanishing external sources.  Because $\Gamma$ is convex, any stationary
point is necessarily a global minimum.  Consequently, $\Gamma$ admits no metastable
vacua and no local positive--energy minima. Extending this to gravity, a de Sitter vacuum would require a stationary configuration of $\Gamma$ with
positive vacuum energy,
\begin{equation}
\frac{\delta \Gamma}{\delta g_{\mu\nu}} = 0,
\qquad
\Gamma[g_{\mu\nu}] > 0,
\end{equation}
together with stability under small fluctuations.  Convexity of $\Gamma$ forbids
such a configuration.  Therefore, if the exact 1PI effective action exists and a
configuration qualifies as a vacuum in the 1PI sense, de Sitter vacua are excluded.  Moreover, in supersymmetric theories such as M-theory, unbroken SUSY together with the existence of the 1PI effective action guarantees Minkowski (or AdS) vacua, but once supersymmetry is broken, convexity alone no longer enforces the existence of any vacuum at all, typically leading instead to runaway behavior while still forbidding de Sitter vacua.

There is however a limitation of the aforementioned conclusion in quantum gravity.
The above conclusion applies only to vacua defined as extrema of $\Gamma$.  It does
not forbid time--dependent or excited configurations.  Moreover, in quantum
gravity the assumptions underlying the construction of $\Gamma$ are nontrivial:
there is no gauge--invariant local source for the metric, no standard thermodynamic
limit, and cosmological observables are naturally formulated in the in--in
(Schwinger--Keldysh) framework rather than in the in--out formalism underlying the
1PI effective action.

Therefore the convexity of the 1PI effective action forbids de Sitter \emph{vacua}, but it does
not forbid de Sitter \emph{physics}.  In string theory and quantum gravity, de Sitter
space may arise as a time--dependent configuration or as a coherent (Glauber--
Sudarshan) excited state whose expectation values follow semiclassical equations of
motion, without extremizing the 1PI effective action.

\subsection{No Wilsonian renormalization in accelerating backgrounds \label{sec2.2}}

Even if we ignore the expectations from imposing the 1PI effective action, Wilsonian renormalization creates more problems.
In flat spacetime, Wilsonian renormalization relies on a clean and time--independent
separation of scales.  Energy and momentum are conserved, and modes classified as
``ultraviolet'' remain ultraviolet under time evolution.  One may therefore define
a Wilsonian effective action $S_\Lambda$ by integrating out all modes with momenta
$|p|>\Lambda$, yielding a scale--dependent but time--independent effective theory. Unfortunately, in an expanding or accelerating cosmological background with metric
\begin{equation}
ds^2 = -dt^2 + a^2(t)\, d\vec x^{\,2},
\end{equation}
this separation of scales no longer holds.  A field mode with fixed comoving
momentum $k$ has physical momentum
\begin{equation}
p_{\rm phys}(t) = \frac{k}{a(t)} .
\end{equation}
In an accelerating universe, $a(t)$ grows rapidly, so that modes which are
ultraviolet at early times inevitably redshift and become infrared at late times.
Thus the distinction between UV and IR modes is intrinsically time--dependent.

As a result, the standard Wilsonian procedure of integrating out modes above a fixed
cutoff $\Lambda$ becomes ambiguous.  A mode that is integrated out at one time may
re--enter the low--energy sector at a later time.  The notion of an effective action
$S_\Lambda$ defined once and for all by integrating out high--energy degrees of
freedom therefore ceases to be well defined.  Instead, any coarse--graining
procedure necessarily depends on the choice of time slicing or physical clock.

This issue is already present in quantum field theory on curved spacetime, but it
becomes particularly acute in quantum gravity.  In a diffeomorphism--invariant
theory, there is no preferred notion of time or energy, and hence no invariant
definition of ``high--energy'' versus ``low--energy'' modes.  Moreover, in an
accelerating spacetime such as de Sitter space, the presence of a cosmological
horizon implies that different observers assign different physical meanings to
energy and scale. The problem may be summarized schematically as follows.  In flat space one assumes
\begin{equation}
\text{UV} \;\xrightarrow{\;\text{integrate out}\;}\; \text{IR},
\end{equation}
with the flow irreversible and time independent.  In an accelerating background,
however, the redshift implies
\begin{equation}
\text{UV at early times}
\;\xrightarrow{\;\text{cosmic expansion}\;}
\text{IR at late times},
\end{equation}
so that the Wilsonian hierarchy of scales is continuously reshuffled by the
background dynamics itself. Note however that this obstruction does not imply an inconsistency of quantum gravity, but it signals
that the usual Wilsonian intuition must be 
modified\footnote{In fact an even more acute difficulty follows from the gravitational constraints. The physical
Hilbert space is defined by \eqref{shoottara}
but it does not generally factorize according to the Wilsonian mode split
in any simple gauge-invariant sense. The constraints couple the retained and
integrated-out sectors,
so that tracing over the ``high-energy modes" produces an effective constraint that is generally
nonlocal and state dependent, thus losing the meaning of the Hamiltonian constraints altogether. The infrared sector is consequently an open system rather than a closed
subsystem governed by a time-independent Hamiltonian. Stated alternatively, without an adiabatic or derivative expansion, integrating out the short modes
does not produce a local action. Instead, one obtains nonlocal kernels together with memory, dissipation, noise, and explicit dependence on the
initial state. \label{barthmeys}}. Any effective description in
an accelerating universe is necessarily time dependent and observer dependent, and
the renormalization group must be understood as a coarse--graining with respect to a
chosen physical scale (such as the Hubble scale) rather than as an integration over
fixed momentum shells.  In this sense, the difficulty of defining a universal,
time--independent Wilsonian RG flow in accelerating backgrounds is one of the key
conceptual challenges of quantum gravity and cosmology.

\subsection{Premise and no-go logic for metastable de Sitter EFTs}
\label{sec:premiseP}

We can in fact quantify the aforementioned line of thoughts as 
{\bf Premise P}: ``No quasi-stationary de Sitter resonances with a parametric gap". The statement then is the following.
Assume that in quantum gravity, and in particular in string compactifications with an infinite tower of states, any configuration with strictly positive four-dimensional vacuum energy $V>0$ necessarily violates at least one of the following conditions:
\begin{itemize}
\item Quasi-stationarity on Hubble times.  
\item Parametric longevity compared to the Hubble time.  
\item Parametric separation between the Hubble scale and the lightest tower of heavy states.
\end{itemize}
Equivalently, assume that there exists no regime in which a de Sitter configuration can be simultaneously long-lived, approximately time-independent, and characterized by a hierarchy
$H \ll m_{\rm tower}$
where $m_{\rm tower}$ denotes the mass scale of the lightest mode in any infinite tower (KK, string, or otherwise). Then we show that, under Premise~P, the notion of a metastable de Sitter \emph{vacuum} described by a four-dimensional Wilsonian effective field theory is internally inconsistent.

\subsubsection*{Fatal Point I: Absence of Quasi-Stationarity}
\label{sec:fatal1}

A necessary condition for defining a vacuum effective field theory is the existence of a regime in which the background is approximately stationary over timescales long compared to the Hubble time. Concretely, this requires:
\begin{equation}
\left|\frac{\dot H}{H^2}\right| \ll 1,
\qquad
\left|\frac{\dot \varphi}{H}\right| \ll \Delta\varphi,
\label{eq:quasistationary}
\end{equation}
for all light fields $\varphi$ controlling couplings and mass scales.
If this condition fails, then the masses of heavy states, which generically depend on $\varphi$, vary on Hubble times:
\begin{equation}
\frac{\dot m_{\rm tower}}{m_{\rm tower}} \sim H .
\end{equation}
In this regime, adiabatic decoupling fails. Integrating out heavy modes no longer produces a local Wilsonian action, but instead yields a nonlocal, time-dependent influence functional (see also footnote \ref{barthmeys}). Consequently, the standard notion of a low-energy effective action expanded in local operators suppressed by powers of $m_{\rm tower}$ ceases to exist.
Thus, in the absence of quasi-stationarity, the very split into ``light'' and ``heavy'' degrees of freedom becomes ill-defined.

\subsubsection*{Fatal Point II: Insufficient Lifetime of the de Sitter Configuration}
\label{sec:fatal2}

Even if the classical background admits a critical point with $V>0$, a vacuum EFT description requires that the associated metastable state persist for times parametrically longer than the Hubble time:
\begin{equation}
\Gamma^{-1} \gg H^{-1},
\label{eq:lifetime}
\end{equation}
where $\Gamma$ denotes the decay rate per unit four-volume. If instead the decay time is of order the Hubble time,
\begin{equation}
\Gamma^{-1} \sim H^{-1},
\end{equation}
then there exists no temporal window in which correlation functions approximate those of a de Sitter-invariant state. In such a situation, it is meaningless to speak of a vacuum EFT defined around a local minimum of a potential. The configuration should instead be regarded as a transient cosmological episode, describable only within an explicitly time-dependent in-in or open-system framework.
Therefore, insufficient longevity eliminates the physical regime in which a metastable de Sitter vacuum EFT could apply.

\subsubsection*{Fatal Point III: Tower Descent and Loss of Scale Separation}
\label{sec:fatal3}

A Wilsonian effective field theory requires a cutoff $\Lambda$ parametrically larger than the characteristic physical scale of interest, which in a de Sitter background is set by $H$. In string compactifications, the cutoff is controlled by the mass of the lightest state in an infinite tower:
\begin{equation}
\Lambda \lesssim m_{\rm tower}(\varphi),
\end{equation}
where $\varphi$ collectively denotes higher dimensional fields descending (or compactified) to four-dimensions
If uplifting to positive vacuum energy necessarily induces a descent of an infinite tower such that:
\begin{equation}
m_{\rm tower} \lesssim \mathcal{O}(H),
\end{equation}
then no parametric hierarchy exists between curvature scales and the cutoff. Higher-derivative corrections, stringy effects, and contributions from the full tower become unsuppressed:
\begin{equation}
\frac{H}{m_{\rm tower}} \sim \mathcal{O}(1),
\qquad
\alpha' R \sim \mathcal{O}(1).
\end{equation}
In this regime, the four-dimensional effective description is not controlled, and integrating out the tower is illegitimate. The failure is not quantitative but qualitative: the putative EFT lacks a well-defined domain of validity.

\subsubsection*{Combined Conclusion}
\label{sec:combined}

Under Premise~P, therefore any configuration with $V>0$ necessarily violates at least one of the following:
\begin{itemize}
\item quasi-stationarity required for adiabatic decoupling,
\item longevity required for a vacuum EFT interpretation,
\item parametric separation between the Hubble scale and an infinite tower of states.
\end{itemize}
Violation of any one of these conditions is sufficient to invalidate the notion of a metastable de Sitter vacuum described by a four-dimensional Wilsonian effective field theory. Consequently, many of the recent constructions  are not merely technically incomplete, but rely on assumptions that fail at a structural level once Premise~P is imposed.

Therefore, metastable de Sitter vacua in string theory are not ruled out by a single obstruction, but by the incompatibility of positive vacuum energy with the simultaneous requirements of stationarity, longevity, and scale separation that underlie effective field theory itself.

The claim then is that it is the GS state that overcomes all the aforementioned issues and leads to four-dimensional de Sitter spacetime as an excited state in string theory (or in quantum gravity). However our discussion in section \ref{dubaitag1} related to the 1PI effective action and source $J$ may raise the following question.
Consider now a Glauber--Sudarshan (GS) state defined by:
\begin{equation}
|\sigma\rangle = D(\sigma)\,|\Omega\rangle,
\end{equation}
where $|\Omega\rangle$ is a reference interacting state and $D(\sigma)$ is a
displacement operator. Why does $D(\sigma)$ define an \emph{operator}, despite being expressed as an
integral over the entire spacetime? As we shall explain shortly, its
operatorial character follows from the consistent manner in which it enters
the path-integral formulation as a time-ordered product of repeatedly applied
``kick'' operators. For the time being we will take a simple toy case with a real $\sigma$, and not worry about its integral form, and express
$D^\dagger(\sigma)\,D(\sigma) = \exp\!\big(2\sigma \phi\big)$ such that the expectation values in the GS state may be written as\footnote{This will be modified to a more precise form using in-in Schwinger-Keldysh contours or in-out path-integral format soon.}:
\begin{equation}
\langle O \rangle_\sigma
=
\frac{
\int \mathcal D\phi\;
\exp\!\left(i S[\phi]\right)\,
\exp\!\left(2\sigma \phi\right)\,
O(\phi)
}{
\int \mathcal D\phi\;
\exp\!\left(i S[\phi]\right)\,
\exp\!\left(2\sigma \phi\right)
}.
\label{GS_expectation}
\end{equation}
Formally, the insertion $\exp(2\sigma\phi)$ resembles a source term which is an issue because it raises the concern about the convexity of the 1PI effective action that we discussed in section \ref{dubaitag1}. However, this
similarity is only algebraic and not structural.
The crucial differences between $\exp(2\sigma\phi)$ and a source $J\phi$ are:
\begin{itemize}
\item No Legendre transform is performed with respect to $\sigma$.
\item $\sigma$ is fixed once the GS state is chosen and is not an external variable
      that can be freely varied. Moreover there is no factor of $i = \sqrt{-1}$ with the insertion.
\item The insertion $\exp(2\sigma\phi)$ appears both in the numerator and denominator
      of \eqref{GS_expectation}, so its effect is normalized and does not define a
      free--energy functional.
\item The construction \eqref{GS_expectation} computes expectation values in a
      specific quantum state, rather than probing the response of the system to an
      external force.
\end{itemize}
Because $D^\dagger D$ does \emph{not} act as a source in the sense required for the
Legendre transform, GS states are not constrained by the convexity of the 1PI
effective action.  Convexity applies to vacua defined by extrema of $\Gamma$, whereas
GS states describe excited or time--dependent configurations whose expectation
values follow dynamical evolution rather than energy minimization.  Consequently,
the realization of de Sitter space as a GS coherent state is compatible with the
absence of de Sitter vacua in the exact effective action.

\section{Glauber-Sudarshan states versus the coherent states \label{sec3.0}}

Let us now come to the main theme of the paper: the key differences between the coherent states and the Glauber-Sudarshan (GS) states. Some of the story was already developed in \cite{joydeep, wdwpaper, coherbeta, coherbeta2, borel2, hetborel, dileep, desitter2} but here we want to emphasize on more generic issues pertaining to temporal evolutions and path-integral representations that were not emphasized earlier. We will start by addressing one of the key subtle points, namely the {\it interacting vacuum} in gravity.

\subsection{Interacting vacuum in gravity: bulk constraints and boundary evolution \label{sec3.1}}

One of the key concepts in standard quantum field theory is the
\emph{interacting vacuum}, which incorporates the effects of all the
interactions in the theory. In ordinary quantum field theory, the interacting
vacuum may be obtained by evolving the free vacuum with the full Hamiltonian,
together with the usual $i\epsilon$ prescription:
\begin{equation}
|\Omega\rangle
\propto
\lim_{T\rightarrow\infty(1-i\epsilon)}
e^{-iHT}|0\rangle,
\qquad
\epsilon>0 ,
\label{sidsween1}
\end{equation}
where $H= H_0 + H_{\rm int}$, with $H_0$ being the free part and the $H_{\rm int}$ being the interacting part. This will be elaborated below. 
Equivalently, in the interaction picture, where $H_{\rm int}^{I}(t) \equiv e^{iH_0t}H_{\rm int}e^{-iH_0t}$ is the interaction Hamiltonian, the prescription is:
\begin{equation}
|\Omega\rangle
=
\lim_{T\rightarrow\infty(1-i\epsilon)}
\frac{
{\cal T}
\exp\left[
-i\int_{-T}^{0}dt\,
H_{\rm int}^{I}(t)
\right]
|0\rangle
}{
\langle0|
{\cal T}
\exp\left[
-i\int_{-T}^{0}dt\,
H_{\rm int}^{I}(t)
\right]
|0\rangle
}.
\label{eq:interacting_vacuum_Peskin_form}
\end{equation}
 The Hamiltonian
is a nonvanishing bulk operator that generates physical time translations,
and the deformation
$T\rightarrow T(1-i\epsilon)$ projects the free vacuum onto the interacting
vacuum, provided that $\langle\Omega|0\rangle\neq0$. In canonical gravity the situation is more subtle. The bulk Hamiltonian is a sum of
constraints:
\begin{equation}
H_{\rm bulk}
=
\int d^3x\,
\left(
N\,\mathcal H_\perp
+
N^i\,\mathcal H_i
\right)
\approx 0 ,
\end{equation}
and physical states obey
$\mathcal H_\perp|\Psi\rangle=0$ and 
$\mathcal H_i|\Psi\rangle=0$.
The bulk Hamiltonian therefore generates gauge transformations (time
reparametrizations) rather than physical evolution. This does \emph{not}, however, imply the absence of time evolution altogether.
In asymptotically Minkowski spacetimes the total Hamiltonian contains a boundary
contribution,
\begin{equation}
H_{\rm phys}=H_{\partial\Sigma},
\end{equation}
given by the ADM (or Brown--York) energy at spatial infinity.  
It is this boundary Hamiltonian that generates physical time translations of states.
Accordingly, the interacting vacuum is defined by:
\begin{equation}\label{fitboushona3}
|\Omega\rangle
=
\lim_{T\to\infty(1-i\epsilon)} e^{-i H_{\partial\Sigma} T}|0\rangle ,
\end{equation}
rather than by evolution with a bulk Hamiltonian. The crucial difference from ordinary quantum field theory is that $H_{\partial\Sigma}$
is a nonlocal boundary operator obtained only after solving the bulk constraints.
Consequently there is no simple decomposition of the form $H=H_0+H_{\rm int}$ acting
locally in the bulk, and the standard construction of the interacting vacuum by
exponentiating a bulk interaction Hamiltonian does not directly carry over at the
fully nonperturbative level. Instead, the gravitational interacting vacuum is defined by solving the Hamiltonian
constraints together with appropriate boundary conditions, or equivalently by the
full covariant path integral\footnote{In canonical gravity, a state $|\Psi\rangle$ is represented in the metric
representation by a wavefunctional on a Cauchy slice $\Sigma$ as
$\Psi[\gamma,\Phi]
\equiv
\langle \gamma,\Phi;\Sigma\,|\,\Psi\rangle$
where $\gamma_{ij}$ is the induced spatial metric on $\Sigma$ and $\Phi$ denotes all
non--gravitational fields. This definition
applies to \emph{any} physical state.
For any such state, the wavefunctional must satisfy the quantum constraints
(Wheeler--DeWitt equations):
${\mathcal H}_\perp\,\Psi[\gamma,\Phi]=0$ and 
${\mathcal H}_i\,\Psi[\gamma,\Phi]=0$. (We don't put hats to distinguish the operators from the functions unless it becomes necessary.)
Thus $\Psi$ denotes a generic solution of the constraint equations.
The vacuum state $|\Omega\rangle$ is a particular physical state singled out by
boundary conditions (regularity, asymptotic Minkowski branch, $i\epsilon$
projection, etc.).  Its wavefunctional is
$\Psi_\Omega[\gamma,\Phi]
\equiv
\langle \gamma,\Phi;\Sigma\,|\,\Omega\rangle$.
This is a distinguished solution of WdW equation selected by those boundary conditions. In the presence of multiple Minkowski minima the choice of the interacting minimum becomes more subtle. We will describe this generic story in section \ref{fitboushona}. \label{lucmaxie}},
\begin{equation}\label{fitboushona1}
\Psi_\Omega[g,\Phi]
=
\int_{\substack{g|_{\partial M}=g \\ \Phi|_{\partial M}=\Phi}}
\mathcal D g_{\mu\nu} \, \mathcal D \Phi \;
e^{\, i S[g,\Phi]} ,
\end{equation}
where $S[g,\Phi]$ contains the Einstein--Hilbert term, matter interactions and the
complete tower of higher--curvature corrections, including ghosts and gauge fixing terms. (This will soon be generalized to $S_{\rm tot}$ whose precise form will be described in section \ref{sec4.3}.) In other words, because
the integral over $g_{\mu\nu}$ includes infinitely many diffeomorphism--equivalent
configurations, the proper definition is a gauge--fixed path
integral with the Faddeev--Popov determinant (or BRST form). Concretely one should
write:

{\footnotesize
\begin{equation}
\Psi_\Omega[\gamma,\Phi]
=
\lim_{T\to\infty(1-i\epsilon)}
\int_{\substack{g|_{\Sigma}=\gamma \\ \Phi|_{\Sigma}=\Phi}}
\mathcal D g_{\mu\nu}\,\mathcal D\Phi\,
\mathcal Dc\,\mathcal D\bar c\,\mathcal Db\;
\exp\!\left[
i\Big(
S_{\rm inv}[g,\Phi]
+
S_{\rm gf}[g,\Phi,b]
+
S_{\rm gh}[g,\Phi,c,\bar c]
\Big)
\right],
\label{eq:PI_with_ghosts}
\end{equation}}
where $S_{\rm inv}[g,\Phi]$ is the diffeomorphism--invariant action including
Einstein--Hilbert, matter/flux interactions, and higher--curvature corrections,
together with the required boundary completion (GHY and counterterms) for the chosen
boundary conditions; $S_{\rm gf}$ imposes a gauge condition (often in BRST form using a Nakanishi--Lautrup
field $b$); and $S_{\rm gh}$ is the associated ghost action implementing the Faddeev--Popov determinant.
This is the correct covariant definition of the vacuum wavefunctional.

Time evolution and energy are encoded in boundary terms such as the ADM or
Brown--York Hamiltonian.  In the absence of boundaries the notion of time evolution
is replaced by the Wheeler--DeWitt constraints, and physical states are defined by $\Psi$ annihilated by $\mathcal H_{\perp}$ and $\mathcal H_i$ as shown in footnote \ref{lucmaxie}.
Within this framework a Glauber--Sudarshan coherent state is naturally defined as \cite{joydeep, wdwpaper, coherbeta, coherbeta2, borel2, hetborel, dileep, desitter2}:
\begin{equation}
|\sigma\rangle = D(\sigma) |\Omega\rangle ,
\end{equation}
where $|\Omega\rangle$ is the exact gravitational vacuum determined by the full
interacting action, and $D(\sigma)$ generates coherent excitations above this
nonperturbative background. We will have more to say on the specific form for $D(\sigma)$ soon as it will decide whether we have a conventional coherent state or a Glauber-Sudarshan state.

Now consider a semiclassical regime in which the metric is expanded around a classical
background solution admitting a preferred time function,
\begin{equation}
g_{\mu\nu}
=
g^{(0)}_{\mu\nu}
+
\kappa \, h_{\mu\nu} .
\end{equation}
In this situation the Hamiltonian constraints may be solved perturbatively and the
boundary Hamiltonian reduces to an effective Hamiltonian for physical fluctuations.
After gauge fixing and constraint reduction one obtains a Schr\"odinger equation:

{\footnotesize
\begin{equation}\label{scroood}
i \frac{\partial}{\partial t} \psi[h,\Phi;t]
=
H_{\rm eff}[h,\Phi] \, \psi[h,\Phi;t], ~~~{\rm where}~~ 
\Psi[h_{ij},\Phi; t]
=
\exp\!\left(\frac{i}{\kappa^2} S_0[h_{ij}] \right)\;
\psi[h_{ij},\Phi; t],
\end{equation}}
where we used a WKB (Born--Oppenheimer) ansatz by
separating slow ``heavy'' gravitational degrees of freedom from
``light'' fluctuations. Here $S_0$ encodes the background classical geometry (Hamilton--Jacobi functional), and 
$\psi$ encodes quantum fluctuations of matter and gravitons on that background. (How \eqref{scroood} appears from Wheeler-DeWitt equation, given in say footnote \ref{lucmaxie}, will be elaborated in more details in section \ref{sec8.1}.)
The effective Hamiltonian admits an expansion:
\begin{equation}
H_{\rm eff}
=
H_0
+
H_{\rm int}^{(3)}
+
H_{\rm int}^{(4)}
+
H_{\rm h.c.}
+ \cdots = H_0 + H_{\rm int} ,
\end{equation}
where $H_0$ describes free gravitons and matter on the chosen background,
$H_{\rm int}^{(3)}$ and $H_{\rm int}^{(4)}$ contain the cubic and quartic
nonlinear interactions, and $H_{\rm h.c.}$ denotes contributions induced by
higher-curvature operators. In this semiclassical regime, the interacting
vacuum may be defined in close analogy with ordinary quantum field theory:

{\footnotesize
\begin{equation}
|\Omega\rangle_{g^{(0)}}
=
\lim_{T\rightarrow\infty(1-i\epsilon)}
\frac{
e^{-iH_{\rm eff}T}|0\rangle_{g^{(0)}}
}{
\left\|
e^{-iH_{\rm eff}T}|0\rangle_{g^{(0)}}
\right\|
} =
\lim_{T\rightarrow\infty(1-i\epsilon)}
\frac{
{\cal T}
\exp\left[
-i\int_{-T}^{0}dt\,
H_{\rm int}^{I}(t)
\right]
|0\rangle_{g^{(0)}}
}{
{}_{g^{(0)}}\langle0|
{\cal T}
\exp\left[
-i\int_{-T}^{0}dt\,
H_{\rm int}^{I}(t)
\right]
|0\rangle_{g^{(0)}}
}.
\end{equation}}
where the second equality is in the interaction picture where 
$H_{\rm int}^{I}(t)
=
e^{iH_0t}
H_{\rm int}
e^{-iH_0t}$ as before. The full quantum state therefore incorporates graviton self--interactions,
matter loops and higher--curvature corrections already at the level of the
vacuum.  A Glauber--Sudarshan coherent state may then be defined as:
\begin{equation}
|\sigma\rangle = D(\sigma) \, |\Omega\rangle_{g^{(0)}} ,
\end{equation}
which is exactly what we used in \cite{coherbeta, coherbeta2}. In this language, semiclassical geometries arise as coherent excitations above the exact
interacting vacuum associated with the background geometry $g^{(0)}_{\mu\nu}$.

We can make this correspondence more precise by taking the following steps.
Assume a background geometry $g^{(0)}_{\mu\nu}$ admitting a global time function and
a foliation by spacelike hypersurfaces $\Sigma_t$, so that the spacetime region may
be taken as a time slab
\begin{equation}
M_T = [-T,T]\times\Sigma .
\end{equation}
We assume gauge fixing has been imposed and ghosts included, the semiclassical
expansion is valid, and an $i\epsilon$ prescription (or Euclidean cap) is used to
project onto the vacuum. The covariant path integral on the slab with final boundary data is:
\begin{equation}
\Psi_T[h_f,\Phi_f]
=
\int_{\substack{h(+T)=h_f \\ \Phi(+T)=\Phi_f}}
\mathcal D g_{\mu\nu}\,\mathcal D\Phi\;
e^{\, i S[g,\Phi]},
\end{equation}
where $S[g, \Phi]$ contains the ghosts and the gauge-fixing terms. (Alternatively we could write this in the full-blown way as in 
\eqref{eq:PI_with_ghosts} by explicitly introducing the other DOFs.)
With the usual $i\epsilon$ prescription this projects onto the vacuum as
$T\to\infty(1-i\epsilon)$:
\begin{equation}
\Psi_\Omega[h_f,\Phi_f]
=
\lim_{T\to\infty(1-i\epsilon)}
\langle h_f,\Phi_f; +T \,|\, \Omega\rangle .
\end{equation}
From here it is easy to see how \eqref{sidsween1} and \eqref{eq:PI_with_ghosts} could be related to each other in the presence of a background $g_{\mu\nu}^{(0)}$. Generically however \eqref{eq:PI_with_ghosts} continues to be the correct choice if we want a background independent description, but the story takes a more conventional turn once we expand over the Minkowski background. Thus the logical chain is:
\bg\label{sidsween2}
&&\text{\textcolor{blue}{covariant path integral}}
\Longleftrightarrow
\text{\textcolor{blue}{boundary Hamiltonian evolution}}\nonumber\\
 \Longrightarrow &&
\textcolor{blue}{H_{\rm eff}}\ \text{\textcolor{blue}{for fluctuations}}
\Longrightarrow
\textcolor{blue}{T\to\infty(1-i\epsilon)}\ \text{\textcolor{blue}{projection}}
\Longrightarrow
\textcolor{blue}{|\Omega\rangle_{g_{\mu\nu}^{(0)}}} 
\nd
where $g_{\mu\nu}^{(0)}$ is the background and the connection in \eqref{sidsween2} clarifies the canonical and the path-integral representations presented in \cite{coherbeta, coherbeta2}. 

\begin{figure}[h]
\centering
\begin{tabular}{c}
\includegraphics[width=6in]{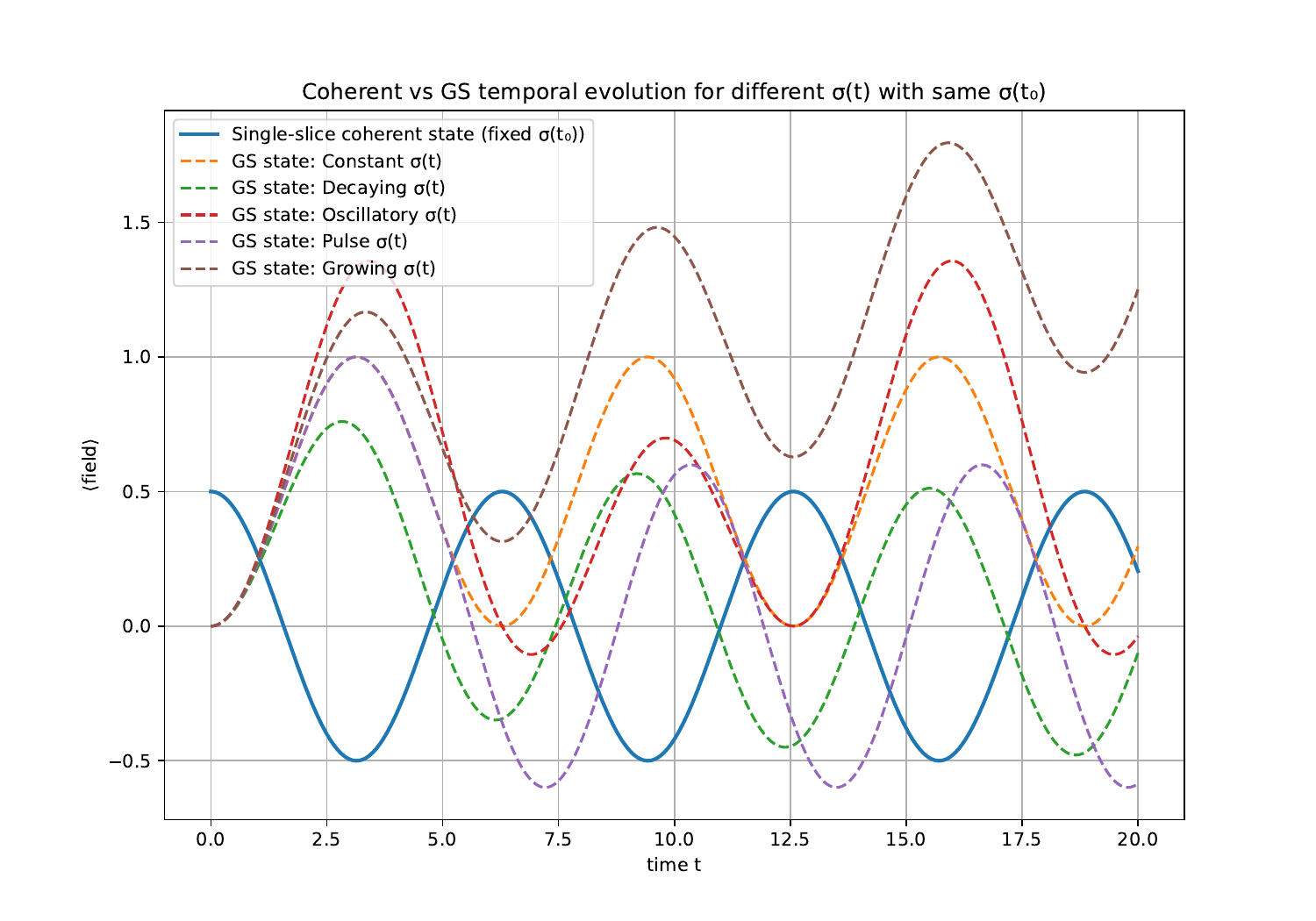}
\end{tabular}
\caption[]{The difference between the temporal evolutions between the coherent states and the Glauber-Sudarshan states.}
\label{gsvscoher}
\end{figure}


\subsection{Coherent states vs Glauber-Sudarshan states under  Hamiltonian actions \label{sec3.2}}

The key distinction between a conventional coherent state \cite{dvali} and a Glauber-Sudarshan state \cite{joydeep, wdwpaper, coherbeta, coherbeta2, borel2, hetborel, dileep, desitter2}
comes from the form of the so-called displacement operator $D(\sigma)$. They may be represented in the following way:

{\footnotesize
\bg\label{sweenseyf}
D(\sigma) = \begin{cases}~~~\exp\!\left[i
\int_{\Sigma_{t_0}} d^3x\;\sqrt{h(t_0,\mathbf x) }~
\sigma^{ij}(t_0,\mathbf x)\,
\hat \pi_{ij}(t_0,\mathbf x)
\right]~~~~~~~~~~ {\rm Coherent~State}\\
~~~ \\
~~\mathcal T\exp\!\left[
\int dt
\int_{\Sigma_t} d^3x\;\sqrt{-g(t,\mathbf x)}~
\sigma^{\mu\nu}(t,\mathbf x)\,
\hat g_{\mu\nu}(t,\mathbf x)
\right] ~~~~~~ {\rm Glauber-Sudarshan~State}  \end{cases}
\nd}
where $\Sigma_{t_n}$ is the Cauchy slice at time $t_n$, $h_{ij}$ is the induced metric on $\Sigma_{t_0}$, $\pi^{ij}$ its conjugate
momentum, $\mathcal T$ is the time-ordering symbol\footnote{We have used $\mathcal T$ instead of $T$ to denote the time ordering because $T$ will be used to denote the temporal coordinate later. We will use hats to denote the operators when the distinction is pertinent to the problem at hand, otherwise both operators and functions will be denoted without hats. Which is which should be clear from the context.}, and $\sigma^{ij}$ a prescribed classical profile (indices raised/lowered
with $h_{ij}$). Similarly $\sigma^{\mu\nu}$ a prescribed classical profile with indices raised or lowered
by $g_{\mu\nu}$. From \eqref{sweenseyf} we can see the key structural difference. A GS insertion is a covariant
\emph{bulk source deformation}, so quantum corrections are naturally organized by the
1PI/2PI effective action. A canonical coherent state (CS) is a \emph{boundary/state choice}
in phase space; it enters as a boundary wavefunctional (or equivalently a boundary term).
Quantum corrections are then incorporated through the same effective-action machinery,
but now with \emph{boundary conditions} rather than a bulk source.
Thus a coherent state is a state created at time $t = t_0$ and then it evolves via the boundary Hamiltonian (as we will elaborate soon), whereas a Glauber-Sudarshan (GS) state is state that is been continuously pushed. In fact
we can consider a GS state to be prepared by repeated action of infinitesimal slice operators
on a reference vacuum $|\Omega\rangle$. The purpose of this subsection is to make
precise how higher-order interactions in the Hamiltonian influence the evolution
of such a state.

To start let us consider the slice operators and their continuum limit. Let $\Sigma_t$ be a Cauchy slice at time $t$, and let $\hat g_{\mu\nu}(t,\mathbf x)$
denote the metric operator in a fixed time gauge adapted to a chosen Minkowski
vacuum with ADM decomposition $(N, N^i, h_{ij})$. Note however the presence of $\sqrt{h}$ and $\sqrt{-g}$ in the definition \eqref{sweenseyf}. This is a subtle point that needs some clarification. In canonical gravity the equal-time commutator is:
\begin{equation}\label{enmill2}
\big[
\hat g_{ij}(\mathbf x),\,
\hat \pi^{kl}(\mathbf y)
\big]
=
i\,\delta_{(i}^{k}\delta_{j)}^{l}\,
\delta^{(3)}(\mathbf x-\mathbf y).
\end{equation}
which is motivated from a similar commutator in non-gravitational QFT, and $(i, j) \in \mathbb{R}^3$. With this premise in mind, let us consider a displacement operator for the coherent state in the 
following way over an initial Cauchy slice:
\begin{equation}\label{enmill1}
D(\sigma)
=
\exp\!\left[
i\int_{\Sigma_{t_0}} d^3x\;
\sigma^{ij}(\mathbf x)\,
\hat \pi_{ij}(\mathbf x)
\right],
\end{equation}
which is similar to \eqref{sweenseyf}, but {\it without} $\sqrt{h}$. The reason for the choice \eqref{enmill1} instead of \eqref{sweenseyf} is because the displacement operator \eqref{enmill1} shifts the metric by a classical tensor $\sigma_{ij}$ in the following way:
\begin{equation}\label{twogonec}
D^\dagger(\sigma)\,
\hat g_{ij}(\mathbf x)\,
D(\sigma)
=
\hat g_{ij}(\mathbf x)
+
\sigma_{ij}(\mathbf x) ,
\end{equation}
where we used the commutation relation \eqref{enmill2}. Question is whether we need $\sqrt{h}$ in the definition of the displacement operator for the coherent state. It turns out that the appearance
of the $\sqrt{h}$ factor in the exponent depends on density
conventions:
\begin{itemize}
\item If $\pi^{ij}$ is a density of weight $+1$, no $\sqrt{h}$ is needed.
\item If $\pi_{ij}$ is treated as a weight-$0$ tensor, then
$\sqrt{h}\,\pi_{ij}$ restores scalar density weight.
\end{itemize}
In the latter case we expect the commutation relation \eqref{enmill2} to change. For example we can define a ``undensitize" momentum operator as $\hat p^{ij}(\mathbf x)\;\equiv\;\frac{1}{\sqrt{\hat h(\mathbf x)}}\,\hat \pi^{ij}(\mathbf x)$ on the initial Cauchy slice. 
For such a case the commutator brackets take the following forms:
\bg\label{enmill3}
&& \big[\hat g_{mn}(\mathbf x),\,\hat p^{ij}(\mathbf y)\big]
=
i\,\delta_{(m}^{i}\delta_{n)}^{j}\,
\frac{1}{\sqrt{\hat h(\mathbf y)}}\,
\delta^{(3)}(\mathbf x-\mathbf y) \nonumber\\
&&  \big[\hat g_{mn}(\mathbf x),\,\hat p_{ij}(\mathbf y)\big]
=
i\,\frac{1}{\sqrt{\hat h(\mathbf y)}}\,
\hat g_{i(m}(\mathbf y)\hat g_{n)j}(\mathbf y)\,
\delta^{(3)}(\mathbf x-\mathbf y), \nd
implying that the commutator becomes \emph{operator-valued} through the factor $1/\sqrt{\hat h}$ and, in addition, explicitly nonlinear in $\hat{g}$ for the second case with all lowered indices. The presence of the inverse metric factor still needs to be addressed. Clearly $\hat{h}({\bf y})$ is a composite operator and if we expand the background over a Minkowski minimum, then $\hat{h}_{ij}({\bf x}) = \delta_{ij} + \delta \hat{h}_{ij}({\bf x})$. In this language the inverse can be Taylor expanded, keeping track of the ordering of the 
$\delta\hat{h}_{ij}({\bf x})$ operators. These subtleties become irrelevant if $\pi^{ij}$ continues to be a density of weight $+1$.
Thus the appearance of $\sqrt{h}$ is not fundamental but convention-dependent. On the other hand, for the GS states we express it explicitly using $\sqrt{-g}$. Again over a Minkowski minimum $-$ which in fact is necessary for the system to have a well-defined boundary Hamiltonian $-$ we can Taylor expand this to write using powers of composite operators with due considerations to their orderings. More appropriately, the quantity
$\sqrt{-\hat g(x)}$
should be understood as \emph{renormalized composite operator densities} in the following sense:
\begin{equation}
\hat{G}_{\mu\nu}(x) = \sqrt{-\hat g(x)}\,\hat g_{\mu\nu}(x)
\equiv
\big(\sqrt{-g(x)}\,g_{\mu\nu}(x)\big)_{\rm ren},
\end{equation}
defined by the same renormalization prescription used for local
composite operators in quantum field theory (e.g.\ point-splitting,
dimensional regularization, or background-field expansion).
In the path-integral formulation, these quantities are ordinary
functions of the integration variable $g_{\mu\nu}(x)$.
Only when translated back into operator language do they require
a renormalized composite-operator definition. Thus to summarize:
The coherent-state operator is canonically well-defined
(up to density conventions), whereas the GS operator involves renormalized composite operators
such as $\sqrt{-\hat g}\,\hat g_{\mu\nu}$.
(Henceforth, to avoid clutter, we will absorb $\sqrt{h({\bf x}, t_0)}$ and $\sqrt{-g({\bf x}, t)}$ in the definition of $h_{ij}({\bf x}, t_0)$ and $g_{\mu\nu}({\bf x}, t)$ respectively.)  Discretize time as $t_n = t + n\Delta t$ and define:
\begin{equation}
D_{t_n}(\sigma)
=
\exp\!\left[
\Delta t
\int_{\Sigma_{t_n}} d^3x\;
\sigma^{\mu\nu}(t_n,\mathbf x)\,
\hat G_{\mu\nu}(t_n,\mathbf x)
\right].
\label{eq:Dt_sigma_again}
\end{equation}
In fact we can fix the initial time as $t = t_0$ and then define $\Delta t$ as $\Delta t = {t_N - t_0\over N}$ where $0 \le n \le N$ and $n \in \mathbb{Z}_+$. (This is independent of $n$ of course.)
Acting on the vacuum, the Glauber--Sudarshan state may be represented as
the chronologically ordered product:
\begin{equation}
|\Psi\rangle
\equiv
|\sigma\rangle
=
\overleftarrow{\prod_{n=0}^{N-1}}
D_{t_n}(\sigma)\,
|\Omega\rangle
=
D_{t_{N-1}}(\sigma)\cdots
D_{t_1}(\sigma)D_{t_0}(\sigma)
|\Omega\rangle ,
\label{eq:state_product}
\end{equation}
where the earliest-time operator $D_{t_0}(\sigma)$ acts first, while the
latest-time operator $D_{t_{N-1}}(\sigma)$ acts last. This
makes the distinction between the GS and the coherent states more transparent. Since the operators $\hat G_{\mu\nu}(t_n,\mathbf x)$ at different times do not
commute, the continuum limit $\Delta t\to 0$ yields a time-ordered exponential:
\begin{equation}
|\Psi\rangle
=
\mathcal T\exp\!\left[
\int dt
\int d^3x\;
\sigma^{\mu\nu}(t,\mathbf x)\,
\hat G_{\mu\nu}(t,\mathbf x)
\right]
|\Omega\rangle ,
\label{eq:state_time_ordered}
\end{equation}
which is exactly the form it appeared in \cite{wdwpaper}. The subtlety now is the temporal evolution of the GS state since we are already integrating over $t$. More precisely the question is as to where does the Hamiltonian show up here. To see this
consider the physical Hamiltonian ({\it i.e.} the boundary Hamiltonian after constraints are solved) be decomposed as:
\begin{equation}
H_{\rm phys}
=
H_0
+
H_{\rm int},
\label{eq:H_split}
\end{equation}
where $H_0$ is the quadratic (free) part and $H_{\rm int}$ contains all
higher-order interactions, including self-interactions, higher-curvature terms,
and stringy corrections. Then the metric operator appearing in \eqref{eq:state_time_ordered} is the Heisenberg
operator evolved with the \emph{full} Hamiltonian\footnote{For any operator $\hat O$ in the physical Hilbert space, the Heisenberg evolution reads
$\hat O_H(t)=e^{\,iH_{\rm phys}(t-t_0)}\,\hat O_H(t_0)\,e^{-\,iH_{\rm phys}(t-t_0)}$.
In particular, once the composite operator is renormalized as
$\hat G_{\mu\nu}(x)
\;\equiv\;
\big(\sqrt{-\hat g(x)}\,\hat g_{\mu\nu}(x)\big)_{\rm ren}$,
it obeys the same evolution law: 
$\hat G_{\mu\nu}(t,\mathbf x)
=
e^{\,iH_{\rm phys}(t-t_0)}\,\hat G_{\mu\nu}(t_0,\mathbf x)\,e^{-\,iH_{\rm phys}(t-t_0)}$.}:
\begin{equation}
\hat G_{\mu\nu}(t,\mathbf x)
=
e^{\,iH_{\rm phys} (t - t_0)}\,
\hat G_{\mu\nu}(t_0,\mathbf x)\,
e^{-\,iH_{\rm phys} (t-t_0)} ,
\label{eq:g_heisenberg}
\end{equation}
where we have defined out metric operator on the initial Cauchy slice at $t= t_0$.
Thus, all interactions enter through the operator time dependence itself, implying that
there is no sense in which the slice operators $D_t(\sigma)$ evolve independently
of $H_{\rm int}$. Moreover, because of interactions, the metric operators at different times fail to commute:
\begin{equation}\label{peribindu1}
\big[
\hat G_{\mu\nu}(t_1,\mathbf x),
\hat G_{\rho\sigma}(t_2,\mathbf y)
\big]
\neq 0,
\end{equation}
with the commutator determined by the full interacting dynamics.
This non-commutativity is precisely why the product
\eqref{eq:state_product} exponentiates to the time-ordered form
\eqref{eq:state_time_ordered}. The Baker--Campbell--Hausdorff corrections arising
from successive slice insertions encode the effects of 
interactions\footnote{With the 
the renormalized composite operator
$\hat G_{\mu\nu}(x)
\;\equiv\;
\big(\sqrt{-\hat g(x)}\,\hat g_{\mu\nu}(x)\big)_{\rm ren}$ implying that 
$\hat G_{\mu\nu}$ is a nonlinear functional of $\hat g_{\alpha\beta}$,
and since the fields do not commute at unequal times, it follows generically that
\eqref{peribindu1} continues to hold for $t_1 \ne t_2$. Now 
consider a discretized sequence of insertions at times $t_n$ with
$\Delta t = (t_f-t_0)/N$ as in \eqref{eq:Dt_sigma_again} then \eqref{eq:state_time_ordered} is completely general for any operator-valued distribution
$\hat O(t)$, because of the following limiting condition:
\begin{equation}
\overleftarrow{\prod_{n=0}^{N-1}} e^{\epsilon \hat O(t_n)}
\;\longrightarrow\;
\mathcal T\exp\!\left(\int dt\,\hat O(t)\right),
\end{equation}
provided the continuum limit is taken carefully with $\epsilon \equiv \Delta t \to 0$ and $N \to \infty$. For the fundamental metric operator $\hat g_{\mu\nu}$,
non-commutativity arises from the canonical algebra and interactions.
For the composite operator $\hat G_{\mu\nu}$ we note the following three points: (1) 
The commutator \eqref{peribindu1} generally contains nonlinear
      interaction-dependent terms; (2) additional local contact terms may appear due to short-distance
      singularities of composite operators; and (3) renormalization can induce operator mixing among operators
      with the same tensor structure.
Nevertheless, none of these effects alters the structural statement
\eqref{eq:state_time_ordered}: the ordered product exponentiates to a
time-ordered exponential, with the Baker--Campbell--Hausdorff corrections
encoding the full interacting dynamics and composite-operator effects.
In other words replacing $\hat g_{\mu\nu}$ by the composite
$\hat G_{\mu\nu}$
does not change the time-ordered exponentiation structure.}.

It is often convenient to rewrite the construction in the interaction picture.
Defining
$\hat G_{\mu\nu}^{I}(t,\mathbf x)
=
e^{\,iH_0 t}\,
\hat G_{\mu\nu}(0,\mathbf x)\,
e^{-\,iH_0 t}$
and the interaction-picture evolution operator as
$U_{\rm int}(t)
=
\mathcal T\exp\!\left(
-\,i\int^t dt'\,H_{\rm int}^{I}(t')
\right)$, the Heisenberg operator is then represented in the standard way as
$\hat G_{\mu\nu}(t,\mathbf x)
=
U_{\rm int}^\dagger(t)\,
\hat G_{\mu\nu}^{I}(t,\mathbf x)\,
U_{\rm int}(t)$. Substituting this into
\eqref{eq:state_time_ordered}, the prepared state may be written as:
\begin{equation}
|\Psi\rangle
=
\mathcal T\exp\!\left[
\int dt
\int d^3x\;
\sigma^{\mu\nu}(t,\mathbf x)\,
U_{\rm int}^\dagger(t)\,
\hat G_{\mu\nu}^{I}(t,\mathbf x)\,
U_{\rm int}(t)
\right]
|\Omega\rangle.
\label{eq:state_interaction_picture}
\end{equation}
This expression makes it explicit that the source prepares a coherent-type
displacement in the free variables, which is then dressed by interactions
through $U_{\rm int}(t)$. The same construction is represented in real time by a Schwinger--Keldysh
path integral in which the operator $D(\sigma)$ appears as a multiplicative
insertion rather than as a deformation of the Lorentzian action.
Explicitly, for a single branch one may write schematically:
\begin{equation}
\int \mathcal D g\,\mathcal D\phi\; \mathcal D({\rm gh})\:
\exp\!\Big[
iS_{\rm EH}[g]
+
iS_{\rm int}[g,\phi, {\rm gh}]
\Big]\,
\exp\!\Big[
\int d^4x\;\sqrt{-g(x)}~
\sigma^{\mu\nu}(x)\,g_{\mu\nu}(x)
\Big],
\label{eq:path_integral_sigma_correct}
\end{equation}
where $S_{\rm int}$ contains all higher-order interaction terms including ghosts and we take $\sigma^{\mu\nu}$ to be a real function.
The second exponential originates directly from the operator definition
of $D(\sigma)$ and therefore does \emph{not} carry an additional factor of $i$.
It multiplies the usual Lorentzian phase $e^{iS_{\rm tot}}$
rather than modifying the action itself. (The precise form for $S_{\rm tot}$ will be described in section \ref{sec4.3}.) Accordingly, the source $\sigma^{\mu\nu}$ biases the saddle point of the
\emph{fully interacting} theory through the insertion of the composite
operator $\sqrt{-g}\,g_{\mu\nu}$.
It does not bypass or replace the interaction dynamics.
The state \eqref{eq:state_time_ordered} should therefore be interpreted as the vacuum
evolved under the full interacting Hamiltonian while being continuously deformed
by this operator insertion.
All interaction effects are already encoded in the Heisenberg evolution
of $\hat g_{\mu\nu}(t,\mathbf x)$ and in the non-commutativity of operators at
different times.
No additional interaction terms need to be inserted into the slice operators
$D_t(\sigma)$; their effects are automatically incorporated through
the full interacting Hamiltonian.


Thus from the above discussion it should be clear 
why the infinite product of slice operators
$D_{t_n}(\sigma)$ from \eqref{eq:Dt_sigma_again} does not require additional factors of
$e^{-iH_{\rm phys}\Delta t}$ between successive slices, as inserting
such factors would amount to double counting the time evolution already present in the Heisenberg operators. To quantify this, let us first propose the following approximate identity  
using the Heisenberg definition \eqref{eq:g_heisenberg}:
\begin{equation}
D_{t_{n+1}}(\sigma)
\approx
e^{\,iH_{\rm phys}\Delta t}\,
D_{t_n}(\sigma)\,
e^{-\,iH_{\rm phys}\Delta t} ,
\label{eq:Dt_time_shift}
\end{equation}
implying that the time dependence of successive slice operators is already encoded
through conjugation by the physical Hamiltonian. The approximate nature of the relation in \eqref{eq:Dt_time_shift} is crucial because of a small subtlety that one should carefully consider. For any operator $\hat X(t)$ in the Heisenberg picture, it is known that 
$\hat X(t+\Delta t)
=
e^{\,iH_{\rm phys}\Delta t}\,
\hat X(t)\,
e^{-\,iH_{\rm phys}\Delta t}$.
Then its exponential satisfies
$e^{\hat X(t+\Delta t)}
=
e^{\,iH_{\rm phys}\Delta t}\,
e^{\hat X(t)}\,
e^{-\,iH_{\rm phys}\Delta t}$. What this really tells us is that there is a comparison between two sets of operators that are not necessarily equivalent:
\begin{equation}
D_{t_{n+1}}(\sigma)
=
\exp\!\left[
\Delta t
\int d^3x\;
\sigma^{\mu\nu}(t_{n+1},\mathbf x)\,
\hat G_{\mu\nu}(t_{n+1},\mathbf x)
\right] \nonumber
\label{eq:Dtnplus1_again}
\end{equation}
\begin{equation}\label{sheeshsami}
~~~~~~~~~~ e^{\,iH_{\rm phys}\Delta t}\,
D_{t_n}(\sigma)\,
e^{-\,iH_{\rm phys}\Delta t}
=
\exp\!\left[
\Delta t
\int d^3x\;
\sigma^{\mu\nu}(t_n,\mathbf x)\,
\hat G_{\mu\nu}(t_{n+1},\mathbf x)
\right].
\end{equation}
Therefore, if one wants to identify these two expressions \emph{without any extra correction term}, one must match the c-number source on the two adjacent slices, {\it i.e.}
$\sigma^{\mu\nu}(t_{n+1},\mathbf x)
=
\sigma^{\mu\nu}(t_n,\mathbf x)$. But for a genuinely continuous time-dependent source, one instead writes:
\begin{equation}
\sigma^{\mu\nu}(t_{n+1},\mathbf x)
=
\sigma^{\mu\nu}(t_n,\mathbf x)
+
\Delta t\,
\partial_t\sigma^{\mu\nu}(t_n,\mathbf x)
+
{\cal O}(\Delta t^2),
\label{eq:sigma_taylor_continuous}
\end{equation}
implying that a relation like \eqref{eq:Dt_time_shift} can only be understood as an approximate stepwise relation. For a continuous source, the simple discrete equality is replaced by the leading Heisenberg time-shift relation together with explicit $\partial_t\sigma^{\mu\nu}$ corrections. Once this is taken into consideration, the identity 
\eqref{eq:Dt_time_shift} explains
why inserting {\it additional} $e^{-iH_{\rm phys}\Delta t}$ factors between the Heisenberg slice operators would be incorrect. If one were to insert explicit evolution operators between slices,
for example:
\begin{equation}
\cdots
e^{-iH_{\rm phys}\Delta t}\,
D_{t_n}(\sigma)\,
e^{-iH_{\rm phys}\Delta t}\,
D_{t_{n+1}}(\sigma)\,
\cdots,
\end{equation}
this would double count time evolution: once through the explicit
$e^{-iH_{\rm phys}\Delta t}$ factors, and once through the Heisenberg time
dependence of $\hat G_{\mu\nu}(t_n,\mathbf x)$. Such a construction is therefore
incorrect. Thus the most precise statement is the following:
\begin{equation}
|\Psi(t)\rangle
=
e^{-\,i H_{\rm phys}(t-t_f)}\;
\mathcal T\exp\!\left[
\int_{t_0}^{t_f} dt'\!
\int d^3x\;
\sigma^{\mu\nu}(t',\mathbf x)\,
\hat G_{\mu\nu}(t',\mathbf x)
\right]
|\Omega\rangle ,
\end{equation}
captures the full temporal evolution because 
$D(\sigma)\equiv
\mathcal T\exp\!\left[\int_{t_0}^{t_f} dt'\!\int d^3x\,
\sigma^{\mu\nu}(t',\mathbf x)\hat G_{\mu\nu}(t',\mathbf x)\right]$
prepares the state during the interval $[t_0,t_f]$, while
$H_{\rm phys}$ generates the subsequent time evolution for $t>t_f$.
There is therefore no double counting: the source operator prepares the state
while it is active, and the physical Hamiltonian governs the evolution after
the source is turned off.

There is however a way to make a relation like \eqref{eq:Dt_time_shift} more precise. For this let us first look into another useful way to interpret the time evolution by defining the \emph{kick} operator to be built instead from fields at the reference time
$t_0$ as:
\begin{equation}
\widetilde D_{t_n}(\sigma)
\equiv
\exp\!\left[
\Delta t
\int_{\Sigma_{t_0}} d^3x\;
\sigma^{\mu\nu}(t_n,\mathbf x)\,
\hat G_{\mu\nu}(t_0,\mathbf x)
\right] ,
\label{eq:Dtilde_def}
\end{equation}
keeping the Cauchy slice $\Sigma_{t_0}$ as the reference slice. This is the Schr\"odinger-picture, or reference-time, description.
Let $H_{\rm phys}$ be the physical Hamiltonian, namely the boundary Hamiltonian after the constraints are solved. Define the unitary evolution operator:
\begin{equation}
U(t',t)
\equiv
e^{-iH_{\rm phys}(t'-t)} ,
\qquad
U(t,t)=\mathbf 1.
\label{eq:U_def}
\end{equation}
Assume for simplicity that $H_{\rm phys}$ is time independent; the time-dependent
generalization is obtained by replacing \eqref{eq:U_def} with a time-ordered
exponential. Then, using the standard Heisenberg operator \eqref{eq:g_heisenberg}, one has the exact conjugation identity:
\begin{equation}
D_{t_n}(\sigma)
=
U^\dagger(t_n,t_0)\,
\widetilde D_{t_n}(\sigma)\,
U(t_n,t_0) ,
\label{eq:Dt_conjugation}
\end{equation}
with $D_{t_n}$ as in \eqref{eq:Dt_sigma_again}. We can also use \eqref{eq:Dt_conjugation} to write the ordered product $\overleftarrow{\prod\limits_{n=0}^{N-1}} D_{t_n}(\sigma)$ in an alternative way. By inserting
$\mathbf 1 = U(t_{n+1},t_n)U^\dagger(t_{n+1},t_n)$ appropriately, a convenient
and standard form is:
\begin{equation}
\overleftarrow{\prod_{n=0}^{N-1}} D_{t_n}(\sigma)
=
U^\dagger(t_N,t_0)\;
\overleftarrow{\prod_{n=0}^{N-1}}
\Big(
U(t_{n+1},t_n)\,
\widetilde D_{t_n}(\sigma)
\Big),
\label{eq:product_with_intermediate_U}
\end{equation}
where $t_N=t_0+N\Delta t$ and the product is ordered with increasing $n$
from right to left. Equation \eqref{eq:product_with_intermediate_U} makes the intermediate
$e^{-iH_{\rm phys}\Delta t}$ factors explicit via
$U(t_{n+1},t_n)=e^{-iH_{\rm phys}\Delta t}$.
In the limit $\Delta t\to 0$, $N\Delta t\to t_f-t_0$, one obtains the
equivalent continuum statement\footnote{This is not hard to show. Define
$X(t)
\equiv
\mathcal T\exp\!\left[
\int_{t_0}^{t}dt'\;
\widetilde V_H(t')
\right]$.
Then
$\frac{dX}{dt}
=
\widetilde V_H(t)\,X(t)$ with 
$X(t_0)=1$.
Now define
$Y(t)
\equiv
U^\dagger(t,t_0)\,
\mathcal T\exp\!\left[
\int_{t_0}^{t}dt'\;
\big(
\widetilde V(t')-iH_{\rm phys}(t')
\big)
\right]$.
Now
$\frac{d}{dt}U^\dagger(t,t_0)
=
i\,U^\dagger(t,t_0)\,H_{\rm phys}(t)$,
and
\begin{equation}
\frac{d}{dt}
\mathcal T\exp\!\left[
\int_{t_0}^{t}dt'\;
\big(
\widetilde V(t')-iH_{\rm phys}(t')
\big)
\right]
=
\big(
\widetilde V(t)-iH_{\rm phys}(t)
\big)
\mathcal T\exp\!\left[
\int_{t_0}^{t}dt'\;
\big(
\widetilde V(t')-iH_{\rm phys}(t')
\big)
\right],\nonumber
\label{eq:dTexp_dt}
\end{equation}
where we note that we have taken a generalized story where the physical Hamiltonian $-$ which is the boundary Hamiltonian for our case $-$ is kept time-dependent. Using the above set of results, one 
finds:
\bg
\frac{dY}{dt}
=
iU^\dagger H_{\rm phys}\,
\mathcal T\exp[\cdots]
+
U^\dagger
\big(
\widetilde V-iH_{\rm phys}
\big)
\mathcal T\exp[\cdots]
=
U^\dagger \widetilde V\,\mathcal T\exp[\cdots]
=
U^\dagger \widetilde V U\,
Y(t)
=
\widetilde V_H(t)\,Y(t) , \nonumber
\nd
where 
$Y(t_0)=1$. Therefore
$Y(t)=X(t)$,
which proves \eqref{eq:continuum_relation}. Note however that, in \eqref{eq:continuum_relation}, $H_{\rm phys}$ is understood as the Schr\"odinger-picture Hamiltonian $H_{\rm phys}(t)$ generating $U(t,t_0)$, and provided the same time ordering $\mathcal T$ is used on the right-hand side.}:
\begin{equation}
\mathcal T\exp\!\left[
\int_{t_0}^{t_f} dt
\int d^3x\;
\sigma^{\mu\nu}(t,\mathbf x)\,
\hat G_{\mu\nu}(t,\mathbf x)
\right]
=
U^\dagger(t_f,t_0)\;
\mathcal T\exp\!\left[
\int_{t_0}^{t_f} dt\;
\left(
\widetilde V(t)-iH_{\rm phys}(t)
\right)
\right],
\label{eq:continuum_relation}
\end{equation}
where we have taken a generalized picture where $H_{\rm phys} = H_{\rm phys}(t)$; and where
the appearance of intermediate factors $U(t_{n+1},t_n)=e^{-iH_{\rm phys}\Delta t}$
in \eqref{eq:product_with_intermediate_U} does not ``interfere'' with the
construction; it is simply the same operator written in a different representation. In the Heisenberg
form, the same evolution is already encoded in the time
dependence of $\hat G_{\mu\nu}(t_n,\mathbf x)$. Equation
\eqref{eq:continuum_relation} makes this evolution explicit by moving
the Heisenberg time dependence into the interleaved $U$ factors, or equivalently into the $-iH_{\rm phys}$ term in the reference-time ordered exponential, while keeping the ``kick''
operators $\widetilde{V}(t)$ built from fields at a fixed reference
time defined on a Cauchy slice $\Sigma_{t_0}$, where:
\begin{equation}
\widetilde V(t)
\equiv
\int_{\Sigma_{t_0}} d^3x\;
\sigma^{\mu\nu}(t,\mathbf x)\,
\hat G_{\mu\nu}(t_0,\mathbf x).
\end{equation}
At first sight, since $\hat G_{\mu\nu}(t_0,\mathbf x)$ is independent of $t$,
one might think that the time ordering in \eqref{eq:continuum_relation}
is trivial and that the exponential reduces to
$\exp\!\left[
\int_{t_0}^{t_f} dt\;\widetilde V(t)
\right]$.
This is incorrect.
The key point is that \eqref{eq:continuum_relation} is obtained by moving the Heisenberg time dependence into the reference-time representation. More precisely,
$\hat G_{\mu\nu}(t,\mathbf x)
=
U^\dagger(t,t_0)\,
\hat G_{\mu\nu}(t_0,\mathbf x)\,
U(t,t_0)$. Therefore the nontrivial ordering in the reference-time representation is governed not only by possible commutators among the $\widetilde V(t)$ themselves, but also by their commutators with $H_{\rm phys}$. In particular, even if
$\big[\widetilde V(t_1),\widetilde V(t_2)\big]$ vanishes in a given sector, the time-ordered exponential in \eqref{eq:continuum_relation} is generally nontrivial because
$\big[H_{\rm phys},\widetilde V(t)\big]\neq 0$.
Although the $\widetilde V(t)$ are built from operators at time $t_0$, their placement inside
\eqref{eq:continuum_relation} is such that they are interleaved with the physical evolution when mapped back to the Heisenberg representation.
More explicitly, the Dyson expansion of the left-hand side of
\eqref{eq:continuum_relation} reads:
\bg
\mathcal T\exp\!\left[
\int dt\;K(t)
\right]
=
\mathbf 1
+
\int dt_1\;K(t_1)
+
\int dt_1 dt_2\;\theta(t_1-t_2)\,K(t_1)K(t_2)
+\cdots,
\nd
where 
$K(t)
=
\int d^3x\;\sigma^{\mu\nu}(t,\mathbf x)\,
\hat G_{\mu\nu}(t,\mathbf x)$. Now using the fact that $\hat{G}_{\mu\nu}(t, {\bf x})$ is related to $\hat{G}_{\mu\nu}(t_0, {\bf x})$ via the conjugation with $U(t, t_0)$ and the composition property
$U(t_1,t_0)U^\dagger(t_2,t_0)=U(t_1,t_2)$,
we obtain
\begin{equation}
K(t_1)K(t_2)
=
U^\dagger(t_1,t_0)\,
\widetilde V(t_1)\,
U(t_1,t_2)\,
\widetilde V(t_2)\,
U(t_2,t_0).
\label{eq:K1K2}
\end{equation}
Hence the time ordering on the left-hand side of
\eqref{eq:continuum_relation} translates into a nontrivial ordering structure
for the $\widetilde V(t)$ operators together with the intervening evolution operators. Putting everything together, 
the reference-time expression is best understood as an interaction-picture evolution operator:
\begin{equation}
U^\dagger(t_f,t_0)\;
\mathcal T\exp\!\left[
\int_{t_0}^{t_f} dt\;
\left(
\widetilde V(t)-iH_{\rm phys}
\right)
\right]
=
\mathcal T\exp\!\left[
\int_{t_0}^{t_f} dt\;
U^\dagger(t,t_0)\,\widetilde V(t)\,U(t,t_0)
\right].
\end{equation}
Thus the time ordering remains essential; the integral does not collapse into a
simple exponential. We therefore conclude that
even though $\hat G_{\mu\nu}(t_0,\mathbf x)$ itself carries no explicit
time dependence, the time-ordered exponential in
\eqref{eq:continuum_relation} is nontrivial. The non-commutativity induced by
the physical Hamiltonian $H_{\rm phys}$ prevents the $dt$ integral from
reducing to a simple ordinary exponential. This result, combined with \eqref{eq:Dtilde_def} and \eqref{eq:Dt_conjugation}, now provides a way to modify the approximate identity in \eqref{eq:Dt_time_shift}. Following the conjugation process, it is easy to show that:
\bg\label{chrisapp}
D_{t_{n+1}}(\sigma)
& = &
U^\dagger(t_n,t_0)\,
U^\dagger_{n+1,n}\,
\widetilde D_{t_{n+1}}(\sigma)\,
U_{n+1,n}\,
U(t_n,t_0) \nonumber\\
& = & U^\dagger(t_n,t_0)\,
U^\dagger_{n+1,n}
\Bigg[
\widetilde D_{t_n}(\sigma)
+
\Delta t^2
\int d^3x\,
\dot\sigma^{\mu\nu}(t_n,\mathbf x)\,
\hat G_{\mu\nu}(t_0,\mathbf x)\,
\widetilde D_{t_n}(\sigma)
\Bigg]\nonumber\\
&\times & U_{n+1,n}\,
U(t_n,t_0)
+
O(\Delta t^3), \nd
where $U_{n+1,n}
\equiv
U(t_{n+1},t_n)$ and we used \eqref{eq:sigma_taylor_continuous} to write the terms inside the square bracket. We should also note that 
possible commutator corrections start at higher order in this
discretized expansion. To write \eqref{chrisapp} cleanly, let us define the short-time evolution operators
transported to the $t_n$ Heisenberg frame by:
\bg\label{marvix}
{\cal U}_{n+1,n}^{(H)}
\equiv
U^\dagger(t_n,t_0)\,
U_{n+1,n}\,
U(t_n,t_0), ~~
\left({\cal U}_{n+1,n}^{(H)}\right)^\dagger
=
U^\dagger(t_n,t_0)\,
U^\dagger_{n+1,n}\,
U(t_n,t_0) ,
\nd
which are not the standard evolution operators but may be viewed as combinations of the evolution operators. In fact the usefulness of these operators becomes clear once we plug in $\widetilde D_{t_n}(\sigma)
=
U(t_n,t_0)\,
D_{t_n}(\sigma)\,
U^\dagger(t_n,t_0)$ into the first term in \eqref{chrisapp}. This term then simply takes the following form:
\begin{equation}
\left({\cal U}_{n+1,n}^{(H)}\right)^\dagger
D_{t_n}(\sigma)~
{\cal U}_{n+1,n}^{(H)} ,
\label{first_term_H_frame}
\end{equation}
which is {\it almost} like \eqref{eq:Dt_time_shift}, except that the ``evolution operator" is ${\cal U}_{n+1,n}^{(H)}$. However we also have the higher-order terms from \eqref{chrisapp}. Could they also be brought in a form similar to \eqref{first_term_H_frame}? To see whether this is possible, let us define a Heisenberg-frame source-variation operator in the following way:
\begin{equation}
{\cal J}_n^{(H)}(t_n, t_0)
\equiv
U^\dagger(t_n,t_0)
\left[
\int d^3x\,
\dot\sigma^{\mu\nu}(t_n,\mathbf x)\,
\hat G_{\mu\nu}(t_0,\mathbf x)
\right]
U(t_n,t_0) ,
\label{Jn_H_def}
\end{equation}
which is written in a suggestive way, but is basically the representation-changing conjugation action. The usefulness of this representation is that the second term in \eqref{chrisapp} can be written as $\Delta t^2 {\cal S}_n^{(H)}$, where 
the source-variation term ${\cal S}_n^{(H)}$ can be written as:
\begin{equation}
\begin{aligned}
{\cal S}_n^{(H)}
&\equiv
U^\dagger(t_n,t_0)\,
U^\dagger_{n+1,n}
\left[
\int d^3x\,
\dot\sigma^{\mu\nu}(t_n,\mathbf x)\,
\hat G_{\mu\nu}(t_0,\mathbf x)\,
\widetilde D_{t_n}(\sigma)
\right]
U_{n+1,n}\,
U(t_n,t_0)
\\
&=
\left({\cal U}_{n+1,n}^{(H)}\right)^\dagger
\left[
{\cal J}_n^{(H)}(t_n, t_0)
D_{t_n}(\sigma)
\right]
{\cal U}_{n+1,n}^{(H)} ,
\end{aligned}
\label{source_variation_H}
\end{equation}
showing again that the combination ${\cal J}_n^{(H)}(t_n, t_0)
D_{t_n}$ is evolved using ${\cal U}_{n+1,n}^{(H)}$. Plugging \eqref{first_term_H_frame} and \eqref{source_variation_H} into \eqref{chrisapp}, it is now easy to show that:
\begin{equation}
D_{t_{n+1}}(\sigma)
=
\left({\cal U}_{n+1,n}^{(H)}\right)^\dagger
\left[
D_{t_n}(\sigma)
+
\Delta t^2\,
{\cal J}_n^{(H)}(t_n, t_0)
D_{t_n}(\sigma)
\right]
{\cal U}_{n+1,n}^{(H)}
+
O(\Delta t^3),
\label{Dtnplus1_correct_relation_boxed}
\end{equation}
where additional commutator contributions enter at higher orders in the short-time expansion. This is the precise short-time expression, order by order in $\Delta t$, that replaces the naive identity \eqref{eq:Dt_time_shift}. However, it can be brought somewhat closer to \eqref{eq:Dt_time_shift} once we ignore the $\Delta t^2$ and higher-order terms and perform the following manipulations:

{\footnotesize
\bg\label{manipull}
&& {\cal U}_{n+1,n}^{(H)}
=
\exp\!\left[
-\,i\Delta t\,\mathcal H_{\rm phys}^{(H)}(t_n)
\right]
+
O(\Delta t^2), ~~~ \mathcal H_{\rm phys}^{(H)}(t_n)
\equiv
U^\dagger(t_n,t_0)\,
H_{\rm phys}(t_n)\,
U(t_n,t_0) \\
&& D_{t_{n+1}}(\sigma)
=
\left({\cal U}_{n+1,n}^{(H)}\right)^\dagger
D_{t_n}(\sigma)~
{\cal U}_{n+1,n}^{(H)}
+
O(\Delta t^2) = e^{\,i\mathcal H_{\rm phys}^{(H)}\Delta t} D_{t_n}(\sigma)~  
e^{-\,i\mathcal H_{\rm phys}^{(H)}\Delta t} +
O(\Delta t^2) . \nonumber \nd}
Here $\mathcal H_{\rm phys}^{(H)}(t_n)$ should be viewed as the physical Hamiltonian transported to the $t_n$ Heisenberg frame. If $H_{\rm phys}$ is time independent, this distinction is immaterial; however, in the more general time-dependent case the transported Hamiltonian is the more accurate object. Thus, to avoid these subtleties, one should regard \eqref{Dtnplus1_correct_relation_boxed} as the precise short-time relation governing the evolution of the displacement operators for the GS states. All of these then bring us to the following distinction between the standard coherent state \cite{dvali} and the Glauber-Sudarshan state \cite{joydeep, wdwpaper, coherbeta, coherbeta2, borel2, hetborel, dileep, desitter2}:

\vskip.2in

\begin{center} 
\begin{tabular}{|c|c|}
\hline
{\bf Standard coherent state} & {\bf Glauber-Sudarshan state} \\ \hline
Prepared at one time & Prepared over time interval \\ \hline
Free evolution after preparation & Continuous deformation + evolution \\ \hline
Classical solution from initial data & Solution with external driving/source \\ \hline
Single displacement operator & Time-ordered product of displacements \\ \hline
\end{tabular}
\end{center}

\vskip.1in

\noindent In other words, a standard coherent state corresponds to initial data on a Cauchy slice
evolved by the boundary Hamiltonian, whereas a GS state corresponds to a spacetime-supported operator that continuously
deforms the geometry.
Thus, a GS state is better viewed as a \emph{driven semiclassical background}
rather than a freely evolving coherent state.

Question then is how natural is the construction of a GS state? More generally, what supplies the drive? We will show in section \ref{sec7.0} that the driving condition, somewhat remarkably, follows a {\it bootstrap equation} \eqref{bootsforwalking} implying that the drive comes from {\it within} the system itself! But before delving into that let us clarify how the wavefunctions for the coherent and the GS states differ. After which we can discuss how the 
path integral structures look like once we use Lorentzian metric. This is where the Schwinger-Keldysh in-in formalism becomes a very useful tool.

\subsection{Wavefunctions for the coherent states vs the Glauber-Sudarshan states \label{sec3.3}}

To study and compare the wavefunctions of the coherent and the GS states we will start with the vacuum wavefunction defined over the interacting Minkowski vacuum $|\Omega\rangle$.
Let $|h\rangle$ denotes an eigenstate of the induced metric on a Cauchy
slice $\Sigma$, and let
$\Psi_\Omega[h] \equiv \langle h | \Omega \rangle$
be the vacuum wavefunctional. For the GS state
$|\sigma\rangle_{\rm GS}
=
D_{\rm GS}[\sigma]|\Omega\rangle$
where $D_{\rm GS}$ is the displacement operator from \eqref{sweenseyf}, 
the corresponding wavefunctional is:
\begin{equation}
\Psi_{\rm GS}[h;\sigma]
=
\langle h,t_f|D_{\rm GS}[\sigma]|\Omega,t_i\rangle ,
\end{equation}
where $|h,t_f\rangle$
is the eigenstate of the induced metric on the final slice $\Sigma_{t_f}$, and the interacting vacuum $|\Omega\rangle$ is defined on the initial Cauchy slice $\Sigma_{t_i}$. 
Since $D_{\rm GS}$ is an operator {\it stretched} for a given interval of time, the wavefunction is not defined on the initial Cauchy slice $\Sigma$. This is not an obstruction. It simply means that the resulting state is defined \emph{after} the evolution generated by $D_{\rm GS}$, and the wavefunctional is evaluated on the final slice.
To make this explicit, we can divide the interval $[t_i,t_f]$ into $N$ slices as 
$t_n=t_i+n\Delta t$, such that 
$\Delta t=\frac{t_f-t_i}{N}$,
with 
$n=0,1,\dots,N$.
Then following \eqref{eq:Dt_sigma_again}, we can express the displacement operator from  \eqref{sweenseyf} as:
\begin{equation}
D[\sigma] \equiv D_{\rm GS}[\sigma]
=
\lim_{N\to\infty}
\prod_{n=N-1}^{0}
\exp\!\left[
\Delta t
\int_{\Sigma_{t_n}} d^3x\;
\sqrt{-\hat{g}(t_n,\mathbf x)}\,
\sigma^{\mu\nu}(t_n,\mathbf x)\,
\hat g_{\mu\nu}(t_n,\mathbf x)
\right] ,
\label{eq:DGS_discrete_reply}
\end{equation}
where we have used the full form of the operator $\hat{G}_{\mu\nu}(t_n, {\bf x})$. Now we can insert complete sets of field eigenstates on every intermediate slice using the resolution of identity as
$\mathbf 1_{\Sigma_{t_n}}
=
\int Dh_n\;
|h_n,t_n\rangle\langle h_n,t_n|$,
with 
$n=1,\dots,N-1$.
Then it is easy to show that:
\begin{equation}
\begin{aligned}
\Psi_{\rm GS}[h_f;\sigma]
&=
\langle h_f,t_f|D_{\rm GS}[\sigma]|\Omega,t_i\rangle
\\
&=
\int \prod_{n=1}^{N-1}Dh_n\;
\prod_{n=0}^{N-1}
\langle h_{n+1},t_{n+1}|
e^{\Delta t\,\hat J_n}
|h_n,t_n\rangle\;
\langle h_0,t_i|\Omega,t_i\rangle,
\end{aligned}
\label{eq:Psi_GS_discrete_chain_reply}
\end{equation}
where
$h_N=h_f$,
and
$\hat J_n
\equiv
\int_{\Sigma_{t_n}} d^3x\;
\sqrt{-\hat{g}(t_n,\mathbf x)}\,
\sigma^{\mu\nu}(t_n,\mathbf x)\,
\hat g_{\mu\nu}(t_n,\mathbf x)$ is the same factor appearing in the exponential of \eqref{eq:DGS_discrete_reply}. In fact, since 
$\hat J_n$ contains no conjugate momenta or time-derivative operators, 
the whole operator $\hat J_n$ acts multiplicatively on $|h_n,t_n\rangle$, {\it i.e.}
$\hat J_n\,|h_n,t_n\rangle
=
J[h_n;t_n]\,
|h_n,t_n\rangle$. Using this on the kets appearing in \eqref{eq:Psi_GS_discrete_chain_reply}, we get:
\bg\label{sapphireelo}
\langle h_{n+1},t_{n+1}|e^{\Delta t\,\hat J_n}|h_n,t_n\rangle
& = &
\langle h_{n+1},t_{n+1}|h_n,t_n\rangle\\
&\times & \exp\!\left[
\Delta t
\int_{\Sigma_{t_n}} d^3x\;
\sqrt{-g[h_n](t_n,\mathbf x)}\,
\sigma^{\mu\nu}(t_n,\mathbf x)\,
g_{\mu\nu}[h_n](t_n,\mathbf x)
\right],\nonumber
\nd
showing that the GS displacement operator cleanly separates out with its eigenvalues leaving us the standard expression $\langle h_{n+1},t_{n+1}|h_n,t_n\rangle$. (A brief caveat is that the eigenvalue equation assumes that $|h_n,t_n\rangle$ is an eigenstate of the full set of equal-time metric components entering $\hat g_{\mu\nu}$ and contains no conjugate momenta or time-derivative operators. This is clearly true for the present case.) Therefore, 
once one restores the ordinary dynamical evolution and passes to the continuum, this becomes precisely the path integral with the GS insertion, namely:
\begin{equation}\label{vanadi1}
\Psi_{\rm GS}[h_f;\sigma]
=
\int_{\Omega}^{\,h_f}
{\cal D}g\,{\cal D}(\text{gh})\;
\exp\!\left(
iS_{\rm tot}[g,\text{gh}]
\right)
\exp\!\left(
\int d^4x\,\sqrt{-g(x)}\,\sigma^{\mu\nu}(x)\,g_{\mu\nu}(x)
\right) ,
\end{equation}
where $S_{\rm tot}[g, {\rm gh}]$ is the total action whose form will be described in section \ref{sec4.3}; and the notation
$\int_{\Omega}^{\,h_f}$
means that the path integral is taken over histories that start in the state $|\Omega,t_i\rangle$ and end on the configuration
$g_{ij}(t_f,\mathbf x)=h_{ij}(\mathbf x)$.
Thus the GS wavefunctional is obtained from the vacuum wavefunctional
by a bulk source-like deformation rather than by a simple translation
of the boundary data. By contrast, for the coherent state
$|\sigma\rangle_{\rm CS}
=
D_{\rm CS}[\sigma]|\Omega\rangle$ with $D_{\rm CS}$ is given in \eqref{sweenseyf} with 
the metric representation of the momentum operator as
$\pi_{ij}(\mathbf x)
=
-\,i\,\frac{\delta}{\delta h_{ij}(\mathbf x)}$, 
shows that $D_{\rm CS}[\sigma]$ acts as a translation operator on the
wavefunctional.  Hence:
\begin{equation}\label{vanadi2}
\Psi_{\rm CS}[h;\sigma]
=
\langle h|D_{\rm CS}[\sigma]|\Omega\rangle
=
\exp\!\left(i
\int d^3x\,\sigma^{ij}(\mathbf x)\,\pi_{ij}(\mathbf x)
\right)\Psi_\Omega[h]
=
\Psi_\Omega[h+\delta_\sigma h],
\end{equation}
with the precise form of $\delta_\sigma h$ depending on the convention
for the displacement operator. The essential distinction is therefore that the coherent-state
wavefunctional is a translated vacuum wavefunctional \eqref{vanadi2}
whereas the GS wavefunctional is a bulk-deformed wavefunctional \eqref{vanadi1}.

The wavefunction \eqref{vanadi1} does capture the full story, but to see how the wavefunction evolves from a given initial Cauchy slice one could reinterpret the GS displacement operator as a {\it multiple kicked} operator. In other words, let us 
suppose that instead of the four-dimensional GS operator, one defines
an instantaneous kick on a fixed Cauchy slice $\Sigma_{t_0}$ by the following variation of \eqref{eq:Dt_sigma_again} :
\begin{equation}
D_\Sigma[\sigma]
=
\exp\!\left(
\int_{\Sigma_{t_0}} d^3x\;
\sqrt{\hat{h}(\mathbf x)}\,
\sigma^{ij}(\mathbf x)\,
\hat{h}_{ij}(\mathbf x)
\right),
\label{eq:instant_GS_kick}
\end{equation}
where $h_{ij}(\mathbf x)$ is the induced spatial metric on the slice
and $\sigma^{ij}(\mathbf x)$ is a prescribed source profile. Note that we are taking only the spatial metric instead of the full metric\footnote{On a Cauchy slice the canonical configuration variable is the induced
spatial metric $h_{ij}$, not the full spacetime metric $g_{\mu\nu}$.
Accordingly, the coherent-state smearing tensor is naturally a
symmetric spatial tensor $\sigma^{ij}$ on $\Sigma_{t_0}$.  Writing
$\sigma^{\mu\nu}h_{\mu\nu}$ would be redundant, since the induced
metric has only tangential components on the slice from the ADM decomposition, and after
projection this reduces precisely to $\sigma^{ij}h_{ij}$.} $g_{\mu\nu}({\bf x}, t)$. Let $|h\rangle$ denote an eigenstate of the induced metric, such that
$\hat h_{ij}(\mathbf x)\,|h\rangle
=
h_{ij}(\mathbf x)\,|h\rangle$.
and 
$\Psi_\Omega[h;t_0]
\equiv
\langle h|\Omega\rangle$
be the vacuum wavefunctional on $\Sigma_{t_0}$. We can now define the instantaneously kicked GS state by:
\begin{equation}\label{vandi0}
|\sigma;t_0^+\rangle
=
D_\Sigma[\sigma]\,
|\Omega\rangle \equiv D_{t_0}[\sigma]|\Omega\rangle ,
\end{equation}
and since in the metric representation the operator
$\hat h_{ij}(\mathbf x)$ acts multiplicatively, the kick operator
\eqref{eq:instant_GS_kick} is diagonal in the $|h\rangle$ basis.
Therefore one immediately obtains the wavefunctional from the state \eqref{vandi0} as:
\bg\label{vandi3}
\Psi_{\rm GS}[h;\sigma,t_0^+]
& \equiv &
\langle h|\sigma;t_0^+\rangle
=
\langle h|D_\Sigma[\sigma]|\Omega\rangle \nonumber\\
& = &
\exp\!\left(
\int_{\Sigma_{t_0}} d^3x\;
\sqrt{h(\mathbf x)}\,
\sigma^{ij}(\mathbf x)\,
h_{ij}(\mathbf x)
\right)
\Psi_\Omega[h;t_0].
\nd
Thus, for an instantaneous GS kick, the wavefunctional is not
translated in configuration space.  Instead, it is obtained by
multiplying the vacuum wavefunctional by the local weight
$\exp\!\left(
\int_{\Sigma_{t_0}} d^3x\;
\sqrt{h}\,\sigma^{ij}h_{ij}
\right)$.
For comparison, a coherent-state kick on the same slice is generated
by the canonical momentum $\hat \pi_{ij}({\bf x})$,
so $D_{\rm CS}[\sigma]$ acts as a translation operator on the
wavefunctional as in \eqref{vanadi2}.
This difference is important.  The coherent-state kick changes the
configuration by shifting the argument of the wavefunctional, whereas
the instantaneous GS kick appears to leave the argument $h_{ij}$ unchanged and
instead changes the relative weight assigned to different metric
configurations.  But as we shall show soon, the multiplicative dressing of the vacuum wavefunctional is {\it equivalent} to a shift.

We can also quantify the time-evolution of the wavefunctional \eqref{vandi0}. In other words, if one wishes to evolve the kicked state to a later time $t>t_0$, the
wavefunctional becomes:

{\footnotesize
\begin{equation}
\Psi_{GS}[h_f;\sigma,t]
=
\int Dh_l\;
K[h_f,t;h_l,t_0]\,
\exp\!\left(
\int_{\Sigma_{t_0}} d^3x\;
\sqrt{h_l(\mathbf x)}\,
\sigma^{ij}(\mathbf x)\,
h_{l,ij}(\mathbf x)
\right)
\Psi_\Omega[h_i;t_0],
\end{equation}}
where $K[h_f,t;h_l,t_0]$ is the appropriate evolution kernel (and we use subscript $l$ to denote the initial state so that $h_{l, ij}$ is the initial metric with components $i, j$).  Thus the
instantaneous GS kick modifies the initial data multiplicatively,
after which the state evolves in the usual way. However the question is to find a clean way to understand the motion induced by repeated
Glauber--Sudarshan (GS) kicks. For this one needs to distinguish between the
\emph{instantaneous action} of each kick on the wavefunctional and the
\emph{resulting motion of the dominant support} of the wavefunctional
in configuration space after repeated kicks and ordinary time
evolution. We can quantify this in the following way.
Let $\Sigma_t$ be a Cauchy slice with induced metric $h_{ij}(\mathbf
x)$, and let $|h\rangle$ denote an eigenstate of the spatial metric,
$\hat h_{ij}(\mathbf x)\,|h\rangle
=
h_{ij}(\mathbf x)\,|h\rangle$. If the state immediately before the kick is
$|\Psi(t^-)\rangle$, 
with wavefunctional
$\Psi[h,t^-]
\equiv
\langle h|\Psi(t^-)\rangle$, 
then immediately after the kick one has the following multicative form:
\begin{equation}
\Psi[h,t^+]
=
\exp\!\left(
\int_{\Sigma_t} d^3x\;
\sqrt{h(\mathbf x)}\,
\sigma^{ij}(\mathbf x,t)\,
h_{ij}(\mathbf x)
\right)
\Psi[h,t^-].
\label{eq:single_GS_kick_wavefunctional}
\end{equation}
Thus a single GS kick does {not} appear to translate the argument
$h_{ij}(\mathbf x)$ of the wavefunctional. Instead, it reweights the
relative amplitudes of the metric configurations. On the other hand, 
we can ask how the wavefunctional evolves under repeated kicks.
Suppose that such kicks occur at a sequence of times:
\begin{equation}\label{vandi7}
t_1 < t_2 < \cdots < t_N ,
\end{equation}
which we take at discrete steps. This is different from the action of the displacement operator $D[\sigma] \equiv D_{\rm GS}[\sigma]$ from \eqref{sweenseyf} which implies {\it continuous} kicks. Our analysis can be easily generalized for the continuous case but, for illustrative purpose, we will take discrete case. 
Then the state after $N$ kicks is:
\begin{equation}
|\Psi(t_N^+)\rangle
=
D_{t_N}[\sigma_N]\,
U(t_N,t_{N-1})\,
D_{t_{N-1}}[\sigma_{N-1}]
\cdots
U(t_2,t_1)\,
D_{t_1}[\sigma_1]\,
|\Omega\rangle,
\label{eq:state_after_many_kicks}
\end{equation}
where $U(t_b,t_a)$ is the ordinary time-evolution operator. Note that the evolution of the wavefunctional goes through two essential steps. For the first step and in the metric representation, the kick at time $t_n$ acts as:
\begin{equation}
\Psi[h,t_n^+]
=
\exp\!\left(
\int_{\Sigma_{t_n}} d^3x\;
\sqrt{h(\mathbf x)}\,
\sigma_n^{ij}(\mathbf x)\,
h_{ij}(\mathbf x)
\right)
\Psi[h,t_n^-] ,
\label{eq:nth_kick_action}
\end{equation}
which is the usual multiplicative action that we analyzed earlier. 
(The superscripts
$t_n^-$ and $t_n^+$
 mean the instants immediately before and immediately after the kick at the same time slice $t_n$, {\it i.e.}
$t_n^- \equiv t_n-\epsilon$ and 
$t_n^+ \equiv t_n+\epsilon$,
for 
$\epsilon\to 0^+$. So 
$\Psi[h,t_n^-]$
is the wavefunctional just before the instantaneous GS kick at time $t_n$, while
$\Psi[h,t_n^+]$
is the wavefunctional immediately after that kick.)
For the second step, the wavefunctional evolves with the ordinary kernel between kicks. In other words:

{\footnotesize
\bg\label{vandi5}
\Psi[h_{n+1}^-,t_{n+1}]
& = &
\int Dh_n\;
K[h_{n+1}^-,t_{n+1};h_n,t^+_n]\,
\Psi[h_n,t_n^+]\\
&= &
\int Dh_n\;
K[h_{n+1}^-,t_{n+1};h_n,t^+_n]\,
\exp\!\left(
\int_{\Sigma_{t_n}} d^3x\;
\sqrt{h_n(\mathbf x)}\,
\sigma_n^{ij}(\mathbf x)\,
h_{n,ij}(\mathbf x)
\right)
\Psi[h_n,t_n^-] , \nonumber \nd}
which shows precisely how the
motion of the state is controlled by the time-dependent profile
$\sigma^{ij}(\mathbf x,t_n)$ where $\sigma^{ij}_n({\bf x})\vert_{{}_{{\Sigma_{t_n}}}} = \sigma^{ij}({\bf x}, t_n)$. 
We can now iterate \eqref{vandi5} from $t^-_0=t_i$ to $t_N=t_f$ to determine the wave-function of the system undergoing repeated kicks interspaced with usual temporal evolution via the boundary Hamiltonian. Doing this one gets the following quantitative result:
\begin{equation}
\begin{aligned}
\Psi_{\rm GS}^{(N)}[h_f;\sigma]
&=
\int \prod_{n=1}^{N-1}Dh_n\;
\Bigg[
\prod_{n=0}^{N-1}
K[h^-_{n+1},t_{n+1};h_n,t^+_n]
\Bigg]
\\
&\qquad\times
\Bigg[
\prod_{n=0}^{N-1}
\exp\!\left(
\int_{\Sigma_{t_n}} d^3x\;
\sqrt{h_n(\mathbf x)}\,
\sigma_n^{ij}(\mathbf x)\,
h_{n,ij}(\mathbf x)
\right)
\Bigg]
\Psi_\Omega[h_0;t_i],
\end{aligned}
\label{eq:iterated_many_kicks}
\end{equation}
with
$h_N=h_f$.
This is the precise repeated-kick wavefunctional.
In fact we can now use \eqref{eq:iterated_many_kicks} to study the continuum limit of the product of kicks. Using
$\sigma_n^{ij}(\mathbf x)
=
\Delta t_n\,\sigma^{ij}(t_n,\mathbf x)$,
the product of multiplicative kick factors becomes:

{\footnotesize
\bg\label{vandiii}
\prod_{n=0}^{N-1}
\exp\!\left(
\int_{\Sigma_{t_n}} d^3x\;
\sqrt{h_n(\mathbf x)}\,
\sigma_n^{ij}(\mathbf x)\,
h_{n,ij}(\mathbf x)
\right)
& = &
\exp\!\left[
\sum_{n=0}^{N-1}
\Delta t_n
\int_{\Sigma_{t_n}} d^3x\;
\sqrt{h_n(\mathbf x)}\,
\sigma^{ij}(t_n,\mathbf x)\,
h_{n,ij}(\mathbf x)
\right] 
\nonumber\\
& \longrightarrow &
\exp\!\left[
\int_{t_i}^{t_f}dt
\int_{\Sigma_t} d^3x\;
\sqrt{h(t,\mathbf x)}\,
\sigma^{ij}(t,\mathbf x)\,
h_{ij}(t,\mathbf x)
\right], \nd}
which is exactly the source-dressing factor appearing in the continuous path integral. Moreover the product of kernels in \eqref{eq:iterated_many_kicks} gives the ordinary path-integral weight:
\begin{equation}
\prod_{n=0}^{N-1}
K[h^-_{n+1},t_{n+1};h_n,t^+_n]
\longrightarrow
\int_{\Omega}^{\,h_f}
{\cal D}g\,{\cal D}(\text{gh})\;
\exp\!\left(
iS_{\rm tot}[g,\text{gh}]
\right),
\label{eq:kernels_to_path_integral}
\end{equation}
with the initial vacuum wavefunctional included in the lower boundary condition. Combining \eqref{vandiii} and \eqref{eq:kernels_to_path_integral}, we obtain \eqref{vanadi1}. Therefore the path-integral wavefunctional \eqref{vanadi1} is exactly the continuum limit of a repeated-kick construction in which, at each time slice $t_n$ the wavefunctional is multiplied by the exponential factor in \eqref{eq:nth_kick_action} and the resulting state is then propagated to the next slice by the ordinary kernel. The continuous GS insertion is therefore the Trotterized limit of infinitely many infinitesimal instantaneous GS kicks.


Let us now clarify how the multiplicative form of the wavefunctional can also be regarded as a {\it shift} of the center $-$ {\it i.e.} the most probable part $-$ of the wavefunctional.
The key point is that an instantaneous GS kick does not rigidly
translate the argument of the wavefunctional, but instead reweights the
relative amplitudes of the configurations.  Repeated kicks therefore do
not generate a simple geometric translation in configuration space;
rather, they continuously displace the \emph{dominant support} of the
state, namely its maximum, saddle, mean, or any other suitable notion
of the effective configuration characterizing the wavefunctional.
To make this precise, let
$\Psi[h,t^-]$
denote the wavefunctional immediately before a kick at time $t$, and
let the kick be
$D_t[\sigma]$.
Immediately after the kick one has
$\Psi[h,t^+]$
given by \eqref{eq:single_GS_kick_wavefunctional}.
If one introduces the normalized probability functional
$P[h,t]$,
then \eqref{eq:single_GS_kick_wavefunctional} implies:
\begin{equation}
P[h,t^+] \equiv
\frac{
|\Psi[h,t^+]|^2
}{
\int Dh\;|\Psi[h,t^+]|^2
}\propto
\exp\!\left(
2\,{\rm Re}\!\int_{\Sigma_t} d^3x\;
\sqrt{h(\mathbf x)}\,
\sigma^{ij}(\mathbf x,t)\,
h_{ij}(\mathbf x)
\right)
P[h,t^-] ,
\label{eq:gskick_repeat2}
\end{equation}
implying that the kick changes the probability landscape in configuration
space.  Configurations that align positively with the profile
$\sigma^{ij}(\mathbf x,t)$ are enhanced, while others are suppressed.
As a result, the dominant configuration of the state is shifted.

Equation \eqref{eq:gskick_repeat2} basically summarizes what we meant by the {\it shift} of the wavefunctional for the GS case: the most probable part of the wavefunctional shifts in the configuration space. We can quantify this further in the following way.
Suppose, for example, that the vacuum wavefunctional is peaked around
a configuration
$h_{ij}^{(0)}(\mathbf x)$.
After a kick the new peak is determined by extremizing:
\begin{equation}\label{vandi6}
\ln P[h,t^+]
=
\ln P[h,t^-]
+
2\,{\rm Re}\!\int_{\Sigma_t} d^3x\;
\sqrt{h(\mathbf x)}\,
\sigma^{ij}(\mathbf x,t)\,
h_{ij}(\mathbf x)
-
\ln {\cal N}(t),
\end{equation}
where ${\cal N}(t)$ is the normalization factor.  The condition for the
new dominant configuration $h_{ij}^\ast(\mathbf x,t)$ is therefore the stationary point of the function \eqref{vandi6}. In other words:
\begin{equation}
\frac{\delta}{\delta h_{ij}(\mathbf x)}
\ln P[h,t^-]
+
2\,\frac{\delta}{\delta h_{ij}(\mathbf x)}
{\rm Re}\!\left(
\int_{\Sigma_t} d^3y\;
\sqrt{h(\mathbf y)}\,
\sigma^{kl}(\mathbf y,t)\,
h_{kl}(\mathbf y)
\right)
=
0,
\end{equation}
evaluated at
$h_{ij}=h_{ij}^\ast(t)$.
This shows explicitly that the displacement of the dominant
configuration is governed by the profile
$\sigma^{ij}(\mathbf x,t)$. In a similar vein if we suppose that kicks occur at a sequence of times \eqref{vandi7}
then the state after $N$ kicks is \eqref{eq:state_after_many_kicks}.
In the metric representation this means that after each time step the
wavefunctional is first reweighted by the instantaneous profile
$\sigma_n^{ij}(\mathbf x)$ and then propagated by the ordinary
evolution kernel.  Consequently, the location of the dominant support
changes from step to step, and the sequence of dominant
configurations,
\begin{equation}
h_{ij}^\ast(t_1),
\qquad
h_{ij}^\ast(t_2),
\qquad
\ldots,
\qquad
h_{ij}^\ast(t_N),
\end{equation}
defines an effective trajectory in configuration space. This motion can be described as \emph{random} once the
profile $\sigma^{ij}(\mathbf x,t)$ itself is chosen randomly.
In other words, for a fixed choice of $\sigma^{ij}(\mathbf x,t)$ the
trajectory is \emph{driven} or \emph{steered} by that profile.
Different choices of $\sigma^{ij}(\mathbf x,t)$ produce different
motions of the dominant configuration.

So far we have been studying discrete kicks followed by intermediate temporal
evolution through the Feynman kernel. We can now discuss what happens in the
continuous limit. In the continuum limit of densely spaced kicks, one may naively think to 
write the wavefunctional in the following way:
\begin{equation}
\partial_t\Psi[h,t]
=
-\,i\,\widehat H_\partial\,\Psi[h,t]
+
\left(
\int_{\Sigma_t} d^3x\;
\sqrt{h(\mathbf x)}\,
\sigma^{ij}(\mathbf x,t)\,
h_{ij}(\mathbf x)
\right)
\Psi[h,t] ,
\label{eq:gs_continuous_drift}
\end{equation}
where note that the second term on the RHS of \eqref{eq:gs_continuous_drift} does not come with a factor of $i = \sqrt{-1}$ because of the way we have defined our instantaneous displacement operator.
Taken literally, however, the second term in
\eqref{eq:gs_continuous_drift} would describe an externally driven system
with an effective generator:
\begin{equation}
\widehat H_{\rm eff}(t)
=
\widehat H_\partial(t)
+
\widehat H_\sigma(t).
\end{equation}
This is not the interpretation appropriate to the present construction\footnote{Moreover, because the infinitesimal GS kicks are nonunitary reweightings, the formal
quantity $\widehat H_\sigma$ obtained by rewriting them in Hamiltonian form
is not a Hermitian physical Hamiltonian. It should therefore not be added to
$\widehat H_\partial$ as an independent generator of temporal evolution.
The operator $D(\sigma)$ prepares the GS state
$|\sigma\rangle=D(\sigma)|\Omega\rangle$ and appears in the path integral as
a state-dependent operator insertion, without modifying the microscopic
action. The resulting state must remain in the Wheeler--DeWitt physical
sector and must evolve according to the original boundary Schr\"odinger
equation generated solely by $\widehat H_\partial$. This is precisely the
interpretation consistent with the later Wheeler--DeWitt plus boundary
Schr\"odinger analysis discussed in section \ref{hwlelo}.}.
Here the densely spaced kicks are not independent external perturbations;
rather, they provide a time-sliced representation of the operator
$D(\sigma)$ acting within one fixed microscopic theory. (This will be elaborated further in section \ref{hwlelo}.) Their continuum
limit must therefore be inherited from, and consistently matched to, the
evolution generated by $\widehat H_\partial$. Equivalently, if
$|\sigma;t\rangle
=
D(\sigma;t)|\Omega;t\rangle$,
then the displacement operator must satisfy, at least on the reference
state the following equation\footnote{Note that if $|\sigma;t\rangle$ denotes the unnormalized state, there is no need to replace it by a normalized state in order to ask whether it satisfies the boundary Schr\"odinger equation. As long as the reference state $|\Omega; t\rangle$ satisfies the boundary Schr\"odinger equation, \eqref{eq:D_intertwining_continuum} follows. Therefore, the nonunitary definition of $D(\sigma)$ is fully compatible with an
unnormalized GS state of finite, nonunit norm. The norm need not equal unity,
and normalization may be imposed separately when computing probabilities or
expectation values. Its time independence, however, does not follow merely
from the exponential form of $D(\sigma)$. It follows when $D(\sigma;t)$
satisfies the boundary-Hamiltonian intertwining condition, so that
$D(\sigma;t)|\Omega;t\rangle$ obeys the ordinary Schr\"odinger equation
generated solely by the Hermitian operator $\widehat H_{\partial}$. Under
this condition $\langle\sigma;t|\sigma;t\rangle$ 
is time independent, even though it
need not equal one. A more quantitative way to see this is the following. Starting from
${\cal N}_\sigma(t)
=
\langle\Omega;t|
D^\dagger D
|\Omega;t\rangle$, 
one obtains:
\begin{equation}
\frac{d{\cal N}_\sigma}{dt}
=
\langle\dot\Omega|
D^\dagger D
|\Omega\rangle
+
\langle\Omega|
\dot D^\dagger D
|\Omega\rangle +
\langle\Omega|
D^\dagger\dot D
|\Omega\rangle
+
\langle\Omega|
D^\dagger D
|\dot\Omega\rangle , \nonumber
\label{eq:direct_norm_derivative}
\end{equation}
where dots denote temporal derivatives. Using
$|\dot\Omega\rangle
=
-i\widehat H_\partial|\Omega\rangle$ and 
$\langle\dot\Omega|
=
i\langle\Omega|\widehat H_\partial$,
together with
$i\dot D|\Omega\rangle
=
\left[
\widehat H_\partial,D
\right]
|\Omega\rangle$
and its Hermitian conjugate, all terms cancel, yielding
$\dot{\cal N}_\sigma(t)=0$.}:
\begin{equation}
\left[
i\frac{\partial D(\sigma;t)}{\partial t}
-
\left[
\widehat H_\partial,
D(\sigma;t)
\right]
\right]
|\Omega;t\rangle
=
0 ,
\label{eq:D_intertwining_continuum}
\end{equation}
where the commutator on the second term appears because we have assumed that both the interacting vacuum $|\Omega; t\rangle$ as well as the GS state $|\sigma; t\rangle$ satisfy the same boundary Schr\"odinger equation. Extending this to the wavefunctional discussion above, we can rewrite this as an
 ordinary boundary
Schr\"odinger equation:
\begin{equation}
\partial_t\Psi[h,t]
=
-\,i\,\widehat H_\partial\Psi[h,t],
\label{eq:gs_physical_schrodinger}
\end{equation}
rather than an independently driven equation generated by
$\widehat H_{\rm eff}$, that the physical GS state must obey. The fact that this whole scenario is also consistent with the Wheeler-DeWitt equation, implying that $\sigma^{\mu\nu}$ cannot be an arbitrary external steering function if
the state is to evolve solely under $\widehat H_{\partial}$. It must belong
to the admissible class:

{\footnotesize
\begin{equation}
{\cal S}_{\rm phys}
=
\Bigg\{
\sigma\;\Bigg|\;
\left[
\widehat{\mathcal H}_{\perp},
D(\sigma)
\right]|\Omega\rangle=0 = 
\left[
\widehat{\mathcal H}_{i},
D(\sigma)
\right]|\Omega\rangle, 
\left[
i\partial_tD(\sigma;t)
-
[\widehat H_{\partial},D(\sigma;t)]
\right]
|\Omega;t\rangle=0
\Bigg\} ,
\end{equation}}
which will be discussed in more detail in section \ref{hwlelo}.
The time-dependent function $\sigma^{ij}(\mathbf x,t)$ may nevertheless be
viewed as parametrizing the manner in which the GS state is assembled from
the repeated kicks. Its influence on the dominant configuration is already
encoded in the state $\Psi_\sigma[h,t]$ and in its initial spectral
coefficients. Accordingly, the evolution of the dominant configuration
should be written schematically as:
\begin{equation}
\dot h^\ast_{ij}(\mathbf x,t)
=
{\cal V}_{ij}
\big[
h^\ast;\Psi_\sigma(t)
\big],
\end{equation}
where the dependence on $\sigma^{ij}(\mathbf x,t)$ is implicit through the
physical GS state, rather than appearing as an additional external drift
term. In particular, $\sigma^{ij}$ does not define an independent force
${\cal F}_{ij}$ added to the boundary-Hamiltonian evolution. Its role is to
specify an admissible GS excitation whose complete time dependence remains
consistent with $\widehat H_\partial$.

This is precisely where the distinction from coherent states becomes
sharp.  For a coherent-state kick generated by the canonical momentum
from the displacement operator
$D_{\rm CS}[\sigma]$ from \eqref{sweenseyf},
the action in the metric representation is translational:
\begin{equation}
\Psi_{\rm CS}[h;\sigma]
\sim
\Psi_\Omega[h+\delta_\sigma h].
\end{equation}
Thus coherent states move through configuration space by shifts
generated by the canonical momentum, and so their motion is tied to a
much more restricted class of directions. By contrast, GS kicks act by multiplicative reweighting:
\begin{equation}
\Psi_{\rm GS}[h;\sigma]
=
\exp\!\left(
\int_{\Sigma_t} d^3x\;
\sqrt{h}\,\sigma^{ij}h_{ij}
\right)
\Psi_\Omega[h].
\end{equation}
Since $\sigma^{ij}(\mathbf x,t)$ is an arbitrary time-dependent
function, the GS construction allows one to continuously reshape the
probability landscape and thereby steer the dominant support of the
state through a far broader region of configuration space.
Under sufficiently favorable conditions, namely if the map:
\begin{equation}
\sigma^{ij}(\mathbf x,t)
\;\longmapsto\;
h^\ast_{ij}(\mathbf x,t)
\end{equation}
is sufficiently non-degenerate, the family of all admissible profiles
$\sigma^{ij}(\mathbf x,t)$ generates an extremely large class of
effective trajectories.  In this sense the GS construction can be used
to approximate a much wider set of on-shell and off-shell trajectories
than is available in the coherent-state construction.

It is therefore more precise to say that coherent states are
associated with a comparatively restricted family of trajectories,
largely tied to canonical translations and hence to a narrower
kinematic sector, whereas GS states, by continuously reweighting the
wavefunctional through arbitrary profiles
$\sigma^{ij}(\mathbf x,t)$, can steer the dominant configuration along
a much broader class of paths in configuration space.

Here is then our final interpretation. The essential distinction may therefore be summarized as follows: (a) {\Su coherent states translate the argument of the wavefunctional},
whereas (b) {\Su GS states deform the probability landscape and thereby move its dominant support.}
For Gaussian vacua these two viewpoints can coincide, because the
multiplicative reweighting may amount to an exact shift of the center.
For generic vacua, however, the GS construction should be understood
not as a rigid translation, but as a controlled displacement of the
maximum, mean, or dominant saddle of the wavefunctional.  This is the
sense in which repeated GS kicks, driven by arbitrary
$\sigma^{ij}(\mathbf x,t)$, can explore a far richer class of
trajectories $-$ both on- and off-shell ones $-$ in configuration space than the coherent-state
construction. These distinctions appear in {\bf figure \ref{gsvscoher}}.

\subsection{SUSY breaking for the coherent vs SUSY breaking for the GS states \label{sec3.4}}

Let us assume that the starting point is a supersymmetric Minkowski vacuum
$|\Omega\rangle$,
constructed over a chosen supersymmetric Minkowski minimum of the exact trans-series action
$S_{\rm tot}$. (Details about these two choices will be elaborated in section \ref{sec5}, and for the present discussion we will not require the subtleties and the nuances explored therein.) By assumption this vacuum preserves supersymmetry, so the supercharges annihilate it:
\begin{equation}
Q_\alpha\,|\Omega\rangle = 0,
\qquad
\bar Q_{\dot\alpha}\,|\Omega\rangle = 0 ,
\label{eq:Q_annihilate_vac}
\end{equation}
where note that we are writing the supercharge operator simply as $Q_\alpha$ without embellishing it with details from possible color degrees of freedom.
Equivalently, in a bosonic Minkowski vacuum the vacuum energy vanishes and the supersymmetry algebra gives:
\begin{equation}
\langle \Omega|\,\{Q_\alpha,Q_\alpha^\dagger\}\,|\Omega\rangle = 0.
\label{eq:susy_alg_vac}
\end{equation}
We can now define the Glauber--Sudarshan type state by acting with $D(\sigma)$ from \eqref{sweenseyf} on the supersymmetric vacuum in the usual way as
$|\sigma\rangle
=
D(\sigma)\,|\Omega\rangle$. One immediate observation is 
that this operator is built from bosonic fields only. Therefore, unless it commutes with all supercharges, the state will not remain supersymmetric. However a more detailed criterion for preserved supersymmetry is the following: The state $|\sigma\rangle$ preserves supersymmetry if and only if it is annihilated by $Q_\alpha$, {\it i.e.}
$Q_\alpha\,|\sigma\rangle = 0$ for all $\alpha$.
We can express this in a more suggestive way by first rewriting this as:
\bg\label{coribhalo1}
Q_\alpha\,|\sigma\rangle
=
Q_\alpha\,D(\sigma)\,|\Omega\rangle
=
D(\sigma)\,
\Big(
D^{-1}(\sigma)\,Q_\alpha\,D(\sigma)
\Big)\,
|\Omega\rangle ,
\nd
where now we can use the strategy that supersymmetry requires the annihilation condition from \eqref{eq:Q_annihilate_vac}, implying that if the terms in the bracket in \eqref{coribhalo1} is proportional to $Q_\alpha$ itself, then using \eqref{eq:Q_annihilate_vac} the state will still be supersymmetric if:
\begin{equation}
D^{-1}(\sigma)\,Q_\alpha\,D(\sigma)
=
Q_\alpha \qquad \implies [Q_\alpha,D(\sigma)] = 0
\label{eq:D_commutes_with_Q}
\end{equation}
when acting on the vacuum, or more generally if the commutator vanishes. Note that the above condition does not rely on which of the two choices of $D(\sigma)$ we take from \eqref{sweenseyf}. For both coherent and the GS states, supersymmetry will be preserved if 
\eqref{eq:D_commutes_with_Q} is satisfied. Thus supersymmetry is broken whenever:
\begin{equation}
[Q_\alpha,D(\sigma)] \neq 0.
\label{eq:broken_if_nonzero}
\end{equation}
Question is does \eqref{eq:broken_if_nonzero} hold for both the displacement operators in \eqref{sweenseyf}? To see this let us start with the commutator with the GS operator, and express the GS operator as:
\begin{equation}
W_\sigma
\equiv
\int d^4x\;
\sqrt{-\hat g(x)}\,
\sigma^{\mu\nu}(x)\,\hat g_{\mu\nu}(x),
\qquad
D(\sigma)= {\cal T}e^{W_\sigma} ,
\label{eq:W_sigma_operator}
\end{equation}
where $\mathcal T$ is the time-ordering. Note that both the metric and the determinant are operators whereas $\sigma^{\mu\nu}(x)$ is a c-number function. 
Then the bracket in \eqref{coribhalo1} becomes:
\begin{equation}
D^{-1}(\sigma)\,Q_\alpha\,D(\sigma)
=
Q_\alpha
+
[Q_\alpha,W_\sigma]
+
\frac{1}{2!}[[Q_\alpha,W_\sigma],W_\sigma]
+\cdots ,
\label{eq:BCH_Q}
\end{equation}
where we used $D(\sigma)$ as a time-ordered exponential of $W_\alpha$ from \eqref{eq:W_sigma_operator} and implemented the Baker-Campbell-Hausdorff (BCH) expansion.  So the question reduces to whether:
\begin{equation}
[Q_\alpha,W_\sigma]
=
\int d^4x\;
\sigma^{\mu\nu}(x)\,
[Q_\alpha,\sqrt{-\hat g}\,\hat g_{\mu\nu}(x)]
\label{eq:QW_comm}
\end{equation}
vanishes. We can simplify the commutator bracket in \eqref{eq:QW_comm}
by expressing the metric operator as a fluctuation over a supersymmetric Minkowski minimum. This will help us to expand $\sqrt{-\hat{g}(x)}$ in a more legitimate way and then work out the commutator bracket. In fact, expanding $\sqrt{-\hat{g}(x)}$ will give us commutators of $Q_\alpha$ with the products of the metric operator $\hat{g}_{\mu\nu}$. 
But supersymmetry transforms the metric into the gravitino. Schematically,
\begin{equation}
[Q_\alpha,\hat g_{\mu\nu}(x)]
\sim  \delta_\epsilon \hat{g}_{\mu\nu}(x)
=
\bar\epsilon\,\Gamma_{(\mu}\psi_{\nu)(x)},
\label{eq:Q_on_g}
\end{equation}
with similar results from products of the metric operators. 
Therefore for a generic nonzero profile $\sigma^{\mu\nu}(x)$, and because $W_\sigma$ is not invariant under supersymmetry, we see that:
\begin{equation}
[Q_\alpha,W_\sigma]
\neq 0 ,
\label{eq:QW_nonzero}
\end{equation}
because it generates fermionic terms proportional to the gravitino. The second commutator bracket in \eqref{eq:BCH_Q} is also non-zero because of \eqref{eq:QW_nonzero} and \eqref{eq:Q_on_g} as well as the fact that $W_\sigma$ allows terms proportional to the metric operators coming from Taylor expanding the determinant operator $\sqrt{-\hat{g}(x)}$.
Hence, in general:
\begin{equation}
Q_\alpha\,|\sigma\rangle \neq 0.
\label{eq:Q_sigma_nonzero}
\end{equation}
This is the precise sense in which the GS state breaks supersymmetry. One would expect similar story for coherent states with $D(\sigma) \equiv D_{\rm coh}(\sigma)$ given by the first relation in \eqref{sweenseyf} as $D_{\rm coh}(\sigma) = \exp\left[\Xi(\sigma)\right]$ where $\Xi_\sigma
=
i\int_{\Sigma_{t_i}} d^3x\;
\sigma^{ij}(\mathbf x)\,\hat\pi_{ij}(\mathbf x)$. Following the same logic as before, we now have, 
using the Baker--Campbell--Hausdorff expansion:
\begin{equation}\label{coribhalo3}
D_{\rm coh}^{-1}(\sigma)\,Q_\alpha\,D_{\rm coh}(\sigma)
=
Q_\alpha
+
[Q_\alpha,\Xi_\sigma]
+
\frac{1}{2!}[[Q_\alpha,\Xi_\sigma],\Xi_\sigma]
+\cdots,
\end{equation}
implying that we have to start by looking at the commutator of the supercharge with $\hat{\pi}_{ij}$, {\it i.e.} $[Q_\alpha, \hat{\pi}_{ij}(x)]$. This is more non-trivial now because in supergravity there are additional fermion bilinears, so more precisely one has:
\begin{equation}
\hat\pi_{ij}(x) \equiv \hat\pi^{ij}_{\rm SUGRA}(x)
=
\sqrt{\hat h(x)}\,
\left(
\hat K^{ij}(x)-\hat h^{ij}(x)\hat K(x)
\right)
+
\hat\pi^{ij}_{\rm ferm}(x),
\label{eq:pi_sugra_def}
\end{equation}
on a given Cauchy slice $\Sigma_{t_i}$ so $t = t_i$, and where $\hat\pi^{ij}_{\rm ferm}(x)$ denotes the gravitino-induced corrections. These terms depend on the precise formulation and gauge choice. We should also really speak about the supersymmetry variation generated by the smeared supercharge:
\begin{equation}
Q[\epsilon]
\equiv
\bar\epsilon^\alpha Q_\alpha,
\label{eq:smeared_supercharge}
\end{equation}
and then the canonical action of supersymmetry on any bosonic operator $\hat{\cal O}$ is
$\delta_\epsilon \hat{\cal O}
=
i\,
[Q[\epsilon],\hat{\cal O}]$.
So the precise objective is to compute
$i\,[Q[\epsilon],\hat\pi^{ij}]
=
\delta_\epsilon \pi^{ij}$,
implying that what one needs is the supersymmetry variation of the canonical momentum $\hat\pi^{ij}$. This becomes:

{\footnotesize
\bg\label{claugarsea1}
i[Q[\epsilon],\pi^{ij}]
 &= &
\delta_\epsilon \pi^{ij}
=
\delta_\epsilon
\left[
\sqrt{h}\,
\left(
K^{ij}-h^{ij}K
\right)
+
\pi^{ij}_{\rm ferm}
\right] \\
& = &
\sqrt{h}\,
\left[
\frac{1}{2N}\,
\partial_t\!\left(
\bar\epsilon\gamma^{(i}\psi^{j)}
\right)
-
\frac{1}{2N}\,
D^{(i}\!\left(
\delta_\epsilon N^{j)}
\right)
+
{\cal A}^{ij}{}_{k}\,
\bar\epsilon\psi^k
+
{\cal B}^{ij}{}_{k\ell}\,
\bar\epsilon D^k\psi^\ell
\right]
+
\delta_\epsilon \pi^{ij}_{\rm ferm}, \nonumber
\nd}
where ${\cal A}$ and ${\cal B}$ are functions of the ADM variables 
$N, N^i, h_{ij}$, and $K_{ij}$. This is still schematic, but it is 
now precise in the sense that it identifies the derivative order and the geometric origin of the various terms. Plugging \eqref{claugarsea1} in \eqref{coribhalo3} would imply:
\bg\label{coribhalo5}
[Q[\epsilon], \Xi_\sigma] \ne 0 \qquad \implies ~~~ Q[\epsilon] |\sigma\rangle_{\rm coh} \ne 0, \nd
because the higher commutator terms in \eqref{coribhalo3} are also non-zero, thus justifying SUSY breaking for the coherent state. Note that in both cases the supersymmetries are broken by the states and {\it not} by the action.
This point is important. The exact trans-series action $S_{\rm tot}$ may still be the supersymmetric action of the theory, and the chosen Minkowski minimum may still be a supersymmetric minimum. The breaking occurs because the \emph{state} $|\sigma\rangle$
is not annihilated by the supercharges, even though the vacuum
$|\Omega\rangle$
is. This implies that the underlying theory may remain supersymmetric, while the chosen states, GS and the coherent states, need not be.
This is exactly analogous to ordinary spontaneous symmetry breaking in quantum mechanics or QFT: the equations and the Hamiltonian may preserve the symmetry, while a particular state does not.

The supersymmetry breaking mechanism for both the coherent and the Glauber-Sudarshan states can also be argued from energy criterion.
Recall that the supersymmetry algebra implies that the supercharges $Q_\alpha$, $Q^\dagger_\beta$ are directly related to the Hamiltonian via:
\begin{equation}
\{Q_\alpha,Q_\beta^\dagger\}
=
2\,\delta_{\alpha\beta}\,H_{\partial\Sigma}
+\text{momentum and central-charge terms} ,
\label{eq:susy_algebra}
\end{equation}
where $H_{\partial\Sigma}$ is the boundary Hamiltonian.
For a translationally invariant Minkowski vacuum with vanishing momentum and no relevant central charges,
\begin{equation}
\langle \Omega|H_{\partial\Sigma}|\Omega\rangle = 0
\qquad\Longleftrightarrow\qquad
Q_\alpha|\Omega\rangle=0.
\label{eq:vacuum_energy_zero}
\end{equation}
For asymptotically flat spacetimes this boundary Hamiltonian is precisely the ADM energy 
$H_{\partial\Sigma}
=
E_{\rm ADM}$, such that the above equation implies the vanishing of the ADM energy $\langle\Omega| E_{\rm ADM} | \Omega\rangle = 0$.
For the GS state:
\begin{equation}\label{coribhalo10}
E_\sigma
\equiv
\frac{\langle \sigma|H_{\partial\Sigma}|\sigma\rangle}{\langle \sigma|\sigma\rangle}
=
\frac{\langle \Omega|D^\dagger(\sigma)\,H_{\partial\Sigma}\,D(\sigma)|\Omega\rangle}
{\langle \Omega|D^\dagger(\sigma)D(\sigma)|\Omega\rangle} =  
\frac{\langle \Omega|D^\dagger(\sigma)\,[E_{\rm ADM}, \,D(\sigma)]|\Omega\rangle}
{\langle \Omega|D^\dagger(\sigma)D(\sigma)|\Omega\rangle} \, > 0,
\end{equation}
where $E_\sigma = 0$ if and only if the commutator in \eqref{coribhalo10} vanishes, {\it i.e.} $[E_{\rm ADM},\, D(\sigma)] = 0$. But this is highly nongeneric. For coherent or GS states, $D(\sigma)$ in \eqref{sweenseyf} is precisely designed to create a nontrivial displaced or excited state, so typically
$[E_{\rm ADM},D(\sigma)]\neq 0$,
and therefore
$E_\sigma > 0$.
Then the supersymmetry algebra implies that the norm of the supercharge variation is nonzero:
\begin{equation}
\sum_\alpha
\frac{
\|Q_\alpha|\sigma\rangle\|^2
}{
\langle \sigma|\sigma\rangle
}
=
\sum_\alpha
\frac{
\langle \sigma|Q_\alpha^\dagger Q_\alpha|\sigma\rangle
}{
\langle \sigma|\sigma\rangle
}
\propto
E_\sigma.
\label{eq:norm_Qsigma}
\end{equation}
So a positive energy density in the GS state is a direct signal that supersymmetry is broken in that state. The story for the coherent state is exactly similar with $D(\sigma) \to D_{\rm coh}(\sigma)$ from \eqref{sweenseyf}.

One may also see the breaking through order parameters. If a bosonic field has a SUSY variation into a fermion, then a nontrivial bosonic expectation value selected by the GS operator generally induces a nontrivial fermionic variation:
\begin{equation}
\delta_\epsilon \langle g_{\mu\nu}(x)\rangle_\sigma
=
\bar\epsilon\,
\langle \Gamma_{(\mu}\psi_{\nu)}(x)\rangle_\sigma.
\label{eq:susy_variation_metric_ev}
\end{equation}
If the state were supersymmetric, the RHS would have to vanish consistently for all preserved supercharges. For a generic GS profile sourcing a nontrivial metric expectation value, it is clear that:
\begin{equation}
\langle g_{\mu\nu}(x)\rangle_\sigma
\neq
\langle g_{\mu\nu}(x)\rangle_\Omega,
\label{eq:metric_shift_sigma}
\end{equation}
the therefore the corresponding fermionic SUSY variations are generally nonzero, showing again that the state is not invariant.

We can therefore conclude the following. 
The fact that the GS state is constructed over a supersymmetric Minkowski minimum does not prevent SUSY breaking generically. It only means that the \emph{reference vacuum}
$|\Omega\rangle$
preserves supersymmetry. Once one acts with
$D(\sigma)$,
one moves to an excited, displaced, or reweighted state in Hilbert space. Unless this displacement lies entirely along a supersymmetric flat direction and is compatible with all Killing-spinor conditions, the new state will not be supersymmetric. In particular, the vacuum may satisfy
$\delta_\epsilon \psi_\mu = 
\delta_\epsilon \lambda =
\delta_\epsilon \chi = 0$,
while the GS-deformed expectation values instead satisfy the following for generic $\sigma$:
\begin{equation}
\delta_\epsilon \langle \psi_\mu\rangle_\sigma \neq 0,
\qquad
\delta_\epsilon \langle \lambda\rangle_\sigma \neq 0,
\qquad
\delta_\epsilon \langle \chi\rangle_\sigma \neq 0 .
\label{eq:sigma_breaks_killing}
\end{equation}
A useful refinement of the above discussion is that the conclusion regarding supersymmetry breaking applies only to the restricted class of \emph{purely bosonic} coherent or GS insertions that we have been considering explicitly. If the displacement operator is built only from bosonic fields, for example from a metric-type kernel of the schematic form 
\eqref{eq:W_sigma_operator}
then one generally has \eqref{eq:QW_nonzero}
for the supercharges $Q_\alpha$ preserved by the original background. In this restricted setting, the deformation is not embedded into a full supersymmetric multiplet, and supersymmetry is therefore generically broken. The same remark applies to the corresponding bosonic coherent-state construction, which should also be viewed as a bosonic truncation of a more general supersymmetric displacement operator.

This conclusion, however, should not be overstated. It does \emph{not} imply that every coherent or GS-type state must break supersymmetry. A nontrivial semiclassical state can preserve part of the supersymmetry provided the source is completed into a full supermultiplet and the resulting expectation values satisfy the usual BPS or Killing-spinor conditions. In other words, the correct requirement is not merely that the state be nontrivial, but that the deformation be supersymmetric in the stronger sense that
\begin{equation}
[Q_\alpha,D_{\rm susy}(\Xi)]\,|\text{phys}\rangle=0
\label{eq:susy_source_condition_appendix}
\end{equation}
for the preserved supercharges, where $D_{\rm susy}(\Xi)$ denotes a displacement operator built from the appropriate bosonic and fermionic source data $\Xi$. Equivalently, the resulting semiclassical expectation values must satisfy conditions of the schematic form
\begin{equation}
\delta_\epsilon\psi_M=0,
\qquad
\delta_\epsilon\lambda=0,
\qquad
\cdots
\label{eq:BPS_conditions_appendix}
\end{equation}
for some nonzero spinor $\epsilon$. An interesting example of the aforementioned procedure is 
furnished by supersymmetric D-brane configurations \cite{dileep}. Although such backgrounds are highly nontrivial, they preserve a fraction of the ambient supersymmetry because the full sourced configuration obeys the corresponding BPS projector conditions, for example:
\begin{equation}
\Gamma_\kappa\,\epsilon=\epsilon,
\label{eq:kappa_projector_appendix}
\end{equation}
together with the ambient Killing-spinor equations. Thus, if a D-brane configuration is realized through a coherent or GS-type state, then that state need not break supersymmetry at all. What matters is not the mere presence of a nontrivial source, but whether the source is completed in a way compatible with the preserved supercharges. Put differently,
\begin{equation}
\text{\textcolor{blue}{bosonic-only source}}
\quad\Longrightarrow\quad
\text{\textcolor{blue}{generic SUSY breaking}} \nonumber
\label{eq:bosonic_only_summary_appendix}
\end{equation}
whereas a fully supersymmetrized source satisfying BPS conditions
implies possible partial SUSY preservation. In fact as we shall discuss in section \ref{sec7.5}, the so-called bootstrap map for the GS states could in-principle lead to fixed points that are supersymmetric.

The correct conclusion is therefore the following. Our earlier argument for supersymmetry breaking should be understood as applying only to the bosonic metric-type coherent or supersymmetry breaking bosonic (and fermionic) GS insertions used in the present construction. More generally, one can certainly imagine coherent or GS states associated with nontrivial semiclassical configurations that preserve supersymmetry, provided the corresponding displacement operator is embedded into a full supermultiplet source and the induced expectation values satisfy the appropriate BPS constraints. Supersymmetric D-branes are the canonical example of this more general possibility.


\section{Coherent states and the Schwinger-Keldysh contours \label{sec4.0}}

First let us ask why and how does the Schwinger-Keldysh in-in formalism appear for our case. The issue appears when we wish to compute the expectation value of the metric operator expressed in the following form:
\begin{equation}\label{philipdal}
\langle g_{\mu\nu}(x)\rangle_\sigma
=
\frac{\bra{\sigma}\, g_{\mu\nu}(x)\,\ket{\sigma}}
{\braket{\sigma}{\sigma}} ,
\end{equation}
where \eqref{philipdal} is meant to compute the expectation value in the state $|\sigma\rangle$. The following analysis is specifically tuned to deal with the coherent states, namely the states defined on an initial Cauchy slice which are then propagated by the {\it boundary} Hamiltonian. (What happens for the Glauber-Sudarshan states will be discussed soon.) So at a face-value this does not appear as a correlation function where we do an in-out computation although, as we shall discuss later, this may also be represented by an in-out correlator.

\subsection{Towards construction of the Schwinger-Keldysh contours \label{sec4.1}}

The steps that we follow now may be started by first fixing $t$ such that $t_i<t<t_f$. The coherent state $|\sigma\rangle$ is defined on the initial Cauchy slice $\Sigma_{t_i} \equiv \Sigma_i$. $\Sigma_f\equiv \Sigma_{t_f}$ an arbitrary late-time slice used to implement the trace, and $t$ the operator insertion time.
Let $\rho_\sigma$ be the density matrix on $\Sigma_i$ (for a pure state,
$\rho_\sigma=\ket{\sigma}\bra{\sigma}$).
Start from the precise trace identity:
\begin{equation}
\mathrm{Tr}\!\left(\rho_\sigma\,\hat O_H(t)\right)
=
\mathrm{Tr}\!\left(
U(t_f,t_i)\,\rho_\sigma\,U^\dagger(t_f,t_i)\;
U(t_f,t)\,\hat O_S(t_i)\,U^\dagger(t_f,t)
\right),
\label{eq:trace_precise}
\end{equation}
where $\hat O_S(t_i)$ is defined on the Schr\"odinger Hilbert space (reference slice $\Sigma_i$), and $U(t_b,t_a) \equiv e^{-iH(t_b - t_a)}$, with $H$ being the boundary Hamiltonian, is the time-evolution operator.
Using $\mathbf 1_{\Sigma_f}=\int Dh_f\,\ket{h_f,t_f}\bra{h_f,t_f}$,
and inserting $\mathbf 1_{\Sigma_i}=\int Dh\,\ket{h,t_i}\bra{h,t_i}$ on both sides of $\rho_\sigma$, \eqref{eq:trace_precise} converts to the following:
\begin{align}
\mathrm{Tr}\!\left(\rho_\sigma\,\hat O_H(t)\right)
&=
\int Dh_f \int Dh_i^+ \int Dh_i^-\;
\bra{h_f,t_f}U(t_f,t_i)\ket{h_i^+,t_i}\;
\bra{h_i^+,t_i}\rho_\sigma\ket{h_i^-,t_i}
\nonumber\\
&\qquad\qquad\times
\bra{h_i^-,t_i}\,
U^\dagger(t_f,t_i)\;U(t_f,t)\,\hat O_S(t_i)\,U^\dagger(t_f,t)\,
\ket{h_f,t_f}.
\label{eq:trace_hi_correct}
\end{align}
The second line in \eqref{eq:trace_hi_correct} is important. This contains the Schr\"odinger operator $\hat{O}_S(t_i) \equiv \hat{O}_S({\bf x}, t_i)$ sandwiched between various evolution operators. This can be simplified in the following way.
Use the composition law
$U^\dagger(t_f,t_i)=U^\dagger(t,t_i)\,U^\dagger(t_f,t)$,
$U^\dagger(t_f,t)=U^\dagger(t,t_f)$, and 
$U^\dagger(t_f,t)\,U(t_f,t)=\mathbf 1$
and inserting a complete set at $\Sigma_t$,
$\mathbf 1_{\Sigma_t}=\int Dh_t\,\ket{h_t,t}\bra{h_t,t}$,
\emph{between} $U(t_f,t)$ and $\hat O_S(t_i)$, we get:
\begin{align}
&\bra{h_i^-,t_i}\,
U^\dagger(t_f,t_i)\;U(t_f,t)\,\hat O_S(t_i)\,U^\dagger(t_f,t)\,
\ket{h_f,t_f}
\nonumber\\
\quad= &
\bra{h_i^-,t_i}\,U^\dagger(t,t_i)
\Big[
U^\dagger(t_f,t)\,U(t_f,t)
\Big]\,
\hat O_S(t_i)\,U^\dagger(t_f,t)\,\ket{h_f,t_f}
\nonumber\\
\quad= &
\bra{h_i^-,t_i}\,U^\dagger(t,t_i)\,
\hat O_S(t_i)\,
U^\dagger(t_f,t)\,\ket{h_f,t_f},
\label{eq:key_simplify}
\end{align}
where we used $U^\dagger(t_f,t)\,U(t_f,t)=\mathbf 1$.
At this stage, to make the operator insertion explicitly occur at the time slice $\Sigma_t$,
we insert $\mathbf 1_{\Sigma_t}$ on both sides of $\hat O_S(t_i)$. This then converts \eqref{eq:key_simplify} to the following:

{\footnotesize
\begin{align}
\bra{h_i^-,t_i}\,U^\dagger(t,t_i)\,\hat O_S(t_i)\,U^\dagger(t_f,t)\ket{h_f,t_f}
&=
\int Dh_t^-\int Dh_t^+\;
\bra{h_i^-,t_i}U^\dagger(t,t_i)\ket{h_t^-,t}\;
\nonumber\\
&\qquad\times
\bra{h_t^-,t}\hat O_S(t_i)\ket{h_t^+,t}\;
\bra{h_t^+,t}U^\dagger(t_f,t)\ket{h_f,t_f}.
\label{eq:insert_ht_correct}
\end{align}}
This is one of the key step because it splits \eqref{eq:key_simplify} into three manageable pieces, whose importance will become clear from the following analysis.
Substituting \eqref{eq:insert_ht_correct} into \eqref{eq:trace_hi_correct} gives the fully consistent Hilbert-space decomposition:

{\footnotesize
\begin{align}
\mathrm{Tr}\!\left(\rho_\sigma\,\hat O_H(t)\right)
&=
\int Dh_f \int Dh_i^+\int Dh_i^-\int Dh_t^+\int Dh_t^-\;
\bra{h_f,t_f}U(t_f,t_i)\ket{h_i^+,t_i}\;
\bra{h_i^+,t_i}\rho_\sigma\ket{h_i^-,t_i}
\nonumber\\
&\qquad\times
\bra{h_i^-,t_i}U^\dagger(t,t_i)\ket{h_t^-,t}\;
\bra{h_t^-,t}\hat O_S(t_i)\ket{h_t^+,t}\;
\bra{h_t^+,t}U^\dagger(t_f,t)\ket{h_f,t_f} ,
\label{eq:trace_split_correct_all}
\end{align}}
which basically brings us right at the point where we can now replace each of the brackets that contains $U(t_b, t_a)$ into the corresponding path integrals. There are however two kernels that do not have $U(t_b, t_a)$ and they are $\bra{h_i^+,t_i}\rho_\sigma\ket{h_i^-,t_i}$ and $\bra{h_t^-,t}\hat O_S(t_i)\ket{h_t^+,t}$. The first one is easy to deal with. 
With pure-state specialization, 
if $\rho_\sigma=\ket{\sigma}\bra{\sigma}$, then:
\begin{equation}\label{wavefunction}
\bra{h_i^+,t_i}\rho_\sigma\ket{h_i^-,t_i}
=
\Psi_\sigma[h_i^+]\Psi_\sigma^*[h_i^-],
\qquad
\Psi_\sigma[h]\equiv \braket{h,t_i}{\sigma},
\end{equation}
which tells us how the wave-functions would enter inside the expectation value. Note that, as expected the wave-functions are defined exactly on the initial Cauchy slice $\Sigma_i$. On the other hand, 
the middle factor:
\begin{equation}
\bra{h_t^-,t}\hat O_S(t_i)\ket{h_t^+,t}
\label{eq:O_matrix_element}
\end{equation}
is the operator insertion at time $t$ in the $h$-basis on $\Sigma_t$.
For a local operator such as $\hat O=g_{\mu\nu}(\mathbf x,t)$, one typically chooses
the insertion on the $+$ (forward) branch, which amounts to replacing
\eqref{eq:O_matrix_element} by a functional of the $+$ history evaluated at time $t$:
\begin{equation}
\bra{h_t^-,t}\hat O_S(t_i)\ket{h_t^+,t}
\;\;\longrightarrow\;\;
O[g_+](\mathbf x,t)\,
\delta[h_t^--h_t^+] ,
\label{eq:O_plus_branch_replacement}
\end{equation}
where note that the $t_i$ information of the original Cauchy slice is no longer needed.
The delta functional enforces that, for the chosen insertion, the two configurations at the
operator time are identified.  More generally, keeping $\bra{h_t^-}\hat O_S\ket{h_t^+}$
explicit generates the full Keldysh operator algebra (time-ordered, anti-time-ordered,
Wightman, etc.), but for the single expectation value we will use \eqref{eq:O_plus_branch_replacement}. 

\begin{figure}[h]
\centering
\begin{tabular}{c}
\includegraphics[width=6in]{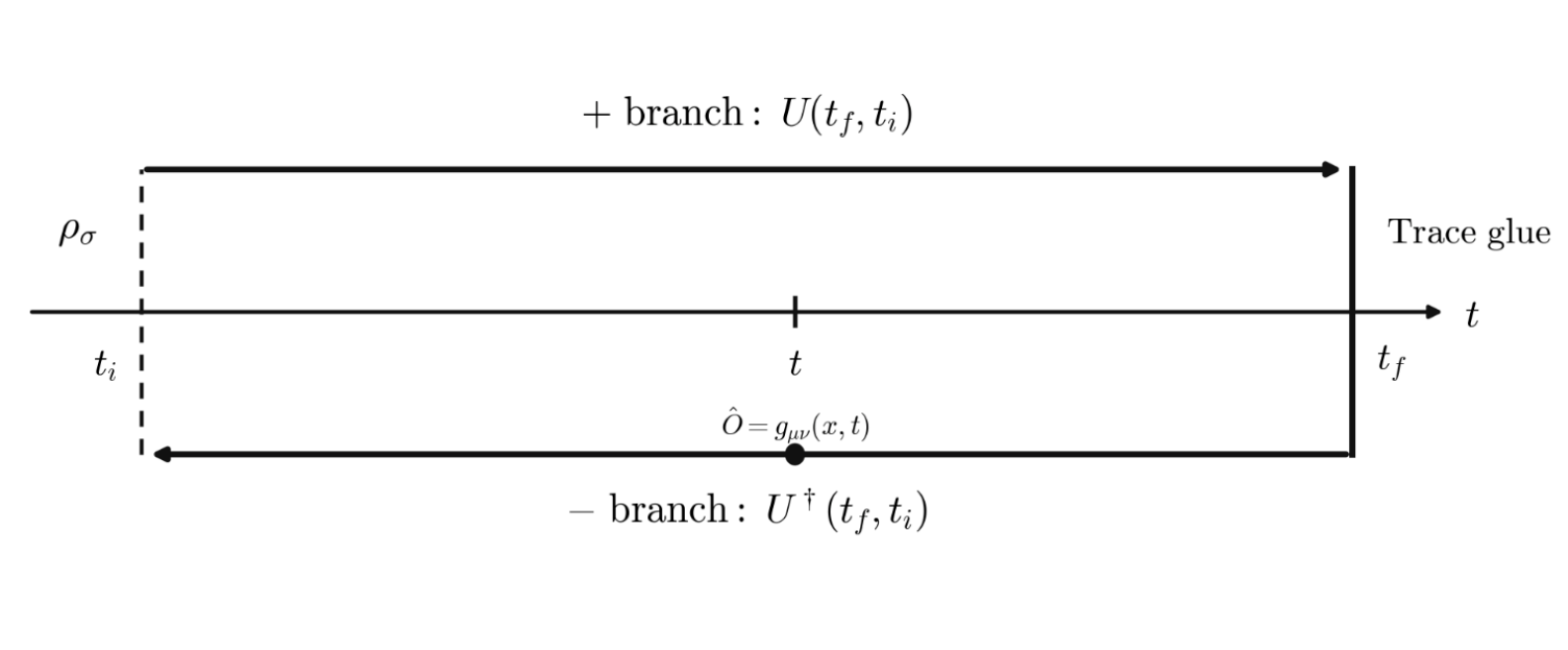}
\end{tabular}
\caption[]{Schwinger-Keldysh contour for the coherent state.}
\label{skcontour1}
\end{figure}

\subsection{Expectation values using the Schwinger-Keldysh contours \label{sec4.2}}

Thus \eqref{wavefunction} and \eqref{eq:O_plus_branch_replacement} took care of the kernels without $U(t_b, t_c)$ in \eqref{eq:trace_split_correct_all}. In the following we will argue that
each kernel $\bra{h_b,t_b}U(t_b,t_a)\ket{h_a,t_a}$ becomes a path integral with weight
$e^{+iS_{\rm tot}}$, and each kernel $\bra{h_a,t_a}U^\dagger(t_b,t_a)\ket{h_b,t_b}$ becomes the complex-conjugate path integral with weight
$e^{-iS_{\rm tot}}$.  This is where the doubled fields $(g_+,g_-)$ and the exponent
$e^{\,iS[g_+]-iS[g_-]}$ arise. Our aim then is the following.
Starting from the correct Hilbert-space decomposition \eqref{eq:trace_split_correct_all} we will convert each time-evolution kernel into a gauge-fixed gravitational path integral.
Throughout, $S_{\rm tot}[g,\text{gh}]$ will denote the full Lorentzian action including
Einstein--Hilbert, gauge-fixing and ghost terms. (Recall that we are ignoring matter DOFs for simplicity. Their inclusion is a straightforward matter to deal here.) For the forward (ket) evolution, we have\footnote{Here the notation $|h,t\rangle$ is used only to indicate that $h$ is a configuration
eigenvalue on the hypersurface $\Sigma_t$. Thus the kernel
$\langle h_f,t_f|U(t_f,t_i)|h^\pm_i,t_i\rangle$ is equivalent to the more economical notation
$\langle h_f|U(t_f,t_i)|h^\pm_i\rangle$, with the understanding that $h^\pm_i$ and $h_f$ live on
$\Sigma_i$ and $\Sigma_f$ respectively. In this sense the explicit time labels on the bras and
kets are slice labels, not an additional time evolution beyond that already encoded in
$U(t_f,t_i)$. Therefore it is perfectly consistent to represent this kernel by the path
integral with boundary conditions fixed on $\Sigma_i$ and $\Sigma_f$. We will continue to follow this convention throughout the paper.}:
\begin{equation}
\bra{h_f,t_f}U(t_f,t_i)\ket{h_i^+,t_i}
=
\int_{g_+|_{\Sigma_i}=h_i^+}^{g_+|_{\Sigma_f}=h_f}
\mathcal D g_+\;\mathcal D(\text{gh}_+)\;
\exp\!\Big(iS_{\rm tot}[g_+,\text{gh}_+]\Big),
\label{eq:kernel_plus_full}
\end{equation}
where ${\rm gh}_+$ is the ghost and the gauge fixing sector, and we are ignoring the matter sector.
For the backward (bra) evolution, we use the complex-conjugate amplitude.
It is convenient to split it at the intermediate time $t$:
\begin{align}
\bra{h_i^-,t_i}U^\dagger(t,t_i)\ket{h_t^-,t}
&=
\int_{g_-|_{\Sigma_i}=h_i^-}^{g_-|_{\Sigma_t}=h_t^-}
\mathcal D g_-^{(1)}\;\mathcal D(\text{gh}_-^{(1)})\;
\exp\!\Big(-iS_{\rm tot}[g_-^{(1)},\text{gh}_-^{(1)}]\Big),
\label{eq:kernel_minus_piece1}
\\[4pt]
\bra{h_t^+,t}U^\dagger(t_f,t)\ket{h_f,t_f}
&=
\int_{g_-|_{\Sigma_t}=h_t^+}^{g_-|_{\Sigma_f}=h_f}
\mathcal D g_-^{(2)}\;\mathcal D(\text{gh}_-^{(2)})\;
\exp\!\Big(-iS_{\rm tot}[g_-^{(2)},\text{gh}_-^{(2)}]\Big),
\label{eq:kernel_minus_piece2}
\end{align}
where the superscripts denote the parametric separation between the $+$ and the $-$ channels. They actually combine in a nice way as may be seen from the following:
Because we integrate over the intermediate boundary data $h_t^\pm$, the two backward
pieces \eqref{eq:kernel_minus_piece1} and \eqref{eq:kernel_minus_piece2} glue into a
single backward history $g_-$ defined on the full interval $[t_i,t_f]$, with the
values on $\Sigma_t$ kept explicit:

{\footnotesize
\bg
&&\int Dh_t^-\; Dh_t^+
\int_{h_i^-}^{h_t^-}\!\mathcal D g_-^{(1)}\;\mathcal D(\text{gh}_-^{(1)})\;\,e^{-iS_{\rm tot}[g_-^{(1)}, {\rm gh}_-^{(1)}]}
\;\;\times\;\;
\int_{h_t^+}^{h_f}\!\mathcal D g_-^{(2)}\;\mathcal D(\text{gh}_-^{(2)})\;\,e^{-iS_{\rm tot}[g_-^{(2)}, {\rm gh}_-^{(2)}]}\;
\delta(h_t^--h_t^+) \nonumber\\
&& \longrightarrow\;\;
\int_{h_i^-}^{h_f}\!\mathcal D g_-\;\mathcal D(\text{gh}_-)\;\,e^{-iS_{\rm tot}[g_-, {\rm gh}_-]},
\label{eq:gluing_statement}
\nd}
with the understanding that the operator insertion is evaluated using the boundary
values on $\Sigma_t$ (see below), and the delta function comes from \eqref{eq:O_plus_branch_replacement}.  This is the standard composition property of the
propagator written in functional-integral language. Now
substituting \eqref{eq:kernel_plus_full}, \eqref{eq:kernel_minus_piece1},
\eqref{eq:kernel_minus_piece2} into \eqref{eq:trace_split_correct_all}, and then using the gluing
property \eqref{eq:gluing_statement}, we obtain:
\bg
\mathrm{Tr}\!\left(\rho_\sigma\,\hat O_H(t)\right)
&=&
\int Dh_f \int Dh_i^+\int Dh_i^-\;
\bra{h_i^+,t_i}\rho_\sigma\ket{h_i^-,t_i}~ O[g_+](\mathbf x,t)\nonumber\\
&\times &
\int_{g_+|_{\Sigma_i}=h_i^+}^{g_+|_{\Sigma_f}=h_f}
\mathcal D g_+\;\mathcal D(\text{gh}_+)\;
\exp\!\Big(iS_{\rm tot}[g_+,\text{gh}_+]\Big)
\nonumber\\
&\times &
\int_{g_-|_{\Sigma_i}=h_i^-}^{g_-|_{\Sigma_f}=h_f}
\mathcal D g_-\;\mathcal D(\text{gh}_-)\;
\exp\!\Big(-iS_{\rm tot}[g_-,\text{gh}_-]\Big), 
\label{eq:SK_numerator_final}
\nd
where the common integration over $h_f$ (the same $h_f$ appears for both $g_+$ and $g_-$)
is precisely the trace/gluing condition at $\Sigma_f$. In a similar vein the denominator is obtained by setting $O[g_+](\mathbf x,t)\to 1$:
\bg
\mathrm{Tr}(\rho_\sigma)
&= &
\int Dh_f \int Dh_i^+\int Dh_i^-\;
\bra{h_i^+,t_i}\rho_\sigma\ket{h_i^-,t_i}
\nonumber\\
&\times &
\int_{g_+|_{\Sigma_i}=h_i^+}^{g_+|_{\Sigma_f}=h_f}
\mathcal D g_+\;\mathcal D(\text{gh}_+)\;
\exp\!\Big(iS_{\rm tot}[g_+,\text{gh}_+]\Big)
\nonumber\\
&\times &
\int_{g_-|_{\Sigma_i}=h_i^-}^{g_-|_{\Sigma_f}=h_f}
\mathcal D g_-\;\mathcal D(\text{gh}_-)\;
\exp\!\Big(-iS_{\rm tot}[g_-,\text{gh}_-]\Big) ,
\label{eq:SK_denominator_final}
\nd
where for both \eqref{eq:SK_numerator_final} and \eqref{eq:SK_denominator_final} the kernel may be read up from \eqref{wavefunction}. Putting everything together, the Schwinger--Keldysh expectation value is: 

{\scriptsize
\bg\label{Sirlauren}
\langle \hat O_H(t)\rangle_\sigma
& = &
\frac{\mathrm{Tr}\!\left(\rho_\sigma\,\hat O_H(t)\right)}{\mathrm{Tr}(\rho_\sigma)}\\
& = &
\frac{
\displaystyle
\int Dh_f \int Dh_i^+\int Dh_i^-\;
\Psi_\sigma[h_i^+]\,\Psi_\sigma^*[h_i^-]\;
\int_{h_i^+}^{h_f}\!\mathcal D g_+\,\mathcal D(\text{gh}_+)\;
\int_{h_i^-}^{h_f}\!\mathcal D g_-\,\mathcal D(\text{gh}_-)\;
O[g_+](\mathbf x,t)\;
e^{\,iS_{\rm tot}[+] - iS_{\rm tot}[-]}
}{
\displaystyle
\int Dh_f \int Dh_i^+\int Dh_i^-\;
\Psi_\sigma[h_i^+]\,\Psi_\sigma^*[h_i^-]\;
\int_{h_i^+}^{h_f}\!\mathcal D g_+\,\mathcal D(\text{gh}_+)\;
\int_{h_i^-}^{h_f}\!\mathcal D g_-\,\mathcal D(\text{gh}_-)\;
e^{\,iS_{\rm tot}[+] - iS_{\rm tot}[-]}
}.\nonumber
\nd}
where $S_{\rm tot}[\pm] \equiv S_{\rm tot}[g_\pm,\text{gh}_\pm]$. Note that 
the doubled weight
$e^{\,iS_{\rm tot}[g_+,\text{gh}_+] - iS_{\rm tot}[g_-,\text{gh}_-]}$
is forced solely by the appearance of $U$ and $U^\dagger$ in the trace.
Gauge-fixing and ghost sectors are duplicated on the two branches in the same way
as the metric field. The table below explains all the parameters appearing in \eqref{Sirlauren}.

\vskip.2in

\begin{center} 
\begin{tabular}{|c|c|}
\hline
{\bf Object} & {\bf Interpretation} \\
\hline
$g_+$ & field in ket evolution \\ \hline
$g_-$ & field in bra evolution \\ \hline
$S[g_+, {\rm gh}_+]$ & forward action \\ \hline
$S[g_-, {\rm gh}_-]$ & backward action \\ \hline
$h_f^+=h_f^-$ & trace over final states \\ \hline
$\Psi_\sigma[h_i^\pm]$ & initial coherent state \\ \hline
\end{tabular}
\end{center}
To see why \eqref{Sirlauren} can provide a non-zero answer for the expectation values, first note that $\Psi_\sigma[h] = \Psi_\Omega[h-\sigma]$.  Substituting into the Schwinger--Keldysh expression gives
us 
$\Psi_\sigma[h_i^+]\,\Psi_\sigma^*[h_i^-]
=
\Psi_\Omega[h_i^+ - \sigma]\,
\Psi_\Omega^*[h_i^- - \sigma]$.
We can then perform a change of variables in the initial data as
$h_i^\pm = \tilde h_i^\pm + \sigma$ which leave the measure invariant.
The expectation value \eqref{Sirlauren} becomes:

{\footnotesize
\begin{align}\label{Sirlauren2}
\langle \hat O_H(t)\rangle_\sigma
=
\frac{
\displaystyle
\int Dh_f \int D\tilde h_i^+\int D\tilde h_i^-\;
\Psi_\Omega[\tilde h_i^+]\,\Psi_\Omega^*[\tilde h_i^-]\;
\int_{\tilde h_i^+ + \sigma}^{h_f}\!\mathcal D g_+\;
\int_{\tilde h_i^- + \sigma}^{h_f}\!\mathcal D g_-\;
O[g_+]\;
e^{iS_{\rm tot}[+] - iS_{\rm tot}[-]}
}{
\displaystyle
\int Dh_f \int D\tilde h_i^+\int D\tilde h_i^-\;
\Psi_\Omega[\tilde h_i^+]\,\Psi_\Omega^*[\tilde h_i^-]\;
\int_{\tilde h_i^+ + \sigma}^{h_f}\!\mathcal D g_+\;
\int_{\tilde h_i^- + \sigma}^{h_f}\!\mathcal D g_-\;
e^{iS_{\rm tot}[+] - iS_{\rm tot}[-]}
} ,
\end{align}}
implying that 
the coherent state is equivalent to the vacuum path integral with
{shifted boundary data} as
$g_\pm(t_i,\mathbf x)=\tilde h_i^\pm(\mathbf x)+\sigma(\mathbf x)$. This is in fact the key point which makes the integral non-zero.

\noindent We can specialize our analysis to computing the expectation of the metric operator on the coherent state. The precise aim is to get the 
coherent state SK expression with explicit $Dh_i^\pm$ and trans-series action. The expectation value takes the form:

{\footnotesize
\bg\label{SKmetric}
\langle g_{\mu\nu}(x)\rangle_\sigma
&= &
\frac{1}{Z_\sigma}
\int Dh_f
\int D\tilde h_i^+\,D\tilde h_i^- \;
\Psi_\Omega[\tilde h_i^+]\,\Psi_\Omega^*[\tilde h_i^-]
\\
&\times &
\int_{g_+(t_i)=\tilde h_i^+ + \sigma}^{g_+(t_f)=h_f}
\!\!\!\!\!\!\!\!\!\!\mathcal D g_+\,\mathcal D(\text{gh}_+)
\int_{g_-(t_i)=\tilde h_i^- + \sigma}^{g_-(t_f)=h_f}
\!\!\!\!\!\!\!\!\!\!\mathcal D g_-\,\mathcal D(\text{gh}_-)
~~g_{+\mu\nu}(x)
\;
\exp\!\Big(
iS_{\rm tot}[g_+, {\rm gh}_+] - iS_{\rm tot}[g_-, {\rm gh}_-]
\Big) \nonumber
\nd}
where the coherent state is equivalent to the vacuum Schwinger--Keldysh
path integral with shifted initial boundary data
$g_\pm(t_i,\mathbf x)
=
\tilde h_i^\pm(\mathbf x) + \sigma(\mathbf x)$.
The SK contour is shown in {\bf figure \ref{skcontour1}} with 
$g_{+\mu\nu}$ on the $+$ branch. The normalization factor is:

{\scriptsize
\begin{equation}
Z_\sigma =
\int Dh_f
\int Dh_i^+\,Dh_i^- \;
\Psi_\sigma[h_i^+]\,\Psi_\sigma^*[h_i^-]
\int_{h_i^+}^{h_f}\!\mathcal D g_+ D(\text{gh}_+)\int_{h_i^-}^{h_f}\!\mathcal D g_- D(\text{gh}_-)\exp\!\Big(
iS_{\rm tot}[g_+, {\rm gh}_+] - iS_{\rm tot}[g_-, {\rm gh}_-]
\Big).
\end{equation}}
This can equivalently be formulated in terms of shifted boundary conditions. Equation \eqref{SKmetric} therefore represents our final expression for the expectation value, up to the resolution of three outstanding issues: (1) what is the typical structure of the total action 
$S_{\rm tot}[g,{\rm gh}]$ that appears in \eqref{SKmetric}? 
(2) why is the quantity in \eqref{SKmetric} manifestly real? and
(3) how is the analysis modified in the Euclidean formulation?

\subsection{The trans-series form of the gauge-fixed action $S_{\rm tot}[g, {\rm gh}]$ \label{sec4.3}}

Let us start by answering the first question on the structure of the action $S_{\rm tot}[g, {\rm gh}]$.
In a gauge theory (including gravity) the Schwinger--Keldysh weight must be built
from the \emph{gauge-fixed} action, i.e.\ the sum of the classical action, a gauge-fixing
term, and the Faddeev--Popov ghost action. On each SK branch we therefore take:
\begin{equation}\label{fargomey}
S_{\rm pert}[\lambda, {\rm g}]
\equiv
S_{\rm cl}[\lambda, {\rm g}]
+
S_{\rm gf}[\lambda, {\rm g}]
+
S_{\rm gh}[\lambda, {\rm g},\bar c,c]
+
S_{\rm ct}[\lambda, {\rm g}]
+\cdots,
\end{equation}
where $S_{\rm ct}$ denotes counterterms (and ``$\cdots$'' denotes any additional fields
or higher-derivative terms); $\lambda$ is the coupling constant; and ${\rm g} = (g, {\rm gh})$.  Beyond perturbation theory the effective infrared action receives
contributions from nonperturbative saddles of the gauge-fixed theory
(such as instantons, wormholes, or other semiclassical configurations).
Consequently the full action is derived over a chosen Minkowski minimum by including non-perturbative saddles implying that 
the \emph{full} gauge-fixed action admits a trans-series organization of the following form \cite{borel2,hetborel,dyson,dorigoni,unsal,grossper,gevrey,borelborel,ecalle,maximE,mirzakhanipaper}: 
\begin{equation}\label{bellcroft}
S_{\rm tot}[\lambda, \text{g}]
=
S_{\rm pert}[\lambda, \text{g}]
+
\sum_I e^{-A_I/\lambda}\,S_I[\lambda, \text{g}]
+
\sum_{I,J} e^{-(A_I+A_J)/\lambda}\,S_{IJ}[\lambda, \text{g}]
+\cdots,
\end{equation}
where $S_{\rm pert}$ contains the perturbative sector (including $S_{\rm gf}$ and
$S_{\rm gh}$) from \eqref{fargomey}, and $S_I, S_{IJ},\ldots$ denote the nonperturbative sectors (instantons,
wormholes, etc.) together with their gauge-fixing and ghost completions\footnote{In other words, non-perturbative saddles and their corresponding fluctuation determinants. Each sector admits its own perturbative expansion in $\lambda$. Such trans-series structures are known to be \emph{resurgent}:
the large-order behavior of the perturbative coefficients is controlled
by the nonperturbative sectors, and the different sectors are related
through resurgence relations \cite{borel2,hetborel,dyson,dorigoni,unsal,grossper,gevrey,borelborel,ecalle,maximE,mirzakhanipaper}.
As a consequence the formal trans-series in \eqref{bellcroft} is
Borel--Écalle summable \cite{borelborel,ecalle} and therefore defines an unambiguous
nonperturbative functional $S_{\rm tot}$ once an appropriate
resummation prescription is chosen.
In what follows we regard $S_{\rm tot}$ as this resummed infrared
effective action.}. (Detailed derivation of \eqref{bellcroft} using Lefschetz thimbles appears in section \ref{sec5}.) Inserting the coupling constant explicitly allows us the following replacement: $S_{\rm tot}[{\rm g}_\pm] \to S_{\rm tot}[\lambda; {\rm g}_\pm]$. The SK exponent is then
$\exp\!\Big(
iS_{\rm tot}[\lambda; g_+,\text{gh}_+]
-
iS_{\rm tot}[\lambda; g_-,\text{gh}_-]
\Big)$,
and inserting the trans-series decomposition yields:

{\footnotesize
\bg\label{malliniamey}
&&\exp\!\Big(
iS_{\rm tot}[\lambda; g_+,\text{gh}_+]
-
iS_{\rm tot}[\lambda; g_-,\text{gh}_-]
\Big) =
\\
&&
\exp\!\Big(
iS_{\rm pert}[\lambda; g_+,\text{gh}_+]
-
iS_{\rm pert}[\lambda; g_-,\text{gh}_-]
\Big)
\prod_I
\exp\!\Big(
i e^{-A_I/\lambda}\,S_I[\lambda; g_+, \text{gh}_+]
-
i e^{-A_I/\lambda}\,S_I[\lambda; g_-, \text{gh}_-]
\Big)
\nonumber\\
&\times &
\prod_{I,J}
\exp\!\Big(
i e^{-(A_I+A_J)/\lambda}\,S_{IJ}[\lambda; g_+, \text{gh}_+]
-
i e^{-(A_I+A_J)/\lambda}\,S_{IJ}[\lambda; g_-, \text{gh}_-]
\Big)
\times\cdots .\nonumber
\nd}
Plugging \eqref{malliniamey} in \eqref{SKmetric} tells us that
the coherent-state expectation value is not merely a contour integral over the doubled
fields, but a contour integral together with the boundary wavefunctional factors.
Suppressing only some inessential notation, the exact expression is therefore given by the following:

{\footnotesize
\bg\label{fitboudahl}
\langle g_{\mu\nu}(x)\rangle_\sigma
 & = &
\frac{N_\sigma(x)}{Z_\sigma(x)} = {1\over Z_\sigma(x)}
\int Dh_f
\int D\tilde h_i^+\,D\tilde h_i^- \;
\Psi_\Omega[\tilde h_i^+]\,\Psi_\Omega^*[\tilde h_i^-]\\
&\times &
\int_{g_+(t_i)=\tilde h_i^+ + \sigma}^{g_+(t_f)=h_f}
\!\!\!\!\!\!\!\!\!\!\mathcal D g_+\,\mathcal D(\text{gh}_+)
\int_{g_-(t_i)=\tilde h_i^- + \sigma}^{g_-(t_f)=h_f}
\!\!\!\!\!\!\!\!\!\!\mathcal D g_-\,\mathcal D(\text{gh}_-)
~~g_{+\mu\nu}(x)
\;
\exp\!\Big(
iS_{\rm tot}[\lambda; {\rm g}_+]
-
iS_{\rm tot}[\lambda;{\rm g}_-]
\Big) , \nonumber
\nd}
where $\mathcal D(\text{gh}_\pm)\equiv \mathcal D\bar c_\pm\,\mathcal Dc_\pm$, ${\rm g}_\pm \equiv (g_\pm, {\rm gh}_\pm)$ and
$S_{\rm tot}$ includes $S_{\rm gf}$ and $S_{\rm gh}$ on each branch.
Since $S_{\rm tot}$ already incorporates the resummed trans-series
structure of the infrared theory, the Schwinger--Keldysh functional
integral automatically includes both perturbative and nonperturbative
contributions. Expanding the doubled fields around a given saddle configuration
$s$ as:
\begin{equation}
g_\pm = g_\pm^{(s)} + \delta g_\pm,
\qquad
{\rm gh}_\pm = {\rm gh}_\pm^{(s)} + \delta{\rm gh}_\pm,
\label{doubledsaddleexpansion}
\end{equation}
the leading contribution to the path integral arises from the classical
value of the action evaluated on that configuration in the following way:
\begin{equation}
S_{\rm eff}[\lambda; {\rm g}^{(s)}_\pm] \equiv
S_{\rm tot}[\lambda;g_+^{(s)}, {\rm gh}_+^{(s)}]
-
S_{\rm tot}[\lambda;g_-^{(s)}, {\rm gh}_-^{(s)}].
\label{saddleactiondifference}
\end{equation}
The remaining integration over fluctuations together with the boundary integrations over
$h_f,\tilde h_i^\pm$ produces the loop expansion. Consequently the numerator and denominator each admit a
sector expansion labelled by the saddles $s$, and the expectation
value may be written schematically as:
\begin{equation}\label{oldbutkid}
\langle g_{\mu\nu}(x)\rangle_\sigma
\sim
\frac{
\sum\limits_s
{\cal W}_\sigma^{(s)}
\exp\!\Big(
iS_{\rm eff}[\lambda;{\rm g}_\pm^{(s)}]
\Big)
\Big[
g_{\mu\nu}^{(s;0)}(x;\sigma)
+
\lambda g_{\mu\nu}^{(s;1)}(x;\sigma)
+
\lambda^2 g_{\mu\nu}^{(s;2)}(x;\sigma)
+\cdots
\Big]
}{
\sum\limits_s
{\cal W}_\sigma^{(s)}
\exp\!\Big(
iS_{\rm eff}[\lambda;{\rm g}_\pm^{(s)}]
\Big)
\Big[
1
+
\lambda Z_\sigma^{(s;1)}
+
\lambda^2 Z_\sigma^{(s;2)}
+\cdots
\Big]
} ,
\end{equation}
where
${\cal W}_\sigma^{(s)}$ denotes the contribution from the boundary wavefunctionals and the corresponding saddle-point evaluation of the integrations over $h_f,\tilde h_i^\pm$. In other words, the factors:
\begin{equation}
\Psi_\Omega[\tilde h_i^+]\,\Psi_\Omega^*[\tilde h_i^-]
\end{equation}
have not disappeared; they are absorbed into the sector weights ${\cal W}_\sigma^{(s)}$ and, more generally, into the saddle coefficients multiplying each sector in \eqref{oldbutkid}. The exponentials in this expansion therefore arise directly from the
value of the trans-series action $S_{\rm tot}$ evaluated on the
corresponding doubled saddle configuration, while the wavefunctional factors determine how strongly a given saddle is sourced by the chosen coherent-state initial data. The quantities
$g_{\mu\nu}^{(s;0)}(x;\sigma),
g_{\mu\nu}^{(s;1)}(x;\sigma),
g_{\mu\nu}^{(s;2)}(x;\sigma)$ et cetera
have the following quantitative meaning. The leading term
$g_{\mu\nu}^{(s;0)}(x;\sigma)$
is the classical value of the metric operator on the $+$ branch evaluated on the saddle $s$:
\begin{equation}
g_{\mu\nu}^{(s;0)}(x;\sigma)
=
g_{+\mu\nu}^{(s)}(x;\sigma).
\label{leadinggsector}
\end{equation}
The higher coefficients
$g_{\mu\nu}^{(s;i)}(x;\sigma)$, with $i\ge 1$,
are the $i$-loop corrections in that same saddle sector. More precisely, they arise from integrating the operator insertion
$g_{+\mu\nu}(x)$
against the fluctuation measure around the doubled saddle
$(g_+^{(s)},{\rm gh}_+^{(s)};g_-^{(s)},{\rm gh}_-^{(s)})$,
including all gauge-fixing, ghost, and interaction vertices, as well as the boundary wavefunctional factors. Thus
$g_{\mu\nu}^{(s;i)}(x;\sigma)$
collects the full $i$-loop contribution to the metric one-point function in the sector $s$. Similarly, the coefficients
$Z_\sigma^{(s;1)},
Z_\sigma^{(s;2)}$ et cetera are the loop corrections from the normalization factor $Z_\sigma$ in the same saddle sector. Concretely,
$Z_\sigma^{(s;i)}$ is the full $i$-loop contribution to the denominator in the saddle sector $s$, again including the fluctuation determinants, higher Feynman graphs, ghost contributions, gauge-fixing terms, and the integrations over the boundary data weighted by the wavefunctionals. Thus $Z_\sigma^{(s;i)}$ is the normalization analogue of $g_{\mu\nu}^{(s;i)}$. Therefore \eqref{oldbutkid} should be interpreted as follows. Each saddle $s$ contributes
(i) a boundary-state weight ${\cal W}_\sigma^{(s)}$, (ii) a semiclassical phase
$\exp\!\Big(
iS_{\rm eff}[\lambda;{\rm g}_\pm^{(s)}]\Big)$,
(iii) a loop expansion of the operator insertion through $g_{\mu\nu}^{(s;i)}$, and
(iv) a loop expansion of the normalization through $Z_\sigma^{(s;i)}$.

In particular, the nonperturbative
weights $e^{-A_I/\lambda}$ appearing in the trans-series
representation of $S_{\rm tot}$ translate into the expected
nonperturbative sectors in the Schwinger--Keldysh expectation value.
Additionally, the parameter $s$ labels the contributing saddles (including nonperturbative ones), and the
loop coefficients $g_{\mu\nu}^{(s;i)}$ include \emph{all} fluctuation determinants and
Feynman graphs built from the gauge-fixed quadratic and interaction vertices,
including the ghost contributions dictated by $S_{\rm gh}$ and coupling constant $\lambda$. In particular, each saddle $s$ is evaluated with the coherent-state-induced shifted
initial data on the SK contour, while the ghost and gauge-fixing sectors are treated
in the standard doubled SK manner on the $+$ and $-$ branches. The coefficients
$Z_\sigma^{(s;i)}$ are defined analogously, but without the insertion of the metric operator.


\subsection{Reality of $\langle g_{\mu\nu}\rangle_\sigma$ on the Schwinger--Keldysh contour \label{sec4.4}}

Our next question is on the reality of the \eqref{SKmetric}, {\it i.e.}
we wish to show that the expectation value of the Hermitian operator
$\hat g_{\mu\nu}(x)$ in a coherent state with 
$\rho_\sigma = |\sigma\rangle\langle\sigma|$
is real when computed using the Schwinger--Keldysh (SK) contour. The operator argument is trivial.
By definition,
\bg\label{veritomey}
\langle g_{\mu\nu}(x)\rangle_\sigma
=
\frac{\mathrm{Tr}(\rho_\sigma\,\hat g_{\mu\nu}(x))}
{\mathrm{Tr}(\rho_\sigma)} ,
\nd
and since $\rho_\sigma$ is Hermitian and $\hat g_{\mu\nu}$ is Hermitian,
$\langle g_{\mu\nu}(x)\rangle_\sigma^*
=
\langle g_{\mu\nu}(x)\rangle_\sigma$.
Thus the result must be real. The SK path integral representation must reproduce this property. In fact it is not too hard to see the reality directly from the Schwinger-Keldysh representation.
The SK expression reads:
\begin{equation}
\begin{aligned}
\langle g_{\mu\nu}(x)\rangle_\sigma
&=
\frac{1}{Z_\sigma}
\int Dh_f
\int Dh_i^+ Dh_i^-
\;
\Psi_\sigma[h_i^+]\,\Psi_\sigma^*[h_i^-]
\\
&\qquad\times
\int_{h_i^+}^{h_f}\!\mathcal D {\rm g}_+
\int_{h_i^-}^{h_f}\!\mathcal D {\rm g}_-
\;
g_{+\mu\nu}(x)
\;
e^{\,iS[{\rm g}_+] - iS[{\rm g}_-]} ,
\end{aligned}
\end{equation}
where ${\rm g}_\pm = (g_\pm, {\rm gh}_\pm)$, and we henceforth suppress $\lambda$ unless mentioned otherwise.
The normalization $Z_\sigma$ has the same structure but without the insertion of $g_{+\mu\nu}$.
Taking the complex conjugate gives the following expression:
\begin{equation}
\begin{aligned}
\langle g_{\mu\nu}(x)\rangle_\sigma^*
&=
\frac{1}{Z_\sigma^*}
\int Dh_f
\int Dh_i^+ Dh_i^-
\;
\Psi_\sigma^*[h_i^+]\,\Psi_\sigma[h_i^-]
\\
&\qquad\times
\int_{h_i^+}^{h_f}\!\mathcal D {\rm g}_+
\int_{h_i^-}^{h_f}\!\mathcal D {\rm g}_-
\;
g_{+\mu\nu}(x)
\;
e^{-\,iS[{\rm g}_+] + iS[{\rm g}_-]} ,
\end{aligned}
\end{equation}
which flips the positions of $\pm$ variables in the wave-functions and in the action. However since they are dummy variables, 
we can exchange the $+$ and $-$ variables:
\begin{equation}
{\rm g}_+ \leftrightarrow {\rm g}_-,
\qquad
h_i^+ \leftrightarrow h_i^- .
\end{equation}
Since the integration measures as well as the one-point functions, measured on either the $+$ or the $-$ branches, are identical, the expression becomes identical to the original one. Therefore
$\langle g_{\mu\nu}(x)\rangle_\sigma^*
=
\langle g_{\mu\nu}(x)\rangle_\sigma$.
The reality therefore follows from the symmetry of the SK contour under interchange of the two branches. One may also verify that for the full action admitting a trans-series expansion as in \eqref{bellcroft},
in the SK weight,
\begin{equation}
e^{\,iS_{\rm tot}[g_+] - iS_{\rm tot}[g_-]},
\end{equation}
each sector appears with opposite signs on the two branches. Under complex conjugation and branch exchange, the structure is preserved. Therefore perturbative loops, instanton sectors, and all resurgent corrections maintain the same reality property\footnote{To see this from a toy example, consider the simple oscillatory integral
$I(\sigma)
=
\int_{-\infty}^{\infty} dx\;|\psi(x)|^2
\int_{x+\sigma}^{\infty} dy\; y\,e^{-i y^2}$.
Introducing the standard $i\epsilon$ prescription,
$e^{-i y^2} \to e^{-i(1-i\epsilon)y^2}$, with $\epsilon>0$,
the inner integral evaluates to
$\int_{a}^{\infty} dy\; y\,e^{-i y^2}
=
\frac{1}{2i}\,e^{-i a^2}$, where 
$a=x+\sigma$.
Thus
$I(\sigma)
=
\frac{1}{2i}
\int_{-\infty}^{\infty} dx\;|\psi(x)|^2
e^{-i(x+\sigma)^2}$ is generically complex because there is no second branch to enforce conjugation symmetry. By contrast, the SK analogue would be
\begin{equation}
I_{\rm SK}
=
\int dx\,dx'\,
|\psi(x)|^2
|\psi(x')|^2
\,
e^{\,i(x+\sigma)^2 - i(x'+\sigma)^2}, \nonumber
\end{equation}
which is manifestly real after exchanging $x \leftrightarrow x'$ under complex conjugation. Moreover this also remains non-zero justifying the non-vanishing of the 1-point function over a coherent state.}. To conclude, 
the single-branch oscillatory integral can be complex.  
The Schwinger--Keldysh doubled contour ensures that expectation values of Hermitian operators remain real because of the symmetry
\begin{equation}
({\rm g}_+,h_i^+) \leftrightarrow ({\rm g}_-,h_i^-).
\end{equation}
This remains true even in the presence of coherent states and full trans-series corrections in the action.

\subsection{Euclidean formulation of coherent-state expectation values \label{sec4.5}}

Finally, let us address the question as how would the expectation value \eqref{SKmetric} change once we go to the Euclidean signature.
In Euclidean signature the time coordinate is rotated as
$t = -\,i\tau$,
and the action becomes:
\bg\label{taghitag}
S_E[{\rm g}] = -\,i\,S_L[{\rm g}]\Big|_{t=-i\tau}, \nd 
where ${\rm g}$ is the generalized field content containing both the metric components and the corresponding Faddeev-Popov ghosts. In other words, ${\rm g} = (g, {\rm gh})$. (Further generalization by incorporating other matter fields and their corresponding ghosts can be made and we shall come back to this later.) 
The Euclidean path integral weight is therefore:
\bg\label{tagondho}
{\rm exp}\Big({-S_E[{\rm g}]}\Big), \nd
which is real (for real saddles) and exponentially damped. Clearly now,
unlike the Lorentzian case, there is no need for a forward/backward
doubling of fields in order to represent $U^\dagger O U$.
Euclidean path integrals naturally compute matrix elements of
$e^{-\tau H}$. This implies that 
the vacuum wavefunctional can be written as a Euclidean path integral
over a half-cylinder:
\begin{equation}\label{vperibindu}
\Psi_\Omega[h_0]
=
\langle h_0 | \Omega \rangle
=
\lim_{\beta\to\infty}
\int_{g(-\beta)=\text{free}}^{g(0)=h_0}
\mathcal D g\,\mathcal D(\text{gh})\;
e^{-S_E[{\rm g}]} ,
\end{equation}
where the Euclidean evolution $e^{-\beta H}$ projects onto the ground state in a somewhat similar way we saw for the Lorentzian case; and 
$\mathcal Dg = \mathcal Dg(-\beta)~ \mathcal Dg_{\rm bulk}$ implying that there is no boundary condition\footnote{This may be alternatively presented in the following way. Using the standard projection identity
$|\Omega\rangle
=
\lim\limits_{\beta\to\infty}
e^{-\beta H}\,|\chi\rangle$
where $|\chi\rangle$ is any state with nonzero overlap with the vacuum,
$\beta>0$, and the 
excited states are suppressed as $e^{-\beta(E_n-E_0)}$; followed by the 
insertion of a complete set of field eigenstates at Euclidean time $-\beta$ as
$\int Dh_{-\beta}\;
|h_{-\beta}\rangle
\langle h_{-\beta}| = 1$, the vacuum wavefunctions takes the form:
\begin{equation}
\Psi_\Omega[h_0]
=
\lim_{\beta\to\infty}
\int Dh_{-\beta}\;
\left(
\int_{g(-\beta)=h_{-\beta}}^{g(0)=h_0}
\mathcal D g\,\mathcal D(\mathrm{gh})\;
e^{-S_{E,\rm tot}[g,\mathrm{gh}]}
\right)
\langle h_{-\beta}|\chi\rangle.\nonumber
\end{equation}
This may look slightly different from \eqref{vperibindu}, but the point is the following. As $\beta\to\infty$,
$e^{-\beta H}$
projects onto the vacuum independently of the initial state $|\chi\rangle$.
The factor $\langle h_{-\beta}|\chi\rangle$ becomes irrelevant up to normalization. Thus effectively one may replace it by $1$, meaning that the initial boundary
metric is integrated over. Thus we express schematically the wavefunction as \eqref{vperibindu} where ``$g(-\beta)=\text{free}$'' means that the induced metric at $\tau=-\beta$
is integrated over rather than fixed.} at $\tau = -\beta$.
Therefore, independently of signature, we expect\footnote{This is not too hard to see from either canonical or path-integral formalism. In the canonical framework, using the Baker--Campbell--Hausdorff identity \eqref{twogonec}, one can easily see that
$D(\sigma)\ket{h}=\ket{h+\sigma}$ and 
$\bra{h}D(\sigma)=\bra{h-\sigma}$, implying
$\Psi_\sigma[h]
=
\bra{h}D(\sigma)\ket{\Omega}
=
\bra{h-\sigma}\Omega\rangle
=
\Psi_\Omega[h-\sigma]$. In the Euclidean path-integral derivation,
the vacuum wavefunctional can be written as a Euclidean path integral \eqref{vperibindu}.The displacement operator shifts the canonical coordinate
$g_{ij}\rightarrow g_{ij}+\sigma_{ij}$.
Acting on the boundary configuration eigenstate therefore shifts the
boundary data
$h_{ij}\rightarrow h_{ij}-\sigma_{ij}$ which amounts to the shifted wavefunction as before. In the functional-derivative derivation, 
the boundary operator can be written as
$D(\sigma)
=
\exp\!\left[
\int d^3x\;
\sigma^{ij}\frac{\delta}{\delta h_{ij}}
\right]$. Acting on the vacuum wavefunctional gives
$\Psi_\sigma[h]
=
\exp\!\left[
\sigma\cdot\frac{\delta}{\delta h}
\right]\Psi_0[h]$. 
Exponentiating the functional derivative translates the argument as
$\exp\!\left[
\sigma\cdot\frac{\delta}{\delta h}
\right]\Psi_\Omega[h]
=
\Psi_\Omega[h-\sigma]$. \label{corirheat}}
$\Psi_\sigma[h]
=
\langle h|\sigma\rangle
=
\Psi_\Omega[h-\sigma]$ and so
the Euclidean expectation value of a local operator
$g_{\mu\nu}(x)$ in the pure state $|\sigma\rangle$ is:

{\footnotesize
\bg\label{indigomara}
\langle g_{\mu\nu}(x)\rangle_{\sigma,E}
&= &
\frac{1}{Z_{\sigma,E}}
\int Dh_0\,Dh_0'
\;
\Psi_\sigma^*[h_0']\,\Psi_\sigma[h_0]
~\int_{g(0^-)=h_0'}^{g(0^+)=h_0}
\mathcal D g\,\mathcal D(\text{gh})\;
g_{\mu\nu}(x)\;
e^{-S_{E,{\rm tot}}[{\rm g}]}\nonumber\\
& = & 
\frac{1}{Z_{\sigma,E}}
\int D\tilde h_0\,D\tilde h_0'
\;
\Psi_\Omega^*[\tilde h_0']\,\Psi_\Omega[\tilde h_0]
~ \int_{g(0^-)=\tilde h_0'+\sigma}^{g(0^+)=\tilde h_0+\sigma}
\mathcal D g\,\mathcal D(\text{gh})\;
g_{\mu\nu}(x)\;
e^{-S_{E, {\rm tot}}[{\rm g}]}.
\nd}
where $Z_{\sigma, E}$ is similar to the numerator but without the source metric insertion; and we have inserted complete sets of configuration eigenstates at $\tau=0$, substituted $\Psi_\sigma[h]=\Psi_0[h-\sigma]$
and changed variables as
$h_0 = \tilde h_0 + \sigma$ and
$h_0' = \tilde h_0' + \sigma$.
Thus the coherent state is equivalent to the Euclidean vacuum path
integral with shifted boundary data
\begin{equation}
g(\tau=0,\mathbf x)
=
\tilde h_0(\mathbf x)+\sigma(\mathbf x).
\end{equation}
Notice that no doubling of fields is required in purely Euclidean signature. We can also see how the saddle and trans-series structure enter the coherent-state in the Euclidean signature.
Following the vacuum wavefunctional prepared by a Euclidean half-cylinder in \eqref{vperibindu}, the coherent wavefunctional 
becomes (see footnote \ref{corirheat}):
\begin{equation}
\Psi_\sigma[h_0]
=
\lim_{\beta\to\infty}
\int_{g(-\beta)=\text{free}}^{g(0)=h_0-\sigma}
\mathcal D g\,\mathcal D(\mathrm{gh})\;
e^{-S_{E,\rm tot}[g,\mathrm{gh}]} ,
\label{eq:coherent_E}
\end{equation}
where note that, unlike the GS case, there is \emph{no additional source term in the bulk action}.
The deformation appears only as a \emph{shift of the boundary data}. Moreover the Euclidean functional integral is evaluated by expanding the fields
around the stationary configurations of the Euclidean action.
Assuming that the infrared effective action admits the following trans-series
representation:
\begin{equation}\label{amaliagorom}
S_{\rm tot}[\lambda, {\rm g}]=
S_{\rm pert}[\lambda, {\rm g}]
+
\sum_I e^{-A_I/\lambda}\,S_I[\lambda, {\rm g}]
+
\sum_{I,J} e^{-(A_I+A_J)/\lambda}\,S_{IJ}[\lambda, {\rm g}]
+\cdots ,
\end{equation}
the path integral receives contributions from the perturbative saddle
and from additional nonperturbative saddles associated with the
sectors $I,J,\ldots$. The Euclidean expectation value is as in \eqref{indigomara}. As before, expanding the fields around a saddle configuration $s$ as
$
{\rm g} = {\rm g}^{(s)}(\sigma) + \delta {\rm g} ,
$
where ${\rm g} = (g, {\rm gh})$,
the leading contribution is determined by the classical value of the
action evaluated on that configuration,
$
S_{\rm tot}[\lambda; {\rm g}^{(s)}].
$
Integrating over the fluctuations $\delta {\rm g}$ produces the usual loop
expansion around the saddle.  Performing the saddle expansion in both
numerator and denominator therefore yields a sector decomposition
of the form:

{\footnotesize
\begin{equation}\label{oldbutkid2}
\langle g_{\mu\nu}(x)\rangle_\sigma
\sim
\frac{
\sum\limits_s
{\cal W}_{\sigma,E}^{(s)}
\exp\Big(-S_{\rm tot}[\lambda; {\rm g}^{(s)}]\Big)
\Big[
g^{(s;0)}_{\mu\nu}(x;\sigma)
+
\lambda g^{(s;1)}_{\mu\nu}(x;\sigma)
+
\lambda^2 g^{(s;2)}_{\mu\nu}(x;\sigma)
+\cdots
\Big]
}{
\sum\limits_s
{\cal W}_{\sigma,E}^{(s)}
\exp\Big(-S_{\rm tot}[\lambda; {\rm g}^{(s)}]\Big)
\Big[
1
+
\lambda Z^{(s;1)}_{\sigma,E}
+
\lambda^2 Z^{(s;2)}_{\sigma,E}
+\cdots
\Big]
} ,
\end{equation}}
which is similar to what we had in \eqref{oldbutkid}. Here
${\cal W}_{\sigma,E}^{(s)}$
denotes the contribution from the boundary wavefunctional product
$\Psi_\sigma^*[h_0']\,\Psi_\sigma[h_0]$
together with the corresponding saddle-point evaluation of the boundary integrations over $h_0$ and $h_0'$. (Or alternatively using the boundary wavefunctional product $\Psi_\Omega^*[\tilde{h}_0']\,\Psi_\Omega[\tilde{h}_0]$ but with shifted $\tilde{h}_0$ and $\tilde{h}'_0$. Either formulation leads to the same expectation value.) Thus the wavefunctional factors do not disappear in \eqref{oldbutkid2}; rather, they are absorbed into the sector weights ${\cal W}_{\sigma,E}^{(s)}$.

The other parameters appearing in \eqref{oldbutkid2} are defined as follows. The coefficient
$g_{\mu\nu}^{(s;0)}(x;\sigma)$
is the leading saddle value of the metric operator evaluated on the configuration ${\rm g}^{(s)}(\sigma)$, while
$g_{\mu\nu}^{(s;1)}(x;\sigma),
g_{\mu\nu}^{(s;2)}(x;\sigma)$ et cetera
are the one-loop, two-loop, and higher-loop corrections in the same saddle sector, obtained by integrating the operator insertion against the fluctuation measure around ${\rm g}^{(s)}(\sigma)$. Similarly,
$Z_{\sigma,E}^{(s;1)},
Z_{\sigma,E}^{(s;2)}$ et cetera
are the corresponding loop corrections from the denominator, i.e.\ from the normalization factor $Z_{\sigma,E}$, in the same saddle sector. Hence \eqref{oldbutkid2} keeps track of four ingredients in each sector $s$: the boundary-state weight ${\cal W}_{\sigma,E}^{(s)}$, the classical saddle action $S_{\rm tot}[\lambda;{\rm g}^{(s)}]$, the loop expansion of the numerator through $g_{\mu\nu}^{(s;i)}$, and the loop expansion of the denominator through $Z_{\sigma,E}^{(s;i)}$.

As in \eqref{oldbutkid}, the conclusion remains identical: 
The exponential weights appearing in this expression are 
determined directly by the value of the trans-series action
$S_{\rm tot}$ evaluated on each saddle configuration.
In particular, the factors $e^{-A_I/\lambda}$ present in the
trans-series structure of $S_{\rm tot}$ generate the expected
nonperturbative contributions in the Euclidean functional integral.

\subsection{An alternative but exact formulation of $\langle g_{\mu\nu}\rangle_\sigma$ \label{sec4.6}}

The structural similarity of \eqref{oldbutkid} and \eqref{oldbutkid2} is expected, but one can also put the above results in a more compact form using Picard--Lefschetz theory. In the process we can also clarify how $Z^{(s,n)}_\sigma$ appear in \eqref{oldbutkid} and \eqref{oldbutkid2}. The only point that must be kept explicit is that the boundary wavefunctional factors do not disappear in this formulation; rather, they are absorbed into the definitions of the saddle coefficients ${\cal Z}_s$ and ${\cal N}_s$ defined in the following discussion, which sets the stage for formulating the functional integrals of previous section in terms of integrals along Lefschetz thimbles.

We will start with the Euclidean story for simplicity (and to avoid the complications of the $+$ and $-$ channels in the Lorentzian SK picture). Expanding the action around the saddle yields:
\begin{equation}
S_{\rm tot}[\lambda; {\rm g}]
=
S_{\rm tot}[\lambda, {\rm g}^{(s)}(\sigma)]
+
\frac12\,\delta {\rm g}\,\mathcal O_s(\sigma, \lambda)\,\delta {\rm g}
+
S_{\rm int}^{(s)}(\lambda, \sigma)[\delta {\rm g}],
\end{equation}
where $\mathcal O_s(\lambda, \sigma) \equiv {\delta^2 S_{\rm tot}[\lambda, {\rm g}]\over \delta{\rm g} ~\delta{\rm g}}\Big\vert_{{\rm g} = {\rm g}^{(s)}}$ is the quadratic fluctuation operator and
$S_{\rm int}^{(s)}(\sigma, \lambda)[\delta {\rm g}]
=
\sum\limits_{n\ge 3}
\frac{1}{n!}\,
{\delta^n S_{\rm tot}(\lambda, {\rm g})\over \delta{\rm g}^n}\Bigg\vert_{{\rm g} = {\rm g}^{(s)}(\sigma)}
(\delta {\rm g})^n$. The Gaussian integral over $\delta {\rm g}$ produces the determinant
$(\det\mathcal O_s)^{-1/2}$, while the interaction term generates
the usual loop expansion. The denominator therefore takes the form
\begin{equation}
Z_\sigma
=
\sum_s
e^{-S_{\rm tot}[\lambda, {\rm g}^{(s)}(\sigma)]}
\,\mathcal Z_s(\lambda, \sigma),
\end{equation}
where $S_{\rm tot}[\lambda;{\rm g}^{(s)}]$ denotes the value of the
gauge-fixed action evaluated on the saddle configuration
${\rm g}^{(s)}$.  The quantity ${\cal Z}_s(\lambda,\sigma)$ arises from integrating over the fluctuations
$\delta{\rm g}$ around this saddle together with the boundary integrations weighted by the appropriate wavefunctionals, and therefore depends on the
coupling $\lambda$, the saddle label $s$, and the coherent-state parameter $\sigma$, but no longer on the
dynamical fields themselves. One may express this as:
\begin{equation}\label{amaliabru2}
\mathcal Z_s(\lambda, \sigma)
=
{\cal W}^{(s)}_{\sigma,E}\,
[\det\mathcal O_s(\lambda, \sigma)]^{-1/2}
\Big[
1+\lambda c^{(s)}_1(\sigma)+\lambda^2 c^{(s)}_2(\sigma)+\cdots
\Big],
\end{equation}
where ${\cal W}^{(s)}_{\sigma,E}$ denotes the contribution of the Euclidean boundary wavefunctional product
$\Psi_\sigma^*[h_0']\,\Psi_\sigma[h_0]$
after saddle expansion of the boundary integrations. From this expression we may identify the denominator coefficients appearing in \eqref{oldbutkid} and \eqref{oldbutkid2} as
$Z^{(s,n)}_{\sigma,E}
=
{\cal W}^{(s)}_{\sigma,E}\,
[\det\mathcal O_s(\lambda, \sigma)]^{-1/2}\,
c_n^{(s)}(\sigma)$, and it is no surprise that the story is quite identical to what we had earlier. This implies that the numerator of the expectation value should also resonate the same staructure that we had before. And indeed the numerator becomes:
\bg\label{amaliabru}
&& N_\sigma(x)
=
\sum_s
\exp\!\left(
-
S_{\rm tot}[\lambda, {\rm g}^{(s)}(\sigma)]
\right)
\mathcal{N}_s(x,\sigma,\lambda),
\\
&& \mathcal{N}_s(x,\sigma,\lambda)
=
{\cal W}^{(s)}_{\sigma,E}\,
\left(\det\mathcal{O}_s(\sigma,\lambda)\right)^{-1/2}
\left(
g^{(s;0)}_{\mu\nu}(x;\sigma)
+
\lambda g^{(s;1)}_{\mu\nu}(x;\sigma)
+
\lambda g^{(s;2)}_{\mu\nu}(x;\sigma)
+
\cdots
\right), \nonumber
\nd
which may be extracted from the path-integral expression presented earlier by expanding over the saddles ${\rm g}^{(s)}(\sigma)$. Here
$g_{\mu\nu}^{(s;0)}(x;\sigma)
=
g_{\mu\nu}^{(s)}(x;\sigma)$
is the classical value of the metric on the saddle, while
$g_{\mu\nu}^{(s;1)}(x;\sigma),
g_{\mu\nu}^{(s;2)}(x;\sigma)$ et cetera
are the one-loop, two-loop, and higher-loop corrections obtained by inserting the metric operator into the fluctuation integral around the same saddle exactly as we saw earlier. Putting everything together, 
the expectation value therefore admits the sector expansion:
\begin{equation}\label{kinmaline}
\langle g_{\mu\nu}(x)\rangle_\sigma
=
\frac{
\sum\limits_s
\exp\!\left[
-
S_{\rm tot}(\lambda, {\rm g}^{(s)}(\sigma))
\right]
\mathcal{N}_s(x,\sigma,\lambda)
}{
\sum\limits_s
\exp\!\left[
-
S_{\rm tot}(\lambda, {\rm g}^{(s)}(\sigma))
\right]
\mathcal{Z}_s(\sigma,\lambda)
},
\end{equation}
where $\mathcal N_s(x, \sigma, \lambda)$ and $\mathcal Z_s(\sigma, \lambda)$ may be read from \eqref{amaliabru2} and \eqref{amaliabru}. Thus the boundary wavefunctional product is still present in \eqref{kinmaline}; it is simply contained inside the coefficients ${\cal Z}_s$ and ${\cal N}_s$ through the factor ${\cal W}^{(s)}_{\sigma,E}$.

For $S_{\rm tot}(\lambda, {\rm g}^{(s)}(\sigma))$, we can use the following procedure. Because the effective action itself admits a trans-series structure as in \eqref{amaliagorom}, 
the exponential weight appearing in the path integral becomes:
\begin{equation}
\exp\!\left(
-
S_{\rm tot}[\lambda;{\rm g}]
\right)
=
\exp\!\left(
-
S_{\rm pert}[\lambda;{\rm g}]
\right)
\exp\!\left(
-
\sum_I e^{-A_I/\lambda} S_I[\lambda;{\rm g}]
-
\cdots
\right),
\end{equation}
where $I$ counts the various non-perturbative saddles (instantons, worm-holes etc.) in the theory. The label $I$ appearing in the trans-series representation of the
effective action should not be confused with the saddle label $s$
arising in the functional integral.  The index $I$ enumerates
nonperturbative sectors (instantons, wormholes, etc.) that contribute
to the construction of the infrared effective action itself, typically
computed around a chosen vacuum background.  By contrast, the index
$s$ labels classical saddle configurations of the full path integral
constructed from this effective action. This may be represented by the following diagram:

{\footnotesize
\begin{equation}
\begin{array}{ccc}
\textbf{Microscopic theory} 
& \longrightarrow &
\textbf{Infrared effective action}
\\[6pt]
\text{UV dynamics (strings, branes, etc.)}
& &
S_{\rm tot}[\lambda;{\rm g}]
=
S_{\rm pert}[\lambda;{\rm g}]
+
\displaystyle\sum_I e^{-A_I/\lambda} S_I[\lambda;{\rm g}]
+\cdots
\\[12pt]
& &
\downarrow
\\[10pt]
\textbf{Saddle expansion} & \longleftarrow &
\textbf{Functional integral}
\\[6pt]
Z_\sigma
=
\displaystyle
\sum_s
\exp\!\left(
-
S_{\rm tot}[\lambda;{\rm g}^{(s)}(\sigma)]
\right)
\,
{\cal Z}_s(\sigma,\lambda)& &
\displaystyle
Z_\sigma
=
\int D{\rm g}\;
\exp\!\left(
-
S_{\rm tot}[\lambda;{\rm g}]
\right)
\\[12pt]
\end{array}
\end{equation}}

\vskip.1in

\noindent The above diagram makes it clear that the index $I$ labels the nonperturbative sectors appearing in the
trans-series representation of the effective action, while the index
$s$ labels classical saddle configurations of the functional integral
constructed from this action.  The two sets of labels correspond to
distinct levels of the construction. Now, because the effective action itself admits a trans-series structure as \eqref{amaliagorom},
the Euclidean weight may be written as:
\begin{equation}
e^{-S_{\rm tot}[\lambda, {\rm g}]}
=
e^{-S_{\rm pert}[\lambda, {\rm g}]}
\left[
1
+
\sum_I e^{-A_I/\lambda}\,{\cal C}_I[\lambda, {\rm g}]
+
\sum_{I,J} e^{-(A_I+A_J)/\lambda}\,{\cal C}_{IJ}[\lambda, {\rm g}]
+\cdots
\right],
\end{equation}
where the coefficients ${\cal C}_I$, ${\cal C}_{IJ}$, $\ldots$ are determined by
expanding the exponential of the full trans-series action. In particular,
${\cal C}_{IJ}$ receives contributions both from the genuine sector
$S_{IJ}$ and from products of lower sectors such as $S_I S_J$, and
similarly for higher orders. One may determine the exact form of these coefficients by introducing the following trick. Let us write the formal nonperturbative contribution as:

{\footnotesize
\begin{equation}
\Delta_{\rm np}[\lambda, {\rm g}]
\equiv
\sum_I e^{-A_I/\lambda}\,S_I[\lambda, {\rm g}]
+
\sum_{I,J} e^{-(A_I+A_J)/\lambda}\,S_{IJ}[\lambda, {\rm g}]
+
\sum_{I,J,K} e^{-(A_I+A_J+A_K)/\lambda}\,S_{IJK}[\lambda, {\rm g}]
+\cdots ,
\end{equation}}
so that the full effective action in \eqref{amaliagorom} may be written as
$S_{\rm tot}[\lambda, {\rm g}]
=
S_{\rm pert}[\lambda, {\rm g}]
+
\Delta_{\rm np}[\lambda, {\rm g}]$.
(This way $\Delta_{\rm np}[\lambda, {\rm g}]$ captures the genuine non-perturbative contributions.) If we now introduce formal variables
$q_I$ as
$q_I \equiv e^{-A_I/\lambda}$,
then:
\bg\label{amaliamey}
&& \exp\!\left(
-
\Delta_{\rm np}[\lambda, {\rm g}]
\right)
=
\sum_{n=0}^{\infty}
\sum_{I_1,\ldots,I_n}
q_{I_1}\cdots q_{I_n}\,
{\cal C}_{I_1\cdots I_n}[\lambda, {\rm g}],
\nonumber\\
\implies ~ &&
{\cal C}_{I_1\cdots I_n}[\lambda, {\rm g}]
=
\frac{1}{n!}
\left.
\frac{\partial^n}{\partial q_{I_1}\cdots \partial q_{I_n}}
\exp\!\left(
-
\Delta_{\rm np}[\lambda, {\rm g}]
\right)
\right|_{q_{I_1} = q_{I_2}= \cdots = q_{I_n}=0} .
\nd
which is the exact answer to any order in $n$. In fact, in this language the coefficient functionals
${\cal C}_{I_1\cdots I_n}$ are defined directly as the coefficients of
the exponential expansion of the full trans-series action. We see that 
they receive contributions both
from the genuine higher nonperturbative sectors already present in the
effective action and from products of lower sectors generated by
expanding the exponential.
The first few coefficients can be easily computed from the derivative formula in \eqref{amaliamey} as: 

{\scriptsize
\bg
&& {\cal C}_I[\lambda, {\rm g}]
=
-\,S_I[\lambda, {\rm g}],\qquad
{\cal C}_{IJ}[\lambda, {\rm g}]
=
-\,S_{IJ}[\lambda, {\rm g}]
+
\frac{1}{2}\,
S_I[\lambda, {\rm g}]\,S_J[\lambda, {\rm g}],
\\
&&  {\cal C}_{IJK}[\lambda, {\rm g}]
=
-\,S_{IJK}[\lambda, {\rm g}]
+
S_I[\lambda, {\rm g}]\,S_{JK}[\lambda, {\rm g}]
+
S_J[\lambda, {\rm g}]\,S_{IK}[\lambda, {\rm g}]
+
S_K[\lambda, {\rm g}]\,S_{IJ}[\lambda, {\rm g}]
-
\frac{1}{6}\,
S_I[\lambda, {\rm g}]\,S_J[\lambda, {\rm g}]\,S_K[\lambda, {\rm g}],\nonumber
\nd}
and similarly at higher orders. Thus the path integral naturally produces contributions proportional to
$e^{-A_I/\lambda}$, $e^{-(A_I+A_J)/\lambda}$ etc.,
which correspond to the expected nonperturbative sectors.

\subsection{Derivation using thimbles and the Picard--Lefschetz theory \label{sec4.7}}

We now reformulate the functional integral in terms of Lefschetz
thimbles\footnote{For details on Lefschetz thimbles, see
\cite{wittenlefschetz}, where the Picard--Lefschetz (PL) theory \cite{picardlefschetz} is
explained in a remarkably transparent manner.  See also
\cite{hetborel} for an application of PL theory to Glauber--Sudarshan
states. Again, we shall refrain from using any stringy inputs in our analysis here.}.  For clarity we first discuss the Euclidean formulation; the
Lorentzian Schwinger--Keldysh construction follows analogously with
the doubling of fields on the $+$ and $-$ branches.

\subsubsection{Euclidean path-integral formulation \label{sec4.7.1}}

In the Euclidean formulation, the expectation value of the metric operator in the coherent state
labelled by $\sigma$ can be written as:
\begin{equation}\label{camekim}
\langle g_{\mu\nu}(x)\rangle_\sigma
=
\frac{1}{Z_\sigma}
\int Dh_0\,Dh_0'\;
\Psi_\sigma^*[h_0']\,\Psi_\sigma[h_0]
\int_{{\cal C}(h_0',h_0)}
{\cal D}{\rm g}\;
g_{\mu\nu}(x)
\exp
\left(
-
S_{\rm tot}[\lambda;{\rm g}]
\right),
\end{equation}
where
${\rm g} \equiv (g,\text{ghosts},\text{matter},\ldots) $
denotes the full multiplet of fields appearing in the gauge--fixed
theory, and where the contour ${\cal C}(h_0',h_0)$ indicates the Euclidean
functional integral with boundary data determined by the coherent-state
wavefunctionals. The normalization factor is defined similarly, but without
the insertion of $g_{\mu\nu}(x)$:
\begin{equation}
Z_\sigma
=
\int Dh_0\,Dh_0'\;
\Psi_\sigma^*[h_0']\,\Psi_\sigma[h_0]
\int_{{\cal C}(h_0',h_0)}
{\cal D}{\rm g}\;
\exp
\left(
-
S_{\rm tot}[\lambda;{\rm g}]
\right).
\end{equation}
We assume that the infrared effective action admits the trans--series
organization \eqref{bellcroft}
where $S_{\rm pert}$ denotes the perturbative gauge--fixed action,
while the functionals $S_I$, $S_{IJ}$, $\ldots$ represent the
nonperturbative sectors of the effective theory, such as instantons
or wormholes together with their fluctuation determinants.

\begin{figure}[h]
\centering
\begin{tabular}{c}
\includegraphics[width=6in]{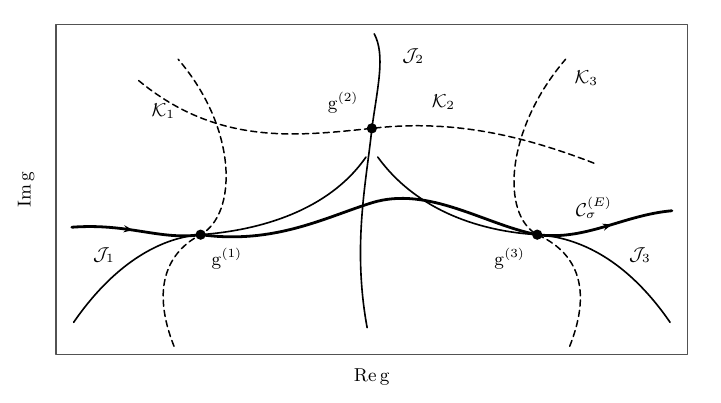}
\end{tabular}
\caption[]{Euclidean Picard--Lefschetz decomposition of the coherent-state path integral in \eqref{camekim}. The thick oriented curve denotes the original integration cycle $\mathcal C_{\sigma}^{(E)}$. The thin solid curves are the Lefschetz thimbles $\mathcal J_s\equiv\mathcal J_s^{(E)}(\sigma)$ attached to the critical configurations ${\rm g}^{(s)}\equiv{\rm g}^{(s)}(\sigma)$, indicated by black points. The dashed curves are the dual cycles $\mathcal K_s\equiv\mathcal K_s^{(E)}(\sigma)$; their oriented intersections with $\mathcal C_{\sigma}^{(E)}$ determine the integers $n_s$ in the decomposition \eqref{abbiverit}. The geometry is schematic and no numerical scale in the complexified field space is implied.}
\label{thimble1}
\end{figure}

The coherent--state parameter $\sigma$ does not appear explicitly as a
source term in the bulk action $S_{\rm tot}$.  Instead it enters
through the boundary conditions that define the coherent state in the
functional integral, equivalently through the boundary wavefunctional
product
$\Psi_\sigma^*[h_0']\,\Psi_\sigma[h_0]$.
Consequently the saddle configurations depend
on $\sigma$ through these boundary data,
\begin{equation}
{\rm g}^{(s)} \;\longrightarrow\; {\rm g}^{(s)}(\sigma).
\end{equation}
In the Picard--Lefschetz formulation the dynamical fields are first
complexified and the thimbles are defined through a gradient flow in
the space of field configurations. Since the dynamical variables are
fields, the flow acts on the full field configuration
\begin{equation}
{\rm g}(x) =
\big(
g_{\mu\nu}(x),
c(x),
\bar c(x),
\text{matter fields}(x),
\ldots
\big).
\end{equation}
The flow therefore evolves an entire field configuration rather than a
single spacetime point. Denoting the flow parameter by $\tau$, the
fields become
${\rm g}(x,\tau)$. The downward Picard--Lefschetz flow equation then takes the form:
\begin{equation}\label{dudhcan1}
\frac{\partial {\rm g}^{\alpha}(x,\tau)}{\partial \tau}
=
\overline{
\frac{\delta S_{\rm tot}[\lambda;{\rm g}]}
{\delta {\rm g}^{\alpha}(x,\tau)}
},
\end{equation}
where the index $\alpha$ labels all components of the dynamical field
multiplet,
${\rm g}^{\alpha} =
\big(
g_{\mu\nu},
c,
\bar c,$ $
\text{matter fields},
\ldots
\big)$, which should now be considered as functions of $(x, \tau)$. More generally one may introduce a metric $G^{\alpha\beta}$ on field
space, leading to:
\begin{equation}
\frac{\partial {\rm g}^{\alpha}(x,\tau)}{\partial \tau}
=
G^{\alpha\beta}
\overline{
\frac{\delta S_{\rm tot}[\lambda;{\rm g}]}
{\delta {\rm g}^{\beta}(x,\tau)}
}.
\end{equation}
In most physical applications the field--space metric is chosen to be
flat, so that 
$G^{\alpha\beta} = \delta^{\alpha\beta}$.
Along this flow the real part of the action increases monotonically, whereas the imaginary part remains constant. In other words\footnote{There is an elegant way to express \eqref{dudhcandi} following \cite{wittenlefschetz} in the following way. Let ${\rm g}^{\alpha}(x,\tau)$ denote the complexified field
configuration, where $\alpha$ labels the components of the full field
multiplet.  The downward Picard--Lefschetz flow is \eqref{dudhcan1}.
Introducing the field--space inner product
$\langle A,B\rangle
=
\int d^dx\,
A^{\alpha}(x)\,\overline{B^{\alpha}(x)}$,
the evolution of the action along the flow may be written as
$\frac{d}{d\tau}S_{\rm tot}
=
\left\langle
\frac{\delta S_{\rm tot}}{\delta {\rm g}},
\frac{\partial {\rm g}}{\partial \tau}
\right\rangle$.
Substituting the flow equation gives
$\frac{d}{d\tau}\,{\rm Re}\,S_{\rm tot}
=
\left\|
\frac{\delta S_{\rm tot}}{\delta {\rm g}}
\right\|^2
\ge 0$, 
where
$\left\|
\frac{\delta S_{\rm tot}}{\delta {\rm g}}
\right\|^2
=
\int d^dx\,
\left|
\frac{\delta S_{\rm tot}}
{\delta {\rm g}(x,\tau)}
\right|^2$.
Similarly, the imaginary part remains constant along the flow:
$\frac{d}{d\tau}\,{\rm Im}\,S_{\rm tot}
=
0$. This is mathematically cleaner and avoids the impression that the LHS of \eqref{dudhcandi} depends on 
$x$.}:
\begin{equation}\label{dudhcandi}
\frac{d}{d\tau}
{\rm Re}\,
S_{\rm tot}[{\rm g}(x,\tau)]
=
\int d^dx
\left|
\frac{\delta S_{\rm tot}}
{\delta {\rm g}(x,\tau)}
\right|^2
\ge 0,~~~~~~~~
\frac{d}{d\tau}
{\rm Im}\,
S_{\rm tot}[{\rm g}(x,\tau)]
=
0 .
\end{equation}
The Lefschetz thimble ${\cal J}_s$ attached to a saddle
${\rm g}^{(s)}(\sigma)$ is defined as the set of all field
configurations that approach this saddle under the downward flow,
\begin{equation}
{\cal J}_s
=
\left\{
{\rm g}(x,0)
\;\middle|\;
\lim_{\tau\to -\infty}
{\rm g}(x,\tau)
=
{\rm g}^{(s)}(x;\sigma)
\right\}.
\end{equation}
The dual cycles ${\cal K}_s$ are defined using the upward flow and
determine the intersection numbers appearing in the decomposition of
the integration contour.
After complexifying the fields in the gauge--fixed theory, the
integration cycle ${\cal C}$ in field space can be decomposed into a
sum of Lefschetz thimbles attached to the critical points of the
action:
\begin{equation}\label{abbiverit}
{\cal C}
=
\sum_s n_s\,{\cal J}_s ,
\end{equation}
where the integers $n_s$ are intersection numbers determined by the
choice of integration contour. This is shown in {\bf figure \ref{thimble1}} alongwith the corresponding saddles. These saddles associated with the thimbles satisfy the stationary
condition:
\begin{equation}\label{sadthimble}
\frac{\delta S_{\rm tot}[\lambda;{\rm g}]}
{\delta {\rm g}}
\Big|_{{\rm g}={\rm g}^{(s)}(\sigma)}
=
0 .
\end{equation}
It is important to emphasize that the coherent--state parameter
$\sigma$ does not modify the functional form of the action
$S_{\rm tot}[\lambda;{\rm g}]$.  Instead it enters through the
boundary data that define the functional integral, or equivalently
through the boundary wavefunctional product
$\Psi_\sigma^*[h_0']\Psi_\sigma[h_0]$. Consequently the
equations of motion obtained from \eqref{sadthimble}
remain unchanged, while the classical saddle configurations depend on
$\sigma$ through the boundary conditions of the path integral.  This
dependence is therefore indicated explicitly by writing the saddle
backgrounds as
${\rm g}^{(s)}(\sigma)$. Thus the parameter $\sigma$ labels different boundary conditions of
the functional integral, and the corresponding thimbles
${\cal J}_s(\sigma)$ are attached to saddle configurations that
depend on this boundary data.
The rest of the story follows in a similar fashion to what we had earlier. For example, on each thimble the fields are expanded around the saddle background as:
\begin{equation}
{\rm g}(x)
=
{\rm g}^{(s)}(x;\sigma)
+
\delta{\rm g}(x) ,
\end{equation}
where $s$ denotes the saddle.
Performing the Gaussian integral over the fluctuations produces the
standard loop expansion that we had encountered earlier. The only
additional point is that the boundary wavefunctional product must be
carried along and is absorbed into the thimble coefficients. Thus
one should write more precisely:

{\footnotesize
\bg
&& \int Dh_0\,Dh_0'\;
\Psi_\sigma^*[h_0']\,\Psi_\sigma[h_0]
\int_{{\cal J}_s(h_0',h_0)}
{\cal D}{\rm g}\;
\exp
\left(
-
S_{\rm tot}[\lambda;{\rm g}]
\right)
=
\exp
\left(
-
S_{\rm tot}[\lambda;{\rm g}^{(s)}(\sigma)]
\right)
{\cal Z}_s(\sigma,\lambda)\\
&& \int Dh_0\,Dh_0'\;
\Psi_\sigma^*[h_0']\,\Psi_\sigma[h_0]
\int_{{\cal J}_s(h_0',h_0)}
{\cal D}{\rm g}\;
g_{\mu\nu}(x)
\exp
\left(
-
S_{\rm tot}[\lambda;{\rm g}]
\right)
=
\exp
\left(
-
S_{\rm tot}[\lambda;{\rm g}^{(s)}(\sigma)]
\right)
{\cal N}_{s,\mu\nu}(x;\sigma,\lambda) ,\nonumber
\nd}
where $\mathcal Z_s(\sigma, \lambda)$ and $\mathcal N_{s, \mu\nu}(x; \sigma, \lambda)$ appearing above are exactly
the fluctuation series introduced earlier around the saddle
configuration ${\rm g}^{(s)}(\sigma)$
in \eqref{amaliabru2} and \eqref{amaliabru} respectively, now understood to include the contribution of the boundary wavefunctional product as well. 
The expectation value therefore takes the thimble--sum form:
\begin{equation}\label{hotatkalu}
\langle g_{\mu\nu}(x)\rangle_\sigma
=
\frac{
\sum\limits_s n_s
\exp
\left(
-
S_{\rm tot}[\lambda;{\rm g}^{(s)}(\sigma)]
\right)
{\cal N}_{s,\mu\nu}(x;\sigma,\lambda)}
{
\sum\limits_s n_s
\exp
\left(
-
S_{\rm tot}[\lambda;{\rm g}^{(s)}(\sigma)]
\right)
{\cal Z}_s(\sigma,\lambda)
} ,
\end{equation}
which should be compared to \eqref{kinmaline}. 
At first sight the Picard--Lefschetz result
appears to differ from the earlier saddle expansion written in
\eqref{kinmaline}, where the coefficients $n_s$ did not appear
explicitly.  The reason is simply that the earlier expression
represented the usual semiclassical saddle expansion in schematic
form, whereas the Picard--Lefschetz analysis keeps track of the
precise decomposition of the original integration cycle.
Recall that, in Picard--Lefschetz theory the functional integration cycle
${\cal C}$ is decomposed as
${\cal C}
=
\sum\limits_s n_s\,{\cal J}_s$,
where ${\cal J}_s$ are the Lefschetz thimbles associated with the
critical points ${\rm g}^{(s)}(\sigma)$ of the action and the integers
$n_s$ are the corresponding intersection numbers with the dual
cycles.  Each saddle contribution is therefore weighted by its
intersection number $n_s$.  In the schematic saddle expansion written
earlier one implicitly assumed $n_s=1$ for the contributing saddles,
or equivalently absorbed these factors into the definition of the
fluctuation series.  The Picard--Lefschetz formulation therefore
provides a more precise version of the same semiclassical expansion.
In other words, 
the Picard--Lefschetz formulation does not alter the structure of
the fluctuation expansion itself; it simply makes explicit that the
semiclassical sum over saddles is weighted by the intersection
numbers $n_s$ determined by the geometry of the integration contour and the saddles.

Let us finish the Euclidean story by making the following observation.
Because the effective action itself contains the trans--series
structure, the saddle weights contain factors of the form:
\begin{equation}
\exp
\left(
-
S_{\rm tot}[\lambda;{\rm g}^{(s)}]
\right)
=
\exp
\left(
-
S_{\rm pert}[\lambda;{\rm g}^{(s)}]
\right)
\exp
\left(
-
\sum_I e^{-A_I/\lambda}
S_I[\lambda;{\rm g}^{(s)}]
-\cdots
\right),
\end{equation}
which we encountered earlier.
Introducing
$\Delta_{\rm np}
=
\sum\limits_I e^{-A_I/\lambda}S_I+\cdots$,
we obtain the identity
$e^{-S_{\rm tot}}
=
e^{-S_{\rm pert}}
e^{-\Delta_{\rm np}}$,
and expanding the second exponential 
generates contributions proportional to
$e^{-A_I/\lambda},
e^{-(A_I+A_J)/\lambda},
e^{-(A_I+A_J+A_K)/\lambda},
\ldots$ et cetera.

\subsubsection{Lorentzian Schwinger--Keldysh formulation \label{sec4.7.2}}

In the Lorentzian formulation the path integral is defined on the
Schwinger--Keldysh (SK) contour and the dynamical fields are doubled.
The expectation value of the metric operator in the coherent state
labelled by $\sigma$ takes the form:

{\footnotesize
\begin{equation}
\langle g_{\mu\nu}(x)\rangle_\sigma
=
\frac{1}{Z_\sigma}
\int Dh_f
\int D\tilde h_i^+\,D\tilde h_i^- \;
\Psi_\Omega[\tilde h_i^+]\,\Psi_\Omega^*[\tilde h_i^-]
\int_{\mathcal S}
{\cal D}{\rm g}_\pm\;
g_{+\mu\nu}(x)
\exp
\left(
iS_{\rm tot}[\lambda;{\rm g}_+]
-
iS_{\rm tot}[\lambda;{\rm g}_-]
\right),
\end{equation}}
where the doubled field multiplet is
${\rm g}_\pm
=
(g_\pm,\text{ghosts}_\pm,\text{matter}_\pm,\ldots)$, and where the
functional integrations over $\mathcal S$ are understood to be taken with the coherent--state
shifted boundary conditions on the SK contour. To simplify the subsequent analysis, 
it is convenient to introduce the Schwinger--Keldysh (SK) action:
\begin{equation}\label{sksimple}
S_{\rm SK}[\lambda;{\rm g}_+,{\rm g}_-]
\equiv
S_{\rm tot}[\lambda;{\rm g}_+]
-
S_{\rm tot}[\lambda;{\rm g}_-].
\end{equation}
After complexification of the doubled field space
$({\rm g}_+,{\rm g}_-)$, the SK integration cycle
${\cal C}_\sigma$ may be decomposed using Picard--Lefschetz theory
into a sum of Lefschetz thimbles:
\begin{equation}
{\cal C}_\sigma
=
\sum_s n_s\,{\cal J}_s(\sigma),
\end{equation}
where $s$ labels the saddle configurations and $n_s$ are the
corresponding intersection numbers. This is much like what we had in \eqref{abbiverit} earlier. The difference is that the saddles correspond to stationary points of the doubled SK action
and therefore satisfy:
\begin{equation}
\frac{\delta S_{\rm SK}}{\delta {\rm g}_+}=0,
\qquad
\frac{\delta S_{\rm SK}}{\delta {\rm g}_-}=0
~~\implies ~~\frac{\delta S_{\rm tot}[\lambda;{\rm g}_+]}{\delta {\rm g}_+}=0,
\qquad
\frac{\delta S_{\rm tot}[\lambda;{\rm g}_-]}{\delta {\rm g}_-}=0,
\end{equation}
where we used the definition of $S_{\rm SK}$ above. The Lefschetz thimbles therefore live in the complexified
configuration space of the doubled variables $({\rm g}_+,{\rm g}_-)$.
Moreover, in parallel with this geometric decomposition of the functional
integral, the effective action itself may admit a trans--series
representation.  The nonperturbative sectors labelled by
$I,J,\ldots$ enter the construction of the effective action through:
\begin{equation}\label{trandhukai}
S_{\rm tot}[\lambda;{\rm g}]
=
S_{\rm pert}[\lambda;{\rm g}]
+
\sum_I e^{-A_I/\lambda}S_I[\lambda;{\rm g}]
+
\sum_{I,J} e^{-(A_I+A_J)/\lambda}S_{IJ}[\lambda;{\rm g}]
+
\cdots .
\end{equation}
Once this effective action is specified, Picard--Lefschetz theory
provides a decomposition of the SK integration cycle into thimbles
attached to classical solutions of the same action.  In the physical
limit where the two branches coincide, the relevant saddles satisfy:
\begin{equation}
{\rm g}_+^{(s)}(x;\sigma)
=
{\rm g}_-^{(s)}(x;\sigma)
=
{\rm g}^{(s)}(x;\sigma),
~~~{\rm with}~~~~~ 
\frac{\delta S_{\rm tot}[\lambda;{\rm g}]}
{\delta {\rm g}}
\Big|_{{\rm g}={\rm g}^{(s)}(\sigma)}
=
0 .
\end{equation}
Thus the trans--series structure of the effective action and the
Picard--Lefschetz decomposition of the path integral represent two
distinct but compatible levels of description.  The relation between
the two may be summarized schematically as:
\begin{equation}
\text{nonperturbative sectors } (I)
\;\Longrightarrow\;
S_{\rm tot}[\lambda;{\rm g}]
\;\Longrightarrow\;
\text{thimble saddles } {\rm g}^{(s)}(\sigma).
\end{equation}
In particular, the exponential factors $e^{-A_I/\lambda}$ originate
from the trans--series structure of the effective action itself,
while the label $s$ enumerates the classical saddle configurations of
the functional integral constructed from this action.

Let us now discuss the consistency between 1PI, Wilsonian, and thimble descriptions. The Lorentzian SK path integral provides a common starting point for the 
three complementary descriptions: the 1PI effective action, the
Wilsonian (or exact renormalization group) effective action, and the
trans--series description based on the Lefschetz thimble decomposition.
The microscopic generating functional of the SK theory is:
\begin{equation}
Z[J_+,J_-]
=
\int {\cal D}{\rm g}_\pm\;
\exp
\left(
iS[{\rm g}_+]
-
iS[{\rm g}_-]
+
iJ_+\cdot{\rm g}_+
-
iJ_-\cdot{\rm g}_-
\right)
\rho_{\rm init},
\end{equation}
where $\rho_{\rm init}$ encodes the initial density matrix.  In the
coherent--state setup considered here the parameter $\sigma$ enters
through the boundary conditions specified by $\rho_{\rm init}$ rather
than through a modification of the bulk action\footnote{From our earlier discussion this is basically \eqref{eq:SK_denominator_final} but with non-zero sources $J_\pm$. In other words, $Z[J_+, J_-] = {\rm Tr}(\rho_\sigma U_{J_+} U^\dagger_{J_-})$, so that when $J_+ = J_- = 0$ we get ${\rm Tr}(\rho_\sigma)$.}. (This is in fact a crucial point and will play an important role when we study the SK path-integral formalism for the Glauber--Sudarshan states. We will come back to this in the next section.)
\begin{figure}[h]
\centering
\begin{tabular}{c}
\includegraphics[width=6in]{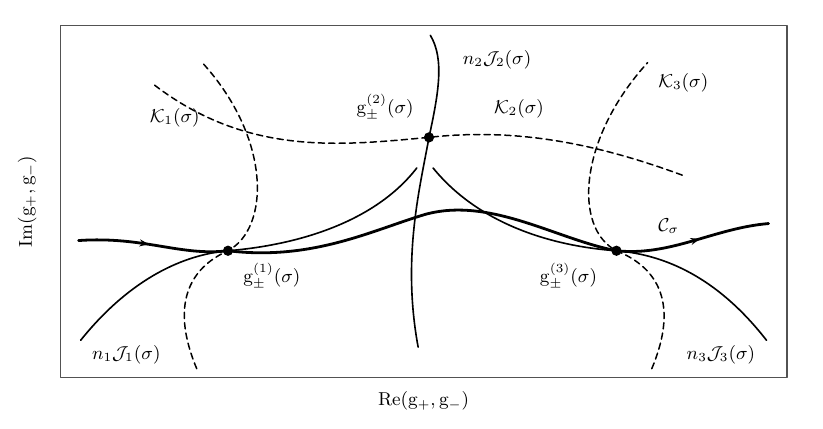}
\end{tabular}
\caption[]{Schematic Picard--Lefschetz decomposition of the Lorentzian coherent-state Schwinger--Keldysh path integral in the complexified doubled configuration space $({\rm g}_+,{\rm g}_-)$. The thick oriented curve is the original SK integration cycle $\mathcal C_\sigma$. The thin solid curves depict representative chain contributions $n_s\mathcal J_s(\sigma)$, where $\mathcal J_s(\sigma)$ are the Lefschetz thimbles and $n_s\in\mathbb Z$ are their intersection-number coefficients. The thimbles are attached to the doubled saddles ${\rm g}^{(s)}_\pm(\sigma)\equiv\big({\rm g}^{(s)}_+(\sigma),{\rm g}^{(s)}_-(\sigma)\big)$, indicated by black points. The dashed curves are the dual cycles $\mathcal K_s(\sigma)$; their oriented intersections with $\mathcal C_\sigma$ determine the coefficients $n_s$ in $\mathcal C_\sigma=\sum_s n_s\mathcal J_s(\sigma)$. Each saddle satisfies $\delta S_{\rm SK}/\delta{\rm g}_+=\delta S_{\rm SK}/\delta{\rm g}_-=0$ and, in the physical limit, ${\rm g}^{(s)}_+={\rm g}^{(s)}_-$. The coherent-state parameter $\sigma$ enters through the boundary conditions encoded by the initial state, $\rho_{\rm init}$.}
\label{thimble2}
\end{figure} 
For the coherent state at hand, 
the Schwinger--Keldysh 1PI effective action is obtained by a Legendre
transform of
\begin{equation}
W[J_+,J_-] = -\,i\ln Z[J_+,J_-] ,
\end{equation}
which is standard. The difference is only in the form of $\rho_{\rm init}$. Keeping that in mind, the Legendre transformation follows the usual route. We can make this clear by first
introducing the classical fields
$\bar{\rm g}_\pm
=
\frac{\delta W}{\delta J_\pm}$,
one finds that $\bar{\rm g}_\pm(x)$ correspond to the expectation values
of the fields on the two SK branches.  For an initial coherent state
$\rho_{\rm init} = \rho_\sigma = |\sigma\rangle\langle\sigma|$ this gives the following form for the {\it classical} fields:
\begin{equation}
\bar{\rm g}_\pm(x)
=
\frac{\langle \sigma |\, {\rm g}_{\pm, H}(x) \,| \sigma \rangle}
{\langle \sigma | \sigma \rangle} ,
\end{equation}
where ${\rm g}_{\pm, H} \equiv g_\pm
=
(g_\pm,\text{ghosts}_\pm,\text{matter}_\pm,\ldots)$ in the Heisenberg picture. 
In the physical limit $J_+ = J_-$ the two branches coincide,
$\bar{\rm g}_+ = \bar{\rm g}_-$, and one recovers the usual expectation
value:
\begin{equation}
\bar{\rm g}(x)
=
\frac{\langle \sigma |\, {\rm g}_H(x) \,| \sigma \rangle}
{\langle \sigma | \sigma \rangle}  \equiv
\frac{\langle \sigma |\, {\rm g}(x) \,| \sigma \rangle}
{\langle \sigma | \sigma \rangle} ,
\end{equation}
thus coinciding with the expectation value \eqref{veritomey} at least for the case where we restrict to the metric degrees of freedom.
Putting everything together, 
the 1PI action becomes:
\begin{equation}
\Gamma_{\rm 1PI}[\bar{\rm g}_+,\bar{\rm g}_-]
=
W[J_+,J_-]
-
J_+\cdot \bar{\rm g}_+
+
J_-\cdot \bar{\rm g}_- .
\end{equation}
The real--time equations of motion follow from
$\frac{\delta \Gamma_{\rm 1PI}}{\delta \bar{\rm g}_\pm}=0$
supplemented by the physical limit
$\bar{\rm g}_+ = \bar{\rm g}_-$. Note that because of $\bar{\rm g}_\pm = {\delta W\over \delta J_{\pm}}$, the sources are now functions of ${\rm g}_\pm$ {\it i.e.} $J_\pm \equiv J_\pm({\rm g}_\pm)$, so the EOMs ${\delta W[J_\pm]\over \delta \bar{\rm g}_\pm} = J_\pm = 0$ do not get trivialized.

Let us now come to the Wilsonian effective action.
The Wilsonian effective action $S_\Lambda$ is obtained by integrating
out modes above a scale $\Lambda$.  In the SK formulation this again
leads to a doubled action:
\begin{equation}
e^{\,iS_\Lambda[{\rm g}_{<,+},{\rm g}_{<,-}]}
=
\int {\cal D}{\rm g}_{>,\pm}\;
\exp
\left(
iS[{\rm g}_{<,+}+{\rm g}_{>,+}]
-
iS[{\rm g}_{<,-}+{\rm g}_{>,-}]
\right)
\rho_{\rm init} ,
\end{equation}
where the modes ${\rm g}_{>, \pm}$ are being integrated away. It is easy to see that 
the Wilsonian action governs the scale dependence of the
low--energy degrees of freedom, while the 1PI action describes the
dynamics of expectation values of the fields.

Finally let us come to the trans--series and thimble description related to the coherent states.
If the infrared effective action admits the trans--series structure
of the form \eqref{trandhukai},
then the doubled SK action becomes 
$S_{\rm tot}^{\rm SK}[\lambda;{\rm g}_+,{\rm g}_-]
=
S_{\rm tot}[\lambda;{\rm g}_+]
-
S_{\rm tot}[\lambda;{\rm g}_-]$ as in \eqref{sksimple}.
After complexification the SK contour decomposes into thimbles,
${\cal C}_\sigma
=
\sum_s n_s\,{\cal J}_s(\sigma)$,
leading to the thimble expansion:
\begin{equation}
Z_\sigma
=
\sum_s n_s
\exp
\left(
iS_{\rm tot}[\lambda;{\rm g}^{(s)}_+(\sigma)]
-
iS_{\rm tot}[\lambda;{\rm g}^{(s)}_-(\sigma)]
\right)
{\cal Z}_s(\sigma,\lambda) ,
\end{equation}
where the fluctuation factors ${\cal Z}_s(\sigma,\lambda)$ are understood to include the contribution of the initial-state boundary factor, equivalently the coherent-state wavefunctional product:
\begin{equation}
\Psi_\Omega[\tilde h_i^+]\,\Psi_\Omega^*[\tilde h_i^-]
\end{equation}
that appeared in the full SK path-integral representation. In other words, the wavefunctional product does not disappear in the thimble decomposition; rather, it is absorbed into the definition of ${\cal Z}_s(\sigma,\lambda)$ and, similarly, into the corresponding numerator coefficients. The structural representation appears in {\bf figure \ref{thimble2}}. 
The three descriptions represent complementary reorganizations of the
same real--time functional integral.  Their consistency requires the following identifications:
\bg\label{brazmukh1}
\bar{\rm g}_{\mu\nu}(x;\sigma)
=
{\rm g}^{(s)}_{\mu\nu}(x;\sigma), ~~~~~
\Gamma^{(2)}_{\rm 1PI}
=
{\cal O}_s
=
S_\Lambda^{(2)} ,
\nd
where the quadratic kernels appearing in the three descriptions of the theory
are given by the second functional derivatives of the corresponding
effective actions. In other words, the kernel of the 1PI effective action is:
\begin{equation}
\Gamma^{(2)}(x,y)
=
\frac{\delta^2 \Gamma_{\rm 1PI}}
{\delta \bar{\rm g}(x)\,\delta \bar{\rm g}(y)},
\end{equation}
which represents the inverse full propagator of the theory. On the other hand, in the thimble (or saddle-point) expansion one writes
${\rm g}(x) = {\rm g}^{(s)}(x) + \delta{\rm g}(x)$,
and the quadratic fluctuation operator is:
\begin{equation}
\mathcal O_s(x,y)
=
\left.
\frac{\delta^2 S_{\rm tot}}
{\delta {\rm g}(x)\,\delta {\rm g}(y)}
\right|_{{\rm g}={\rm g}^{(s)}} ,
\end{equation}
defined over the saddle. Finally, 
in the Wilsonian description the quadratic kernel is obtained from the
Wilsonian effective action $S_\Lambda$ in the following way:
\begin{equation}
S_\Lambda^{(2)}(x,y)
=
\frac{\delta^2 S_\Lambda}
{\delta {\rm g}(x)\,\delta {\rm g}(y)} .
\end{equation}
Thus $\Gamma^{(2)}$, $\mathcal O_s$, and $S_\Lambda^{(2)}$ encode the same
quadratic dynamics viewed respectively in the 1PI, thimble, and Wilsonian
descriptions of the theory. Moreover, and as we also saw earlier,
the nonperturbative sectors
$e^{-A_I/\lambda},
e^{-(A_I+A_J)/\lambda},
\ldots $
appear consistently in the Wilsonian flow, the trans--series action,
and the thimble expansion. Note that the coherent--state parameter $\sigma$ enters all three
descriptions only through the boundary conditions of the SK path
integral and therefore selects the physical branch of solutions
without modifying the local form of the bulk action.
The relation between the different descriptions may be summarized
schematically as:

{\footnotesize
\begin{equation}
\text{microscopic SK theory}
\;\Longrightarrow\;
Z[J_+,J_-;\sigma]
\;\Longrightarrow\;
W[J_+,J_-;\sigma]
\;\Longrightarrow\;
\left\{
\begin{array}{l}
\Gamma_{\rm 1PI}^{\rm SK}[\bar g_+,\bar g_-] \\[6pt]
S_\Lambda^{\rm SK}[g_{<,+},g_{<,-}] \\[6pt]
\displaystyle \sum_s n_s {\cal J}_s(\sigma)
\end{array}
\right.
\end{equation}}
It is no surprise that all three descriptions must reproduce the same physical saddles,
causal kernels, and nonperturbative sectors for the system to make sense.



\section{Trans-series structure and the fate of the Minkowski minimum \label{sec5}}


Now that we have done some details of the Picard--Lefschetz theory \cite{picardlefschetz,wittenlefschetz,hetborel}, it is time to revisit the trans-series decomposition of the total action $S_{\rm tot}$ {\transpapers} from \eqref{bellcroft} and ask the following two generic questions: one, whether this can be derived from the short-distance dynamics of the theory, and two, whether this decides the fate of the chosen Minkowski minimum. First however it is useful to clarify in what sense a trans-series decomposition of the form:
\begin{equation}
{\cal O}[J;\lambda]
\sim
\sum_s
n_s\,
{\cal O}_s[J;\lambda]
\label{genericsectorstart1}
\end{equation}
may arise from a well-defined underlying theory, and how this in turn leads to a trans-series form of the infrared effective action. The correct statement is not that such a decomposition follows automatically from integrating out heavy modes. Rather, it follows when the exact reduced functional obtained after integrating out the heavy modes admits a decomposition into several relevant saddles, Lefschetz thimbles, or nonperturbative sectors. This formulation is sufficiently general to apply to ordinary quantum field theory, effective field theory, and also to string- or M-theoretic settings whenever a suitable reduced functional description exists. One may then turn to the question of the fate of the particular Minkowski minimum around which the trans-series structure has been constructed.

\subsection{Sector decomposition and revisiting the trans-series form of $S_{\rm tot}$ \label{sec4.7.3}}

Let the microscopic theory contain light fields $\phi$ and heavy degrees of freedom $\Psi_{\rm heavy}$, and let $\lambda$ denote the parameter controlling the asymptotic expansion of interest. Depending on the problem, $\lambda$ may be the renormalized IR coupling at scale $\mu$, the string coupling $g_s$, a large-$N$ parameter such as $1/N$, or some other small quantity. Then an exact reduced functional for the light fields may be defined by the following standard prescription:
\begin{equation}
\exp\!\left(
-\,S_{\rm red}[\phi;\lambda]
\right)
\equiv
\int {\cal D}\Psi_{\rm heavy}\;
\exp\!\left(
-\,S_{\rm UV}[\phi,\Psi_{\rm heavy};\lambda]
\right),
\label{defineSred_clean1}
\end{equation}
where ``heavy'' should be understood in a broad sense: it may refer to genuinely heavy field species, to high-momentum modes above a Wilsonian scale $\mu$, or to both. Thus the light fields are genuinely light and are all defined within the Wilsonian scale $\mu$. If the observable of interest is written in the microscopic theory as
\begin{equation}
{\cal O}_{\rm UV}[J;\lambda]
=
\int {\cal D}\phi\,{\cal D}\Psi_{\rm heavy}\;
\exp\!\left(
-\,S_{\rm UV}[\phi,\Psi_{\rm heavy};\lambda]
+
\int d^4x \sqrt{-g}\,J\!\cdot\!\phi
\right),
\label{OUVclean1}
\end{equation}
where $J$ is the set of source terms attached to the light degrees of freedom $\phi$ such that $J\!\cdot\!\phi$ denotes the appropriate tensor contractions, then, after integrating out the heavy degrees of freedom, the same exact observable may be written as:
\begin{equation}\label{coridahl10}
{\cal O}_{\rm red}[J;\lambda]
=
\int {\cal D}\phi\;
\exp\!\left(
-\,S_{\rm red}[\phi;\lambda]
+
\int d^4x \sqrt{-g}\,J\!\cdot\!\phi
\right),
\end{equation}
with
${\cal O}_{\rm UV}[J;\lambda]
=
{\cal O}_{\rm red}[J;\lambda]$.
For notational simplicity one may thereafter denote this common exact observable simply by ${\cal O}[J;\lambda]$. Equation \eqref{coridahl10} is the exact reduced description in terms of the light fields alone. Up to this point no trans-series structure has yet been assumed. Integrating out heavy modes produces the reduced functional $S_{\rm red}[\phi;\lambda]$, but not by itself a sector decomposition.

At this stage one should also clarify the physical picture behind the later trans-series form of the IR action. In many familiar field-theoretic examples, the reduced functional $S_{\rm red}[\phi;\lambda]$ admits several semiclassical minima, often including Minkowski vacua or near-Minkowski vacua. One then chooses one such vacuum as the reference perturbative sector, expands perturbatively around it, and interprets the remaining sectors as tunnelling or instanton contributions associated with the other minima. In this situation the trans-series action $S_{\rm IR}$ derived below is precisely the effective action around the chosen reference vacuum, corrected by the nonperturbative contributions of the other sectors. This is the standard intuition coming from tunnelling between neighboring minima. However, for full generality one should not restrict the discussion only to other Minkowski minima. Depending on the theory, the additional sectors may also come from complex saddles, bounce-like solutions, Lefschetz thimbles of the complexified reduced action, Euclidean brane sectors, worldsheet sectors, or other nonperturbative configurations. Thus the logic is more general than tunnelling among ordinary minima, although that case remains an especially important and intuitive realization of the general framework.

The sector decomposition appears if the reduced functional integral \eqref{coridahl10} has several relevant contributions associated with distinct saddles or sectors. Denoting these sectors by $s$, one may then write schematically
\begin{equation}
{\cal O}[J;\lambda]
\sim
\sum_s
n_s\,
{\cal O}_s[J;\lambda],
\label{sectorobservable1}
\end{equation}
where $n_s$ are trans-series parameters or sector weights and ${\cal O}_s$ is the contribution of the sector $s$. The label $s$ may refer to a perturbative vacuum sector, instanton sectors, multi-instanton sectors, complex saddles, Euclidean brane sectors, worldsheet sectors, or any other nonperturbative configurations relevant to the problem. The precise interpretation of the sectors depends on the theory, but the structural statement \eqref{sectorobservable1} is generic.

When the integration cycle of the reduced functional integral can be complexified, one may refine this statement further using Picard--Lefschetz theory. In that case the original integration cycle ${\cal C}$ may be decomposed into a sum of Lefschetz thimbles ${\cal J}_\sigma$ associated with critical points $\phi_\sigma$ of the complexified reduced action as:
\begin{equation}
{\cal C}
=
\sum_\sigma
n_\sigma\,{\cal J}_\sigma ,
\end{equation}
where $n_\sigma$ denotes the integer intersection number between the original integration cycle and the dual upward-flow cycle associated with the saddle $\phi_\sigma$; it determines whether the thimble ${\cal J}_\sigma$ contributes and with what orientation.
(In the physics language used here, $n_\sigma$ is what ultimately fixes the sector weights in the thimble decomposition of the contour.)
The reduced observable then takes the form:
\begin{equation}
{\cal O}[J;\lambda]
=
\sum_\sigma
n_\sigma
\int_{{\cal J}_\sigma}
{\cal D}\phi\;
\exp\!\left(
-\,S_{\rm red}[\phi;\lambda]
+
J\!\cdot\!\phi
\right).
\label{thimbleobservable1}
\end{equation}
Thus the sector label $s$ in \eqref{sectorobservable1} may in favorable cases be identified with a thimble label $\sigma$, although more generally it may also denote sectors not captured by a simple finite-dimensional thimble picture. In this sense the language of saddles, thimbles, and nonperturbative sectors should be understood as nested levels of refinement rather than mutually exclusive alternatives. We can then denote the sector label by $s$ generically, and for each sector one expects an asymptotic contribution of the 
form:
\begin{equation}
{\cal O}_s[J;\lambda]
=
\exp\!\left(
-\frac{A_s[J]}{\lambda^{\alpha_s}}
\right)
\Phi_s[J;\lambda],
\label{sectorformgeneric1}
\end{equation}
where $A_s[J]$ is the sector action, $\alpha_s>0$ is the dominant scaling exponent appropriate to that sector, and $\Phi_s[J;\lambda]$ is the asymptotic fluctuation series around it. If we do not isolate the dominant scaling $\alpha_s$, then in principle one should sum over the possible scaling exponents associated with the sector. We will however avoid these nuances here. In the simplest situations the fluctuation factor admits an expansion:
\begin{equation}
\Phi_s[J;\lambda]
\sim
\sum_{n=0}^{\infty}
\lambda^{\beta_{s,n}}\,
\phi_{s,n}[J],
\label{sectorflucgeneric1}
\end{equation}
although the detailed form of the powers $\beta_{s,n}$ may vary from one context to another. In particular, different sectors may come with different exponents $\alpha_s$. For example, in string theory a D-brane instanton sector may scale as ${\rm exp}\left(-{{\rm A}\over g_s}\right)$ while an NS5-type sector may scale as ${\rm exp}\left(-{{\rm B}\over g_s^2}\right)$, so the full answer may naturally contain several distinct exponential scales. Thus, the decomposition:
\begin{equation}
{\cal O}[J;\lambda]
\sim
\sum_s
n_s\,
\exp\!\left(
-\frac{A_s[J]}{\lambda^{\alpha_s}}
\right)
\Phi_s[J;\lambda]
\label{fullsectorobservable1}
\end{equation}
does admit a derivation from a well-defined UV theory, but only after two logically distinct steps. {\Su First}, one integrates out the heavy modes to obtain the exact reduced functional $S_{\rm red}[\phi;\lambda]$. {\Su Second}, one analyzes the reduced functional integral and identifies its relevant saddles, thimbles, or sectors. The first step yields the exact IR description; the second yields the trans-series decomposition.

This distinction is important because the trans-series structure is not created by the elimination of heavy modes alone. Rather, it reflects the existence of multiple semiclassical or nonperturbative sectors in the exact theory. In many ordinary quantum field theories these sectors may be interpreted as perturbative vacua, instantons, multi-instantons, complex saddles, or, when the contour is complexified, as distinct Lefschetz thimbles of the reduced IR action. In particularly simple situations they may be viewed as contributions from tunnelling between the chosen reference minimum and neighboring minima of the reduced potential. In string-theoretic settings, depending on what formulation is available, they may instead correspond to worldsheet sectors, Euclidean D-branes, NS-type sectors, flux sectors, or other genuinely stringy nonperturbative objects. The generic structure \eqref{fullsectorobservable1} is therefore more fundamental than any one particular realization.

To obtain an IR action as a functional of the light field, one should now switch from the fully integrated observable \eqref{coridahl10} to a background-field formulation. The point is that if the light field is fully integrated over, as in \eqref{coridahl10}, the resulting quantity is naturally a functional of $J$, not of $\phi$. If instead one wants an effective action depending on a light-field configuration, then one should not integrate over that final field argument. Rather, one splits the light field into a background plus fluctuations:
\begin{equation}
\phi
=
\varphi+\eta,
\label{splitfield_clean1}
\end{equation}
where $\varphi$ is the background or classical light field and $\eta$ is the fluctuation to be integrated over. The reduced functional integral is then performed over $\eta$, not over $\varphi$:
\begin{equation}
\exp\!\left(
-\,S_{\rm IR}[\varphi;\lambda]
\right)
\equiv
\sum_\sigma
n_\sigma
\int_{{\cal J}_\sigma(\varphi)}
{\cal D}\eta\;
\exp\!\left(
-\,S_{\rm red}[\varphi+\eta;\lambda]
\right).
\label{defineSIR_background1}
\end{equation}
Now the dependence on $\varphi$ is perfectly consistent, because $\varphi$ was never integrated out. It is an external background argument of the effective functional. One should also stress that $\varphi$ is, in general, an \emph{off-shell} background field. Only after the effective action has been defined does one impose its equations of motion. Thus it is perfectly consistent later to choose a reference Minkowski minimum $\phi_a$ and write:
\begin{equation}
\varphi=\phi_a+\delta\phi,
\label{varphiphiasplit1}
\end{equation}
where $\delta\phi$ is an arbitrary background deviation from the chosen minimum. As long as no effective equation of motion is imposed, both $\delta\phi$ and hence $\varphi$ are generically off-shell. This is a crucial point. Combining \eqref{splitfield_clean1} and \eqref{varphiphiasplit1}, one may think of the full light field as:
\begin{equation}
\phi=\phi_a+\delta\phi+\eta,
\label{fullsplit_clean1}
\end{equation}
where $\phi_a$ fixes the perturbative reference sector, $\delta\phi$ is the off-shell background deformation, and $\eta$ is the quantum fluctuation integrated out in the path integral. With this understood, one may now formulate the discussion directly at the level of the light fields. Let ${\cal O}[\varphi;\lambda]$ denote an exact reduced quantity depending on the background light field $\varphi$ and on the expansion parameter $\lambda$. Then, whenever the exact reduced theory admits a sector decomposition, one may write:
\begin{equation}\label{coridahl11}
{\cal O}[\varphi;\lambda]
\sim
\sum_s
n_s\,
{\cal O}_s[\varphi;\lambda],
\qquad
{\rm with}
\qquad
{\cal O}_s[\varphi;\lambda]
=
\exp\!\left(
-\frac{A_s[\varphi]}{\lambda^{\alpha_s}}
\right)
\Phi_s[\varphi;\lambda].
\end{equation}
Here $A_s[\varphi]$ is the sector action or sector weight controlling the dominant exponential suppression, $\alpha_s>0$ is the scaling exponent appropriate to that sector, and $\Phi_s[\varphi;\lambda]$ is the fluctuation series around that sector. The fluctuation factor $\Phi_s[\varphi;\lambda]$ is itself asymptotic and may be expanded as:
\begin{equation}
\Phi_s[\varphi;\lambda]
\sim
\sum_{n=0}^{\infty}
\lambda^{\beta_{s,n}}\,
\phi_{s,n}[\varphi],
\label{sectorfluct1}
\end{equation}
where the sequence of powers $\beta_{s,n}$ depends on the problem. In the simplest cases one has $\beta_{s,n}=n$, but in string theory or in more general resurgence problems the sectorial fluctuation expansion may be organized by genus, by loop order around an instanton, by fractional powers, or by logarithmic corrections. The trans-series structure does not depend on the detailed form of these powers; it depends only on the existence of distinct sectors of the form \eqref{coridahl11}. Combining \eqref{coridahl11} and \eqref{sectorfluct1}, the exact observable takes the schematic trans-series form:
\begin{equation}
{\cal O}[\varphi;\lambda]
\sim
\sum_s
n_s\,
\exp\!\left(
-\frac{A_s[\varphi]}{\lambda^{\alpha_s}}
\right)
\sum_{n=0}^{\infty}
\lambda^{\beta_{s,n}}\,
\phi_{s,n}[\varphi].
\label{fullobservabletrans1}
\end{equation}
This is the fundamental trans-series structure. It is obtained directly from the existence of multiple sectors, without requiring any detailed assumption about the form of an exact UV action. The merit of this formulation is that it does not rely on detailed knowledge of the microscopic UV action beyond the existence of a well-defined exact theory and of a reduced light-field description. It therefore applies equally well to ordinary QFT and, more abstractly, to any framework in which the exact observable can be represented as a sum over perturbative and nonperturbative sectors.

The next step is to convert this trans-series structure of observables into a trans-series form for the effective infrared action. Let us define the exact IR effective action $S_{\rm IR}[\varphi;\lambda]$ through the standard exponential relation:
\begin{equation}
{\cal O}[\varphi;\lambda]
=
\exp\!\left(
-\,S_{\rm IR}[\varphi;\lambda]
\right),
\label{defineSIR1}
\end{equation}
where ${\cal O}$ may be interpreted, depending on context, as a partition function, a wavefunctional, a reduced density element, a vacuum functional, or some other quantity from which the effective action is extracted. Substituting \eqref{fullobservabletrans1} into \eqref{defineSIR1}, one obtains
\begin{equation}
\exp\!\left(
-\,S_{\rm IR}[\varphi;\lambda]
\right)
\sim
\sum_s
n_s\,
\exp\!\left(
-\frac{A_s[\varphi]}{\lambda^{\alpha_s}}
\right)
\Phi_s[\varphi;\lambda].
\label{SIRtransbasic1}
\end{equation}
This equation is already the desired trans-series representation of the IR theory at the level of the exponentiated effective action.

To make the structure more explicit, let us isolate a reference sector, usually the perturbative vacuum sector, which we denote by $s=0$. In many field-theoretic situations this reference sector may be chosen to be one particular Minkowski minimum of the reduced action. Concretely, one chooses one exact Minkowski minimum $\phi_a$ as the reference perturbative sector and then studies the effective action for backgrounds $\varphi$ in its neighborhood, namely $\varphi=\phi_a+\delta\phi$. If this sector has action $A_0[\varphi]$ and exponent $\alpha_0$, then one may factor out its leading contribution:

{\footnotesize
\begin{equation}
\exp\!\left(
-\,S_{\rm IR}[\varphi;\lambda]
\right)
\sim
\exp\!\left(
-\frac{A_0[\varphi]}{\lambda^{\alpha_0}}
\right)
\Phi_0[\varphi;\lambda]
\left[
1+
\sum_{s\neq 0}
\widetilde n_s\,
\exp\!\left(
-\frac{A_s[\varphi]}{\lambda^{\alpha_s}}
+
\frac{A_0[\varphi]}{\lambda^{\alpha_0}}
\right)
\frac{\Phi_s[\varphi;\lambda]}{\Phi_0[\varphi;\lambda]}
\right],
\label{factorrefsector1}
\end{equation}}
where the constants $\widetilde n_s$ absorb appropriate normalizations. (If we take $n_0=1$, then $\widetilde n_s=n_s$.) We have also suppressed the subscript $a$ from the choice of the Minkowski vacuum $\phi_a$, but \eqref{factorrefsector1} should be understood as the IR action defined over a chosen Minkowski minimum. When all sectors share the same exponent $\alpha_s=\alpha$, this simplifies to:

{\footnotesize
\begin{equation}
\exp\!\left(
-\,S_{\rm IR}[\varphi;\lambda]
\right)
\sim
\exp\!\left(
-\frac{A_0[\varphi]}{\lambda^{\alpha}}
\right)
\Phi_0[\varphi;\lambda]
\left[
1+
\sum_{s\neq 0}
\widetilde n_s\,
\exp\!\left(
-\frac{A_s[\varphi]-A_0[\varphi]}{\lambda^{\alpha}}
\right)
\frac{\Phi_s[\varphi;\lambda]}{\Phi_0[\varphi;\lambda]}
\right].
\label{samealpha1}
\end{equation}}
Note that for the generic scenario where we sum over all $s$ as well as over the set $\{\alpha_s\}$, implementing the choice $\alpha_s=\alpha$ can become tricky but can nevertheless be done with some careful ordering of the terms. Since we are avoiding these nuances, we will not worry too much about them here. Defining the relative sector actions by:
\begin{equation}
\Delta A_s[\varphi]
\equiv
{\lambda^{\alpha_0}A_s[\varphi]-\lambda^{\alpha_s}A_0[\varphi]\over \lambda^{\alpha_0}},
\end{equation}
with the assumption that now we are resorting back to the case where each $\alpha_s$ remains distinct,
and then taking the logarithm now yields the IR effective action itself:
\begin{align}
S_{\rm IR}[\varphi;\lambda]
&=
\frac{A_0[\varphi]}{\lambda^{\alpha_0}}
-
\log\, \Phi_0[\varphi;\lambda]
-
\log\!\left[
1+
\sum_{s\neq 0}
\widetilde n_s\,
\exp\!\left(
-\frac{\Delta A_s[\varphi]}{\lambda^{\alpha_s}}
\right)
\frac{\Phi_s[\varphi;\lambda]}{\Phi_0[\varphi;\lambda]}
\right],
\label{SIRbeforeexpand1}
\end{align}
which is {\it almost} what we are after but not quite in the trans-series form \eqref{bellcroft} yet. However note that, if the nonperturbative sectors are exponentially suppressed relative to the reference sector, the last logarithm may be expanded:
\begin{align}
S_{\rm IR}[\varphi;\lambda]
&\sim
\frac{A_0[\varphi]}{\lambda^{\alpha_0}}
-
\log\, \Phi_0[\varphi;\lambda]
-
\sum_{s\neq 0}
\widetilde n_s\,
\exp\!\left(
-\frac{\Delta A_s[\varphi]}{\lambda^{\alpha_s}}
\right)
\frac{\Phi_s[\varphi;\lambda]}{\Phi_0[\varphi;\lambda]}
\nonumber\\
&\qquad
+
\frac{1}{2}
\left(
\sum_{s\neq 0}
\widetilde n_s\,
\exp\!\left(
-\frac{\Delta A_s[\varphi]}{\lambda^{\alpha_s}}
\right)
\frac{\Phi_s[\varphi;\lambda]}{\Phi_0[\varphi;\lambda]}
\right)^2
-\cdots,
\label{SIRexpanded1}
\end{align}
where the dotted terms can be easily worked out to any order we want. The result we have after expansion is quite remarkable: it shows explicitly that the IR effective action itself inherits a trans-series structure. This can be expressed quantitatively as:
\bg\label{SIRtransseries1}
S_{\rm IR}[\varphi;\lambda]
& \sim &
S_{\rm IR}^{(0)}[\varphi;\lambda]
+
\sum_{s\neq 0}
\exp\!\left(
-\frac{\Delta A_s[\varphi]}{\lambda^{\alpha_s}}
\right)
S_{\rm IR}^{(s)}[\varphi;\lambda] \nonumber\\
&& ~~~~~~~~~~~ +
\sum_{s,t\neq 0}
\exp\!\left(
-\frac{\Delta A_s[\varphi]}{\lambda^{\alpha_s}}
-
\frac{\Delta A_t[\varphi]}{\lambda^{\alpha_t}}
\right)
S_{\rm IR}^{(s,t)}[\varphi;\lambda]
+\cdots,
\nd
where the series expansion simplifies slightly once we choose $\alpha_s=\alpha_t=\alpha$ but even in this limit we do not expect $\Delta A_s[\varphi]$ to be equal to $\Delta A_t[\varphi]$.
The leading perturbative part in \eqref{SIRtransseries1} is simply the first two terms from \eqref{SIRexpanded1}, namely
\begin{equation}
S_{\rm IR}^{(0)}[\varphi;\lambda]
=
\frac{A_0[\varphi]}{\lambda^{\alpha_0}}
-
\log\, \Phi_0[\varphi;\lambda],
\label{SIRpert1}
\end{equation}
while the sectorial corrections $S_{\rm IR}^{(s)}[\varphi;\lambda]$, $S_{\rm IR}^{(s,t)}[\varphi;\lambda]$, and so on are determined by the logarithmic expansion in \eqref{SIRexpanded1}. The result \eqref{SIRtransseries1} matches what we proposed earlier in \eqref{bellcroft} once we identify the sectors $s,t,\ldots$ etc with $I,J,\ldots$ etc from \eqref{bellcroft}. However the derivation is intentionally general. In our derivation it is {\it not} necessary to assume that the sectors arise from a known UV spacetime action despite our emphasis on it earlier. It requires only that the exact theory admit several contributing sectors, each with a characteristic exponential weight and an associated asymptotic fluctuation series. This makes the derivation suitable not only for ordinary quantum field theory but also for string theory, where one may know the perturbative genus expansion and the relevant nonperturbative objects without knowing an exact UV action.

Although not important for the present discussion, let us ask what happens in string theory. In the string-theoretic case one should therefore start from the actual perturbative data. For example, a perturbative closed-string quantity typically has the genus 
expansion:
\begin{equation}
{\cal A}_{\rm pert}(g_s)
\sim
\sum_{h=0}^{\infty}
g_s^{2h-2}\,
{\cal A}_h.
\label{stringgenusstart1}
\end{equation}
If the coefficients ${\cal A}_h$ grow factorially at large genus, then this perturbative series is asymptotic and generally incomplete. The large-order behavior, together with independent physical knowledge of D-brane sectors, worldsheet instantons, NS5 sectors, or other nonperturbative effects, indicates that the exact answer must be enlarged to a trans-series of the schematic form:
\begin{equation}
{\cal A}(g_s)
\sim
\Phi_{\rm pert}(g_s)
+
\sum_I
{\rm exp}\left(-{A_I\over g_s}\right)\,
\Phi_I(g_s)
+
\sum_J
{\rm exp}\left(-{B_J\over g_s^2}\right)\,
\Psi_J(g_s)
+\cdots,
\label{stringexampletrans1}
\end{equation}
where the dotted terms could in principle incorporate other non-perturbative effects like exotic instantons, bound states of branes et cetera. Defining the effective IR action by
${\cal A}(g_s)
=
\exp\!\left[
-\,S_{\rm IR}(g_s)
\right]$,
one immediately obtains:
\begin{equation}
\exp\!\left[
-\,S_{\rm IR}(g_s)
\right]
\sim
\Phi_{\rm pert}(g_s)
+
\sum_I
{\rm exp}\left(-{A_I\over g_s}\right)\,
\Phi_I(g_s)
+
\sum_J
{\rm exp}\left(-{B_J\over g_s^2}\right)\,
\Psi_J(g_s)
+\cdots,
\label{stringSIRexp1}
\end{equation}
and hence, after taking the logarithm and expanding around the perturbative sector, an IR trans-series action of precisely the same structure as \eqref{SIRtransseries1}\footnote{See also \cite{coherbeta,coherbeta2,borel2,hetborel} where similar trans-series action for a stringy/M-theory set-up is presented.}. The important point is that this derivation uses only the existence of the perturbative and nonperturbative sectors themselves. It does not require one to postulate a detailed exact UV spacetime action. The logic then is:
\begin{equation}
S_{\rm UV}
\longrightarrow
S_{\rm red}
\longrightarrow
\text{sector decomposition}
\longrightarrow
S_{\rm tot}
\equiv
S_{\rm IR},
\end{equation}
and it is this final effective trans-series action $S_{\rm tot}$ that appears in the Euclidean path integral for the one-point function in say \eqref{camekim}\footnote{A question could arise at this stage: if $S_{\rm tot}$ is already in trans-series form, why do we need thimble decomposition? The answer is that the thimbles are no longer needed to derive the fact that $S_{\rm tot}$ has a trans-series decomposition. Instead, they play a different role. First, they tell us how to evaluate the path integral consistently. Even with a trans-series effective action, the contour ${\cal C}$ may still need to be decomposed into relevant thimbles for an actual semiclassical evaluation. Secondly, they determine the Stokes data and sector weights. The coefficients $n_s$ in \eqref{abbiverit} and \eqref{hotatkalu} are not arbitrary. Thimble structure tells us which sectors contribute on a given contour and how they jump across Stokes walls. So once $S_{\rm tot}$ is already known to be trans-series-valued, the thimbles are not there to replace it, but to explain and organize its sector content and contour dependence.}.

The conceptual conclusion is therefore straightforward. A decomposition such as \eqref{coridahl11} can indeed arise from a well-defined UV theory after integrating out heavy modes, but not merely because the heavy modes were integrated out. What is essential is that the resulting exact reduced functional, or equivalently the full theory itself, possesses multiple relevant saddles, thimbles, or nonperturbative sectors. The trans-series structure of the IR theory is then the natural reflection of that underlying sectorial decomposition. Equivalently, once the exponentiated IR object is written as the sum over these sectors, the IR effective action obtained by taking the logarithm necessarily inherits a trans-series structure. This argument is sufficiently general to apply equally well to ordinary field theory and to string theory.

\subsection{Localized vacua, instanton mixing, and the trans-series interpretation \label{fitboushona}}

The physical picture behind the trans-series action may be understood most transparently by starting from the case in which the reduced theory admits several classically degenerate Minkowski minima. The basic intuition is the familiar one from the double-well system, but it should be formulated somewhat carefully in the present field-theoretic/gravity setting.

Suppose that the reduced or microscopic theory has several degenerate minima, and let $\phi_a$ be one chosen minimum. Then the perturbative expansion around $\phi_a$ defines a natural reference sector. However, the exact low-energy spectrum is not, in general, built from states that remain forever localized in a single minimum. Rather, the true stationary states of the full Hamiltonian are typically delocalized linear combinations spread over the various minima, while a state localized near $\phi_a$ is obtained by taking an appropriate superposition of those exact eigenstates. In this sense the state
$|\Omega\rangle_a$ should be viewed not as an exact eigenstate of the full theory, but as a quasi-localized or approximate vacuum associated with the chosen minimum $\phi_a$. Consequently it is not perfectly stationary and will tunnel into the neighboring minima. This is precisely the phenomenon encoded by the instanton sectors. The chain of ideas may therefore be summarized schematically as:

{\footnotesize
\begin{equation}
\text{\textcolor{blue}{multiple minima}}
\;\Longrightarrow\;
\text{\textcolor{blue}{localized quasi-vacua}}\nonumber
\end{equation}
\begin{equation}
\;\Longrightarrow\;
\text{\textcolor{blue}{tunnelling between minima}}
\;\Longrightarrow\;
\text{\textcolor{blue}{instanton contributions}}
\;\Longrightarrow\;
\text{\textcolor{blue}{trans-series structure}}
\label{logic_chain_multi}\nonumber
\end{equation}}
Two distinctions are important here. First, one should distinguish the exact stationary states from the localized vacua. In the simplest two-well case the exact eigenstates are the analogs of the symmetric and antisymmetric combinations, while the localized vacuum near one well is a suitable linear combination of those exact eigenstates. More generally, with finitely many degenerate minima one obtains a finite multiplet of split states, while only an infinite periodic family of minima leads to a genuine Bloch-band picture. Thus the expression ``band structure'' should be used literally only in the periodic case; more generally, the exact stationary states are delocalized combinations, while the localized vacua are approximate states constructed from them.

Second, one should distinguish the existence of the instanton saddle from the choice of localized state. The instanton is a property of the theory itself, namely of its Euclidean saddle structure. In the simplest case it interpolates between two minima, while in more general settings it belongs to a larger nonperturbative sector. The localized state does not create the instanton. Rather, once one chooses to work with a vacuum localized near one minimum, the effect of that instanton saddle appears as the exponentially suppressed mixing between the chosen minimum and the others. In this sense the localized-vacuum picture is a physical interpretation of the instanton effects, while the instanton itself is more fundamentally a saddle of the Euclidean path integral.

Accordingly, the effective action around the chosen minimum is not merely the purely perturbative action around $\phi_a \equiv (g, {\rm matter})_a$, but rather a trans-series-corrected action that incorporates the contributions of the other saddles:
\begin{equation}
S_{\rm tot}[\lambda;{\rm g}]
=
S_{\rm pert}^{(a)}[\lambda;{\rm g}]
+
\sum_I
\exp\!\left(
-\frac{A_I[{\rm g}]}{\lambda^{\alpha_I}}
\right)
S_I^{(a)}[\lambda;{\rm g}]
+\cdots .
\label{Stot_multi_min}
\end{equation}
where ${\rm g} = (g, {\rm gh}, {\rm matter})$; $S_{\rm pert}^{(a)}$ describes fluctuations around the chosen minimum $\phi_a$, while the exponentially suppressed sectors encode tunnelling to the other minima or, more generally, contributions of the additional relevant saddles. This same statement may be rephrased in terms of the localized vacuum wavefunctional. If one chooses boundary conditions selecting the minimum $\phi_a$, then:
\begin{equation}
\langle h|\Omega\rangle_a
\sim
\int_{\text{b.c. selecting }a}
{\cal D}{\rm g}\;
\exp\!\left(
-\,S_{\rm tot}[\lambda;{\rm g}]
\right) ,
\label{wf_localized}
\end{equation}
which is what we had in \eqref{fitboushona1} and \eqref{eq:PI_with_ghosts} earlier in the Lorentzian setting although there are some subtleties that will become clear soon.
Thus the perturbative and nonperturbative sectors are not separate ad hoc ingredients, but are all encoded in the same trans-series action.

The relation between the localized-vacuum picture and the instanton picture may then be stated more sharply. If one chooses the quasi-localized state $|\Omega\rangle_a$, then there are two complementary ways to describe its mixing with neighboring minima. The first is the Lorentzian picture, in which one studies the real-time evolution
$|\Omega(t)\rangle_a
=
e^{-iHt}\,|\Omega\rangle_a$, 
or, in gravity with suitable asymptotics:
\begin{equation}
|\Omega(t)\rangle_a
=
e^{-iH_{\partial\Sigma}t}\,|\Omega\rangle_a ,
\label{real_time_grav_localized}
\end{equation}
with $H_{\partial\Sigma}$ being the boundary Hamiltonian.
Since the localized state is not an exact eigenstate, this evolution causes it to spread into the neighboring minima. The second is the Euclidean instanton picture, in which one computes the exponentially small tunnelling matrix elements or level splittings from the relevant Euclidean saddles:
\begin{equation}
\Delta E
\sim
\exp\!\left(
-\frac{A_{\rm inst}}{\lambda^\alpha}
\right)
\times
(\text{sectorial fluctuations}).
\label{instanton_splitting}
\end{equation}
Thus the Euclidean computation does not describe a different physical process. Rather, it computes the nonperturbative mixing data responsible for the real-time spreading:
\begin{equation}
\text{\textcolor{blue}{instanton amplitude}}
\;\Longrightarrow\;
\text{\textcolor{blue}{energy splitting or tunnelling matrix element}}\nonumber
\end{equation}
\begin{equation}
\;\Longrightarrow\;
\text{\textcolor{blue}{real-time spreading of the localized state}}
\label{instanton_to_realtime}\nonumber
\end{equation}
The two descriptions are therefore complementary: the real-time evolution of a localized state {\it implies} Euclidean instanton-induced level mixing.

In describing the interacting vacua $|\Omega\rangle_a$ we have kept a subtlety under the rug. This corresponds to the difference between $|\Omega\rangle_a$ used above and $|\Omega\rangle$ in \eqref{fitboushona3}. The latter corresponds to exact interacting vacuum and therefore one must distinguish it from the localized quasi-vacuum $|\Omega\rangle_a$. The formula:
\begin{equation}
|\Omega\rangle \equiv |\Omega_{\rm exact}\rangle
=
\lim_{T\to\infty(1-i\epsilon)}
e^{-iH_{\partial\Sigma}T}\,|0\rangle
\label{exact_projection}
\end{equation}
projects onto an exact eigenstate of the full Hamiltonian. If the full theory has several minima but instanton mixing is included, then \eqref{exact_projection} selects a delocalized exact vacuum rather than a state strictly localized near one chosen minimum. Thus, in general,
\begin{equation}
|\Omega_{\rm exact}\rangle
\neq
|\Omega\rangle_a
\qquad
\text{in the presence of instanton mixing.}
\label{exact_vs_localized}
\end{equation}
The state $|\Omega\rangle_a$ is therefore best understood as a sector-selected quasi-vacuum, prepared by boundary conditions selecting the minimum $\phi_a$, whereas the $i\epsilon$ projection prepares the true exact vacuum of the full theory. This distinction becomes especially important when one considers expectation values. For a quasi-localized vacuum associated with a chosen Minkowski minimum, the in-in expectation value discussed in the numerator of say \eqref{philipdal} should more appropriately be denoted via:
\begin{equation}
{}_a\langle \Omega|
D^\dagger(\sigma)\,
{\cal O}_H\,
D(\sigma)
|\Omega\rangle_a
\label{localized_SK_exp}
\end{equation}
and is naturally represented by a Schwinger--Keldysh contour anchored by the initial state selecting that minimum. In this contour representation the same trans-series action \eqref{Stot_multi_min} appears, and the nonperturbative sectors of $S_{\rm tot}$ precisely encode the mixing of the chosen quasi-vacuum with the other relevant saddles. In the GS case, where:
\begin{equation}
D(\sigma)
=
\mathcal {T}\exp\!\left(
\int d^4x\,
\sqrt{-g}\,
\sigma^{\mu\nu}(x)\,
g_{\mu\nu}(x)
\right),
\label{GS_case}
\end{equation}
the operator is naturally represented as a source insertion along the contour. In the coherent-state case, where:
\begin{equation}
D(\sigma)
=
\exp\!\left(
\int d^3x\,
\sigma^{ij}(x)\,
\pi_{ij}(x)
\right),
\label{coherent_case}
\end{equation}
the operator is more naturally represented as a boundary shift of the initial state. In either case, the same trans-series effective action governs the perturbative and nonperturbative contributions. 

The situation is different if the theory has a unique Minkowski vacuum. (This would correspond more appropriately to \eqref{fitboushona3} with the interacting vacuum $|\Omega\rangle$.) In that case one no longer has the double-well-type interpretation in which a localized state tunnels to neighboring minima. Instead the state:
\begin{equation}
|\Omega\rangle \equiv |\Omega_{\rm exact}\rangle
=
\lim_{T\to\infty(1-i\epsilon)}
e^{-iH_{\partial\Sigma}T}\,|0\rangle
\label{unique_exact_vac}
\end{equation}
is the exact vacuum of the full theory. The absence of multiple minima, however, does \emph{not} eliminate the possibility of a trans-series. It only changes the physical origin of the nonperturbative sectors. In string/M theory these sectors may arise from other Euclidean saddles, such as wrapped brane instantons on internal cycles. In other words, while 
{unique Minkowski vacuum}
implies {no vacuum-mixing interpretation}, the presence of 
{Euclidean brane saddles} imply {new nonperturbative sectors} which in turn imply the
trans-series structure\footnote{Alternatively, we can state this in the following way. The existence of brane instantons is controlled primarily by the geometry, topology, fluxes, and zero-mode structure of the compactification, whereas the number of Minkowski minima is controlled by the effective potential. The two are not generically in one-to-one correspondence, although brane instantons can certainly modify the effective potential and hence affect the vacuum structure. So there are really two different roles instantons can play:
(i) connect different vacua / induce tunnelling,
and
(ii) produce NP corrections within one vacuum.
Brane instantons are often of type (ii), though in some setups they may also be related to vacuum transitions. Either or both lead to the trans-series structure for the effective action at the IR.}.
If an Euclidean $p$-brane wraps an internal cycle $\Sigma_p$, then its contribution is weighted schematically by
$\exp\!\left(
-\,S_{\rm inst}^{(p)}
\right)$, where 
$S_{\rm inst}^{(p)}
\sim
T_p\,{\rm Vol}(\Sigma_p)$
with {couplings to background potentials and fluxes} added in.
In M-theory one may similarly have sectors from Euclidean M2- and M5-branes:
\begin{equation}
S_{M2}
\sim
T_{M2}\,{\rm Vol}(\Sigma_3)+\cdots,
\qquad
S_{M5}
\sim
T_{M5}\,{\rm Vol}(\Sigma_6)+\cdots ,
\label{M2M5_actions}
\end{equation}
with $T_{Mp}$ denoting the tension of the $Mp$-brane and the dotted terms are additional couplings to the background fields.
After rewriting these in terms of the relevant low-energy expansion parameter, the resulting nonperturbative sectors again take the generic form:
\begin{equation}
\exp\!\left(
-\frac{A_I}{\lambda^{\alpha_I}}
\right).
\label{generic_brane_sector}
\end{equation}
Accordingly, in the unique-vacuum case the trans-series action should be interpreted not as a combination of the 
{perturbation theory around one minimum}
and {tunnelling to other minima},
but rather as
{perturbation theory around the unique vacuum} {\it plus} the contributions from the {nonperturbative wrapped-brane sectors}.
One may therefore write schematically:
\begin{equation}
S_{\rm tot}[\lambda;{\rm g}]
\sim
S_{\rm pert}[\lambda;{\rm g}]
+
\sum_I
\exp\!\left(
-\frac{A_I[{\rm g}]}{\lambda^{\alpha_I}}
\right)
S_I[\lambda;{\rm g}]
+\cdots,
\label{unique_vac_trans}
\end{equation}
where now the sectors $I$ are labeled by Euclidean wrapped-brane saddles or other nonperturbative configurations rather than by neighboring Minkowski minima. The role of the nonperturbative sectors is correspondingly different: they no longer encode vacuum mixing, but rather nonperturbative corrections to observables \emph{within the same vacuum}, such as corrections to superpotentials, couplings, moduli-space metrics, higher-derivative terms, or correlation functions. Thus in the unique-vacuum case the natural expectation value is:
\begin{equation}
\langle\Omega_{\rm exact}|
D^\dagger(\sigma)\,
{\cal O}_H({\bf x},t)\,
D(\sigma)
|\Omega_{\rm exact}\rangle,
\label{unique_vac_exp}
\end{equation}
where $|\Omega_{\rm exact}\rangle$ is the exact vacuum selected by \eqref{unique_exact_vac}. The corresponding path integral may still involve the trans-series action \eqref{unique_vac_trans}, but now the nonperturbative terms arise from Euclidean brane instantons or other wrapped saddles rather than from vacuum tunnelling.

The main message is therefore the following. When several degenerate minima are present, the trans-series action describes perturbation theory around a chosen minimum together with the instanton sectors that mix it with the others. In that case the localized quasi-vacuum picture, the instanton picture, and the Schwinger--Keldysh description are all complementary formulations of the same nonperturbative physics. When the vacuum is unique, the trans-series does not disappear; rather, its nonperturbative sectors are reinterpreted as contributions of other Euclidean saddles, such as wrapped brane instantons, within the same exact vacuum. In either case, the trans-series action provides the unified framework in which perturbative and nonperturbative effects are simultaneously encoded.


\subsection{On-shell shift and the fate of the chosen Minkowski minimum \label{sec5.3}}

In section \ref{sec4.7.3} we used the background field method to derive the IR action $S_{\rm IR}[\varphi;\lambda]\equiv S_{\rm tot}[\varphi;\lambda]$ in \eqref{SIRtransseries1}, where $\varphi=(g,{\rm gh},{\rm matter})$. Our strategy was to use an off-shell background field $\varphi(x)$ and integrate the fluctuations $\eta$ around it using \eqref{splitfield_clean1} to obtain the action\footnote{There is, however, a small but important point in the use of the
background field method. Since the background $\varphi(x)$ is kept
off-shell, the fluctuation expansion around it generically contains
terms linear in the quantum fluctuation $\eta$. Explicitly, expanding
the action around $\varphi$ gives:
\begin{equation}
S[\varphi+\eta]
=
S[\varphi]
+
\int d^d x\,
\left.
\frac{\delta S}{\delta\phi(x)}
\right|_{\phi=\varphi}
\eta(x)
+
\frac{1}{2}
\int d^d x\,d^d y\,
\eta(x)
\left.
\frac{\delta^2 S}{\delta\phi(x)\delta\phi(y)}
\right|_{\phi=\varphi}
\eta(y)
+\cdots .\nonumber
\label{action_expansion_offshell}
\end{equation}
The term linear in $\eta$ is the tadpole term:
$S_{\rm tad}[\varphi,\eta]
=
\int d^d x\,
\left.
\frac{\delta S}{\delta \phi(x)}
\right|_{\phi=\varphi}
\eta(x)$.
This term vanishes only if the background satisfies the classical
equation of motion,
$\left.
\frac{\delta S}{\delta\phi(x)}
\right|_{\phi=\varphi}
=0$.
Therefore, if $\varphi(x)$ is a genuinely off-shell background, one
should expect tadpoles. The background field method does not require the tadpole to vanish.
Rather, it is designed precisely to allow $\varphi(x)$ to be arbitrary.
The path integral over fluctuations is then schematically
$\exp\left(i\Gamma[\varphi]\right)
=
\int {\cal D}\eta\,
\exp\left(
iS[\varphi+\eta]
\right)$, where the fluctuation action contains the linear term
$S_{\rm tad}[\varphi,\eta]$. Thus the tadpole is not an inconsistency;
it is the signal that $\varphi$ is not a stationary point of the
classical action but a generic off-shell field configuration.} in terms of a generic off-shell field $\varphi(x)$. We then parameterized the background field around a chosen on-shell Minkowski minimum $\phi_a$ as in \eqref{varphiphiasplit1},
\begin{equation}
\varphi=\phi_a+\delta\phi,
\label{varphiarounda_here}
\end{equation}
and studied the trans-series action \eqref{SIRtransseries1} to quantify the dynamics of the {\it off-shell} fluctuation $\delta\phi$. The question that we want to ask now concerns the {\it on-shell} value of the same fluctuation $\delta\phi$ around the minimum $\phi_a$. Can we quantify it? In the process we may also summarize the relation between the microscopic theory $S_{\rm UV}$, the exact reduced theory $S_{\rm red}$, and the trans-series effective action $S_{\rm tot}\equiv S_{\rm IR}$, together with the corresponding notions of vacuum, localized state, and physical observable. The central point is that one must distinguish carefully between the exact theory, the perturbative organization of that theory around a chosen minimum, and the choice of quantum state used in an expectation value.

Let $S_{\rm UV}$ denote the microscopic action, and let $S_{\rm red}$ be the exact reduced functional obtained by integrating out heavy modes without discarding any light field that distinguishes the relevant vacua. This procedure preserves the full physical vacuum structure of the UV theory, but not necessarily as a one-to-one set of ordinary minima in the reduced light-field description. If the UV theory has only Minkowski vacua, the exact IR theory should not generate genuinely new physical de-Sitter or Anti de-Sitter vacua, although the vacuum structure may appear in a more complicated branched or nonlocal way.
 However, this statement must be formulated carefully. Since the microscopic action is really of the form
$S_{\rm UV}
=
S_{\rm UV}[\phi,\Psi_{\rm heavy};\lambda]$,
the exact vacua of the UV theory are not specified by the light field $\phi$ alone. Rather, they are full static field 
configurations\footnote{The notation $ib_i$ is very convenient but not necessarily essential. If the intention is simply to say that the $i$-th light field is accompanied by
$b_i$
heavy fields, then writing
$\Psi_{{\rm heavy},\,b_i}$
is still not ideal, because $b_i$ is a \emph{number}, not a field label. The cleanest notation is to introduce a separate index for the heavy fields, for example
$\Psi_{{\rm heavy},\,i\alpha}$,
with 
$\alpha=1,\ldots,b_i$,
so that the full static vacuum configuration is written as:
\begin{equation}
(\phi,\Psi_{\rm heavy})
=
\Big\{
\big(
\phi_i,\,
\Psi_{{\rm heavy},\,i\alpha}
\big)
\; ;\;
\alpha=1,\ldots,b_i
\Big\}. \nonumber
\label{eq:fullvacconfig_clean}
\end{equation}
If one wishes to suppress the explicit heavy-field index, then one may simply write
$(\phi,\Psi_{\rm heavy})
=
\{(\phi_i,\Psi_{{\rm heavy},\,i})\}$,
with the understanding that
$\Psi_{{\rm heavy},\,i}
=
\{\Psi_{{\rm heavy},\,i\alpha}\}_{\alpha=1}^{b_i}$. This is of course better but introduces too much clutter, and therefore we shall stick with the $ib_i$ notation.}:
\begin{equation}
(\phi,\Psi_{\rm heavy})
=
\{(\phi_i,\Psi_{{\rm heavy},ib_i})\},
\label{fullvacconfig_local}
\end{equation}
that solve the coupled UV equations of motion
$\frac{\delta S_{\rm UV}}{\delta\phi}=0$ and 
$\frac{\delta S_{\rm UV}}{\delta\Psi_{\rm heavy}}=0$
together with the requirement that the corresponding vacuum be Minkowski; and the subscript $b_i$ denotes $b_i$ number of heavy fields associated with the $i$-th light field. (To comply with this notation, and if $1 \le i \le I$, for $i \ge I+1$ we will simply assume that $(\phi_i, \Psi_{{\rm heavy},ib_i}) \equiv \Psi_{{\rm heavy}, i}$ thus making $\phi_i$ an empty set and removing the multiplicity $b_i$.) Therefore the fully explicit UV statement may be expressed in the following way:

{\footnotesize
\begin{equation}
S_{\rm UV}[\phi,\Psi_{\rm heavy};\lambda]
\;\Longrightarrow\;
\left\{
(\phi_1,\Psi_{{\rm heavy},1b_1}),
(\phi_2,\Psi_{{\rm heavy},2b_2}),
\ldots
\right\},
\qquad
\{|\psi^{(b_1)}_{\rm 1H}\rangle,|\psi^{(b_2)}_{\rm 2H}\rangle,\ldots\},
\label{fullyexplicit_local}
\end{equation}}
where the first set denotes the full UV vacuum configurations and the second set denotes the exact stationary spectrum of the microscopic theory. At this stage we should emphasize that the reduced functional $S_{\rm red}$ is obtained \emph{exactly} by integrating out the heavy fields:
\begin{equation}
\exp\!\left(
-\,S_{\rm red}[\phi;\lambda]
\right)
\equiv
\int {\cal D}\Psi_{\rm heavy}\;
\exp\!\left(
-\,S_{\rm UV}[\phi,\Psi_{\rm heavy};\lambda]
\right) ,
\label{Sredexactdef_local}
\end{equation}
in an Euclidean framework where the convergence is easily visible.
Only in this exact sense does $S_{\rm red}$ retain the same reduced physical content as the microscopic theory, but the vacua counting could in-principle be more complicated. For example  
several distinct UV minima from \eqref{fullyexplicit_local}
may project to the same light-field value
$\phi_a=\phi_b$,
in which case the reduced IR description does not distinguish them as separate minima in the $\phi$-space description.
So, in general,
$N_{\rm vac}^{\rm IR}
\neq
N_{\rm vac}^{\rm UV}$
as a naive counting statement in reduced field space.
What \emph{is} true is that no genuinely new physical vacua are created or destroyed by an exact change of variables or by exact integration over heavy fields. The full physical vacuum structure is preserved, but its representation in the reduced theory can be more complicated.
In particular, the matching of the reduced vacuum structure and of the exact stationary spectrum relies on the exact functional integration \eqref{Sredexactdef_local}, not merely on a classical elimination of the heavy fields. Indeed, if one only solves the heavy-field equations formally as:
\begin{equation}
\frac{\delta S_{\rm UV}}{\delta\Psi_{\rm heavy}}=0
\qquad
\Longrightarrow
\qquad
\Psi_{\rm heavy}
=
\Psi_{\rm heavy}^{\rm cl}[\phi;\lambda],
\label{heavyonshell_local}
\end{equation}
and substitutes the result back into the UV action, one obtains only the tree-level or saddle-point reduced action of the following form:
\begin{equation}
S_{\rm red}^{\rm tree}[\phi;\lambda]
=
S_{\rm UV}\!\left[\phi,\Psi_{\rm heavy}^{\rm cl}[\phi;\lambda];\lambda\right],
\label{Sredtree_local}
\end{equation}
whereas the exact reduced functional \eqref{Sredexactdef_local} also contains the heavy-field fluctuation determinants and all higher corrections. Thus the exact matching of vacua and stationary spectral data between the UV and reduced descriptions relies on \eqref{Sredexactdef_local}.

Equivalently, one may think of the reduced vacuum problem in the following way. Since the heavy fields have been integrated out exactly, the vacua of the reduced theory are determined by
$\frac{\delta S_{\rm red}}{\delta\phi}=0$,
and the corresponding solutions are precisely, upto constraints, the light-field components
$(\phi_1,\phi_2,\ldots)$
of the full UV vacua. Thus the schematic notation:
\begin{equation}
S_{\rm UV}
\;\Longrightarrow\;
\{\phi_1,\phi_2,\ldots, \phi_k\}\big\vert_{\Psi^{b_1\to b_k}_{\rm heavy}},
\qquad
\{|\psi_1\rangle,|\psi_2\rangle,\ldots, |\psi_k\rangle\}\big\vert_{\Psi^{b_1\to b_k}_{\rm heavy}}
\label{UVshorthand_local}
\end{equation}
should always be understood as a shorthand for the more explicit statement \eqref{fullyexplicit_local}, followed by projection to the light-field sector after the heavy fields have been integrated out exactly. In this notation one may summarize the UV and the reduced descriptions as
\bg\label{lenulila1}
S_{\rm UV}
\;&\Longrightarrow& \;
\left\{
(\phi_1,\Psi_{{\rm heavy},1b_1}), \ldots ,
(\phi_k,\Psi_{{\rm heavy},kb_k})
\right\},
\qquad
\{|\psi^{(b_1)}_{\rm 1H}\rangle, \ldots , |\psi^{(b_k)}_{k{\rm H}}\rangle\}
\nonumber\\
&\Longrightarrow & \;
\{\phi_1, \ldots,\phi_k\}\big\vert_{\Psi^{b_1\to b_k}_{\rm heavy}},
\qquad
\{|\psi_1\rangle, \ldots , |\psi_k\rangle\}\big\vert_{\Psi^{b_1\to b_k}_{\rm heavy}}
\nonumber\\[3pt]
S_{\rm red}
&\Longrightarrow &\;
\{\phi_1, \ldots ,\phi_l\},
\qquad
\{|\psi_1\rangle,\ldots ,|\psi_l\rangle\} ,
\nd
where $k \ge l$.
Here the set $\{\phi_i\}$ denotes the light-field vacuum data extracted from the full UV vacua, whereas the set $\{|\psi_n\rangle\}$ denotes the exact stationary spectrum as represented in the reduced theory. Now choose one exact Minkowski minimum $\phi_a$ only as a \emph{reference perturbative sector} and rewrite the infrared theory in trans-series form, following the same logic as in \eqref{splitfield_clean1}, \eqref{varphiphiasplit1} and \eqref{fullsplit_clean1}:
\begin{equation}
S_{\rm tot}[\varphi; \lambda]\equiv S_{\rm IR}[\varphi; \lambda].
\label{Stotdef}
\end{equation}
If $S_{\rm tot}$ is the \emph{full exact} trans-series completion, i.e. it includes all relevant nonperturbative sectors --- ordinary tunnelling saddles, wrapped brane instantons, Lefschetz-thimble sectors, and any other required Euclidean saddles --- then it describes the same exact theory as $S_{\rm UV}$ and $S_{\rm red}$. 
This implies that the exact IR reduction should not create new physical vacua with different cosmological constant out of nothing.
But this does \emph{not} imply that the reduced effective potential written in terms of the surviving variables must display the same minima in a one-to-one way as ordinary extrema.
Hence we expect for $k$ Minkowski minima in the UV theory:
\begin{equation}
S_{\rm tot}
\;\Longrightarrow\;
\{\phi_1,\phi_2,\ldots, \phi_l\},
\qquad
\{|\psi_1\rangle,|\psi_2\rangle,\ldots, |\psi_l\rangle\},
\label{Stotspectrum}
\end{equation}
with $k \ge l$
provided the following three conditions hold: (i) $S_{\rm tot}$ is exact, (ii) no relevant light field has been discarded, and 
(iii) the same field variable $\phi$, or more appropriately $\varphi$ here, is used in all descriptions.
Thus the clean exact statement is
$S_{\rm UV}, S_{\rm red}$ and $S_{\rm tot}$
all describe the same exact theory,
and therefore:
\begin{equation}
\{|\psi^{(b_n)}_{n{\rm H}}\rangle\}_{\rm UV}
\ge
\{|\psi_n\rangle\}_{\rm red}
=
\{|\psi_n\rangle\}_{\rm tot},
\qquad
\{\phi_i\}_{\rm UV}
\ge
\{\phi_i\}_{\rm red}
=
\{\phi_i\}_{\rm tot}.
\label{boxsame2}
\end{equation}
In particular, in the strict exact sense one should not introduce genuinely new exact vacuum locations $\phi_{i\star}$ distinct from the original $\phi_i$. Rather,
$\phi_{i\star}=\phi_i$
{for the exact theory in the same field variable.} This is a subtle point because one might worry that the potential $V_{\rm tot}$ coming from the trans-series action $S_{\rm tot}$ might actually lead to the on-shell condition where the Minkowski minimum is shifted slightly from the original choice of $\varphi = \phi_a$ to $\phi_\ast$ where:
\begin{equation}
\phi_\star=\phi_a+\delta\phi_a = \phi_a -
\frac{
\left.
\frac{\partial V_{\rm NP}}{\partial\phi}
\right|_{\phi=\phi_a}
}{
\left.
\frac{\partial^2 V_{\rm pert}}{\partial\phi^2}
\right|_{\phi=\phi_a}
}\label{phistar}
\end{equation}
where $V_{\rm tot} \equiv V_{\rm pert} + V_{\rm NP}$ is in the trans-series form. Our discussion above, and especially in \eqref{boxsame2}, guarantees that this is {\it not} the case. Any on-shell shift has to vanish implying $\delta\phi_a = 0$. The shift $\delta\phi_a$ 
is useful only in a weaker sense, namely:
(i) as local bookkeeping around one perturbative branch,
(ii) as an approximation in a truncated trans-series,
or (iii) as a consequence of a field redefinition or a change of variables.
In the strict exact sense, if $\phi_a$ is already one of the exact vacua of the full theory, then the on-shell condition for the exact trans-series action is satisfied exactly at
$\varphi=\phi_a$ because 
$\frac{\partial V_{\rm tot}}{\partial\phi}
\big|_{\phi=\phi_a}=0$, including 
$\frac{\partial V_{\rm pert}}{\partial\phi}
\big|_{\phi=\phi_a}=0$ and 
$\frac{\partial V_{\rm NP}}{\partial\phi}
\big|_{\phi=\phi_a}=0$ in the strict exact sense. A nonzero $\delta\phi_a$ is therefore meaningful only as an approximate, local, or coordinate-dependent notion. Put differently, for the off-shell background parameterization:
\begin{equation}
\varphi=\phi_a+\delta\phi,
\label{offshellparam}
\end{equation}
the fluctuation $\delta\phi$ is generically nonzero off-shell, but vanishes on-shell whenever one uses the exact trans-series action and $\phi_a$ is already one of the exact vacua in the same field 
variable\footnote{Unless there is a moduli space of vacua here. We are assuming that any such moduli space is lifted by quantum effects. In the absence of supersymmetry this is a reasonable assumption, although nowhere close to being a proof. We will not worry about these subtleties here.}.

\subsection{Persistence of the UV Minkowski vacuum sectors in the far IR \label{sec5.4}}

Our aim in this section is to justify the following statement in a form that is compatible with the exact saddle decomposition of the heavy-field path integral: If the heavy fields are integrated out exactly, then even in the presence of multiple heavy saddle branches
$\Psi_{{\rm heavy},ab_a}$
over a given light-field configuration $\phi_a$, the exact reduced theory cannot generate genuinely new physical vacuum sectors other than those already present in the full UV theory. In particular, if the full UV theory contains only Minkowski vacuum sectors and no de Sitter or anti-de Sitter vacuum sectors, then the exact reduced theory cannot produce genuinely new physical de Sitter, anti-de Sitter, or additional Minkowski vacuum sectors that were absent in the UV theory.
(A key point is that the heavy-field functional integral is organized by \emph{saddles}, not merely by minima. Therefore the correct language throughout is the language of saddle sectors or thimbles, with exact Minkowski vacua appearing as a distinguished subset of those saddle sectors.)

Let us start by following the nomenclature developed in the earlier section, namely that the UV theory contain light fields $\phi$ and heavy fields $\Psi_{\rm heavy}$, with exact action
$S_{\rm UV}[\phi,\Psi_{\rm heavy}]$. We will also 
assume, as before, that the exact stationary vacua of the full theory are labeled by pairs
$(\phi_a,\Psi_{{\rm heavy},ab_a})$, with 
$1\le a \le I$, and 
$1\le b_a \le B_a$, 
where for each light-field vacuum location $\phi_a$ there are $B_a$ heavy branches. Also as mentioned earlier, we could 
allow additional purely heavy vacua, not coupled to the light fields in the relevant sense. Let us denote these by
$\Psi_a$, with 
$a\ge I+1$. We will assume all these vacua to be Minkowski, and denote 
the full set of exact UV vacua by $\mathcal V_{\rm UV}$. They are given by:
\begin{equation}
\Gamma_{\rm UV}\big[\phi_a,\Psi_{{\rm heavy},ab_a}\big]=0,
\qquad
\Gamma_{\rm UV}[\Psi_a]=0 \nonumber
\label{eq:all_Minkowski_assumption}
\end{equation}
\begin{equation}
~~~{\cal V}_{\rm UV}
=
\left\{
(\phi_a,\Psi_{{\rm heavy},ab_a})
\right\}_{1\le a\le I,\ 1\le b_a\le B_a}
\cup
\left\{
\Psi_a
\right\}_{a\ge I+1} ,
\label{eq:full_vacuum_set}
\end{equation}
where $\Gamma_{\rm UV}[\phi_a,\Psi_{{\rm heavy}, ab_a}]$
denotes the exact quantum effective action of the full UV theory, obtained as the Legendre transform of the connected generating functional, and its stationary points determine the exact quantum vacuum 
sectors\footnote{Recall that, and as mentioned above,
$\Gamma_{\rm UV}$
denotes the \emph{exact quantum effective action} of the ultraviolet theory. More explicitly, if the UV theory has fields collectively denoted by
$\Phi_{\rm UV}
=
(\phi,\Psi_{\rm heavy})$,
then
$\Gamma_{\rm UV}[\phi,\Psi_{\rm heavy}]$
is the Legendre transform of the connected generating functional of the UV theory. To see this, start with the UV partition function in the presence of sources:
\begin{equation}
Z_{\rm UV}[J_\phi,J_\Psi]
=
\int {\cal D}\phi\,{\cal D}\Psi_{\rm heavy}\;
\exp\!\left(
iS_{\rm UV}[\phi,\Psi_{\rm heavy}]
+
i\int J_\phi\,\phi
+
i\int J_\Psi\,\Psi_{\rm heavy}
\right). \nonumber
\label{eq:ZUV_definition}
\end{equation}
Define the connected generating functional by
$W_{\rm UV}[J_\phi,J_\Psi]
=
-\,i\log Z_{\rm UV}[J_\phi,J_\Psi]$.
Then the expectation values of the fields in the presence of sources are
$\phi_c
=
\frac{\delta W_{\rm UV}}{\delta J_\phi}$,
and 
$\Psi_c
=
\frac{\delta W_{\rm UV}}{\delta J_\Psi}$.
The exact UV effective action is the Legendre transform:
\begin{equation}
\Gamma_{\rm UV}[\phi_c,\Psi_c]
=
W_{\rm UV}[J_\phi,J_\Psi]
-
\int J_\phi\,\phi_c
-
\int J_\Psi\,\Psi_c. \nonumber
\label{eq:GammaUV_definition}
\end{equation}
Usually one then drops the subscript $c$ and writes simply
$\Gamma_{\rm UV}[\phi,\Psi_{\rm heavy}]$. The importance of $\Gamma_{\rm UV}$ is that its stationary points determine the exact quantum vacua of the UV theory, {\it i.e.}
$\frac{\delta \Gamma_{\rm UV}}{\delta \phi}=0$, and 
$\frac{\delta \Gamma_{\rm UV}}{\delta \Psi_{\rm heavy}}=0$.
So when we write
$\Gamma_{\rm UV}\big[\phi_a,\Psi_{{\rm heavy},ab_a}\big]=0$,
the meaning is that
$(\phi_a,\Psi_{{\rm heavy},ab_a})$
is an exact Minkowski vacuum sector of the full quantum theory.
It is also important not to confuse
$S_{\rm UV}[\phi,\Psi_{\rm heavy}]$
with
$\Gamma_{\rm UV}[\phi,\Psi_{\rm heavy}]$.
The bare or microscopic action $S_{\rm UV}$ is the action we write in the path integral. By contrast, $\Gamma_{\rm UV}$ is the exact quantum effective action obtained after including all loop and nonperturbative corrections. In this sense $S_{\rm red}$ and $S_{\rm tot}$ are more powerful because they are generated by exactly integrating out the heavy modes.}. The set of equations in \eqref{eq:full_vacuum_set} is arranged 
such that there are no other exact stationary vacua of de-Sitter or anti de-Sitter type. In fact we can define the exact reduced action by integrating out the heavy fields 
as in \eqref{Sredexactdef_local}. 
Equivalently, for sources $J$ coupled only to the light fields, the total partition function in the Lorentzian language becomes:
\begin{equation}
Z_{\rm red}[J]
=
\int {\cal D}\phi\;
e^{\,iS_{\rm red}[\phi]+i\int J\phi}
=
\int {\cal D}\phi\,{\cal D}\Psi_{\rm heavy}\;
e^{\,iS_{\rm UV}[\phi,\Psi_{\rm heavy}]+i\int J\phi}
=
Z_{\rm UV}[J] ,
\label{eq:Zred_equals_ZUV_corrected}
\end{equation}
from where it is easy to see that 
the exact reduced theory reproduces exactly the light-field physics of the UV theory. Both \eqref{eq:Zred_equals_ZUV_corrected} and \eqref{Sredexactdef_local} are useful because one may show,
for fixed $\phi$, the heavy-field functional integral is organized as a sum over saddle sectors:
\begin{equation}
\int {\cal D}\Psi_{\rm heavy}\;
e^{\,iS_{\rm UV}[\phi,\Psi_{\rm heavy}]}
\sim
\sum_{\alpha\in{\cal B}(\phi)}
{\cal Z}_\alpha[\phi],
\label{eq:saddle_decomposition_corrected}
\end{equation}
where ${\cal B}(\phi)$ denotes the exact set of heavy-field saddle sectors contributing at the given light-field value. Note that 
the partition function is organized by stationary points of the action, {\it i.e.} saddles, because these are the configurations around which the phase or weight is locally extremal. Minima are only one special class of stationary points, so a sum over minima is in general incomplete. This means each contributing sector may be written schematically as:
\begin{equation}
{\cal Z}_\alpha[\phi]
=
{\cal N}_\alpha[\phi]\,
e^{\,i\Gamma_\alpha[\phi]},
\label{eq:sector_contribution_corrected}
\end{equation}
where $\alpha$ labels a saddle sector rather than necessarily a minimum. Among these saddle sectors, those descending from exact UV Minkowski vacua satisfy, at the corresponding reduced light-field value $\phi=\phi_a$, the following equation:
\begin{equation}
\left.
\frac{\delta \Gamma_{\alpha(a,b_a)}[\phi]}{\delta\phi}
\right|_{\phi=\phi_a}
=
0,
\qquad
\Gamma_{\alpha(a,b_a)}[\phi_a]=0 ,
\label{eq:branch_stationarity_corrected}
\end{equation}
implying that 
every exact UV Minkowski vacuum sector contributes a corresponding exact stationary saddle sector to the reduced theory.
This proves persistence of the original Minkowski sectors, namely, 
every exact UV Minkowski vacuum sector survives in the exact reduced theory.

One may also ask as to why do no genuinely new physical vacuum sectors can appear. To answer this let us suppose, for contradiction, that the exact reduced theory has a genuinely new vacuum sector at some light-field configuration $\phi_\star$ that does not descend from any UV vacuum sector. This means that:
\begin{equation}
\left.
\frac{\delta \Gamma_{\rm red}[\phi]}{\delta\phi}
\right|_{\phi=\phi_\star}
=
0,
\label{eq:new_IR_stationary_corrected}
\end{equation}
and either
$\Gamma_{\rm red}[\phi_\star]=0$
for a new Minkowski sector, or
$\Gamma_{\rm red}[\phi_\star]\neq 0$
for an exotic de-Sitter or anti de-Sitter sector. But \eqref{eq:Zred_equals_ZUV_corrected} shows that the exact reduced theory is physically equivalent to the exact UV theory in all light-field observables. Therefore any exact stationary vacuum sector of the reduced theory must correspond to some exact stationary sector of the UV theory.

Hence a genuinely new reduced vacuum sector at $\phi_\star$ would imply the existence of an additional exact UV vacuum sector with the same light-field observables and the same vacuum energy. By assumption, no such additional UV sector exists. This is a contradiction. Therefore
the exact reduced theory cannot create genuinely new physical vacuum sectors that were absent in the exact UV theory.

What happens once we include the decoupled heavy vacuum sectors? These decoupled fields appear from the additional vacuum sectors $\Psi_a$ with $a\ge I+1$, which are decoupled from the light fields.
There are now two possibilities. The first one is when we have the truly decoupled spectator sectors, {\it i.e.}
these sectors do not couple to the light fields at all. In that case their contribution to the heavy-field integral factorizes:

{\footnotesize
\begin{equation}
\int {\cal D}\Psi_{\rm heavy}\;
e^{\,iS_{\rm UV}[\phi,\Psi_{\rm heavy}]}
=
\left(
\int {\cal D}\Psi_{\rm coupled}\;
e^{\,iS_{\rm UV}^{\rm coupled}[\phi,\Psi_{\rm coupled}]}
\right)
\left(
\int {\cal D}\Psi_{\rm dec}\;
e^{\,iS_{\rm UV}^{\rm dec}[\Psi_{\rm dec}]}
\right) ,
\label{eq:factorized_decoupled_corrected}
\end{equation}}
where $\Psi_{\rm dec}$ is associated with the saddles from the decoupled sector such that $\{\Psi_a\} \subset \Psi_{\rm dec}$. 
The second factor is independent of $\phi$, so it contributes only an overall normalization:
\begin{equation}
S_{\rm red}[\phi]
=
S_{\rm red}^{\rm coupled}[\phi]
+
{\rm constant} ,
\label{eq:constant_shift_only_corrected}
\end{equation}
implying that the IR reduced action only comes from those heavy fields that are directly coupled to the light fields. This is of course intuitively expected, and 
therefore it cannot create new stationary points in the light-field sector, because:
\begin{equation}
\frac{\delta S_{\rm red}[\phi]}{\delta\phi}
=
\frac{\delta S_{\rm red}^{\rm coupled}[\phi]}{\delta\phi} ,
\label{eq:no_new_extrema_from_constant_corrected}
\end{equation}
implying further that 
truly decoupled heavy vacuum sectors cannot generate new light-field vacuum sectors. However a more interesting question is to ask as to what happens when multiple heavy saddle branches over the same or different light vacua. If several heavy saddle branches sit above a given $\phi_a$, then the exact reduced action contains the sum over those sectors:
\begin{equation}
e^{\,iS_{\rm red}[\phi]}
\sim
\sum_{\alpha\in{\cal B}_{\rm coupled}(\phi)}
{\cal Z}_\alpha[\phi]
+
\sum_{a\ge I+1}
{\cal Z}^{\rm dec}_a,
\label{eq:sum_over_all_branches_corrected}
\end{equation}
where the decoupled terms ${\cal Z}^{\rm dec}_a$ are $\phi$-independent, while the coupled sector terms are precisely those descending from the UV saddle sectors. Taking the log and then extremising over $\phi$ we see that no new physical vacuum sectors are created; the reduced theory simply reorganizes the already existing UV sectors. Putting everything together our arguments boil down to the following three conclusions:
\begin{itemize}
\item{$S_{\rm red}[\phi]$ is obtained by exact integration over the heavy fields.}
\item{The full UV theory has only the exact Minkowski vacuum sectors listed in ${\cal V}_{\rm UV}$}
\item{There are no additional exact UV de-Sitter or anti de-Sitter vacuum sectors.}
\end{itemize}
\noindent A contradiction proof following \eqref{eq:new_IR_stationary_corrected} reveals the same outcome, namely that, because $\Gamma_{\rm red}$ is exact and physically equivalent to the UV theory in light observables, this vacuum sector must correspond to some exact stationary UV sector. But by assumption, no such additional UV sector exists, implying that \eqref{eq:new_IR_stationary_corrected} contradicts itself {\it unless} $\phi_\ast = \phi_a$, {\it i.e.} $\phi_\ast$ equals to one of the existing Minkowski minimum. This again means that 
no new exact vacuum sector can be generated by exact integrating out, even in the presence of multiple heavy saddle branches or decoupled heavy vacuum sectors. 

What \emph{can} happen is not creation of new physical vacuum sectors, but rather a more complicated representation of the old ones. For example (a) 
distinct UV vacuum sectors may project to the same $\phi_a$,
(b) the reduced theory may become multibranched, nonlocal, or sectorial,
and/or (c) approximate truncations may display spurious extra stationary points.
But none of these are genuinely new exact physical vacuum sectors, and our conclusions remain unchanged.

There is however one more thing.
Although the exact vacua and exact stationary eigenstates satisfy the aforementioned conditions in all three exact descriptions, namely the microscopic, the far IR and the trans-series reorganization, a state constructed by choosing one perturbative vacuum sector is not, in general, one of the exact stationary eigenstates. This is the real nontrivial distinction. To see this, let $\phi_a$ be one of the exact vacua that survives in the IR. One may build a localized or sector-selected state
$|\Omega\rangle_a$, 
centered on the perturbative sector around $\phi_a$. This state is analogous to the localized state $|L\rangle$ in the ordinary double-well system (see below). It is physically meaningful, but it need not be an exact stationary eigenstate of the full Hamiltonian. Rather, the exact stationary states are the eigenstates
$|\psi_1\rangle,|\psi_2\rangle,\ldots.$
So the fundamental distinction is not
$\phi_a$ {versus} 
$\phi_{\star}$,
but rather
$|\Omega\rangle_a$ {versus}
$|\psi_n\rangle$.

This is exactly what happens in the symmetric double-well problem. The classical potential has minima at $\phi=\pm a$, but the exact stationary states are the symmetric and antisymmetric combinations
$|+\rangle$ and $|-\rangle$, 
with energies
$E_\pm
=
E_0
\pm
\frac{\Delta E}{2}$,
where $\Delta E\neq 0$. The localized state
$|L\rangle
\sim
\frac{1}{\sqrt{2}}
\left(
|+\rangle+|-\rangle
\right)$
is not an eigenstate, and evolves as:
\begin{equation}
|L(t)\rangle
=
e^{-iHt}|L\rangle
\sim
\frac{1}{\sqrt{2}}
\left(
e^{-iE_+t}|+\rangle
+
e^{-iE_-t}|-\rangle
\right),
\label{Ltime}
\end{equation}
so it exhibits spreading or oscillatory leakage into the other well. This oscillatory leakage is responsible for converting the action to the requisite trans-series form from the point of view of (say) the left well. Likewise, the state $|\Omega\rangle_a$ is a sector-selected quasi-vacuum, not in general an exact eigenstate. The same applies to GS or coherent states built over it:
\begin{equation}
|\sigma\rangle_a
=
D(\sigma)\,|\Omega\rangle_a.
\label{GSstate}
\end{equation}
These are useful displaced states probing the sector around $\phi_a$, but they are not new stationary vacua. Compared to the double well example above, now there are multiple such Minkowski minima and precisely due to the fact that the quasi-localized interacting vacuum $|\Omega\rangle_a$ around the vacuum $\phi_a$ is {\it not} an eigenstate of the total Hamiltonian, provides the reason why the dynamics is controlled by the exact trans-series action $S_{\rm tot}[\varphi; \lambda]$.
For the quasi-localized sector around $\phi_a$, the natural in-in expectation value is:
\begin{equation}
{}_a\langle\Omega|
D^\dagger(\sigma)\,
{\cal O}_H({\bf x}, t)\,
D(\sigma)
|\Omega\rangle_a.
\label{quasiexp}
\end{equation}
Its path-integral representation is naturally written on a Schwinger--Keldysh contour using the trans-series action $S_{\rm tot}$, because $S_{\rm tot}$ encodes the perturbative sector around $\phi_a$ together with the nonperturbative sectors that mix it with the other saddles. Thus $S_{\rm tot}$ governs the dynamics of the quasi-localized state even though $|\Omega\rangle_a$ is not an exact eigenstate.

Although not important for the present work, 
if one is instead interested in the exact stationary sector of the same theory, then one should work in one of the exact eigenstates $|\psi_n\rangle$ already contained in the spectrum \eqref{boxsame2}. In particular, for the exact vacuum state one may write:
\begin{equation}
\langle\psi_0|
D^\dagger(\sigma)\,
{\cal O}_H({\bf x}, t)\,
D(\sigma)
|\psi_0\rangle,
\label{exactexp}
\end{equation}
where $|\psi_0\rangle$ is the exact ground state of the full theory. This state is stationary, so 
$H_{\rm tot}|\psi_0\rangle
=
E_0|\psi_0\rangle$,
and, by exact equivalence, the same state also belongs to the spectra of the Hamiltonians derived from $S_{\rm UV}$ and $S_{\rm red}$.
Therefore, the same exact action
$S_{\rm tot}=S_{\rm IR}$
may underlie both \eqref{quasiexp} and \eqref{exactexp} despite the fact that for the case \eqref{exactexp} we are not restricted to any vacua. However we do not expect \eqref{quasiexp} and \eqref{exactexp} to match as the expectation values differ because the states differ:
\begin{equation}
{}_a\langle\Omega|
D^\dagger(\sigma)\,
{\cal O}_H({\bf x}, t)\,
D(\sigma)
|\Omega\rangle_a
\neq
\langle\psi_0|
D^\dagger(\sigma)\,
{\cal O}_H({\bf x}, t)\,
D(\sigma)
|\psi_0\rangle
\label{differentexps}
\end{equation}
in general. Thus the action alone does not determine the observable; one must also specify the state. For us \eqref{exactexp} is not useful and as it does not lead to any observable consequence. The only useful expectation values are the ones computed from the quasi-localized vacua $|\Omega\rangle_a$ which henceforth we will simply denote by $|\Omega\rangle$. After the dust settles, the logically clean picture then is the following.

\begin{enumerate}

\item The microscopic action $S_{\rm UV}$, the exact reduced action $S_{\rm red}$ obtained by integrating out the heavy fields, and the exact trans-series completion $S_{\rm tot}$ are three equivalent descriptions of the same exact theory, provided they are written in terms of the same light variables and no operator content is discarded.

\item More precisely, the microscopic theory determines full vacuum configurations given via
$(\phi_i,\Psi_{{\rm heavy},ib_i})$,
where $\phi_i$ denotes the light-field component and $\Psi_{{\rm heavy},ib_i}$ collectively denotes the heavy-sector data with multiplicity $b_i$. The reduced and trans-series descriptions are written only in terms of the light variables, so they encode the vacuum structure through the corresponding reduced data $\{\phi_i\}$, but not necessarily as a naive one-to-one list of minima in the reduced field space.

\item The reduced functional reproduces the exact light-field dynamics only because it is obtained by exact functional integration over the heavy fields, as in \eqref{Sredexactdef_local},
and not merely by substituting the classical equations of motion of the heavy modes.

\item In that exact sense, no genuinely new physical vacua are created by passing from $S_{\rm UV}$ to $S_{\rm red}$ or $S_{\rm tot}$. However, the reduced description need not display the vacuum structure as the same naive set of ordinary minima in the reduced variables, because distinct UV vacua may project to the same $\phi_i$, and the exact reduced theory may represent the vacuum structure through branches, nonlocalities, or different trans-series sectors.

\item Accordingly, one should not claim that exact reduction produces genuinely new exact vacuum locations $\phi_{i\star}$ that were absent in the UV theory. If the same field variables are used and one is already sitting on an exact vacuum branch, then on-shell there is no further exact shift of that vacuum location. Nevertheless, off-shell variations $\delta\phi$ are of course nonzero in general, and approximate truncations may display spurious shifted minima that are not exact vacua of the full theory.

\item A useful and genuinely nontrivial distinction is instead between an exact stationary eigenstate $|\psi_n\rangle$ of the full theory and a localized, sector-selected state $|\Omega\rangle_a$ associated to one chosen vacuum sector labeled by $\phi_a$.

\item The state $|\Omega\rangle_a$ is the natural object for organizing the perturbative expansion, the local trans-series completion, and the Schwinger--Keldysh observables around the chosen vacuum sector $\phi_a$. In general it should be viewed as a sector-adapted state rather than as a globally exact stationary eigenstate of the full theory.

\item Exact physical vacuum observables are obtained by evaluating operators in the exact stationary vacuum state $|\psi_0\rangle$, or more generally in the exact stationary eigenstates $|\psi_n\rangle$. By contrast, observables computed in $|\Omega\rangle_a$ describe the physics of the chosen local vacuum sector and need not coincide with globally exact vacuum observables unless the sector-selected state itself is exact.

\item The trans-series action $S_{\rm tot}$ does not represent a different theory from $S_{\rm UV}$ or the exact $S_{\rm red}$. Rather, it is the same exact dynamics reorganized asymptotically around a chosen saddle or vacuum sector. What changes from one description to another is not the exact theory itself, but the state, the sector being emphasized, and the asymptotic organization of the computation.

\end{enumerate}
\noindent Thus the correct exact outcome is that the vacua and exact stationary spectrum match across $S_{\rm UV}$, $S_{\rm red}$, and $S_{\rm tot}$, while the physically nontrivial distinction lies in the difference between exact eigenstates and localized sector-selected states. It is only the latter that has any physically observable consequence for the system.

\section{Glauber-Sudarshan states and the Schwinger-Keldysh contours \label{sec6.0}}

A central advantage of the Glauber--Sudarshan states over ordinary
coherent states is that they provide access to a much larger class of
trajectories in configuration space, including both on-shell and
off-shell configurations, whereas coherent states are largely confined
to a more restricted set of predominantly on-shell trajectories. This
difference is already apparent in {\bf figures ~1--4} of \cite{borelborel} and in {\bf figure \ref{gsvscoher}},
where the relevant configuration-space trajectories are shown
explicitly. Standard eigenstates are even more constrained, with a
reach in configuration space that is narrower than that of either
coherent or Glauber--Sudarshan states.

Their wider kinematic reach is, however, only part of the story. As we
will demonstrate below, the Glauber--Sudarshan formalism also allows
quantum corrections to be incorporated into expectation values in a far
more direct way than is possible for coherent states. In addition,
these states reveal striking connections to bootstrap structures and to
the vertex-operator framework of string theory. In the following we turn to a full
quantitative development of these features.

\subsection{Operator algebras for the coherent and the GS states \label{sec6.1}}

We will start by elaborating on the operator algebra for the coherent and the GS states. In fact we briefly discussed this, at least for the coherent states, in \eqref{twogonec} using the displacement operator from \eqref{sweenseyf}. In the following we will elaborate the story slightly more, but with the scalar fields, first for the coherent states and then for the GS states. We start by 
define the coherent-state displacement operator at a fixed time slice $\Sigma_{t_0}$ by:
\begin{equation}
D(\sigma)
=
\exp\!\left(
i\int_{\Sigma_{t_0}} d^3x\;\sigma(\mathbf x)\,\pi(\mathbf x,t_0)
\right),
\end{equation}
where note the absence of $\sqrt{h}$ and our choice of the conjugate momentum for the scalar field $\phi({\bf x}, t_0)$ localized on an initial Cauchy slice. The field and the conjugate momentum satisfy the 
standard canonical equal-time commutator
$[\phi(\mathbf x,t_0),\pi(\mathbf y,t_0)]
=
i\,\delta^{(3)}(\mathbf x-\mathbf y)$. Using the Baker-Campbell-Hausdorff rule $e^{-A}Be^{A}=B+[B,A]+\frac12[[B,A],A]+\cdots$ with $A$ defined as
$A=i\int d^3y\,\sigma(\mathbf y)\pi(\mathbf y,t_0)$ gives us the following commutation algebra:

{\footnotesize
\begin{align}
[\phi(\mathbf x,t_0),A]
&=
i\int d^3y\;\sigma(\mathbf y)\,[\phi(\mathbf x,t_0),\pi(\mathbf y,t_0)]
=
i\int d^3y\;\sigma(\mathbf y)\,i\delta^{(3)}(\mathbf x-\mathbf y)
=
-\sigma(\mathbf x) ,
\end{align}}
showing that the final result is expectedly a $c$-number concentrated on the initial Cauchy slice. Moreover 
since this is a $c$-number, all higher nested commutators vanish. This would imply the following (see also \eqref{twogonec}): 
\begin{equation}\label{ancooper}
D^\dagger(\sigma)\,\phi(\mathbf x,t_0)\,D(\sigma)
=
\phi(\mathbf x,t_0)+\sigma(\mathbf x) ,
\end{equation}
justifying the shift of the field by the function $\sigma({\bf x})$ localized on the initial Cauchy slice. The relation \eqref{ancooper}
holds in the Schr\"odinger picture at the initial time slice
$t=t_0$.  To obtain the corresponding statement at a later time, one
must evolve the shifted operator with the time-evolution operator.
Accordingly, and defining the evolution appropriately, we get:
\begin{equation}
\phi_H^{(\sigma)}(\mathbf x,t)
\equiv
U^\dagger(t,t_0)\,
\Big(
D^\dagger(\sigma)\,\phi(\mathbf x,t_0)\,D(\sigma)
\Big)\,
U(t,t_0)=
\phi_H(\mathbf x,t)+\phi_\sigma(\mathbf x,t), 
\end{equation}
where we substituted the shifted Schr\"odinger-picture relation into the first
expression and then used the conjugation to get the Heisenberg picture. We have also defined:
\begin{equation}
\phi_\sigma(\mathbf x,t)
\equiv
U^\dagger(t,t_0)\,\sigma(\mathbf x)\,U(t,t_0).
\end{equation}
Note that the later-time shift does not follow by directly replacing
$\phi(\mathbf x,t_0)$ with $\phi_H(\mathbf x,t)$ inside the original
conjugation formula.  Rather, one first shifts the Schr\"odinger
operator at the initial time and then evolves the resulting operator
to time $t$. Moreover, since $\sigma(\mathbf x)$ is treated as a fixed $c$-number profile,
it commutes with the time-evolution operator and one simply has
$\phi_\sigma(\mathbf x,t)=\sigma(\mathbf x)$.


The story so far is rather straightforward and the result is an expected one. However once we take the displacement operator associated with the GS states, the analysis gets bit more non-trivial. Part of the reason is the form of the displacement operator in \eqref{sweenseyf}: it is a time-ordered integrated operator over all space and time instead of being localized on an initial Cauchy slice. To see how the story unfolds now, 
let us repeat the discussion for the GS-type operator written in the
form:
\begin{equation}
D[\sigma]
\equiv
\mathcal T \exp\!\left(
\int d^4x\;\sqrt{-g(x)}\,\sigma(x)\,\phi(x)
\right),
\end{equation}
where $\mathcal T$ denotes time ordering with respect to the Lorentzian
time coordinate $x^0$. Interestingly, even if we choose $\sigma(x)$ to be of the form $\sigma({\bf x})\delta(t - t_0)$ so as to localize the operator on a Cauchy slice, this would be different from the displacement operator for the coherent state because of the presence of the field $\phi(x)$ instead of its conjugate momentum $\pi(x)$. With this in mind, we would like to understand how the conjugation
of the scalar operator $\phi(y)$ changes under the operation:
\begin{equation}
D[\sigma]^\dagger\,\phi(y)\,D[\sigma],
\qquad
y\equiv (y^0,\mathbf y).
\end{equation}
From the get go one might worry because of the presence of $\sqrt{-g(x)}$. For scalar field $\phi(x)$ this is not a problem:
as long as $\sqrt{-g(x)}\,\sigma(x)$ is
treated as a prescribed $c$-number function\footnote{In other words taking the metric as the background metric and ignoring the fluctuations. Alternatively in the operator representation $\hat{g}_{\mu\nu}(x) = g^{(0)}_{\mu\nu}(x) + \delta\hat{g}_{\mu\nu}(x)$ consider only the $g^{(0)}_{\mu\nu}(x)$ piece for simplicity.}, the previous argument goes
through essentially unchanged, except that the source profile
$\sigma(x)$ is replaced everywhere by the weighted profile
$\sqrt{-g(x)}\,\sigma(x)$.  The result is therefore a nonlocal shift
weighted by the spacetime volume element.  If, however, the metric
itself is promoted to an operator, then additional operator-ordering
effects appear and the argument becomes more involved. For the simple case if we define for convenience
$J_g(x)\equiv \sqrt{-g(x)}\,\sigma(x)$, 
then expanding in Dyson series gives
\begin{equation}
D[\sigma]
=
1
+
\int d^4x\;J_g(x)\,\phi(x)
+
\frac{1}{2}
\int d^4x\,d^4x'\;
\mathcal T\!\Big[
J_g(x)\phi(x)\,J_g(x')\phi(x')
\Big]
+\cdots ,
\end{equation}
where the time ordering shows up from the second term onwards. (The way we wrote this is by keeping the ordering of $J_g(x)$ intact, but for the $c$-number functions this is not important.)
To first order in $\sigma$ one finds:
\bg\label{HKPGKM}
D[\sigma]^\dagger\,\phi(y)\,D[\sigma]
&= &
\phi(y)
+
\int d^4x\;J_g(x)\,[\phi(y),\phi(x)]
+\mathcal O(\sigma^2) \nonumber\\
& = & \phi(y)
+
i\int d^4x\;\sqrt{-g(x)}\,\sigma(x)\,\Delta(y,x)
+
\mathcal O(\sigma^2) ,
\nd
where, going from the first line to the next, 
the only change at this stage is the replacement
$\sigma(x)\;\longrightarrow\;\sqrt{-g(x)}\,\sigma(x)$ and the usage of the Pauli--Jordan function $\Delta(y, x)$ which, 
for a free (or asymptotic) scalar field, comes from 
\begin{equation}
[\phi(y),\phi(x)]
=
i\,\Delta(y,x) .
\end{equation}
In flat space this
reduces to $\Delta(y-x)$, whereas in a general background it depends on
the two spacetime points separately.  It still vanishes for spacelike
separation, {\it i.e.}
$\Delta(y,x)=0$
{for spacelike-separated $x$ and $y$, 
but it is generically nonzero when the two points are timelike or null
related. Thus the result in \eqref{HKPGKM} is again a shift, but it is not a local shift
$\phi(y)\;\to\;\phi(y)+\sigma(y)$, 
as in the coherent state case. 
Rather, it is a spacetime-smeared shift controlled by the Pauli--Jordan
kernel and weighted by the invariant measure. Moreover, for a free field the commutator is a $c$-number, so all higher nested
commutators vanish:
\begin{equation}
\Big[\,[\phi(y),\phi(x)],\,\phi(x')\Big]
=
\Big[\,i\Delta(y-x)\mathbf 1,\,\phi(x')\Big]
=0.
\end{equation}
Hence, in the free theory, the transformation truncates exactly at
first order as \eqref{HKPGKM} with $\Delta(y, x) = \Delta(y-x)$ with no $\mathcal O(\sigma^2)$ term.
By contrast, in an interacting theory the commutator
$[\phi(y), \phi(x)] = i{\cal C}(y,x)$ is itself an operator $-$ which still satisfies microcausality,
${\cal C}(y,x)=0$
{for spacelike separation $-$ so higher nested commutators
need not vanish:
\begin{equation}
\Big[\,[\phi(y),\phi(x)],\,\phi(x')\Big]
=
\big[\,i{\cal C}(y,x),\,\phi(x')\big]
\neq 0 ,
\end{equation}
in general.
Therefore the conjugation by the GS operator generates an infinite
series of nested commutators, and there is no simple exact shift
identity analogous to the free-field result.

The discussion becomes more subtle if one promotes the metric to an
operator.  In that case the insertion is not simply
$\sqrt{-g(x)}\,\sigma(x)\,\phi(x)$
with $\sqrt{-g(x)}$ a $c$-number, but rather something of the form
$\sqrt{-\hat g(x)}\,\sigma(x)\,\phi(x)$.
Then $\sqrt{-\hat g(x)}$ need not commute with the other operator
insertions, and the first commutator term becomes:
\begin{equation}\label{chukmukha}
\int d^4x\;
\left[
\phi(y),
\sqrt{-\hat g(x)}\,\sigma(x)\,\phi(x)
\right] ,
\end{equation}
which could probably be simplified a bit by taking $\hat{g}_{\mu\nu}(x) = g^{(0)}_{\mu\nu}(x) + \delta \hat{g}_{\mu\nu}(x)$ and then Taylor expanding over $\delta\hat{g}_{\mu\nu}(x)$ by keeping track of the operator orderings. In general however \eqref{chukmukha} takes the form:
\begin{equation}
\int d^4x\;
\left(
[\phi(y),\sqrt{-\hat g(x)}]\,\sigma(x)\,\phi(x)
+
\sqrt{-\hat g(x)}\,\sigma(x)\,[\phi(y),\phi(x)]
\right).
\end{equation}
Thus even the first correction need no longer be a $c$-number, and the
higher nested commutators will generally fail to vanish.  In such a
situation one no longer has a simple exact shift formula.

The scenario does not improve once we take the displacement operator  to study the shift of a metric field. Suppose now that the GS operator is built directly from the metric, as in the second relation from \eqref{sweenseyf},
and we wish to study the conjugation of the metric operator itself, {\it i.e.}:
\begin{equation}
D^\dagger[\sigma]\,
\hat g_{\alpha\beta}(y)\,
D[\sigma].
\end{equation}
This is qualitatively more complicated than the scalar example because
the operator appearing in the exponential is now a nonlinear
functional of the same operator $\hat g_{\mu\nu}$ that we are
conjugating.  In particular, the insertion is no longer of the simple
form
$J(x)\,\hat{\cal O}(x)$
with $J(x)$ a prescribed $c$-number source.  Rather, the entire
quantity
$\mathcal V(x) \equiv  \sqrt{-\hat g(x)}\,\sigma^{\mu\nu}(x)\,\hat g_{\mu\nu}(x)$
is operator valued.
Expanding in Dyson series gives:
\begin{equation}
D[\sigma]
=
1
+
\int d^4x\;{\cal V}(x)
+
\frac{1}{2}
\int d^4x\,d^4x'\;
\mathcal T\!\Big[
{\cal V}(x)\,{\cal V}(x')
\Big]
+\cdots ,
\end{equation}
where $\mathcal T$ is the time-ordering and, as before, it appears from the second term onwards. (We also don't put a hat over $\mathcal V(x)$, but it should be understood as an operator.)
To first order in $\sigma$, the conjugated metric becomes:
\bg
D[\sigma]^\dagger\,
\hat g_{\alpha\beta}(y)\,
D[\sigma]
& = &
\hat g_{\alpha\beta}(y)
+
\int d^4x\;
\Big[
\hat g_{\alpha\beta}(y),\,{\cal V}(x)
\Big]
+
{\cal O}(\sigma^2)\nonumber\\
& = &
\hat g_{\alpha\beta}(y)
+
\int d^4x\;
\Big[
\hat g_{\alpha\beta}(y),\,
\sqrt{-\hat g(x)}\,
\sigma^{\mu\nu}(x)\,
\hat g_{\mu\nu}(x)
\Big]
+
{\cal O}(\sigma^2). \nonumber\\
&= &
\hat g_{\alpha\beta}(y)
+
\int d^4x\;
\sigma^{\mu\nu}(x)\,
[\hat g_{\alpha\beta}(y),\sqrt{-\hat g(x)}]\,
\hat g_{\mu\nu}(x) \nonumber\\
&+ &
\int d^4x\;
\sqrt{-\hat g(x)}\,
\sigma^{\mu\nu}(x)\,
[\hat g_{\alpha\beta}(y),\hat g_{\mu\nu}(x)]
+
{\cal O}(\sigma^2) ,
\nd
where, since $\sigma^{\mu\nu}(x)$ is an external $c$-number profile, we have
pulled it outside the commutator. The result is the direct analogue of the scalar-field formula, except that
both terms are now genuinely operator valued. However
for an ordinary free scalar field with a linear source insertion, the
basic commutator is a $c$-number, and therefore higher nested
commutators vanish.  Here the situation is very different because (a) 
the operator $\sqrt{-\hat g(x)}$ is a nonlinear functional of the
metric operator; (b) 
the commutator
$[\hat g_{\alpha\beta}(y),\hat g_{\mu\nu}(x)]$
is not, in general, a $c$-number; and (c)
the commutator
$[\hat g_{\alpha\beta}(y),\sqrt{-\hat g(x)}]$
is itself operator valued. Therefore even the first correction is already an operator-valued
expression, and the higher nested commutators generically do not
vanish:
\begin{equation}
\Big[
\Big[
\hat g_{\alpha\beta}(y),{\cal V}(x)
\Big],
{\cal V}(x')
\Big]
\neq 0 ,
\end{equation}
in general. One should expect an infinite BCH/Dyson series rather than a
simple exact shift. However there does exist a way to tackle the situation by expansion around a background metric. In other words, 
a useful way to organize the computation is to expand the metric
operator around a background,
$\hat g_{\mu\nu}(x)
=
g^{(0)}_{\mu\nu}(x)
+
\delta\hat g_{\mu\nu}(x)$ 
where $g^{(0)}_{\mu\nu}(x)$ is a $c$-number background and
$\delta\hat g_{\mu\nu}(x)$ denotes the fluctuation operator.
Then substituting the determinant operator\footnote{The determinant can be expanded as $
\sqrt{-\hat g(x)}
=
\sqrt{-g^{(0)}(x)}
+
\frac{1}{2}\sqrt{-g^{(0)}(x)}\,
g_{(0)}^{\mu\nu}(x)\,
\delta\hat g_{\mu\nu}(x)
+
{\cal O}\!\left((\delta\hat g)^2\right)$.} into the insertion gives:

{\scriptsize
\bg\label{ennaboro}
{\cal V}(x)
&= &
\sqrt{-\hat g(x)}\,
\sigma^{\mu\nu}(x)\,
\hat g_{\mu\nu}(x) =  {\cal V}_0(x)
+
{\cal V}_1(x)
+\cdots \\
&= &
\underbrace{\sqrt{-g^{(0)}(x)}\,
\sigma^{\mu\nu}(x)\,
g^{(0)}_{\mu\nu}(x)}_{\mathcal V_0(x)}
+
\underbrace{\sqrt{-g^{(0)}(x)}\,
\sigma^{\mu\nu}(x)\,
\delta\hat g_{\mu\nu}(x)
+
\frac{1}{2}\sqrt{-g^{(0)}(x)}\,
g_{(0)}^{\rho\sigma}(x)\,
\delta\hat g_{\rho\sigma}(x)\,
\sigma^{\mu\nu}(x)\,
g^{(0)}_{\mu\nu}(x)}_{\mathcal V_1(x)}
+
\cdots , \nonumber
\nd}
where 
the first term, $\mathcal V_0(x)$, is a pure $c$-number and therefore drops out of the
commutator.  The leading nontrivial operator terms are linear in the
fluctuation, and is given by $\mathcal V_1(x)$ in \eqref{ennaboro}. 
Therefore, to leading order in fluctuations,
\begin{equation}
D^\dagger[\sigma]\,
\hat g_{\alpha\beta}(y)\,
D[\sigma]
=
\hat g_{\alpha\beta}(y)
+
\int d^4x\;
\Big[
\hat g_{\alpha\beta}(y),{\cal V}_1(x)
\Big]
+\cdots ,
\end{equation}
which makes it explicit that the first correction is controlled by the
commutator of the metric fluctuation with a linearized source-like
operator built from $\delta\hat g_{\mu\nu}$. Note that
even if one imposes canonical equal-time commutation relations on the
basic ADM variables, the insertion in $D[\sigma]$ is integrated over
all times.  Hence the first correction samples
$[\hat g_{\alpha\beta}(y^0,\mathbf y),\hat g_{\mu\nu}(x^0,\mathbf x)]$
with $x^0\neq y^0$ in general, and there is no reason for such a
commutator to vanish.  Thus, just as in the scalar case, equal-time
commutativity does not imply that the first correction is zero. 
However,
unlike the free scalar case, there is in general no simple
exact shift identity.  The nonlinear dependence of
$\sqrt{-\hat g}$ on the same metric operator and the fact that the
metric commutators are not $c$-numbers imply that higher nested
commutators generically survive, so the full conjugation must be
treated as an infinite operator series. 
Expanding around a background metric helps in organizing the result order by
order in the fluctuation operator $\delta\hat g_{\mu\nu}$. The analysis nevertheless remains technically involved, since the
higher-order corrections are considerably more intricate.  A more
practical route is therefore to adopt the path-integral formulation,
which has the additional advantage of making the incorporation of
quantum corrections systematic and transparent.  This framework also
allows one to exhibit more explicitly the distinction between the
results obtained from coherent states and those arising from the
Glauber--Sudarshan states.  We turn to these issues, together with
several related questions, in the following discussion.

\subsection{The Lorentzian in-in Schwinger-Keldysh representation of GS states \label{sec6.2}}

In Lorentzian signature, real-time expectation values are generally
represented on the Schwinger--Keldysh (SK) contour. We now derive the
SK path-integral representation of the GS-state expectation value in a
way parallel to the coherent-state derivation, but keeping track of the
additional GS insertions explicitly.

We will start with the usual gauge-fixed action and doubled fields.
Let the full gauge-fixed action $S_{\rm tot}$ take the trans-series form \eqref{bellcroft} with $S_{\rm pert}$ as in \eqref{fargomey}.
Let ${\rm g}$ denote the full field multiplet of the gauge-fixed
theory, such that on the
Schwinger--Keldysh contour the fields are doubled in the following way:
\begin{equation}
{\rm g} \; \longrightarrow\; ({\rm g}_+, {\rm g}_-)~~~~ {\rm with}~~~
{\rm g}
\equiv
(g,\bar c,c,\text{matter},\ldots) ,
\end{equation}
where $(c, \bar{c})$ denote the ghosts and all other degrees of freedom are grouped under matter. Thus 
${\rm g}_\pm
=
(g_\pm,\bar c_\pm,c_\pm,\text{matter}_\pm,\ldots)$ will henceforth denote the degrees of freedom at the two branches of the SK contour such that the corresponding contour weight is:
\begin{equation}
\exp\!\Big(
iS_{\rm tot}[{\rm g}_+]
-
iS_{\rm tot}[{\rm g}_-]
\Big).
\label{eq:SK_weight_detailed2}
\end{equation}
So far this is similar to what we saw for the coherent state case. The difference appears when we look at the operator definition of the GS expectation value. To quantify this, let us take the displacement operator for the GS state from \eqref{sweenseyf}, {\it i.e.}:
\begin{equation}\label{vandikok}
D[\sigma]
=
\mathcal T \exp\!\left(
\int d^4x\;
\sqrt{-\hat g(x)}\,
\sigma^{\mu\nu}(x)\,
\hat g_{\mu\nu}(x)
\right),
\end{equation}
where $\mathcal T$ is the time-ordering symbol and hats denote the operators. In the way we wrote $D[\sigma]$, all the metric components appearing in \eqref{vandikok} are operators. However, to avoid unnecessary operator-ordering ambiguities, we treat
$\sigma^{\mu\nu}(x)$ as a prescribed external contravariant source,
independent of the metric operator $\hat g_{\mu\nu}(x)$.  Only the
factors $\sqrt{-\hat g(x)}$ and $\hat g_{\mu\nu}(x)$ are therefore
operator valued in the GS insertion\footnote{If we do not make this assumption then the insertion should really be written differently, because if $\sigma_{\mu\nu}$
 is metric-raised, we are hiding extra operator dependence. In that case one should write the source in a lower-index form from the start, for example:
\begin{equation}
D[\sigma] = \mathcal T \exp\!\left(
\int d^4x\;
\sqrt{-\hat g(x)}\,
\sigma_{\mu\nu}(x)\,
\hat g^{\mu\nu}(x)
\right),
\end{equation}
or something equivalent, and then handle the operator ordering carefully. To avoid all these ambiguities we will henceforth define $\sigma^{\mu\nu}(x)$ be be a $c$-number function from the get go.}. We will let the GS density matrix on the initial slice be:
\begin{equation}
\rho_\sigma
=
D[\sigma]\,
\rho_\Omega\,
D^\dagger[\sigma],
\label{eq:rho_sigma_def_detailed2}
\end{equation}
where $\rho_\Omega \equiv |\Omega\rangle\langle\Omega |$ is the vacuum density matrix with $|\Omega\rangle$ denoting the  vacuum over which we are constructing the GS states, \emph{e.g.} an interacting supersymmetric Minkowski vacuum in string theory. Then the
expectation value of a local operator $\hat O_H(x)$ in the GS state is:
\begin{equation}
\langle O(x)\rangle_\sigma
=
\frac{
{\rm Tr}\!\left(
\rho_\sigma\,\hat O_H(x)
\right)
}{
{\rm Tr}\!\left(
\rho_\sigma
\right)
} = \frac{{\rm Tr}\!\left(
\rho_\Omega\,
D^\dagger[\sigma]\,
\hat O_H(x)\,
D[\sigma]
\right)}{\mathrm{Tr}\!\left(
\rho_\Omega\,
D^\dagger[\sigma]\,
D[\sigma]
\right)} ,
\label{eq:GS_expect_start2}
\end{equation}
where we used the cyclicity of the trace in the intermediate step. Note that, since
$D[\sigma]$ is itself a four-dimensional time-ordered exponential, it
is not an operator localized on a single Cauchy slice.  Therefore one
must not treat it as if it were a Schr\"odinger-picture operator at
$t_i$ as we did for the equivalent coherent state case.  The Schwinger--Keldysh contour must instead be derived directly
from the trace for both the numerator and the denominator. The numerator takes the form:
\begin{equation}
N_\sigma(x)
=
\mathrm{Tr}\!\left(
\rho_\Omega\,
D^\dagger[\sigma]\,
\hat O_H(x)\,
D[\sigma]
\right)=
\mathrm{Tr}\!\left(D[\sigma]\,
\rho_\Omega\,
D^\dagger[\sigma]\,
\hat O_H(x)
\right) ,
\label{eq:GS_num_start}
\end{equation}
where we can write the trace by inserting a complete basis at the return time $t_f$. In other words, choosing a return time $t_f>t$, and letting $|h_f,t_f\rangle$ denote a
complete basis of metric eigenstates on the hypersurface $\Sigma_f$, we get the following 
using the definition of the trace:
\begin{equation}
N_\sigma(x)
=
\int Dh_f\;
\langle h_f,t_f| D[\sigma]\,\rho_\Omega\, 
D^\dagger[\sigma]\,
\hat O_H(x)
|h_f,t_f\rangle.
\label{eq:trace_with_hf}
\end{equation}
At this stage the contour is already implicit: the same final state
$|h_f,t_f\rangle$ appears both as bra and ket, which is precisely what
will force the forward and backward histories to meet at $t_f$. We can make this a bit more explicit by choosing $\Sigma_i$ as the initial hypersurface at time $t_i$, and then inserting 
$\mathbf 1_{\Sigma_i}
=
\int Dh^\pm_i\;|h^\pm_i,t_i\rangle\langle h^\pm_i,t_i|$
on both sides of $\rho_\Omega$ appropriately. This converts \eqref{eq:trace_with_hf} to take the form:
\begin{equation}
\begin{aligned}
N_\sigma(x)
&=
\int Dh_f
\int Dh_i^+
\int Dh_i^-\;
\langle h_i^+,t_i|\rho_\Omega|h_i^-,t_i\rangle
\\
&\qquad\times
\langle h_f,t_f|D[\sigma]|h_i^+,t_i\rangle
\;
\langle h_i^-,t_i|D^\dagger[\sigma]\hat O_H(x)|h_f,t_f\rangle ,
\end{aligned}
\label{eq:num_with_initial_basis}
\end{equation}
where the SK contour becomes more explicit. 
The doubled structure is also evident:
(a) $h_i^+$ {belongs to the forward branch,} and (b) 
$h_i^-$ {belongs to the backward branch.}

We can now interpret the two matrix elements as forward and backward contour amplitudes in the following way.
Because $D[\sigma]$ is time ordered over the Lorentzian contour, the
matrix element:
\begin{equation}
\langle h_f,t_f|D[\sigma]|h_i^+,t_i\rangle
\end{equation}
is represented by a forward path integral from $h_i^+$ at $t_i$ to
$h_f$ at $t_f$, with the GS insertion living on the forward branch.
Similarly, since:
\begin{equation}
D^\dagger[\sigma]
=
\bar{\mathcal T}
\exp\!\left(
\int d^4x\;
\sqrt{-\hat g(x)}\,
\bar\sigma^{\mu\nu}(x)\,
\hat g_{\mu\nu}(x)
\right),
\label{eq:Ddagger_GS}
\end{equation}
where $\bar{\mathcal T}$ is the anti time-ordering symbol with $\bar{\sigma}^{\mu\nu}(x)$ being the corresponding complex conjugate function that remains equal to $\sigma^{\mu\nu}(x)$ for real function, 
the matrix element:
\begin{equation}
\langle h_i^-,t_i|D^\dagger[\sigma]\hat O_H(x)|h_f,t_f\rangle
\end{equation}
is represented by a backward path integral from $h_f$ at $t_f$ back to
$h_i^-$ at $t_i$, with the anti-time-ordered GS insertion living on
the backward branch, together with the operator insertion $\hat O_H(x)$. Such an observation also helps us to convert the two branches into path integrals. Taking ${\rm g}$ to denote the full field multiplet of the gauge-fixed
theory as
${\rm g}
\equiv
(g,\bar c,c,\text{matter},\ldots)$ with $S_{\rm tot}[{\rm g}]$ being the full trans-series action \eqref{amaliagorom}, then 
 the forward and the backward matrix elements become:
\bg\label{coridahl}
&& \langle h_f,t_f|D[\sigma]|h_i^+,t_i\rangle
=
\int_{h_i^+}^{h_f}
{\cal D}{\rm g}_+\;
\exp\!\left(
iS_{\rm tot}[{\rm g}_+]
\right)
\,
\mathcal I_+[g_+;\sigma] \\
&& \langle h_i^-,t_i|D^\dagger[\sigma]\hat O_H(x)|h_f,t_f\rangle
=
\int_{h_i^-}^{h_f}
{\cal D}{\rm g}_-\;
\exp\!\left(
-iS_{\rm tot}[{\rm g}_-]
\right)
\,
\mathcal I_-[g_-;\sigma]\,
O[g_-](x), \nonumber \nd
where we see that the operator insertion is done on the $-$ branch. 
For a Hermitian local operator one may, after the standard contour
identification, equally place the insertion on the $+$ branch, so the
final normalized expectation value may be written with $O[g_+](x)$. The other parameters appearing in \eqref{coridahl} are defined as:
\bg\label{coridahl2}
&& \mathcal I_+[g_+;\sigma]
\equiv
\exp\!\left[
\int d^4x\;
\sqrt{-g_+(x)}\,
\sigma^{\mu\nu}(x)\,
g_{+\mu\nu}(x)
\right] \nonumber\\
&& \mathcal I_-[g_-;\sigma]
\equiv
\exp\!\left[
\int d^4x\;
\sqrt{-g_-(x)}\,
\bar\sigma^{\mu\nu}(x)\,
g_{-\mu\nu}(x)
\right] ,
\nd
where for real source profiles one may simply identify the two branch
sources,
$\bar\sigma^{\mu\nu}(x)=\sigma^{\mu\nu}(x)$,
so that the distinction between the $+$ and $-$ branches lies only in
their contour ordering and in the overall SK phase
$\exp\!\big(iS_+-iS_-\big)$. The only thing remaining to complete the story is the insertion of the vacuum density matrix from \eqref{eq:num_with_initial_basis}. 
On the initial slice this takes the form:
\begin{equation}
\langle h_i^+,t_i|\rho_\Omega|h_i^-,t_i\rangle
=
\Psi_\Omega[h_i^+]\,
\Psi_\Omega^\ast[h_i^-] ,
\label{coridahl4}
\end{equation}
which is represented by the product of two vacuum wavefunctions on the two branches respectively. Plugging \eqref{coridahl} and \eqref{coridahl4} into \eqref{eq:num_with_initial_basis}, we get:

{\footnotesize
\bg\label{coridahl6}
N_\sigma(x)
&= &
\int Dh_f
\int Dh_i^+\,Dh_i^-\;
\Psi_\Omega[h_i^+]\,
\Psi_\Omega^\ast[h_i^-]\\
&\times &
\int_{h_i^+}^{h_f}
{\cal D}{\rm g}_+\;
\exp\!\left(
iS_{\rm tot}[{\rm g}_+]
\right)
\mathcal I_+[g_+;\sigma]~
\int_{h_i^-}^{h_f}
{\cal D}{\rm g}_-\;
\exp\!\left(
-iS_{\rm tot}[{\rm g}_-]
\right)
\mathcal I_-[g_-;\sigma]\,
O[g_-](x) , \nonumber
\nd}
justifying how the products of the vacuum wavefunctions as well as the double copies of the action on the two  branches fit together to provide the SK in-in path integral formulation of the expectation value. In a similar vein, the denominator is obtained by removing the operator insertion from \eqref{coridahl6} as:

{\footnotesize
\bg\label{tricksbrke}
Z_\sigma
&=&
\int Dh_f
\int Dh_i^+\,Dh_i^-\;
\Psi_\Omega[h_i^+]\,
\Psi_\Omega^\ast[h_i^-]\\
&\times &
\int_{h_i^+}^{h_f}
{\cal D}{\rm g}_+\;
\exp\!\left(
iS_{\rm tot}[{\rm g}_+]
\right)
\mathcal I_+[g_+;\sigma]~
\int_{h_i^-}^{h_f}
{\cal D}{\rm g}_-\;
\exp\!\left(
-iS_{\rm tot}[{\rm g}_-]
\right)
\mathcal I_-[g_-;\sigma] , \nonumber
\nd}
giving us the required expression in terms of the two branches. The full expectation value of the operator $\hat{O}(x)$ would simply be the ratio of \eqref{coridahl6} and \eqref{tricksbrke}. This takes the following form:

{\footnotesize
\bg\label{coridahl7}
\langle \hat{O}(x)\rangle_\sigma
 & = & {N_\sigma \over Z_\sigma} = 
\frac{1}{Z_\sigma}
\int Dh_f
\int Dh_i^+\,Dh_i^-\;
\Psi_\Omega[h_i^+]\,
\Psi_\Omega^\ast[h_i^-] \\
&& ~~ \times \int_{h_i^+}^{h_f}
{\cal D}{\rm g}_+\;
\int_{h_i^-}^{h_f}
{\cal D}{\rm g}_-\;
O[g_-](x)\,
e^{\,iS_{\rm tot}[{\rm g}_+]-iS_{\rm tot}[{\rm g}_-]}\,
\mathcal I_+[g_+;\sigma]\,
\mathcal I_-[g_-;\sigma] , \nonumber
\nd}
thus providing the correct
Schwinger--Keldysh representation of the genuine GS 
operator\footnote{It is also easy to see how $Z_\sigma$ and $N_\sigma$ are real. The Schwinger--Keldysh partition function $Z_\sigma$ is not invariant under the bare exchange of the two contour branches by itself; rather, its relevant reality property is obtained by combining complex conjugation with the dummy-variable exchange $+\leftrightarrow -$. Under this combined operation, and using the fact that the backward-branch insertion is the Hermitian conjugate of the forward-branch insertion, one recovers the original functional, so that $Z_\sigma^\ast=Z_\sigma$. The same logic applies to the one-point function $N_\sigma(x)$, except that under complex conjugation the operator insertion is carried from one contour branch to the other. Since the Schwinger--Keldysh formalism represents the same physical in--in one-point function whether the operator is inserted on the forward or backward branch, the complex-conjugated expression maps back to the same observable, and therefore $N_\sigma(x)$ is real as well. Thus the reality of both $Z_\sigma$ and $N_\sigma$ is a consequence not of a bare branch exchange alone, but of the combined Schwinger--Keldysh involution consisting of complex conjugation together with exchange of the two dummy contour branches.}.  
This expectation value should be compared to the ones we got in \eqref{Sirlauren} and \eqref{Sirlauren2} for the coherent states. In \eqref{Sirlauren}, we see that the knowledge of the wavefunction of the coherent state is necessary. Alternatively, in \eqref{Sirlauren2}, we can keep the wavefunctions simple, but change the boundary conditions. Whereas for the GS states both the boundary conditions and the wavefunctions are simple, and all information of the GS state goes to the contour insertions that in turn may be absorbed into deformations of
the branch actions as:
\bg\label{coridahl9}
&& \exp\!\left(
iS_{\rm tot}[{\rm g}_+]
\right)\,
\mathcal I_+[g_+;\sigma]
=
\exp\!\left(
iS_{\rm tot}[{\rm g}_+]
+
\int d^4x\;
\sqrt{-g_+}\,
\sigma^{\mu\nu}\,
g_{+\mu\nu}
\right) \nonumber\\
&& \exp\!\left(
-iS_{\rm tot}[{\rm g}_-]
\right)\,
\mathcal I_-[g_-;\sigma]
=
\exp\!\left(
-iS_{\rm tot}[{\rm g}_-]
+
\int d^4x\;
\sqrt{-g_-}\,
\bar\sigma^{\mu\nu}\,
g_{-\mu\nu}
\right) ,
\nd
after which one may insert the operator in either $+$ or the $-$ branch. (For the analysis that we performed above, the operator insertion was on the $-$ branch; whereas for the coherent state, the operator insertion was in the $+$ branch. This difference is just convention and doesn't alter the main conclusions for either cases.) Moreover, we can reinterpret \eqref{coridahl9} equivalently as shifts of the doubled actions in the following way: 
\bg\label{coridahl101}
&& S_{\rm tot}[{\rm g}_+]
\;\to\;
S_{\rm tot}[{\rm g}_+]
-
i\int d^4x\;
\sqrt{-g_+}\,
\sigma^{\mu\nu}\,
g_{+\mu\nu} \nonumber\\
&& S_{\rm tot}[{\rm g}_-]
\;\to\;
S_{\rm tot}[{\rm g}_-]
+
i\int d^4x\;
\sqrt{-g_-}\,
\bar\sigma^{\mu\nu}\,
g_{-\mu\nu} ,
\nd
where note the relative appearance of the $i = \sqrt{-1}$. It should also be clear that the contour structure in \eqref{coridahl9} follows directly from the trace because 
the same final state $h_f$ appears on both branches,
so the forward and backward histories are sewn together at $t_f$.  The
operator $D[\sigma]$ lives on the forward branch, the operator
$D^\dagger[\sigma]$ lives on the backward branch, and the resulting
doubled functional integral is precisely the SK contour. This is described in detail in {\bf figure \ref{skcontourGS}}.

\begin{figure}[h]
\centering
\begin{tabular}{c}
\includegraphics[width=6in]{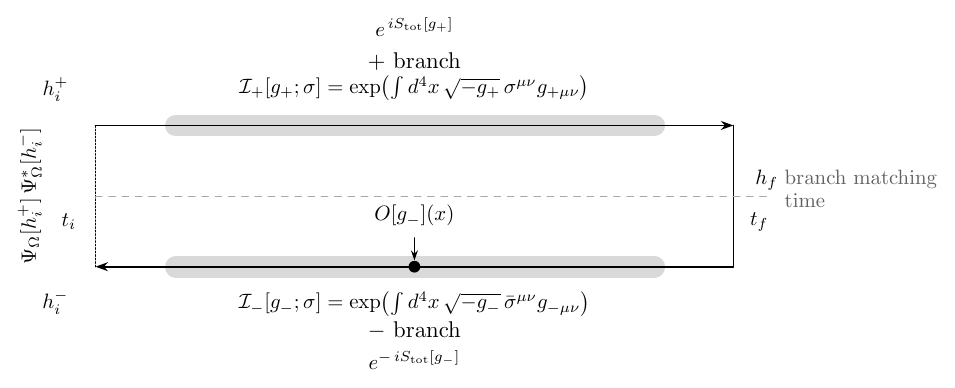}
\end{tabular}
\caption[]{The Schwinger--Keldysh contour associated with the Lorentzian Glauber--Sudarshan state. The gray intervals denote the continuous insertions $\mathcal I_+[g_+;\sigma]$ and $\mathcal I_-[g_-;\sigma]$ on the forward and backward branches, respectively. The two histories are sewn at the common final configuration $h_f$, while $\Psi_\Omega[h_i^+]\Psi_\Omega^*[h_i^-]$ supplies the initial boundary data. The operator $O[g_-](x)$ is inserted on the backward branch. The complete functional-integral expression represented by the contour is given in Eq.~\eqref{coridahl7} of the main text. One should also compare the SK contour with the one for the coherent state in {\bf figure \ref{skcontour1}}.}
\label{skcontourGS}
\end{figure}


\subsubsection{SK contour and deriving the first equation in \eqref{coridahl} \label{sec6.333}}

The result presented in \eqref{coridahl} provides the precise way to convert some of the key parts of \eqref{eq:num_with_initial_basis} into path-integral format. However the derivation of \eqref{coridahl} is more subtle and in the following we will see how this may come about by first going to the corresponding Schr\"odinger picture and then converting it back to the path-integral form. If the Schr\"odinger picture is anchored on the initial slice $\Sigma_i\equiv \Sigma_{t_i}$, then the correct relation is:
\begin{equation}
\hat g_{H\,\mu\nu}(t,\mathbf x)
=
U^\dagger(t,t_i)\,
\hat g_{S\,\mu\nu}(t_i,\mathbf x)\,
U(t,t_i),
\label{eq:gH_gS_correct_rewrite2}
\end{equation}
where $H$ and $S$ denote the Heisenberg and the Schr\"odinger pictures respectively. Now comes our first crucial point. 
The GS density inserted at time $t$, for the Heisenberg and the Schr\"odinger pictures are defined by the following forms:
\bg\label{lenulilaboro}
&& \hat J_H(t)
=
\int d^3x\;
\sqrt{-\hat g_H(t,\mathbf x)}\,
\sigma^{\mu\nu}(t,\mathbf x)\,
\hat g_{H\,\mu\nu}(t,\mathbf x) \nonumber\\
&& \hat J_S(t_i;t)
\equiv
\int d^3x\;
\sqrt{-\hat g_S(t_i,\mathbf x)}\,
\sigma^{\mu\nu}(t,\mathbf x)\,
\hat g_{S\,\mu\nu}(t_i,\mathbf x), \nd
where 
the corresponding Schr\"odinger operator is defined on the reference slice $t_i$. The crucial observation is then the following.
The $t$-dependence here comes only from the c-number source profile
$\sigma^{\mu\nu}(t,\mathbf x)$; the operator part itself lives on the
reference Hilbert space at time $t_i$; {\it i.e.} on the initial Cauchy slice defined at $t_i$. Thus the correct relation is:
\begin{equation}
\hat J_H(t)
=
U^\dagger(t,t_i)\,
\hat J_S(t_i;t)\,
U(t,t_i),
\label{eq:JH_JS_correct_rewrite2}
\end{equation}
which converts the GS density operator from one picture to another. The actual GS operator from say \eqref{sweenseyf} is now easy to write using \eqref{lenulilaboro} and \eqref{eq:JH_JS_correct_rewrite2} in the following way:
\bg\label{lenlil01}
D[\sigma]
=
{\cal T}\exp\!\left(
\int_{t_i}^{t_f} dt\; \hat J_H(t)
\right) = {\cal T}\exp\!\left(
\int_{t_i}^{t_f} dt\;
U^\dagger(t,t_i)\,
\hat J_S(t_i;t)\,
U(t,t_i)
\right) ,
\nd
defined on the interval $[t_i,t_f]$. However what we need is actually the discretized form of \eqref{lenlil01}. For that we can discretize the interval as
$t_i=t_0<t_1<\cdots<t_N=t_f$ such that a small interval $\Delta t$ may be defined as $\Delta t=\frac{t_f-t_i}{N}$.
Then the time-ordered exponential \eqref{lenlil01} takes the following form:

{\footnotesize
\begin{equation}\label{lenlin02}
D[\sigma]
=
\lim_{N\to\infty}
\prod_{n=N-1}^{0}
\exp\!\left(
\Delta t\,
\hat J_H(t_n)
\right) = \lim_{N\to\infty}
\prod_{n=N-1}^{0}
\left[
U(t_{n+1},t_n)\,
\exp\!\left(
\Delta t\,\hat J_S(t_i;t_n)
\right)
\right],
\end{equation}}
with later times placed to the left. The above can be easily derived by decomposing the unitary operator $U(t_n, t_i)$ appropriately and then using the standard Dyson/Trotter rearrangement up to the usual ${\cal O}(\Delta t^2)$ ordering ambiguities that vanish in the continuum limit\footnote{In fact the second equality in \eqref{lenlin02} follows by expressing each Heisenberg insertion as a conjugate Schr\"odinger-picture insertion, using $e^{U^\dagger XU}=U^\dagger e^X U$ and then collapsing adjacent evolution operators with the composition law for $U(t_b,t_a)$.}. We can now plug in the expansion on the RHS of \eqref{lenlin02} to evaluate 
our first matrix element from \eqref{coridahl}, namely:
\bg\label{lenu3}
\langle h_f,t_f|D[\sigma]|h_i^+,t_i\rangle, \nd
where we can define a set of $h_n$ with $h_0 \equiv h_i^+$ and $h_N \equiv h_f$ forming 
complete sets of configuration eigenstates $|h_n, t_n\rangle$ with 
$\mathbf 1_n
=
\int Dh_n\,|h_n,t_n\rangle\langle h_n,t_n|$ and $n=1,\ldots,N-1$. We can then insert the identity operator 
at each intermediate slice of $D[\sigma]$ from \eqref{lenlin02} in \eqref{lenu3}. This gives us the following form of the matrix element:

{\footnotesize
\begin{align}
\langle h_f,t_f|D[\sigma]|h_i^+,t_i\rangle
&=
\lim_{N\to\infty}
\int \prod_{m=1}^{N-1} Dh_m\;
\prod_{n=0}^{N-1}
\langle h_{n+1},t_{n+1}|
U(t_{n+1},t_n)\,
e^{\,\Delta t\,\hat J_S(t_i;t_n)}
|h_n,t_n\rangle,
\label{eq:first_discrete_ME_rewrite2}
\end{align}}
where the integration measure runs from $m = 1$ to $m = N-1$ because we do not integrate over $h_0 \equiv h_i^+$. (Here the second line is valid because, after the intermediate states are inserted, each
factor is a $c$-number matrix element, so the resulting product may be written with the
index $n$ increasing from $0$ to $N-1$ without affecting the value.) At this stage, to justify the factorization properly, insert one more identity on the same slice $t_n$ using 
$\mathbf 1_{t_n}
=
\int Dh_m\; |h_m,t_n\rangle\langle h_m,t_n|$.
Then:
\begin{align}\label{lenu5}
&\langle h_{n+1},t_{n+1}|
U(t_{n+1},t_n)\,
e^{\,\Delta t\,\hat J_S(t_i;t_n)}
|h_n,t_n\rangle
\nonumber\\
&\qquad=
\int Dh_m\;
\langle h_{n+1},t_{n+1}|U(t_{n+1},t_n)|h_m,t_n\rangle\;
\langle h_m,t_n|e^{\,\Delta t\,\hat J_S(t_i;t_n)}|h_n,t_n\rangle\nonumber\\
&\qquad = \int Dh_m\;
\langle h_{n+1},t_{n+1}|U(t_{n+1},t_n)|h_m,t_n\rangle\;
e^{\,\Delta t\,J[h_n;t_n]}\,
\delta[h_m-h_n] \nonumber\\
&\qquad=
\langle h_{n+1},t_{n+1}|U(t_{n+1},t_n)|h_n,t_n\rangle\;
e^{\,\Delta t\,J[h_n;t_n]}
\end{align}
where in the first two lines of \eqref{lenu5}
$\hat J_S(t_i;t_n)$ is the operator in the Schr\"odinger representation and is 
multiplicative in the configuration basis at the slice $t_n$. In the third line we used the eigenvalue equation:
\begin{equation}\label{lenu6}
\hat J_S(t_i;t_n)\,|h_n,t_n\rangle
=
J[h_n;t_n]\,
|h_n,t_n\rangle,
\end{equation}
giving us the eigenvalue $J[h_n; t_n]$ and the delta functional $\delta[h_m-h_n]$. Note that while the operator $\hat{J}_S[t_i; t_n)$ is a function of the initial slice at $t_i$, the eigenvalue is only a function of $t_n$ and $h_n$ because it acts on an eigenstate defined on the Cauchy slice at $t_n$. The form of the eigenvalue may be derived from \eqref{lenulilaboro} and takes the form:
\begin{equation}\label{lenu7}
J[h_n;t_n]
=
\int d^3x\;
\sqrt{-g_n(\mathbf x)}\,
\sigma^{\mu\nu}(t_n,\mathbf x)\,
g_{n, \mu\nu}(\mathbf x) ,
\end{equation}
alongwith the delta functional removing the $Dh_m$ integral to set
$h_m=h_n$, reproduces the requisite result on the fourth line of \eqref{lenu5}. Here $g_{n,\mu\nu}(\mathbf x)$ is simply shorthand for the metric evaluated on the time slice $t=t_n$, namely $g_{\mu\nu}(t_n,\mathbf x)$. Likewise $ \sqrt{-g_n(\mathbf x)}
\equiv
\sqrt{-g(t_n,\mathbf x)}$ fixes the determinant\footnote{There is however one important conceptual caveat.
If $|h_n,t_n\rangle$ is strictly an eigenstate of the induced spatial metric on the Cauchy slice $\Sigma_{t_n}$, then what is directly fixed by the state is
$h_{n,ij}(\mathbf x)
=
h_{ij}(t_n,\mathbf x)$,
not necessarily the full spacetime metric components
$g_{\mu\nu}(t_n,\mathbf x)$.
So the formula
$g_{n,\mu\nu}(\mathbf x)=g_{\mu\nu}(t_n,\mathbf x)$
is correct as notation, but whether all components are really determined by the eigenstate depends on what basis of states we are using. For the present case, if we view the GS state as repeated kicks, we can arrange the basis of states to reflect the full metric components. This way, being on any Cauchy slice, we do not lose any information.}. After the dust settles we get:

{\footnotesize
\bg\label{lenu9}
\langle h_f,t_f|D[\sigma]|h_i^+,t_i\rangle
=
\lim_{N\to\infty}
\int \prod_{m=1}^{N-1} Dh_m\;
\prod_{n=0}^{N-1}
\langle h_{n+1},t_{n+1}|U(t_{n+1},t_n)|h_n,t_n\rangle
\;
e^{\,\Delta t\,J[h_n;t_n]} ,
\nd}
where note two important developments: one, we now have $U(t_{n+1}, t_n)$ sandwiched between a bra and a ket, and two, the appearance of the eigenvalue $J[h_n; t_n]$; both of them for all $n$ from $n = 0$ to $n = N-1$. The product of short-time propagators and the product of the GS insertions become:

{\footnotesize
\bg\label{lenu10}
&& \prod_{n=0}^{N-1}
\langle h_{n+1},t_{n+1}|U(t_{n+1},t_n)|h_n,t_n\rangle
\;\longrightarrow\;
\int_{h_i^+}^{h_f}
{\cal D}{\rm g}_+\;
e^{\,iS_{\rm tot}[{\rm g}_+]}\\
&&\prod_{n=0}^{N-1} e^{\,\Delta t\,J[h_n;t_n]}
=
\exp\!\left(
\sum_{n=0}^{N-1}\Delta t\,J[h_n;t_n]
\right)
\longrightarrow
\exp\!\left[
\int_{t_i}^{t_f} dt\int d^3x\;
\sqrt{-g_+(t,\mathbf x)}\,
\sigma^{\mu\nu}(t,\mathbf x)\,
g_{+\mu\nu}(t,\mathbf x)
\right] \nonumber
\nd}
yielding respectively the usual forward Lorentzian path integral and $\mathcal I_+[g_+; \sigma]$ from \eqref{coridahl2}. In the second line of \eqref{lenu10}, since we are dealing with $c$-number eigenvalues, all the exponential together produce $\mathcal I_+[g_+; \sigma]$. Putting everything together we reproduce the first relation in \eqref{coridahl}, namely:
\begin{equation}
\langle h_f,t_f|D[\sigma]|h_i^+,t_i\rangle
=
\int_{h_i^+}^{h_f}
{\cal D}{\rm g}_+\;
\exp\!\left(
iS_{\rm tot}[{\rm g}_+]
\right)
\,
\mathcal I_+[g_+;\sigma].
\label{eq:first_target_derived_rewrite2}
\end{equation}

\subsubsection{SK contour and deriving the second equation in \eqref{coridahl} \label{6.444}}

Our aim now is to derive the second relation in \eqref{coridahl}, and the technique that we shall adopt here will be akin to what we did in section \ref{sec6.333} above. More concretely, 
we want to derive the backward-branch matrix element:
\begin{equation}
\langle h_i^-,t_i|D^\dagger[\sigma]\hat O_H(x)|h_f,t_f\rangle,
\qquad
x=({\bf x},t),
\qquad
t_i<t<t_f,
\label{eq:goal_second_ME}
\end{equation}
and show that it admits the path-integral representation presented in \eqref{coridahl}. The cleanest derivation is indeed to place the operator insertion at the discrete time slice that contains the physical time $t$, and to keep the anti-time-ordered product explicit until the end. The actual implementation of this is a bit more subtle so let us elaborate the steps carefully. First 
we discretize the interval:
\begin{equation}
t_i=t_0<t_1<\cdots<t_N=t_f,
\qquad
\Delta t=\frac{t_f-t_i}{N},
\label{eq:time_discretization_second}
\end{equation}
and choose an integer $k$ such that
$t \equiv t_k$. (If $t$ does not coincide exactly with a lattice point, one simply chooses the nearest slice and takes the continuum limit at the end.) The anti-time-ordered GS operator is:
\begin{equation}
D^\dagger[\sigma]
=
\bar{\cal T}\exp\!\left(
\int_{t_i}^{t_f} dt'\;\hat J_H(t')
\right) ,
\label{eq:Ddagger_start}
\end{equation}
where $\hat J_H(t')$ is defined in \eqref{eq:JH_JS_correct_rewrite2}. We can also replace $t'$ by $t_n$ and define $\hat J_H(t_n)$ accordingly. Such a notation is useful because, 
using
$\hat J_H(t_n)$ the discretized anti-time-ordered product takes the following form:
\begin{equation}
D^\dagger[\sigma]
=
\lim_{N\to\infty}
\prod_{n=0}^{N-1}
\left[
e^{\,\Delta t\,\hat J_S(t_i;t_n)}\,
U^\dagger(t_{n+1},t_n)
\right],
\label{eq:Ddagger_discrete_S}
\end{equation}
where the order now runs from early to late times, as appropriate for $\bar{\cal T}$. One may also compare \eqref{eq:Ddagger_discrete_S} with \eqref{lenlin02} where the time ordering is opposite. For the local Heisenberg operator we use:
\begin{equation}
\hat O_H(x) \equiv \hat O_H({\bf x}, t_k)
=
U^\dagger(t_k,t_i)\,
\hat O_S({\bf x},t_i)\,
U(t_k,t_i) ,
\label{eq:OH_relation_second}
\end{equation}
which, in our notation, is defined at $t = t_k$. This implies that 
the correct discrete representation of the matrix element
$\langle h_i^-,t_i|D^\dagger[\sigma]\hat O_H(x)|h_f,t_f\rangle$
is obtained by placing the Schr\"odinger-picture operator at the slice $t_k$. Thus:
\begin{align}
&\langle h_i^-,t_i|D^\dagger[\sigma]\hat O_H(x)|h_f,t_f\rangle
\nonumber\\
&\qquad=
\lim_{N\to\infty}
\int \prod_{m=1}^{N-1}Dh_m
\Bigg[
\prod_{n=0}^{k-1}
\langle h_n,t_n|
e^{\,\Delta t\,\hat J_S(t_i;t_n)}
U^\dagger(t_{n+1},t_n)
|h_{n+1},t_{n+1}\rangle
\Bigg]
\nonumber\\
&\qquad\qquad\times
\langle h_k,t_k|
e^{\,\Delta t\,\hat J_S(t_i;t_k)}
\hat O_S({\bf x},t_i)\,
U^\dagger(t_{k+1},t_k)
|h_{k+1},t_{k+1}\rangle
\nonumber\\
&\qquad\qquad\times
\Bigg[
\prod_{n=k+1}^{N-1}
\langle h_n,t_n|
e^{\,\Delta t\,\hat J_S(t_i;t_n)}
U^\dagger(t_{n+1},t_n)
|h_{n+1},t_{n+1}\rangle
\Bigg],
\label{eq:second_discrete_ME_correct}
\end{align}
with
$h_0=h_i^-$ and 
$h_N=h_f$. This is the correct way of representing the product of two operators: the insertion sits at one distinguished slice $t_k$ (albeit dressed by the operator $e^{\Delta t \hat J_S(t_i; t_k)}$), not inside a repeated factor, and the products are split into the parts before and after that insertion\footnote{The reason the other slices in \eqref{eq:second_discrete_ME_correct} are untouched is that the Heisenberg operator insertion is localized at the single time $t_k$. Therefore only the factor corresponding to that time slice is modified. All other slices still carry the usual factors
$U^\dagger(t_n,t_i)\,
e^{\Delta t\,\hat J_S(t_i;t_n)}\,
U(t_n,t_i)$,
for
$n\neq k$. So the modified chain is:
\begin{equation}
\begin{aligned}
&\Big[
U^\dagger(t_{N-1},t_i)\,
e^{\Delta t\,\hat J_S(t_i;t_{N-1})}\,
U(t_{N-1},t_i)
\Big]
\cdots
\Big[
U^\dagger(t_{k+1},t_i)\,
e^{\Delta t\,\hat J_S(t_i;t_{k+1})}\,
U(t_{k+1},t_i)
\Big]
\\
&\qquad\times
\Big[
U^\dagger(t_k,t_i)\,
\hat O_S({\bf x},t_i)\,
e^{\Delta t\,\hat J_S(t_i;t_k)}\,
U(t_k,t_i)
\Big]
\\
&\qquad\times
\Big[
U^\dagger(t_{k-1},t_i)\,
e^{\Delta t\,\hat J_S(t_i;t_{k-1})}\,
U(t_{k-1},t_i)
\Big]
\cdots
\Big[
U^\dagger(t_0,t_i)\,
e^{\Delta t\,\hat J_S(t_i;t_0)}\,
U(t_0,t_i)
\Big]. \nonumber
\end{aligned}
\label{eq:modified_full_chain}
\end{equation}
This is the correct chain with a single insertion at the $k$-th slice. Taking the Hermitian conjugate of the full chain, including the modified insertion, one indeed reproduces the kind of sequence that appears in the discretized expression for the bra-side matrix element from \eqref{eq:second_discrete_ME_correct}. The reason why this works 
is not because $D[\sigma]$ itself is unchanged, but because one has replaced the $k$-th slice by the factor appropriate to the operator-inserted matrix element. An alternative way to see this is the following. Expand the anti-time-ordered exponential as:
\begin{equation}
D^\dagger[\sigma]
=
1
+
\int_{t_i}^{t_f}dt_1\;\hat J_H(t_1)
+
\frac{1}{2!}
\int_{t_i}^{t_f}dt_1dt_2\;
\bar{\cal T}\!\left[
\hat J_H(t_1)\hat J_H(t_2)
\right]
+\cdots ,\nonumber
\label{eq:Ddag_expand}
\end{equation}
where $\bar{\cal T}$ orders all operator insertions from \emph{earlier} times to the left and \emph{later} times to the right.
Now inserting $\hat O({\bf x},t_k)$ inside the anti-time-ordered product gives us:
\bg
D^\dagger[\sigma]\hat O({\bf x},t_k)
&= &
\hat O({\bf x},t_k)
+
\int_{t_i}^{t_f}dt_1\;
\hat J_H(t_1)\hat O({\bf x},t_k)
+
\frac{1}{2!}
\int_{t_i}^{t_f}dt_1dt_2\;
\bar{\cal T}\!\left[
\hat J_H(t_1)\hat J_H(t_2)
\right]
\hat O({\bf x},t_k)
+\cdots \nonumber\\
&\equiv &
\hat O({\bf x},t_k)
+
\int_{t_i}^{t_f}dt_1\;
\bar{\cal T}\!\left[
\hat J_H(t_1)\hat O({\bf x},t_k)
\right]
+
\frac{1}{2!}
\int_{t_i}^{t_f}dt_1dt_2\;
\bar{\cal T}\!\left[
\hat J_H(t_1)\hat J_H(t_2)\hat O({\bf x},t_k)
\right]
+\cdots ,\nonumber
\nd
where the second equality makes manifest the fact that the backward path-integration forces the operator $\hat{O}({\bf x}, t_k)$ to appear in \eqref{eq:second_discrete_ME_correct} with the same ordering imposed by $\bar{\cal T}$.
This is exactly the expansion of
$\bar{\cal T}\!\left[
\exp\!\left(
\int_{t_i}^{t_f}dt'\;\hat J_H(t')
\right)
\hat O({\bf x},t_k)
\right] =\bar{\cal T}\!\left[
\left(
\prod\limits_{n=0}^{N-1}
\exp\!\left(
\Delta t\,\hat J_H(t_n)
\right)
\right)
\hat O({\bf x},t_k)
\right]   $, 
implying that the anti-time-ordering places the operator at the unique position corresponding to the time $t_k$ in either continuous or discretized version.\label{shirtagra}}.

The next step is similar to what we did earlier: we insert an additional resolution of the identity on each relevant slice using 
$\mathbf 1_{t_n}
=
\int Dh_m^{(n)}\;
|h_m^{(n)},t_n\rangle\langle h_m^{(n)},t_n|$,
so that for every $n\neq k$, we get the following:
\begin{align}
&\langle h_n,t_n|
e^{\,\Delta t\,\hat J_S(t_i;t_n)}
U^\dagger(t_{n+1},t_n)
|h_{n+1},t_{n+1}\rangle
\nonumber\\
&\qquad=
\int Dh_m^{(n)}\;
\langle h_n,t_n|
e^{\,\Delta t\,\hat J_S(t_i;t_n)}
|h_m^{(n)},t_n\rangle\;
\langle h_m^{(n)},t_n|
U^\dagger(t_{n+1},t_n)
|h_{n+1},t_{n+1}\rangle ,
\label{eq:generic_slice_identity}
\end{align}
thus explicitly excluding the operator $\hat O(x)$. To see how the operator insertion would work at the insertion slice $n = k$, we can use the resolution of identity between the dressing $e^{\Delta t \hat J_S(t_i; t_k)}$ and the operator $\hat O_S(t_i; t_k)$ as:
\begin{align}\label{biralaador1}
&\langle h_k,t_k|
e^{\,\Delta t\,\hat J_S(t_i;t_k)}
\hat O_S({\bf x},t_i)\,
U^\dagger(t_{k+1},t_k)
|h_{k+1},t_{k+1}\rangle\\
&\qquad=
\int Dh_m^{(k)}\;
\langle h_k,t_k|
e^{\,\Delta t\,\hat J_S(t_i;t_k)}
|h_m^{(k)},t_k\rangle\;
\langle h_m^{(k)},t_k|
\hat O_S({\bf x},t_i)\,
U^\dagger(t_{k+1},t_k)
|h_{k+1},t_{k+1}\rangle.\nonumber
\end{align}
At this stage we can use diagonality of the GS factor:
Because $\hat J_S(t_i;t_n)$ is multiplicative in the configuration basis on the slice $t_n$, we have the eigenvalue equation \eqref{lenu6}
with the eigenvalue taking the form \eqref{lenu7}. Therefore
\begin{equation}
\langle h_n,t_n|
e^{\,\Delta t\,\hat J_S(t_i;t_n)}
|h_m^{(n)},t_n\rangle
=
e^{\,\Delta t\,J[h_n;t_n]}\,
\delta[h_n-h_m^{(n)}].
\label{eq:diag_expJ_second}
\end{equation}
Substituting \eqref{eq:diag_expJ_second} into \eqref{eq:generic_slice_identity} and \eqref{biralaador1} collapses all the extra $Dh_m^{(n)}$ integrals and sets
$h_m^{(n)}=h_n$ for all $n$ reminiscent of what we had in section \ref{sec6.333}.
Thus for $n\neq k$ we get:
\bg\label{biralaador2}
&&\langle h_n,t_n|
e^{\,\Delta t\,\hat J_S(t_i;t_n)}
U^\dagger(t_{n+1},t_n)
|h_{n+1},t_{n+1}\rangle \nonumber\\
&& \qquad =
e^{\,\Delta t\,J[h_n;t_n]}\,
\langle h_n,t_n|
U^\dagger(t_{n+1},t_n)
|h_{n+1},t_{n+1}\rangle ,
\nd
where in the second line we have a clean separation between the eigenvalue $J[h_n; t_n]$ and the evolution operator $U^\dagger(t_{n+1}, t_n)$. On the other hand for $n = k$, {\it i.e.} at the insertion slice, we have:
\bg\label{biralaador3}
&&\langle h_k,t_k|
e^{\,\Delta t\,\hat J_S(t_i;t_k)}
\hat O_S({\bf x},t_i)\,
U^\dagger(t_{k+1},t_k)
|h_{k+1},t_{k+1}\rangle \nonumber\\
&& \qquad =
e^{\,\Delta t\,J[h_k;t_k]}\,
\langle h_k,t_k|
\hat O_S({\bf x},t_i)\,
U^\dagger(t_{k+1},t_k)
|h_{k+1},t_{k+1}\rangle, 
\nd
where now in the second line we still have the Schr\"odinger operator $\hat O_S({\bf x}, t_i)$ inside the matrix element. To isolate the operator insertion we can insert one more complete set on the same slice $t_k$, {\it i.e.}
$\mathbf 1_{t_k}
=
\int Dh_k^+\;
|h_k^+,t_k\rangle\langle h_k^+,t_k|$.
Then:
\begin{align}
&\langle h_k,t_k|
\hat O_S({\bf x},t_i)\,
U^\dagger(t_{k+1},t_k)
|h_{k+1},t_{k+1}\rangle
\nonumber\\
&\qquad=
\int Dh_k^+\;
\langle h_k,t_k|\hat O_S({\bf x},t_i)|h_k^+,t_k\rangle\;
\langle h_k^+,t_k|
U^\dagger(t_{k+1},t_k)
|h_{k+1},t_{k+1}\rangle ,
\label{eq:operator_identity_insert}
\end{align}
which isolates the Schr\"odinger operator $\hat O_S({\bf x}, t_i)$ and the unitary evolution operator $U^\dagger(t_{k+1}$, $t_k)$ into two separate matrix elements. 
For a local multiplicative operator on the chosen branch one uses the standard replacement:
\begin{equation}
\langle h_k,t_k|\hat O_S({\bf x},t_i)|h_k^+,t_k\rangle
\;\longrightarrow\;
O[h_k]({\bf x}, t_k)\,
\delta[h_k-h_k^+] ,
\label{eq:operator_local_replacement}
\end{equation}
which produces the requisite delta functional $\delta[h_k - h_k^+]$ that identifies $h_k^+$ with $h_k$, and therefore removes the functional integral over $Dh_k^+$ in \eqref{eq:operator_identity_insert}. Note also that the Schr\"odinger operator $\hat O_S({\bf x}, t_i)$ converts to the eigenvalue $O[h_k]({\bf x}, t_k)$ defined exactly on the Cauchy slice $\Sigma_{t_k}$.
Hence:
\begin{align}
&\langle h_k,t_k|
\hat O_S({\bf x},t_i)\,
U^\dagger(t_{k+1},t_k)
|h_{k+1},t_{k+1}\rangle
\nonumber\\
&\qquad=
O[h_k]({\bf x}, t_k)\,
\langle h_k,t_k|
U^\dagger(t_{k+1},t_k)
|h_{k+1},t_{k+1}\rangle ,
\label{eq:operator_factorized}
\end{align}
producing the requisite eigenvalue $O[h_k]({\bf x}, t_k)$ and a matrix element defined completely using the unitary evolution operator $U^\dagger(t_{k+1}, t_k)$ over the Cauchy slice $\Sigma_{t_k}$.
Substituting this into \eqref{biralaador3} gives:
\begin{align}\label{biralaador6}
&\langle h_k,t_k|
e^{\,\Delta t\,\hat J_S(t_i;t_k)}
\hat O_S({\bf x},t_i)\,
U^\dagger(t_{k+1},t_k)
|h_{k+1},t_{k+1}\rangle
\nonumber\\
&\qquad=
e^{\,\Delta t\,J[h_k;t_k]}\,
O[h_k](x)\,
\langle h_k,t_k|
U^\dagger(t_{k+1},t_k)
|h_{k+1},t_{k+1}\rangle ,
\end{align}
thus isolating the matrix element and the eigenvalues appropriately. We can now reconstruct the full discrete expression by plugging \eqref{biralaador2} and \eqref{biralaador6} in \eqref{eq:second_discrete_ME_correct} to get the following form:
\begin{align}
&\langle h_i^-,t_i|D^\dagger[\sigma]\hat O_H(x)|h_f,t_f\rangle
\nonumber\\
&\qquad=
\lim_{N\to\infty}
\int \prod_{n=1}^{N-1}Dh_n\;
\prod_{n=0}^{N-1}
e^{\,\Delta t\,J[h_n;t_n]}\,
\prod_{n=0}^{N-1}
\langle h_n,t_n|
U^\dagger(t_{n+1},t_n)
|h_{n+1},t_{n+1}\rangle
\;
O[h_k](x) ,
\label{eq:full_discrete_reconstructed}
\end{align}
where $x \equiv ({\bf x}, t_k)$, and we notice that the products of the matrix elements and the displacement operators are naturally ordered from $n = 0$ to $n = N-1$. However \eqref{eq:full_discrete_reconstructed} is still not in the path-integral format. This may be brought to the requisite form by making the following two observations: {\Su one},  
the product of backward kernels is the standard backward branch path integral:
\begin{equation}\label{biralaador11}
\prod_{n=0}^{N-1}
\langle h_n,t_n|
U^\dagger(t_{n+1},t_n)
|h_{n+1},t_{n+1}\rangle
\;\longrightarrow\;
\int_{h_i^-}^{h_f}
{\cal D}{\rm g}_-\;
\exp\!\left(
-iS_{\rm tot}[{\rm g}_-]
\right) ,
\end{equation}
which clarifies precisely why we have $g_-$ and the trans-series action $S_{\rm tot}$ appearing with an opposite sign\footnote{An alternative way would be to write the backward product as the Hermitian conjugate of the forward product. However, this implies the complex conjugate of the forward path integral, not an additional sign flip beyond complex conjugation. Therefore the continuum limit is \eqref{biralaador11}, which is equivalently $\int_{{\cal C}_-}{\cal D}g_-\,e^{\,iS_{\rm tot}[g_-]\vert_{{\cal C}_-}}$ when the contour is written with reversed orientation. Thus there is no contradiction with the alternative notation where one writes everything with $e^{+\,iS}$ along an oriented contour. In that language one rewrites:
\begin{equation}
e^{-\,iS[g_-;t_i\to t_f]}
=
e^{\,iS[g_-;t_f\to t_i]}. \nonumber
\label{eq:reverse_orientation_equiv}
\end{equation}
So the same backward branch can be described in two equivalent ways:
one using the branch notation, {\it i.e.}
$e^{-\,iS[g_-;t_i\to t_f]}$,
or two, using the oriented contour notation, {\it i.e.}
$e^{\,iS[g_-;t_f\to t_i]}$.
The formula \eqref{biralaador11} is using the first convention. We will elaborate further on the other convention a bit later.} And {\Su two},  
the product of GS factors exponentiates as:
\begin{align}\label{biralaador12}
\prod_{n=0}^{N-1}
e^{\,\Delta t\,J[h_n;t_n]}
&=
\exp\!\left(
\sum_{n=0}^{N-1}\Delta t\,J[h_n;t_n]
\right)\\
&\longrightarrow
\exp\!\left[
\int_{t_i}^{t_f}dt\int d^3x\;
\sqrt{-g_-(t,{\bf x})}\,
\sigma^{\mu\nu}(t,{\bf x})\,
g_{-\mu\nu}(t,{\bf x})
\right] \equiv \mathcal I_-[g_-; \sigma] , \nonumber
\end{align}
where note that in the continuum limit, the discrete history 
$\{h_n\}$ becomes the backward branch field configuration $g_-(t,{\bf x})$, and the Riemann sum in the exponential turns into the corresponding time integral. In a similar vein:
\begin{equation}\label{biralaador13}
O[h_k]({\bf x}, t_k)
\;\longrightarrow\;
O[g_-]({\bf x}, t)
\end{equation}
in the continuum limit, since the configuration $h_k$ becomes the value of the backward history $g_-$ at the time slice $t=t_k$.
It is also important to note that the minus sign in the backward branch comes from the complex-conjugate evolution kernel,
$U^\dagger(..)
\longrightarrow
e^{-\,iS_{\rm tot}[g_-]}$
not from the GS insertion itself. The GS factor is not part of the Lorentzian phase $iS_{\rm tot}[g_\pm]$; it is an ordinary multiplicative operator insertion. Therefore its discrete contribution remains
$e^{\,\Delta t\,J[h_n;t_n]}$
on both branches in this coherent-state-style derivation. This also confirms that the factor $\mathcal I_-[g_-;\sigma]$
appears on the $-$ branch for exactly the same reason that 
$\mathcal I_+[g_+;\sigma]$
appears on the $+$ branch: each time slice contributes one diagonal GS factor. The only difference between the two branches is in the dynamical phase:
$e^{\,iS_{\rm tot}[g_+]}$
versus 
$e^{-\,iS_{\rm tot}[g_-]}$.
The GS multiplicative factors exponentiate in the appropriate way on both branches. Therefore, after plugging \eqref{biralaador11}, \eqref{biralaador12} and \eqref{biralaador13} in \eqref{eq:full_discrete_reconstructed}, we reproduce the second equation from \eqref{coridahl}, namely:
\begin{equation}
\langle h_i^-,t_i|D^\dagger[\sigma]\hat O_H(x)|h_f,t_f\rangle
=
\int_{h_i^-}^{h_f}
{\cal D}{\rm g}_-\;
\exp\!\left(
-iS_{\rm tot}[{\rm g}_-]
\right)
\,
\mathcal I_-[g_-;\sigma]\,
O[g_-](x).
\end{equation}

\subsection{Euclidean path-integral representation for the GS state \label{sec6.3}}

The Euclidean case is structurally simpler than the Lorentzian
Schwinger--Keldysh construction because there is no need to double the
fields into $+$ and $-$ branches.  Instead, one has a single Euclidean
functional integral, and the GS insertion appears as a source-like
deformation of the Euclidean action.

For starters, we can quantify the Euclidean continuation and the vacuum projector in the following way. Let $\tau$ denote Euclidean time, obtained from Lorentzian time by
$t=-\,i\tau$. In the Euclidean theory, propagation over a long interval projects
onto the interacting vacuum:
\begin{equation}
e^{-H(\tau_f-\tau_i)}
\;\xrightarrow[\tau_f-\tau_i\to\infty]{}\;
|\Omega\rangle\langle\Omega|.
\end{equation}
Accordingly, Euclidean vacuum expectation values may be represented by
a single path integral over a Euclidean manifold with appropriate
boundary conditions at $\tau_i$ and $\tau_f$.

The Euclidean GS operator can now be arranged in the following way.
The Lorentzian GS operator was defined as a time-ordered exponential of
a source-like metric insertion as in \eqref{sweenseyf}.  Its Euclidean analogue is naturally
defined as:
\begin{equation}
D_E[\sigma_E]
=
{\cal T}_\tau
\exp\!\left(
\int d^dx_E\;
\sqrt{g_E(x)}\,
\sigma_E^{\mu\nu}(x)\,
g^E_{\mu\nu}(x)
\right),
\label{eq:DE_def}
\end{equation}
where ${\cal T}_\tau$ denotes Euclidean time ordering, $g^E_{\mu\nu}$
is the Euclidean metric, $\sigma^{\mu\nu}_E(x)$ is the Euclidean source, and $d^dx_E$ denotes the Euclidean volume
element.  When the metric is treated as a configuration variable in
the path integral, the insertion in \eqref{eq:DE_def} becomes an
ordinary multiplicative functional of the Euclidean field history\footnote{With the choice \eqref{eq:DE_def}, we are in fact hiding one subtlety. If one performs a \emph{literal} Wick rotation of the Lorentzian GS
operator while keeping the same source symbol
$\sigma_L^{\mu\nu}$, then the exponent acquires an explicit factor of
$-i$ because of our choice $t = -i\tau$ to go to the Euclidean space in the following way:
\begin{equation}
D_L[\sigma_L]
\;\longrightarrow\;
{\cal T}_\tau
\exp\!\left(
-\,i
\int d^dx_E\;
\sqrt{g_E}\,
\sigma_L^{\mu\nu}\,
g^E_{\mu\nu}
\right) , \nonumber
\end{equation}
where $\sigma^{\mu\nu}_L(x) = \sigma^{\mu\nu}(x)$ is the Lorentzian source in \eqref{sweenseyf}.
If, however, one introduces the Euclidean source by defining
$\sigma_E^{\mu\nu}=-\,i\,\sigma_L^{\mu\nu} = - i\sigma^{\mu\nu}(x)$,
then the Euclidean GS operator takes the simpler form:
\begin{equation}
D_E[\sigma_E]
=
{\cal T}_\tau
\exp\!\left(
\int d^dx_E\;
\sqrt{g_E}\,
\sigma_E^{\mu\nu}\,
g^E_{\mu\nu}
\right). \nonumber
\end{equation}
In this convention there is no explicit $i$ in the Euclidean exponent,
because it has been absorbed into the definition of the Euclidean
source. This is entirely analogous to the familiar continuation of an ordinary
source term.  In Lorentzian signature one often writes
$Z_L[J_L]
=
\int {\cal D}\phi\;
\exp\!\left(
iS_L[\phi]
+
i\int d^dx_L\,J_L(x)\phi(x)
\right)$.
After Wick rotation this becomes
$Z_E[J_E]
=
\int {\cal D}\phi_E\;
\exp\!\left(
-\,S_E[\phi_E]
+
\int d^dx_E\,J_E(x)\phi_E(x)
\right)$,
with the Euclidean source defined appropriately.  The GS insertion is
the same phenomenon, except that the Lorentzian operator was written
without an explicit factor of $i$ to begin with.}. With this in mind, and for $\hat O_E(x)$ to be an Euclidean local operator, the normalized GS
expectation value is:
\begin{equation}
\langle O_E(x)\rangle_\sigma
=
\frac{
\langle \Omega|\,D_E^\dagger[\sigma]\,
\hat O_E(x)\,
D_E[\sigma]\,|\Omega\rangle
}{
\langle \Omega|\,D_E^\dagger[\sigma]\,
D_E[\sigma]\,|\Omega\rangle
} ,
\label{eq:OE_sigma_operator}
\end{equation}
where, for real $\sigma^{\mu\nu}$, one may simply take
$D_E^\dagger[\sigma]=D_E[\sigma]$. The next set of steps is an obvious follow-up of the steps leading to the in-in Schwinger-Keldysh formulation. 
Let $|h,\tau\rangle$ denote an eigenstate of the induced Euclidean
metric on a constant-$\tau$ slice, and let
\begin{equation}
\mathbf 1_{\Sigma_\tau}
=
\int Dh\;|h,\tau\rangle\langle h,\tau| ,
\end{equation}
so that we can use this to bring \eqref{eq:OE_sigma_operator} into a more manageable form. 
Using the long Euclidean evolution to project onto the vacuum and the fact that the resolution of identity can be done for both initial and final states, the
numerator of \eqref{eq:OE_sigma_operator} may be written as:
\begin{equation}
\begin{aligned}
N_\sigma^E(x)
&=
\langle \Omega|\,D_E^\dagger[\sigma]\,
\hat O_E(x)\,
D_E[\sigma]\,|\Omega\rangle = {\rm Tr}\left(\rho_\Omega\,D^\dagger_E[\sigma]\, \hat O_E(x) \, D_E[\sigma]\right) 
\\
&=
\int Dh_f\,
\langle h_f,\tau_f| D_E[\sigma]\, \rho_\Omega\,
D_E^\dagger[\sigma]\,
\hat O_E(x)\,
|h_f,\tau_f\rangle,
\end{aligned}
\label{eq:NE_before_PI}
\end{equation}
where in the second line we have used the cyclicity of the trace, and $\rho_\Omega = |\Omega \rangle\langle\Omega|$ is the interacting vacuum density matrix which will eventually give us the Euclidean vacuum wavefunctional $\Psi_\Omega^E[h]
\equiv
\langle h,\tau|\Omega\rangle$. The appearance of the vacuum wavefunction $\Psi^E_\Omega[h]$ and not the wavefunction $\Psi^E_\sigma[h]$ of the GS state is again the familiar advantage we get for using the GS states over the coherent states. Similarly, the
denominator is:
\begin{equation}
\begin{aligned}
Z_\sigma^E
&=
\langle \Omega|\,D_E^\dagger[\sigma]\,
D_E[\sigma]\,|\Omega\rangle = {\rm Tr}\left(\rho_\Omega\, D_E^\dagger[\sigma]\,
D_E[\sigma]\right)\\
&=
\int Dh_f\,
\langle h_f,\tau_f|D_E[\sigma]\, \rho_\Omega\,
D_E^\dagger[\sigma]
|h_f,\tau_f\rangle.
\end{aligned}
\label{eq:ZE_before_PI}
\end{equation}
Our aim now is to put both \eqref{eq:NE_before_PI} and \eqref{eq:ZE_before_PI} in the path integral form. For this, let us start by denoting ${\rm g}_E$ as the full Euclidean field multiplet,
${\rm g}_E
\equiv
(g_E,\bar c,c,\text{matter},\ldots)$,
such that the the full Euclidean gauge-fixed action be:
\begin{equation}
S_{\rm tot}^E[\lambda;{\rm g}_E]
=
S_{\rm pert}^E[\lambda;{\rm g}_E]
+
\sum_I e^{-A_I/\lambda}\,
S_I^E[\lambda;{\rm g}_E]
+
\sum_{I,J} e^{-(A_I+A_J)/\lambda}\,
S_{IJ}^E[\lambda;{\rm g}_E]
+\cdots ,
\label{eq:SE_transseries}
\end{equation}
taking the expected trans-series form that we studied earlier (but now in the Euclidean language). The first piece
$S_{\rm pert}^E$ contain the
classical, gauge-fixing, ghost, matter, and counterterm pieces expressed using higher order perturbative expansions. Then the Euclidean vacuum matrix elements with GS insertions become:

{\footnotesize
\bg\label{lenulila6}
Z_\sigma^E
&=&
\int Dh_f\,Dh_i\, Dh'_i\;
\Psi_\Omega^{\ast E}[h'_i]\,
\Psi_\Omega^E[h_i]
\int_{h_i, h'_i}^{h_f}
{\cal D}{\rm g}_E\;
\exp\!\left(
-\,S_{\rm tot}^E[\lambda;{\rm g}_E]
\right)
\,
{\cal I}_E^\dagger[{\rm g}_E;\sigma]\,
{\cal I}_E[{\rm g}_E;\sigma] \\
 N_\sigma^E(x)
&=&
\int Dh_f\,Dh_i\, Dh'_i\;
\Psi_\Omega^{\ast E}[h'_i]\,
\Psi_\Omega^E[h_i]
\int_{h_i, h'_i}^{h_f}
{\cal D}{\rm g}_E\;
\exp\!\left(
-\,S_{\rm tot}^E[\lambda;{\rm g}_E]
\right)
\,
{\cal I}_E^\dagger[{\rm g}_E;\sigma]\,
O_E[{\rm g}_E](x)\,
{\cal I}_E[{\rm g}_E;\sigma],\nonumber
\nd}
which may now be compared to \eqref{coridahl6} and \eqref{tricksbrke}. Clearly, and as in the Euclidean coherent state case, there are no double branches but the differences from \eqref{indigomara} should be clear: the boundaries are not shifted and instead an additional set of source terms $\mathcal I_E[{\rm g}_E; \sigma]$ and its complex conjugate enter the path integral,  
where:
\begin{equation}
{\cal I}_E[{\rm g}_E;\sigma]
\equiv
\exp\!\left[
\int d^dx_E\;
\sqrt{g_E(x)}\,
\sigma^{\mu\nu}(x)\,
g^E_{\mu\nu}(x)
\right] ,
\label{eq:IE_def}
\end{equation}
which when inserted in \eqref{lenulila6} reproduces the required path integral form for the expectation value of any operator $O_E[{\rm g}_E](x)$. We can further simplify this once we take real $\sigma_{\mu\nu}(x)$. 
For real $\sigma^{\mu\nu}(x)$ one has
${\cal I}_E^\dagger[{\rm g}_E;\sigma]
=
{\cal I}_E[{\rm g}_E;\sigma]$,
and therefore
the Euclidean GS expectation value may be written as

{\scriptsize
\begin{equation}
\langle O_E(x)\rangle_\sigma
=
\frac{
\int Dh_f\,Dh\;
\Psi_\Omega^{\ast E}[h'_i]\,
\Psi_\Omega^E[h_i]
\int_{h}^{h_f}
{\cal D}{\rm g}_E\;
\exp\!\left(
-\,S_{\rm tot}^E[\lambda;{\rm g}_E]
+
2\int d^dx_E\;
\sqrt{g_E(x)}\,
\sigma^{\mu\nu}(x)\,
g^E_{\mu\nu}(x)
\right)
O_E[{\rm g}_E](x)
}{
\int Dh_f\,Dh\;
\Psi_\Omega^{\ast E}[h'_i]\,
\Psi_\Omega^E[h_i]
\int_{h}^{h_f}
{\cal D}{\rm g}_E\;
\exp\!\left(
-\,S_{\rm tot}^E[\lambda;{\rm g}_E]
+
2\int d^dx_E\;
\sqrt{g_E(x)}\,
\sigma^{\mu\nu}(x)\,
g^E_{\mu\nu}(x)
\right)
},
\label{eq:Euclidean_GS_master}
\end{equation}}
which is precisely the form it appeared in {\gspapers} earlier in the Euclidean set-up (modulo some minor subtleties that we will clarify in section \ref{biralador}); and $h \equiv (h_i, h'_i)$ with $Dh \equiv Dh_i Dh'_i$. We can further rearrange the expectation value \eqref{eq:Euclidean_GS_master} in a condensed form by absorbing the GS insertion into a deformation of the Euclidean
action:
\begin{equation}
S_{\rm eff,\sigma}^E[\lambda;{\rm g}_E]
\equiv
S_{\rm tot}^E[\lambda;{\rm g}_E]
-
2\int d^dx_E\;
\sqrt{g_E(x)}\,
\sigma^{\mu\nu}(x)\,
g^E_{\mu\nu}(x) ,
\label{eq:SE_sigma_eff}
\end{equation}
implying the existence of an {\it effective} action $S^E_{{\rm eff}, \sigma}[\lambda; {\rm g}_E]$. Recall that in the Lorentzian case \eqref{coridahl101} there is no real ({\it i.e.} only complex) shift of the action because of the relative $i = \sqrt{-1}$ factor, but in both cases the shifts are only for the gauge-fixed actions. By construction there are no gauge invariant ways of expressing the shifts for both cases as they only appear as source terms in the path integrals. Putting everything together converts \eqref{eq:Euclidean_GS_master} to the following condensed form:
\begin{equation}
\langle O_E(x)\rangle_\sigma
=
\frac{
\int Dh_f\,Dh_i\, Dh'_i\;
\Psi_\Omega^{\ast E}[h'_i]\,
\Psi_\Omega^E[h_i]
\int_{h_i, h'_i}^{h_f}
{\cal D}{\rm g}_E\;
e^{-S_{\rm eff,\sigma}^E[\lambda;{\rm g}_E]}\,
O_E[{\rm g}_E](x)
}{
\int Dh_f\,Dh_i\, Dh'_i\;
\Psi_\Omega^{\ast E}[h'_i]\,
\Psi_\Omega^E[h_i]
\int_{h_i, h'_i}^{h_f}
{\cal D}{\rm g}_E\;
e^{-S_{\rm eff,\sigma}^E[\lambda;{\rm g}_E]}
}.
\label{eq:Euclidean_sigma_eff_ratio}
\end{equation}
This is basically our final expression, and now we can easily contrast with the Lorentzian SK story from section \ref{sec6.2}. In the Lorentzian theory the normalized expectation value requires the closed Schwinger--Keldysh contour, so the fields are doubled:
\begin{equation}
{\rm g}
\;\longrightarrow\;
{\rm g}_+,\qquad {\rm g}_-,
\end{equation}
and the GS insertions appear separately on the two branches as shown in \eqref{coridahl101}. In the Euclidean theory there is only a single branch, so the same GS construction appears simply as a source-like deformation of the Euclidean weight:
\begin{equation}
e^{-S_{\rm tot}^E}
\;\longrightarrow\;
e^{-S_{\rm tot}^E + 2\int \sqrt{g_E}\,\sigma^{\mu\nu}g^E_{\mu\nu}}.
\end{equation}
Thus the Euclidean path integral is simpler, but it encodes the same basic physical idea: the GS operator reweights the path integral by a functional linear in the metric. This basically resonates with what we said earlier, but now we can also compare the thimble decomposition in the Euclidean theory for the GS case with the one from the coherent state case studied in section \ref{sec4.7.1}.

\subsubsection{Single contour representation and a proof of \eqref{lenulila6} \label{biralador}}

The clean way to handle the Euclidean operator insertions is exactly parallel to the Lorentzian discussion from sections \ref{sec6.333} and \ref{6.444}, except that there is now only a \emph{single} contour and the evolution operator is Euclidean:
\begin{equation}
U_E(\tau_b,\tau_a)
=
e^{-H(\tau_b-\tau_a)}.
\label{eq:UE_basic}
\end{equation}
Accordingly, there is no doubling into $+$ and $-$ branches. Both 
$D_E[\sigma]$ and $D_E^\dagger[\sigma]$ are inserted on the same Euclidean contour between $\tau_i$ and $\tau_f$. We start from \eqref{eq:NE_before_PI} and \eqref{eq:ZE_before_PI} and clarify how the path-integral structure in \eqref{lenulila6} may be motivated. In the process we can also see how the Euclidean case differs from the Lorentzian case studied earlier.  The key issue is how to represent the matrix elements:
\begin{equation}
\langle h_f,\tau_f|
D_E^\dagger[\sigma]\,
\hat O_E(x)\,
D_E[\sigma]
|h_i,\tau_i\rangle
~~~{\rm and} ~~~
\langle h_f,\tau_f|
D_E^\dagger[\sigma]\,
D_E[\sigma]
|h_i,\tau_i\rangle
\label{eq:key_Eucl_matrix_element_den}
\end{equation}
as a single Euclidean path integral. For this we need to first define the Euclidean GS operator and then its Schr\"odinger form. 
Let:

{\footnotesize
\begin{equation}
D_E[\sigma]
=
{\cal T}_\tau
\exp\!\left(
\int_{\tau_i}^{\tau_f} d\tau\;
\hat J_H^E(\tau)
\right)
~~~ {\rm where}~~~
\hat J_H^E(\tau)
=
\int d^{d-1}x\;
\sqrt{\hat g_E(\tau,{\bf x})}\,
\sigma^{\mu\nu}(\tau,{\bf x})\,
\hat g^E_{\mu\nu}(\tau,{\bf x}) ,
\label{eq:JHE_def}
\end{equation}}
denote the displacement operator in the Euclidean format.
As before, the Schr\"odinger picture is anchored on the initial slice 
$\Sigma_{\tau_i}$. Therefore the Euclidean Heisenberg-to-Schr\"odinger relation may be expressed as the following set\footnote{In any picture change, the operator transforms by similarity, that is, by $U^{-1}(\cdots)U$.
In Lorentzian time, unitary evolution implies $U^{-1}=U^\dagger$, so one often writes $U^\dagger(\cdots)U$.
In Euclidean time, $U_E$ is not unitary, so $U_E^{-1}\neq U_E^\dagger$. Hence one must use $U_E^{-1}$, not $U_E^\dagger$.}:
\bg\label{biral2}
&& \hat J_H^E(\tau)
=
U_E^{-1}(\tau,\tau_i)\,
\hat J_S^E(\tau_i;\tau)\,
U_E(\tau,\tau_i) \nonumber\\
&& \hat J_S^E(\tau_i;\tau)
\equiv
\int d^{d-1}x\;
\sqrt{\hat g_S^E(\tau_i,{\bf x})}\,
\sigma^{\mu\nu}(\tau,{\bf x})\,
\hat g^E_{S\,\mu\nu}(\tau_i,{\bf x}) ,
\nd
where the $\tau$-dependence comes only from the c-number source profile $\sigma^{\mu\nu}(\tau,{\bf x})$.
Similarly, for a local operator inserted at Euclidean time $\tau_k$,
and $x=({\bf x},\tau_k)$,
we write:
\begin{equation}
\hat O_E(x)
=
U_E^{-1}(\tau_k,\tau_i)\,
\hat O_S^E({\bf x},\tau_i)\,
U_E(\tau_k,\tau_i).
\label{eq:OE_Heis_to_Schr}
\end{equation}
Our next step would be to 
discretize the interval as
$\tau_i=\tau_0<\tau_1<\cdots<\tau_N=\tau_f$ with 
$\Delta\tau=\frac{\tau_f-\tau_i}{N}$ and 
choose $k$ such that
$\tau_k=\tau_x$,
where $\tau_x$ is the Euclidean time coordinate of the insertion point $x$. Then the Euclidean time-ordered exponential becomes:
\bg\label{biral3}
D_E[\sigma]
 &= &
\lim_{N\to\infty}
\prod_{n=N-1}^{0}
\exp\!\left(
\Delta\tau\,\hat J_H^E(\tau_n)
\right)\nonumber\\
& = &
\lim_{N\to\infty}
\prod_{n=N-1}^{0}
\left[
U_E(\tau_{n+1},\tau_n)\,
\exp\!\left(
\Delta\tau\,\hat J_S^E(\tau_i;\tau_n)
\right)
\right] \nonumber\\
& = & \mathcal B_{N-1}\, \mathcal B_{N_2}\, \mathcal B_{N-3}\, ....... \mathcal B_2\, \mathcal B_1\, \mathcal B_0\nonumber\\
D_E^\dagger[\sigma] 
& = &
\lim_{N\to\infty}
\prod_{n=0}^{N-1}
\exp\!\left(
\Delta\tau\,\hat J_H^{E\,\dagger}(\tau_n)
\right)\nonumber\\
& = &
\lim_{N\to\infty}
\prod_{n=0}^{N-1}
\left[
\exp\!\left(
\Delta\tau\,\hat J_S^{E\,\dagger}(\tau_i;\tau_n)
\right)\,
U_E(\tau_{n+1},\tau_n)
\right]\nonumber\\
& = & \mathcal A_0\, \mathcal A_1\, \mathcal A_2\, ..... \, \mathcal A_{N-2}\, \mathcal A_{N-1} = \mathcal B^\dagger_0\, \mathcal B^\dagger_1\, \mathcal B^\dagger_2\, ..... \, \mathcal B^\dagger_{N-2}\, \mathcal B^\dagger_{N-1} , 
\nd
where $\mathcal B_{N-1} =U_E(\tau_{N},\tau_{N-1})\,
\exp\!\left(
\Delta\tau\,\hat J_S^E(\tau_i;\tau_{N-1})
\right)$; and 
we used the Euclidean Dyson/Trotter rearrangement for the time ordered product of the exponential from \eqref{eq:JHE_def}. For $D^\dagger_E[\sigma]$ the order is reversed because the Hermitian conjugate reverses the time ordering similar to the Lorentzian case. For real source profiles and Hermitian metric operators one may take:
\begin{equation}
\hat J_S^{E\,\dagger}(\tau_i;\tau_n)
=
\hat J_S^E(\tau_i;\tau_n).
\label{eq:JSE_herm}
\end{equation}
Let us now discuss the 
one-contour representation of the numerator \eqref{eq:NE_before_PI}. 
The point is that,
because the Euclidean contour is single-branched, the operator product
given by $D_E^\dagger[\sigma]\,
\hat O_E(x)\,
D_E[\sigma]$
is represented by \emph{one} ordered product along the same contour, with $D_E[\sigma]$ inserted on the ket side, $\hat O_E(x)$ inserted at $\tau_k$, and $D_E^\dagger[\sigma]$ inserted on the bra side. To see how this works, let us place $D_E[\sigma]$ on the ket side. Expanding the first relation in \eqref{biral3} between $|h_i,\tau_i\rangle$ and the insertion slice $\Sigma_{\tau_k}$, one gets the standard forward chain:
\begin{align}
&\langle h_f,\tau_f|
D_E[\sigma]
|h_i,\tau_i\rangle
\nonumber\\
&\qquad=
\lim_{N\to\infty}
\int \prod_{m=1}^{N-1}Dh_m\;
\prod_{n=0}^{N-1}
\langle h_{n+1},\tau_{n+1}|
U_E(\tau_{n+1},\tau_n)\,
e^{\,\Delta\tau\,\hat J_S^E(\tau_i;\tau_n)}
|h_n,\tau_n\rangle,
\label{eq:DE_on_ket_chain}
\end{align}
with
$h_0=h_i$ and 
$h_N=h_f$.
The reason why we do not restrict ourselves to some intermediate state $|h_k^+, \tau_k\rangle$ will become clear soon. Before going into that, let us 
place $D_E^\dagger[\sigma]$ on the bra side. Then the second relation in \eqref{biral3} gives:
\begin{align}
&\langle h_f,\tau_f|
D_E^\dagger[\sigma]
|h_i,\tau_i\rangle
\nonumber\\
&\qquad=
\lim_{N\to\infty}
\int \prod_{m=1}^{N-1}Dh_m\;
\prod_{n=0}^{N-1}
\langle h_{n+1},\tau_{n+1}|
e^{\,\Delta\tau\,\hat J_S^{E\,\dagger}(\tau_i;\tau_n)}\,
U_E(\tau_{n+1},\tau_n)
|h_n,\tau_n\rangle,
\label{eq:DEdagger_on_bra_chain}
\end{align}
with
$h_0=h_i$ and 
$h_N=h_f$. The important thing to note here is that the $D_E^\dagger$ factors should not be mixed into a single product that looks identical to the ket-side chain. The bra-side chain starts at $\tau_f$ slice and runs to $\tau_i$ in the reverse order, while the ket-side chain starts at $\tau_i$ and runs to $\tau_f$.

What happens at the insertion slice? 
At the insertion slice $\tau_k$, we insert the local Euclidean operator. To do this one way would be to allow for a cleaner split at $\tau_k$. In other words we may want to organize the Euclidean numerator by splitting the contour at the operator insertion time
$x=({\bf x},\tau_k)$ for 
$\tau_i<\tau_k<\tau_f$,
and to treat the three ingredients separately:
(i) the part of $D_E[\sigma]$ from $\tau_i$ to $\tau_k$,
{(ii) the local insertion } $D_E^\dagger[\sigma]\hat O_E(x)$ { at } $\tau_k$,
and 
{(iii) the remaining part of } $D_E^\dagger[\sigma]$ { from } $\tau_{k+1}$ { to } $\tau_f$.
This would then appear to avoid carrying the full $D_E^\dagger[\sigma]$ through the whole interval before isolating the insertion. 

While the above is an attractive strategy, the actual implementation of such a procedure is cumbersome because of its difficulty to justify the {\it absence} of the pieces of the displacement operators beyond the Cauchy slice $\Sigma_{\tau_f}$. A better strategy is to take the $N_\sigma^E(x)$ defined in the first line of \eqref{eq:NE_before_PI} and, using cyclicity of the trace, rewrite it as:
\bg\label{kimpacha}
N_\sigma^E(x)
 &= &
{\rm Tr}\!\left(
D_E[\sigma]\,
\rho_\Omega\,
D_E^\dagger[\sigma]\,
\hat O_E(x)
\right)\nonumber\\
& = &
\int Dh_f\,
\langle h_f,\tau_f|
D_E[\sigma]\,
\rho_\Omega\,
D_E^\dagger[\sigma]\,
\hat O_E(x)\,
|h_f,\tau_f\rangle ,
\nd
by inserting a complete basis of metric eigenstates on a final Euclidean slice $\Sigma_f$ at time $\tau_f$ using 
$\mathbf 1_{\Sigma_f}
=
\int Dh_f\;
|h_f,\tau_f\rangle\langle h_f,\tau_f|$.
The crucial point is that in Euclidean signature this expression describes a \emph{single} matrix element around one Euclidean contour. There is no independent forward and backward branch. The same final configuration $h_f$ appears only because of the trace, and it closes a single Euclidean time contour rather than sewing together two independent Lorentzian histories.

Let us illustrate this by first 
inserting an initial basis, isolating the density matrix, and then constructing the requisite one-sided contour. For the initial basis, 
let us choose an initial Euclidean slice $\Sigma_i$ at time $\tau_i$, and insert complete sets of states on the two sides of $\rho_\Omega$ as
resolutions of identities via
$\mathbf 1_{\Sigma_i}
=
\int Dh_i\;
|h_i,\tau_i\rangle\langle h_i,\tau_i|$
and 
$\mathbf 1_{\Sigma_i}
=
\int Dh_i'\;
|h_i',\tau_i\rangle\langle h_i',\tau_i|$.
Then \eqref{eq:NE_before_PI} takes the following form:
\begin{equation}
\begin{aligned}
N_\sigma^E(x)
&=
\int Dh_f\,Dh_i\,Dh_i'\;
\langle h_f,\tau_f|D_E[\sigma]|h_i,\tau_i\rangle
\langle h_i,\tau_i|\rho_\Omega|h_i',\tau_i\rangle
\\
&\qquad\qquad\times
\langle h_i',\tau_i|
D_E^\dagger[\sigma]\,
\hat O_E(x)
|h_f,\tau_f\rangle ,
\end{aligned}
\label{eq:NE_initial_density_onecontour}
\end{equation}
where $D_E[\sigma]$ takes the ket state at the Cauchy slice defined at $\tau_i$ and drags it all the way to the Cauchy slice at $\tau_f$; whereas $D^\dagger_E[\sigma]$ takes the bra state defined at $\tau_i$ and drags it all the way to $\tau_f$. However,
at this stage the expression still looks like it contains two pieces, but in Euclidean signature these two pieces are simply two segments of the \emph{same} Euclidean contour. The density matrix supplies the initial Euclidean boundary conditions, while the common final state $|h_f,\tau_f\rangle$ closes the trace. To see this we will first 
discretize $D_E[\sigma]$ and $D_E^\dagger[\sigma]$ by 
letting
$\tau_n \equiv \tau_i + n\Delta\tau$,
such that 
$\Delta\tau = \frac{\tau_f-\tau_i}{N}$ with 
$n=0,1,\dots,N$.
We can then define the following variables:
$U_n \equiv U_E(\tau_{n+1},\tau_n),
J_n \equiv \hat J_S^E(\tau_i;\tau_n)$
and $J_n^\dagger \equiv \hat J_S^{E\dagger}(\tau_i;\tau_n)$
such that $D_E[\sigma]$ and $D^\dagger_E[\sigma]$ from \eqref{biral3} may be expressed using the following ordered-product notation as:
\bg\label{friendwatch}
&&D_E[\sigma]
=
\overleftarrow{\prod}_{n=0}^{N-1}
\left(
U_n e^{\Delta\tau J_n}
\right)
=
\left(
U_{N-1}e^{\Delta\tau J_{N-1}}
\right)\cdots
\left(
U_0e^{\Delta\tau J_0}
\right)\nonumber\\
&& D_E^\dagger[\sigma]
=
\overrightarrow{\prod}_{n=0}^{N-1}
\left(
e^{\Delta\tau J_n^\dagger}U_n
\right)
=
\left(
e^{\Delta\tau J_0^\dagger}U_0
\right)\cdots
\left(
e^{\Delta\tau J_{N-1}^\dagger}U_{N-1}
\right), \nd
where the arrows denote the ordering of the operators. (This is similar to the ordering of the $\mathcal B_{N}$ and $\mathcal A_N$ operators in \eqref{biral3}.) It's also clear that, using the above ordering and then
inserting complete sets of states at every intermediate Euclidean slice via 
$\mathbf 1_{\Sigma_{\tau_n}}
=
\int Dh_n\;
|h_n,\tau_n\rangle\langle h_n,\tau_n|$
for 
$n=1,\dots,N-1$, we reproduce \eqref{eq:DE_on_ket_chain} and \eqref{eq:DEdagger_on_bra_chain} as expected but also the following  
matrix element:

{\footnotesize
\begin{equation}
\begin{aligned}
\langle h_i',\tau_i|
D_E^\dagger[\sigma]\,
\hat O_E(x)
|h_f,\tau_f\rangle
&=
\int \prod_{m=1}^{N-1}D\widetilde h_m\;
\prod_{n=0}^{N-1}
\langle \widetilde h_n,\tau_n|
e^{\Delta\tau J_n^\dagger}U_n
|\widetilde h_{n+1},\tau_{n+1}\rangle
\;
O_E[\widetilde h_k](x),
\end{aligned}
\label{eq:second_piece_factorized_onecontour}
\end{equation}}
with
$\widetilde h_0=h_i'$
and $\widetilde h_N=h_f$, whose justification will be provided soon. Now, because the same final configuration $h_f$ is integrated over in both pieces in \eqref{kimpacha}, the trace identifies the end of the first Euclidean segment with the end of the second Euclidean segment. In Euclidean signature this does \emph{not} represent two independent histories, but rather a single contour closed by the trace. Equivalently, one can regard the matrix element
from \eqref{kimpacha}
as one Euclidean evolution kernel with an operator insertion and periodic gluing at $\tau_f$. The density matrix $\rho_\Omega$ only fixes the initial Euclidean state and does not generate a second real-time branch. If the vacuum is represented by a Euclidean wavefunctional, one may write:
\begin{equation}
\langle h_i,\tau_i|\rho_\Omega|h_i',\tau_i\rangle
=
\Psi_\Omega^E[h_i]\,
\Psi_\Omega^{\ast E}[h_i'].
\label{eq:rho_wavefunctional_onecontour}
\end{equation}
Then the two Euclidean segments are sewn through the common integrations over $h_i$, $h_i'$, and $h_f$, producing a single Euclidean functional integral.

Now consider the short-time matrix elements from \eqref{eq:DE_on_ket_chain} and \eqref{eq:DEdagger_on_bra_chain}, {\it i.e.} the ones lying between the temporal interval $\tau_n$ and $\tau_{n+1}$ given by
$\langle h_{n+1},\tau_{n+1}|U_n e^{\Delta\tau J_n}|h_n,\tau_n\rangle$ and its hermitian conjugate respectively.
We can insert a same-slice identity via the resolution of identity
using the ket states $|\hat{h}_n, \tau_n\rangle$ as $\mathbf 1_{\Sigma_{\tau_n}}
=
\int D\widehat h_n\;
|\widehat h_n,\tau_n\rangle\langle \widehat h_n,\tau_n|$
to obtain:
\bg\label{amytagra}
\langle h_{n+1},\tau_{n+1}|U_n e^{\Delta\tau \hat J_n}|h_n,\tau_n\rangle
&=&
\int D\widehat h_n\;
\langle h_{n+1},\tau_{n+1}|U_n|\widehat h_n,\tau_n\rangle
\langle \widehat h_n,\tau_n|e^{\Delta\tau \hat J_n}|h_n,\tau_n\rangle\nonumber\\
& = & \langle h_{n+1},\tau_{n+1}|U_n|h_n,\tau_n\rangle
e^{\Delta\tau J_E[h_n;\tau_n]} \nonumber\\
\langle h_n,\tau_n|e^{\Delta\tau \hat J_n^\dagger}U_n|h_{n+1},\tau_{n+1}\rangle
& = & \int D\widehat h_n\,\langle h_{n+1},\tau_{n+1}|
e^{\,\Delta\tau\,\hat J_n^{\dagger}}\,|\widehat h_n,\tau_n\rangle 
\langle \widehat h_n, \tau_n | U_n
|h_n,\tau_n\rangle \nonumber\\
& = &
e^{\Delta\tau J_E^\dagger[h_n;\tau_n]}
\langle h_n,\tau_n|U_n|h_{n+1},\tau_{n+1}\rangle , \nd
which pulls out the eigenvalues $J_E[h_n; \tau_n]$ and its hermitian conjugate and keep only the corresponding unitary operators as matrix elements. This is because  
the ket $|h_n , \tau_n\rangle$ is an eigenstate of $\hat J_n$, {\it i.e.} $\hat J_n |h_n, \tau_n\rangle = J_E[h_n; \tau_n] |h_n, \tau_n\rangle$ and using the fact that:
\begin{equation}
\langle \widehat h_n,\tau_n|e^{\Delta\tau J_n}|h_n,\tau_n\rangle
=
e^{\Delta\tau J_E[h_n;\tau_n]}\,
\delta[\widehat h_n-h_n] ,
\label{eq:Jdiag_B_onecontour}
\end{equation}
we can get the results as depicted in \eqref{amytagra}. This is good because it implies that both the source insertions reduce to multiplicative factors along the \emph{same} Euclidean chain. However question still remains about the operator insertion. Since the operator is inserted on the slice $\tau_k$, the cleanest way is to first insert a complete basis on that slice via
$\mathbf 1_{\Sigma_{\tau_k}}
=
\int Dh_k\;
|h_k,\tau_k\rangle\langle h_k,\tau_k|$, 
and then express the matrix element as:
\begin{equation}
\begin{aligned}
\langle h_i',\tau_i|D_E^\dagger[\sigma]\hat O_E(x)|h_f,\tau_f\rangle
&=
\int Dh_k\;
\langle h_i',\tau_i|D_E^\dagger[\sigma]|h_k,\tau_k\rangle
\langle h_k,\tau_k|\hat O_E(x)|h_f,\tau_f\rangle ,
\end{aligned}
\label{eq:bra_side_with_k_insertion}
\end{equation}
which naturally splits it up to two manageable pieces. As before, 
it is more convenient to discretize the entire chain first and then isolate the operator on the slice $\tau_k$. Using
$D_E^\dagger[\sigma]$ defined in \eqref{friendwatch}, 
together with complete sets on all intermediate slices, one finds:

{\footnotesize
\begin{equation}
\begin{aligned}
\langle h_i',\tau_i|D_E^\dagger[\sigma]\hat O_E(x)|h_f,\tau_f\rangle
&=
\int \prod_{m=1}^{N-1}D\widetilde h_m\;
\prod_{n=0}^{N-1}
\langle \widetilde h_n,\tau_n|
e^{\Delta\tau \hat J_n^\dagger}U_n
|\widetilde h_{n+1},\tau_{n+1}\rangle
\;
O_E[\widetilde h_k](x),
\end{aligned}
\label{eq:bra_side_discretized}
\end{equation}}
with
$\widetilde h_0=h_i'$ and 
$\widetilde h_N=h_f$, where we see that the operator is naturally isolated from the matrix element. The reason the operator becomes multiplicative is that it is local on the slice $\tau_k$:
\begin{equation}
\langle \widetilde h_k,\tau_k|\hat O_E(x)|\widetilde h_k,\tau_k\rangle
=
O_E[\widetilde h_k](x) ,
\label{eq:local_O_multiplicative}
\end{equation}
so there is no need to go to the intermediate Schr\"odinger picture although we could also follow that route.  This justifies \eqref{eq:second_piece_factorized_onecontour}.  At the special slice $\tau_k$, the relevant factor in the bra-side chain is:
\begin{equation}
\langle h_k,\tau_k|
e^{\Delta\tau \hat J_k^\dagger}U_k
|h_{k+1},\tau_{k+1}\rangle
\,O_E[h_k](x) = e^{\Delta\tau J_E^\dagger[h_k;\tau_k]}\,
\langle h_k,\tau_k|U_k|h_{k+1},\tau_{k+1}\rangle
\,O_E[h_k](x)\label{eq:k_slice_factor_bra} ,
\end{equation}
where we used \eqref{amytagra} (one may also follow the procedure outlined in footnote \ref{shirtagra} starting with \eqref{eq:bra_side_discretized} to reach \eqref{eq:k_slice_factor_bra}). Equivalently, if one prefers to place the operator directly between equal-time states, one may insert the same-slice identity at $\tau_k$ and use:
\begin{equation}
\langle h_k,\tau_k|\hat O_E(x)|\widehat h_k,\tau_k\rangle
=
O_E[h_k](x)\,
\delta[h_k-\widehat h_k].
\label{eq:O_diagonal_same_slice}
\end{equation}
Either way, the net result is that the local operator is inserted multiplicatively at the slice $\tau_k$. After the dust settles, we can combine the short-time kernels \eqref{amytagra} and \eqref{eq:k_slice_factor_bra} into the two matrix elements from \eqref{eq:NE_initial_density_onecontour} to get:

{\footnotesize
\bg\label{amyphule}
&& \langle h_f,\tau_f|D_E[\sigma]|h_i,\tau_i\rangle
=
\int \prod_{m=1}^{N-1}Dh_m\;
\Bigg[
\prod_{n=0}^{N-1}
\langle h_{n+1},\tau_{n+1}|U_n|h_n,\tau_n\rangle
\Bigg]
\Bigg[
\prod_{n=0}^{N-1}
e^{\Delta\tau J_E[h_n;\tau_n]}
\Bigg] \\
&& \langle h_i',\tau_i|D_E^\dagger[\sigma]\hat O_E(x)|h_f,\tau_f\rangle
=
\int \prod_{m=1}^{N-1}D\widetilde h_m\;
\Bigg[
\prod_{n=0}^{N-1}
e^{\Delta\tau J_E^\dagger[\widetilde h_n;\tau_n]}
\Bigg]
\Bigg[
\prod_{n=0}^{N-1}
\langle \widetilde h_n,\tau_n|U_n|\widetilde h_{n+1},\tau_{n+1}\rangle
\Bigg]
O_E[\widetilde h_k](x) , \nonumber
\nd}
where we see that the displacement operators $D_E[\sigma]$ and $D^\dagger_E[\sigma]$ as well as the insertion operator $\hat O_E(x)$ neatly isolate out of the matrix elements as multiplicative factors. We can now
combine the two Euclidean segments into one contour to get the following form for $N_\sigma^E(x)$ from \eqref{eq:NE_initial_density_onecontour}:
\bg\label{meypechap}
N_\sigma^E(x)
&= &
\int Dh_f\,Dh_i\,Dh_i'\;
\langle h_i,\tau_i|\rho_\Omega|h_i',\tau_i\rangle
\\
&\times &
\int \prod_{m=1}^{N-1}Dh_m\;
\Bigg[
\prod_{n=0}^{N-1}
\langle h_{n+1},\tau_{n+1}|U_n|h_n,\tau_n\rangle
\Bigg]
\Bigg[
\prod_{n=0}^{N-1}
e^{\Delta\tau J_E[h_n;\tau_n]}
\Bigg]\nonumber\\
&\times &
\int \prod_{p=1}^{N-1}D\widetilde h_p\;
\Bigg[
\prod_{n=0}^{N-1}
e^{\Delta\tau J_E^\dagger[\widetilde h_n;\tau_n]}
\Bigg]
\Bigg[
\prod_{n=0}^{N-1}
\langle \widetilde h_n,\tau_n|U_n|\widetilde h_{n+1},\tau_{n+1}\rangle
\Bigg]
O_E[\widetilde h_k](x) , \nonumber
\nd
where, at this stage the one-contour nature is expressed by the fact that the same final configuration $h_f$ appears in both chains and is integrated over. In Euclidean signature this is not a forward/backward Schwinger--Keldysh doubling. Rather, the trace closes a single Euclidean contour. From \eqref{meypechap}, we see that the continuum kernel
$\exp\!\left(
-\,S_{\rm tot}^E[g_E]
\right)$
arises from the standard short-time composition law:
\begin{equation}
\prod_{n=0}^{N-1}
\langle h_{n+1},\tau_{n+1}|U_n|h_n,\tau_n\rangle
\longrightarrow
\exp\!\left(
-\,S_{\rm tot}^E[g_E]
\right),
\label{eq:shorttime_to_action_filled}
\end{equation}
providing precisely the reason behind the one-contour representation of the path-integral. On the other hand, the GS factors exponentiate according to:
\bg
&& \prod_{n=0}^{N-1}
e^{\Delta\tau J_E[h_n;\tau_n]}
\longrightarrow
{\cal I}_E[g_E;\sigma]  =
\exp\!\left[
\int d^dx\;
\sqrt{g_E(x)}\,
\sigma^{\mu\nu}(x)\,
g_{\mu\nu}^E(x)
\right]\nonumber\\
&& \prod_{n=0}^{N-1}
e^{\Delta\tau J_E^\dagger[h_n;\tau_n]}
\longrightarrow
{\cal I}_E^\dagger[g_E;\sigma] =
\exp\!\left[
\int d^dx\;
\sqrt{g_E(x)}\,
\sigma^{\mu\nu\ast}(x)\,
g_{\mu\nu}^E(x)
\right], \nd
where we take $\sigma^{\mu\nu *} = \sigma^{\mu\nu}$, although we could easily handle the complex case. It is also important to note that, while we have used only the displacement operators for the metric degrees of freedom, we can easily extend this to include other matter (and even supergravity) degrees of freedom. Finally, 
the local insertion simply becomes:
\begin{equation}
O_E[h_k](x)\longrightarrow O_E[g_E](x).
\label{eq:O_continuum_filled}
\end{equation}
One may therefore combine the two segments into a single Euclidean functional integral with boundary data encoded by the initial density matrix:
\begin{equation}
\begin{aligned}
N_\sigma^E(x)
&=
\int Dh_f\,Dh_i\,Dh_i'\;
\langle h_i,\tau_i|\rho_\Omega|h_i',\tau_i\rangle
\\
&\qquad\times
\int_{h_i,h_i'}^{h_f}
{\cal D}g_E\;
\exp\!\left(
-\,S_{\rm tot}^E[g_E]
\right)\,
{\cal I}_E[g_E;\sigma]\,
{\cal I}_E^\dagger[g_E;\sigma]\,
O_E[g_E](x).
\end{aligned}
\label{eq:NE_discrete_to_continuum_onecontour}
\end{equation}
Equivalently, once the initial density matrix is absorbed into the Euclidean preparation of the vacuum state, this becomes the standard one-contour Euclidean path integral:
\begin{equation}
N_\sigma^E(x)
=
\int {\cal D}g_E\;
\exp\!\left(
-\,S_{\rm tot}^E[g_E]
\right)\,
{\cal I}_E^\dagger[g_E;\sigma]\,
O_E[g_E](x)\,
{\cal I}_E[g_E;\sigma] ,
\label{eq:NE_one_contour_final}
\end{equation}
where
${\cal I}_E[g_E;\sigma]$ (and its hermitian conjugate $
{\cal I}_E^\dagger[g_E;\sigma]$) are defined in \eqref{eq:IE_def}. 
The essential reason \eqref{eq:NE_discrete_to_continuum_onecontour} or \eqref{eq:NE_one_contour_final} is a \emph{one-contour} Euclidean path integral is because
in Euclidean signature there is no independent forward and backward time branch.
The trace only identifies the final Euclidean configuration $h_f$ and closes a single Euclidean contour.
After time slicing and inserting complete sets, all kernels compose into one Euclidean chain weighted by $e^{-S_{\rm tot}^E}$.
A compact final statement is therefore:
\bg\label{anytagbhalo}
&& N_\sigma^E(x)
=
{\rm Tr}\!\left(
\rho_\Omega\,D_E^\dagger[\sigma]\,\hat O_E(x)\,D_E[\sigma]
\right) \nonumber\\
\Longrightarrow &&
N_\sigma^E(x)
=
\int {\cal D}g_E\;
e^{-S_{\rm tot}^E[g_E]}\,
{\cal I}_E^\dagger[g_E;\sigma]\,
O_E[g_E](x)\,
{\cal I}_E[g_E;\sigma],\nd
which is a path integral over a \emph{single Euclidean contour}. One can now follow the same computational procedure to compute $Z_\sigma^E$
from \eqref{biralador} which we won't do here but leave it as an exercise for our diligent reader.

\subsection{Derivation using thimbles and the Picard-Lefschetz theory \label{sec6.4}}

Having presented the Euclidean path-integral and the Lorentzian Schwinger--Keldysh (SK) formulation for the
Glauber--Sudarshan (GS) state, we can now rewrite it in Picard--Lefschetz form \cite{wittenlefschetz,picardlefschetz,hetborel}.
The discussion parallels the coherent--state case, but there is one
important structural difference: for the GS state the parameter
$\sigma^{\mu\nu}(x)$ enters as a \emph{bulk} insertion on the contour rather
than as a shift of the initial boundary data.  Consequently the SK action
itself is deformed by the GS source, whereas the initial density matrix
may still be taken to be the vacuum density matrix.

\subsubsection{Euclidean path-integral formulation \label{sec6.4.1}}

Because the Euclidean path integral has a single exponential weight
$e^{-S_{\rm eff,\sigma}^E}$, it is particularly natural to analyze it using Picard--Lefschetz theory. Upon complexifying the Euclidean field space, the integration cycle may be decomposed as
${\cal C}
=
\sum\limits_s n_s\,{\cal J}_s$,
where the thimbles ${\cal J}_s$ are attached to saddles satisfying:
\begin{equation}
\frac{\delta S_{\rm eff,\sigma}^E[\lambda;{\rm g}_E]}
{\delta {\rm g}_E}
\Bigg|_{{\rm g}_E={\rm g}_E^{(s)}(\sigma)}
=
0,
\end{equation}
and $n_s$ tells us whether the thimble ${\cal J}_s$ contributes and with what orientation. Equivalently, $n_s$ is the intersection number between the original contour and the dual upward-flow cycle attached to the saddle ${\rm g}_E^{(s)}(\sigma)$. Thus
$n_s\in \mathbb Z$,
and in particular
$n_s=0$ implies that 
${\cal J}_s$ { does not contribute.} At this stage one should note that the boundary wavefunctionals
$\Psi_\Omega^{\ast E}[h_f]\,
\Psi_\Omega^E[h_i]$
do not disappear in the thimble decomposition. Rather, they are absorbed into the definitions of the corresponding thimble factors for the denominator and numerator. Then we can define the partition function and the operator insertion respectively as:

{\footnotesize
\bg\label{coridahl350}
Z_\sigma^E
&=&
\sum_s n_s\,
e^{-S_{\rm eff,\sigma}^E[\lambda;{\rm g}_E^{(s)}(\sigma)]}\,
{\cal Z}_s^E(\sigma,\lambda),
\nonumber\\
N_\sigma^E(x)
&=&
\sum_s n_s\,
e^{-S_{\rm eff,\sigma}^E[\lambda;{\rm g}_E^{(s)}(\sigma)]}\,
{\cal N}_{s}^E(x;\sigma,\lambda),
\nd}
where the quantity
${\cal Z}_s^E(\sigma,\lambda)$
is the fluctuation factor associated with the denominator on the thimble ${\cal J}_s$, {including the integrations over the boundary data weighted by the Euclidean vacuum wavefunctionals}. More explicitly, after expanding around the saddle
${\rm g}_E^{(s)}(\sigma)$,
one may write schematically:

{\scriptsize
\begin{equation}
{\cal Z}_s^E(\sigma,\lambda)
\sim
\int Dh_f\,Dh_i\;
\Psi_\Omega^{\ast E}[h_f]\,
\Psi_\Omega^E[h_i]
\int_{{\cal J}_s(h_i,h_f)}
{\cal D}(\delta{\rm g}_E)\;
\exp\!\left(
-\frac{1}{2}\delta{\rm g}_E\cdot
\frac{\delta^2 S_{\rm eff,\sigma}^E}{\delta{\rm g}_E^2}
\Bigg|_{{\rm g}_E^{(s)}}
\cdot\delta{\rm g}_E
+\cdots
\right),
\label{Zs_with_boundary}
\end{equation}}
where the dots denote cubic and higher terms in the fluctuation expansion. Thus ${\cal Z}_s^E$ contains not only the Gaussian determinant around the saddle and the higher fluctuation corrections, but also the boundary-state overlap encoded by
$\Psi_\Omega^{\ast E}[h_f]\Psi_\Omega^E[h_i]$. Similarly the quantity
${\cal N}_s^E(x;\sigma,\lambda)$ is the analogous fluctuation factor for the numerator with the operator insertion, again including the same boundary integrations. Schematically:

{\scriptsize
\begin{equation}
{\cal N}_s^E(x;\sigma,\lambda)
\sim
\int Dh_f\,Dh_i\;
\Psi_\Omega^{\ast E}[h_f]\,
\Psi_\Omega^E[h_i]
\int_{{\cal J}^{(if)}_s}
{\cal D}(\delta{\rm g}_E)\;
\exp\!\left(
-\frac{1}{2}\delta{\rm g}_E\cdot
\frac{\delta^2 S_{\rm eff,\sigma}^E}{\delta{\rm g}_E^2}
\Bigg|_{{\rm g}_E^{(s)}}
\cdot\delta{\rm g}_E
+\cdots
\right)
\,O_E[\tilde{\rm g}_E^{(s)}](x),
\end{equation}}
where $\tilde{\rm g}_E^{(s)} \equiv {\rm g}_E^{(s)}+\delta{\rm g}_E$ and ${\cal J}^{(if)}_s \equiv {\cal J}_s(h_i,h_f)$.
So ${\cal N}_s^E$ differs from ${\cal Z}_s^E$ precisely by the operator insertion. At leading semiclassical order one often has
${\cal N}_s^E(x;\sigma,\lambda)
\approx
O_E[{\rm g}_E^{(s)}(\sigma)](x)\,
{\cal Z}_s^E(\sigma,\lambda)$,
while beyond leading order the operator itself also fluctuates and contributes additional corrections. Therefore, the thimble-decomposed partition function $Z_\sigma^E$ and numerator $N_\sigma^E$ then provide the following expression for the expectation value:
\begin{equation}\label{coridahlfitboushona}
\langle O_E(x)\rangle_\sigma
=
\frac{
\sum\limits_s n_s\,
e^{-S_{\rm eff,\sigma}^E[\lambda;{\rm g}_E^{(s)}(\sigma)]}\,
{\cal N}_s^E(x;\sigma,\lambda)
}{
\sum\limits_s n_s\,
e^{-S_{\rm eff,\sigma}^E[\lambda;{\rm g}_E^{(s)}(\sigma)]}\,
{\cal Z}_s^E(\sigma,\lambda)
},
\end{equation}
from where the following clean interpretation arises:
$n_s$ selects the contributing thimbles and their orientation,
${\cal Z}_s^E$
is the fluctuation determinant and higher corrections for the denominator on ${\cal J}_s$ together with the boundary-state overlap,
and
${\cal N}_s^E$
is the corresponding fluctuation factor for the numerator with the insertion of $O_E(x)$, again including the same boundary-state factors. In other words, $n_s$ tells us which saddle sectors contribute, ${\cal Z}_s^E$ tells us how each saddle contributes to the partition function, and ${\cal N}_s^E$ tells us how each saddle contributes to the operator expectation value. These details are shown in {\bf figure \ref{skeuclibeta}}.

Therefore \eqref{coridahlfitboushona} is the natural Euclidean analogue of the Lorentzian GS story. The main difference is that in the Euclidean case the contour is single-branched and the GS insertion simply deforms the Euclidean action, whereas in the Lorentzian case the same physics is encoded on the doubled Schwinger--Keldysh contour.

\begin{figure}[h]
\centering
\begin{tabular}{c}
\includegraphics[width=6in]{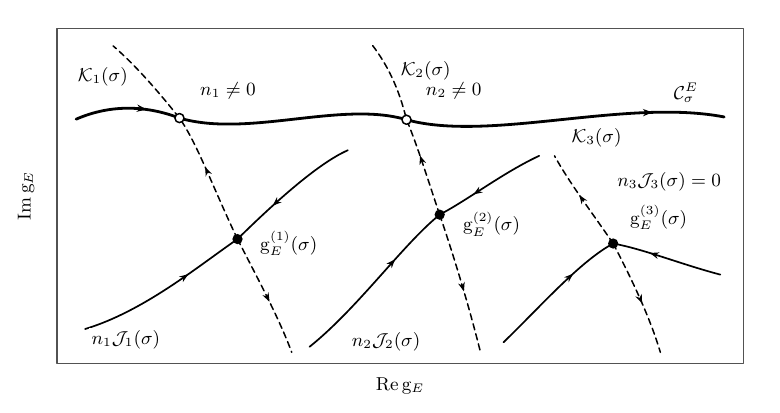}
\end{tabular}
\caption[]{Schematic two-dimensional projection of the Picard--Lefschetz decomposition of the single-contour Euclidean GS path integral. The thick solid curve is the original cycle $\mathcal C_\sigma^E$. The thin solid curves are the stable thimbles $\mathcal J_s(\sigma)$, with arrows directed toward the corresponding saddles, whereas the dashed curves are the dual upward-flow cycles $\mathcal K_s(\sigma)$, with arrows directed away from the saddles. In this decomposition, the arrows indicate the gradient flow used to define the Lefschetz thimbles, namely the downward flow of the real part of the complexified Euclidean action, namely
$\frac{d\phi^I}{du}
=
-\,\overline{\frac{\partial {\cal S}_E}{\partial \phi^I}}$,
or schematically
$\frac{d}{du}\,{\rm Re}\,{\cal S}_E \le 0$. The filled circles denote the source-dependent saddle configurations $\mathrm g_E^{(s)}(\sigma)$ of $S_{\mathrm{eff},\sigma}^E$. The open circles mark intersections of $\mathcal C_\sigma^E$ with the dual cycles and therefore determine the integers $n_s=\langle\mathcal C_\sigma^E,\mathcal K_s(\sigma)\rangle$.
In the schematic example shown, $n_1$ and $n_2$ are nonzero, while $\mathcal K_3(\sigma)$ does not intersect the original cycle and hence $n_3=0$; consequently, $\mathcal C_\sigma^E=n_1\mathcal J_1(\sigma)+n_2\mathcal J_2(\sigma)$ for the sectors displayed.}
\label{skeuclibeta}
\end{figure}

\subsubsection{Lorentzian Schwinger-Keldysh formulation \label{sec6.4.2}}

To study the Lorentzian Schwinger-Keldysh formulation, we can go back to our Lorentzian in-in computation from section \ref{sec6.2} and redefine some of the parameters. We can start by defining:
\begin{equation}\label{dixierupor}
S_{\rm SK}[\lambda;{\rm g}_+,{\rm g}_-]
\equiv
S_{\rm tot}[\lambda;{\rm g}_+]
-
S_{\rm tot}[\lambda;{\rm g}_-],
\end{equation}
where as before ${\rm g}_\pm = (g_\pm, {\rm gh}_\pm, {\rm matter}_\pm)$ with $\pm$ denoting the two branches. It is now convenient to absorb the GS insertion into a deformed SK action, and therefore let us define the GS--deformed SK action from \eqref{dixierupor} by:
\bg\label{dixtucker}
S_{{\rm SK},\sigma}[\lambda;{\rm g}_+,{\rm g}_-]
 & \equiv &
S_{\rm tot}[\lambda;{\rm g}_+]
-
S_{\rm tot}[\lambda;{\rm g}_-] \\
&+ & i\int d^4x\,
\sqrt{-g_+(x)}\,
\sigma^{\mu\nu}(x)\,
g_{+\mu\nu}(x)
-i\int d^4x\,
\sqrt{-g_-(x)}\,
\sigma^{\mu\nu}(x)\,
g_{-\mu\nu}(x) ,\nonumber
\nd
where notice the appearance of relative factor of $i = \sqrt{-1}$. Using \eqref{dixtucker}, the weight appearing in the SK path integral may be written simply as
$\exp\!\Big(
iS_{{\rm SK},\sigma}[\lambda;{\rm g}_+,{\rm g}_-]
\Big)$. With this notation the numerator and denominator of \eqref{eq:GS_expect_start2} become:
\bg\label{dixtucker2}
Z_\sigma
&=&
\int Dh_f
\int Dh_i^+\,Dh_i^- \;
\Psi_\Omega[h_i^+]\,\Psi_\Omega^*[h_i^-]
\int_{h_i^+}^{h_f}\!{\cal D}{\rm g}_+
\int_{h_i^-}^{h_f}\!{\cal D}{\rm g}_-\;
e^{\,iS_{{\rm SK},\sigma}[\lambda;{\rm g}_+,{\rm g}_-]}\\
N_\sigma(x)
&= &
\int Dh_f
\int Dh_i^+\,Dh_i^- \;
\Psi_\Omega[h_i^+]\,\Psi_\Omega^*[h_i^-]
\int_{h_i^+}^{h_f}\!{\cal D}{\rm g}_+
\int_{h_i^-}^{h_f}\!{\cal D}{\rm g}_-\;
{\cal O}[{\rm g}_+](x)\,
e^{\,iS_{{\rm SK},\sigma}[\lambda;{\rm g}_+,{\rm g}_-]} \nonumber
\nd
which is the basic Lorentzian SK formulation for the GS state. We can use this to study the Picard--Lefschetz decomposition of the doubled contour. For this, we now complexify the doubled field space:
\begin{equation}
({\rm g}_+,{\rm g}_-)
\;\longrightarrow\;
({\rm g}_+^{\mathbb C},{\rm g}_-^{\mathbb C}),
\end{equation}
and decompose the SK contour into Lefschetz thimbles
${\cal C}_\sigma
=
\sum_s n_s\,{\cal J}_s(\sigma)$,
where $n_s\in \mathbb Z$ are the intersection numbers with the dual upward-flow
cycles.  The relevant saddles are critical points of the GS--deformed doubled
action, namely:
\begin{equation}
\frac{\delta S_{{\rm SK},\sigma}}{\delta {\rm g}_+}=0,
\qquad
\frac{\delta S_{{\rm SK},\sigma}}{\delta {\rm g}_-}=0.
\label{GS_saddle_eqs_doubled}
\end{equation}
Because the GS parameter enters as a bulk insertion, these equations are
\emph{not} the same as the undeformed equations of motion. In fact this is where the significant difference from the coherent state appears. More explicitly:
\bg\label{dixtucker3}
&& \frac{\delta S_{\rm tot}[\lambda;{\rm g}_+]}{\delta {\rm g}_+}
+
i\,\frac{\delta}{\delta {\rm g}_+}
\left(
\int d^4x\,
\sqrt{-g_+}\,
\sigma^{\mu\nu}g_{+\mu\nu}
\right)
=0 \nonumber\\
&& \frac{\delta S_{\rm tot}[\lambda;{\rm g}_-]}{\delta {\rm g}_-}
+
i\,\frac{\delta}{\delta {\rm g}_-}
\left(
\int d^4x\,
\sqrt{-g_-}\,
\sigma^{\mu\nu}g_{-\mu\nu}
\right)
=0,
\nd
where, and as mentioned above, the GS
state shifts the saddle equations themselves, unlike the coherent state where
the action was unchanged and only the boundary data were shifted. We can now let
$\bigl({\rm g}_+^{(s)}(\sigma),{\rm g}_-^{(s)}(\sigma)\bigr)$
denote a saddle of the doubled GS action.  On each thimble one expands
${\rm g}_\pm
=
{\rm g}_\pm^{(s)}(\sigma)
+
\delta{\rm g}_\pm$. Then the doubled action expands as:
\begin{align}
S_{{\rm SK},\sigma}[\lambda;{\rm g}_+,{\rm g}_-]
&=
S_{{\rm SK},\sigma}[\lambda;{\rm g}_+^{(s)}(\sigma),{\rm g}_-^{(s)}(\sigma)]
\nonumber\\
&\qquad
+\frac12
\begin{pmatrix}
\delta{\rm g}_+ & \delta{\rm g}_-
\end{pmatrix}
{\cal O}_s^{\rm SK}(\sigma,\lambda)
\begin{pmatrix}
\delta{\rm g}_+ \\ \delta{\rm g}_-
\end{pmatrix}
+
S_{{\rm int},\sigma}^{(s)}[\delta{\rm g}_+,\delta{\rm g}_-],
\label{GS_SK_quad_expand}
\end{align}
where the linear order terms vanish due to the saddle constraint \eqref{dixtucker3}. The form \eqref{GS_SK_quad_expand} is an exact answer because:
\begin{equation}
{\cal O}_s^{\rm SK}(\sigma,\lambda)
=
\left.
\frac{\delta^2 S_{{\rm SK},\sigma}}
{\delta({\rm g}_+,{\rm g}_-)\,\delta({\rm g}_+,{\rm g}_-)}
\right|_{({\rm g}_+^{(s)}(\sigma),{\rm g}_-^{(s)}(\sigma))}
\label{GS_SK_kernel}
\end{equation}
is the doubled quadratic fluctuation operator, and all the infinite series of higher order corrections are inserted in $S_{{\rm int},\sigma}^{(s)}[\delta{\rm g}_+,\delta{\rm g}_-]$. Performing the Gaussian integrals and then the higher fluctuation expansion yields
\bg\label{dixtucker5}
Z_\sigma
 &= &
\sum_s
n_s\,
\exp\!\left(
iS_{{\rm SK},\sigma}[\lambda;{\rm g}_+^{(s)}(\sigma),{\rm g}_-^{(s)}(\sigma)]
\right)
{\cal Z}_s(\sigma,\lambda) \nonumber\\
N_\sigma(x)
& = &
\sum_s
n_s\,
\exp\!\left(
iS_{{\rm SK},\sigma}[\lambda;{\rm g}_+^{(s)}(\sigma),{\rm g}_-^{(s)}(\sigma)]
\right)
{\cal N}_s(x;\sigma,\lambda),
\nd
where the coefficient ${\cal Z}_s(\sigma,\lambda)$ includes the doubled
fluctuation determinant, all higher loop corrections, and the contribution of
the initial boundary wavefunctional product
$\Psi_\Omega[h_i^+]\,\Psi_\Omega^*[h_i^-]$
and ${\cal N}_s(x;\sigma,\lambda)$ is the corresponding fluctuation series
with the operator insertion on the $+$ branch. Putting everything together, the GS expectation value takes the exact thimble form:
\begin{equation}
\langle {\cal O}_H(x)\rangle_\sigma
=
\frac{
\sum\limits_s
n_s\,
\exp\!\left(
iS_{{\rm SK},\sigma}[\lambda;{\rm g}_+^{(s)}(\sigma),{\rm g}_-^{(s)}(\sigma)]
\right)
{\cal N}_s(x;\sigma,\lambda)
}{
\sum\limits_s
n_s\,
\exp\!\left(
iS_{{\rm SK},\sigma}[\lambda;{\rm g}_+^{(s)}(\sigma),{\rm g}_-^{(s)}(\sigma)]
\right)
{\cal Z}_s(\sigma,\lambda)
} ,
\label{GS_thimble_ratio}
\end{equation}
which may now to compared to \eqref{coridahl6}, \eqref{tricksbrke} and especially to \eqref{coridahl7}. The results are expectedly similar but \eqref{GS_thimble_ratio} is more {\it accurate} because it precisely captures which thimbles contribute to the expectation value. 
For the metric operator, \eqref{GS_thimble_ratio} becomes:
\begin{equation}
\langle g_{\mu\nu}(x)\rangle_\sigma
=
\frac{
\sum\limits_s
n_s\,
\exp\!\left(
iS_{{\rm SK},\sigma}[\lambda;{\rm g}_+^{(s)}(\sigma),{\rm g}_-^{(s)}(\sigma)]
\right)
{\cal N}_{s,\mu\nu}(x;\sigma,\lambda)
}{
\sum\limits_s
n_s\,
\exp\!\left(
iS_{{\rm SK},\sigma}[\lambda;{\rm g}_+^{(s)}(\sigma),{\rm g}_-^{(s)}(\sigma)]
\right)
{\cal Z}_s(\sigma,\lambda)
} ,
\label{GS_metric_thimble_ratio}
\end{equation}
which is our final answer from the Picard-Lefschetz theory. In writing \eqref{GS_thimble_ratio} and \eqref{GS_metric_thimble_ratio}, we should note that 
the nonperturbative sectors of the effective action are labelled by
$I,J,\ldots$ through:
\begin{equation}
S_{\rm tot}[\lambda;{\rm g}]
=
S_{\rm pert}[\lambda;{\rm g}]
+
\sum_I e^{-A_I/\lambda}S_I[\lambda;{\rm g}]
+
\sum_{I,J} e^{-(A_I+A_J)/\lambda}S_{IJ}[\lambda;{\rm g}]
+\cdots ,
\end{equation}
but these labels should not be confused with the thimble label $s$.
The index $I$ labels nonperturbative sectors in the \emph{construction of the
effective action}, whereas the index $s$ labels critical points of the
\emph{doubled SK functional integral} built from the GS--deformed action.
Schematically:

{\footnotesize
\begin{equation}
\text{nonperturbative sectors }(I)
\;\Longrightarrow\;
S_{\rm tot}[\lambda;{\rm g}]
\;\Longrightarrow\;
S_{{\rm SK},\sigma}[\lambda;{\rm g}_+,{\rm g}_-]
\;\Longrightarrow\;
\text{thimble saddles }s ,
\label{GS_sector_to_thimble}
\end{equation}}
implying that
the trans--series and the thimble decomposition are distinct but fully
compatible layers of description. In fact this also directly relates to the 1PI effective action. To see the connection, we can define:
\begin{align}
Z[J_+,J_-;\sigma]
&=
\int Dh_f
\int Dh_i^+\,Dh_i^- \;
\Psi_\Omega[h_i^+]\,\Psi_\Omega^*[h_i^-]\\
&\qquad\times
\int_{h_i^+}^{h_f}\!{\cal D}{\rm g}_+
\int_{h_i^-}^{h_f}\!{\cal D}{\rm g}_-\;
\exp\!\Big(
iS_{{\rm SK},\sigma}[\lambda;{\rm g}_+,{\rm g}_-]
+
iJ_+\cdot{\rm g}_+
-
iJ_-\cdot{\rm g}_-
\Big). \nonumber
\label{GS_ZJ}
\end{align}
which forms the microscopic SK generating functional with external sources $J_\pm$. We can also define the connected generating functional in the following standard way:
\begin{equation}
W[J_+,J_-;\sigma]
=
-\,i\ln Z[J_+,J_-;\sigma].
\label{GS_Wdef}
\end{equation}
The classical doubled fields are then
$\bar{\rm g}_\pm
= \pm
\frac{\delta W}{\delta J_\pm}$ with the minus sign for the $-$ 
branch\footnote{The minus sign in the second relation is the standard SK convention coming from the source term
$iJ_+\!\cdot{\rm g}_+ - iJ_-\!\cdot{\rm g}_-$.
If one uses a different convention for the signs of the sources, then the sign in the classical field expectation value changes accordingly. In many physics texts this sign is absorbed into the definition of the $-$ branch source, but conceptually nothing changes.}.
Because the GS insertion is built into the SK action itself, these classical
fields are expectation values in the GS state. In the physical limit
$J_+=J_-$ one recovers the physical one-point function:
\begin{equation}
\bar{\rm g}(x;\sigma)
=
\frac{\langle \sigma|\,{\rm g}_H(x)\,|\sigma\rangle}
{\langle \sigma|\sigma\rangle} ,
\label{GS_bar_g_physical}
\end{equation}
where ${\rm g}_H$ is in the Heisenberg form. Note however that, while we write ${\rm g} = (g, {\rm gh}, {\rm matter})$, the expectation value \eqref{GS_bar_g_physical} is only defined over $(g, {\rm matter})$ and not the ghosts. The SK 1PI effective action is obtained by the usual Legendre transform:
\begin{equation}
\Gamma_{\rm 1PI}^{\rm SK}[\bar{\rm g}_+,\bar{\rm g}_-;\sigma]
=
W[J_+,J_-;\sigma]
-
J_+\cdot\bar{\rm g}_+
+
J_-\cdot\bar{\rm g}_- 
\label{GS_1PI}
\end{equation}
with the understanding that after the differentiation $\bar{\rm g}_\pm = \pm {\partial W\over \partial J_\pm}$ one solves for the sources as functionals of the classical fields 
$J_+ = J_+[\bar{\rm g}_+,\bar{\rm g}_-;\sigma]$ and 
$J_- = J_-[\bar{\rm g}_+,\bar{\rm g}_-;\sigma]$.
After substituting these back into \eqref{GS_1PI}, the 1PI action becomes a functional of $\bar{\rm g}_\pm$ and $\sigma$ alone. The 
equations of motion and the quadratic 1PI kernel are:
\begin{equation}
\frac{\delta \Gamma_{\rm 1PI}^{\rm SK}}{\delta \bar{\rm g}_\pm}=0,
\qquad  \Gamma_{\rm 1PI}^{(2)}(x,y;\sigma)
=
\frac{\delta^2 \Gamma_{\rm 1PI}^{\rm SK}}
{\delta \bar{\rm g}(x)\,\delta \bar{\rm g}(y)}, 
\end{equation}
in the physical limit $\bar{\rm g}_+=\bar{\rm g}_-$. Thus ${\rm g}_\pm$
are the microscopic path-integral variables, whereas 
$\bar{\rm g}_\pm
=\pm
\frac{\delta W}{\delta J_\pm}$
are the classical fields entering $\Gamma_{\rm 1PI}^{\rm SK}$.
In other words, the logical chain is:

{\footnotesize
\begin{equation}
({\rm g}_+,{\rm g}_-)
\quad\stackrel{\text{integrate over}}{\longrightarrow}\quad
Z[J_+,J_-;\sigma]
\quad\longrightarrow\quad
W[J_+,J_-;\sigma]
\quad\stackrel{\delta/\delta J_\pm}{\longrightarrow}\quad
(\bar{\rm g}_+,\bar{\rm g}_-),
\end{equation}}
and only after this does one define
$\Gamma_{\rm 1PI}^{\rm SK}[\bar{\rm g}_+,\bar{\rm g}_-;\sigma]$. In a similar vein one may relate this to the Wilsonian effective action.
The Wilsonian effective action in the doubled SK theory is obtained by splitting
modes into low and high frequency pieces:
\begin{equation}
{\rm g}_\pm
=
{\rm g}_{<,\pm}+{\rm g}_{>,\pm},
\end{equation}
for both the $+$ and $-$ branches. We can now 
integrate out the high frequency modes $-$ which exist because we are over a Minkowski minimum and therefore do not suffer the same problems that we discussed in section \ref{sec2.2} $-$ in the presence of the same GS deformation:

{\footnotesize
\bg\label{dixtucker8}
e^{\,iS_{\Lambda,\sigma}^{\rm SK}[{\rm g}_{<,+},{\rm g}_{<,-}]}
=
\int {\cal D}{\rm g}_{>,+}\,{\cal D}{\rm g}_{>,-}\;
\exp\!\Big(
iS_{{\rm SK},\sigma}[\lambda;{\rm g}_{<,+}+{\rm g}_{>,+},
{\rm g}_{<,-}+{\rm g}_{>,-}]
\Big) \rho_{\rm init}.
\nd}
\begin{figure}[h]
\centering
\begin{tabular}{c}
\includegraphics[width=6in]{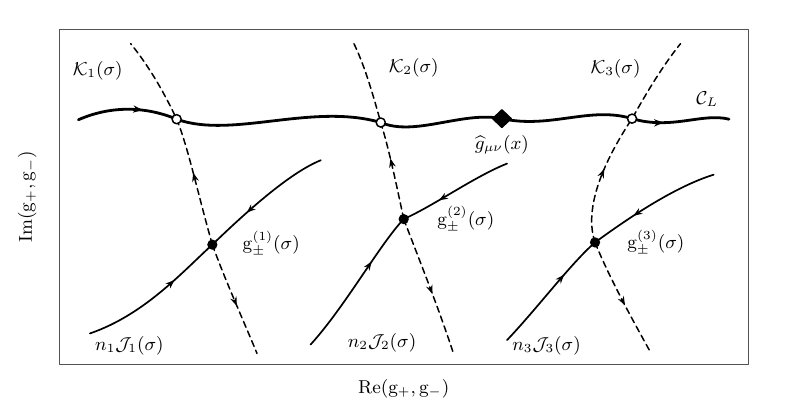}
\end{tabular}
\caption[]{Schematic two-dimensional projection of the Picard--Lefschetz decomposition of the complexified doubled Schwinger--Keldysh field space for the Lorentzian Glauber--Sudarshan state. The thick solid curve is the original contour $\mathcal C_L$. The thin solid curves are the descent thimbles $\mathcal J_s(\sigma)$, oriented toward the saddle points $({\rm g}_+^{(s)}(\sigma),{\rm g}_-^{(s)}(\sigma))$, which satisfy the GS-deformed saddle equations~\eqref{GS_saddle_eqs_doubled}. The dashed curves show the dual upward cycles $\mathcal K_s(\sigma)$, with arrows directed away from the saddles; the open circles mark their intersections with $\mathcal C_L$, so that $n_s=\langle\mathcal C_L,\mathcal K_s\rangle$ and $\mathcal C_L=\sum_s n_s\mathcal J_s(\sigma)$. The diamond denotes the metric insertion $\widehat g_{\mu\nu}(x)$ on the original cycle; it appears the numerator of Eq.~\eqref{GS_metric_thimble_ratio} but does not define an additional saddle or thimble. The doubled Lorentzian exponent retains the forward/backward signs $+iS_{\rm tot}[{\rm g}_+]$ and $-iS_{\rm tot}[{\rm g}_-]$, while the GS bulk deformation is included in $S_{{\rm SK},\sigma}$ and therefore shifts the saddles themselves. The corresponding thimble sums and metric expectation value are given in Eqs.~\eqref{dixtucker5} and \eqref{GS_metric_thimble_ratio}. }
\label{skloren2}
\end{figure} 
Thus the Wilsonian description inherits both the SK doubling and the GS bulk
deformation. Its quadratic kernel therefore is:
\begin{equation}
S_{\Lambda,\sigma}^{(2)}(x,y)
=
\frac{\delta^2 S_{\Lambda,\sigma}^{\rm SK}}
{\delta {\rm g}(x)\,\delta {\rm g}(y)}.
\label{GS_Wilsonian_kernel}
\end{equation}
The three descriptions $-$ 1PI, Wilsonian, trans--series and thimble descriptions $-$ are complementary reorganizations of the same underlying
real--time functional integral.  Their consistency implies that the same physical
background and the same quadratic dynamics are being described in different
languages.  Schematically one may write
\begin{equation}
\bar{\rm g}_{\mu\nu}(x;\sigma)
\;\sim\;
{\rm g}^{(s)}_{\mu\nu}(x;\sigma),
\qquad
\Gamma_{\rm 1PI}^{(2)}
\;\sim\;
{\cal O}_s^{\rm SK}
\;\sim\;
S_{\Lambda,\sigma}^{(2)},
\label{GS_identifications}
\end{equation}
where the symbol ``$\sim$'' means ``encodes the same quadratic physical
dynamics,'' not literal equality as bare functionals. More explicitly the three functions:
\bg\label{dixtucker00}
&& {\cal O}_s^{\rm SK}(x,y;\sigma,\lambda)
=
\left.
\frac{\delta^2 S_{{\rm SK},\sigma}}
{\delta({\rm g}_+,{\rm g}_-)\,\delta({\rm g}_+,{\rm g}_-)}
\right|_{({\rm g}_+^{(s)}(\sigma),{\rm g}_-^{(s)}(\sigma))}\nonumber\\
&& \Gamma_{\rm 1PI}^{(2)}(x,y;\sigma)
=
\frac{\delta^2 \Gamma_{\rm 1PI}^{\rm SK}}
{\delta \bar{\rm g}(x)\,\delta \bar{\rm g}(y)}, \qquad S_{\Lambda,\sigma}^{(2)}(x,y)
=
\frac{\delta^2 S_{\Lambda,\sigma}^{\rm SK}}
{\delta {\rm g}(x)\,\delta {\rm g}(y)}, \nd
are respectively the quadratic fluctuation operator on the thimble saddle, the inverse exact propagator of the theory, and the Wilsonian quadratic kernel. The nonperturbative sectors
$e^{-A_I/\lambda},
e^{-(A_I+A_J)/\lambda}$, et cetera
appear consistently in all descriptions: in the trans--series action
$S_{\rm tot}$, in the Wilsonian flow after integrating out high modes, in the
1PI vertices after resummation, and in the thimble expansion of the GS--deformed
SK contour. Therefore the logical relation among the various objects in the GS case may be summarized
as:

{\footnotesize
\begin{equation}
\text{microscopic SK theory}
\;\Longrightarrow\;
Z[J_+,J_-;\sigma]
\;\Longrightarrow\;
W[J_+,J_-;\sigma]
\;\Longrightarrow\;
\left\{
\begin{array}{l}
\Gamma_{\rm 1PI}^{\rm SK}[\bar g_+,\bar g_-;\sigma] \\[6pt]
S_{\Lambda,\sigma}^{\rm SK}[g_{<,+},g_{<,-}] \\[6pt]
\displaystyle \sum_s n_s {\cal J}_s(\sigma)
\end{array}
\right. ,
\label{GS_master_summary}
\end{equation}}
while the trans--series structure enters through $S_{\rm tot}[\lambda; {\rm g}_\pm]$ and the GS state modifies the doubled action itself according to \eqref{dixtucker}. More details appear in {\bf figures \ref{skloren2}} and {\bf \ref{sklorenthimble}}.

Thus, for the GS state, the 1PI, Wilsonian, trans--series, and
Picard--Lefschetz descriptions remain mutually consistent, but the role of
$\sigma$ is conceptually different from the coherent--state case: it does not
merely select a boundary condition, but directly deforms the doubled SK action
and hence shifts the classical saddle equations themselves.

\begin{figure}[h]
\centering
\begin{tabular}{c}
\includegraphics[width=6in]{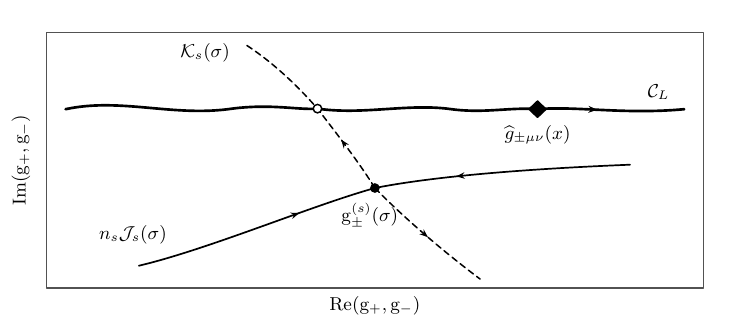}
\end{tabular}
\caption[]{Branch-resolved metric insertion in a representative saddle sector of the Lorentzian Glauber--Sudarshan thimble decomposition. The diamond on the original doubled Lorentzian cycle $\mathcal C_L$ denotes $\widehat g_{\pm\mu\nu}(x)$, where $+$ selects an insertion on the forward Schwinger--Keldysh branch and $-$ selects an insertion on the backward branch. The thin solid curve is the descent thimble $\mathcal J_s(\sigma)$, with arrows directed toward the filled saddle $({\rm g}+^{(s)}(\sigma),{\rm g}-^{(s)}(\sigma))$; the dashed curve is its dual upward cycle $\mathcal K_s(\sigma)$, with arrows directed away from that saddle. The open circle marks $\mathcal C_L\cap\mathcal K_s(\sigma)$, which fixes $n_s$. Note that, in the Lorentzian case the {thimble-defining auxiliary flows} are not the same Euclidean gradient flows of ${\rm Re}\,S_E$ that we saw in {\bf figure \ref{skeuclibeta}}. Instead they are the Picard--Lefschetz flows associated with the complexified Lorentzian exponent. If
${\cal I}_L[g_+,g_-;\sigma]
=
iS_{\rm tot}[g_+]
-
iS_{\rm tot}[g_-]
+
W_+[g_+;\sigma]
+
W_-[g_-;\sigma]$,
then the Lorentzian thimbles are defined by downward flow of the Morse function
$h[g_+,g_-]
=
{\rm Re}\,{\cal I}_L[g_+,g_-;\sigma]$.
Accordingly, the auxiliary Picard--Lefschetz flow equations are schematically
$\frac{d\phi^I}{du}
=
-\,\overline{
\frac{\partial {\cal I}_L}{\partial \phi^I}
}$ with 
$\phi^I\in\{g_+,g_-\}_{\mathbb C}$,
so that
$\frac{d}{du}\,{\rm Re}\,{\cal I}_L
\le 0$.
Thus the descent and the ascent flows in the Lorentzian thimble picture can be interpreted as {auxiliary Picard--Lefschetz flows in complexified doubled field space}. (HB: Note that metric operator insertions do not change the contour nor the thimble decomposition, regardless of the branch where they appear! So, I don't think we need this figure, all the relevant info is already in the main text and summarized in Figure 7. Please feel free to drop Figure 8.)}
\label{sklorenthimble}
\end{figure}

\subsection{In-in versus in-out: A comparative analysis of the expectation values \label{sec6.5}}

A useful way to phrase the issue is that the choice between an
in--out and an in--in description is not determined solely by the
local operator identity. It is also determined by how one completes
the contour with vacuum wavefunctionals and by what physical quantity
one ultimately wishes to compute. To see this, consider first the vacuum two-point function with:
\begin{equation}
t_2>t_1>t_0,
\qquad
x=({\bf x},t_2),
\qquad
y=({\bf y},t_1) ,
\end{equation}
where the temporal distinction is provided by the two Cauchy slices at the two points and, because the background is Minkowski $-$ recall that we are always at a chosen Minkowski minimum $-$ there is a specific meaning of time here.  In this set-up and in the interaction picture one may write:
\begin{equation}
\mathcal T\phi(x)\phi(y)
=
U^\dagger(t_2,t_0)\,
\phi_I({\bf x},t_2)\,
U(t_2,t_1)\,
\phi_I({\bf y},t_1)\,
U(t_1,t_0) ,
\end{equation}
where $\mathcal T$ is the time ordering symbol. Note that we are using the interaction picture instead of the Schr\"odinger picture to be consistent with the QFT computations. 
This operator identity already contains the local sequence:
\begin{equation}
t_0\to t_1\to t_2\to t_0,
\end{equation}
so at the level of local time evolution it has the structure of a
closed contour. The contour structure is clearly coming from the identity $U^\dagger(t_2, t_0) = U(t_0, t_2)$ so is already built in the definition of the correlation function itself. Equivalently,
\begin{equation}
\langle \Omega|\,\mathcal T\phi(x)\phi(y)\,|\Omega\rangle
=
{\rm Tr}\!\left(
\rho_\Omega\,\mathcal T\phi(x)\phi(y)
\right),
\qquad
\rho_\Omega=|\Omega\rangle\langle\Omega|,
\end{equation}
which suggests a Schwinger--Keldysh closed contour interpretation. However, the same correlator also admits the usual in--out vacuum
completion if we can appropriately define the out vacuum. For example if:
\begin{equation}
|\Omega\rangle_{\rm in}
=
U(t_0,-T)\,|0\rangle,
\qquad
{}_{\rm out}\langle\Omega|
=
\langle 0|\,U(+T,t_0),
\end{equation}
where $|0\rangle$ denotes the free vacuum;
then one obtains the standard in--out contour completion from
$-T$ to $+T$. In this sense the ordinary Feynman vacuum correlator is
naturally treated as an in--out object, even though it also admits a
closed-contour representation.

Now consider the GS construction. Let $D(\sigma)$ be denoted by \eqref{vandikok} and $\rho_\sigma$ by \eqref{eq:rho_sigma_def_detailed2}.
The exact physical quantity of interest is the normalized real-time
expectation value:
\begin{equation}
\langle \hat{\mathcal O}_H({\bf x}, t)\rangle_\sigma
=
\frac{
{\rm Tr}\!\left(
\rho_\sigma\,\hat{\mathcal O}_H({\bf x}, t)
\right)
}{
{\rm Tr}(\rho_\sigma)
},
\label{exact_inin_GS_ratio_local}
\end{equation}
which, for the metric itself, is represented by the replacement
$\hat{\mathcal O}_H(x) \to \hat{g}_{\mu\nu}(x)$.
This quantity is intrinsically an in--in observable and is exactly
represented by the closed Schwinger--Keldysh contour. Nevertheless, the same local operator structure may also be completed
into an in--out contour. This is the point we want to make precise
below. However before going into that let us clarify how a seemingly in--out type contour appears once the $-$ branch of the SK contour is integrated out.

\subsubsection{Effective decoupling of the backward branch and reality
\label{dixiruporfatieachi}}

Having established the Lorentzian Schwinger--Keldysh description of the Glauber--Sudarshan state, it is useful to record the exact real-time one-point function in a form that makes three points manifest: first, the exact role of the GS deformation in the closed-time-path functional integral; second, in what precise sense the $-$ branch can and cannot be eliminated; and third, why the final answer remains real even after rewriting the formula in a $+$ branch language with an influence functional.

Recall that with the GS-deformed density matrix 
$\rho_\sigma
\equiv
D(\sigma)\,\rho_\Omega\,D^\dagger(\sigma)$,
where $\rho_\Omega$ is the density matrix associated with the reference vacuum state and $D(\sigma)$ is the GS operator, the exact in--in expectation value of the Heisenberg metric operator at the spacetime point
$x=(x^0,\mathbf x)$ takes the following standard Schwinger--Keldysh representation:

{\scriptsize
\begin{align}
\langle \hat g_{\mu\nu}(x)\rangle_\sigma
& = \frac{
\mathrm{Tr}\!\left(
\rho_\sigma\,\hat g_{\mu\nu}(x)
\right)
}{
\mathrm{Tr}(\rho_\sigma)
}\nonumber\\
&=
\frac{
\displaystyle
\int Dh_f
\int Dh_i^\pm\;
\rho_\Omega[h_i^\pm]\;
\int_{h_i^+}^{h_f}
{\cal D}{\rm g}_+\;
\int_{h_i^-}^{h_f}
{\cal D}{\rm g}_-\;
g_{+\mu\nu}(x)\;
e^{\,iS_{\rm tot}[{\rm g}_+] - iS_{\rm tot}[{\rm g}_-]}\;
e^{\,W_\sigma[g_+] + W_\sigma[g_-]}
}{
\displaystyle
\int Dh_f
\int Dh_i^\pm
\;
\rho_\Omega[h_i^\pm]\;
\int_{h_i^+}^{h_f}
{\cal D}{\rm g}_+\;
\int_{h_i^-}^{h_f}
{\cal D}{\rm g}_-\;
e^{\,iS_{\rm tot}[{\rm g}_+] - iS_{\rm tot}[{\rm g}_-]}\;
e^{\,W_\sigma[g_+] + W_\sigma[g_-]}
},
\label{eq:exact_GS_SK_h_full}
\end{align}}
where ${\rm g}_\pm = (g_\pm, {\rm gh}_\pm, {\rm matter}_\pm)$ (although we don't always consider the matter part); 
$Dh_i^\pm = Dh_+ Dh_-$ integrate over the initial induced three-metrics on the initial hypersurface at
$t=t_i$, while
$Dh_f$ integrates over the common final induced three-metric on the gluing hypersurface at $t=t_f$.
The constrained Lorentzian path integrals are therefore taken over four-metrics whose induced boundary data satisfy:
\begin{equation}
g_\pm(t_i,\mathbf x)\big|_{\rm induced}=h_i^\pm(\mathbf x),
\qquad
g_\pm(t_f,\mathbf x)\big|_{\rm induced}=h_f(\mathbf x) ,
\label{eq:SK_induced_bc_exact}
\end{equation}
which are defined on the initial and the final Cauchy slices respectively. All these slices are over a supersymmetric Minkowski minimum, so are well defined. The GS deformation appears through the source functional:
\begin{equation}
W_\sigma[g]
\equiv
\int_{t_i}^{t_f}dt\int d^3x\;
\sqrt{-g(t,\mathbf x)}\;
\sigma^{\alpha\beta}(t,\mathbf x)\,
g_{\alpha\beta}(t,\mathbf x).
\label{eq:W_sigma_def_exact}
\end{equation}
Thus, unlike the coherent-state case where the deformation is implemented through shifted boundary data, the GS state modifies the bulk weight directly and does so separately on the two branches of the contour:
\begin{equation}
e^{\,iS_{\rm tot}[g_+,\mathrm{gh}_+] - iS_{\rm tot}[g_-,\mathrm{gh}_-]}
\quad\longrightarrow\quad
e^{\,iS_{\rm tot}[g_+,\mathrm{gh}_+] - iS_{\rm tot}[g_-,\mathrm{gh}_-]}\;
e^{\,W_\sigma[g_+] + W_\sigma[g_-]}.
\end{equation}
Note that \eqref{eq:exact_GS_SK_h_full} is exact. No saddle-point approximation has been used, no semiclassical limit has been assumed, and no truncation of the ghost or matter sectors has been made. All perturbative corrections, all nonperturbative sectors already encoded in the exact trans-series action $S_{\rm tot}$, and the full measure data are included in the functional integrations.

At this point
one might ask whether the $-$ branch can be removed so that the one-point function is represented by a single forward path integral. For a genuine in--in observable, the answer is negative in general. The reason is structural: the backward branch implements the return evolution required by the trace, enforces the SK normalization
$Z[J,J]=1$,
and is responsible for the standard reality and causality properties of the expectation value. Therefore there is no exact algebraic elimination of the $-$ branch that preserves the same Lorentzian one-point function for a generic initial density matrix $\rho_\Omega$ and a generic spacetime-dependent GS deformation $\sigma(x)$.

This statement is not changed by the fact that the operator insertion appears only on the $+$ branch. The $-$ branch is still required to complete the closed-time contour and to provide the conjugate amplitude needed for the trace. Thus one should distinguish carefully between
{inserting the operator only on the $+$ branch}
and {eliminating the $-$ branch altogether}.
The former is standard and exact; the latter is impossible for a generic in--in observable.

There are, however, several useful weaker statements. {\Su First}, if one chooses the final time $t_f$ to coincide with the largest operator time, then one obtains a minimal closed contour. This removes any unnecessary late-time segment, but it does not remove the backward branch itself. {\Su Second}, in special nongeneric limits where the initial density kernel factorizes and the final-time gluing is removed or replaced by fixed data, the backward functional integral may collapse to an overall factor that cancels between numerator and denominator. This is not the physically relevant situation for a normalizable vacuum density matrix of the form
$\rho_\Omega[h_i^+,h_i^-]
=
\Psi_\Omega[h_i^+]\,
\Psi_\Omega^*[h_i^-]$.
{\Su Third}, and most importantly, one may integrate out the $-$ branch \emph{as a definition} and thereby repackage its contribution into an influence functional. This is exact, but it is not the same as removing the $-$ branch physically.

We can follow the third argument and try to integrate out the $-$ branch. This way we can express \eqref{eq:exact_GS_SK_h_full} as an
exact rewriting in terms of an {\it influence functional}. To see this, let us define the influence functional by integrating over the backward branch together with the initial $-$ boundary data:

{\footnotesize
\begin{align}
{\cal F}_\sigma[h_i^+,h_f]
&\equiv
\int Dh_i^-\;
\rho_\Omega[h_i^+,h_i^-]\;
\int_{h_i^-}^{h_f}
{\cal D}g_-\,
{\cal D}(\mathrm{gh}_-)\;
\exp\!\Big(
-iS_{\rm tot}[g_-,\mathrm{gh}_-] + W_\sigma[g_-]
\Big) ,
\label{eq:influence_def_exact}
\end{align}}
which is by construction a function of $h_i^+$ and $h_f$ and is generically a complex function (this will be useful soon).
With this definition the exact one-point function becomes:

{\footnotesize
\begin{align}
\langle \hat g_{\mu\nu}(x)\rangle_\sigma
&=
\frac{
\displaystyle
\int Dh_f
\int Dh_i^+\;
\int_{h_i^+}^{h_f}
{\cal D}{\rm g}_+\;
g_{+\mu\nu}(x)\;
\exp\!\Big(
iS_{\rm tot}[{\rm g}_+]+W_\sigma[g_+]
\Big)\;
{\cal F}_\sigma[h_i^+,h_f]
}{
\displaystyle
\int Dh_f
\int Dh_i^+\;
\int_{h_i^+}^{h_f}
{\cal D}{\rm g}_+\;
\exp\!\Big(
iS_{\rm tot}[{\rm g}_+]+W_\sigma[g_+]
\Big)\;
{\cal F}_\sigma[h_i^+,h_f]
}.
\label{eq:single_plus_with_influence_exact}
\end{align}}
This is an exact rewriting of the original SK expression. However, it must be interpreted correctly. The object
${\cal F}_\sigma[h_i^+,h_f]$
is not a simple multiplicative constant and not a local functional of the bulk $+$ history alone. Rather, it is a complicated nonlocal boundary functional that contains all the information of the backward branch. In other words, the $-$ branch has not disappeared; it has merely been repackaged into
${\cal F}_\sigma[h_i^+,h_f]$.
Thus the precise statement is that the 
exact algebraic elimination of the backward branch is impossible,
but the exact repackaging of the backward branch into an influence functional is certainly possible.

However the repackaging in \eqref{eq:single_plus_with_influence_exact}
raises the following question: how can we show that it is real as we expect 
the exact expectation value of the Hermitian metric operator to be real? In the full SK representation this is automatic, because the numerator is a contraction of a forward amplitude with its conjugate backward amplitude\footnote{The reality can be seen in the following way. In the Schwinger--Keldysh formalism, the insertion of a Hermitian operator on the plus or minus branch gives the same physical expectation value once the contour is closed and normalized. Concretely, define:

{\scriptsize
\bg
&& N_{\sigma,+}(x)
\equiv
\int Dh_f
\int Dh_i^\pm\;
\rho_\Omega[h_i^\pm]\,
\int_{h_i^+}^{h_f}{\cal D}g_+\,
\int_{h_i^-}^{h_f}{\cal D}g_-\,
O[g_+](x)\,
e^{\,iS_{\rm tot}[g_+]-iS_{\rm tot}[g_-]}\,
{\cal I}_+[g_+;\sigma]\,
{\cal I}_-[g_-;\sigma]\nonumber\\
&& N_{\sigma,-}(x)
\equiv
\int Dh_f
\int Dh_i^\pm\;
\rho_\Omega[h_i^\pm]\,
\int_{h_i^+}^{h_f}{\cal D}g_+\,
\int_{h_i^-}^{h_f}{\cal D}g_-\,
O[g_-](x)\,
e^{\,iS_{\rm tot}[g_+]-iS_{\rm tot}[g_-]}\,
{\cal I}_+[g_+;\sigma]\,
{\cal I}_-[g_-;\sigma] ,\nonumber
\nd}
where $Dh_i^\pm = Dh_i^+ Dh_i^-$ and $\rho_\Omega[h_i^\pm] = \rho_\Omega[h_i^+, h_i^-]$.
Then using the fact that we can rename dummy variables by exchanging the two branches as 
$h_i^+\leftrightarrow h_i^-$ and 
$g_+\leftrightarrow g_-$, one may easily show that
$N^\ast_{\sigma,-}(x)=N_{\sigma,+}(x)$.
But for a Hermitian operator in the in-in formalism, the two contour placements are equal, implying that 
$N_{\sigma,+}(x)=N_{\sigma,-}(x)$.
Hence
$N^\ast_\sigma(x)=N_\sigma(x)$ in \eqref{coridahl6}. In a similar vein, using the exchange of the dummy variables, one may show that 
$Z^\ast_\sigma=Z_\sigma$ in \eqref{tricksbrke}. This proves the reality of \eqref{coridahl7}.}. The same reality must persist after rewriting the formula in terms of the influence functional, and it is useful to see explicitly how this happens.

We will first assume that the metric operator is Hermitian, {\it i.e.}
$\hat g_{\mu\nu}^\dagger(x)=\hat g_{\mu\nu}(x)$.
We will also assume that the initial density matrix is Hermitian, {\it i.e.}
$\rho_\Omega[h_i^+,h_i^-]^*
=
\rho_\Omega[h_i^-,h_i^+]$.
Finally we can assume the usual complex-conjugation property of the gauge-fixed Lorentzian path integral with the standard $i\epsilon$ prescription. This means that the forward and backward branch kernels are conjugates of one another. To make this explicit, let us define the branch kernels without insertion as:
\bg  \label{eq:Kminus_def_exact}
&&{\cal K}_{+,\sigma}[h_f,h_i^+]
\equiv
\int_{h_i^+}^{h_f}
{\cal D}g_+\,
{\cal D}(\mathrm{gh}_+)\;
\exp\!\Big(
iS_{\rm tot}[g_+,\mathrm{gh}_+]+W_\sigma[g_+]
\Big) \nonumber\\
&&{\cal K}_{-,\sigma}[h_f,h_i^-]
\equiv
\int_{h_i^-}^{h_f}
{\cal D}g_-\,
{\cal D}(\mathrm{gh}_-)\;
\exp\!\Big(
-iS_{\rm tot}[g_-,\mathrm{gh}_-]+ W_\sigma[g_-]
\Big).
\nd
Then one has
${\cal K}_{-,\sigma}[h_f,h_i]
=
\Big(
{\cal K}_{+,\sigma}[h_f,h_i]
\Big)^*$, with $+ \leftrightarrow -$, implying that the two branch kernels are complex conjugates of each other.
In terms of these kernels the influence functional is simply given by 
$\mathcal K_{-,\sigma}[h_f, h_i^-]$ as:
\begin{equation}
{\cal F}_\sigma[h_i^+,h_f]
=
\int Dh_i^-\;
\rho_\Omega[h_i^+,h_i^-]\;
{\cal K}_{-,\sigma}[h_f,h_i^-] ,
\label{eq:F_as_contraction_exact}
\end{equation}
which, as we pointed out earlier, is a complex function. The usefulness of this observation appears when we take the complex conjugate of \eqref{eq:F_as_contraction_exact}. Using the fact that the two branch kernels in \eqref{eq:Kminus_def_exact} are complex conjugates of each other, gives:
\begin{align}
{\cal F}_\sigma[h_i^+,h_f]^*
&=
\int Dh_i^-\;
\rho_\Omega[h_i^+,h_i^-]^*\;
{\cal K}_{-,\sigma}[h_f,h_i^-]^*
\nonumber\\
&=
\int Dh_i^-\;
\rho_\Omega[h_i^-,h_i^+]\;
{\cal K}_{+,\sigma}[h_f,h_i^-].
\label{eq:F_conj_explicit_exact}
\end{align}
Thus ${\cal F}_\sigma[h_i^+, h_f]$ is not arbitrary. It is precisely the object that supplies the conjugate branch contribution required by the closed-time-path construction. If we now define the forward kernel with metric insertion:
\begin{equation}
{\cal K}_{+\mu\nu,\sigma}[h_f,h_i^+;x]
\equiv
\int_{h_i^+}^{h_f}
{\cal D}g_+\,
{\cal D}(\mathrm{gh}_+)\;
g_{+\mu\nu}(x)\;
\exp\!\Big(
iS_{\rm tot}[g_+,\mathrm{gh}_+]+W_\sigma[g_+]
\Big) ,
\label{eq:Kplus_insert_def_exact}
\end{equation}
which is possible because we have defined the metric insertion only on the $+$ branch; and $W_\sigma[g_+]$ from \eqref{eq:W_sigma_def_exact} also in the same branch.
Then the numerator and denominator of \eqref{eq:single_plus_with_influence_exact} may be written in the following way:
\begin{align}
{\cal D}[x]
&=
\int Dh_f
\int Dh_i^+\;
{\cal F}_\sigma[h_i^+,h_f]\;
{\cal K}_{+,\sigma}[h_f,h_i^+]\nonumber\\[4pt]
{\cal N}_{\mu\nu}[x]
&=
\int Dh_f
\int Dh_i^+\;
{\cal F}_\sigma[h_i^+,h_f]\;
{\cal K}_{+\mu\nu,\sigma}[h_f,h_i^+;x],
\label{eq:N_K_form_exact}
\end{align}
Taking complex conjugates and using the Hermiticity of the metric insertion together with the conjugation properties above reproduces the original SK complex-conjugated expression. Since the original expectation value is that of a Hermitian operator in a Hermitian density matrix, one obtains:
\begin{equation}
{\cal N}^*={\cal N},
\qquad
{\cal D}^*={\cal D},
\qquad
\Longrightarrow
\qquad
\langle \hat g_{\mu\nu}(x)\rangle_\sigma\in\mathbb R.
\label{eq:reality_ratio_exact}
\end{equation}
This is the correct functional-integral version of the familiar statement that an in--in expectation value is a sum of amplitudes times conjugate amplitudes.

It is important to phrase this carefully. One should \emph{not} say that the numerator is literally a real insertion multiplied by a single positive measure of the form
$|\mathcal A|^2$. That would be too naive, because the insertion changes the forward kernel. The correct statement is instead that the full numerator is a contraction of an inserted forward kernel with the conjugate branch structure encoded in the influence functional. This is the functional analogue of the ordinary quantum-mechanical structure:
\begin{equation}
\sum_f A_f^{(O)}\,A_f^*,
\end{equation}
rather than a bare
$\sum\limits_f |A_f|^2$. Therefore keeping in mind the trans-series and thimble analysis, the essential conclusions may therefore be summarized as follows. {\Su First}, the exact GS one-point function is given by the closed-time-path expression \eqref{eq:exact_GS_SK_h_full}, in which the GS operator appears as a branch-dependent bulk reweighting through:
\begin{equation}
W_\sigma[g_+] + W_\sigma[g_-].
\end{equation}
{\Su Second}, there is no exact removal of the $-$ branch for a genuine in--in one-point function. The best exact reformulation is the influence-functional representation \eqref{eq:single_plus_with_influence_exact}. {\Su Third}, the reality of the resulting one-point function follows from:
\begin{equation}
\rho_\Omega^\dagger=\rho_\Omega,
\qquad
\hat g_{\mu\nu}^\dagger=\hat g_{\mu\nu},
\qquad
{\cal K}_{-,\sigma}=
{\cal K}_{+,\sigma}^*
\end{equation}
with the standard contour and $i\epsilon$ prescription. Therefore the exact Lorentzian GS one-point function satisfies
$\langle \hat g_{\mu\nu}(x)\rangle_\sigma
\in
\mathbb R$.
Finally, this discussion fits naturally with the trans-series viewpoint adopted earlier. The exact action
$S_{\rm tot}[g,\mathrm{gh}]$
already contains the perturbative and nonperturbative sectors of the gauge-fixed infrared theory, while the closed-time-path structure ensures the correct real-time interpretation of the expectation value. Thus the GS one-point function in Lorentzian signature is governed simultaneously by two exact structures: (1)
{the trans-series completion of the effective action}; and (2)
{the forward/backward SK contour required by in--in dynamics.}
The former determines the dynamical sector content of the theory, while the latter determines how those sectors contribute to a physically meaningful real-time expectation value.

\subsubsection{In-out expectation value from the in-in SK path integral \label{sec6.5.2}}

The expectation value derived in \eqref{eq:single_plus_with_influence_exact} only has the $+$ branch and resembles somewhat the expectation value studied in {\gspapers} but differs crucially by the presence of the influence function $\mathcal F_\sigma[h_i^+, h_f]$ which is a complex function.
Thus this is not exactly the in--out picture that we have in mind. 

Instead our aim is to present the following picture. 
Suppose that the support of the GS operator \eqref{vandikok} lies between
$t_0$ and $t_1$, and consider an insertion time $t$ with
$t_0<t<t_1$. Then the unnormalized matrix element:
\begin{equation}
\langle\Omega|\,D^\dagger(\sigma)\,\hat{\mathcal O}_H({\bf x}, t)\,D(\sigma)\,|\Omega\rangle
\end{equation}
has the same local sequence
$t_0\to t\to t_1\to t_0$
that may be embedded either into a closed contour or into an
in--out contour by appropriately choosing the out vacuum. For example, 
if one chooses the in--out completion:
\begin{equation}
|\Omega\rangle_{\rm in}
=
U(t_0,-T)\,|0\rangle,
\qquad
{}_{\rm out}\langle\Omega|
=
\langle 0|\,U(+T,t_0),
\label{inout_completion_local}
\end{equation}
then one may define the corresponding in--out numerator and denominator
by $N_\sigma^{\rm in\text{-}out}({\bf x}, t)$ and $Z_\sigma^{\rm in\text{-}out}({\bf x}, t)$ respectively; and using them rewrite the expectation value formally in the following way:
\begin{equation}
\langle \hat{\mathcal O}_H({\bf x}, t)\rangle_\sigma^{\rm in\text{-}out}
=
\frac{
N_\sigma^{\rm in\text{-}out}({\bf x}, t)
}{
Z_\sigma^{\rm in\text{-}out}
} \equiv  \frac{{}_{\rm out}\langle\Omega|\,
D^\dagger(\sigma)\,\hat{\mathcal O}_H({\bf x}, t)\,D(\sigma)\,
|\Omega\rangle_{\rm in}}{{}_{\rm out}\langle\Omega|\,
D^\dagger(\sigma)\,D(\sigma)\,
|\Omega\rangle_{\rm in}} .
\label{inout_ratio_local}
\end{equation}
It is important to emphasize that \eqref{inout_ratio_local} is an
\emph{in--out contour completion} of the same operator structure.
It is not, in general, identical to the exact in--in expectation value
\eqref{exact_inin_GS_ratio_local}. The latter is the genuine physical
GS expectation value. The former is a different contour completion of
the same local operator insertion. We will discuss the possible equivalence in section \ref{sec6.5.3}.

We can now use \eqref{inout_ratio_local} to derive the 
single-contour in--out path-integral representation. We start by 
introducing the asymptotic vacuum wavefunctionals:
\begin{equation}
\Psi_0(g_i,-T)=\langle g_i,-T|0\rangle,
\qquad
\Psi_0^\ast(g_f,+T)=\langle 0|g_f,+T\rangle ,
\end{equation}
and derive the path-integral form for the numerator and the denominator of \eqref{inout_ratio_local} separately. 
We begin with the in--out matrix element $N_\sigma^{\rm in\text{-}out}({\bf x}, t)$ from the numerator of \eqref{inout_ratio_local}, {\it i.e.}
$N_\sigma^{\rm in\text{-}out}({\bf x}, t)
=
{}_{\rm out}\langle \Omega|
\,D^\dagger(\sigma)\,\hat{\mathcal O}_H({\bf x}, t)\,D(\sigma)\,
|\Omega\rangle_{\rm in}$.
Here the asymptotic in/out vacua are taken to be the ones from 
\eqref{inout_completion_local},
so that $N_\sigma^{\rm in\text{-}out}({\bf x}, t)$ becomes:
\begin{equation}
N_\sigma^{\rm in\text{-}out}({\bf x}, t)
=
\langle 0|\,
U(+T,t_0)\,
D^\dagger(\sigma)\,\hat{\mathcal O}_H({\bf x}, t)\,D(\sigma)\,
U(t_0,-T)\,
|0\rangle ,
\label{eq:Ninout_with_U}
\end{equation}
where $|0\rangle$ is the free vacuum for a given choice of Minkowski minimum as discussed in section \ref{sec5}. The next step is to insert complete sets of field eigenstates at the
initial and final times.  Writing
$\int {\cal D}g_f\;
|g_f,+T\rangle\langle g_f,+T| = \mathbf{1}$ and 
$\int {\cal D}g_i\;
|g_i,-T\rangle\langle g_i,-T| = \mathbf{1}$
we obtain:

{\footnotesize
\bg\label{mzdanirupor}
N_\sigma^{\rm in\text{-}out}({\bf x}, t)
&= &
\int {\cal D}g_i\,{\cal D}g_f\;
\langle \Omega|g_f,+T\rangle\,
\langle g_f,+T|
D^\dagger(\sigma)\,\hat{\mathcal O}_H({\bf x}, t)\,D(\sigma)
|g_i,-T\rangle\,
\langle g_i,-T|\Omega\rangle \nonumber\\
&=&
\int {\cal D}g_i\,{\cal D}g_f\;
\Psi_\Omega^\ast(g_f,+T)\,
\Psi_\Omega(g_i,-T)\,
\langle g_f,+T|
D^\dagger(\sigma)\,\hat{\mathcal O}_H({\bf x}, t)\,D(\sigma)
|g_i,-T\rangle,
\nd}
where we defined the asymptotic vacuum wavefunctionals in the standard way using $|\Omega\rangle$ as
$\Psi_\Omega(g_i,-T)
\equiv
\langle g_i,-T|\Omega\rangle$ and 
$\Psi_\Omega^\ast(g_f,+T)
\equiv
\langle \Omega|g_f,+T\rangle$. Notice that we have traded the free vacuum $|0\rangle$ with the interacting vacuum $|\Omega\rangle$ but kept the information about the asymptotic boundaries using different bra and ket (this way we can avoid using the subscript ``in" and ``out"). In fact that is all there is to it because now everything reduces to understanding the kernel:
\begin{equation}
K_\sigma[g_f,+T;g_i,-T]
\equiv
\langle g_f,+T|
D^\dagger(\sigma)\,\hat{\mathcal O}_H({\bf x}, t)\,D(\sigma)
|g_i,-T\rangle ,
\label{eq:kernel_def}
\end{equation}
in \eqref{mzdanirupor}. The story now is slightly different from what we encountered earlier when using the $+$ and $-$ branches of the SK contour because we now expect both $D(\sigma)$ and $D^\dagger(\sigma)$ to lie on the same branch. To quantify the behavior, suppose the full contour runs from $-T$ to $+T$, and choose a time-slicing:
\begin{equation}
-T=t_{(0)}<t_{(1)}<\cdots<t_{(N)}=+T.
\end{equation}
Then the time-ordered exponential \eqref{vandikok} and its Hermitian conjugate are represented as the limit of an ordered product over only those slices lying inside the support of the source as the following time-ordered products of the operators:
\bg\label{cocovan2}
&&D(\sigma) =
\prod_{t_{(k)}\in[t_0,t_1]}^{\leftarrow}
\exp\!\left(
\Delta t\int d^3x\;
\sqrt{-g(t_{(k)},\mathbf x)}\,
\sigma^{\alpha\beta}(t_{(k)},\mathbf x)\,
g_{\alpha\beta}(t_{(k)},\mathbf x)
\right) \nonumber\\
&& D^\dagger(\sigma) =
\prod_{t_{(k)}\in[t_0,t_1]}^{\rightarrow}
\exp\!\left(
\Delta t\int d^3x\;
\sqrt{-g(t_{(k)},\mathbf x)}\,
\sigma^{\alpha\beta}(t_{(k)},\mathbf x)\,
g_{\alpha\beta}(t_{(k)},\mathbf x)
\right),
\nd
which is what we studied earlier in \eqref{eq:Dt_sigma_again} and in section \ref{sec3.2}. The arrows in \eqref{cocovan2} indicate opposite orderings, corresponding respectively to $\mathcal T$ and $\bar{\mathcal T}$. Thus the insertion
$D^\dagger(\sigma)\,\hat{\mathcal O}_H({\bf x}, t)\,D(\sigma)$
is not concentrated at a single time slice. It is an extended ordered insertion spread over the interval $[t_0,t_1]$, with $\hat{\mathcal O}_H({\bf x}, t)$ inserted at its own time $t$.  The support matters because the relative position of $t$ with respect to $[t_0,t_1]$ determines whether the metric operator lies: (i) before the support of $D(\sigma)$, (ii) inside the support of $D(\sigma)$,
or (iii) after the support of $D(\sigma)$. These cases are different.

\paragraph{Case 1: $t<t_0$.}

If the insertion time lies before the support of the GS operator, then $\hat{\mathcal O}_H({\bf x}, t)$ is earlier than all factors entering both $D(\sigma)$ and $D^\dagger(\sigma)$. In the time-sliced product, the operator sits entirely to one side of the GS factors. In this case the ordered structure simplifies considerably.

\paragraph{Case 2: $t_0<t<t_1$.}

If the insertion time lies inside the support of $D(\sigma)$, then some of the GS factors occur at earlier times than $t$, while others occur at later times than $t$. In that case the operator is sandwiched nontrivially inside the ordered exponential, and one must keep the full ordered expression. This is the most delicate case.

\paragraph{Case 3: $t>t_1$.}

If the insertion time lies after the support of the GS operator, then $\hat{\mathcal O}_H({\bf x}, t)$ is later than all the factors in $D(\sigma)$ and $D^\dagger(\sigma)$, and the ordered structure again simplifies.

This means that the support of $D(\sigma)$ matters because it controls the relative time-ordering between 
$\hat{\mathcal O}_H({\bf x}, t)$ and the GS insertion. However, the key point is to identify what remains true independently of the support. In particular, what does \emph{not} change is the formal existence of the single-contour kernel representation:
\begin{equation}
\langle g_f,+T|
D^\dagger(\sigma)\,\hat{\mathcal O}_H({\bf x}, t)\,D(\sigma)
|g_i,-T\rangle
=
\int_{g(-T)=g_i}^{g(+T)=g_f}
{\cal D}{\rm g}\;
e^{\,iS_{\rm tot}[{\rm g}]}\,
\Big[
D^\dagger(\sigma)\,{\mathcal O}_H({\bf x}, t)\,D(\sigma)
\Big]_{\rm g} ,
\label{eq:single_contour_still_true}
\end{equation}
providing the path-integral representation of the kernel \eqref{eq:kernel_def}, and ${\rm g}(x)$ is as usual defined as ${\rm g}(x) = (g(x), {\rm gh}(x), {\rm matter}(x))$ although we are typically suppressing the matter part.
This formula is always valid as a time-sliced operator/path-integral identity. What changes is only the detailed content of following operator insertion:
\begin{equation}\label{emmagulab}
\Big[
D^\dagger(\sigma)\,\mathcal O_H({\bf x}, t)\,D(\sigma)
\Big]_{\rm g} ,
\end{equation}
where the subscript means that the operator insertion is first represented in the usual time-sliced path-integral construction, and is then evaluated on the same c-number history
${\rm g}(x) \equiv (g(x), {\rm gh}(x), {\rm matter}(x))$
over which the functional integral is performed. The notation \eqref{emmagulab} should be interpreted carefully.  It does {\it not} mean that the
operator expression has somehow been replaced by an arbitrary function
by hand.  Rather, it is a shorthand for the result of the standard
time-sliced insertion prescription in the single-history path integral.
In particular, if $D(\sigma)$ and $D^\dagger(\sigma)$ denote respectively $\mathcal T$ and $\hat{\mathcal T}$ ordered product in \eqref{vandikok}, then for a purely multiplicative insertion such as
$\hat{\mathcal O}_H({\bf x}, t)=\hat g_{\mu\nu}({\bf x}, t)$,
the bracket notation corresponds schematically to\footnote{Let us make the following clean notational choice. The bracket
$[\cdots]_{\rm g}$
denotes the c-number insertion obtained from the corresponding operator expression after time slicing and evaluation on the field history ${\rm g}(x)$. This way no hats appear on the RHS of \eqref{eq:single_contour_still_true}. \label{conjugvis}}:
\begin{equation}
\Big[
D^\dagger(\sigma)\, g_{\mu\nu}(t)\,D(\sigma)
\Big]_{\rm g}
=
\bar{\mathcal T}\exp\!\left(
W_\sigma[g]
\right)\,
g_{\mu\nu}(t)\,
\mathcal T\exp\!\left(
W_\sigma[g]
\right),
\label{eq:bracket_metric_explicit}
\end{equation}
where $W_\sigma[g]$ is given by \eqref{eq:W_sigma_def_exact}.
If the insertion is more complicated and involves canonical momenta,
time derivatives, or composite operators, then the bracket notation
must be understood with the appropriate discretization prescription.
In such cases contact terms and ordering subtleties can arise, so one
should keep the ordered operator form until the insertion has been
treated carefully. After the dust settles, the numerator \eqref{mzdanirupor} and the denominator from \eqref{inout_ratio_local}, may be succinctly presented as:

{\footnotesize
\bg\label{jailmomvis}
&& Z_\sigma^{\rm in\text{-}out}
=
\int {\cal D}g_i\,{\cal D}g_f\;
\Psi_\Omega^\ast(g_f,+T)\,
\Psi_\Omega(g_i,-T)
\int_{g(-T)=g_i}^{g(+T)=g_f}
{\cal D}{\rm g}\;
\exp\!\left(
iS_{\rm tot}[{\rm g}]
\right) \Big[
D^\dagger(\sigma)\,D(\sigma)
\Big]_{\rm g} \\
&& N_\sigma^{\rm in\text{-}out}({\bf x}, t)
=
\int {\cal D}g_i\,{\cal D}g_f\;
\Psi_\Omega^\ast(g_f,+T)\,
\Psi_\Omega(g_i,-T)
\int_{g(-T)=g_i}^{g(+T)=g_f}
{\cal D}{\rm g} \; \exp\!\left(
iS_{\rm tot}[{\rm g}]\right)
\Big[
D^\dagger(\sigma)\,\mathcal O_H({\bf x}, t)\,D(\sigma)
\Big]_{\rm g}, \nonumber \nd}
where ${\rm g} = (g, {\rm gh}, {\rm matter})$ and the contour is from $-T$ to $+T$ with $-T < t < +T$. The bracket $[\cdots]_{\rm g}$ has the same interpretation as in footnote \ref{conjugvis}. Plugging \eqref{jailmomvis} in \eqref{inout_ratio_local},
the formal in--out contour completion of the GS ratio takes the form:

{\footnotesize
\begin{equation}
\begin{aligned}
\langle \hat{\mathcal O}_H({\bf x}, t)\rangle_\sigma^{\rm in\text{-}out}
&=
\frac{
\int {\cal D}g_i\,{\cal D}g_f\;
\Psi_\Omega^\ast(g_f,+T)\,
\Psi_\Omega(g_i,-T)
\int_{g(-T)=g_i}^{g(+T)=g_f}
{\cal D}{\rm g}\;
e^{\,iS_{\rm tot}[{\rm g}]}
\,
\bar{\mathcal T} e^{W_\sigma[g]}
\,
\mathcal O_H({\bf x}, t)
\,
\mathcal T e^{W_\sigma[g]}
}{
\int {\cal D}g_i\,{\cal D}g_f\;
\Psi_\Omega^\ast(g_f,+T)\,
\Psi_\Omega(g_i,-T)
\int_{g(-T)=g_i}^{g(+T)=g_f}
{\cal D}{\rm g}\;
e^{\,iS_{\rm tot}[{\rm g}]}
\,
\bar{\mathcal T} e^{W_\sigma[g]}
\,
\mathcal T e^{W_\sigma[g]}
},
\end{aligned}
\label{ratio_inout_compact_local}
\end{equation}}
where $W_\sigma[g]$ is given by \eqref{eq:W_sigma_def_exact}. This is the generic result for the expectation value of any local operator
$\hat{\mathcal O}_H({\bf x}, t)$ over a given GS state, and is represented by a single contour going from $-T$ to $+T$.

There is however a deeper level of subtlety when implementing operator product \eqref{emmagulab} to the path-integral \eqref{ratio_inout_compact_local} via \eqref{eq:bracket_metric_explicit}. The question is when can the exponentials be combined? This is where the support matters most.
If the operator insertion is purely multiplicative and its time lies completely outside the support of $D(\sigma)$, then one may often combine the two exponentials into a single factor:
\begin{equation}
\bar{\mathcal T}\exp\!\left(W_\sigma[\hat{g}]\right)\,
\hat{\mathcal O}_H({\bf x}, t)\,
\mathcal T\exp\!\left(W_\sigma[\hat{g}]\right)
\;\longrightarrow\;
\hat{\mathcal O}_H({\bf x}, t)\,
\exp\!\left(
2W_\sigma[\hat{g}]
\right),
\label{eq:combine_exponentials_safe}
\end{equation}
with
$W_\sigma[\hat{g}]
=
\int_{t_0}^{t_1} dt\int d^3x\;
\sqrt{-\hat{g}}\,
\sigma^{\alpha\beta}\hat{g}_{\alpha\beta}$, and $t > t_1$. On the other hand, if the insertion time satisfies
$t_0 < t < t_1$,
then the operator $\hat{\mathcal O}_H({\bf x}, t)$
sits inside the support of the time-ordered and anti-time-ordered exponentials. In that situation, the operator identity  \eqref{eq:combine_exponentials_safe}
is in general \emph{not} justified, because operator ordering and possible contact terms matter. So at the operator level one should keep the ordered form:
\begin{equation}
D^\dagger(\sigma)\,\hat{\mathcal O}_H({\bf x}, t)\,D(\sigma)
=
\bar{\mathcal T}\exp\!\left(W_\sigma[\hat g]\right)\,
\hat{\mathcal O}_H({\bf x}, t)\,
\mathcal T\exp\!\left(W_\sigma[\hat g]\right) ,
\label{eq:ordered_operator_true}
\end{equation}
which is the safest procedure in the language of operators. However
once the matrix element has been converted into a path integral over a single c-number history ${\rm g}(x)$, the situation changes. One then deals with an insertion of the form \eqref{emmagulab}
which is the c-number functional obtained from the operator insertion after time slicing and evaluation on the path-integral history ${\rm g}(x)$. If the insertion is \emph{purely multiplicative}, meaning that it depends only on the field values on the history and involves no canonical momenta, no extra time derivatives requiring discretization choices, and no singular composite-operator renormalization subtleties, then inside the path integral the two exponentials do indeed become ordinary commuting functions of the same history $g$. In that case one may write

{\footnotesize
\begin{equation}
\Big[
D^\dagger(\sigma)\,\hat{\mathcal O}_H({\bf x}, t)(t)\,D(\sigma)
\Big]_{\rm g} \xrightarrow{\rm PI} 
\bar{\mathcal T}\exp\!\left(W_\sigma[g]\right)\,
\mathcal O_H({\bf x}, t)\,
\mathcal T\exp\!\left(W_\sigma[g]\right) = \mathcal O_H({\bf x}, t)\,
\exp\!\left(2W_\sigma[g]\right),
\label{eq:PI_insertion_multiplicative}
\end{equation}}
since all factors on the right are now c-number functionals of the same history, they commute. Therefore at the level of the c-number path-integral integrand, and for a purely multiplicative insertion, once the operator matrix element has been converted to a single-history path integral, the two GS exponentials may be combined into $\exp\left({2W_\sigma[g]}\right)$.
even if the time $t$ lies in the interval
$t_0<t<t_1$. Accordingly:

{\footnotesize
\begin{equation}
\begin{aligned}
\langle \hat{\mathcal O}_H({\bf x}, t)\rangle_\sigma^{\rm in\text{-}out}
&=
\frac{
\int {\cal D}g_i\,{\cal D}g_f\;
\Psi_\Omega^\ast(g_f,+T)\,
\Psi_\Omega(g_i,-T)
\int_{g(-T)=g_i}^{g(+T)=g_f}
{\cal D}{\rm g}\;
\exp\!\left(
iS_{\rm tot}[{\rm g}]+2W_\sigma[g]
\right)
\mathcal O_H({\bf x}, t)
}{
\int {\cal D}g_i\,{\cal D}g_f\;
\Psi_\Omega^\ast(g_f,+T)\,
\Psi_\Omega(g_i,-T)
\int_{g(-T)=g_i}^{g(+T)=g_f}
{\cal D}{\rm g}\;
\exp\!\left(
iS_{\rm tot}[{\rm g}]+2W_\sigma[g]
\right)
} ,
\end{aligned}
\label{inout_ratio_combined_local}
\end{equation}}
giving us the final simplified expression for the in--out expectation value from the generic result in \eqref{ratio_inout_compact_local}, where
$W_\sigma[g]$ is given by \eqref{eq:W_sigma_def_exact} with $t_i = t_0$ and $t_f = t_1$
However, this simplification should be used with care. It is reliable
only when $\mathcal{O}_H({\bf x}, t)$ is a purely multiplicative field
insertion. If the inserted operator contains canonical momenta, time
derivatives, or composite operators that require regularization, then
the explicit $\mathcal T$ and $\bar{\mathcal T}$ ordering should be retained.

\subsubsection{Single contour representation of the in-out kernel revisited \label{sec6.5.3}}

The kernel \eqref{eq:kernel_def}, while leads to a reasonably consistent in-out path-integral structure for the matrix element, hides the actual temporal evolution of a state from the initial Cauchy slice at $-T$ to the final Cauchy slice at $+T$. 

We now give a careful derivation of the in--out matrix element in a notation that keeps the operator ordering explicit and does not truncate the displacement operators at the insertion time. The key point is that the source-supported evolution contains a forward excursion from $t_r$ to $t_s$ generated by $D(\sigma)$ and a backward excursion from $t_s$ to $t_r$ generated by $D^\dagger(\sigma)$, with the local operator inserted at an intermediate time
$t_r<t_k<t_s$. We take the in--out numerator to be:
\bg\label{ivanakhuleche}
N_\sigma^{\rm in\text{-}out}({\bf x},t_k)
 &= &
\langle 0|\,
U(+T,t_r)\,
D^\dagger_{[t_r,t_s]}(\sigma)\,
\hat{\mathcal O}_H({\bf x},t_k)\,
D_{[t_r,t_s]}(\sigma)\,
U(t_r,-T)\,
|0\rangle \nonumber\\
 & = &
\int Dh_f\,Dh_i\;
\Psi_0^\ast[h_f,+T]\,
\Psi_0[h_i,-T]\,
{\cal M}[h_f,h_i], \nd
where $|0\rangle$ is the free Minkowski vacuum. In going from first line to the second in \eqref{ivanakhuleche} we have 
inserted complete bases at the outer endpoints using
$\mathbf 1_{\Sigma_{+T}}
=
\int Dh_f\;
|h_f,+T\rangle\langle h_f,+T|$,
and 
$\mathbf 1_{\Sigma_{-T}}
=
\int Dh_i\;
|h_i,-T\rangle\langle h_i,-T|$. Such a procedure produces two vacuum wavefunctionals from the two endpoints that we define as 
$\Psi_0^\ast[h_f,+T]
\equiv
\langle 0|h_f,+T\rangle$, and 
$\Psi_0[h_i,-T]
\equiv
\langle h_i,-T|0\rangle$. Once these wavefunctionals are isolated, the remain piece is the amplitude ${\cal M}[h_f,h_i]$ that we define in the following way:
\begin{equation}
{\cal M}[h_f,h_i]
\equiv
\langle h_f,+T|\,
U(+T,t_r)\,
D^\dagger_{[t_r,t_s]}(\sigma)\,
\hat{\mathcal O}_H({\bf x},t_k)\,
D_{[t_r,t_s]}(\sigma)\,
U(t_r,-T)\,
|h_i,-T\rangle.
\label{eq:M_definition}
\end{equation}
The question that we want to ask now is whether we can simplify the amplitude ${\cal M}[h_f,h_i]$ to get a handle on the in-out contour. For this, a first step would be to have a proper 
time-slicing strategy and full definitions of $D$ and $D^\dagger$. We can start by discretizing the source-supported interval $[t_r,t_s]$ as
$t_n=t_r+n\Delta t$ such that 
$\Delta t=\frac{t_s-t_r}{N}$, where 
$n=0,1,\dots,N$
with
$t_0=t_r,
t_N=t_s$, 
$t_k=t_r+k\Delta t$ and 
$0<k<N$. Using our slicing strategy, we can now define three parameters 
$U_n\equiv U(t_{n+1},t_n),
J_n\equiv \hat J_S(t_r;t_n)$ and 
$J_n^\dagger\equiv \hat J_S^\dagger(t_r;t_n)$, such that they may be used to define two important functions:
\begin{equation}
B_n\equiv U_n\,e^{\Delta t\,J_n},
\qquad
A_n\equiv e^{\Delta t\,J_n^\dagger}\,U_n^\dagger ,
\label{eq:An_Bn_defs}
\end{equation}
which one may recognize as the functions that enter the definitions of $D[\sigma]$ and $D^\dagger[\sigma]$ respectively once we express them as multiple kicked operators. Saying differently, we can express 
the full source-supported displacement operators in a way where the split at the insertion slice $t_k$ become visible. In other words, we express the displacement operators in the following suggestive way:
\bg\label{ennakoigelo}
&& D_{[t_r,t_s]}(\sigma) =D_{\ge}D_{<}
=
\underbrace{B_{N-1}B_{N-2}\cdots B_k}_{D_{\ge}}\underbrace{B_{k-1} B_{k-2} \cdots B_0}_{D_<} \nonumber\\
&&  D^\dagger_{[t_r,t_s]}(\sigma) = D_{<}^\dagger D_{\ge}^\dagger
=
\underbrace{A_0A_1\cdots A_{k-2} A_{k-1}}_{D^\dagger_<} \underbrace{A_k A_{k+1} \cdots A_{N-2}A_{N-1}}_{D^\dagger_{\ge}} ,
\nd
with no $(A_N, B_N)$ factors because the short-time operators are indexed by the \emph{intervals} between slices, not by the slices themselves. Note that 
these definitions are exact and do not truncate the full operators; they only factorize them. The split at $t_k$ is made so as to insert the source operator there as we shall soon see. Meanwhile, using 
\eqref{ennakoigelo}, the matrix element \eqref{eq:M_definition} takes the form:
\begin{equation}
{\cal M}[h_f,h_i]
=
\langle h_f,+T|\,
U(+T,t_r)\,
D_{<}^\dagger D_{\ge}^\dagger\,
\hat{\mathcal O}_H({\bf x},t_k)\,
D_{\ge} D_{<}\,
U(t_r,-T)\,
|h_i,-T\rangle ,
\label{eq:M_with_split_Ds}
\end{equation}
which is not the full story because we haven't placed the source operator $\hat{\mathcal O}_H({\bf x},t_k)$ in an appropriate time-ordered form yet. To perform the task efficiently, we will
insert complete sets at the times $t_r$, $t_s$, and on both sides of the operator at $t_k$ using:
\begin{equation}
\mathbf 1_{\Sigma_{t_r}}
=
\int Dh_r^-\;
|h_r^-,t_r\rangle\langle h_r^-,t_r|,
\qquad
\mathbf 1_{\Sigma_{t_r}}
=
\int Dh_r^+\;
|h_r^+,t_r\rangle\langle h_r^+,t_r|\nonumber
\label{eq:id_tr_pm}
\end{equation}
\begin{equation}
\mathbf 1_{\Sigma_{t_s}}
=
\int Dh_s\;
|h_s,t_s\rangle\langle h_s,t_s|\nonumber
\label{eq:id_ts}
\end{equation}
\begin{equation}
~~~~~~~~~\mathbf 1_{\Sigma_{t_k}}
=
\int Dh_k^-\;
|h_k^-,t_k\rangle\langle h_k^-,t_k|,
\qquad
\mathbf 1_{\Sigma_{t_k}}
=
\int Dh_k^+\;
|h_k^+,t_k\rangle\langle h_k^+,t_k|.
\label{eq:id_tk_pm}
\end{equation}
Before proceeding however we need a 
few more details associated with time-ordering operators. We discussed this briefly in the footnote \ref{shirtagra}, and here we will elaborate this according to our need. First, recall that 
if the displacement operator is defined as a time-ordered exponential as in \eqref{lenlil01} but now the limit goes from $t_r$ to $t_s$,
then:
\begin{equation}
\mathcal T\!\left\{
D[\sigma]\,
\hat{\mathcal O}_H({\bf x},t_k)
\right\} = 
\mathcal T\!\left\{
\exp\!\left(
\int_{t_r}^{t_s} dt'\; \hat J_H(t')
\right)
\hat{\mathcal O}_H({\bf x},t_k)
\right\}  ,
\label{eq:T_D_O_basic_reply}
\end{equation}
implying that any operator defined on a Cauchy slice at $t_k$, when conjugated with the displacement operator can be time-ordered in the way shown above because of how they are arranged in the forward branch. In other words, 
this means that the operator
$\hat{\mathcal O}_H({\bf x},t_k)$
is time-ordered together with all insertions of $\hat J_H(t')$ appearing inside the exponential.
Expanding the exponential, one gets the following expression:
\begin{equation}
\begin{aligned}
\mathcal T\!\left\{
D[\sigma]\,
\hat{\mathcal O}_H({\bf x},t_k)
\right\}
&=
\hat{\mathcal O}_H({\bf x},t_k)
+
\int_{t_r}^{t_s}dt_1\;
\mathcal T\!\left\{
\hat J_H(t_1)\,
\hat{\mathcal O}_H({\bf x},t_k)
\right\}
\\
&\qquad
+
\frac{1}{2!}
\int_{t_r}^{t_s}dt_1dt_2\;
\mathcal T\!\left\{
\hat J_H(t_1)\hat J_H(t_2)\,
\hat{\mathcal O}_H({\bf x},t_k)
\right\}
+\cdots ,
\end{aligned}
\label{eq:T_D_O_expanded_reply}
\end{equation}
which is similar to what we had in footnote \ref{shirtagra}. Note that 
the simplest way to understand the time-ordering in the first equality of \eqref{eq:T_D_O_basic_reply} is that the time-ordering is not created by a mysterious extra step; it is already built into the definition of the displacement operator itself. The operation 
${\cal T}$ orders everything according to the time labels. Since $\hat O_H({\bf x},t_k)$ has the definite time label $t_k$, its place inside the ordered product is fixed uniquely, {\it i.e.}
all operators at times $t'>t_k$ go to the left of $\hat O_H({\bf x},t_k)$, and 
all operators at times $t'<t_k$ go to the right. In other words, a useful explicit form is:
\begin{equation}
\mathcal T\!\left\{
D[\sigma]\,
\hat{\mathcal O}_H({\bf x},t_k)
\right\}
=
\left[
\mathcal T\exp\!\left(
\int_{t_k}^{t_s}dt'\;\hat J_H(t')
\right)
\right]
\hat{\mathcal O}_H({\bf x},t_k)
\left[
\mathcal T\exp\!\left(
\int_{t_r}^{t_k}dt'\;\hat J_H(t')
\right)
\right],
\label{eq:T_D_O_split_reply}
\end{equation}
up to the usual convention about operators exactly at equal time $t_k$. This simply means that $\hat{\mathcal O}_H({\bf x},t_k)$ is inserted at its proper place in the time-ordered string. This shows that 
conjugation by the displacement operator does not create time-ordering anew; it only reveals the ordering that was already built into ${\cal T}$ or $\bar{\cal T}$ from the start.

The above formalism still doesn't say exactly what happens at the Cauchy slice $t_k$. If the insertion at $t_k$ is included in the source support, then we can rewrite \eqref{eq:T_D_O_split_reply} using 
$D[\sigma]
=
D_{>}D_kD_{<}$ as the following time-ordered product:
\begin{equation}
\mathcal T\!\left\{
D[\sigma]\,
\hat{\mathcal O}_H({\bf x},t_k)
\right\}
=
D_{>}\,
\mathcal T\!\left\{
D_k\,\hat{\mathcal O}_H({\bf x},t_k)
\right\}
D_{<} 
\label{eq:T_D_O_with_Dk_reply}
\end{equation}
\begin{equation}
D_{>}
=
\prod_{n>k}
\exp\!\left(
\Delta t\,\hat J_H(t_n)
\right),
\qquad
D_k
=
\exp\!\left(
\Delta t\,\hat J_H(t_k)
\right),
\qquad
D_{<}
=
\prod_{n<k}
\exp\!\left(
\Delta t\,\hat J_H(t_n)
\right) , \nonumber
\label{eq:Dgreater_Dk_Dless_reply}
\end{equation}
which simplifies to \eqref{eq:T_D_O_split_reply} only if $\hat J_H(t_k)$ commutes with $\hat{\mathcal O}_H({\bf x},t_k)$ so that it is simply inserted at the appropriate time location inside the time-ordered exponential. Such a simplification should not be viewed as generic. In fact 
if
$\hat{\mathcal O}_H({\bf x},t_k)$
is itself one of the supergravity field operators, such as a metric component, flux component, dilaton, \emph{etc.}, then whether it commutes with
$\hat J_H(t_k)$
depends on the precise operator content of $\hat J_H$.
A careful statement is that
$[\hat{\mathcal O}_H({\bf x},t_k),\hat J_H(t_k)]
\neq 0$ in general. This typically happens if $\hat J_H(t_k)$ contains conjugate momenta, time derivatives, covariant time derivatives, or any operator not diagonal in the same configuration basis.
For example, if
$\hat{\mathcal O}_H({\bf x},t_k)=\hat h_{ij}({\bf x},t_k)$
but $\hat J_H(t_k)$ contains the conjugate momentum
$\hat\pi^{mn}({\bf y},t_k)$,
then because:
\begin{equation}
[\hat h_{ij}({\bf x},t_k),\hat\pi^{mn}({\bf y},t_k)]
=
i\,\delta_{(i}^{\,m}\delta_{j)}^{\,n}\,\delta^{(d-1)}({\bf x}-{\bf y}),
\label{eq:h_pi_commutator}
\end{equation}
the commutator with $\hat J_H$ is generically nonzero.
Likewise, if $\hat{\mathcal O}_H$ is a flux operator involving both coordinates and momenta, or if $\hat J_H$ is nonlinear and contains noncommuting composite operators, then the commutator need not vanish.
On the other hand, if $\hat J_H(t_k)$ is built only from the same equal-time configuration operators and contains no conjugate momenta or time derivatives, then it is multiplicative in the configuration basis. For example, if:
\begin{equation}
\hat J_H(t_k)
=
\int d^{d-1}y\;
\sigma^{A}({\bf y},t_k)\,
\hat\Xi_A({\bf y},t_k),
\label{eq:JH_linear_field_source}
\end{equation}
and if
$\hat{\mathcal O}_H({\bf x},t_k)
=
\hat\Xi_B({\bf x},t_k)$,
with the equal-time field commutation relation between
$\hat\Xi_A({\bf y},t_k)$ and $\hat\Xi_B({\bf x},t_k)$
then
$\hat{\mathcal O}_H({\bf x},t_k)$ will commute with $\hat J_H(t_k)]$.
In that case:
\begin{equation}
\left[
\hat{\mathcal O}_H({\bf x},t_k),
e^{\Delta t\,\hat J_H(t_k)}
\right]
=0 ,
\label{eq:O_expJ_commute_special}
\end{equation}
which is the situation when both operators are functions only of commuting equal-time bosonic fields.
In the GS construction discussed here, the source is typically taken to be linear in the bosonic fields themselves, schematically taking the form \eqref{eq:JH_linear_field_source}, 
so if
$\hat{\mathcal O}_H({\bf x},t_k)=\hat\Xi_B({\bf x},t_k)$
is another bosonic field operator at equal time, then one usually has
vanishing commutator\footnote{But this is not a completely general theorem about arbitrary supergravity operators.}.
Under this assumption, the operator at the insertion slice may be moved past the equal-time source insertion, so that
$[\hat{\mathcal O}_H({\bf x},t_k), D_{\ge}] = 0$
inside the discretized matrix element. Consequently,
\begin{equation}
\begin{aligned}
\langle h_s,t_s|\hat{\mathcal O}_H({\bf x},t_k)\,D_{\ge}|h_k^+,t_k\rangle
&=
\int Dh_k'\;
\langle h_s,t_s|D_{\ge}|h_k',t_k\rangle\,
\langle h_k',t_k|\hat{\mathcal O}_H({\bf x},t_k)|h_k^+,t_k\rangle ,
\end{aligned}
\label{eq:matrix_element_rewritten_final}
\end{equation}
which would provide a way to handle the factorization for the matrix element in \eqref{eq:M_with_split_Ds}. In fact since the 
derivation of the in--out path-integral form from the fully split matrix element can become relatively technical, we will start by considering the fully split expression:
\begin{equation}
\begin{aligned}
{\cal M}[h_f, h_i]
&=
\int Dh_r^-\,Dh_r^+\,Dh_k^-\,Dh_k^+\,Dh_s\,Dh_k'
\;
\langle h_f,+T|U(+T,t_r)|h_r^-,t_r\rangle
\langle h_r^-,t_r|D_<^\dagger|h_k^-,t_k\rangle
\\
&\qquad\times
\langle h_k^-,t_k|D_\ge^\dagger|h_s,t_s\rangle
\langle h_s,t_s|D_\ge|h_k',t_k\rangle
\langle h_k',t_k|\hat{\mathcal O}_H({\bf x},t_k)|h_k^+,t_k\rangle
\\
&\qquad\times
\langle h_k^+,t_k|D_<|h_r^+,t_r\rangle
\langle h_r^+,t_r|U(t_r,-T)|h_i,-T\rangle ,
\end{aligned}
\label{eq:M_full_split_start}
\end{equation}
that keeps all the operators appearing in \eqref{eq:M_with_split_Ds} sandwiched between states on the corresponding Cauchy slices. This will help us to convert each factor into a time-sliced functional integral and then combine the pieces. To facilitate this let the source-supported interval be
$[t_r,t_s]$
with
$t_r<t_k<t_s$,
and discretize it as before, {\it i.e.} as
$t_n=t_r+n\Delta t$ where
$\Delta t=\frac{t_s-t_r}{N}$ for 
$n=0,1,\dots,N$
with
$t_0=t_r,
t_N=t_s,
t_k=t_r+k\Delta t$
and $0<k<N$. The rest of the parameters appearing in \eqref{eq:M_full_split_start} are defined in \eqref{eq:An_Bn_defs} and \eqref{ennakoigelo}. With these, the two ordinary in--out segments outside the source support are easy to work out. For example, 
the first and last factors in \eqref{eq:M_full_split_start} are ordinary in--out propagation pieces such that 
their standard time-sliced representations are:
\bg\label{missenna}
&&\langle h_f,+T|U(+T,t_r)|h_r^-,t_r\rangle
=
\int_{h_r^-}^{h_f}{\cal D}g_f\;
\exp\!\left(
iS[g_f;\,t_r\to +T]
\right) \nonumber\\
&& \langle h_r^+,t_r|U(t_r,-T)|h_i,-T\rangle
=
\int_{h_i}^{h_r^+}{\cal D}g_i\;
\exp\!\left(
iS[g_i;\,-T\to t_r]
\right) ,
\nd
converting them directly to their path-integral formats with the temporal flow $-T \to t_r \to +T$ although at $t = t_r$ the two set of Cauchy slices are not exactly identical or overlapping. This is essential to accommodate all the pieces of the matrix elements from 
\eqref{eq:M_full_split_start}. For example, we have five such matrix elements, namely:
\begin{equation}
\langle h_k',t_k|\hat{\mathcal O}_H({\bf x},t_k)|h_k^+,t_k\rangle\nonumber
\end{equation}
\begin{equation}\label{shudhuaador}
\langle h_r^-,t_r|D_<^\dagger|h_k^-,t_k\rangle,~~~ \langle h_k^-,t_k|D_\ge^\dagger|h_s,t_s\rangle, ~~~
\langle h_s,t_s|D_\ge|h_k',t_k\rangle, ~~~
\langle h_k^+,t_k|D_<|h_r^+,t_r\rangle ,
\end{equation}
that need to be represented in path-integral formats. We will start with the time-sliced form of the lower forward source-supported segment, namely $\langle h_k^+,t_k|D_<|h_r^+,t_r\rangle$. This may be brought in the path-integral form by first
introducing the intermediate configurations
$h_n^+$ with 
$n=1,\dots,k-1$
and 
$h_0^+=h_r^+$,
$h_k^+=h_k^+$
and then expressing it in the following suggestive way:
\bg\label{ennergondho}
\langle h_k^+,t_k|D_<|h_r^+,t_r\rangle
 & = &
\int \prod_{m=1}^{k-1}Dh_m^+\;
\prod_{n=0}^{k-1}
\langle h_{n+1}^+,t_{n+1}|B_n|h_n^+,t_n\rangle \\
& = & \int \prod_{m=1}^{k-1}Dh_m^+\;
\prod_{n=0}^{k-1} D\widehat h_n^+\;
\langle h_{n+1}^+,t_{n+1}|U_n|\widehat h_n^+,t_n\rangle
\langle \widehat h_n^+,t_n|e^{\Delta t\,J_n}|h_n^+,t_n\rangle\nonumber\\
& = &
\int \prod_{m=1}^{k-1}Dh_m^+\;
\Bigg[
\prod_{n=0}^{k-1}
\langle h_{n+1}^+,t_{n+1}|U_n|h_n^+,t_n\rangle
\Bigg]
\Bigg[
\prod_{n=0}^{k-1}
e^{\Delta t\,J[h_n^+;t_n]}
\Bigg] ,\nonumber
\nd
where going from the first to the second line we applied the resolution of identity
$\mathbf 1_{\Sigma_{t_n}}
=
\int D\widehat h_n^+\;
|\widehat h_n^+,t_n\rangle\langle \widehat h_n^+,t_n|$ $k$ times and used $B_n$ from \eqref{eq:An_Bn_defs}. Since $J_n$ satisfies the eigenvalue equation
$J_n|h_n^+,t_n\rangle
=
J[h_n^+;t_n]\,
|h_n^+,t_n\rangle$
the intermediate matrix element on the second line simplifies as 
$\langle \widehat h_n^+,t_n|e^{\Delta t\,J_n}|h_n^+,t_n\rangle
=
e^{\Delta t\,J[h_n^+;t_n]}\,
\delta[\widehat h_n^+-h_n^+]$
providing the necessary delta functions. The two terms in the square brackets on the third line take the following standard forms:
\begin{equation}
\int \prod_{m=1}^{k-1}Dh_m^+\;
\prod_{n=0}^{k-1}
\langle h_{n+1}^+,t_{n+1}|U_n|h_n^+,t_n\rangle
\longrightarrow
\int_{h_r^+}^{h_k^+}{\cal D}g_<^+\;
\exp\!\left(
iS[g_<^+;\,t_r\to t_k]
\right)
\label{eq:kernel_to_path_integral}
\end{equation}
\begin{equation}
\prod_{n=0}^{k-1}
e^{\Delta t\,J[h_n^+;t_n]}
=
\exp\!\left[
\sum_{n=0}^{k-1}
\Delta t\,
J[h_n^+;t_n]
\right] \longrightarrow
\exp\!\left[
\int_{t_r}^{t_k}dt\int d^{d-1}x\;
J[g_<^+(t,{\bf x});t]
\right]
\equiv
{\cal I}_<[g_<^+;\sigma] ,\nonumber
\label{eq:prod2sum}
\end{equation}
implying that the discrete expression becomes the continuum path integral because the first product gives the usual propagator and the second product exponentiates into the continuum source functional.
After the dust settles, and 
in the continuum limit, the matrix element becomes:
\begin{equation}
\langle h_k^+,t_k|D_<|h_r^+,t_r\rangle
=
\int_{h_r^+}^{h_k^+}{\cal D}g_<^+\;
\exp\!\left(
iS[g_<^+;\,t_r\to t_k]
\right)\,
{\cal I}_<[g_<^+;\sigma] \nonumber
\label{eq:Dless_continuum}
\end{equation}
\begin{equation}
~~~~{\cal I}_<[g_<^+;\sigma]
\equiv
\exp\!\left[
\int_{t_r}^{t_k}dt\int d^{d-1}x\;
J[g_<^+(t,{\bf x});t]
\right] ,
\label{patho4}
\end{equation}
which is the requisite path-integral form. In \eqref{patho4}, 
$g_<^+$
denotes the {plus-branch field history restricted to the lower time segment}
$[t_r,t_k]$.
So it is not a different field from $g_+$. Rather, it is the same plus-branch configuration, but only on the portion of the contour between the initial time $t_r$ and the insertion time $t_k$. More explicitly, one may write
$g_<^+
\equiv
g_+\big|_{[t_r,t_k]}$.
Accordingly, the functional integral
$\int_{h_r^+}^{h_k^+}{\cal D}g_<^+$
means that we
integrate over all plus-branch histories $g_+(t,\mathbf x)$ on the interval $t_r\le t \le t_k$
such that 
$g_+(t_r,\mathbf x)=h_r^+(\mathbf x)$ and 
$g_+(t_k,\mathbf x)=h_k^+(\mathbf x)$.
Similarly,
$S[g_<^+;\,t_r\to t_k]$
means the action evaluated only on that restricted history segment, namely on the portion of the trajectory from $t_r$ to $t_k$; and 
${\cal I}_<[g_<^+;\sigma]$
means the source insertion restricted to the same lower segment.

The time-sliced form of the upper forward source-supported segment, namely the matrix element $\langle h_s,t_s|D_\ge|h_k',t_k\rangle$ from the list \eqref{shudhuaador} may be dealt in the following way. We first 
introduce the intermediate configurations
$\widehat h_n^+$ for 
$n=k+1,\dots,N-1$
with
$\widehat h_k^+=h_k'$ and 
$\widehat h_N^+=h_s$.
Then the matrix element takes the following form:
\bg\label{ennatumik}
\langle h_s,t_s|D_\ge|h_k',t_k\rangle
& = &
\int \prod_{m=k+1}^{N-1}D\widehat h_m^+\;
\prod_{n=k}^{N-1}
\langle \widehat h_{n+1}^+,t_{n+1}|B_n|\widehat h_n^+,t_n\rangle \\
& = &
\int \prod_{m=k+1}^{N-1}D\widehat h_m^+\;
\Bigg[
\prod_{n=k}^{N-1}
\langle \widehat h_{n+1}^+,t_{n+1}|U_n|\widehat h_n^+,t_n\rangle
\Bigg]
\Bigg[
\prod_{n=k}^{N-1}
e^{\Delta t\,J[\widehat h_n^+;t_n]}
\Bigg] ,\nonumber
\nd
where we used similar reduction scheme as in \eqref{ennergondho}; and we see that the relevant parts of the displacement operator cleanly separate out. Following similar strategy as \eqref{eq:prod2sum}, the path-integral structure for \eqref{ennatumik} then takes the following form:
\begin{equation}
\langle h_s,t_s|D_\ge|h_k',t_k\rangle
=
\int_{h_k'}^{h_s}{\cal D}g_\ge^+\;
\exp\!\left(
iS[g_\ge^+;\,t_k\to t_s]
\right)\,
{\cal I}_\ge[g_\ge^+;\sigma] \nonumber
\label{eq:Dgreater_continuum}
\end{equation}
\begin{equation}
~~~ {\cal I}_\ge[g_\ge^+;\sigma]
\equiv
\exp\!\left[
\int_{t_k}^{t_s}dt\int d^{d-1}x\;
J[g_\ge^+(t,{\bf x});t]
\right] ,
\label{patho3}
\end{equation}
which provides the answer that we seek. In our notation,
$g_\ge^+$
denotes the plus-branch field history restricted to the upper time segment
$[t_k,t_s]$.
So, just like
$g_<^+
\equiv
g_+\big|_{[t_r,t_k]}$,
the object
$g_\ge^+
\equiv
g_+\big|_{[t_k,t_s]}$
is the same plus-branch configuration $g_+$, but restricted to the later portion of the contour, namely from the insertion time $t_k$ up to the upper time $t_s$. Thus the difference is simply the time interval on which the history is defined, namely
$g_<^+$ lives on 
$[t_r,t_k]$,
and 
$g_\ge^+$ lives on 
$[t_k,t_s]$.
Correspondingly, the boundary conditions are different.
For the lower segment,
$\int_{h_r^+}^{h_k^+}{\cal D}g_<^+$
means integrating over all plus-branch histories satisfying
$g_<^+(t_r,\mathbf x)=h_r^+(\mathbf x)$ and 
$g_<^+(t_k,\mathbf x)=h_k^+(\mathbf x)$.
For the upper segment,
$\int_{h_k'}^{h_s}{\cal D}g_\ge^+$
means integrating over all plus-branch histories satisfying
$g_\ge^+(t_k,\mathbf x)=h_k'(\mathbf x)$ and 
$g_\ge^+(t_s,\mathbf x)=h_s(\mathbf x)$.
Likewise,
${\cal I}_\ge[g_\ge^+;\sigma]$
is just the source insertion restricted to that upper segment\footnote{Note that we are not using more nuanced symbol like $[t_r, t_k)$ and $[t_k, t_s]$ to distinguish \eqref{patho4} and \eqref{patho3} respectively. These distinctions are unnecessary because of almost overlapping Cauchy slices at $t_k$.}.

The time-sliced form of the upper backward source-supported segment, namely the matrix element $\langle h_k^-,t_k|D_\ge^\dagger|h_s,t_s\rangle$, is interesting because this is where we study the parts of $D^\dagger[\sigma]$. The procedure however is similar to what we encountered earlier. First we
introduce the intermediate configurations
$\widehat h_n^-$ for $n=k+1,\dots,N-1$
with
$\widehat h_k^-=h_k^-$ and 
$\widehat h_N^-=h_s$, 
and then express the matrix element in the following way:
\bg\label{alemink}
\langle h_k^-,t_k|D_\ge^\dagger|h_s,t_s\rangle
& = &
\int \prod_{m=k+1}^{N-1}D\widehat h_m^-\;
\prod_{n=k}^{N-1}
\langle \widehat h_n^-,t_n|A_n|\widehat h_{n+1}^-,t_{n+1}\rangle\nonumber\\
& = & \int \prod_{m=k+1}^{N-1}D\widehat h_m^-\;
\prod_{n=k}^{N-1} \int D\bar h_n^-\;
\langle \widehat h_n^-,t_n|e^{\Delta t\,J_n^\dagger}|\bar h_n^-,t_n\rangle
\langle \bar h_n^-,t_n|U_n^\dagger|\widehat h_{n+1}^-,t_{n+1}\rangle \nonumber\\
& = & \int \prod_{m=k+1}^{N-1}D\widehat h_m^-\;
\Bigg[
\prod_{n=k}^{N-1}
e^{\Delta t\,J^\dagger[\widehat h_n^-;t_n]}\,
\langle \widehat h_n^-,t_n|U_n^\dagger|\widehat h_{n+1}^-,t_{n+1}\rangle
\Bigg], \nd
where going from the first to the second line we 
inserted a same-slice identity 
$\mathbf 1_{\Sigma_{t_n}}
=
\int D\bar h_n^-\;
|\bar h_n^-,t_n\rangle\langle \bar h_n^-,t_n|$, or the resolution of identity, and defined $A_n$ from \eqref{eq:An_Bn_defs}. Going from the second to the third line we used the eigenvalue equation on the bra 
$\langle \widehat h_n^-,t_n|J_n^\dagger
=
J^\dagger[\widehat h_n^-;t_n]\,
\langle \widehat h_n^-,t_n|$, which pulls out the necessary eigenvalues. Now using the matrix element:
\bg\label{avaloalex}
\langle \widehat h_n^-,t_n|e^{\Delta t\,J_n^\dagger}|\bar h_n^-,t_n\rangle
=
e^{\Delta t\,J^\dagger[\widehat h_n^-;t_n]}\,
\delta[\widehat h_n^- - \bar h_n^-], 
\nd
we can express the third line 
in the product form  {\it without} integrals over $D\bar{h}_n^-$ because of the delta functions arising from \eqref{avaloalex}. The terms in the square bracket are in product forms and therefore can be rewritten as:
\bg\label{miramink}
\langle h_k^-,t_k|D_\ge^\dagger|h_s,t_s\rangle
= \int \prod_{m=k+1}^{N-1}D\widehat h_m^-\;
\Bigg[
\prod_{n=k}^{N-1}
\langle \widehat h_n^-,t_n|U_n^\dagger|\widehat h_{n+1}^-,t_{n+1}\rangle
\Bigg] \Bigg[
\prod_{n=k}^{N-1}
e^{\Delta t\,J^\dagger[\widehat h_n^-;t_n]}\Bigg], \nonumber\\ \nd
as they are eigenvalues and c-number pieces but not operators. This is important because the placement of $J^\dagger[\widehat h_n^-;t_n]$ doesn't matter now. After the dust settles the two individual product terms convert to the following path-integral form:
\begin{equation}
\langle h_k^-,t_k|D_\ge^\dagger|h_s,t_s\rangle
=
\int_{h_s}^{h_k^-}{\cal D}g_\ge^-\;
\exp\!\left(
-\,iS[g_\ge^-;\,t_k\to t_s]
\right)\,
{\cal I}_\ge^\dagger[g_\ge^-;\sigma] \nonumber
\label{eq:Dgreaterdag_continuum}
\end{equation}
\begin{equation}
~~~{\cal I}_>^\dagger[g_>^-;\sigma]
\equiv
\exp\!\left[
\int_{t_k}^{t_s}dt\int d^{d-1}x\;
J^\dagger[g_>^-(t,{\bf x});t]
\right] ,
\label{patho2}
\end{equation}
following similar strategy\footnote{Notice however the sign flip for the action term. This will be recurrent for the $-$ branch, and we will come back to this and explain the convention soon.} as in \eqref{eq:prod2sum}. In fact similar argument also goes through in an identical way for the 
time-sliced form of the lower backward source-supported segment, namely $\langle h_r^-,t_r|D_<^\dagger|h_k^-,t_k\rangle$. As above, we now introduce the intermediate configurations
$h_n^-$
for
$n=1,\dots,k-1$
with
$h_0^-=h_r^-$
and 
$h_k^-=h_k^-$.
Then the matrix element becomes:
\bg\label{lonava}
\langle h_r^-,t_r|D_<^\dagger|h_k^-,t_k\rangle
& = &
\int \prod_{m=1}^{k-1}Dh_m^-\;
\prod_{n=0}^{k-1}
\langle h_n^-,t_n|A_n|h_{n+1}^-,t_{n+1}\rangle \\
& = &
\int \prod_{m=1}^{k-1}Dh_m^-\;
\Bigg[
\prod_{n=0}^{k-1}
e^{\Delta t\,J^\dagger[h_n^-;t_n]}\,
\langle h_n^-,t_n|U_n^\dagger|h_{n+1}^-,t_{n+1}\rangle
\Bigg]\nonumber\\
& = & \int \prod_{m=1}^{k-1}Dh_m^-\;
\Bigg[
\prod_{n=0}^{k-1}
\langle h_n^-,t_n|U_n^\dagger|h_{n+1}^-,t_{n+1}\rangle
\Bigg]\Bigg[\prod_{n=0}^{k-1}
e^{\Delta t\,J^\dagger[h_n^-;t_n]}\Bigg], \nonumber
\nd
where $A_n$ appears from \eqref{eq:An_Bn_defs}, and going from the first to the second line we used the eigenvalue equation for $J_n^\dagger$ which produces as c-number. In fact this helps us to split the product in the way shown in the third line. After the dust settles, and using \eqref{eq:prod2sum}, we get the following form:
\begin{equation}
\langle h_r^-,t_r|D_<^\dagger|h_k^-,t_k\rangle
=
\int_{h_k^-}^{h_r^-}{\cal D}g_<^-\;
\exp\!\left(
-\,iS[g_<^-;\,t_r\to t_k]
\right)\,
{\cal I}_<^\dagger[g_<^-;\sigma],\nonumber
\label{eq:Dlessdag_continuum}
\end{equation}
\begin{equation}
~~~{\cal I}_<^\dagger[g_<^-;\sigma]
\equiv
\exp\!\left[
\int_{t_r}^{t_k}dt\int d^{d-1}x\;
J^\dagger[g_<^-(t,{\bf x});t]
\right] ,
\label{patho1}
\end{equation}
which provides the requisite path-integral structure using metric parameters $g_<^-(t, {\bf x})$ defined in a similar way as before. With this we are basically done with the $D[\sigma]$ and $D^\dagger[\sigma]$ insertions and their path-integral representations. The last remaining one is the insertion of the source term. This is the local operator matrix element represented by
$\langle h_k',t_k|\hat{\mathcal O}_H({\bf x},t_k)|h_k^+,t_k\rangle$.
If $\hat{\mathcal O}_H$ is multiplicative in the configuration basis, then clearly:
\begin{equation}
\langle h_k',t_k|\hat{\mathcal O}_H({\bf x},t_k)|h_k^+,t_k\rangle
=
{\mathcal O}[h_k^+]({\bf x},t_k)\,
\delta[h_k'-h_k^+] ,
\label{patho5}
\end{equation}
which collapses
$h_k'=h_k^+$.
If one also chooses the left and right configurations to be identified by the local insertion, one may further write
$h_k^- = h_k^+ \equiv h_k$. The analysis in \eqref{patho5} is done using directly the Heisenberg representation of the operator. What happens once we use the Schr\"odinger picture following the insertion method outlined in \eqref{biralaador1} to \eqref{biralaador6}? Do we get the same result? The answer is yes as may be seen in the following way. Writing $D[\sigma] = D_> B_k D_<$, where $B_k$ is from \eqref{eq:An_Bn_defs}, our concern is only with the intermediate insertion of $B_k$ with the operator defined at $t_k$. This produces:
\bg\label{sarakaalu}
&&\langle h_{k+1},t_{k+1}|
U(t_{k+1},t_k)\,
e^{\,\Delta t\,\hat J_S(t_i;t_k)}\,
\hat O_S({\bf x},t_i)
|h'_k,t_k\rangle\nonumber\\
&&\qquad=
\int Dh_k''\;
\langle h_{k+1},t_{k+1}|
U(t_{k+1},t_k)
|h_k'',t_k\rangle\,
\langle h_k'',t_k|
e^{\,\Delta t\,\hat J_S(t_i;t_k)}\,
\hat O_S({\bf x},t_i)
|h'_k,t_k\rangle \nonumber\\
&&\qquad=
\int Dh_k''\;
\langle h_{k+1},t_{k+1}|
U(t_{k+1},t_k)
|h_k'',t_k\rangle\,
e^{\,\Delta t\,J[h'_k;t_k]}\,
O[h'_k](x)\,
\delta[h_k''-h'_k]\nonumber\\
&&\qquad=
e^{\,\Delta t\,J[h'_k;t_k]}\,
O[h'_k](x)\,
\langle h_{k+1},t_{k+1}|
U(t_{k+1},t_k)
|h'_k,t_k\rangle ,
\nd
which precisely gives us back the c-number form of the operator as in \eqref{patho5}, provided we identify $h_k^+ = h'_k$. The remaining piece of the displacement operator along with the matrix element defined on the Cauchy slice $t_k$ can be combined with the other matrix elements appropriately. The operator ordering on the given Cauchy slice doesn't matter. For example if instead we want the operator to appear to the left of the source factor on the same slice, namely:
\begin{equation}
\langle h_{k+1},t_{k+1}|
U(t_{k+1},t_k)\,
\hat O_S({\bf x},t_i)\,
e^{\,\Delta t\,\hat J_S(t_i;t_k)}
|h'_k,t_k\rangle,
\label{eq:using_ODk_start}
\end{equation}
then the same derivation gives identical result provided $[\hat O_S({\bf x},t_i),\hat J_S(t_i;t_k)]=0$. This is clearly guaranteed for our case. We can also see how this may work for the case when we replace $D_\ge$ by $D_>$ in \eqref{ennatumik}. For such a scenario, and by definition, the strict upper piece starts after the insertion slice:
\bg\label{ennadekao}
\langle h_s,t_s|D_>|h_{k+1},t_{k+1}\rangle
& = & 
\int \prod_{m=k+2}^{N-1}D\widehat h_m^+\;
\prod_{n=k+1}^{N-1}
\langle \widehat h_{n+1}^+,t_{n+1}|B_n|\widehat h_n^+,t_n\rangle\\
& = &
\int \prod_{m=k+2}^{N-1}D\widehat h_m^+\;
\Bigg[
\prod_{n=k+1}^{N-1}
\langle \widehat h_{n+1}^+,t_{n+1}|U_n|\widehat h_n^+,t_n\rangle
\Bigg]
\Bigg[
\prod_{n=k+1}^{N-1}
e^{\,\Delta t\,J[\widehat h_n^+;t_n]}
\Bigg] , \nonumber
\nd
where we used 
$B_n
=
U_n\,e^{\,\Delta t\,J_n}$ from \eqref{eq:An_Bn_defs} with
$U_n\equiv U(t_{n+1},t_n)$ and the following definitions: $\widehat{h}_N^+ = h_s$ and $\widehat{h}^+_{k+1} = h_{k+1}$. The missing slice is the Cauchy slice at $t_k$ containing precisely the matrix element \eqref{eq:using_ODk_start}. This provides a multiplicative piece of the form:
\bg\label{vixmary}
&& \int Dh_{k+1} \langle h_{k+1},t_{k+1}|
U(t_{k+1},t_k)\,
\hat O_S({\bf x},t_i)\,
e^{\,\Delta t\,\hat J_S(t_i;t_k)}
|h'_k,t_k\rangle \nonumber\\
= && \int D\widehat{h}_{k+1} \langle \widehat{h}_{k+1},t_{k+1}|
U(t_{k+1},t_k)\,
\hat O_S({\bf x},t_i)\,
e^{\,\Delta t\,\hat J_S(t_i;t_k)}
|\widehat{h}^+_k,t_k\rangle ,
\nd
which fits well because $\widehat{h}_k^+ = h'_k$. It is now easy to see that combining \eqref{ennadekao} and \eqref{vixmary} produces \eqref{ennatumik} and the result from \eqref{patho5}. This justifies the similarity between the two approaches. It should also be clear that $D^\dagger_\ge$ doesn't need to be split because the source operator is never conjugated with reverse temporal evolution controlled by $D^\dagger$. Therefore to summarize: the strict upper chain $D_>$ starts at $t_{k+1}$, so adjoining the reduced $k$-th slice factor precisely upgrades it to $D_\ge$, while the local insertion contributes the extra multiplicative factor $O[h'_k]({\bf x}, t_k)$.

With these, we are ready to address the exact continuum in--out path-integral structure. Our first step would be to assimilate all the ingredients appearing in the matrix element $\mathcal M$ from \eqref{eq:M_full_split_start} to see how the path-integral structure looks like. 
Substituting
\eqref{missenna},
\eqref{patho4},
\eqref{patho3},
\eqref{patho2},
\eqref{patho1} and
\eqref{patho5}
into \eqref{eq:M_full_split_start}, we obtain:
\bg \label{eq:M_continuum_exact_final}
{\cal M}
&=&
\int Dh_r^\pm\,Dh_s\,Dh_k^\pm\,Dh'_k\;
\Bigg[
\int_{h_r^-}^{h_f}{\cal D}g_f\;
e^{\,iS[g_f;\,t_r\to +T]}
\Bigg]
\Bigg[
\int_{h_k^-}^{h_r^-}{\cal D}g_<^-\;
e^{-\,iS[g_<^-;\,t_r\to t_k]}
{\cal I}_<^\dagger[g_<^-;\sigma]
\Bigg]\nonumber\\
&\times &
\Bigg[
\int_{h_s}^{h_k^-}{\cal D}g_\ge^-\;
e^{-\,iS[g_\ge^-;\,t_k\to t_s]}
{\cal I}_\ge^\dagger[g_\ge^-;\sigma]
\Bigg]
\Bigg[
\int_{h'_k}^{h_s}{\cal D}g_\ge^+\;
e^{\,iS[g_\ge^+;\,t_k\to t_s]}
{\cal I}_\ge[g_\ge^+;\sigma]
\Bigg]
\\
&\times &
\langle h_k',t_k|\hat{\mathcal O}_H({\bf x},t_k)|h_k^+,t_k\rangle
\Bigg[
\int_{h_r^+}^{h_k^+}{\cal D}g_<^+\;
e^{\,iS[g_<^+;\,t_r\to t_k]}
{\cal I}_<[g_<^+;\sigma]
\Bigg]
\Bigg[
\int_{h_i}^{h_r^+}{\cal D}g_i\;
e^{\,iS[g_i;\,-T\to t_r]}
\Bigg] ,\nonumber
\nd
where $Dh_a^\pm \equiv Dh_a^+ Dh_a^-$. Note that
we have retained the $D_\ge$ and $D^\dagger_\ge$ operator pieces so that the source operator can appear in the matrix form as in the LHS of \eqref{patho5}. (As mentioned above, we could have used only $D_>$ and $D^\dagger_>$, but then the result would have been exactly the same.) For a multiplicative operator, using \eqref{patho5}, this becomes:
\bg \label{eq:M_continuum_exact_local}
{\cal M}
&= &
\int Dh_r^\pm\,Dh_s\,Dh_k^\pm\;
\Bigg[
\int_{h_r^-}^{h_f}{\cal D}g_f\;
e^{\,iS[g_f;\,t_r\to +T]}
\Bigg]
\Bigg[
\int_{h_k^-}^{h_r^-}{\cal D}g_<^-\;
e^{-\,iS[g_<^-;\,t_r\to t_k]}
{\cal I}_<^\dagger[g_<^-;\sigma]
\Bigg]\nonumber\\
&\times &
\Bigg[
\int_{h_s}^{h_k^-}{\cal D}g_\ge^-\;
e^{-\,iS[g_\ge^-;\,t_k\to t_s]}
{\cal I}_\ge^\dagger[g_\ge^-;\sigma]
\Bigg]
\Bigg[
\int_{h_k^+}^{h_s}{\cal D}g_\ge^+\;
e^{\,iS[g_\ge^+;\,t_k\to t_s]}
{\cal I}_\ge[g_\ge^+;\sigma]
\Bigg]
\\
&\times &
{\mathcal O}[h_k^+]({\bf x},t_k)\,
\Bigg[
\int_{h_r^+}^{h_k^+}{\cal D}g_<^+\;
e^{\,iS[g_<^+;\,t_r\to t_k]}
{\cal I}_<[g_<^+;\sigma]
\Bigg]
\Bigg[
\int_{h_i}^{h_r^+}{\cal D}g_i\;
e^{\,iS[g_i;\,-T\to t_r]}
\Bigg] ,\nonumber
\nd
which is the exact in--out path-integral structure. The equation \eqref{eq:M_continuum_exact_final} captures everything but there is a subtlety about the temporal ordering that we should clear up. The 
point is this, 
$\int_{h_k^-}^{h_r^-}{\cal D}g_<^-$ and 
$\int_{h_s}^{h_k^-}{\cal D}g_\ge^-$
already indicate that the minus-branch histories are being traversed in the reverse direction, namely
$t_k\to t_r$ and 
$t_s\to t_k$ respectively.
Therefore, if one writes the action as
$e^{-\,iS[g_<^-;\,t_r\to t_k]}$
or
$e^{-\,iS[g_>^-;\,t_k\to t_s]}$,
then one is using the convention that
$S[g;\,t_r\to t_k]$
is the ordinary forward-oriented action, and the explicit minus sign in the exponential compensates for the fact that the contour runs backward\footnote{This is easy to see from the following arguments. Suppose a contour has a forward segment
$t:\ a\to b$
and then a backward segment
$t:\ b\to a$.
If the Lagrangian is $L[q,\dot q]$, then the actions on the forward and the backward segments are:
\begin{equation}
S_{a\to b}[q_+]
=
\int_a^b dt\;L[q_+(t),\dot q_+(t)]\nonumber
\label{eq:S_forward}
\end{equation}
\begin{equation}
S_{b\to a}[q_-]
=
\int_b^a dt\;L[q_-(t),\dot q_-(t)] =
-\int_a^b dt\;L[q_-(t),\dot q_-(t)] , \nonumber
\label{eq:S_backward_oriented}
\end{equation}
implying that 
$S_{b\to a}[q_-]
=
-\,S_{a\to b}[q_-]$.
So nothing mysterious happens: the minus sign is absorbed into the reversed orientation of the contour.
Hence
\begin{equation}
e^{\,iS_{a\to b}[q_+]}\,
e^{-\,iS_{a\to b}[q_-]}
=
e^{\,iS_{a\to b}[q_+]}\,
e^{\,iS_{b\to a}[q_-]}.
\label{eq:key_identity_simple}
\end{equation}
This is the precise reason why a product of a forward weight and a backward conjugate weight can be written as one contour weight. One might however worry that such a way of writing would masquerade the  reality of the expectation values, but this need not be the case. We will discuss a bit more on the issue of reality in the next sub-section.}. This is the convention that we used here and is perfectly consistent. However we can entertain two equivalent conventions to write the minus branch. The $\color{blue}{\rm first}$ one is exactly what we have followed so far, namely, 
keep the action written with forward time labels, and put an explicit minus sign in the exponent. This is what our formula is doing:
\begin{equation}
\int_{h_k^-}^{h_r^-}{\cal D}g_<^-\;
e^{-\,iS[g_<^-;\,t_r\to t_k]},~~~~~~
\int_{h_s}^{h_k^-}{\cal D}g_\ge^-\;
e^{-\,iS[g_\ge^-;\,t_k\to t_s]} ,
\label{eq:convention_A_upper}
\end{equation}
where the symbol
$S[g;\,t_r\to t_k]$
means the ordinary action integrated with increasing time, while the overall factor
$-\,i$ implements the backward branch. In the $\color{blue}{\rm second}$ convention, we write the action with the actual backward orientation, and then use $e^{+iS}$. In other words 
one may equivalently write:
\begin{equation}
\int_{h_k^-}^{h_r^-}{\cal D}g_<^-\;
e^{\,iS[g_<^-;\,t_k\to t_r]},~~~~~
\int_{h_s}^{h_k^-}{\cal D}g_\ge^-\;
e^{\,iS[g_\ge^-;\,t_s\to t_k]} ,
\label{eq:convention_B_upper}
\end{equation}
which are equivalent because
$S[g;\,t_a\to t_b]
=
-S[g;\,t_b\to t_a]$,
implying that the minus branch is reversed in orientation, but our formula is already encoding that reversal through the combination of reversed boundary data and the factor
$e^{-\,iS[\cdots]}$. The second convention, {\it i.e.} \eqref{eq:convention_B_upper} looks more transparent because it keeps one specific temporal ordering. Therefore, an alternative way of rewriting \eqref{eq:M_continuum_exact_local} is then the following:
\bg \label{eq:M_continuum_exact_local_transparent}
{\cal M}
&= &
\int Dh_r^\pm\,Dh_s\,Dh_k^\pm\;
\Bigg[
\int_{h_r^-}^{h_f}{\cal D}g_f\;
e^{\,iS[g_f;\,t_r\to +T]}
\Bigg]
\Bigg[
\int_{h_k^-}^{h_r^-}{\cal D}g_<^-\;
e^{\,iS[g_<^-;\,t_k\to t_r]}
{\cal I}_<^\dagger[g_<^-;\sigma]
\Bigg]\nonumber\\
&\times &
\Bigg[
\int_{h_s}^{h_k^-}{\cal D}g_\ge^-\;
e^{\,iS[g_\ge^-;\,t_s\to t_k]}
{\cal I}_\ge^\dagger[g_\ge^-;\sigma]
\Bigg]
\Bigg[
\int_{h_k^+}^{h_s}{\cal D}g_\ge^+\;
e^{\,iS[g_\ge^+;\,t_k\to t_s]}
{\cal I}_\ge[g_\ge^+;\sigma]
\Bigg]\\
&\times &
{\mathcal O}[h_k^+]({\bf x},t_k)\,
\Bigg[
\int_{h_r^+}^{h_k^+}{\cal D}g_<^+\;
e^{\,iS[g_<^+;\,t_r\to t_k]}
{\cal I}_<[g_<^+;\sigma]
\Bigg]
\Bigg[
\int_{h_i}^{h_r^+}{\cal D}g_i\;
e^{\,iS[g_i;\,-T\to t_r]}
\Bigg] ,\nonumber
\nd
which is of course 
equivalent to our original formula, but the orientation of the minus branch is now explicit. In fact the structure \eqref{eq:M_continuum_exact_local_transparent} corresponds to the following temporal ordering and contour representation:
\begin{equation}
-T\to t_r\to t_k\to t_s\to t_k\to t_r\to +T \nonumber
\label{eq:full_contour_shape}
\end{equation}
\begin{equation}
~~~{\cal C}_{\rm in\text{-}out}
=
[-T,t_r]
\cup
[t_r,t_s]_+
\cup
[t_s,t_r]_-
\cup
[t_r,+T] ,
\label{eq:compact_contour_definition}
\end{equation}
telling us precisely that the contour starts at $-T$ then goes forward all the way till $t_s$, turns back to the original Cauchy slice $t_r$ and then moves forward from $t_r$ to $+T$. This is represented by {\bf figure \ref{inoutskcontour}}.  
Thus one may summarize the result as:
\begin{equation}
{\cal M}
=
\int_{{\cal C}_{\rm in\text{-}out}}{\cal D}g\;
\exp\!\left(
iS_{\rm tot}[g]\big|_{{\cal C}_{\rm in\text{-}out}}
\right)\,
{\cal I}^\dagger[g_-;\sigma]\,
{\mathcal O}[g_+(t_k)]({\bf x},t_k)\,
{\cal I}[g_+;\sigma],
\label{eq:compact_contour_result}
\end{equation}
with the understanding that the plus and minus histories are glued at the turning point $t_s$ and at the insertion point when the local operator is diagonal. Finally, the full in--out numerator takes the following form:
\begin{equation}
N_\sigma^{\rm in\text{-}out}({\bf x},t_k)
=
\int Dh_f\,Dh_i\;
\Psi_0^\ast[h_f,+T]\,
\Psi_0[h_i,-T]\,
{\cal M}[h_f,h_i],
\label{eq:full_Nsigma_result}
\end{equation}
with ${\cal M}[h_f,h_i]$ given exactly by \eqref{eq:M_continuum_exact_local_transparent}, or in compact contour form by \eqref{eq:compact_contour_result}. (The denominator $Z_\sigma^{\rm in-out}$ can be easily extracted from $\mathcal M$ by removing the operator source term ${\mathcal O}[g_+(t_k)]({\bf x},t_k)$.)
The path-integral structure emerges by converting each exact matrix element in \eqref{eq:M_full_split_start} into a time-sliced functional integral: ordinary propagation outside the source support gives the segments $[-T,t_r]$ and $[t_r,+T]$,
while
$D_<,\ D_{\ge}$ and 
$D_<^\dagger,\ D_{\ge}^\dagger$
generate the forward and backward source-supported excursions on:
\begin{equation}
[t_r,t_k],\ [t_k,t_s],\ [t_s,t_k],\ [t_k,t_r].
\label{eq:summary_inner_segments}
\end{equation}
Thus the exact in--out matrix element is represented by a contour with a forward and backward excursion inserted between the ordinary in and out propagation segments.

\begin{figure}[h]
\centering
\begin{tabular}{c}
\includegraphics[width=6in]{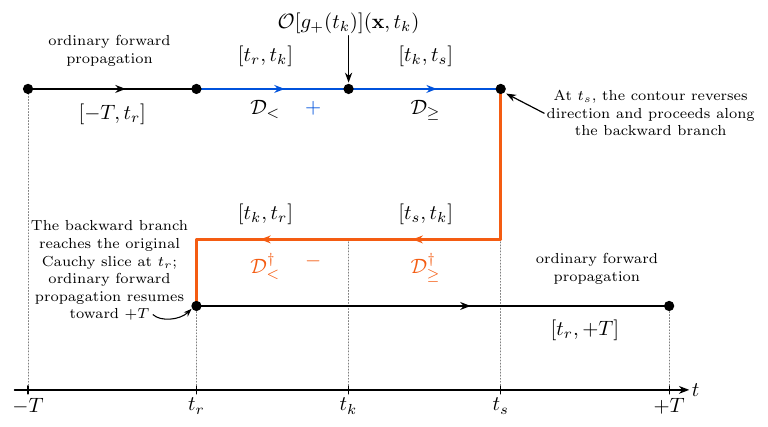}
\end{tabular}
\caption[]{The contour associated with the in--out expectation value. The leftmost black line denote ordinary forward propagation. The \textcolor{blue}{blue} and \textcolor{orange}{orange} lines denote the parts of the forward $(+)$ and backward $(-)$ branches where the displacement operator is supported, respectively. After the backward evolution that ends on $t_r$, the ordinary forward propagation happens again. The parts of the displacement operator before and after the hypersurface of the operator insertion are denoted by $\mathcal{D_<}$ and $\mathcal{D}_\ge$, respectively. The vertical distances are a bookkeeping conventions to separate the forward and backward pieces of the contour and so have no physical meaning.}
\label{inoutskcontour}
\end{figure}

\subsubsection{Interpretation, caveat and when is the in--out contour useful \label{sec6.5.4}}

In the previous section, where reasonably rigorous treatment of the in--out contour was done, we gleamed over two topics a bit too fast. The first one is the {\it orientation} of the in--out contour, and the second one is the connection between the in--in and the in--out formalisms. In the following we will elaborate on them starting with the connection between the in--in and the in--out contours.
The in--out contour completion written above is a mathematically valid
single-contour representation of the same local operator structure.
But one should not confuse it with the exact GS in--in observable. The exact physical quantity remains:
\begin{equation}
\langle \hat{\mathcal O}_H({\bf x}, t)\rangle_\sigma
=
\frac{
{\rm Tr}\!\left(
D(\sigma)\,\rho_\Omega\,D^\dagger(\sigma)\,\hat{\mathcal O}_H({\bf x}, t)
\right)
}{
{\rm Tr}\!\left(
D(\sigma)\,\rho_\Omega\,D^\dagger(\sigma)
\right)
},
\label{exact_GS_repeat_local}
\end{equation}
which is computed exactly by the Schwinger--Keldysh contour and the path-integral form appears in \eqref{coridahl7}. The source insertion can be either in the $+$ or the $-$ branch, although here we take it in the $-$ branch. On the other hand, 
the in--out ratio:
\begin{equation}\label{duimecalgari}
\langle \hat{\mathcal O}_H({\bf x}, t)\rangle_\sigma^{\rm in\text{-}out}
=
\frac{
{}_{\rm out}\langle\Omega|\,
D^\dagger(\sigma)\,\hat{\mathcal O}_H({\bf x}, t)\,D(\sigma)\,
|\Omega\rangle_{\rm in}
}{
{}_{\rm out}\langle\Omega|\,
D^\dagger(\sigma)\,D(\sigma)\,
|\Omega\rangle_{\rm in}
}
\end{equation}
is instead an in--out completion of the same local operator insertion whose simplified form appears in \eqref{inout_ratio_combined_local}.
In general it is a different contour completion and therefore a
different object. Thus the correct conclusion is that
the GS numerator and denominator admit both in--in and in--out contour completions at the operator level,
but
the exact normalized real-time expectation value in the prepared GS state is the in--in Schwinger--Keldysh quantity. Thus generically one would expect this in--out representation to be viewed as a contour
completion of the same local operator structure, not as the exact
replacement of the original in--in Schwinger--Keldysh observable.
In this precise sense, the GS construction admits both an in--in and
an in--out contour completion, even though the physically relevant
normalized expectation value is most naturally and exactly described by
the in--in Schwinger--Keldysh formalism.

The expression for $\mathcal M$ in \eqref{eq:M_continuum_exact_local_transparent} already makes the difference clear. The pieces
$g_<^+,\ g_\ge^+$
and
$g_<^-,\ g_\ge^-$
appear because we split the operator insertion and the displacement operator into segments around the time $t_k$. But this is not yet the same thing as the SK closed-time-path identification. In the SK in-in functional integral, the two branches are glued at the final time through a common integration over
$h_f$.
In our in-out amplitude, by contrast, the plus and minus pieces arise as a factorization device for the transition amplitude, with additional intermediate gluing data such as
$h_r^\pm,\ h_s,\ h_k^\pm$.
So the resemblance is structural, but it is not yet an equality.

When can they coincide? They can coincide only in special cases. The  $\color{blue}{\rm first}$ one would be with a trivial final-state projection.
If the out-state projection is replaced by a trace over final states, then the in-out amplitude turns into an in-in object. This is exactly what happens when one inserts
$\int Dh_f\;
|h_f,t_f\rangle\langle h_f,t_f|
=
\mathbf 1$
at the return time and sums over final configurations. Then the structure becomes the usual SK contour. So the SK formalism may be obtained from an in-out-looking derivation only after replacing the final-state projection by a trace. In other words, in-out becomes in-in only after summing over final states and closing the contour.
The $\color{blue}{\rm second}$ one would be with a tree level or classical saddle approximation. In some semiclassical limits, especially at tree level or when a single saddle dominates, the numerical values of in-in and in-out one-point functions may coincide to leading order. In that case both are controlled by the same classical solution:
\begin{equation}
\langle \hat O(x)\rangle_{\rm in\text{-}in}
\approx
O[g_{\rm cl}](x)
\approx
\frac{\langle {\rm out}|\hat O_H(x)|{\rm in}\rangle}{\langle {\rm out}|{\rm in}\rangle} ,
\label{eq:classical_coincidence}
\end{equation}
where $O[g_{\rm cl}](x)$ denotes the c-number functional obtained from the operator $\hat O(x)$ by replacing the quantum field $\hat g$ with the classical background $g_{\rm cl}$.
But this is only an approximation, not an exact identity.
The $\color{blue}{\rm third}$ one would be with a special adiabatic vacuum situations.
For example, if the initial and final states coincide and there is no particle production or branch sensitivity, then the distinction may collapse. But again this is a special case. So it appears that none of the arguments are compelling enough to declare the equality between \eqref{exact_GS_repeat_local} and \eqref{duimecalgari}.

The point is that the combination of
$e^{\,iS[\text{forward segment}]}$
and 
$e^{-\,iS[\text{backward segment}]}$
is exactly what one gets by writing a \emph{single} contour integral
$e^{\,iS[g]\big|_{{\cal C}_{\rm in\text{-}out}}}$
on a contour whose backward part has the opposite orientation.
In our formula, the forward-oriented pieces were
$e^{\,iS[g_i;\,-T\to t_r]},
e^{\,iS[g_<^+;\,t_r\to t_k]}$, and 
$e^{\,iS[g_>^+;\,t_k\to t_s]}$, whereas the backward-oriented pieces were
$e^{-\,iS[g_>^-;\,t_k\to t_s]}$ and 
$e^{-\,iS[g_<^-;\,t_r\to t_k]}$
together with
$e^{\,iS[g_f;\,t_r\to +T]}$ for the evolution $t > t_r$.
Now using the orientation identity discussed before, one can show that
$e^{-\,iS[g_>^-;\,t_k\to t_s]}
=
e^{\,iS[g_>^-;\,t_s\to t_k]}$
and
$e^{-\,iS[g_<^-;\,t_r\to t_k]}
=
e^{\,iS[g_<^-;\,t_k\to t_r]}$.
So the product of all exponentials becomes:

{\footnotesize
\bg\label{psantosam}
e^{\,iS[g]\big|_{{\cal C}_{\rm in\text{-}out}}}
& = & e^{\,iS[g_i;\,-T\to t_r]}
e^{\,iS[g_<^+;\,t_r\to t_k]}
e^{\,iS[g_\ge^+;\,t_k\to t_s]}
e^{-\,iS[g_\ge^-;\,t_k\to t_s]}
e^{-\,iS[g_<^-;\,t_r\to t_k]}
e^{\,iS[g_f;\,t_r\to +T]} \nonumber\\
&= &
e^{\,iS[g_i;\,-T\to t_r]}
e^{\,iS[g_<^+;\,t_r\to t_k]}
e^{\,iS[g_\ge^+;\,t_k\to t_s]}
e^{\,iS[g_\ge^-;\,t_s\to t_k]}
e^{\,iS[g_<^-;\,t_k\to t_r]}
e^{\,iS[g_f;\,t_r\to +T]} ,\nonumber\\
\nd}
which is exactly the exponential of the action integrated along the contour \eqref{eq:compact_contour_definition}.
So the contour notation is just a compact rewriting of the same product so that following one specific direction of the contour amounts to {\it adding up} all the intermediate action pieces.
This is completely analogous to the usual Schwinger--Keldysh identity
equivalently written as a contour action:
\begin{equation}
e^{\,iS[q_+; t_r \to t_s]}\,e^{-\,iS[q_-; t_r \to t_s]}
=
e^{\,i\left(S[q_+; t_r \to t_s] +S[q_-; t_s \to t_r]\right)} = 
e^{\,iS[q]\big|_{{\cal C}_{\rm in-in}}} .
\label{eq:SK_standard}
\end{equation}
The only difference here is that our in--out contour is not the standard closed time path from $-\infty$ to $+\infty$ and back, but the more specialized in--out contour with the source-supported excursion
$t_r\to t_s\to t_r$
inserted between the outer propagation pieces. In fact this is the key point that helps us to connect to the earlier path-integral formulation for the expectation value developed in \cite{joydeep, coherbeta, coherbeta2, wdwpaper, borelborel}. The contour studied there should now be interpreted as ${\cal C}_{\rm in-in}$ and {\it not} ${\cal C}_{\rm in-out}$. In other words:
\bg\label{comatopam}
\langle \hat O[g](x,t_k) \rangle_\sigma & = &
{\langle \sigma|\hat O[g](x,t_k)|\sigma\rangle\over \langle\sigma |\sigma\rangle}
=
{1\over Z_\sigma^{\rm in-in}} \langle \Omega|
D^\dagger[\sigma]\,
\hat O_H(x,t_k)\,
D[\sigma]
|\Omega\rangle \nonumber\\
& = &  {1\over Z_\sigma^{\rm in-in}} \int {\cal D}g_+\,{\cal D}g_-\;
e^{\,iS[g_+] - iS[g_-]}\,
{\cal I}[g_+;\sigma]\,
{\cal I}^\dagger[g_-;\sigma]\,
O[g_+](x,t_k) \nonumber\\
 & = &
{1\over Z_\sigma^{\rm in-in}}
\int_{{\cal C}_{\rm in-in}} {\cal D}g\;
e^{\,iS[g]\big|_{{\cal C}_{\rm in-in}}}\,
{\cal I}[g_+;\sigma]\,
{\cal I}^\dagger[g_-;\sigma]\,
O[g_+](x,t_k),
\nd
providing both the connection to our earlier works {\gspapers}, as well as the reason why such an expectation value should be {\it real}. On the other hand, the in-out contour description doesn't have to be real. Only in very special cases the in-out path-integral become real. This may be easily demonstrated but we will not do so here and leave it as an exercise for our diligent readers. Instead we want to ask under what conditions do the in--out path-integral become relevant. The answer is depicted in {\bf figures \ref{inoutsourcemoved1}} and {\bf \ref{inoutsourcemoved2}}, and in the following let us provide a brief elaborations on them. 

\begin{figure}[h]
\centering
\begin{tabular}{c}
\includegraphics[width=6in]{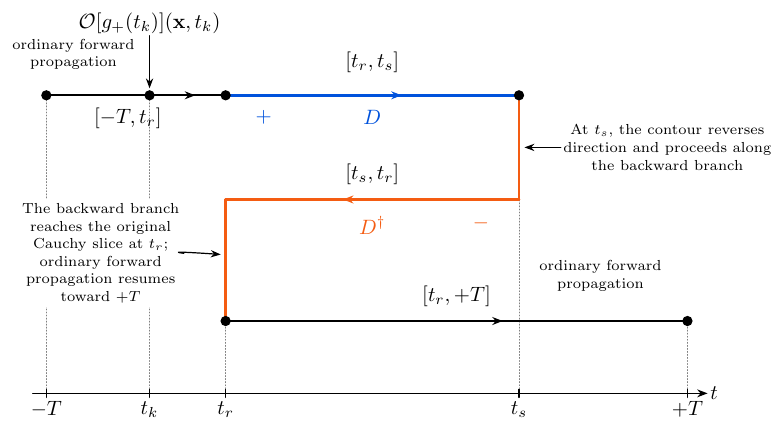}
\end{tabular}
\caption[]{The in--out contour representation similar to {\bf figure \ref{inoutskcontour}} but now the source operator is added before the displacement operators are switched on. The displacement operators continue to be active for $t_r \le t \le t_s$, such that the source is now at $-T < t_k < t_r$. We have denoted the source operator using the $+$ branch. The \textcolor{orange}{orange} line denotes the reverse propagation of $D^\dagger$.}
\label{inoutsourcemoved1}
\end{figure}

\begin{figure}[h]
\centering
\begin{tabular}{c}
\includegraphics[width=6in]{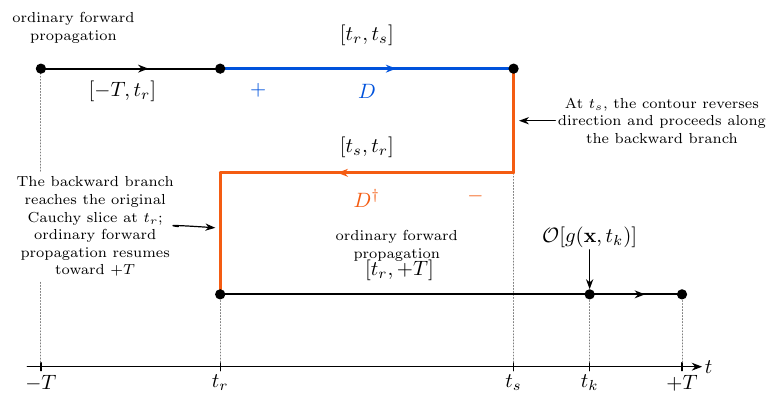}
\end{tabular}
\caption[]{The in--out contour representation with the source now placed after the actions of the displacement operators $D$ and $D^\dagger$. Most of the details appearing here are similar to what we had in {\bf figure \ref{inoutsourcemoved1}}, with the exception that the source operator is now placed at $t_s < t < +T$. We have also denoted the metric in the outer branch using $g$ and not $g_+$ or $g_-$.}
\label{inoutsourcemoved2}
\end{figure}

Let us start by recalling that the in--out path integral is preferred in much of standard QFT because it directly computes vacuum transition amplitudes, $S$-matrix elements, and time-ordered Green's functions, which are the central observables in scattering theory and perturbation theory. It is technically simpler than the in--in formalism because it does not require doubling the fields or introducing a closed time contour. However, it is not the appropriate formalism for general real-time expectation values, nonequilibrium systems, or cosmology, where the in--in Schwinger--Keldysh contour is the correct framework. For the present case, 
the in--out-type contour becomes relevant when one studies matrix elements of the form:
\begin{equation}
\langle 0|
U(+T,t_r)\,
D^\dagger[\sigma]\,
\hat O_H(x,t_k)\,
D[\sigma]\,
U(t_r,-T)
|0\rangle,
\label{eq:inout_matrix_element_basic}
\end{equation}
and the insertion time $t_k$ lies \emph{outside} the support of $D$ and $D^\dagger$, where recall that we have assumed that $D[\sigma]$ remains active in the temporal domain $[t_r, t_s]$, whereas $D^\dagger[\sigma]$ remains active in the reverse domain $[t_s, t_r]$. With this in mind, there are then two distinct possibilities:

\paragraph{Case A: $t_k<t_r$}

The operator is inserted before the contour reaches the source-supported excursion. Then the contour looks schematically like:
\begin{equation}
-T\to t_k\to t_r\to t_s\to t_r\to +T.
\label{eq:caseA_contour}
\end{equation}
Here the operator lies on the outer incoming leg, whereas the pair $D,D^\dagger$ forms an excursion later in time. In this case the object is naturally viewed as an in--out matrix element with an inserted closed excursion.

\paragraph{Case B: $t_k>t_s$}

The operator is inserted after the contour has already completed the source-supported excursion and returned to the outer leg. Then the contour looks schematically like:
\begin{equation}
-T\to t_r\to t_s\to t_r\to t_k\to +T.
\label{eq:caseB_contour}
\end{equation}
Now the operator lies on the final outgoing leg, again outside the support of the $D,D^\dagger$ excursion.
In both cases the operator is not inserted on the internal forward/backward pair of branches generated by the source-supported excursion. Instead, it sits on one of the outer asymptotic legs. That is exactly why the in--out interpretation becomes useful.

So we see the following distinction:
If the operator insertion lies within the time interval where the source-supported forward/backward excursion is active, then the exact object is naturally an in--in Schwinger--Keldysh expectation value.
But
if the operator insertion lies outside that interval, either before $D[\sigma]$ or after $D^\dagger[\sigma]$, then it is often more natural to describe the resulting matrix element as an in--out contour with an inserted closed excursion.

Even in the two ``outer insertion'' cases, the $D[\sigma], D^\dagger[\sigma]$ part itself still carries a forward/backward excursion. So one should not say that the Schwinger--Keldysh structure disappears entirely. Rather, what changes is where the operator sits relative to that excursion. Thus a better explanation is the following.
The internal source-supported part still has a doubled-contour structure, but the full matrix element is most conveniently organized as an in--out contour with an embedded forward/backward excursion.
In other words, the expectation value $\langle \sigma|\hat O[g](x,t_k)|\sigma\rangle$ is intrinsically an in--in Schwinger--Keldysh object, because it is an expectation value in one and the same state and therefore requires a closed time contour. However, when the operator insertion time $t_k$ lies outside the support interval of $D[\sigma]$ and $D^\dagger[\sigma]$, namely for $t_k<t_r$ or $t_k>t_s$, the operator sits on one of the outer asymptotic legs. In those cases it is often more natural to organize the result as an in--out contour from $-T \to +T$ with an embedded forward/backward excursion on $[t_r,t_s]$. Thus the Schwinger--Keldysh structure remains present inside the excursion, but the full matrix element is most transparently viewed in an in--out form when the operator lies before or after the $D[\sigma], D^\dagger[\sigma]$ action.

\section{Bootstrapping the Glauber-Sudarshan state and consistency \label{sec7.0}}

To see how and where bootstrapping enters the story, we will have to first work out the Schwinger-Dyson equations carefully. In the process 
we will dispel some of the common mistakes one does when interpreting the behavior of the GS states. First, however, let us ask as to why we expect bootstrapping to control the dynamics of the GS states. The answer is simple but instructive.
A GS state is not merely an arbitrary perturbation of initial data. Rather, it is defined by a spacetime-supported deformation of the in--in Schwinger--Keldysh contour, and the same deformation that defines the state also changes the mean-field equations whose solution determines the state. Therefore the state and the dynamics are tied together self-consistently. A clean way to say this is the following.

For an ordinary coherent state, one specifies initial data on an initial Cauchy slice and then evolves with the usual {\it boundary} Hamiltonian or, equivalently, with the undeformed in--in effective action. In that case naively the dynamics is governed by the homogeneous mean-field equation:
\begin{equation}
\frac{\delta \Gamma_{\rm SK}[g_r,g_a]}{\delta g_r^{\mu\nu}(x)}
\Bigg|_{g_a=0}
=
0 ,
\label{eq:coherent_homogeneous_boot}
\end{equation}
where $\Gamma_{\rm SK}$ is the effective action defined using Legendre transformation of the connected functional $W_{\rm SK}$ $-$ derived from the partition functional $Z_{\rm SK}$ $-$ that appears in the Schwinger-Keldysh path-integral formulation, and $g^{\mu\nu}_{r,a} = {1\over 2}(g^{\mu\nu}_+ \pm g^{\mu\nu}_-)$. (These will be defined more carefully a little later.)
The coherent-state profile changes the initial condition, but it does not directly deform the bulk mean-field equation.
By contrast, for a GS state one inserts branch functionals
$W_+[g_+;\sigma]$ and 
$W_-[g_-;\sigma]$
into the Schwinger--Keldysh contour, so the generating functional itself changes as
$Z_{\rm SK}
\longrightarrow
Z_{\rm SK}^{(\sigma)}$.
Consequently the connected functional changes,
$W_{\rm SK}
\longrightarrow
W_{\rm SK}^{(\sigma)}$,
and therefore the effective action changes as well:
$\Gamma_{\rm SK}
\longrightarrow
\Gamma_{\rm SK}^{(\sigma)}$. However the physical mean-field equation becomes:
\begin{equation}
{1\over \sqrt{-g_c}}
\frac{\delta \Gamma_{\rm phys}^{(\sigma)}[g_c]}
{\delta g_c^{\mu\nu}(x)}
=
0 ,
\label{eq:GS_deformed_mean_field}
\end{equation}
where $g_c^{\mu\nu} = \langle g_\pm^{\mu\nu} \rangle_\sigma$ and $\Gamma_{\rm phys}^{(\sigma)}[g_c]$, which is a reduced single-field functional, will be defined in \eqref{katadelga}.
Thus the same object $\sigma$ that defines the GS state also modifies the equation that determines the mean geometry. This is already the basic reason bootstrapping appears.

But why is the GS dynamics self-referential? 
The point is that the GS deformation is not external in the same way as a fixed classical source in a textbook generating functional. Rather, once one asks for a self-consistent semiclassical branch, the deformation must be compatible with the geometry and stress tensor that it itself produces. So the deformation does not merely \emph{drive} the dynamics. It also gets \emph{renormalized and constrained} by the dynamics. This is the hallmark of bootstrap structure. In fact the term ``bootstrap'' is appropriate because the unknown quantity $-$ here $\sigma$  $-$ reappears inside the condition that determines it. Alternatively, the physical reason bootstrapping controls GS states is therefore simple to state:

\vskip.1in

\noindent $\bullet$ The GS deformation changes the state.

\vskip.1in

\noindent $\bullet$ The changed state changes the in--in effective action.

\vskip.1in

\noindent $\bullet$ The changed effective action changes the mean-field equations.

\vskip.1in

\noindent $\bullet$ The resulting mean fields determine whether the original GS deformation was consistent.

\vskip.1in

\noindent So one expects bootstrap control precisely because the GS state is not just propagated by the dynamics; it is partly defined by the same dynamics. In other words, bootstrapping is expected to control the dynamics of GS states because a GS state is defined by a spacetime-supported deformation of the Schwinger--Keldysh contour, and that same deformation modifies the effective action whose variation determines the mean fields. Therefore the GS profile, the mean geometry, and the stress tensor must satisfy a self-consistency condition rather than an ordinary initial-value problem.
Unlike an ordinary coherent state, which changes only the initial data, a GS state changes the bulk in--in contour itself. This means that the same profile $\sigma$ that defines the GS state also changes the effective action and hence the mean-field equations. The resulting dynamics is therefore self-referential: the state determines the geometry, and the geometry constrains the state as shown in {\bf figure \ref{bootstrapping3}}. This is precisely why a bootstrap condition is expected to control the dynamics of GS states.

\begin{figure}[h]
\centering
\begin{tabular}{c}
\includegraphics[width=5in]{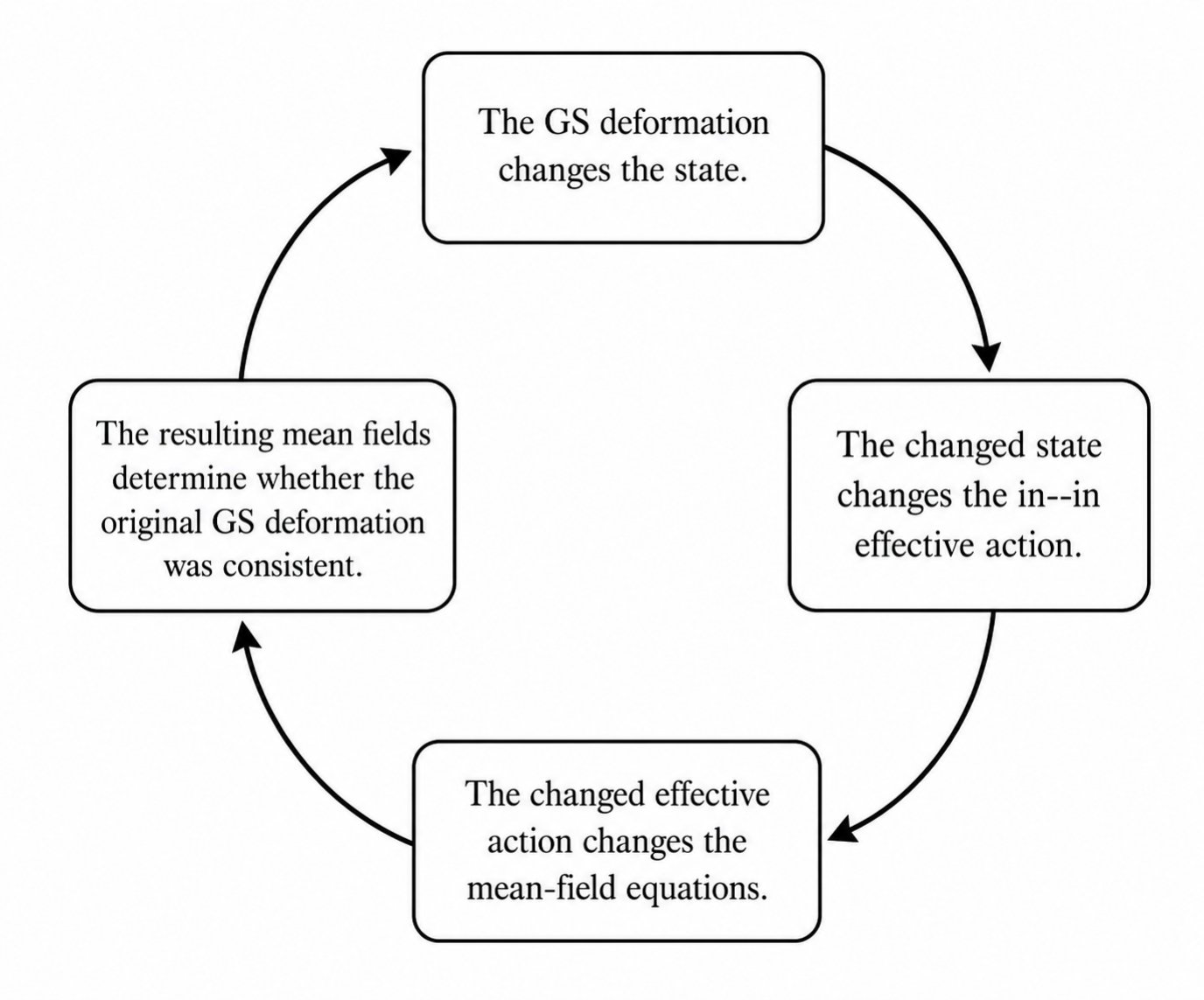}
\end{tabular}
\caption[]{The bootstrapping diagram for the GS states.}
\label{bootstrapping3}
\end{figure}

A closely related question is how the GS bootstrap program differs from the bootstrap methods commonly used to constrain the dynamics of a conformal field theory. The simplest answer is that the two bootstrap programs are quite different in both aim and structure. In the standard CFT bootstrap, one studies an abstract conformal field theory by imposing general consistency conditions such as conformal symmetry, crossing symmetry, unitarity, and the operator product expansion. The basic unknowns are the scaling dimensions and OPE coefficients, and the bootstrap equations constrain these data directly:
\begin{equation}
\sum_{\cal O}
\lambda_{12{\cal O}}\lambda_{34{\cal O}}\,
F_{\cal O}(u,v)
=
\sum_{\cal O}
\lambda_{14{\cal O}}\lambda_{23{\cal O}}\,
F_{\cal O}(v,u) ,
\label{eq:CFT_bootstrap_summary}
\end{equation}
where the quantities $\lambda_{12{\cal O}}$ and $\lambda_{34{\cal O}}$ are the operator-product-expansion coefficients that appear when the operators ${\cal O}_1,{\cal O}_2$ or ${\cal O}_3,{\cal O}_4$ are expanded into intermediate primary operators ${\cal O}$ and their descendants. The function $F_{\cal O}(u,v)$ is the conformal block (or conformal partial wave) associated with the exchange of the primary operator ${\cal O}$, and it depends on the conformal cross-ratios:
\begin{equation}
u
=
\frac{x_{12}^2 x_{34}^2}{x_{13}^2 x_{24}^2},
\qquad
v
=
\frac{x_{14}^2 x_{23}^2}{x_{13}^2 x_{24}^2},
\label{eq:cross_ratios_uv}
\end{equation}
where
$x_{ij}^2 \equiv (x_i-x_j)^2$.
Thus $u$ and $v$ are the two independent conformally invariant combinations of the four insertion points, while $F_{\cal O}(u,v)$ encodes the kinematical contribution of the exchanged conformal family, and the coefficients $\lambda_{ij{\cal O}}$ encode the dynamical OPE data of the theory.
Therefore the CFT bootstrap is a nonperturbative consistency program for the operator algebra of a conformal theory.

By contrast, the GS bootstrap that we are discussing is not a bootstrap of operator dimensions or OPE coefficients. It is instead a self-consistency problem for a family of Glauber--Sudarshan states and their backreacted expectation values. The basic unknowns are the GS labels, the displacement operator data, and the semiclassical expectation values
$\langle \Xi\rangle_\sigma$,
and the bootstrap equations require that these quantities satisfy the coupled Schwinger--Dyson and backreaction conditions. In other words, the GS bootstrap does not solve for the operator algebra of a CFT; it selects those GS states for which the induced semiclassical configuration is self-consistent. Therefore the clean distinction may be presented in the following way.
The CFT bootstrap constrains the intrinsic operator data of a conformal field theory, whereas the GS bootstrap constrains the state-dependent semiclassical configurations and displacement data of Glauber--Sudarshan states. Their goals are different: the CFT bootstrap constrains the quantum correlation functions whereas the GS bootstrap fixes the semiclassical backgrounds.

\subsection{The Schwinger-Dyson equations with GS sources \label{sec7.1}}

To avoid confusion, let us first separate the partition function from the numerator with an operator insertion.  
The genuine in--in GS partition function is:
\begin{equation}
Z_\sigma
=
\int Dh_f
\int Dh_i^+\,Dh_i^-\;
\Psi_\Omega[h_i^+]\,
\Psi_\Omega^\ast[h_i^-]\,
K_{+,\sigma}[h_f,h_i^+]\,
K_{-,\sigma}[h_f,h_i^-],
\label{eq:Zsigma_start_correct}
\end{equation}
which appeared earlier in \eqref{dixtucker2}, \eqref{tricksbrke} and 
\eqref{coridahl7}
with 
$K_{+,\sigma}[h_f,h_i^+]$ and 
$K_{-,\sigma}[h_f,h_i^-]$
defined as in \eqref{eq:Kminus_def_exact}.
The numerator with an insertion of $\hat O(x)$ is 
$N_\sigma[O](x)$ given in \eqref{tricksbrke} or \eqref{coridahl7}
and the expectation value is
$\langle \hat O(x)\rangle_\sigma
=
\frac{N_\sigma[O](x)}{Z_\sigma}$. We can also define the GS insertions carefully using $\mathcal I_\pm$ or $W_\pm[g_\pm; \sigma]$ as in \eqref{eq:exact_GS_SK_h_full} and then express as:
\bg\label{eq:Zsigma_exponentiated}
Z_\sigma
 &= &
\int Dh_f
\int Dh_i^+\,Dh_i^-\;
\Psi_\Omega[h_i^+]\,
\Psi_\Omega^\ast[h_i^-] \\
& \times & \int_{h_i^+}^{h_f}{\cal D}g_+\;
\int_{h_i^-}^{h_f}{\cal D}g_-\;
\exp\!\left(
iS_{\rm tot}[g_+]
-
iS_{\rm tot}[g_-]
+
W_+[g_+;\sigma]
+
W_-[g_-;\sigma]
\right) ,\nonumber
\nd
where in the convention used in our time-sliced derivation, the $+$ and $-$ branch GS functionals appear with the same sign in the exponent, while the dynamical action appears as
$iS_{\rm tot}[g_+] - iS_{\rm tot}[g_-]$.
There is however an important sign remark:
one should \emph{not} try to absorb $W_\pm$ into $iS_{\rm tot}$ unless one is willing to define a complexified branch action.  It is safer to keep the GS functionals separate.  Then there is no ambiguity: the forward branch carries $+\,iS_{\rm tot}$ and the backward branch carries $-\,iS_{\rm tot}$, while the GS insertions appear as the additional real or complex functionals $W_\pm$.

The partition function $Z_\sigma$ in \eqref{eq:Zsigma_exponentiated} is all we need to derive the Schwinger-Dyson equation. The following derivation avoids certain pitfalls that we shall elaborate a little later. First however we need to introduce 
$g_r$ and $g_a$, and the meaning of derivatives with respect to them. They may be defined in the following way:
\begin{equation}
g_{r\,\mu\nu}
\equiv
\frac{1}{2}\big(g_{+\mu\nu}+g_{-\mu\nu}\big),
\qquad
g_{a\,\mu\nu}
\equiv \frac{1}{2}\big(
g_{+\mu\nu}-g_{-\mu\nu}\big) \nonumber
\label{eq:ra_defs_final}
\end{equation}
\begin{equation}
g_{+\mu\nu}
=
g_{r\,\mu\nu}
+
g_{a\,\mu\nu},
\qquad
g_{-\mu\nu}
=
g_{r\,\mu\nu}
-
g_{a\,\mu\nu} 
\label{eq:pm_from_ra_final}
\end{equation}
\begin{equation}
\frac{\delta}{\delta g_{r\,\mu\nu}(x)}
=
\frac{\delta}{\delta g_{+\mu\nu}(x)}
+
\frac{\delta}{\delta g_{-\mu\nu}(x)}, ~~~~
\frac{\delta}{\delta g_{a\,\mu\nu}(x)}
=
\frac{\delta}{\delta g_{+\mu\nu}(x)}
-
\frac{\delta}{\delta g_{-\mu\nu}(x)} , \nonumber
\label{eq:da_relation_final}
\end{equation}
where for the derivative we kept $g_a$ fixed to define $\frac{\delta}{\delta g_{r\,\mu\nu}(x)}$, and kept $g_r$ fixed to define $\frac{\delta}{\delta g_{a\,\mu\nu}(x)}$. With these we are ready to work out the real contour-combined Schwinger--Dyson equation. This may be easily derived\footnote{The way to derive the Schwinger-Dyson equation from a given partition function is standard so we will simply outline the steps. Vary the integration variables by a common shift:
\begin{equation}
g_{+\mu\nu}(x)\to g_{+\mu\nu}(x)+\varepsilon_{\mu\nu}(x),
\qquad
g_{-\mu\nu}(x)\to g_{-\mu\nu}(x)+\varepsilon_{\mu\nu}(x),\nonumber
\label{eq:gr_shift_SD}
\end{equation}
which is precisely a shift of $g_r$ at fixed $g_a$. Since the measure is invariant, one can easily show that this leads to \eqref{eq:SD_gr_start_final}.}
from \eqref{eq:Zsigma_exponentiated} by varying with respect to $g_r$:
\begin{equation}
0
=
\int {\cal D}g_+\,{\cal D}g_-\;
\frac{\delta}{\delta g_{r\,\mu\nu}(x)}
\exp\!\left(
iS_{\rm tot}[g_+]
-
iS_{\rm tot}[g_-]
+
W_+[g_+;\sigma]
+
W_-[g_-;\sigma]
\right) ,
\label{eq:SD_gr_start_final}
\end{equation}
where notice that we have not included the wave-function pieces, and nor have we included the integration over the $(h_i^\pm, h_f)$. These are not necessary at least to derive the Schwinger-Dyson equations from the partition function.
Using \eqref{eq:pm_from_ra_final}, this becomes:
\begin{equation}
\left\langle
i\,
\left[
\frac{\delta S_{\rm tot}[g_+]}{\delta g_{+\mu\nu}(x)}
-
\frac{\delta S_{\rm tot}[g_-]}{\delta g_{-\mu\nu}(x)}
\right]
+
\left[
\frac{\delta W_+[g_+;\sigma]}{\delta g_{+\mu\nu}(x)}
+
\frac{\delta W_-[g_-;\sigma]}{\delta g_{-\mu\nu}(x)}
\right]
\right\rangle_\sigma
=
0 ,
\label{eq:SD_gr_expanded_final}
\end{equation}
which is 
the correct real contour-combined Schwinger--Dyson equation in our convention. One may also verify that,
under complex conjugation together with interchange of branches:
\begin{equation}
g_+
\longleftrightarrow
g_-,
\label{eq:branch_swap_final}
\end{equation}
the first bracket in \eqref{eq:SD_gr_expanded_final}  changes sign, but so does the explicit factor of $i$, while the second bracket is mapped into itself provided the backward branch insertion is the appropriate conjugate counterpart of the forward one. Thus \eqref{eq:SD_gr_expanded_final} is the natural real combination.
It is therefore convenient to rewrite it as:
\begin{equation}
\left\langle
\frac{\delta S_{\rm tot}[g_+]}{\delta g_{+\mu\nu}(x)}
-
\frac{\delta S_{\rm tot}[g_-]}{\delta g_{-\mu\nu}(x)}
\right\rangle_\sigma
=
i\,
\left\langle
\frac{\delta W_+[g_+;\sigma]}{\delta g_{+\mu\nu}(x)}
+
\frac{\delta W_-[g_-;\sigma]}{\delta g_{-\mu\nu}(x)}
\right\rangle_\sigma ,
\label{eq:real_SD_master}
\end{equation}
which will be helpful to extract the bootstrap equation. Note that we could also combine the action, by choosing an appropriate contour, to replace the relative minus sign on the LHS of \eqref{eq:real_SD_master} by one action functional. This is not necessary but nevertheless useful to reorganize the Schwinger-Dyson equation in a more manageable format. We will take advantage of this soon.

\subsection{The GS-deformed Schwinger--Keldysh generating functional \label{sec7.2}}

We can also get another form of the  Schwinger-Dyson equation using the generating function in a slightly different way. 
Switching on external sources ${\cal J}_\pm^{\mu\nu}$ for the two branches, the GS-deformed generating functional may be expressed as:
\begin{align}
Z_{\rm SK}^{(\sigma)}[{\cal J}_+,{\cal J}_-]
&=
\int {\cal D}g_+\,{\cal D}g_-\,
{\cal D}({\rm gh}_+)\,{\cal D}({\rm gh}_-)\;
\rho_i[g_i^+,g_i^-]
\nonumber\\
&\qquad\times
\exp\!\left(
iS_{\rm tot}[g_+,{\rm gh}_+]
-
iS_{\rm tot}[g_-,{\rm gh}_-]
\right)
{\cal I}_+[g_+;\sigma]\,
{\cal I}_-[g_-;\sigma]
\nonumber\\
&\qquad\times
\exp\!\left(
i\int d^4x\;{\cal J}_+^{\mu\nu}(x)\,g_{+\mu\nu}(x)
-
i\int d^4x\;{\cal J}_-^{\mu\nu}(x)\,g_{-\mu\nu}(x)
\right) ,
\label{eq:ZSK_sigma_full}
\end{align}
where the relative signs for the two sources are motivated from the relative signs of the actions on the two SK branches, and $\rho_i[g_i^+, g_i^-]$ is the wavefunctional product. The GS insertions 
${\cal I}_\pm[g_\pm;\sigma]$  are defined as before in terms of 
$\exp\!\left(
W_\pm[g_\pm;\sigma]
\right)$ in the usual way. We can therefore rewrite \eqref{eq:ZSK_sigma_full} as:
\begin{align}
Z_{\rm SK}^{(\sigma)}[{\cal J}_+,{\cal J}_-]
&=
\int {\cal D}g_+\,{\cal D}g_-\,
{\cal D}({\rm gh}_+)\,{\cal D}({\rm gh}_-)\;
\rho_i[g_i^+,g_i^-]
\nonumber\\
&\qquad\times
\exp\!\left(
i\int d^4x\;{\cal J}_+^{\mu\nu}g_{+\mu\nu}
-
i\int d^4x\;{\cal J}_-^{\mu\nu}g_{-\mu\nu}
\right)
\nonumber\\
&\qquad\times
\exp\!\left(
iS_{\rm tot}[g_+,{\rm gh}_+]
-
iS_{\rm tot}[g_-,{\rm gh}_-]
+
W_+[g_+;\sigma]
+
W_-[g_-;\sigma]
\right) ,
\label{eq:ZSK_sigma_with_W}
\end{align}
where we can see how the GS deformations $W_\pm[g_\pm; \sigma]$ differ from the source insertions. This implies that the connected functional will get deformed by both $\sigma$ and $\mathcal J_\pm$, and from there we can derive the deformed classical metrics. All these may be collected together as:
\begin{equation}
g_{c\,\mu\nu}^+(x)
=
\frac{\delta W_{\rm SK}^{(\sigma)}[{\cal J}_+,{\cal J}_-]}{\delta {\cal J}_+^{\mu\nu}(x)},
\qquad
g_{c\,\mu\nu}^-(x)
=
-\,\frac{\delta W_{\rm SK}^{(\sigma)}[{\cal J}_+,{\cal J}_-]}{\delta {\cal J}_-^{\mu\nu}(x)} \nonumber
\label{eq:gcpm_sigma_def}
\end{equation}
\begin{equation}
~~W_{\rm SK}[{\cal J}_+,{\cal J}_-]
\quad\longrightarrow\quad
W_{\rm SK}^{(\sigma)}[{\cal J}_+,{\cal J}_-]
=
-\,i\ln Z_{\rm SK}^{(\sigma)}[{\cal J}_+,{\cal J}_-] ,
\label{eq:WSK_sigma_def}
\end{equation}
Thus the GS deformation changes the relation between the external sources and the mean fields, because the path integral being differentiated is no longer the undeformed one. This means the GS-deformed Schwinger--Keldysh effective action generates a correspondingly deformed map between sources and branchwise classical fields, and hence a modified effective dynamics. The correct effective action for the GS state is therefore:

{\footnotesize
\bg\label{calgarytagra}
\Gamma_{\rm SK}^{(\sigma)}[g_c^+,g_c^-]
& = &
W_{\rm SK}^{(\sigma)}[{\cal J}_+,{\cal J}_-]
-
\int d^4x\;{\cal J}_+^{\mu\nu}(x)\,g_{c\,\mu\nu}^+(x)
+
\int d^4x\;{\cal J}_-^{\mu\nu}(x)\,g_{c\,\mu\nu}^-(x)\\
& = & 
W_{\rm SK}^{(\sigma)}
\Big[
{\cal J}_+[g_c^+,g_c^-],
{\cal J}_-[g_c^+,g_c^-]
\Big]
-
\int d^4x\;
{\cal J}_+^{\mu\nu}[g_c^+,g_c^-]\,
g_{c\,\mu\nu}^+
+
\int d^4x\;
{\cal J}_-^{\mu\nu}[g_c^+,g_c^-]\,
g_{c\,\mu\nu}^- , \nonumber
\nd}
where ${\cal J}_\pm$ are re-expressed in terms of $g_c^\pm$ through \eqref{eq:gcpm_sigma_def}. (The procedure is again standard: first compute $g_c^\pm$ by differentiating $W_{\rm SK}^{(\sigma)}$ with respect to ${\cal J}_\pm$.
Then invert those relations to solve for ${\cal J}_\pm$ as functionals of $g_c^\pm$.
Finally substitute those solutions into the Legendre transform.) At this stage it is also useful to define the corresponding source combinations:
\begin{equation}
{\cal J}_{a}^{\mu\nu}
\equiv
{\cal J}_+^{\mu\nu}
-
{\cal J}_-^{\mu\nu},
\qquad
{\cal J}_{r}^{\mu\nu}
\equiv
{\cal J}_+^{\mu\nu}
+
{\cal J}_-^{\mu\nu} \nonumber
\label{eq:Jra_defs_SD}
\end{equation}
\begin{equation}
~~~~\int d^4x\;{\cal J}_+^{\mu\nu}g_{+\mu\nu}
-
\int d^4x\;{\cal J}_-^{\mu\nu}g_{-\mu\nu}
= 
\int d^4x\;{\cal J}_a^{\mu\nu}g_{r\,\mu\nu}
+ 
\int d^4x\;{\cal J}_r^{\mu\nu}g_{a\,\mu\nu} ,
\label{eq:Jg_rewrite_SD}
\end{equation}
such that the second line quantifies the currents into the symmetric and the anti-symmetric combinations much like $g_r$ and $g_a$. It also quantifies the following reality operation, that is defined as a 
complex conjugation followed by interchange of the two Schwinger--Keldysh branches in the following way:
\begin{equation}
{\cal C}:
\qquad
g_+
\longleftrightarrow
g_-,
\qquad
{\cal J}_+
\longleftrightarrow
{\cal J}_-,
\qquad
W_+[g_+;\sigma]^*
=
W_-[g_-;\sigma] ,
\label{eq:C_operation_SD}
\end{equation}
as we had discussed earlier below \eqref{eq:SD_gr_expanded_final}, but now including the sources. The operation $\mathcal C$ in \eqref{eq:C_operation_SD} is more elaborate but is a natural involution for the system to maintain reality.
Under this combined operation, the average and difference variables should transform in a way such that the conjugation and the branch interchange would compensate each other. What really happens is the following:
\begin{equation}
g_r
\longrightarrow
g_r,
\qquad
g_a
\longrightarrow
-\,g_a, \qquad {\cal J}_r
\longrightarrow
{\cal J}_r,
\qquad
{\cal J}_a
\longrightarrow
-{\cal J}_a \nonumber
\label{eq:ra_C_transform_SD}
\end{equation}
\begin{equation}
\int d^4x\;
\Big[
{\cal J}_a g_r + {\cal J}_r g_a
\Big]
\longrightarrow
-\int d^4x\;
\Big[
{\cal J}_a g_r + {\cal J}_r g_a
\Big] ,
\label{eq:source_term_flips_sign}
\end{equation}
implying that without an $i$, the source term is {\it odd} under $\mathcal C$, not even.  The standard Schwinger--Keldysh way to make the source sector transform properly is to include an explicit factor of $i$ as shown in \eqref{eq:ZSK_sigma_with_W}. This way the two minus signs cancel, and we get:
\begin{equation}
\left[
i\int d^4x\;
\Big(
{\cal J}_a g_r + {\cal J}_r g_a
\Big)
\right]^{\cal C}
=
i\int d^4x\;
\Big(
{\cal J}_a g_r + {\cal J}_r g_a
\Big) ,
\label{eq:source_invariant_with_i}
\end{equation}
so that reality is maintained. However if one insists on writing the source term without an $i$, 
then to keep the exponent invariant under ${\cal C}$ one would need the sources themselves to pick up an extra minus sign under complex conjugation, {\it i.e.}
${\cal J}_r^*
=
-\,{\cal J}_r$ and 
${\cal J}_a^*
=
{\cal J}_a$.
But this is not the standard SK convention for real physical sources.
So the clean and standard choice is to keep the factor of $i$ in the source coupling. Note however that, due to the underlying Legendre transformation, there is no $i$ in \eqref{calgarytagra}. The sequence of transformation is:
\bg\label{pharmatagra}
&& Z_{\rm SK}^{(\sigma)}[{\cal J}_+,{\cal J}_-]^{\mathcal C}
=
Z_{\rm SK}^{(\sigma)}[{\cal J}_-,{\cal J}_+] ~~\implies ~~
W_{\rm SK}^{(\sigma)}[{\cal J}_+,{\cal J}_-]^{\mathcal C}
=
-\,W_{\rm SK}^{(\sigma)}[{\cal J}_-,{\cal J}_+]\nonumber\\
\implies ~~ && \Gamma_{\rm SK}^{(\sigma)}[g_c^+,g_c^-]^{\mathcal C}
=
-\,\Gamma_{\rm SK}^{(\sigma)}[g_c^-,g_c^+] ~~\implies ~~
\Gamma_{\rm SK}^{(\sigma)}[g_r,g_a]^{\mathcal C}
=
-\,\Gamma_{\rm SK}^{(\sigma)}[g_r,-g_a] ,
\nd
so that the complex conjugate is equal to minus the effective action evaluated at the branch-swapped configuration, not minus itself at the same arguments. As a special case on the physical diagonal, where 
$g_a=0$ and 
$g_c^+=g_c^-=g_c$, equation
\eqref{pharmatagra} reduces to
$\Gamma_{\rm SK}^{(\sigma)}[g_r,0]^*
=
-\,\Gamma_{\rm SK}^{(\sigma)}[g_r,0]$
which implies\footnote{The proof is simple. The condition $\Gamma_{\rm SK}^{(\sigma)}[g_r,0]^*
=
-\,\Gamma_{\rm SK}^{(\sigma)}[g_r,0]$ implies that the diagonal value is purely imaginary, {\it i.e.}
$\Re\,
\Gamma_{\rm SK}^{(\sigma)}[g_r,0]
=
0$. The basic microscopic statement is that when the two external sources are equal on the two branches, the closed-time-path functional is normalized, {\it i.e.}
$Z_{\rm SK}^{(\sigma)}[{\cal J},{\cal J}]
=
1$.
This is the usual Schwinger--Keldysh unitarity statement: identical sources on the forward and backward branches cancel in the trace, so the closed contour gives unity. Equivalently, in $r/a$ language this means 
${\cal J}_a=0
\Longrightarrow
Z_{\rm SK}^{(\sigma)}[{\cal J}_r,{\cal J}_a=0]
=
1$. Since
$W_{\rm SK}^{(\sigma)}[{\cal J}_+,{\cal J}_-]
=
-\,i\ln Z_{\rm SK}^{(\sigma)}[{\cal J}_+,{\cal J}_-]$,
it follows immediately that
$W_{\rm SK}^{(\sigma)}[{\cal J},{\cal J}]
=
0$,
or equivalently
$W_{\rm SK}^{(\sigma)}[{\cal J}_r,{\cal J}_a=0]
=
0$. Plugging this in the deformed Schwinger-Keldysh effective action \eqref{calgarytagra}, and in the limit ${\cal J}_+^{\mu\nu}={\cal J}_-^{\mu\nu}\equiv{\cal J}^{\mu\nu}$ and $g_c^+=g_c^-\equiv g_c$, we get:
\bg
\Gamma_{\rm SK}^{(\sigma)}[g_c,g_c]
=
0 \qquad \Longrightarrow \qquad
\Gamma_{\rm SK}^{(\sigma)}[g_r,0]
=
0, \nonumber \nd
justifying the vanishing for all $g_r$ with $g_a = 0$. Thus the $\mathcal C$ operation only shows that the diagonal value of the Schwinger--Keldysh effective action is purely imaginary. By itself it does not imply vanishing. Once combined with SK normalization, or the unitarity condition, for $Z_{\rm SK}^{(\sigma)}$, the vanishing of 
$\Gamma_{\rm SK}^{(\sigma)}[g_r,0]$ appears. \label{sammibhalo}} 
$\Gamma_{\rm SK}^{(\sigma)}[g_r,0]=0$. This is consistent with the SK formalism,
so on the diagonal this is trivially satisfied. After the dust settles, we can conclude that the Schwinger--Keldysh effective action is not simply anti-Hermitian at fixed arguments. Its correct reality property is branch-swapped anti-Hermiticity as explained in the last stage of the sequence \eqref{pharmatagra}.

\subsection{The Schwinger-Dyson equations with GS and $\mathcal J_{\pm\mu\nu}$ sources \label{sec7.3}}

Our first attempt to derive the Schwinger-Dyson equation in \eqref{eq:real_SD_master} revealed an interesting interplay between $S_{\rm tot}[g_\pm]$ and the GS source terms $W_\pm[g_\pm; \sigma]$. But \eqref{eq:real_SD_master} is not the full story because the $\mathcal J_{\pm\mu\nu}$ sources could also play a role. In the following we will derive the exact microscopic Schwinger--Dyson identities in the presence of both the GS and the $\mathcal J_{\pm \mu\nu}$ sources by shifting the integration variables. The only assumption is that the functional measure is invariant under the infinitesimal changes of variables used below. Assuming the measure allows no anomalous Jacobian, we can start with the fundamental branchwise identities by varying $g_\pm$ in two possible ways:
\bg\label{tagradekhche1}
&& g_{+\mu\nu}(x)\to g_{+\mu\nu}(x)+\delta g_{+\mu\nu}(x),
\qquad
g_{-\mu\nu}(x)\to g_{-\mu\nu}(x) \nonumber\\
&& g_{+\mu\nu}(x)\to g_{+\mu\nu}(x),
\qquad
g_{-\mu\nu}(x)\to g_{-\mu\nu}(x)+\delta g_{-\mu\nu}(x) ,
\nd
which is an allowed set of variations because the 
Schwinger--Dyson identities come from the statement that the path integral is invariant under \emph{any} infinitesimal change of integration variables. Therefore one may choose different infinitesimal changes separately and obtain different valid identities.
In fact imposing \eqref{tagradekhche1}, we get the folloiwng two set of equations:
\bg\label{tagdekh2}
&& \left\langle
\frac{\delta S_{\rm tot}[g_+]}{\delta g_{+\mu\nu}(x)}
+
{\cal J}_+^{\mu\nu}(x)
\right\rangle_\sigma
=
i
\left\langle
\frac{\delta W_+[g_+;\sigma]}{\delta g_{+\mu\nu}(x)}
\right\rangle_\sigma \nonumber\\
&& \left\langle
\frac{\delta S_{\rm tot}[g_-]}{\delta g_{-\mu\nu}(x)}
+
{\cal J}_-^{\mu\nu}(x)
\right\rangle_\sigma
=
-\,i
\left\langle
\frac{\delta W_-[g_-;\sigma]}{\delta g_{-\mu\nu}(x)}
\right\rangle_\sigma. 
\nd
These two equations are the fundamental microscopic branchwise Schwinger--Dyson identities. They are not individually ordinary real equations. Rather, they are exchanged into one another by the full reality operation ${\cal C}$. This is consistent because the Schwinger--Keldysh measure is complex. One could however write \eqref{tagdekh2} in a $\mathcal C$-covariant form by making the following shifts:
\bg\label{tagdekh3}
&& g_{+\mu\nu}(x)\to g_{+\mu\nu}(x)+\varepsilon_{\mu\nu}(x),
\qquad
g_{-\mu\nu}(x)\to g_{-\mu\nu}(x)+\varepsilon_{\mu\nu}(x) \nonumber\\
&& g_{+\mu\nu}(x)\to g_{+\mu\nu}(x)+\eta'_{\mu\nu}(x),
\qquad
g_{-\mu\nu}(x)\to g_{-\mu\nu}(x)-\eta'_{\mu\nu}(x), \nd
where the first is precisely a shift of $g_r$ at fixed $g_a$ because
$g_r\to g_r+\varepsilon$ but 
$g_a\to g_a$, whereas the second is a shift of $g_a$ at fixed $g_r$ because
$g_r\to g_r$ but
$g_a\to g_a+\eta'$, where $\eta'_{\mu\nu}$ is a generic small shift. They lead, respectively, to the following two set of equations:
\bg\label{tagdekh4}
&& \left\langle
\frac{\delta S_{\rm tot}[g_+]}{\delta g_{+\mu\nu}(x)}
-
\frac{\delta S_{\rm tot}[g_-]}{\delta g_{-\mu\nu}(x)}
+
{\cal J}_a^{\mu\nu}(x)
\right\rangle_\sigma
=
i
\left\langle
\frac{\delta W_+[g_+;\sigma]}{\delta g_{+\mu\nu}(x)}
+
\frac{\delta W_-[g_-;\sigma]}{\delta g_{-\mu\nu}(x)}
\right\rangle_\sigma \nonumber\\
&& \left\langle
\frac{\delta S_{\rm tot}[g_+]}{\delta g_{+\mu\nu}(x)}
+
\frac{\delta S_{\rm tot}[g_-]}{\delta g_{-\mu\nu}(x)}
+
{\cal J}_r^{\mu\nu}(x)
\right\rangle_\sigma
=
i
\left\langle
\frac{\delta W_+[g_+;\sigma]}{\delta g_{+\mu\nu}(x)}
-
\frac{\delta W_-[g_-;\sigma]}{\delta g_{-\mu\nu}(x)}
\right\rangle_\sigma , 
\nd
which may now be compared to \eqref{eq:real_SD_master}. Both of these set of equations are ${\cal C}$-covariants: in the first equation, the left bracket is odd under branch exchanges, so the prefactor $i$ in right bracket makes it odd; whereas in the second equation, the left bracket is even under branch exchanges, while the $i$ on the right bracket makes it also even. 

Let us also make the following comment on the relation among the four microscopic identities. The $g_r$ and $g_a$ identities are just linear combinations of the branchwise identities \eqref{tagdekh2}. Indeed, adding and subtracting those two branchwise equations reproduces \eqref{tagdekh4}. This is not an inconsistency. It simply means that the microscopic doubled system may be presented either in branchwise form or in $r/a$-combined form. However what must \emph{not} be done is to impose ordinary single-branch reality on one branchwise equation in isolation. The expectation values are taken with respect to the complex Schwinger--Keldysh weight, so an expectation value of a formally real branch functional need not itself be real. In the source-free problem one sets
${\cal J}_\pm^{\mu\nu}=0$
which implies
${\cal J}_r^{\mu\nu}=0 = 
{\cal J}_a^{\mu\nu}$.
With the contour interpretation adopted here, the natural exact microscopic Schwinger--Dyson identity is the contour-combined equation:
\begin{equation}
\left\langle
i
\left[
\frac{\delta S_{\rm tot}[g_+]}{\delta g_{+\mu\nu}(x)}
-
\frac{\delta S_{\rm tot}[g_-]}{\delta g_{-\mu\nu}(x)}
\right]
+
\left[
\frac{\delta W_+[g_+;\sigma]}{\delta g_{+\mu\nu}(x)}
+
\frac{\delta W_-[g_-;\sigma]}{\delta g_{-\mu\nu}(x)}
\right]
\right\rangle_\sigma
=
0 ,
\label{eq:microscopic_contour_identity_corrected_section}
\end{equation}
where we note that the action terms come with a relative minus sign whereas the GS source terms come with a relative plus sign. The relative signs are fixed as follows. The dynamical action appears on the forward and backward branches with opposite contour orientation, so that we can express the difference succinctly as:
\begin{equation}
S_{\rm tot}[g_+]-S_{\rm tot}[g_-]
=
\int_{t_i}^{t_f} dt\;L[g_+]
+
\int_{t_f}^{t_i} dt\;L[g_-],
\label{eq:action_closed_contour_corrected_section}
\end{equation}
which is precisely the full action on the closed time contour. (The backward contour contribution cancels the forward one only when the two branch histories coincide. In the Schwinger--Keldysh path integral, however, $g_+$ and $g_-$ are independent dummy variables before the physical limit is imposed, so $S_{\rm tot}[g_+]-S_{\rm tot}[g_-]$ is generically nonzero\footnote{ Moreover, the integral vanishes only for a {\it trivial path} that stays at a given point. A nontrivial closed loop beginning and ending at a point is not the trivial path, so its integral need not vanish. Stated differently,
a contour integral is not determined only by the initial and final points unless the integrand is an exact differential. In general:   
\begin{equation}
\int_a^b \omega
+
\int_b^c \omega
+
\int_c^a \omega
=
\int_{\zeta_1} \omega
+
\int_{\zeta_2} \omega + \int_{\zeta_3} \omega
=
\int_{\zeta_1\circ \zeta_2\circ\zeta_3}\omega =  
\oint_C \omega \ne \int_a^a \omega , \nonumber
\label{eq:concatenation_true}
\end{equation}
where $\zeta_1\circ\zeta_2\circ \zeta_3$ means the concatenated path. Thus only if we integrate the {\it same exact differential} along successive segments, then we may concatenate the segments into one path. But for a general contour integral, the value depends on the whole path, not just on its endpoints. In other words if $\omega = dF$ for some form $F$, the integral is determined by the end points and it vanishes. Otherwise in general it doesn't.}.) This is why the action sector enters through the difference:
\begin{equation}
\frac{\delta S_{\rm tot}[g_+]}{\delta g_{+\mu\nu}}
-
\frac{\delta S_{\rm tot}[g_-]}{\delta g_{-\mu\nu}}.
\label{eq:action_difference_corrected_section}
\end{equation}
By contrast, the GS insertion sector is not a contour action of the same kind. It arises from the ordered pair of operators $D(\sigma)$ and $ D^\dagger(\sigma)$, with the latter already anti-time-ordered on the backward branch. Hence its branch structure is already built into the definition of the two functionals $W_\pm[g_\pm;\sigma]$, and the correct microscopic combination is therefore:
\begin{equation}
\frac{\delta W_+[g_+;\sigma]}{\delta g_{+\mu\nu}}
+
\frac{\delta W_-[g_-;\sigma]}{\delta g_{-\mu\nu}},
\label{eq:W_sum_corrected_section}
\end{equation}
rather than a difference. This is crucial and lies at the heart of our construction. Moreover  \eqref{eq:microscopic_contour_identity_corrected_section} is the Lorentzian contour analogue of the Euclidean Schwinger--Dyson identity:
\begin{equation}
\left\langle
\frac{\delta S_E[g]}{\delta g_{\mu\nu}(x)}
-
\frac{\delta W_E[g;\sigma]}{\delta g_{\mu\nu}(x)}
\right\rangle_\sigma
=
0,
\label{eq:Euclidean_SD_corrected_section}
\end{equation}
up to the usual convention-dependent Euclidean sign choices. In this sense \eqref{eq:microscopic_contour_identity_corrected_section} is the microscopic identity that is both contour-real and smoothly compatible with the Euclidean description. It is important, however, not to confuse this exact microscopic identity with the effective mean-field equation obtained after the Legendre transform. Once the Schwinger--Keldysh effective action is introduced, one must carefully identify which variable plays the role of the physical metric and which variation reproduces the contour-combined equation above. To this effect, let us define: \begin{equation} g_{r\,\mu\nu} = \frac{1}{2} \big( g_{+\mu\nu}+g_{-\mu\nu} \big), \qquad g_{a\,\mu\nu} = \frac{1}{2} \big( g_{+\mu\nu}-g_{-\mu\nu} \big), \label{eq:ra_defs_corrected_section} \end{equation} so that the physical limit is $g_{a\,\mu\nu}=0$ which would imply that $g_{+\mu\nu}=g_{-\mu\nu}\equiv g_{c\,\mu\nu}$. One may also use the $g_r$ and $g_a$ metric factors to rewrite the source term $\mathcal J^{\mu\nu}_\pm$ appearing in \eqref{eq:ZSK_sigma_full} in the following suggestive way: \begin{equation} \int d^4x\; {\cal J}_+^{\mu\nu}g_{+\mu\nu} - \int d^4x\; {\cal J}_-^{\mu\nu}g_{-\mu\nu} = \int d^4x\; {\cal J}_a^{\mu\nu}g_{r\,\mu\nu} + \int d^4x\; {\cal J}_r^{\mu\nu}g_{a\,\mu\nu}, \label{eq:source_rewrite_corrected_section} \end{equation} where $\mathcal J_r^{\mu\nu} = \mathcal J_+^{\mu\nu} + \mathcal J_-^{\mu\mu}$ and $\mathcal J_a^{\mu\nu} = \mathcal J_+^{\mu\nu} - \mathcal J_-^{\mu\mu}$ are the symmetric and the anti-symmetric combination of sources respectively. In fact we can use these redefined sources to neatly pack the Schwinger-Dyson equations \eqref{tagdekh4} in terms of the Legendre transform $\Gamma^{(\sigma)}_{\rm SK}[g_r, g_a]$ in the following way: \begin{equation} \frac{\delta \Gamma_{\rm SK}^{(\sigma)}[g_r,g_a]} {\delta g_r^{\mu\nu}(x)} = -\,{\cal J}_{a\,\mu\nu}(x), \qquad \frac{\delta \Gamma_{\rm SK}^{(\sigma)}[g_r,g_a]} {\delta g_a^{\mu\nu}(x)} = -\,{\cal J}_{r\,\mu\nu}(x). \label{eq:Legendre_relations_corrected_section} \end{equation}
Since the microscopic contour identity \eqref{eq:microscopic_contour_identity_corrected_section} is precisely the one accompanied by ${\cal J}_a$, one might naively conclude that its proper effective-action counterpart is obtained by varying with respect to $g_r$, not with respect to $g_a$. In fact from the two allowed possibilities:
\begin{equation}
\left.
\frac{\delta \Gamma_{\rm SK}^{(\sigma)}[g_r,g_a]}
{\delta g_r^{\mu\nu}(x)}
\right|_{g_a=0}
=
0, \qquad {\rm or} \qquad
\left.
\frac{\delta \Gamma_{\rm SK}^{(\sigma)}[g_r,g_a]}
{\delta g_a^{\mu\nu}(x)}
\right|_{g_r=0}
=
0 ,
\label{eq:naive_ga_not_eom_cleaned}
\end{equation}
one should not conclude that the physical mean-field equation is obtained directly from {\it either} of the two equations in \eqref{eq:naive_ga_not_eom_cleaned}. This is because 
the doubled Schwinger--Keldysh effective action satisfies the following identity:
\begin{equation}
\Gamma_{\rm SK}^{(\sigma)}[g_r,0]=0, 
~~~ \forall g_r \qquad \implies ~~~~ 
\left.
\frac{\delta \Gamma_{\rm SK}^{(\sigma)}[g_r,g_a]}
{\delta g_r^{\mu\nu}(x)}
\right|_{g_a=0}
=
0 ,
\label{eq:dr_zero_identically_cleaned}
\end{equation}
holds identically. This is a tangential derivative along the vanishing locus
$g_a=0$, so it carries no dynamical information. On the other hand, while the derivative with respect to $g_a$ is transverse to the diagonal and need not vanish identically, it also does not preserve the contour-oriented difference structure of:
\begin{equation}
S[g_+]-S[g_-],
\label{eq:contour_oriented_action_cleaned}
\end{equation}
which, in the present prescription, is the fundamental object representing the dynamical action on the closed time contour. Thus neither of the two equations in \eqref{eq:naive_ga_not_eom_cleaned} should be taken as the primary physical Einstein equation.
The correct physical equation is instead the contour-combined single-field equation extracted from the exact microscopic identity:
\begin{equation}
\left\langle
i
\left[
\frac{\delta S_{\rm tot}[g_+]}{\delta g_{+\mu\nu}(x)}
-
\frac{\delta S_{\rm tot}[g_-]}{\delta g_{-\mu\nu}(x)}
\right]
+
\left[
\frac{\delta W_+[g_+;\sigma]}{\delta g_{+\mu\nu}(x)}
+
\frac{\delta W_-[g_-;\sigma]}{\delta g_{-\mu\nu}(x)}
\right]
\right\rangle_\sigma
\Bigg|_{\langle g_+\rangle_\sigma =\langle g_-\rangle_\sigma =g_c}
=
0 ,
\label{eq:microscopic_E_functional_cleaned}
\end{equation}
where the condition
$\langle g_+\rangle_\sigma
=
\langle g_-\rangle_\sigma
=
g_c$ means only that the \emph{one-point functions} on the two branches coincide. It does not mean that all higher-point functions or all nonlinear functionals coincide branchwise in the naive way. Note that the equality of the branchwise mean fields does not collapse nonlinear or trans-series expectation values to the same functional evaluated at $g_c$: correlators and higher moments still survive. Keeping this in mind, \eqref{eq:microscopic_E_functional_cleaned} is the exact microscopic contour identity appropriate to the present prescription: the action sector appears as a difference because it already carries the closed-contour orientation, while the insertion sector appears as a sum because the operator ordering of $\mathbb D(\sigma)$ and $\mathbb D^\dagger(\sigma)$ is already built into the definitions of $W_+$ and $W_-$. It is this equation, rather than the naive doubled-field variations \eqref{eq:naive_ga_not_eom_cleaned}, that matches naturally onto the Euclidean picture. It is therefore convenient to introduce a reduced single-field functional
$\Gamma_{\rm phys}^{(\sigma)}[g_c]$.
A clean way to do this is to first introduce:
\begin{equation}
{\cal E}^{\mu\nu}_{\rm SK}[g_+,g_-;\sigma](x)
\equiv
\left\langle
i
\left[
\frac{\delta S_{\rm tot}[g_+]}{\delta g_{+\mu\nu}(x)}
-
\frac{\delta S_{\rm tot}[g_-]}{\delta g_{-\mu\nu}(x)}
\right]
+
\left[
\frac{\delta W_+[g_+;\sigma]}{\delta g_{+\mu\nu}(x)}
+
\frac{\delta W_-[g_-;\sigma]}{\delta g_{-\mu\nu}(x)}
\right]
\right\rangle_\sigma ,
\label{eq:E_SK_general_best}
\end{equation}
which is the most general object one can define without making any closure or factorization assumption. In particular, it remains meaningful even when the expectation values contain polynomial, nonlocal, or trans-series dependence on the branch metrics through higher correlators such as
$\langle g_+^2\rangle_\sigma,
\langle g_+g_-\rangle_\sigma,
\langle g_+^3\rangle_\sigma$, et cetera.
The physical diagonal sector is then selected \emph{afterwards} by imposing equality of the branchwise mean fields 
$g^\pm_{c\,\mu\nu}(x)
\equiv
\langle g_{\pm\mu\nu}(x)\rangle_\sigma$
followed by
$g^+_{c\,\mu\nu}(x)
=
g^-_{c\,\mu\nu}(x)
=
g_{c\,\mu\nu}(x)$.
Thus the physical equation is best written as:
\begin{equation}
{\cal E}^{\mu\nu}_{\rm SK}[g_+,g_-;\sigma](x)
\Big|_{\langle g_+\rangle_\sigma=\langle g_-\rangle_\sigma=g_c}
=
0 ,
\label{eq:E_SK_physical_best}
\end{equation}
which avoids any potential inconsistency, because one is not assuming that the equation must be the functional derivative of an action depending only on $g_c$. Instead, \eqref{eq:E_SK_physical_best} keeps the full doubled structure intact and only then selects the physical mean-field sector. One can now define $\Gamma_{\rm phys}^{(\sigma)}[g_c]$
implicitly by the condition:
\bg\label{katadelga}
\frac{1}{\sqrt{-g_c(x)}}
\frac{\delta \Gamma_{\rm phys}^{(\sigma)}[g_c]}
{\delta g_{c\, \mu\nu}(x)}
 = 
{\cal E}^{\mu\nu}[g_c;\sigma] 
& \equiv & 
\mathcal E_{\rm SK}^{\mu\nu}[g_+, g_-, W_\pm = 0; \sigma] \Big|_{\langle g_+\rangle_\sigma=\langle g_-\rangle_\sigma=g_c}^{\langle {X}^n\rangle_\sigma \approx \langle{ X}\rangle^n_\sigma} \\
& = & 
\left\langle
i
\left[
\frac{\delta S_{\rm tot}[g_+]}{\delta g_{+\mu\nu}(x)}
-
\frac{\delta S_{\rm tot}[g_-]}{\delta g_{-\mu\nu}(x)}
\right]
\right\rangle_\sigma
\Bigg|^{\langle {X}^n\rangle_\sigma \approx \langle{ X}\rangle^n_\sigma}_{\langle g_+\rangle_\sigma =\langle g_-\rangle_\sigma =g_c}, \nonumber
\nd
which provides the necessary physical mean-field equation as the vanishing of the LHS of \eqref{katadelga}, where 
$\langle {X}^n\rangle_\sigma \approx \langle{ X}\rangle^n_\sigma$ implies that we are neglecting the connected higher correlators at the mean-field level. The equation \eqref{katadelga} also justifies that $\Gamma_{\rm phys}^{(\sigma)}[g_c]$
is not the exact doubled quantum effective action; it is the reduced mean-field effective action obtained after diagonal reduction and factorization.
Note that once the Schwinger--Keldysh sources are introduced, and as evident from above, the branchwise classical fields are precisely the branchwise one-point functions:
\begin{equation}
g_{c\,\mu\nu}^+(x)
=
\langle g_{+\mu\nu}(x)\rangle_{\sigma,{\cal J}_+,{\cal J}_-},
\qquad
g_{c\,\mu\nu}^-(x)
=
\langle g_{-\mu\nu}(x)\rangle_{\sigma,{\cal J}_+,{\cal J}_-},
\label{eq:gcpm_are_branchwise_means_final}
\end{equation}
with the physical diagonal metric obtained in the limit
$g_{c\,\mu\nu}^+(x)=g_{c\,\mu\nu}^-(x)=g_{c\,\mu\nu}(x)$.
Thus there are really {\it three} distinct objects in the discussion:
(i) the microscopic integration variables $g_{\pm\mu\nu}$,
(ii) the branchwise mean fields $g^\pm_{c\,\mu\nu}$ obtained from $W_{\rm SK}^{(\sigma)}$, and
(iii) the single diagonal mean field $g_{c\,\mu\nu}$ obtained by imposing
$g^+_{c\,\mu\nu}=g^-_{c\,\mu\nu}=g_{c\,\mu\nu}$.
So the correct statement is that
$g_{c\,\mu\nu}^\pm(x)$ appearing in \eqref{eq:WSK_sigma_def}
are the branchwise expectation values, whereas the object obtained by simply setting the microscopic variables equal, {\it i.e.}
$g_{+\mu\nu}=g_{-\mu\nu}$,
defines only the diagonal subspace of the doubled microscopic configuration space and is not by itself a derivative of $W_{\rm SK}^{(\sigma)}$. The distinction is therefore between the microscopic contour variables and the derived branchwise mean fields\footnote{Inside the path integral one integrates over two independent dummy variables
$g_{+\mu\nu}(x)$ and
$g_{-\mu\nu}(x)$.
These are not expectation values, but integration variables.
By contrast,
$g_{c\,\mu\nu}^\pm(x)$
are derived quantities obtained from
$W_{\rm SK}^{(\sigma)}$
through differentiation with respect to the sources, and are therefore equal to the branchwise one-point functions
$\langle g_{\pm\mu\nu}(x)\rangle_{\sigma,{\cal J}_+,{\cal J}_-}$.
The physical diagonal condition should therefore be imposed at the level of these branchwise mean fields,
$g^+_{c\,\mu\nu}=g^-_{c\,\mu\nu}=g_{c\,\mu\nu}$,
rather than directly on the microscopic integration variables themselves.}. Putting everything together, the proper mean-field-action version of the contour-combined microscopic Schwinger--Dyson identity is then the following:
\begin{equation}
\frac{1}{\sqrt{-g_c(x)}}
\frac{\delta \Gamma_{\rm phys}^{(\sigma)}[g_c]}
{\delta g_c^{\mu\nu}(x)}
=
0 ,
\label{eq:physical_mean_field_equation_cleaned}
\end{equation}
which, as we shall see later, will help us to determine the semiclassical trajectories.
Only after this reduced physical equation has been identified does it become natural to define the GS-induced source term. 
With this in mind, the physical equation may be rewritten in the sourced form:
\begin{equation}
\frac{1}{\sqrt{-g_c(x)}}
\frac{\delta \Gamma^{(\sigma)}_{\rm phys}[g_c]}
{\delta g_c^{\mu\nu}(x)}
=
\Sigma_{\mu\nu}^{(\sigma)}(x) ,
\label{eq:actual_SD_with_source_master}
\end{equation}
which basically tells us what contributions do the displacement operators ${D}[\sigma]$ and ${D}^\dagger[\sigma]$ as well as other off-diagonal pieces have on the single-field functional $\Gamma_{\rm phys}^{(\sigma)}[g_c]$. Thus $\Sigma^{(\sigma)}_{\mu\nu}$ is not introduced independently at the microscopic level. Rather, it is extracted only after the doubled contour problem has been reduced to the single-field physical equation.
For gravity, the undeformed variation in \eqref{eq:actual_SD_with_source_master} should yield some equivalent of the ordinary Einstein tensor, but the story gets a bit non-trivial because of the presence of $i$ in \eqref{eq:microscopic_E_functional_cleaned}, although the terms with $i$ and the terms without $i$ are {\it both real}\footnote{This is easy to see from what we discussed earlier: a $\ast$ operation will flip the sign of the $i$ terms, but then we can exchange the dummy variables $g_{+\,\mu\nu} \leftrightarrow g_{-\,\mu\nu}$ to restore the sign. The $i$ independent terms are real, so the dummy variable exchange does nothing.}. To avoid unnecessary complications, we can go to the Euclidean signature and express the Schwinger-Dyson equation in the following suggestive way:
\begin{equation}
G_{\mu\nu}[\langle {g}\rangle_\sigma]
=
8\pi G\,
\langle T_{\mu\nu}[g]\rangle_\sigma
+
\langle\widetilde\Sigma_{\mu\nu}^{(\sigma)}[g]\rangle_\sigma,
\label{eq:Einstein_with_deformed_Gamma_source}
\end{equation}
where $\langle {g}_{\mu\nu}\rangle_\sigma \equiv {g_{c\,\mu\nu}} = \langle {g}_{+\,\mu\nu}\rangle_\sigma = \langle {g}_{-\,\mu\nu}\rangle_\sigma$; and  we used $\langle\widetilde\Sigma_{\mu\nu}^{(\sigma)}[g]\rangle_\sigma$ instead of 
$\Sigma_{\mu\nu}^{(\sigma)}$. They are related, but the story is a bit more subtle than what appears at the face-value from \eqref{eq:Einstein_with_deformed_Gamma_source}. We will come back to this when we study the bootstrap equations in the next section. Meanwhile,  the expectation value
$\langle T_{\mu\nu}[g]\rangle_\sigma$
is the genuine in--in stress tensor in the GS-deformed state that will depend on matter, ghosts and the trans-series form of the action. 
It is also important to emphasize that, in general,
$\widetilde\Sigma_{\mu\nu}^{(\sigma)}$
should \emph{not} be identified naively with a purely microscopic branchwise expression such as:
\begin{equation}
\frac{i}{\sqrt{-g_c(x)}}
\left(
\frac{\delta W_+[g_+;\sigma]}{\delta g^{\mu\nu}_+(x)}
+
\frac{\delta W_-[g_-;\sigma]}{\delta g^{\mu\nu}_-(x)}
\right),
\label{eq:Sigma_not_naive_W_cleaned}
\end{equation}
with $g_c$ defined above,
because the exact object entering the physical mean-field equation is not obtained by simply reading off the branchwise $W_\pm$ variation. A formula of the form \eqref{eq:Sigma_not_naive_W_cleaned} may at best be regarded as a semiclassical or tree-level proxy. The precise GS-induced source term is defined only after one has passed through the doubled Schwinger--Keldysh construction and reduced it to the single-field physical equation.
Accordingly, the correct logical chain may be expressed in the following way:
\begin{equation}
Z_{\rm SK}^{(\sigma)}
\;\longrightarrow\;
W_{\rm SK}^{(\sigma)}
\;\longrightarrow\;
\Gamma_{\rm SK}^{(\sigma)}
\;\longrightarrow\;
{\cal E}_{\mu\nu}[g_c;\sigma]
\;\longrightarrow\;
\Gamma_{\rm phys}^{(\sigma)}[g_c]
\;\longrightarrow\;
\frac{\delta \Gamma_{\rm phys}^{(\sigma)}}{\delta g^{\mu\nu}_c}
=\sqrt{-g_c}\, \Sigma^{(\sigma)}_{\mu\nu},
\label{eq:correct_logic_chain_cleaned}
\end{equation}
where ${\cal E}_{\mu\nu}[g_c;\sigma]$ is the contour-combined equation-of-motion functional defined in \eqref{katadelga}. In this way the doubled Schwinger--Keldysh formalism remains the correct microscopic organizing framework, but the actual physical equation is formulated only after reduction to the single-field functional whose variation reproduces the contour-combined identity.

\subsection{The bootstrapping procedure and the bootstrap equation \label{sec7.4}}

The Schwinger-Dyson equation \eqref{eq:actual_SD_with_source_master} studied in the previous sub-section led us to the EOM of the form \eqref{eq:Einstein_with_deformed_Gamma_source}. On the other hand, in the absence of sources, one could also get the branchwise SD equations that take the form:
\begin{equation}
\left\langle
i\frac{\delta S_{\rm tot}[g_+]}{\delta g_{+\mu\nu}(x)}
+
\frac{\delta W_+[g_+;\sigma]}{\delta g_{+\mu\nu}(x)}
\right\rangle_\sigma
=
0 =
\left\langle
-\,i\frac{\delta S_{\rm tot}[g_-]}{\delta g_{-\mu\nu}(x)}
+
\frac{\delta W_-[g_-;\sigma]}{\delta g_{-\mu\nu}(x)}
\right\rangle_\sigma .
\label{eq:minus_branchwise_old}
\end{equation}
These are correct as formal identities, but by themselves they are not the final real physical equation.  The reason is what we pointed out earlier: they mix real and imaginary pieces branchwise, although this is compensated by the measure.
By contrast, the $g_r$ and $g_a$ variations produce the combined equations \eqref{tagdekh4}, in the presence of the $\mathcal J_\pm$ sources. They are both contour-real objects and are therefore the correct starting point for the physical mean-field and bootstrap analysis in this convention. (A careful analysis however suggests that the $g_r$ variation leading to \eqref{eq:microscopic_E_functional_cleaned} does provide the correct contour-consistent equations coinciding with the ones we get from the Euclidean formalism, so we will stick with the $g_r$ variation.) With this in mind, let us 
split the stress tensor expectation value in the following way:
\begin{equation}
\langle T_{\mu\nu}[g]\rangle_\sigma
=
T_{\mu\nu}^{\rm non\!-\!pert}({g}_c)
+
\Delta T_{\mu\nu}^{(\sigma)}({g}_c, \phi^\pm_c),
\end{equation}
where we have ignored the ghosts and the matter contributions so that the energy-momentum
tensor is from the gravitational sector, namely, the possible higher-derivative corrections.
We have also defined:
\begin{equation}
{g}_c = \langle {g}_{+\mu\nu} \rangle_\sigma = \langle g_{-\mu\nu}\rangle_\sigma ,
\qquad
\phi^\pm_c= {\langle \sigma' | {g}_{\pm\mu\nu} |\sigma \rangle \over \sqrt{\langle\sigma'|\sigma\rangle \langle\sigma |\sigma\rangle}},
\label{eq:target_fields_final}
\end{equation}
with $\sigma' \ne \sigma$. Thus $|\sigma'\rangle$ is another GS state that appear as intermediate state that do not produce on-shell states. They typically give rise to the off-shell fields\footnote{As in \cite{hetborel}, we need to carefully distinguish between off-shell states/DOFs and off-shell fields. For example $\varphi$ is an off-shell state/DOF if $\langle {\varphi} \rangle_\sigma = 0$, whereas if $\varphi$ appears as a generic field configuration in the path-integral then it is an off-shell field. We will provide a concrete example soon.}.
We have also defined $\Delta T_{\mu\nu}^{(\sigma)}(g_c,\phi^\pm_c)$ such that $ \Delta T_{\mu\nu}^{(\sigma)}(g_c,0) = 0$. Therefore putting everything together, the bootstrap equations are:
\bg\label{bootsara}
G_{\mu\nu}[{g}_c]
=
8\pi G\,
T_{\mu\nu}^{\rm non\!-\!pert}({g}_c), ~~~ \langle\widetilde\Sigma^{(\sigma)}_{\mu\nu}(x)\rangle_\sigma = -8\pi G\, \Delta T^{(\sigma)}_{\mu\nu}\left({g}_c,\phi^\pm_c\right)(x) , 
\nd
We now rewrite the bootstrap discussion in a way that is consistent with the in--in formalism developed earlier, with the corrected distinction between off-shell {\it states} and ordinary off-shell {\it fields}, and with the corrected split of the Schwinger--Dyson relations for the off-shell-state sector and the on-shell sector. 
The basic dynamical mean-field equation is the GS-deformed in--in field 
equation given by:
\begin{equation}
{1\over \sqrt{-g_c}}
\frac{\delta \Gamma_{\rm phys}^{(\sigma)}[\langle\Xi\rangle_\sigma]}
{\delta g_c^{\mu\nu}(x)}
=
0,
\label{eq:bootstrap_master_Gamma}
\end{equation}
which is basically \eqref{eq:physical_mean_field_equation_cleaned} but now written in terms of additional matter and ghosts, {\it i.e.}
$\Xi(x)=
(g_{\mu\nu}(x),\text{matter}(x), \Upsilon_g)$ and
$\Upsilon_g
\equiv
(\text{ghosts})$, although the ghost contributions are only implicit through SK path-integral weights because $\langle\Upsilon_g\rangle_\sigma = 0$.
Thus there is no need to write an additional pair $(g_r,g_a)$ variables used earlier as they are now part of the definition of $g_c$ and $\Xi(x)$.  
The role of the bootstrap equations is to express the content of
\eqref{eq:bootstrap_master_Gamma}
after separating the mean fields into two classes:
\begin{equation}
\Xi_{\rm on}
\subset \Xi
\qquad\text{with}\qquad
\langle \Xi_{\rm on}\rangle_\sigma \neq 0,
\label{eq:Xi_on_final}
\end{equation}
\begin{equation}
\Xi_{\rm off}
\not\subset \Xi
\qquad\text{with}\qquad
\langle \Xi_{\rm off}\rangle_\sigma = 0 ,
\label{eq:Xi_off_final}
\end{equation}
where the phrase ``off-shell states'' refers only to the second set, namely those components whose {\it one-point functions} vanish in the chosen GS state. This should not be confused with the ordinary off-shell fields of the functional integral, which remain present in the usual way before one integrates them out.

A key structural point is that expectation values of composite operators in the GS state naturally split into a diagonal piece plus a sum over intermediate channels. Schematically, if $X$ is any operator, then one may write:
\begin{equation}
\langle X^2\rangle_\sigma
=
\langle X\rangle_\sigma^2
+
\sum_{\sigma' \neq \sigma}
\langle X^2\rangle_{(\sigma'|\sigma)},
\qquad
\langle X^2\rangle_{(\sigma'|\sigma)}
\equiv
\frac{
\langle \sigma|X|\sigma'\rangle
\langle \sigma'|X|\sigma\rangle
}{
\langle \sigma|\sigma\rangle
\langle \sigma'|\sigma'\rangle
} + ... ,
\label{eq:X2_offdiag_def_final}
\end{equation}
where the dotted terms incorporate the additional corrections to the standard resolution of identity as shown in \cite{hetborel}. (These subtle nuances, while important for a full quantitative analysis of the bootstrap equations, will not be explicitly used here.)
The logic \eqref{eq:X2_offdiag_def_final} applies to higher powers and, more generally, to nonlinear functionals of the fields. Therefore the mean-field equations extracted from the in--in effective action naturally separate into a part depending only on
$\langle \Xi\rangle_\sigma$
and a part controlled by intermediate channels, which we shall denote using the symbol 
$(\sigma'|\sigma)$. The last statement however involves more subtleties that will become clear as we develop the story further. 
Additionally, equation \eqref{eq:X2_offdiag_def_final} reflects the general decomposition of operator expectation values into the diagonal GS contribution and the off-diagonal GS transitions. This is the origin of the two-part bootstrap split we propose in this section.

There is however a point that we should repeat for clarity.
The bootstrap equations \eqref{bootsara} that appear from \eqref{eq:E_SK_physical_best} are {\it not} expected to also appear from \eqref{katadelga} and \eqref{eq:bootstrap_master_Gamma}. In fact \eqref{katadelga} and \eqref{eq:bootstrap_master_Gamma} are arranged to reproduce only the {\it diagonal} parts of the bootstrap equations. This is simply because (a) we have constrained the dynamics to lie only in the mean-field sector because of our assumption $\langle X^n\rangle_\sigma \approx \langle X\rangle^n_\sigma$, so the off-diagonal terms are invisible in this formalism; and (b) introduced $W_\pm=0$
inside the definition of ${\cal E}_{\rm SK}^{\mu\nu}[g_+, g_-;\sigma]$ so that the source part is separated off and reappears as the other bootstrap equation. This way the mean-field equation is determined by the action sector, while the GS insertion is determined by the companion bootstrap condition. 

Therefore if we want to study the full bootstrap equations, a generic effective action is called for. 
To proceed, let the GS-deformed Schwinger--Keldysh effective action be reduced, after integrating out the ordinary off-shell fields that do not survive as mean fields, to a functional of the surviving variables:
\begin{equation}
\Gamma_{\rm SK}^{(\sigma)}[\Xi]
=
\Gamma_{\rm loc}^{(\sigma)}[\Xi]
+
\Gamma_{\rm nloc}^{(\sigma)}[\Xi]
+
\Gamma_{\rm ghost}^{(\sigma)}[\Xi] ,
\label{eq:Gamma_split_final}
\end{equation}
where both the local and the non-local sectors that include individual ghost structures form their respective trans-series extensions. These may be easily inferred from the full trans-series form of the total action $S_{\rm tot}$.
Then the physical equation
\eqref{eq:bootstrap_master_Gamma}, that capture the diagonal parts,
may be written schematically as:
\begin{equation}
{\cal E}_{\mu\nu}^{\rm loc}(\langle \Xi\rangle_\sigma)
+
{\cal E}_{\mu\nu}^{\rm nloc}(\langle \Xi\rangle_\sigma)
=
0 ,
\label{eq:E_total_from_Gamma_final}
\end{equation}
with no {\it explicit} contributions from the ghost part. 
The point of the bootstrap discussion is to add in the off-diagonal pieces, ghost contributions as well as the Schwinger-Dyson equations corresponding to the off-shell states/DOFs carefully  so that the system naturally splits into 
two sectors
$\Xi_{\rm off}$
and 
$\Xi_{\rm on}$,
and then to resolve each of them into its diagonal and off-diagonal GS pieces. This is a subtle process, so one needs to be careful to avoid certain pitfalls. In the following let us turn to this first.

\subsubsection{How {\it not} to bootstrap using the GS states \label{sec7.4.1}}

To see how and where bootstrapping enters the story, we will first need to dispel some of the common mistakes one does when interpreting the behavior of the GS states. Consider a quantum field (or graviton fluctuation) operator $\hat g_{\mu\nu}$ and an
interacting vacuum $|\Omega\rangle$ defined on a Minkowski branch.  As elaborated before, a standard
coherent state is created by a displacement operator acting at one time slice as given by the first relation in \eqref{sweenseyf}.
The state then evolves freely under the physical Hamiltonian (in gravity, the
boundary Hamiltonian).
Thus an ordinary coherent state describes a freely evolving classical perturbation.
Its energy density redshifts and generically decays.  In particular, it cannot
generate a sustained positive vacuum energy capable of producing a de Sitter
geometry. Saying differently, a standard coherent state provides initial data only.  
It does not continuously inject energy into the system. In fact this is where it appears to differ from the GS state.
A GS state is instead generated by an operator supported over a spacetime region as shown in the second relation in \eqref{sweenseyf},
so that the state is produced by a continuous sequence of ``kicks''.
The expectation value satisfies the Schwinger--Dyson equation in the presence of
a source as shown in \eqref{eq:microscopic_contour_identity_corrected_section}, or more specifically 
\eqref{eq:E_SK_physical_best}.
Thus the classical profile obeys a driven Einstein equation which may be derived from \eqref{eq:Einstein_with_deformed_Gamma_source}, and takes the form:
\bg\label{vivluna}
&& \langle {G}_{\mu\nu}\rangle_\sigma
=
8\pi G\,\langle {T}_{\mu\nu}\rangle_\sigma
+
2\langle{\Sigma}_{\mu\nu}\rangle_\sigma \nonumber\\
&& \langle{\Sigma}_{\mu\nu}\rangle_\sigma = 2\sigma^{MN} \Big[\langle \sqrt{-{g}}~ {g}_{\mu M} {g}_{\nu N}\rangle_\sigma + {1\over 2} \langle \sqrt{-{g}}~ {g}_{\mu\nu} {g}_{M N}\rangle_\sigma\Big], 
\nd
where the factor of 2 comes from ${D}^\dagger(\sigma) {D}(\sigma)$ entering the path-integral \cite{coherbeta, coherbeta2} or alternatively, from the positive and the negative branches of the Schwinger-Keldysh (SK) contour.
(Note that $\langle\Sigma_{\mu\nu}\rangle_\sigma$ appearing here is related to $\langle\widetilde\Sigma^{(\sigma)}_{\mu\nu}\rangle_\sigma$ appearing in 
\eqref{eq:Einstein_with_deformed_Gamma_source} which may be determined easily but for the present discussion this is not necessary.)
If $\langle{\Sigma}_{\mu\nu}\rangle_\sigma$ has approximately vacuum-energy form:
\begin{equation}\label{crepcrawlrani}
\langle{\Sigma}_{\mu\nu}\rangle_\sigma
\approx
-{1\over 2}\Lambda_{\rm eff}\,\langle {g}_{\mu\nu}\rangle_\sigma,~~~~~ \implies~~
\langle {G}_{\mu\nu}\rangle_\sigma
+
\Lambda_{\rm eff}\,\langle {g}_{\mu\nu}\rangle_\sigma
=
8\pi G\,\langle {T}_{\mu\nu}\rangle_\sigma,
\end{equation}
which admits quasi--de Sitter solutions, at least in the mean field approximation, where $\Lambda_{\rm eff}$ acts as an effective dark energy that may or may not be a constant with respect to time. Crucially, the effective vacuum energy is maintained by the continuous action of
the GS operator rather than by a true potential minimum.  
Therefore, in this construction, the geometry might 
correspond to a transient excited state above the Minkowski branch.

Unfortunately, the ansatz \eqref{crepcrawlrani} does \emph{not} capture the structure we seek\footnote{This is despite the fact that one may entertain the following attractive conclusion from \eqref{crepcrawlrani}: A single-time coherent state evolves according to
$\frac{\delta \Gamma_{\rm SK}[g]}{\delta g^{\mu\nu}(x)}=0$
with no source term. Therefore a generic single-time coherent state built from ordinary local excitations does not naturally behave as a sustained cosmological-constant source. Its energy density will in general evolve nontrivially and may redshift, clump, or collapse rather than remain of exact vacuum-energy form. Nevertheless, this does not amount to a no-go theorem: in principle, a specially chosen coherent state could mimic an effective $\Lambda g_{\mu\nu}$ source if its initial data and backreaction satisfy the constraints and preserve the required stress-tensor structure. By contrast, a GS state satisfies
$\frac{\delta \Gamma^{(\sigma)}_{\rm phys}[g_c]}{\delta g_c^{\mu\nu}(x)}
=\sqrt{-g_c(x)}\, \Sigma^{(\sigma)}_{\mu\nu}(x)$ generically,
so the background is continuously driven thus providing the requisite positive energy. This is the pitfall, alluded to earlier, that we want to avoid.}. While one may adopt an assumption of the form \eqref{crepcrawlrani} to motivate the existence of a quasi--de Sitter solution, such a choice is effectively indistinguishable from introducing an explicit $\Lambda_{\rm eff}$ term directly into the Euclidean action. In that case the resulting background is not determined dynamically by the underlying theory, but rather by an externally specified parameter. Consequently, $\Lambda_{\rm eff}$ would function as a tunable input rather than an emergent quantity tied to the dynamics, and this undermines the goal of obtaining a self-consistent quasi--de Sitter solution from the intrinsic degrees of freedom of the system.

\subsubsection{Bootstrap equations for the off-shell-state sector \label{sec7.4.2}}

Having clarified this point,
we will start by analyzing the equation \eqref{eq:E_total_from_Gamma_final} for the off-shell states (or alternatively, the off-shell DOFs).
Consider first a component in the vanishing-one-point 
sector\footnote{If the trans-series action $S_{\rm tot}(\Xi)$ is the exact quantum effective action and
the Minkowski/warped-Minkowski soliton with its supporting fields is an exact stationary
solution, then all 1PI tadpoles vanish on that background. For the
quantum state whose mean fields coincide with this exact soliton, the
one-point functions of the fluctuations vanish in the absence of any GS or coherent states.}, for example a representative metric component of the form\footnote{To be consistent with the earlier works {\gspapers} we will henceforth denote 
$g_{c\,\mu\nu}$ by ${\bf g}_{MN}$ and the generic background metric  ${\rm g}_{c\,\mu\nu} = \langle g_{+\,\mu\nu}\rangle_\sigma = \langle g_{-\mu\nu}\rangle_\sigma$ by $\langle \hat{\bf g}_{MN}\rangle_\sigma$. The operators will be denoted by hats as usual. \label{corirani}}:
\begin{equation}
{\bf g}_{A'B'}
\in
\Xi_{\rm off},
\qquad
\langle \hat{\bf g}_{A'B'}\rangle_\sigma = 0 ,
\label{eq:g0Aprime_off_final}
\end{equation}
which as defined above and also in \cite{hetborel}, is called an off-shell state (or an off-shell DOF) because it does not appear as background for the specific choice of the slicing. Generically, this component is slicing-specific and does not have an invariant meaning as one can go from one slicing to another by coordinate transformations. Nevertheless \eqref{eq:g0Aprime_off_final} provides a simple way to choose a specific slicing from the excited state point of view that is consistent with the underlying coordinate transformations. 
The correct statement then is that the bootstrap equation for this sector splits into two parts: 
The \textcolor{blue}{first} is the diagonal mean-field condition:
\begin{equation}
{\bf R}_{A'B'}(\langle \Xi\rangle_\sigma)
-
{\bf T}^{\rm np}_{A'B'}(\langle \Xi\rangle_\sigma)
-
{\bf T}^{\rm nloc}_{A'B'}(\langle \Xi\rangle_\sigma)
=
0 ,
\label{eq:offshell_boot_diag_final}
\end{equation}
where ``np" is the local non-perturbative piece and ``nloc" is the non-local non-perturbative piece. Both appear from the trans-series form of the action discussed earlier.
This equation is the part obtained from the diagonal GS contribution of the reduced in--in effective action. It should be interpreted as the condition that the reduced mean field be consistent with the vanishing of the one-point function in the off-shell-state direction.
Note that this equation is expressed using the on-shell DOFs $\langle\Xi\rangle_\sigma$.
The \textcolor{blue}{second} is the off-diagonal consistency relation that may be expressed in the following way:
\begin{equation}
\begin{aligned}
&\sum_{\sigma' \neq \sigma}
\Big[
\langle \lambda_3(\Xi,\Upsilon_g)\,\hat{\bf R}_{A'B'}(\Xi)\rangle_{(\sigma|\sigma')}
-
\langle \lambda_3(\Xi,\Upsilon_g)\,\hat{\bf T}_{A'B'}(\Xi)\rangle_{(\sigma|\sigma')}
-
\langle \lambda_3(\Xi,\Upsilon_g)\,\hat{\bf T}^{\rm nloc}_{A'B'}(\Xi)\rangle_{(\sigma|\sigma')}
\Big]\\
&+
\langle
\lambda_3(\Xi,\Upsilon_g)\,
[\hat{\bf T}^{\rm nloc}_{\rm ghost}]_{A'B'}(\Xi,\Upsilon_g)
\rangle_\sigma
=
0 ,
\end{aligned}
\label{eq:offshell_boot_offdiag_final}
\end{equation}
where $\lambda_3$ is the remnant of the determinant of the four-dimensional metric after integrating out the off-shell DOFs, and hats distinguish operators from the functions (compare \eqref{eq:offshell_boot_diag_final}).
The equation \eqref{eq:offshell_boot_offdiag_final} is not a second independent mean-field equation for a new background component. Rather, it is the consistency condition that fixes the induced non-local and ghost sector associated with the off-shell-state direction. In other words, the off-shell-state sector contributes to the bootstrap problem through a pair of conditions: (a) 
diagonal mean-field consistency, and (b) 
off-diagonal induced-sector consistency. 
Thus the complete off-shell-state bootstrap problem is:
\begin{equation}
\left\{
\begin{aligned}
&{\bf R}_{A'B'}(\langle \Xi\rangle_\sigma)
-
{\bf T}_{A'B'}(\langle \Xi\rangle_\sigma)
-
{\bf T}^{\rm nloc}_{A'B'}(\langle \Xi\rangle_\sigma)
=
0,
\\
&\sum_{\sigma' \neq \sigma}
\Big[
\langle \lambda_3\hat{\bf R}_{A'B'}\rangle_{(\sigma|\sigma')}
-
\langle \lambda_3\hat{\bf T}_{A'B'}\rangle_{(\sigma|\sigma')}
-
\langle \lambda_3\hat{\bf T}^{\rm nloc}_{A'B'}\rangle_{(\sigma|\sigma')}
\Big]
+
\langle \lambda_3[\hat{\bf T}^{\rm nloc}_{\rm ghost}]_{A'B'}\rangle_\sigma
=
0 ,
\end{aligned}
\right.
\label{eq:offshell_boot_pair_final}
\end{equation}
where the first equation becomes necessary for the internal consistency especially in higher dimensions where we can have time-dependent warp-factors {\gspapers}. Note that for ${\bf g}_{A'B'}$ not satisfying 
\eqref{eq:g0Aprime_off_final} and consequently \eqref{eq:offshell_boot_pair_final}, the general equation \eqref{tagdekh4}, or more specifically \eqref{eq:microscopic_contour_identity_corrected_section}, is of course always satisfied. For more details about this in the full eleven-dimensional supergravity setting, the readers may refer to section 7.1 of \cite{hetborel}.

\subsubsection{Bootstrap equations for the on-shell-state sector \label{sec7.4.3}}

The on-shell states (or the on-shell DOFs) from the GS state point of view are recovered from integrating over all the off-shell fields in the SK path-integral. They are originally denoted by $g_{c\,\mu\nu}^\pm = \langle g_{\pm\,\mu\nu}\rangle_\sigma$ in say \eqref{eq:gcpm_sigma_def}, and we are looking at the limit where $g_{c\, \mu\nu}^+ = g_{c\,\mu\nu}^- = {\rm g}_{c\,\mu\nu}$, although for this discussion we will drop the subscript $c$, replace $(\mu, \nu) \to (A, B)$ and follow the convention outlined in footnote \ref{corirani}. Therefore an on-shell state would be  
a component in the surviving sector, for example:
\begin{equation}
{\bf g}_{AB}
\in
\Xi_{\rm on},
\qquad
\langle \hat{\bf g}_{AB}\rangle_\sigma \neq 0.
\label{eq:gAB_on_final}
\end{equation}
Here the split differs from the off-shell-state case because the GS operator contributes explicitly through the functional derivative of:
\begin{equation}
\log\Big({D}^\dagger(\sigma){D}(\sigma)\Big) ,
\label{eq:DdagD_log_symbol}
\end{equation}
which is precisely the contribution from $\widetilde{\Sigma}^{(\sigma)}_{AB}$ appearing in  \eqref{eq:Einstein_with_deformed_Gamma_source}.
The corrected split is therefore again in two parts. The \textcolor{blue}{first} one is the 
diagonal mean-field equation:
\begin{equation}
{\bf R}_{AB}(\langle \Xi\rangle_\sigma)
-
\frac{1}{2}
{\bf g}_{AB}
{\bf R}(\langle \Xi\rangle_\sigma)
-
{\bf T}^{\rm np}_{AB}(\langle \Xi\rangle_\sigma)
-
{\bf T}^{\rm nloc}_{AB}(\langle \Xi\rangle_\sigma)
=
0 ,
\label{eq:onshell_boot_diag_final}
\end{equation}
where ``np" and ``nloc" are again the non-perturbative local and non-local contributions respectively. 
This is the corrected on-shell equation of motion from \eqref{bootsara}
where we can easily identify $G_{AB}[{\rm g}_c]$ with the first two terms in 
\eqref{eq:onshell_boot_diag_final}, and the $T_{AB}^{\rm non-pert}({\rm g}_c)$ to the  local and the non-local pieces of \eqref{eq:onshell_boot_diag_final}. The \textcolor{blue}{second} one is the companion off-diagonal relation, and is given by the following:

{\footnotesize
\bg\label{sakaluheat}
-
\left\langle
\frac{\delta}{\delta {\bf g}^{AB}}
\log\Big({D}^\dagger(\sigma){D}(\sigma)\Big)
\right\rangle_\sigma
&= &
\sum_{\sigma' \neq \sigma}
\Big[
\langle \lambda_3(\Xi,\Upsilon_g)\,\hat{\bf R}_{AB}(\Xi)\rangle_{(\sigma|\sigma')}
-
\frac{1}{2}
\langle \lambda_3(\Xi,\Upsilon_g)\,{\bf g}_{AB}\hat{\bf R}(\Xi)\rangle_{(\sigma|\sigma')}
\nonumber\\
&&\qquad
-
\langle \lambda_3(\Xi,\Upsilon_g)\,\hat{\bf T}_{AB}(\Xi)\rangle_{(\sigma|\sigma')}
-
\langle \lambda_3(\Xi,\Upsilon_g)\,\hat{\bf T}^{\rm nloc}_{AB}(\Xi)\rangle_{(\sigma|\sigma')}\Big]
\nonumber\\
&& \qquad
-
\langle \lambda_3(\Xi,\Upsilon_g)\,[\hat{\bf T}^{\rm nloc}_{\rm ghost}]_{AB}(\Xi,\Upsilon_g)\rangle_\sigma ,
\nd}
which is a more detailed version of the second equation in \eqref{bootsara} as is evident from the form of $\Delta T^{(\sigma)}_{\mu\nu}({\rm g}_c, \phi_c)$ given in terms of various components in \eqref{sakaluheat}, with hats distinguishing operators from the functions.
This relation is also different in structure from the off-shell-state case precisely because the GS operator acts directly on the surviving sector. Hence the derivative of
$\log\Big({D}^\dagger{D}\Big)$
must remain explicit in the second equation of the on-shell pair.
Accordingly, the complete on-shell bootstrap system is:
\bg\label{bootsforwalking}
\begin{cases}
& {\bf R}_{AB}(\langle \Xi\rangle_\sigma)
-
\frac{1}{2}
{\bf g}_{AB}
{\bf R}(\langle \Xi\rangle_\sigma)
-
{\bf T}^{\rm np}_{AB}(\langle \Xi\rangle_\sigma)
-
{\bf T}^{\rm nloc}_{AB}(\langle \Xi\rangle_\sigma)
=
0 \\
& ~~~~~~\\
&
\left\langle
\frac{\delta}{\delta {\bf g}^{AB}}
\log\Big({D}^\dagger(\sigma){D}(\sigma)\Big)
\right\rangle_\sigma
=
\sum\limits_{\sigma' \neq \sigma}
\Big[-
\langle \lambda_3\hat{\bf R}_{AB}\rangle_{(\sigma|\sigma')}
+
\frac{1}{2}
\langle \lambda_3{\bf g}_{AB}\hat{\bf R}\rangle_{(\sigma|\sigma')}
\\
& \qquad\qquad\qquad\qquad ~~~~~~~~~~~~~~
+
\langle \lambda_3\hat{\bf T}_{AB}\rangle_{(\sigma|\sigma')}
+
\langle \lambda_3\hat{\bf T}^{\rm nloc}_{AB}\rangle_{(\sigma|\sigma')}\Big]
+
\langle \lambda_3[\hat{\bf T}^{\rm nloc}_{\rm ghost}]_{AB}\rangle_\sigma ,
\end{cases}
\nd
where the first equation fixes the form of the background from where we can determine the on-shell GS state. The second equation then tells us how other off-shell GS states participate to support the on-shell background that we want. 
The bootstrap procedure, however, proceeds somewhat differently. Once a GS state is specified, it determines the left-hand side of the second equation in \eqref{bootsforwalking}. Provided that the state is sufficiently sharply peaked, this information also indicates how large a neighborhood in the space of GS states must be explored in evaluating the right-hand side of that equation. One may then compute the relevant curvature and ghost contributions. The second equation subsequently determines the {\it mean-field} expectation values of the energy--momentum tensors.
Recall that the energy--momentum tensors receive contributions from several sources, including instantons, multi-instantons, nonlocal effects, and other nonperturbative sectors. The computed values from the second equation therefore help identify the underlying microscopic contributions, although such an identification is not essential for implementing the bootstrap\footnote{This is precisely where a direct computation of the perturbative and nonperturbative contributions to the energy--momentum tensor differs from the bootstrap procedure. The former is considerably more demanding and, by itself, does not necessarily isolate which microscopic sectors provide the relevant contributions. The bootstrap approach is more targeted: one of its principal aims is to iteratively identify and converge toward the specific energy--momentum contributions required for a self-consistent solution.}.
The resulting mean-field energy--momentum tensors are then substituted into the first equation in \eqref{bootsforwalking}, which determines an updated GS state. If this output state does not coincide with the original input state, the procedure must be iterated. A self-consistent solution is reached when the iteration converges to a fixed point, so that the GS state used as input is reproduced by the bootstrap equations.
This is precisely the bootstrapping structure that we were aiming for. Note again that,
for ${\bf g}_{AB}$ not satisfying 
\eqref{bootsforwalking}, the general equation \eqref{tagdekh4}, or more specifically \eqref{eq:microscopic_contour_identity_corrected_section}, is of course always satisfied.

Alternatively, we can also state the precise relation with the GS-deformed in--in effective action. The single master equation \eqref{eq:bootstrap_master_Gamma}
contains, upon decomposing the operator expectation values into diagonal and off-diagonal GS sectors, two different kinds of bootstrap pairs: (a) 
for off-shell states we have 
\eqref{eq:offshell_boot_pair_final},
and (b) for on-shell states/DOFs we have 
\eqref{bootsforwalking}.
Thus the bootstrap analysis should not be viewed as though the same split applies uniformly to all sectors. The off-shell-state sector and the on-shell sector differ precisely by whether the explicit GS-source derivative
\begin{equation}
\left\langle
\frac{\delta}{\delta \Xi}
\log\Big({D}^\dagger(\sigma){D}(\sigma)\Big)
\right\rangle_\sigma
\label{eq:explicit_GS_derivative_symbol}
\end{equation}
appears in the second equation of the pair, where $\Xi$ is the set of all on-shell degrees of freedom. Again for more details on \eqref{bootsforwalking}, the readers may refer to section 7.1 of \cite{hetborel}.

\subsubsection{Towards solving the bootstrap equations consistently \label{sec7.4.4}}

A systematic way to solve the pair of equations in \eqref{bootsforwalking} is to treat them as a coupled bootstrap problem rather than as two unrelated equations\footnote{Similar story can be developed for the pair of equations in \eqref{eq:offshell_boot_pair_final}, but we will not do so here.}. Our brief discussion below \eqref{bootsforwalking} outlined the procedure. In the following we will elaborate more on this. Let us start by 
identifying the unknowns.
They are not only the mean fields
$\langle \Xi \rangle_\sigma$,
but also the GS data hidden in
$\mathbb D(\sigma)$,
the off-diagonal matrix elements
$\langle \cdots \rangle_{(\sigma|\sigma')}$,
including the local quantity $T^{\rm np}_{AB}$ as well as the induced non-local quantities
${\bf T}^{\rm nloc}_{AB}$ and $
[\hat{\bf T}^{\rm nloc}_{\rm ghost}]_{AB}$.
So one is solving simultaneously for the background and for the state-dependent source data.
It is then useful to denote the first equation by:
\begin{equation}
{\cal E}_{AB}[\langle \Xi\rangle_\sigma]
\equiv
{\bf G}_{AB}(\langle \Xi\rangle_\sigma)
-
{\bf T}^{\rm np}_{AB}(\langle \Xi\rangle_\sigma)
-
{\bf T}^{\rm nloc}_{AB}(\langle \Xi\rangle_\sigma)
=
0,
\label{eq:EAB_def}
\end{equation}
where
${\bf G}_{AB}
=
{\bf R}_{AB}
-
\frac{1}{2}
{\bf g}_{AB}{\bf R}$ is the Einstein tensor but now expressed in terms of the emergent quantity $\langle \Xi\rangle_\sigma$ which here is the metric $\langle\hat{\bf g}_{\mu\nu}\rangle_\sigma$.
Similarly we can define the second equation in the following way:
\begin{equation}
{\cal B}_{AB}[\sigma,\langle \Xi\rangle_\sigma]
\equiv
\left\langle
\frac{\delta}{\delta {\bf g}^{AB}}
\log\Big({D}^\dagger(\sigma){D}(\sigma)\Big)
\right\rangle_\sigma
-
{\cal S}_{AB}[\sigma,\langle \Xi\rangle_\sigma]
=
0,
\label{eq:BAB_def}
\end{equation}
which relates the GS data, namely from the displacement operator, to other intermediate off-shell GS states. Note that, while writing \eqref{eq:BAB_def}, we are only taking the metric derivative but this may not be generic. One could take derivative with respect to the whole on-shell sector $\Xi_{\rm on} \subset \Xi$. For just the metric derivative, $\mathcal S_{AB}$ is defined as:
\begin{align}
{\cal S}_{AB}[\sigma,\langle \Xi\rangle_\sigma]
&=
\sum_{\sigma' \neq \sigma}
\Big[
-
\langle \lambda_3\hat{\bf R}_{AB}\rangle_{(\sigma|\sigma')}
+
\frac{1}{2}
\langle \lambda_3{\bf g}_{AB}\hat{\bf R}\rangle_{(\sigma|\sigma')}
\nonumber\\
&\qquad
+
\langle \lambda_3\hat{\bf T}^{\rm np}_{AB}\rangle_{(\sigma|\sigma')}
+
\langle \lambda_3\hat{\bf T}^{\rm nloc}_{AB}\rangle_{(\sigma|\sigma')}
\Big]
+
\langle \lambda_3[\hat{\bf T}^{\rm nloc}_{\rm ghost}]_{AB}\rangle_\sigma\nonumber\\
& \equiv  \sum_{\sigma' \neq \sigma} {\cal K}_{AB}(\sigma,\sigma') + 
\langle \lambda_3[\hat{\bf T}^{\rm nloc}_{\rm ghost}]_{AB}\rangle_\sigma ,
\label{eq:SAB_def}
\end{align}
where the sum is over all GS states $|\sigma'\rangle \ne |\sigma\rangle$. The last term with non-local ghost contribution can also be split into diagonal and off-diagonal parts but the difference is that both of them contribute to this equation as we do not expect to see the ghost contribution directly through  \eqref{eq:EAB_def}.
Moreover, the sum over $\sigma'$ may actually be converted to an integral 
if the labels $\sigma'$ form a continuous basis of intermediate states.
Then one may write:
\begin{equation}
\sum_{\sigma' \neq \sigma}
{\cal K}_{AB}(\sigma,\sigma')
\quad\longrightarrow\quad
\int_{{\cal M}_\sigma} d\mu(\sigma')\;
{\cal K}_{AB}(\sigma,\sigma'),
\label{eq:sum_to_integral_sigma_prime}
\end{equation}
where $d\mu(\sigma')$ is the appropriate measure on the space of intermediate states, and the domain ${\cal M}_\sigma$ excludes the diagonal point $\sigma'=\sigma$. In either case, what we want is the solution to the following two equations:
\begin{equation}
{\cal E}_{AB}=0,
\qquad
{\cal B}_{AB}=0.
\label{eq:E_B_system}
\end{equation}
In full generality the system is impossible to solve exactly, so one must choose a controlled truncation. Typical choices are: (a) 
symmetry reduction of the mean fields, (b) truncation of the intermediate-state sum over $\sigma'$, and 
(c) controlled expansion for the energy-momentum tensors.
For example, if the mean background is assumed to respect a slicing symmetry, one first writes:
\begin{equation}
\langle \Xi\rangle_\sigma
=
\Xi_{\rm ansatz}(u_1,\dots,u_n),
\label{eq:Xi_ansatz}
\end{equation}
where $u_i$ are finitely many unknown functions or constants. Then \eqref{eq:EAB_def} becomes a finite set of reduced equations.

Operationally, the second equation is best viewed as determining the state-induced source sector. That is, one uses
${\cal B}_{AB}=0$
to solve for one of the following:
(i) the derivative 
$\left\langle
\frac{\delta}{\delta {\bf g}^{AB}}
\log(D^\dagger D)
\right\rangle_\sigma$, 
(ii) the intermediate-channel sum 
${\cal S}_{AB}$,
or
(iii) the induced local and non-local tensor 
together with the ghost term.
In practice, one often parameterizes the energy-momentum tensors 
in terms of a finite set of unknown coefficients, and then fix them from \eqref{eq:BAB_def}.
Thus the second equation is not usually solved for the mean geometry directly; it is used to determine the state-dependent effective source entering the first equation.

Once the non-local and state-dependent sector has been determined from the second equation, we can substitute the result into the first equation, namely \eqref{eq:EAB_def} forming 
a closed effective equation for the mean fields.
So the practical order is: (a) solve for 
${\cal B}_{AB}=0$, (b) this would fix the input energy-momentum tensors
${\bf T}^{\rm nloc}_{AB}$ and ${\bf T}^{\rm np}_{AB}$, and (c) 
solve for 
${\cal E}_{AB}=0$. However
because ${\cal B}_{AB}$ itself depends on
$\langle \Xi\rangle_\sigma$,
the procedure must be iterative. A standard bootstrap loop then is the following\footnote{It is worth emphasizing that, if the system of equations in
\eqref{bootsforwalking} could be solved exactly, the iterative
procedure would be unnecessary. In the present case, however, the
equations depend crucially on the trans-series structure of the action,
making it exceedingly difficult to anticipate, let alone construct,
their exact solutions. Under such circumstances, the bootstrap
procedure adopted here provides the most practical, and potentially the
only systematic framework for analyzing the system.}:
\begin{itemize}
\item Step 1: Choose a trial background $\langle \Xi\rangle_\sigma^{(0)}$.
\item Step 2: Evaluate the RHS of the second equation \eqref{eq:BAB_def} on $\langle \Xi\rangle_\sigma^{(0)}$.
\item Step 3: Extract ${\bf T}^{\rm nloc}_{AB}{}^{(0)}$ and ${\bf T}^{\rm np(0)}_{AB}$.
\item Step 4: Solve the first equation \eqref{eq:EAB_def} for an updated $\langle \Xi\rangle_\sigma^{(1)}$.
\item Step 5: Repeat until 
$\langle \Xi\rangle_\sigma^{(n+1)}=\langle \Xi\rangle_\sigma^{(n)}$
within tolerance.
\end{itemize}
\noindent We should also make sure that a candidate solution must satisfy at least the following checks: (i) covariance and index consistency,
(ii) conservation of the total effective source {\it i.e.}
$\nabla^A
\Big(
{\bf T}^{\rm np}_{AB}
+
{\bf T}^{\rm nloc}_{AB}
\Big)
=
0$\footnote{It should, however, be emphasized that conservation of the effective source stress-energy tensor does not necessarily imply that the source fields satisfy an independent set of equations of motion. In some cases, particularly when the coupling is topological or when conservation follows from an off-shell differential identity, the associated stress-energy tensor may be conserved without imposing separate source-field equations. For generic non-topological couplings, by contrast, the divergence of the stress-energy tensor is proportional to the source equations of motion, so that stress-energy conservation generally enforces at least some combination of those equations. We will avoid delving into these subtleties here.}, (iii) compatibility of the second equation with the same background used in the first, and 
(iv) stability of the iterative map or perturbative expansion.
The conservation condition is especially important, because otherwise the first equation cannot be compatible with the contracted Bianchi identity. A self-consistent solution is therefore a fixed point of the following map:
\begin{equation}
\langle \Xi\rangle_\sigma
\longmapsto
{\cal B}_{AB}
\longmapsto
\left({\bf T}^{\rm nloc}_{AB}, {\bf T}^{\rm np}_{AB}\right)
\longmapsto
{\cal E}_{AB}
\longmapsto
\langle \Xi\rangle_\sigma ,
\label{eq:fixed_point_map}
\end{equation}
which amounts to saying that 
the second equation \eqref{eq:BAB_def} determines the GS-induced energy-momentum tensors, 
the first equation \eqref{eq:EAB_def} then determines the mean fields in that induced source background,
and the whole procedure is iterated until the background used to compute the source matches the background produced by the source. This is the requisite bootstrap procedure as also elucidated in {\bf figure \ref{bootstraploop}}.

\begin{figure}[h]
\centering
\begin{tabular}{c}
\includegraphics[width=5in]{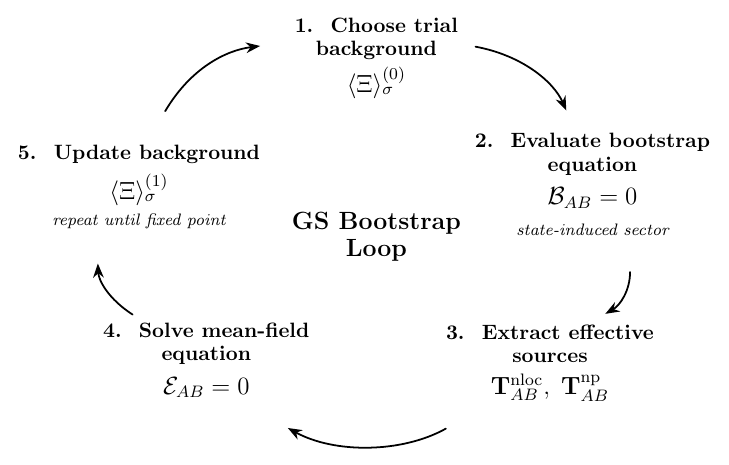}
\end{tabular}
\caption[]{This is an illustration of the bootstrap loop for the GS states, corresponding to equation \eqref{eq:fixed_point_map}.  A candidate solution must satisfy covariance and index consistency, conservation of the total effective source, namely 
$\nabla^{A}\!\left(
T^{\mathrm{np}}_{AB}
+T^{\mathrm{nloc}}_{AB}
\right)=0$, 
compatibility of $\cal E_{AB}$ and $\cal B_{AB}$, and stability of the iterative map or perturbative expansion. The conservation condition is required for consistency with the contracted Bianchi identity.}
\label{bootstraploop}
\end{figure}

\subsection{Nonuniqueness, bootstrap fixed points and basins of attraction \label{sec7.5}}

The bootstrap equations that we developed in \eqref{eq:offshell_boot_pair_final} and  \eqref{bootsforwalking} and as illustrated in {\bf figure \ref{bootstraploop}} now raises the following question: does this procedure provide a {\it unique} solution? The answer should clearly be no, but then we need to reconcile with all possible allowed fixed points of the bootstrap map.

The bootstrap procedure should then be viewed as defining a nonlinear map on the space of admissible Glauber--Sudarshan states. The correct question is therefore not whether the bootstrap equations admit a unique formal solution, but rather how many fixed points the associated bootstrap map possesses, whether those fixed points are stable, and how the space of initial GS data is partitioned into basins of attraction. In the following we will analyze the solutions.

\subsubsection{Bootstrap map, nonuniqueness and fixed points \label{sec7.5.1}}

Let ${\cal S}_{\rm GS}$ denote the space of admissible GS labels $\sigma$. For each state $|\sigma\rangle$, the second bootstrap equation in \eqref{bootsforwalking} determines the induced sources
$\sigma
\longmapsto
\Big(
{\bf T}^{\rm np}_{AB}[\sigma],
{\bf T}^{\rm nloc}_{AB}[\sigma]
\Big)$
which by itself is of course a non-trivial exercise because of the way the energy-momentum tensors appear in \eqref{bootsforwalking}. Nevertheless this could be done with some effort, 
and the first bootstrap equation then determines the corresponding mean field
$\Big(
{\bf T}^{\rm np}_{AB}[\sigma],
{\bf T}^{\rm nloc}_{AB}[\sigma]
\Big)
\longmapsto
{\cal G}[\sigma]
\equiv
\langle \Xi\rangle_\sigma$. 
Finally, from the resulting mean field one reconstructs the GS state label compatible with it:
\begin{equation}
{\cal G}[\sigma]
\longmapsto
{\cal R}\big({\cal G}[\sigma]\big).
\label{eq:geometry_to_state_reordered}
\end{equation}
This step is not very hard following the path-integral representation of the expectation value as in \eqref{coridahl7}: once the expectation value is independently determined from the first bootstrap equation in \eqref{bootsforwalking}, we can ask what GS state is responsible for it. 
Thus the full bootstrap defines a map:
\begin{equation}
{\cal F}:{\cal S}_{\rm GS}\to{\cal S}_{\rm GS},
\qquad
{\cal F}(\sigma)
=
{\cal R}\circ{\cal G}(\sigma) ,
\label{eq:bootstrap_map_reordered}
\end{equation}
because, from the aforementioned comment, we will need multiple iteration of the procedure to finally attain the precise GS state. This means, and as mentioned earlier, 
a self-consistent GS configuration is a fixed point of this map:
\begin{equation}
{\cal F}(\sigma_\ast)=\sigma_\ast ,
\label{eq:fixed_point_reordered}
\end{equation}
which is exactly where we expect our procedure to terminate. The number of iterations would of course depend on how far are we from the fixed point, but this is a subtle point if there exists other basins of attractors nearby. So we will need to tread carefully here as we shall elaborate soon.
Equivalently, the fixed-point set is:
\begin{equation}
{\rm Fix}({\cal F})
\equiv
\{\sigma\in{\cal S}_{\rm GS}\,|\,{\cal F}(\sigma)=\sigma\} ,
\label{eq:fixed_point_set_reordered}
\end{equation}
which provides a meaning to the region in the neighborhood of the point. The fixed point is the exact self-consistent solution of the coupled bootstrap equations. The iterative loop:
\begin{equation}
\sigma_{n+1}={\cal F}(\sigma_n)
\label{eq:iterative_loop_reordered}
\end{equation}
is only a constructive scheme for approximating such a solution. Therefore the physical content lies in \eqref{eq:fixed_point_reordered}, whereas the iteration is merely a method for finding a point in ${\rm Fix}({\cal F})$.

The bootstrap equations do not generically determine a unique fixed point. Indeed, from the existence of one orbit:
\begin{equation}
\sigma_0\longrightarrow \sigma_1\longrightarrow \sigma_2\longrightarrow \cdots
\label{eq:single_orbit_reordered}
\end{equation}
where $\sigma_0$ signifies the initial GS state $|\sigma_0\rangle$,
one cannot infer uniqueness of the endpoint. Different initial ans\"atze, or seeds, may converge to different fixed points:
\begin{equation}
\sigma_0^{(1)}
\longrightarrow
\sigma_\ast^{(1)},
\qquad
\sigma_0^{(2)}
\longrightarrow
\sigma_\ast^{(2)},
\qquad
\sigma_\ast^{(1)}\neq \sigma_\ast^{(2)}.
\label{eq:different_seeds_different_fixed_points}
\end{equation}
Moreover, even existence of a convergent orbit requires additional assumptions, such as contractivity of the map or sufficient compactness of the state space. Thus the correct general statement is
that the bootstrap problem is generically nonunique, because it is a fixed-point problem for a nonlinear map on ${\cal S}_{\rm GS}$.
A sufficient condition for nonuniqueness is the existence of disjoint invariant sectors. Suppose:
\begin{equation}
{\cal S}_{\rm GS}
=
{\cal S}_1\cup{\cal S}_2,
\qquad
{\cal S}_1\cap{\cal S}_2=\varnothing \nonumber
\label{eq:disjoint_union_reordered}
\end{equation}
\begin{equation}
{\cal F}({\cal S}_1)\subseteq{\cal S}_1,
\qquad
{\cal F}({\cal S}_2)\subseteq{\cal S}_2 ,
\label{eq:invariant_sectors_reordered}
\end{equation}
where $\mathcal F(\mathcal S)$ is the bootstrap map. Then the precise statement is the following.
If there exists a fixed point in each sector, then there are at least two distinct self-consistent GS states:
\begin{equation}
\sigma_\ast^{(1)}\in{\cal S}_1,
\qquad
\sigma_\ast^{(2)}\in{\cal S}_2,
\qquad
\sigma_\ast^{(1)}\neq \sigma_\ast^{(2)} ,
\label{eq:two_distinct_fixed_points_reordered}
\end{equation}
with $\sigma_\ast^{(i)}$ denoting the $i$-th fixed point sector. In the language of attractors, we can say that there are at least two basins of attractors here. 
Hence multiplicity follows as soon as the bootstrap map preserves a nontrivial sector decomposition.

Existence of fixed points also lead to question of the local stability near any given fixed point. To study this,  
let $\sigma_a$ be a fixed point satisfying 
${\cal F}(\sigma_a)=\sigma_a$.
Consider a nearby initial state
$\sigma_0=\sigma_a+\delta \sigma_0$.
Assuming ${\cal F}$ to be differentiable near $\sigma_a$, then the first-order Taylor expansion of ${\cal F}$ around $\sigma_a$ is:
\begin{equation}
{\cal F}(\sigma_a+\delta \sigma_0)
=
{\cal F}(\sigma_a)
+
D{\cal F}\big|_{\sigma_a}\cdot \delta \sigma_0
+
{\cal O}( \delta \sigma_0^2) =
\sigma_a
+
D{\cal F}\big|_{\sigma_a}\cdot \delta \sigma_0
+
{\cal O}(\delta \sigma_0^2) ,
\label{eq:Taylor_map_derivation}
\end{equation}
where
$D{\cal F}\big|_{\sigma_a}$
is the derivative, or Jacobian, of the bootstrap map at the fixed point. It is the linear operator that tells us how an infinitesimal perturbation is mapped after one bootstrap step; and we have used the fixed-point condition $\mathcal F(\sigma_a) = \sigma_a$ to express the second equality in \eqref{eq:Taylor_map_derivation}.
We can also define the perturbation at the $n$-th step in the following way.
Let
$\sigma_n=\sigma_a+\delta \sigma_n$.
Then one bootstrap step gives:
\begin{equation}
\sigma_{n+1}
=
{\cal F}(\sigma_n)
=
{\cal F}(\sigma_a+\delta \sigma_n) =
\sigma_a
+
D{\cal F}\big|_{\sigma_a}\cdot \delta \sigma_n
+
{\cal O}(\delta \sigma_n^2) ,
\label{eq:iterate_sigma_n}
\end{equation}
from where we can extract the leading order solution in the following way. First define $\delta\sigma_{n+1} = \sigma_{n+1} - \sigma_a$, providing the $n$-th order deviation from the fixed point. Aplying this to \eqref{eq:iterate_sigma_n}, we get:
\begin{equation}
\delta \sigma_{n+1}
=
D{\cal F}\big|_{\sigma_a}\cdot \delta \sigma_n
+
{\cal O}(\delta \sigma_n^2) \approx
D{\cal F}\big|_{\sigma_a}\cdot \delta \sigma_n , 
\label{eq:delta_recursion_full_derivation}
\end{equation}
where, for the second relation we have assumed the perturbation to be sufficiently small so that the quadratic correction may be neglected to first approximation. 
Applying this repeatedly,
and by induction, this leads to the recursion formula:
\begin{equation}
\delta \sigma_n
\approx
\Big(D{\cal F}\big|_{\sigma_a}\Big)^n\delta \sigma_0 ,
\label{eq:delta_n_by_induction}
\end{equation}
which is sufficient to lead the dynamics of the bootstrap path, but can be brought in a more manageable form by converting it into an eigenvalue equation. In fact this will tell us what happens when the iteration becomes very large.
To understand the large-$n$ behavior, we can decompose the initial perturbation into eigenvectors of the linearized operator. Suppose
\begin{equation}
D{\cal F}\big|_{\sigma_a} v_i=\lambda_i v_i ,
\label{eq:eigenvector_equation_derivation}
\end{equation}
where $\lambda_i$ form the eigenvalues with the choice of eigenvectors 
given via $v_i$. In fact the eigenvalues can also tell us how to control stability. For example, if:
\begin{equation}
\delta \sigma_0=\sum_i c_i v_i ~~ \implies ~~ \delta \sigma_n
\approx
\sum_i c_i \lambda_i^n v_i ,
\label{eq:delta0_eigenexpansion}
\end{equation}
then the behavior of each component is determined by $\lambda_i^n$. There are five possible outcomes: (1) 
if $0 < |\lambda_i|<1$ $\forall i$, then this implies 
$\lambda_i^n\to 0$. For such a case $\sigma_a$ is a {\it sink}, or attractor. This is the case represented in {\bf figure \ref{bootstrapbasins}};
(2) if $|\lambda_i|>1$ $\forall i$, then this implies 
$\lambda_i^n$ grows unboundedly, and $\sigma_a$ is a {\it source}: nearby states are repelled away from it under forward bootstrap iteration. This is shown in {\bf figure \ref{bootsinksourcesaddle}}; 
(3) if some eigenvalues satisfy
$|\lambda_i|<1$
while others satisfy
$|\lambda_j|>1$, 
then $\sigma_a$ is a {\it saddle}. In that case the fixed point attracts along some directions and repels along others as shown in {\bf figure \ref{bootsinksourcesaddle}}; 
(4) if 
$|\lambda_i|=1$, 
we expect a marginal behavior\footnote{For example if there is a complex conjugate pair
$\lambda_\pm
=
e^{\pm i\theta}$,
with 
$0<\theta<\pi$,
then in the corresponding two-dimensional real subspace the map acts as a rotation, {\it i.e.}
$\delta \sigma_n
=
\Big(
D{\cal F}\big|_{\sigma_a}
\Big)^n
\delta \sigma_0
\sim
R(n\theta)\,\delta \sigma_0$,
where $R(n\theta)$ is the usual rotation matrix by angle $n\theta$.
In that case the orbit of a nearby point can circle around the fixed point without approaching it and without running away from it. A unit-modulus eigenvalues signal that linear analysis alone is inconclusive about asymptotic stability.} as shown in {\bf figure \ref{sssrotate}}; and (5) if $\lambda_i = 0$, this corresponds to a maximally contracting, or super-attractive, direction of the bootstrap map. The perturbation along that eigendirection vanishes after one iteration at linear order\footnote{A fixed point with some zero eigenvalues may still be a sink, provided all other eigenvalues also satisfy
$|\lambda_j|<1$.
If instead some directions have
$|\lambda_j|>1$,
then the fixed point is still a saddle, with the $\lambda_i=0$ directions being especially strongly attractive.}.
So if all eigenvalues satisfy the following conditions:
\begin{equation}
0 \le |\lambda_i|<1 ~~~~ \implies ~~~~ \delta \sigma_n\to 0 ,
\label{eq:all_eigs_less_than_1}
\end{equation}
then every component of $\delta \sigma_n$ decays to zero, some even in just one iteration if the corresponding eigenvalue vanishes, and hence
$\sigma_n=\sigma_a+\delta \sigma_n\to \sigma_a$.
This is exactly what is meant by local stability of the fixed point, although one may make the argument a bit more precise in the following way. 
Because the Taylor expansion is valid only when $\delta \sigma_n$ stays sufficiently small, one needs the initial perturbation to lie in some sufficiently small neighborhood of $\sigma_a$. If the linearized operator is strictly contractive, then there exists an open set
$U_a\ni \sigma_a$
such that every initial point in $U_a$ stays in that neighborhood and converges to the fixed point:
\begin{equation}
\sigma_0\in U_a
\quad\Longrightarrow\quad
{\cal F}^n(\sigma_0)\to \sigma_a ,
\label{eq:local_stability_neighborhood}
\end{equation}
which then provides the precise content of the local stability statement. However one could still ask as to 
why nearby initial data do not generically go to $\sigma_a+\epsilon$. This is in fact easy to see from the following argument. 
Suppose we start instead at
$\sigma_0=\sigma_a+\epsilon$.
Then under one step, we get:
\begin{equation}
{\cal F}(\sigma_a+\epsilon)
=
\sigma_a
+
D{\cal F}\big|_{\sigma_a}\cdot \epsilon
+
{\cal O}(\epsilon^2).
\label{eq:F_sigmaa_plus_eps_expand}
\end{equation}
For this to equal
$\sigma_a+\epsilon$
one would need
$D{\cal F}\big|_{\sigma_a}\cdot \epsilon
=
\epsilon$
at linear order. That is, $\epsilon$ would have to lie in an eigen-direction with eigenvalue $1$. This is the marginal case.
So generically, if the fixed point is stable, then
$\sigma_a+\epsilon
\longrightarrow
\sigma_a$
and not to a shifted fixed point. Therefore our above analysis provides the following precise stability criterion for the bootstrap map: 
 If all eigenvalues of $D{\cal F}|_{\sigma_a}$ have modulus smaller than one, the perturbation decays and all sufficiently nearby initial GS states flow back to the same fixed point. Moreover, for a stable fixed point, nearby initial data do {\it not} generically flow to a shifted fixed point $\sigma_a+\epsilon$. Rather, they flow back to $\sigma_a$ itself. Alternatively, a perturbation of the initial GS state leads to a nearby shifted fixed point only if the fixed-point set contains a marginal family. Otherwise nearby seeds either flow back to the same fixed point or are repelled from it.

\begin{figure}[h]
\centering
\begin{tabular}{c}
\includegraphics[width=6.3in]{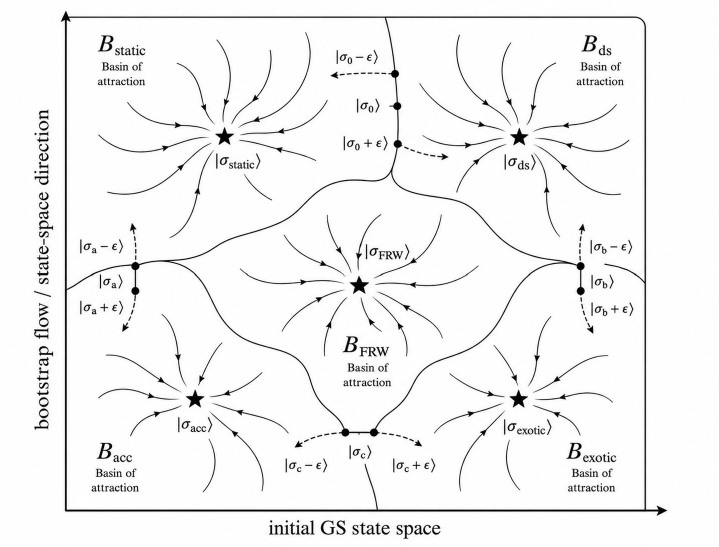}
\end{tabular}
\caption[]{The basins of attractors for the bootstrap maps ${\cal F}$, corresponding to equations \eqref{eq:offshell_boot_pair_final} and \eqref{bootsforwalking}. 
The figure is intended to display only the attracting fixed points of the bootstrap map, namely those self-consistent GS configurations reached under forward iteration from generic initial data. The solid starred points denote such attractors. Black dots indicate sample initial GS states, while dashed
arrows illustrate their flow. A more complete dynamical-systems portrait could also contain repulsive fixed points and saddles. Thus the present figure should be understood as a basin-of-attraction diagram rather than as the full phase portrait of the bootstrap flow. }
\label{bootstrapbasins}
\end{figure}

\subsubsection{Basins of attraction and other fixed points \label{sec7.5.2}}

The correct way to formulate the global picture is in terms of basins of attraction, repulsion and saddle, which in fact provide a much cleaner picture to study the stability criterion as well as to see how the neighborhood of the various fixed points behave. In the following however we will only concentrate on the basins of attraction as they are most interesting aspect of the GS construction. For each fixed point $\sigma_\alpha$, we can define the basin of attractors in the following way:
\begin{equation}
{\cal B}_\alpha
\equiv
\Big\{
\sigma_0\in{\cal S}_{\rm GS}
\;\Big|\;
{\cal F}^n(\sigma_0)
\text{ converges to }
\sigma_\alpha
\text{ under the chosen convergence criterion}
\Big\}.
\label{eq:basin_of_attraction_reordered}
\end{equation}
as depicted in {\bf figure \ref{bootstrapbasins}}.
Such a definition 
avoids the misleading impression that one must literally perform an infinite number of bootstrap loops physically; the point is only that the iterative scheme approaches the exact fixed point to the desired tolerance. Thus if
$\sigma_0\in{\rm int}({\cal B}_\alpha)$,
then there exists an $\epsilon_0>0$ such that:
\begin{equation}
\|\delta \sigma\|<\epsilon_0
\quad\Longrightarrow\quad
\sigma_0+\delta \sigma\in{\cal B}_\alpha.
\label{eq:small_perturbation_same_basin_reordered}
\end{equation}
Therefore all sufficiently nearby seeds flow to the same fixed point. This is the mathematically precise sense in which a whole neighborhood of GS states may determine the same emergent cosmology.

The above discussion accommodates naturally the possibility that the same bootstrap equations admit several distinct cosmological branches. Suppose the fixed-point set contains:
\begin{equation}
\sigma_{\rm static},
\qquad
\sigma_{\rm dS},
\qquad
\sigma_{\rm FRW},
\qquad
\dots
\label{eq:cosmological_fixed_points_reordered}
\end{equation}
corresponding respectively to a static branch with the corresponding GS state $|\sigma_{\rm static}\rangle$, a de Sitter branch with the corresponding GS state $|\sigma_{\rm dS}\rangle$, a generic FRW branch with the corresponding GS state $|\sigma_{\rm FRW}\rangle$, and possibly more exotic accelerating branches. Then one may have:
\begin{equation}
\sigma_{\rm static}^0\in{\cal B}_{\rm static},
\qquad
\sigma_{\rm dS}^0\in{\cal B}_{\rm dS},
\qquad
\sigma_{\rm FRW}^0\in{\cal B}_{\rm FRW},
\label{eq:seeds_in_basins_reordered}
\end{equation}
which implies, for example  $\sigma^0_{\rm dS}$ lies in the basin of attractor $\mathcal B_{\rm dS}$ from \eqref{eq:basin_of_attraction_reordered} such that $\sigma^0_{\rm dS}$ lies in the close neighborhood of the fixed point $\sigma_{\rm dS}$ with the corresponding GS state $|\sigma_{\rm dS}\rangle$ with all trajectories leading to the fixed point; with similar structure for the other cases. Alternatively:
\begin{equation}
\sigma_{\rm static}^0\longrightarrow \sigma_{\rm static},
\qquad
\sigma_{\rm dS}^0\longrightarrow \sigma_{\rm dS},
\qquad
\sigma_{\rm FRW}^0\longrightarrow \sigma_{\rm FRW}.
\label{eq:flows_to_branches_reordered}
\end{equation}
Hence the same pair of bootstrap equations may generate several inequivalent cosmologies. The distinction does not lie in the formal structure of the equations, but in the existence of several stable fixed points and their corresponding basins.
If, in addition, the fixed points $\sigma_{\rm static}, \sigma_{\rm FRW}$, $\sigma_{\rm dS}$ et cetera are locally stable, then there exist open neighborhoods:
\begin{equation}
U_{\rm static}\subset{\cal B}_{\rm static},
\qquad
U_{\rm dS}\subset{\cal B}_{\rm dS}, \qquad U_{\rm FRW} \subset \mathcal B_{\rm FRW}, ~~~~~ ....
\label{eq:local_neighborhoods_reordered}
\end{equation}
where by an {\it open neighborhood} $U_{\rm cosmology}$ of $\sigma_{\rm cosmology}$ one simply means a small region of the GS-state space around the fixed point $\sigma_{\rm cosmology}$ such that every state inside that region is still in the same basin of attraction. In other words, if the initial GS state is chosen sufficiently close to $\sigma_{\rm cosmology}$, then the bootstrap flow will still converge to $\sigma_{\rm cosmology}$. In other words, 
\begin{equation}
\sigma_0\in U_{\rm cosmology}
\quad\Longrightarrow\quad
{\cal F}^n(\sigma_0)\to \sigma_{\rm cosmology},
\label{eq:local_static_convergence_reordered}
\end{equation}
where the subscript ``cosmology" denotes either static, dS, FRW or any other exotic cosmologies.
This is the precise mathematical sense in which a whole neighborhood of GS states may lead to the same cosmological background.

One other issue is the question of countability versus continuity of the fixed-point set. In fact the question of how many fixed points exist is separate from the question of nonuniqueness. The bootstrap equations alone do not determine the cardinality of the fixed-point set ${\rm Fix}({\cal F})$. In principle one may have:
\begin{equation}
{\rm Fix}({\cal F}): ~
\text{ finite, countably infinite, or continuously infinite.} \nonumber
\label{eq:cardinality_options_reordered}
\end{equation}
where generically the first bootstrap equation in \eqref{bootsforwalking}, without additional restrictions, typically admits infinitely many solutions. (Even after quotienting by diffeomorphisms, however, the remaining moduli space is usually still infinite unless strong restrictions are imposed.) So typically a finite fixed-point set is only possible 
only after imposing strong extra conditions such as symmetry, topology, boundary data, and matter content. On the other hand, 
if the admissible GS states form a discrete countable set, {\it i.e.}
\begin{equation}
{\cal S}_{\rm GS}=\{\sigma_n\}_{n\in\mathbb N} ~~~\implies ~~~
{\rm Fix}({\cal F})\subseteq{\cal S}_{\rm GS} ,
\label{eq:countable_sigma_space_reordered}
\end{equation}
is automatically at most countable. In that case one may obtain a countably infinite family of fixed points if the space decomposes into infinitely many invariant sectors each containing at least one fixed point.
However, if the GS labels depend on continuous parameters, then the fixed-point set may be continuous. Therefore countability is {\it not} a consequence of the bootstrap equations alone. It requires an additional discreteness assumption on the GS-state space. After the dust settles, 
the correct conceptual picture is therefore the following.
\begin{itemize}
\item The bootstrap equations define a nonlinear map on the space of admissible GS states. Their self-consistent solutions are fixed points of this map.
\item Nonuniqueness is generic: different initial GS seeds may converge to different fixed points whenever the map has multiple basins of attraction or multiple invariant sectors.
\item Static, de Sitter, FRW, and more exotic accelerating cosmologies may all arise as distinct fixed points of the same bootstrap equations, with different neighborhoods of GS data flowing to different branches as depicted in {\bf figure \ref{bootstrapbasins}}.
\item The bootstrap map may contain sinks, sources, and saddles. Sinks ``attract" nearby GS states and therefore generate basins of attraction; sources ``repel" nearby states and are unstable under forward iteration; saddles ``attract" along some directions and ``repel" along others, and thus organize the separatrices between neighboring basins. This is depicted in {\bf figure \ref{bootsinksourcesaddle}}.
\item The bootstrap equations alone do not determine whether the set of fixed points is finite, countable, or continuous; this depends additionally on the structure of the GS-state space.
\end{itemize}

\begin{figure}[h]
\centering
\begin{tabular}{c}
\includegraphics[width=6.3in]{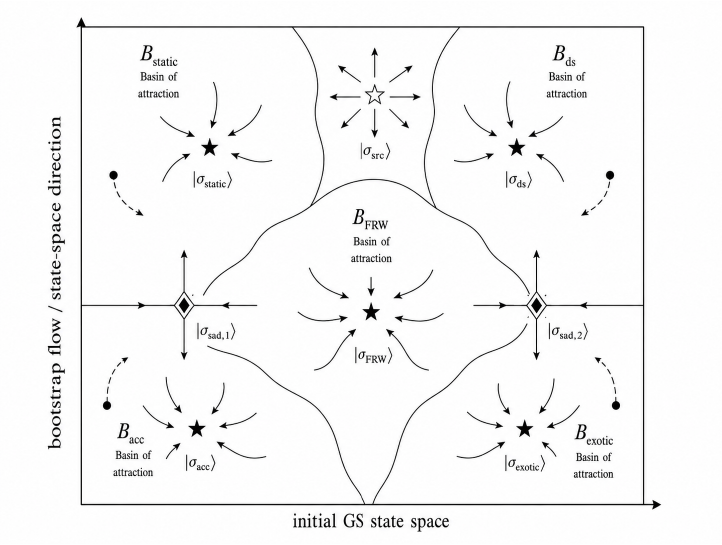}
\end{tabular}
\caption[]{The figure depicts the qualitative phase portrait of the bootstrap map ${\cal F}$ on the space of admissible GS states. Regions represent basins of attraction of stable fixed points, or sinks, corresponding to self-consistent GS backgrounds reached under forward bootstrap iteration. The solid starred points denote such attractors. The repelling fixed point, or source, is illustrated by non-solid star, and represents an unstable self-consistent configuration from which nearby GS states flow away. The diamond-shaped points denote saddles, which are stable along some directions and unstable along others, and therefore lie on the separatrices that organize the boundaries between neighboring basins. Black dots indicate sample initial GS states, while dashed arrows illustrate their flow under repeated application of the bootstrap map. Thus the figure should be read as a state-space portrait of bootstrap dynamics, rather than as a spacetime evolution diagram.}
\label{bootsinksourcesaddle}
\end{figure}

\subsubsection{Bootstrap fixed points, local stability, and de Sitter \label{sec7.5.3}}

In the discussion above, as well as in {\bf figures \ref{bootstrapbasins}} and {\bf \ref{bootsinksourcesaddle}}, we implicitly treated four-dimensional de Sitter space as a sink of the bootstrap map. This, however, is by no means guaranteed {\it a priori}. It is therefore important to revisit this point more carefully and distinguish the mere existence of a de Sitter fixed point from its local stability properties under the bootstrap iteration. To this effect, 
recall that ${\cal F}$ denotes the bootstrap map on the space of admissible GS states such that 
$\sigma_{n+1}
=
{\cal F}(\sigma_n)$. A self-consistent GS configuration is a fixed point of this map, {\it i.e.}
${\cal F}(\sigma_\ast)=\sigma_\ast$.
If the associated mean field is four-dimensional de Sitter, then
$\langle \Xi\rangle_{\sigma_{\rm dS}}
=
\Xi_{\rm dS}$,
with the following Schwinger-Dyson EOMs:
\begin{equation}
{\cal E}_{AB}[\Xi_{\rm dS}]=0,
\qquad
{\cal B}_{AB}[\sigma_{\rm dS},\Xi_{\rm dS}]=0 ,
\label{eq:dS_bootstrap_summary}
\end{equation}
appearing from \eqref{eq:EAB_def} and \eqref{eq:BAB_def}.
This proves only that de Sitter is a fixed point of the bootstrap map. It does not yet imply that de Sitter is an attractor.
To test local stability, perturb the GS state near the fixed point as
$\sigma_n
=
\sigma_{\rm dS}
+
\delta \sigma_n$.
Assuming ${\cal F}$ is differentiable near $\sigma_{\rm dS}$, one has the following relation:
\begin{equation}
\mathcal F(\sigma_n) = {\cal F}(\sigma_{\rm dS}+\delta \sigma)
=
{\cal F}(\sigma_{\rm dS})
+
D{\cal F}\big|_{\sigma_{\rm dS}}\cdot \delta \sigma
+
{\cal O}(\delta \sigma^2),
\label{eq:linearized_F_summary}
\end{equation}
that follows from Taylor expansion and the precise form for $\mathcal D\mathcal F$ will be shown below. Now using the fact that $\sigma_{n+1} = \mathcal F(\sigma_n)$ we can easily argue, as before, that:
\begin{equation}
\delta \sigma_{n+1}
=
D{\cal F}\big|_{\sigma_{\rm dS}}\cdot \delta \sigma_n
+
{\cal O}(\delta \sigma_n^2)~~~ \implies ~~~
\delta \sigma_n
\approx
\Big(
D{\cal F}\big|_{\sigma_{\rm dS}}
\Big)^n
\delta \sigma_0 ,
\label{eq:leading_order_delta_summary}
\end{equation}
where $\delta\sigma_{n+1} = \sigma_{n+1} - \sigma_{\rm dS}$.
If $\lambda_i$ are the eigenvalues of the linearized bootstrap operator
$D{\cal F}\big|_{\sigma_{\rm dS}}$,
then the local classification is:
\begin{equation}
|\lambda_i|<1
\quad
\text{for all }i
\qquad
\Longrightarrow
\qquad
\sigma_{\rm dS}
\text{ is a sink} \nonumber
\label{eq:sink_condition_summary}
\end{equation}
\begin{equation}
|\lambda_i|>1
\quad
\text{for some }i
\qquad
\Longrightarrow
\qquad
\sigma_{\rm dS}
\text{ is not a sink} \nonumber
\label{eq:not_sink_condition_summary}
\end{equation}
\begin{equation}
|\lambda_i|=1
\quad
\text{for some }i
\qquad
\Longrightarrow
\qquad
\text{the linear analysis is inconclusive} \nonumber
\label{eq:marginal_condition_summary}
\end{equation}
where the second one includes both the {\it source} ($|\lambda_i| > 1$ $\forall i)$ and the {\it saddle} ($|\lambda_i| > 1$ for some $i$ and $|\lambda_j| < 1$ for some $j$ with $i+j$ set of eigenvectors).
Thus de Sitter is an attractor of the bootstrap map if and only if the corresponding GS fixed point is a locally stable sink. In other words
de Sitter is a bootstrap attractor implies that 
$|\lambda_i|<1$
for all eigenvalues of 
$D{\cal F}\big|_{\sigma_{\rm dS}}$.
The map ${\cal F}$ is defined only implicitly by the coupled equations, so $D{\cal F}|_{\sigma_{\rm dS}}$ must be extracted by linearizing the full bootstrap loop. Writing:
\begin{equation}
{\cal F}
=
{\cal R}\circ{\cal U}\circ{\cal S} ~~~ \implies ~~ \mathcal F: \sigma
\stackrel{S}{\longmapsto}
{\bf T}
\stackrel{U}{\longmapsto}
\Xi
\stackrel{R}{\longmapsto}
\sigma' ,
\label{eq:F_composition_summary}
\end{equation}
where ${\cal S}$ denotes source extraction from ${\cal B}_{AB}=0$, ${\cal U}$ denotes the mean-field update from ${\cal E}_{AB}=0$, and ${\cal R}$ reconstructs the GS label from the updated mean field, one finds:
\begin{equation}
D{\cal F}\big|_{\sigma_{\rm dS}}
=
D{\cal R}\big|_{\Xi_{\rm dS}}
\circ
D{\cal U}\big|_{{\bf T}_{\rm dS}}
\circ
D{\cal S}\big|_{\sigma_{\rm dS}} ,
\label{eq:DF_chain_rule_summary}
\end{equation}
provides one way to quantify the form of $\mathcal D \mathcal F\big|_{\sigma_{\rm dS}}$. Note that $\mathcal F$ is a composition of maps, not a product of ordinary functions. Therefore one uses the chain rule, not the Leibniz product rule to write \eqref{eq:DF_chain_rule_summary}\footnote{There is an easy way to justify the form \eqref{eq:DF_chain_rule_summary}. Suppose
$s:\mathbb R^n\to\mathbb R^m,
u:\mathbb R^m\to\mathbb R^p$, and 
$r:\mathbb R^p\to\mathbb R^q$.
Then
$f=r\circ u\circ s$
means
$f(x)=r(u(s(x)))$.
If one perturbs $x\to x+\delta x$, then:
\begin{equation}
\delta s
=
Ds|_x\cdot \delta x,~~~~
\delta u
=
Du|_{s(x)}\cdot \delta s,~~~~
\delta f
=
Dr|_{u(s(x))}\cdot \delta u , \nonumber
\label{eq:delta_f_finite}
\end{equation}
where the transformations are arranged so that the nested structure remains intact. This is the key point that distinguishes it from the usual product of functions. Therefore, 
combining them gives:
\begin{equation}
\delta f
=
Dr|_{u(s(x))}
\circ
Du|_{s(x)}
\circ
Ds|_x
\cdot
\delta x~~~ \implies ~~~
Df|_x
=
Dr|_{u(s(x))}
\circ
Du|_{s(x)}
\circ
Ds|_x ,
\label{eq:Df_chain_rule}
\end{equation}
which is exactly the same structure as in the bootstrap problem. Therefore the transformation rule in \eqref{eq:DF_chain_rule_summary} should be understood in exactly the same way.}.
To see this more concretely, perturb both the GS data and the mean fields as 
$\sigma=\sigma_{\rm dS}+\delta \sigma$ and 
$\langle\Xi\rangle_\sigma=\Xi_{\rm dS}+\delta \langle\Xi\rangle_\sigma$ respectively. This perturbed configuration is generically off-shell, so one should {\it not} assume that it still satisfies the coupled bootstrap equations exactly. Instead, the role of the linearized equations is to determine the first-order correction map near the fixed point. This may be seen as follows. Expanding the second bootstrap equation \eqref{eq:BAB_def} around the de Sitter fixed point gives:
\begin{align} 
\delta {\cal B}_{AB}
&=
\frac{\delta {\cal B}_{AB}}{\delta \sigma}\Big|_{\rm dS}\,\delta \sigma
+
\frac{\delta {\cal B}_{AB}}{\delta \langle\Xi\rangle_\sigma }\Big|_{\rm dS}\,\delta \langle\Xi\rangle_\sigma
+
\frac{\delta {\cal B}_{AB}}{\delta {\bf T}^{\rm np}_{CD}}\Big|_{\rm dS}\,\delta {\bf T}^{\rm np}_{CD}
+
\frac{\delta {\cal B}_{AB}}{\delta {\bf T}^{\rm nloc}_{CD}}\Big|_{\rm dS}\,\delta {\bf T}^{\rm nloc}_{CD} ,
\label{eq:deltaB_full_corrected}
\end{align}
where $\langle\Xi\rangle_\sigma$ generically denotes the full collective mean-field variable.
The first-order bootstrap update is defined by requiring that the corrected effective sources satisfy the linearized source relation at the fixed point. Equivalently, one solves the linearized second equation:
\begin{equation}
\delta {\cal B}_{AB}=0
\label{eq:deltaB_zero_corrected}
\end{equation}
as a relation among the source variations and the perturbed trial data. This is {\it not} the statement that the original perturbed configuration is already on-shell; it is the linearized equation that determines how the source sector must be adjusted at first order.
Thus one obtains a linear relation of the form:
\begin{equation}
\sum_{i = 1}^2{\cal M}^{(i)}_{AB}{}^{CD}\,
\delta {\bf T}^{(i)}_{CD}
=
{\cal N}_{AB}\,\delta \sigma
+
{\cal P}_{AB}\,\delta \langle\Xi\rangle_\sigma ,
\label{eq:linear_relation_sources_corrected}
\end{equation}
where
$\Big(\delta {\bf T}^{(1)}_{CD}, \delta {\bf T}^{(2)}_{CD}\Big)
\equiv
\Big(
\delta {\bf T}^{\rm np}_{CD},
\delta {\bf T}^{\rm nloc}_{CD}
\Big)$, which appears in this form because the second bootstrap equation determines at least {\it two} relevant source variations:
$\delta {\bf T}^{\rm np}_{AB}$ and 
$\delta {\bf T}^{\rm nloc}_{AB}$.
Assuming the relevant linear operator is invertible, {\it i.e.} assuming invertibility in this enlarged source space, this becomes:
\begin{equation}
\delta {\bf T}^{(i)}_{AB}
=
{\cal A}^{(i)}_{AB}\,\delta \sigma
+
{\cal B}^{(i)}_{AB}{}^{CD}\,\delta \langle\Xi_{CD}\rangle_\sigma ,
\label{eq:deltaT_from_delta_sigma_deltaXi_corrected}
\end{equation}
for $i = 1, 2$ and we have put in the components assuming that 
$ \langle\Xi_{CD}\rangle_\sigma \propto \langle g_{CD}\rangle_\sigma$ (although the story could be easily generalized for any on-shell DOFs).
Therefore the linearized second bootstrap equation determines how the effective source sector responds to a small perturbation of the GS label and the mean fields.

We can now ask what is the linearized mean-field update from the first bootstrap equation. The first bootstrap equation is \eqref{eq:EAB_def}.
At the exact fixed point this vanishes, but for a nearby trial configuration it is generically nonzero. The linearized first equation therefore determines the first-order mean-field correction produced by the source variations. Expanding around the de Sitter background gives:
\begin{equation}
\delta {\cal E}_{AB}
=
\delta {\bf G}_{AB}
-
\delta {\bf T}^{\rm np}_{AB}
-
\delta {\bf T}^{\rm nloc}_{AB}.
\label{eq:deltaE_corrected}
\end{equation}
Since the Einstein tensor is a functional of the mean fields, its first variation is linear in the perturbation, {\it i.e.}
$\delta {\bf G}_{AB}
=
{\cal L}_{AB}{}^{CD}\,\delta \langle\Xi_{CD}\rangle_\sigma$, 
where ${\cal L}_{AB}{}^{CD}$ is the linearized Einstein operator on the de Sitter background.
Substituting \eqref{eq:deltaT_from_delta_sigma_deltaXi_corrected} into \eqref{eq:deltaE_corrected}, one finds:
\begin{equation}
\delta {\cal E}_{AB}
=
{\cal L}_{AB}{}^{CD}\,\delta \langle\Xi_{CD}\rangle_\sigma
-
{\cal A}_{AB}\,\delta \sigma
-
{\cal B}_{AB}{}^{CD}\,\delta \langle\Xi_{CD}\rangle_\sigma ,
\label{eq:deltaE_substituted_corrected}
\end{equation}
where $\mathcal A_{AB} = \mathcal A^{(1)}_{AB} + \mathcal A^{(2)}_{AB}$ and $ {\cal B}_{AB}{}^{CD} = {\cal B}^{(1)}_{AB}{}^{CD} + {\cal B}^{(2)}_{AB}{}^{CD}$ from \eqref{eq:deltaT_from_delta_sigma_deltaXi_corrected}. (Note that in the variation of ${\bf G}_{AB}$ only its metric component enters directly. Thus the notation $\delta \langle\Xi_{CD}\rangle_\sigma$ is simply shorthand for the corresponding metric perturbation $\delta \langle g_{CD}\rangle_\sigma$.)
The linearized bootstrap update is obtained by requiring that the updated mean field cancel the first-order residual, namely:
\begin{equation}
\delta {\cal E}_{AB}=0.
\label{eq:deltaE_zero_corrected}
\end{equation}
Again, this is not the statement that the original perturbed trial configuration was already on-shell. It is the equation that determines the first-order correction of the mean field.
Thus:
\begin{equation}
\Big(
{\cal L}_{AB}{}^{CD}
-
{\cal B}_{AB}{}^{CD}
\Big)\delta \langle\Xi_{CD}\rangle_\sigma
=
{\cal A}_{AB}\,\delta \sigma ~~~ \implies ~~~
\delta \langle\Xi_{AB}\rangle_\sigma
=
{\cal Q}_{AB}\,\delta \sigma ,
\label{eq:deltaXi_from_delta_sigma_corrected}
\end{equation}
where we assumed that the $\mathcal L_{AB}{}^{CD} - \mathcal B_{AB}{}^{CD}$ operator has an inverse to write the second equation.
This is the linearized one-step update of the mean field induced by a small perturbation of the GS input. This is in fact a useful relation that will help us to quantify the reconstruction of the updated GS state. The point is that the bootstrap map acts on GS labels, so one finally reconstructs the updated GS state from the updated mean field. For example, if
$\sigma'={\cal R}(\langle\Xi'\rangle_\sigma)$,
then at linear order:
\begin{equation}
\delta \sigma'
=
D{\cal R}\big|_{\Xi_{\rm dS}}\cdot \delta \langle\Xi \rangle_\sigma
=
D{\cal R}\big|_{\Xi_{\rm dS}}\cdot {\cal Q}\,\delta \sigma, 
\end{equation}
where we used \eqref{eq:deltaXi_from_delta_sigma_corrected} to write the second relation. Notice that now there is a direct map between $\delta\sigma'$ and $\delta\sigma$, so we can utilize \eqref{eq:leading_order_delta_summary} and get the requisite map.
Therefore the linearized bootstrap map is:
\begin{equation}
D{\cal F}\big|_{\sigma_{\rm dS}}
=
D{\cal R}\big|_{\Xi_{\rm dS}}\circ {\cal Q} ,
\label{eq:DF_final_composition_corrected}
\end{equation}
which is the precise sense in which the practical procedure determines
$D{\cal F}\big|_{\sigma_{\rm dS}}$. Note that in deriving \eqref{eq:DF_final_composition_corrected} we did not
assume that a generic perturbed configuration has remained on-shell. Rather, the linearized second and first bootstrap equations determine the first-order source and mean-field corrections, and their composition yields the linearized update of the GS label itself. The eigenvalues of the operator \eqref{eq:DF_final_composition_corrected} then determine whether we have a sink, a source or a saddle.

The sink/source/saddle language should however be interpreted carefully. It does not refer to physical time evolution in spacetime, since the bootstrap map carries no physical time parameter. Rather, it refers to the discrete iteration flow on the space of trial GS states. In particular, nearby perturbations of an exact fixed point are generically off-shell, {\it i.e.}
$\sigma_\ast+\delta \sigma$
is typically not an exact solution.
Nevertheless, the bootstrap map still acts on such trial data. Thus:
\begin{equation}
\text{\textcolor{blue}{sink}}
\quad
\Longrightarrow
\quad
\text{\textcolor{blue}{small off-shell errors are damped}} \nonumber
\label{eq:sink_meaning_summary}
\end{equation}
\begin{equation}
\text{\textcolor{blue}{source}}
\quad
\Longrightarrow
\quad
\text{\textcolor{blue}{small off-shell errors are amplified}} \nonumber
\label{eq:source_meaning_summary}
\end{equation}
\begin{equation}
\text{\textcolor{blue}{saddle}}
\quad
\Longrightarrow
\quad
\text{\textcolor{blue}{some perturbations are damped and others are amplified}}\nonumber
\label{eq:saddle_meaning_summary}
\end{equation}
So the classification measures the local robustness of the self-consistency procedure, not the Lorentzian dynamics of the perturbed spacetime.
This distinction is essential when discussing the physical significance of de Sitter. The existence of an exact de Sitter GS solution and its stability under bootstrap iteration are logically distinct. If:
\begin{equation}
{\cal E}_{AB}[\Xi_{\rm dS}]=0,
\qquad
{\cal B}_{AB}[\sigma_{\rm dS},\Xi_{\rm dS}]=0,
\label{eq:exact_existence_summary}
\end{equation}
then de Sitter exists as an exact self-consistent branch. Even if the corresponding fixed point is a source, as shown in {\bf figure \ref{sssrotate}}, that does not invalidate the existence claim. It implies only that de Sitter is bootstrap-unstable and not selected from generic nearby trial GS data. Equivalently, 
an exact de Sitter GS solution is not ruled out by being a source of the bootstrap map; it is merely non-generic or unstable under bootstrap iteration.

Accordingly, the sink/source/saddle classification does not primarily describe nearby exact on-shell solutions. Rather, it describes how nearby off-shell trial GS configurations are mapped relative to a given exact fixed point. Other exact fixed points may certainly exist elsewhere in state space, but that is a separate question concerning the global structure of
\begin{equation}
{\rm Fix}({\cal F})
=
\{\sigma\in{\cal S}_{\rm GS}\,|\,{\cal F}(\sigma)=\sigma\}.
\label{eq:FixF_summary}
\end{equation}
In this sense, the classification is dispensable for the bare existence of a solution, but essential for understanding the local organization and robustness of the bootstrap solution space.

\begin{figure}[h]
\centering
\begin{tabular}{c}
\includegraphics[width=6in]{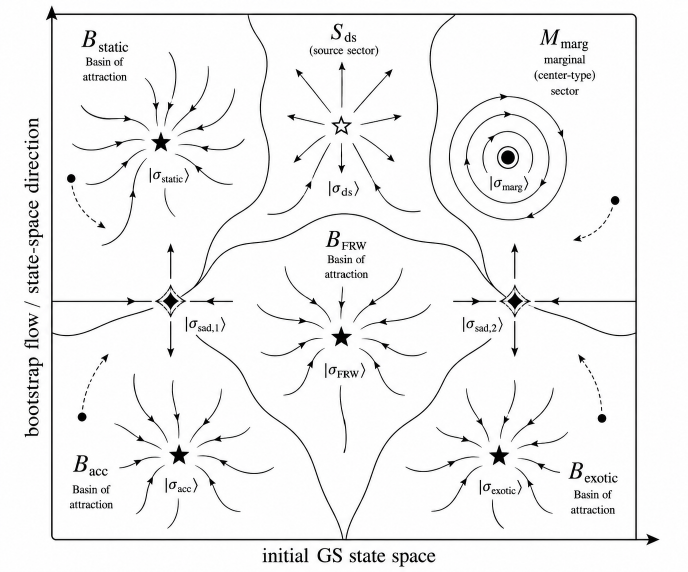}
\end{tabular}
\caption[]{Besides sinks, sources, and saddles, the linearized bootstrap map may also admit center-type behavior. The big solid circle represents the marginal (center-type) sector. This occurs when some eigenvalues lie on the unit circle, for example  $\lambda_\pm=e^{\pm i\theta}$ such that $|\lambda_i| = 1$. In that case nearby trajectories can rotate around the fixed point without either converging to it or diverging away from it. Such directions are marginal and require nonlinear analysis to determine the ultimate behavior. We have also exhibited the possibility that, in the neighborhood of a de Sitter fixed point, the bootstrap flow is predominantly repulsive.}
\label{sssrotate}
\end{figure}

\subsubsection{Unsplit vs split equations and the role of bootstrap \label{sec7.5.44}}

Before ending this section let us elaborate on the aforementioned procedure a bit more which would lead to an important clarification of what the bootstrap procedure is actually doing. This will also tie-up nicely with the cautionary statements we made in section \ref{sec7.4.1}. 

The generic Schwinger--Dyson equation \eqref{vivluna}
which we shall call the \emph{unsplit} equation, is the universal GS Schwinger--Dyson identity. It should hold for any admissible GS state $|\sigma\rangle$, irrespective of whether that state admits a semiclassical interpretation.
By contrast, the \emph{split} equations \eqref{bootsforwalking}
are much more restrictive. They are not expected to hold for arbitrary $|\sigma\rangle$, but only for those special GS states for which the expectation values define a self-consistent semiclassical branch and the GS data satisfy the associated bootstrap fixed-point condition. In other words, the split equations are stronger conditions selecting the special semiclassical/bootstrap-fixed-point sector.

With this in mind, the bootstrap procedure based on the split equations is not solving the most general GS system. Rather, it is solving the \emph{semiclassical self-consistency problem}. In other words, it is designed to answer the following question:
For which choices of $|\sigma\rangle$ do the expectation values 
$\langle \Xi\rangle_\sigma$
define a self-consistent semiclassical background?
That is a much narrower question than the question answered by the unsplit equation, namely: What exact Schwinger--Dyson identity is satisfied by a generic GS state?
Therefore it makes perfect sense to use only the split equations in the bootstrap procedure provided one is explicit that the goal is to identify the semiclassical/bootstrap-fixed-point sector, not the full GS state space. Stated alternatively,
it is consistent to use only the split equations for the bootstrap procedure, because that procedure is not meant to classify all GS states. It is meant to isolate the special GS states that admit a semiclassical interpretation and solve the self-consistency conditions

The reason the unsplit equation alone is not sufficient for the bootstrap is that it does not separate the problem into 
(i) a background equation for $\langle\Xi\rangle_\sigma$,
and (ii) a consistency equation for the GS insertion itself. Certain aspects of this issue were already pointed out in section \ref{sec7.4.1} wherein we cautioned against the usage of the unsplit equation for determining the semiclassical solutions.
In the unsplit form, both effects are entangled inside the exact expectation values. That equation is true for all admissible GS states, including highly off-shell or non-semiclassical ones. Therefore, by itself, it does not tell us which states correspond to self-consistent semiclassical configurations.

The split equations do precisely this extra job: they reorganize the exact identity into a form where one can look for fixed points in the space of GS labels.
So it is not that the split equations are more fundamental than the unsplit equation. They are in fact a stronger reorganization of it, adapted specifically to the search for semiclassical bootstrap fixed points.
The proper hierarchy is therefore:
\begin{equation}
\text{\textcolor{blue}{generic admissible GS state }}\textcolor{blue}{|\sigma\rangle}
\quad\Longrightarrow\quad
\text{\textcolor{blue}{unsplit SD equation \eqref{vivluna}}} \nonumber
\label{eq:hierarchy_unsplit}
\end{equation}
\begin{equation}
\text{\textcolor{blue}{bootstrap-selected semiclassical GS state }}|\textcolor{blue}{\sigma\rangle}
\quad\Longrightarrow\quad
\text{\textcolor{blue}{split equations \eqref{bootsforwalking}}} \nonumber
\label{eq:hierarchy_split}
\end{equation}
Equivalently, one may say that the split equations define a subset
${\cal S}_{\rm sc}
\subset
{\cal S}_{\rm GS}$,
where ${\cal S}_{\rm GS}$ is the full space of admissible GS states satisfying the unsplit equation, and ${\cal S}_{\rm sc}$ is the distinguished subset of GS states that are semiclassical bootstrap fixed points. Thus the bootstrap procedure is naturally a search inside ${\cal S}_{\rm sc}$, not inside all of ${\cal S}_{\rm GS}$. Therefore we can summarize the story in the following way: it makes sense to use only the split equations for the bootstrap procedure, but only because the bootstrap is not intended to solve the full GS Schwinger--Dyson problem for all $|\sigma\rangle$. Rather, it is intended to find the special semiclassical/bootstrap-fixed-point states. The unsplit equation remains the more general identity valid for arbitrary GS states, while the split equations define the distinguished semiclassical sector.



\section{WdW equations, world-sheet vertex operators and the GS states \label{sec8.0}}

For this section we will change the gear a little bit to show the connection between three seemingly disparate topics: the Wheeler--DeWitt equations, the bulk GS operators (that we studied so far), and the vertex operators in the worldsheet representation of string theory. For the present construction, since supersymmetry plays no role, we will not worry too much about the worldsheet fermions or the corresponding GSO projections. These details are important for a more quantitative study, but maybe not so much for the analysis that we have in mind here. We also do not claim to provide a complete account here. Our objective is more modest: to highlight several suggestive connections and structural parallels, while leaving a fuller and more systematic investigation to future work.

We can start by summarizing the salient features of what we have so far. Let us denote the on-shell fields by
$\Xi(x)
=
\Big(
g_{\mu\nu}(x), \Phi(x), \Psi(x),
{\rm ghosts}(x)
\Big)$ where $x= ({\bf x}, t)$,
that include the full set of bosonic, fermionic, and ghost fields, and let
$S_{\rm tot}[\Xi]$
be the full trans-series effective action, including perturbative, non-perturbative, and possibly non-local contributions. This could either be in the Euclidean and the Lorentzian forms $-$ unless mentioned explicitly $-$ and we will make a more informed choice a little later.  The corresponding classical, quantum, canonical, and GS-deformed structures may then be summarized as follows.

\paragraph{Classical saddle-point equation.}

They typically appear from extremizing $S_{\rm tot}[\Xi]$, with $\Xi(x)$ denoting the on-shell fields and $x = ({\bf x}, t)$. 
The stationary configurations of the full trans-series action are determined by:
\begin{equation}
\frac{\delta S_{\rm tot}[\Xi]}{\delta \Xi(x)}
=
0.
\label{eq:classical_saddle_transseries}
\end{equation}
This is the classical equation of motion of the full effective system. In a trans-series setting, its solutions should be understood as the saddle points of the resummed effective action, and may include non-perturbative saddles in addition to the usual perturbative extrema.

\paragraph{Vacuum Schwinger--Dyson equation.}

For the undeformed vacuum theory, assuming an invariant functional measure and no additional insertions, the exact Schwinger--Dyson identity is given by:
\begin{equation}
\left\langle
\frac{\delta S_{\rm tot}[\Xi]}{\delta \Xi(x)}
\right\rangle_0
=
0 ,
\label{eq:vacuum_SD_transseries}
\end{equation}
where the subscript $0$ denote the vacuum.
This is the quantum equation of motion in the vacuum sector. It is weaker than the classical saddle-point equation \eqref{eq:classical_saddle_transseries}, since it is an expectation-value identity rather than a pointwise statement in field space.

\paragraph{Canonical Hamiltonian and constraints.}

Once $S_{\rm tot}$ is specified, we can extract the Lagrangian $L_{\rm tot}$, and from there $H_{\rm tot}$ could be easily evaluated after a 3+1 split on a globally hyperbolic spacetime.
The total Hamiltonian admits the standard decomposition:
\begin{equation}
H_{\rm tot} = \Pi_\Xi \dot{\Xi} - L_{\rm tot}
=
H_{\rm bulk}
+
H_{\partial},\qquad
H_{\rm bulk}
=
\int_\Sigma d^{d-1}x\;
\Big(
N\,{\cal H}_\perp
+
N^i\,{\cal H}_i
\Big).
\label{eq:Hbulk_constraints_formal}
\end{equation}
Here ${\cal H}_\perp$ and ${\cal H}_i$ are the Hamiltonian and momentum constraints of the full gravity-plus-matter-plus-ghost system; and we have imposed the standard ADM formalism. These are related to the metric Euler--Lagrange equations through:
\begin{equation}
E_{\mu\nu}(x)
\equiv
\frac{\delta S_{\rm tot}[\Xi]}{\delta g^{\mu\nu}(x)} 
=
\frac{1}{2\kappa}\sqrt{-g(x)}
\Big(G_{\mu\nu}(x)
-
\kappa T_{\mu\nu}^{\rm tot}(x)\Big),
\label{eq:Emunu_definition_formal}
\end{equation}
for an Einstein-type system,
up to sign conventions and possible higher-derivative, non-local, or genuinely trans-series corrections. Introducing the future-directed unit normal $n^\mu$ to the spatial slice $\Sigma$, together with the induced metric
$h_{\mu\nu}
=
g_{\mu\nu}
+
n_\mu n_\nu$,
the constraint equations are obtained as the normal-tangential and normal-normal projections of \eqref{eq:Emunu_definition_formal} that take the familiar geometric form :
\bg\label{eq:Hconstraints_as_projections}
&& {\cal H}_i
\propto
E_{\mu\nu}n^\mu h^\nu{}_i =
D_j
\Big(
K^j{}_i-\delta^j{}_iK
\Big)
-
j_i^{\rm tot},
\qquad
j_i^{\rm tot}
=
-\,T_{\mu\nu}^{\rm tot}n^\mu h^\nu{}_i \nonumber\\
&& {\cal H}_\perp
\propto
E_{\mu\nu}n^\mu n^\nu =
\frac{1}{2}
\Big(
{}^{(d-1)}R
+
K^2
-
K_{ij}K^{ij}
\Big)
-
\rho_{\rm tot}, ~~~ \rho_{\rm tot}
=
T_{\mu\nu}^{\rm tot}n^\mu n^\nu ,
\nd
respectively
up to overall convention-dependent factors. In the ADM decomposition, the proportionality constants are just ${2\over N}$ and $-{2\over N}$ respectively.
In writing \eqref{eq:Hconstraints_as_projections}, we have assumed Einstein gravity coupled to the full matter sector.
On shell one has:
\begin{equation}
{\cal H}_\perp=0,
\qquad
{\cal H}_i=0,
\label{eq:constraints_vanish_nice}
\end{equation}
so that the bulk Hamiltonian vanishes on physical configurations and the only nontrivial canonical generator that remains is the boundary Hamiltonian $H_\partial$.

\paragraph{Wheeler--DeWitt constraint and boundary time evolution.}

Now comes the crucial observation. Even if we express the total Hamiltonian as in \eqref{eq:Hbulk_constraints_formal}, the WdW equations only come from the bulk parts whereas the Schr\"odinger equation comes from the boundary part. This is most obvious once we use the Weiss variation\footnote{For more details on the Weiss variation \cite{weiss} and the subsequent derivation of the WdW equation from there, the readers may look up the interesting series of papers by Feng and Matzner \cite{feng}.} of the Feynman path integral \cite{feng} (see also \cite{wdwpaper}). 
The bulk WdW equations are obtained by quantizing the Hamiltonian constraint supplemented by the momentum constraints:
\begin{equation}
\widehat{\cal H}_\perp\,\Psi[\Xi({\bf x}); t]
=
0,~~~~~~~~
\widehat{\cal H}_i\,\Psi[\Xi({\bf x}); t]
=
0 ,
\label{eq:momentum_constraints_quantum_nice}
\end{equation}
respectively, both acting on the wavefunctional $\Psi[\Xi({\bf x}); t]$.
If a boundary notion of time exists, then the residual time dependence of the wavefunctional is generated by the boundary Hamiltonian according to:
\begin{equation}
i\frac{\partial \Psi[\Xi({\bf x}); t]}{\partial t}
=
\widehat H_\partial\,\Psi[\Xi({\bf x}); t].
\label{eq:boundary_schrodinger_nice}
\end{equation}
Thus the Wheeler--DeWitt equations and the boundary Schr\"odinger equation play distinct roles: the former imposes the bulk gravitational constraint, while the latter governs the nontrivial boundary evolution.

\paragraph{GS-deformed Schwinger--Dyson equation.}

Let $|\sigma\rangle$ denote a GS state generated by the operator $D(\sigma)$. Then the vacuum Schwinger--Dyson identity \eqref{eq:vacuum_SD_transseries} is replaced by the sourced equation:
\begin{equation}
\left\langle
\frac{\delta S_{\rm tot}[\Xi]}{\delta \Xi(x)}
\right\rangle_\sigma
=
\left\langle
\frac{\delta}{\delta \Xi(x)}
\log\Big(
D^\dagger(\sigma)D(\sigma)
\Big)
\right\rangle_\sigma,
\label{eq:GS_deformed_SD_nice}
\end{equation}
up to convention-dependent factors of $i$ in Lorentzian signature or $\pm$ in the Euclidean signature. This is the precise sense in which the GS state induces a source for the quantum equations of motion of the full trans-series system. From here the actual set of equations are the two sets of the bootstrap equations \eqref{eq:offshell_boot_pair_final} and \eqref{bootsforwalking}.

\subsection{Boundary time, WdW time, and Schr\"odinger evolution \label{sec8.1}}

In a diffeomorphism-invariant theory of gravity coupled to matter, fermions, and ghosts, the canonical description naturally contains two distinct structures: the bulk Wheeler--DeWitt constraints and the boundary Schr\"odinger evolution. It is useful to formulate these carefully in a unified framework.
The full set of canonical variables be denoted collectively by 
$\Xi
=
\Big(
g_{\mu\nu},
\Phi,
\Psi_{\rm f},
{\rm ghosts}
\Big)$, and 
the full wavefunctional depends on all canonical data and on the boundary time parameter as $\Psi \equiv \Psi[\Xi({\bf x}); t]$. 
The exact theory therefore contains the bulk constraints \eqref{eq:momentum_constraints_quantum_nice}
together with the exact boundary-time evolution equation \eqref{eq:boundary_schrodinger_nice}.
Thus the exact canonical theory contains both the WdW constraints and a genuine Schr\"odinger equation with respect to the asymptotic boundary time\footnote{We assume a chosen spacelike boundary hypersurface together with a timelike congruence of geodesics orthogonal to that boundary, and we use the corresponding affine parameter to define the temporal coordinate $t \equiv t_\partial$. In an asymptotically flat spacetime, this choice may be made canonically in the asymptotic region, where $t$ is identified with the proper time measured along a congruence of asymptotically inertial geodesics normal to the boundary.} $t$.
The bulk WdW operator may be split roughly into a purely gravitational part and a predominantly non-gravitational part in the following way:
\begin{equation}
\widehat{\cal H}_{\rm bulk}
=
\widehat{H}_g
+
\widehat{H}_{\rm mat+gh} =
\int_\Sigma d^{d-1}x\;
\Big(
N\,\widehat{\mathcal H}_\perp
+
N^i\,\widehat{\mathcal H}_i
\Big) ,
\label{eq:Hbulk_split_gravity_matter_boundary_time}
\end{equation}
where $\widehat{\cal H}_{\rm mat+gh}$ contains matter, fermionic, and ghost contributions, but is also coupled to the metric due to general covariance of the system. This is important because in a system with dynamical gravity, which we are interested in, a pure split into gravitational and non-gravitational parts is not 
possible\footnote{Henceforth by {\it non-gravitational} we will always mean predominantly non-gravitational unless mentioned otherwise.}. 
The exact WdW equation then reads:
\begin{equation}
\Big(
\widehat{H}_g
+
\widehat{H}_{\rm mat+gh}
\Big)\Psi
=
0~~~ \implies ~~~
\widehat{H}_{\rm mat+gh}\Psi
=
-\,\widehat{H}_g\Psi.
\label{eq:WdW_rearranged_boundary_time}
\end{equation}
This rewriting is exact, but is also presented in a suggestive way. However, one should not immediately interpret the right-hand side as a fundamental time derivative in the following sense:
\begin{equation}
-\,\widehat{H}_g\Psi
\stackrel{?}{=}
i\frac{\partial \Psi}{\partial t'}.
\label{eq:naive_tprime_wrong_boundary_time}
\end{equation}
In general, $\widehat{H}_g$ is a second-order functional differential operator in the metric variables, not a first-order derivative operator generating evolution in some exact time $t'$. Therefore the WdW equation is not itself a Schr\"odinger equation.
A Schr\"odinger-type time emerges only after a semiclassical treatment of the gravitational sector. To make this precise, assume that the bulk WdW operator takes the standard schematic form:
\begin{equation}
\widehat{\cal H}_{\rm bulk}
=
\underbrace{-\frac{1}{2M_{\rm P}^2}\,
G_{ijkl}(h)\,
\frac{\delta^2}{\delta h_{ij}\delta h_{kl}}
+
M_{\rm P}^2\,U[h]}_{\widehat{H}_g}
+
\widehat H_{\rm mat+gh}[\Xi],
\label{eq:WdW_operator_explicit_boundary_time}
\end{equation}
where $G_{ijkl}(h)$ is the DeWitt supermetric on superspace, $U[h]$ is the gravitational potential term, and $\widehat H_{\rm mat+gh}[h]$ is the Hamiltonian operator for matter, fermions, and ghosts on the spatial slice. 
We can now make a more precise account of how the exact boundary-time evolution and the emergent semiclassical time extracted from the bulk Wheeler--DeWitt equation fit together. The essential point is that these are, in general, two distinct structures: one is exact and tied to the existence of a nonvanishing boundary Hamiltonian, while the other is emergent and arises only after a semiclassical treatment of the gravitational sector.
Let the full wavefunctional depend on the canonical data on a spatial slice:
\begin{equation}
\Psi[\Xi({\bf x}); t]
=
\Psi\Big[
h_{ij}(\mathbf x),
\Phi(\mathbf x),
\Psi_{\rm f}(\mathbf x),
{\rm ghosts}(\mathbf x);
t
\Big] ,
\label{eq:full_wavefunctional_exact}
\end{equation}
where note that the wavefunction is expressed using bulk fields localized on a given Cauchy slice defined at time $t$. There is an explicit dependence on time $t$ because the evolution is governed by a {\it boundary} Hamiltonian.
As discussed earlier, the total Hamiltonian is assumed to take the standard form
$\widehat{\mathcal H}_{\rm tot}
=
\widehat{\mathcal H}_{\rm bulk}
+
\widehat H_{\partial}$,
with the bulk part given by \eqref{eq:Hbulk_split_gravity_matter_boundary_time}, and the boundary part is $\widehat{H}_\partial$ which in fact governs the temporal evolution. This means the wavefunction \eqref{eq:full_wavefunctional_exact} has to satisfy:
\begin{equation}
\widehat{\mathcal H}_\perp\,\Psi
=
0,
\qquad
\widehat{\mathcal H}_i\,\Psi
=
0, \qquad
i\frac{\partial}{\partial t}\Psi
=
\widehat H_{\partial}\,\Psi ,
\label{eq:boundary_SE_exact}
\end{equation}
the first two being the bulk constraints and the third one is the exact boundary-time evolution. The first two equations can also be expressed as \eqref{eq:WdW_rearranged_boundary_time}, which is what we shall use here. In this sense, as mentioned earlier, 
the exact canonical theory contains both the bulk Wheeler--DeWitt constraints and a genuine Schr\"odinger equation with respect to the external or asymptotic time $t$. Moreover, 
to obtain a Schr\"odinger-type evolution equation, one must treat the gravitational sector semiclassically. We therefore introduce the Born--Oppenheimer or WKB ans\"atze:

{\footnotesize
\begin{equation}
\Psi\Big[
h_{ij}(\mathbf x),
\Phi(\mathbf x),
\Psi_{\rm f}(\mathbf x),
{\rm ghosts}(\mathbf x);
t
\Big]
=
A[h;t]\,
\exp\!\big(
iM_{\rm P}^2 S_0[h;t]
\big)\,
\psi\Big[
h_{ij}(\mathbf x),
\Phi(\mathbf x),
\Psi_{\rm f}(\mathbf x),
{\rm ghosts}(\mathbf x);
t
\Big],
\label{eq:BO_WKB_ansatz_revised}
\end{equation}}
where $A[h;t]$ is a slowly varying prefactor, $S_0[h;t]$ is the leading gravitational phase, and $\psi$ is the remaining wavefunctional. Again, $\psi$ needs to be a functional of all the fields on the Cauchy slice with an additional explicit dependence on time $t$. For definiteness, suppose the bulk WdW operator takes the schematic form \eqref{eq:WdW_operator_explicit_boundary_time}
where we can assume that all the higher order terms from the pure  gravitational part of the action (that appears from the trans-series form of the action) are to be inserted in $U(h)$.
Substituting \eqref{eq:BO_WKB_ansatz_revised} into
$\widehat{\mathcal H}_{\rm bulk}\Psi=0$ we get the following equation:
\bg\label{eq:rescaled_WdW_equation_simple}
0
&= & 
\frac{\textcolor{blue}{M_{\rm P}^4}}{2}\,
G_{ijkl}
\frac{\delta S_0}{\delta h_{ij}}
\frac{\delta S_0}{\delta h_{kl}}\,
A\psi
+
\textcolor{blue}{M_{\rm P}^4} \,U[h]\,
A\psi -\frac{1}{2}\,
G_{ijkl}
\frac{\delta^2 (A\psi)}{\delta h_{ij}\delta h_{kl}}
\nonumber\\
&- &
\frac{i \textcolor{red}{M_{\rm P}^2}}{2}\,
G_{ijkl}
\frac{\delta^2 S_0}{\delta h_{ij}\delta h_{kl}}\,
A\psi
-
i \textcolor{red}{M_{\rm P}^2} \,
G_{ijkl}
\frac{\delta S_0}{\delta h_{ij}}
\frac{\delta (A\psi)}{\delta h_{kl}}
+
\textcolor{red}{M_{\rm P}^2} \,
\widehat H_{\rm mat+gh}[\Xi]\,
A\psi ,
\nd
where notice that we have $\mathcal O(M_p^4), \mathcal O(M_p^2)$ and $\mathcal O(1)$ terms, assuming that the gravitational potential $U[h]$ does not have any $M_p$ dependent terms. This is of course very naive, but let us still carry on. (We will generalize the story in section \ref{sec8.1.7}.) Now expanding in powers of $M_{\rm P}^2$ yields, at leading order, the gravitational Hamilton--Jacobi equation:
\begin{equation}
\frac{1}{2}
G_{ijkl}
\frac{\delta S_0}{\delta h_{ij}}
\frac{\delta S_0}{\delta h_{kl}}
+
U[h]
=
0 ,
\label{eq:gravitational_HJ_revised}
\end{equation}
where it is assumed that we have extracted the relevant powers of $M_p$ from $U[h]$. (For the time being however we will avoid these nuances, because the story gets much more involved as we shall soon see.)
At the next order, the story is a bit more subtle because there are terms that involve the prefactor $A[h; t]$, and terms that do not involve the prefactor. This is where we can choose $A[h; t]$ to eliminate some of the terms and bring the resulting set of terms in a more suggestive form\footnote{It is interesting that the $\mathcal O(M_p^0)$ term also comes with the same factor of $A$ but in the form
$-\frac{1}{2}\,
G_{ijkl}
\frac{\delta^2(A\psi)}{\delta h_{ij}\delta h_{kl}}$. So {\it naively} one would think that the vanishing of the $\mathcal O(M_p^0)$ term should be equivalent to the condition required to get \eqref{eq:pre_schrodinger_WKB_revised}. This is {\it not} the case because the condition on $A$ to get \eqref{eq:pre_schrodinger_WKB_revised} is actually the following:
\begin{equation}
G_{ijkl}
\frac{\delta S_0}{\delta h_{ij}}
\frac{\delta A}{\delta h_{kl}}
+
\frac{1}{2}\,
G_{ijkl}
\frac{\delta^2 S_0}{\delta h_{ij}\delta h_{kl}}\,A
=
0 , \nonumber
\label{eq:transport_A_again}
\end{equation}
which is not the same as the vanishing of the $\mathcal O(M_p^0)$ term. The reason is that 
the ans\"atze \eqref{eq:BO_WKB_ansatz_revised} is only intended to solve the equation through ${\cal O}(M_{\rm P}^2)$. If we want the ${\cal O}(M_{\rm P}^0)$ terms to cancel as well, then we must enlarge the ans\"atze itself, for example schematically to:
\begin{equation}
\Psi
=
\exp\!\Big(
iM_{\rm P}^2S_0
+
iS_1
+
\frac{i}{M_{\rm P}^2}S_2
+\cdots
\Big)
\Big(
A_0+\frac{1}{M_{\rm P}^2}A_1+\cdots
\Big)
\Big(
\psi_0+\frac{1}{M_{\rm P}^2}\psi_1+\cdots
\Big). \nonumber
\label{eq:extended_ansatz_needed}
\end{equation}
Then the $\mathcal O(M_{\rm P}^0)$ equation is no longer just
$-\frac{1}{2}\,
G_{ijkl}
\frac{\delta^2(A\psi)}{\delta h_{ij}\delta h_{kl}}=0$, 
but instead receives additional contributions from
$S_1,
S_2,
\psi_1$
{etc.} In the same vein, the higher-order corrections will {\it modify} the Hamilton--Jacobi/WKB hierarchy, but this does \emph{not} mean that the leading branch
$h_{ij}(\mathbf x;t_{\rm WKB})$
ceases to be on shell. Rather, it means that its on-shell character is only \emph{leading-order} on-shell, and that at the next orders one obtains a corrected notion of semiclassical trajectory. This may be easily quantified by replacing $S_0 \to S_{\rm eff}$, $U[h] \to U_{\rm eff}[h], G_{ijkl} \to G_{ijkl}^{\rm eff}$, etc. to the Hamilton-Jacobi and the flow equations such that the on-shell/semiclassical branch $h_{ij}({\bf x}, t_{\rm WKB}) \to h_{ij}^{\rm eff}({\bf x}, t_{\rm WKB})$.
The algebra leading to the aforementioned conclusion is quite easy to tackle so we leave this as an exercise for our diligent readers. \label{lenapyar}}. After the dust settles, and after choosing the prefactor $A[h;t]$, the resulting set of terms satisfy the standard transport equation of the form:
\begin{equation}
-\,i
\int d^{d-1}x\;
G_{ijkl}(x)\,
\frac{\delta S_0}{\delta h_{ij}(x)}
\frac{\delta}{\delta h_{kl}(x)}
\psi[\Xi({\bf x}); t]
+
\widehat H_{\rm mat+gh}[\Xi]\,
\psi[\Xi({\bf x}); t]
=
0 ,
\label{eq:pre_schrodinger_WKB_revised}
\end{equation}
where $\psi[\Xi({\bf x}); t] \equiv \psi\Big[
h_{mn}(\mathbf x),
\Phi(\mathbf x),
\Psi_{\rm f}(\mathbf x),
{\rm ghosts}(\mathbf x);
t
\Big]$. Notice the presence of $-i$ on one of the term, which is one of the main prerequisites to convert it into a Schr\"odinger-like form. But it is not the whole story. The full condition is that the operator multiplying $-\,i$ should define a real directional derivative along a real WKB flow, and that the reduced matter Hamiltonian should have the usual Hermiticity properties. One may easily check that this is the case so long as $G_{ijkl}(x)\in\mathbb R$ and 
$\frac{\delta S_0}{\delta h_{ij}(x)}\in\mathbb R$. The first condition is standard for the DeWitt metric in a real Lorentzian or Euclidean canonical formulation. The second means that the leading Hamilton--Jacobi functional $S_0[h]$ is real along the branch being considered. Both are true for the present case, and therefore 
this motivates the definition of the WKB time derivative:
\begin{equation}
\frac{\partial}{\partial t_{\rm WKB}}
\equiv
\int d^{d-1}x\;
G_{ijkl}(x)\,
\frac{\delta S_0}{\delta h_{ij}(x)}
\frac{\delta}{\delta h_{kl}(x)} ,
\label{eq:WKB_time_def_revised}
\end{equation}
which naively appears to make sense because of $t_{\rm WKB}$ is a real parameter and if one uses the usual $L^2$ inner product in that parameter, then $i\frac{\partial}{\partial t_{\rm WKB}}$
is Hermitian provided the boundary terms in $t_{\rm WKB}$ vanish or are controlled appropriately. Therefore,
in terms of this derivative, \eqref{eq:pre_schrodinger_WKB_revised} becomes:

{\footnotesize
\begin{equation}
i\frac{\partial}{\partial t_{\rm WKB}}
\psi\Big[
h_{ij}(\mathbf x),
\Phi(\mathbf x),
\Psi_{\rm f}(\mathbf x),
{\rm ghosts}(\mathbf x);
t
\Big]
=
\widehat H_{\rm matt+gh}[\Xi]\,
\psi\Big[
h_{ij}(\mathbf x),
\Phi(\mathbf x),
\Psi_{\rm f}(\mathbf x),
{\rm ghosts}(\mathbf x);
t
\Big].
\label{eq:emergent_WKB_SE_revised}
\end{equation}}
Equation \eqref{eq:emergent_WKB_SE_revised} is the approximate Schr\"odinger equation for the non-gravitational sector that emerges from the bulk WdW constraint in the semiclassical regime. However the story is much more subtle than what appears at the face-value. \textcolor{blue}{First}, ${\partial\over \partial t_{\rm WKB}}$ is not originally defined as an abstract self-adjoint or anti-self-adjoint operator on the full Hilbert space of wavefunctionals. It is a {\it directional derivative on superspace along the WKB flow implicitly defined by \eqref{eq:gravitational_HJ_revised}}. Thus \eqref{eq:emergent_WKB_SE_revised} is a reduced semiclassical evolution equation, not a fundamental operator identity. This is why the naive identification
$\widehat H_{\rm mat+gh}
=
i\frac{\partial}{\partial t_{\rm WKB}}$
may be slightly misleading. A more accurate statement is that 
the operator $\widehat H_{\rm mat+gh}$ generates evolution of the reduced matter state with respect to the parameter $t_{\rm WKB}$ along a chosen semiclassical branch. Moreover \eqref{eq:WKB_time_def_revised} 
depends on the chosen Hamilton–Jacobi functional $S_0[h]$. If one moves
to a different WKB branch, then one gets a different derivative operator. Thus the relation \eqref{eq:emergent_WKB_SE_revised} holds only branch by branch. This is another reason why one should not elevate \eqref{eq:WKB_time_def_revised} to a universal operator identity. \textcolor{blue}{Secondly}, the Schr\"odinger equation \eqref{eq:emergent_WKB_SE_revised} is defined using derivative of $t_{\rm WKB}$. However the argument of the wavefunctional $\psi\Big[
h_{ij}(\mathbf x),
\Phi(\mathbf x),
\Psi_{\rm f}(\mathbf x),
{\rm ghosts}(\mathbf x);
t
\Big]$ appears to only depend on $({\bf x}, t)$ and the bulk fields. Where and how do the dependence on $t_{\rm WKB}$ appears in the wavefunction? \textcolor{blue}{Thirdly}, the story leading to the Schr\"odinger equation
cannot be quite as straightforward as the discussion above might suggest.
Throughout this paper, the fundamental object has been the trans-series
action, and therefore consistency requires that the entire derivation be
formulated within the same trans-series framework. In other words, it is
not sufficient to resum only the action while retaining conventional notions
of time evolution and quantum dynamics. Rather, every ingredient entering
the derivation of the Schr\"odinger equation should itself inherit the
trans-series structure of the underlying theory.

This observation has several immediate consequences. First, the temporal
parameter that emerges from the semiclassical expansion can no longer be
assumed to coincide with the standard WKB time variable. Since the background
itself receives trans-series corrections, the notion of time extracted from
the gravitational sector should also be modified by corresponding
non-perturbative contributions. Second, the Hamilton--Jacobi equation, which
ordinarily provides the bridge between the gravitational background and the
emergent temporal coordinate, must be replaced by its trans-series
counterpart. Such a generalized Hamilton--Jacobi equation would encode not
only the leading semiclassical dynamics but also the complete hierarchy of
exponentially suppressed sectors that contribute to the full quantum state.
Finally, the Schr\"odinger equation governing the non-gravitational degrees
of freedom should likewise be expected to acquire trans-series corrections.
Consequently, the standard linear time-evolution operator may represent only
the leading approximation to a more elaborate structure in which different
trans-series sectors are coupled.

These considerations raise a deeper conceptual issue concerning the meaning
of ``time'' itself. In the conventional Born--Oppenheimer or WKB analysis,
time emerges from the gravitational 
sector\footnote{Of course, both gravitational and matter sectors are inherently related. By assuming an ans\"atze like \eqref{eq:BO_WKB_ansatz_revised}, we are not decoupling them because the gravitational prefactor $A[h;t]$ satisfy equations that also depends on the matter wavefunction $\psi[\Xi;t]$ and the WKB phase $S_0[h;t]$.} and subsequently serves as the
evolution parameter for matter fields. However, once the gravitational
background is promoted to a trans-series object, the emergent notion of time
may itself become sector-dependent. It is then no longer obvious how the
predominantly non-gravitational sector should interpret temporal evolution,
or even whether a single universal time parameter continues to exist.
Addressing this question requires a reformulation of the semiclassical
expansion in which the notions of time, Hamilton--Jacobi dynamics, and
Schr\"odinger evolution are all treated on the same trans-series footing. A complete analysis of this issue is clearly beyond the scope of the present paper. Therefore, in the following subsection, we will first discuss how the emergent WKB time $t_{\rm WKB}$ may be reconciled with the actual boundary time $t$. We will then provide a brief discussion in section \ref{sec8.1.7} of the corresponding trans-series formulation of the problem, after tackling a few other relevant topics on the way.

\subsubsection{Schr\"odinger equation and what $t_{\rm WKB}$ means \label{sec8.1.1}}

The trans-series story that we alluded to above is an added complication that may be relaxed if one wishes to study the theory in the so-called zero instanton sector. However the question related to $t_{\rm WKB}$ and $t$ is unavoidable and much more subtle. Therefore let us deal with it first. We can phrase the question in two parts: (1) what is the meaning of having two different times $t_{\rm WKB}$ and $t$?; and (2) does the expression \eqref{eq:WKB_time_def_revised} imply that we have somehow replaced the metric components on a Cauchy slice by $t_{\rm WKB}$? 
At this stage one must be careful. The notation in \eqref{eq:WKB_time_def_revised} or in \eqref{eq:emergent_WKB_SE_revised} does not mean that all metric components of the following form:
\begin{equation}
h_{11}(\mathbf x),\;
h_{22}(\mathbf x),\;
h_{12}(\mathbf x),\;
\dots
\label{eq:metric_components_example_revised}
\end{equation}
have somehow been replaced by a single variable $t_{\rm WKB}$. That would be incorrect. The proper meaning is that the Hamilton--Jacobi functional $S_0[h]$ defines a vector field on superspace\footnote{Not to be confused with the notion of ``superspace" in supersymmetry! In quantum gravity, \emph{superspace} means the configuration space of all spatial metrics $h_{ij}(\mathbf x)$ modulo spatial diffeomorphisms, and more generally, when matter is included, the enlarged configuration space of three-geometries together with matter-field configurations. Thus a Wheeler--DeWitt wavefunctional $\Psi[h_{ij},\Phi,\ldots]$ is a functional on this infinite-dimensional space. By contrast, in supersymmetry the word \emph{superspace} refers to an extension of spacetime itself by additional anticommuting Grassmann coordinates, for example $(x^\mu,\theta^\alpha,\bar\theta^{\dot\alpha})$, so that superfields provide a unified description of bosons and fermions. Therefore, superspace in quantum gravity is a \emph{space of configurations}, whereas superspace in supersymmetry is an \emph{extended spacetime or coordinate space}. These are, however, two conceptually distinct notions and should not be conflated.}:
\begin{equation}
V_{ij}(\mathbf x;h)
=
G_{ijkl}(\mathbf x;h)\,
\frac{\delta S_0[h]}{\delta h_{kl}(\mathbf x)},
\label{eq:Vij_on_superspace}
\end{equation}
and $t_{\rm WKB}$ is a parameter along the integral curves of this vector field. In other words, one considers a one-parameter family of metrics
$h_{ij}(\mathbf x;t_{\rm WKB})$
satisfying:
\begin{equation}
\frac{\partial h_{ij}(\mathbf x;t_{\rm WKB})}{\partial t_{\rm WKB}}
=
G_{ijkl}(\mathbf x;h)\,
\left.
\frac{\delta S_0[h]}{\delta h_{kl}(\mathbf x)}
\right|_{h=h(\cdot;t_{\rm WKB})} ,
\label{eq:metric_flow_equation_revised}
\end{equation}
where
$h(\cdot;t_{\rm WKB})$
means the full spatial configuration
$\big\{h_{ij}(\mathbf y;t_{\rm WKB})
\big\}_{\mathbf y\in\Sigma}$
at the branch parameter value $t_{\rm WKB}$. The equation \eqref{eq:metric_flow_equation_revised} is the Hamilton--Jacobi form of the first Hamilton equation after the momentum has been replaced by the functional derivative of $S_0$. By itself it is not the full canonical dynamics. Rather, the pair consisting of the Hamilton--Jacobi equation \eqref{eq:gravitational_HJ_revised}
 and the flow equation \eqref{eq:metric_flow_equation_revised} is equivalent, at leading WKB order, to the classical gravitational equations of motion in canonical form, and hence to the Einstein equations or their effective corrected counterpart.
Thus, for example:
\begin{equation}
h_{ij}(\mathbf x)
\quad\longrightarrow\quad
h_{ij}(\mathbf x;t_{\rm WKB})
\label{eq:h12_tWKB_revised}
\end{equation}
means that one solves the WKB flow equation for the full metric and then reads off its $(i,j)$ component along the chosen branch. This is the meaning of restricting the metric components to the semi-classical branch. By contrast, the non-gravitational fluctuation variables
$\Phi(\mathbf x), \Psi_{\rm f}(\mathbf x)$ and ${\rm ghosts}(\mathbf x)$
remain as ordinary configuration-space arguments of the reduced wavefunctional and are not, at this stage, replaced by semiclassical trajectories. Thus they are truly off-shell. 
More explicitly:
\begin{equation}
h_{ij}(\mathbf x;t_{\rm WKB})
=
h_{ij}^{(0)}(\mathbf x)
+
\int_0^{t_{\rm WKB}} ds\;
G_{ijkl}\!\Big(\mathbf x;h(\cdot;s)\Big)\,
\left.
\frac{\delta S_0[h]}{\delta h_{kl}(\mathbf x)}
\right|_{h=h(\cdot;s)} ,
\label{eq:h12_integral_form_revised}
\end{equation}
$h=h(\cdot;s)$ signifies the same restriction to a given Cauchy slice as above.
The correct statement is therefore not that the metric dependence disappears, but that the dependence on the metric is reorganized in coordinates adapted to the WKB flow:
\begin{equation}
\{h_{ij}(\mathbf x)\}
\quad\longrightarrow\quad
\big(
t_{\rm WKB},\,y^\alpha
\big),
\label{eq:adapted_coordinates_superspace_revised}
\end{equation}
where $t_{\rm WKB}$ labels motion along the semiclassical branch and the $y^\alpha$ denote the remaining transverse gravitational directions. Accordingly, the more precise form of the reduced wavefunctional is:
\bg\label{mizmialola}
\psi[\Xi({\bf x}); t] &\equiv & \psi\Big[
h_{ij}(\mathbf x),
\Phi(\mathbf x),
\Psi_{\rm f}(\mathbf x),
{\rm ghosts}(\mathbf x);
t
\Big] 
=  
\psi\Big[
h_{ij}(\mathbf x;t_{\rm WKB}),
\Phi(\mathbf x),
\Psi_{\rm f}(\mathbf x),
{\rm ghosts}(\mathbf x);
t
\Big] \nonumber\\
& \equiv & \psi\Big[h_{ij}(\mathbf x;t_{\rm WKB}), \widetilde{\Xi}({\bf x}); t\Big] = 
\psi\Big[
t_{\rm WKB},
y^\alpha, \widetilde{\Xi}({\bf x});t
\Big] ,
\nd
with $\widetilde\Xi({\bf x})$ denoting the non-gravitational degrees of freedom, {\it i.e.} the bosonic and the fermionic matter and ghosts. In other words: $ \widetilde\Xi({\bf x}) =  \big(\Phi(\mathbf x),
\Psi_{\rm f}(\mathbf x),
{\rm ghosts}(\mathbf x)\big)$. Interestingly, in \eqref{mizmialola}, 
only if one neglects the transverse gravitational fluctuations $y^\alpha$ does one arrive at the simplified notation:
\begin{equation}
\psi\Big[\Xi({\bf x});
t
\Big]
\approx
\psi\Big[
h_{ij}(\mathbf x;t_{\rm WKB}), \widetilde\Xi({\bf x});
t
\Big]
\approx
\psi\Big[
t_{\rm WKB}, \widetilde\Xi({\bf x});
t
\Big].
\label{eq:psi_simplified_branch_revised}
\end{equation}
This simplification however is uncalled for as it is rarely possible to neglect the transverse gravitational fluctuations even if we restrict ourselves to the on-shell (or more appropriately, the semi-classical) branch.
Accordingly, the more precise WKB Schr\"odinger equation is:

{\footnotesize
\begin{equation}
i\frac{\partial}{\partial t_{\rm WKB}}
\psi\Big[
h_{ij}(\mathbf x;t_{\rm WKB}),\,
\widetilde\Xi(\mathbf x);\,
t
\Big]
=
\int_\Sigma d^{d-1}y\;
\widehat{\mathcal H}_{\rm mat+gh}\!\Big[
h(\mathbf y;t_{\rm WKB}),\,
\widetilde\Xi(\mathbf y)
\Big]\,
\psi\Big[
h_{ij}(\mathbf x;t_{\rm WKB}),\,
\widetilde\Xi(\mathbf x);\,
t
\Big] ,
\label{eq:emergent_WKB_SE_precise}
\end{equation}}
where $\widehat{\mathcal H}_{\rm mat+gh}$ is the Hamiltonian 
density\footnote{However, there is one small refinement one could implement to \eqref{eq:emergent_WKB_SE_precise}. Strictly speaking, the Hamiltonian density
$\widehat{\mathcal H}_{\rm matt+gh}$
is usually a local operator density involving not only the fields at $\mathbf y$, but also their canonical momenta and possibly spatial derivatives. So the most precise form for the Schr\"odinger equation is something like:
\begin{equation}
i\frac{\partial}{\partial t_{\rm WKB}}
\psi\Big[
h_{ij}(\mathbf x;t_{\rm WKB}),\,
\widetilde\Xi(\mathbf x);\,
t
\Big]
=
\int_\Sigma d^{d-1}y\;
\widehat{\mathcal H}_{\rm matt+gh}\!\Big[
h(\mathbf y;t_{\rm WKB}),\,
\{\widetilde\Xi(\mathbf y)\}
\Big]\,
\psi\Big[
h_{ij}(\mathbf x;t_{\rm WKB}),\,
\widetilde\Xi(\mathbf x);\,
t
\Big] , \nonumber
\label{eq:reduced_SE_density_more_precise}
\end{equation}
where $\{\widetilde\Xi({\bf y})\} = \Big(\widetilde\Xi(\mathbf y),\,
\partial \widetilde\Xi(\mathbf y),\,
\Pi_{\widetilde\Xi}(\mathbf y),\,
\ldots\Big)$.
If one does not want to display all of this structure explicitly, then the form \eqref{eq:emergent_WKB_SE_precise} is perfectly fine as schematic notation, and is understood that the added complications exist but are not displayed explicitly. Unless mentioned otherwise, we will follow this strategy to avoid unnecessary clutter. \label{kanobile}}; and  note that the equation depends on two time parameters: one is the genuine time $t$, and the other is the emergent time $t_{\rm WKB}$ appearing from the Hamilton-Jacobi flow \eqref{eq:metric_flow_equation_revised}. The puzzle, mentioned earlier, is not just that, but also the fact that the derivative in \eqref{eq:emergent_WKB_SE_precise} is with respect to $t_{\rm WKB}$ and not $t$. 

The resolution to this conundrum is slightly more non-trivial so we will go in small steps. First recall that, due to the presence of a boundary, we do have a genuine Schr\"odinger equation \eqref{eq:boundary_SE_exact} that tells us how the full wavefunctional evolves with respect to the time $t$. The relevant operator therein is the boundary Hamiltonian $\widehat{H}_\partial$. But in general this Hamiltonian could be a complicated one that depends on all the DOFs. Therefore, 
suppose first that the boundary Hamiltonian depends only on the metric sector, {\it i.e.}
$\widehat H_\partial
=
\widehat H_\partial[h]$.
Then the full state obeys the exact boundary Schr\"odinger equation:
\begin{equation}
i\frac{\partial}{\partial t}
\Psi\Big[ h_{ij}(\mathbf x),
\Phi(\mathbf x),
\Psi_{\rm f}(\mathbf x),
{\rm ghosts}(\mathbf x);
t
\Big]
=
\widehat H_\partial[h]\,
\Psi\Big[
h_{ij}(\mathbf x),
\Phi(\mathbf x),
\Psi_{\rm f}(\mathbf x),
{\rm ghosts}(\mathbf x);
t
\Big] ,
\label{eq:boundary_SE_metric_only_revised}
\end{equation}
where typically all field components appearing in the argument of the wavefunctional are generically off-shell as they should be. Notice also that the wavefunctional is defined with respect to the genuine time $t$, although we could restrict this to the semi-classical branch of the metric. 
Substituting the WKB ansatz \eqref{eq:BO_WKB_ansatz_revised} into \eqref{eq:boundary_SE_metric_only_revised} gives
\begin{equation}
-\,M_{\rm P}^2
\frac{\partial S_0}{\partial t}\,
\psi
+
i\frac{\partial \psi}{\partial t}
+
i\frac{1}{A}
\frac{\partial A}{\partial t}\,
\psi
=
A^{-1}
e^{-\,iM_{\rm P}^2S_0}
\widehat H_\partial[h]\,
A\,e^{\,iM_{\rm P}^2S_0}\,
\psi ,
\label{eq:boundary_equation_after_WKB_metric_only_revised}
\end{equation}
which is still an equation in the full configuration space. The WKB ansatz itself does not impose $h_{ij}(\mathbf x)=h_{ij}(\mathbf x;t_{\rm WKB})$. It is first written on the full off-shell configuration space. The restriction to the semiclassical branch is made only later, after $S_0$ has been shown to satisfy the Hamilton--Jacobi equation and one chooses to evaluate the reduced state along that branch. What the WKB ansatz does is something different. It separates the wavefunctional into
a rapidly varying gravitational phase $e^{iM_{\rm P}^2S_0[h;t]}$,
together with
a slower prefactor $A[h;t]$
and a reduced factor $\psi[h,\Phi,\Psi_{\rm f},{\rm ghosts};t]$.
Only after substituting this ansatz into the Wheeler--DeWitt equation and expanding in the semiclassical parameter does one obtain the Hamilton--Jacobi equation for $S_0$, {\it i.e.} \eqref{eq:gravitational_HJ_revised}. Then, using this $S_0$, one defines the semiclassical branch by the Hamilton--Jacobi flow \eqref{eq:metric_flow_equation_revised}. One may now restrict the metric fluctuations to the semi-classical branch. So the logical order is:

{\scriptsize
\begin{equation}
\text{\textcolor{blue}{WKB ansatz}}
~\Longrightarrow~
\text{\textcolor{blue}{Hamilton--Jacobi equation for} }\textcolor{blue}{S_0}
~\Longrightarrow ~
\text{\textcolor{blue}{definition of the semiclassical branch} }\textcolor{blue}{h_{ij}(\mathbf x;t_{\rm WKB})} \nonumber
\label{eq:logical_order_WKB_branch}
\end{equation}}
The important point however is not the restriction to some semi-classical sector, but the fact that the equation \eqref{eq:boundary_SE_metric_only_revised}
is exact. It shows that the external boundary time $t$ governs the full wavefunctional. Even if the boundary Hamiltonian acts only on the metric sector, the full wavefunctional evolves, including its matter, fermionic, and ghost dependence, because the entire state remains subject to the coupled Wheeler--DeWitt constraints.
There are therefore two conceptually distinct notions of time:
$t$ is the exact external boundary time generated by 
$\widehat H_\partial$,
whereas
$t_{\rm WKB}$
is the emergent intrinsic time extracted from the semiclassical gravitational phase $S_0[h;t]$.
The boundary time appears in the exact evolution law \eqref{eq:boundary_SE_metric_only_revised} for the full state. The WKB time appears in the approximate matter Schr\"odinger equation \eqref{eq:emergent_WKB_SE_precise} obtained from the bulk Wheeler--DeWitt constraint.

One might think that this distinction goes away once one takes the full trans-series form of the action. This is not the case and it will be an error to think along that line! (We will discuss the story soon.) Instead 
the logic is as follows.
If the full trans-series action is written in its most general form, then the boundary Hamiltonian must be derived from the full action, and one should expect, in principle,
$H_\partial
=
H_\partial
\Big[\Xi
\Big]$
rather than a purely gravitational ADM term alone. More explicitly, let us consider the following form of the total action:
\begin{equation}
S_{\rm tot}[\Xi]
=
S_{\rm grav}[\Xi]
+
S_{\rm mat}[\Xi]
+
S_{\rm ferm}[\Xi]
+
S_{\rm gh}[\Xi]
+
S_{\rm np}[\Xi]
+
S_{\rm nloc}[\Xi],
\label{eq:Stot_full_transseries_revised}
\end{equation}
where this may be alternatively arranged in the trans-series form as we saw earlier. Clearly, the total Hamiltonian will change from the simple form that we discussed earlier, but we expect it to retain the basic ADM structure, namely the split into bulk and the boundary pieces. This is indeed the case, and the decomposition now becomes:

{\footnotesize
\bg\label{miaaaria}
&& H_{\rm tot}
=
\int_\Sigma d^{d-1}x\;
\Big(
N{\cal H}_\perp
+
N^i{\cal H}_i
\Big)
+
H_\partial\\
&& {\cal H}_\perp
=
{\cal H}_\perp^{\rm grav}
+
{\cal H}_\perp^{\rm mat}
+
{\cal H}_\perp^{\rm ferm}
+
{\cal H}_\perp^{\rm gh}
+
{\cal H}_\perp^{\rm np}, ~~~
H_\partial
=
H_\partial^{\rm grav}
+
H_\partial^{\rm mat}
+
H_\partial^{\rm ferm}
+
H_\partial^{\rm gh}
+
H_\partial^{\rm np}, \nonumber
\nd}
and similarly for ${\cal H}_i$. In writing the boundary Hamiltonian we have assumed that
the canonical analysis produces genuine boundary contributions from all sectors. This may {\it not} always be the case but the conclusion remains unaffected for any choice of the boundary Hamiltonian. Despite these complications,  
on shell the bulk constraints {\it always} vanish, {\it i.e.}
${\cal H}_\perp=0$ and 
${\cal H}_i=0$ (setting constraints on the wavefunctional), 
so the nontrivial canonical generator reduces to the boundary term:
$H_{\rm phys}
=
H_\partial[\Xi]$.
In that case the exact boundary Schr\"odinger equation becomes:
\begin{equation}
i\frac{\partial}{\partial t}
\Psi\Big[\Xi({\bf x});
t
\Big]
=
\Big(
\widehat H_\partial^{\rm grav}
+
\widehat H_\partial^{\rm mat}
+
\widehat H_\partial^{\rm ferm}
+
\widehat H_\partial^{\rm gh}
+
\widehat H_\partial^{\rm np}
\Big)
\Psi\Big[\Xi({\bf x});
t
\Big].
\label{eq:boundary_schrodinger_sector_split_revised}
\end{equation}
Then all sectors evolve directly with respect to the asymptotic time $t$. If, on the other hand, some sectors do not generate independent boundary terms under the chosen asymptotic conditions, then those sectors continue to appear in the bulk Wheeler--DeWitt constraints but not explicitly in $H_\partial$. In that case the exact boundary-time evolution remains
$i\partial_t\Psi
=
\widehat H_\partial\Psi$,
but the non-gravitational dependence of $\Psi$ is propagated only indirectly through the coupled bulk constraints
$\widehat{\cal H}_\perp\Psi=0$,
and 
$\widehat{\cal H}_i\Psi=0$.

\begin{figure}[h]
\centering
\begin{tabular}{c}
\includegraphics[width=6in]{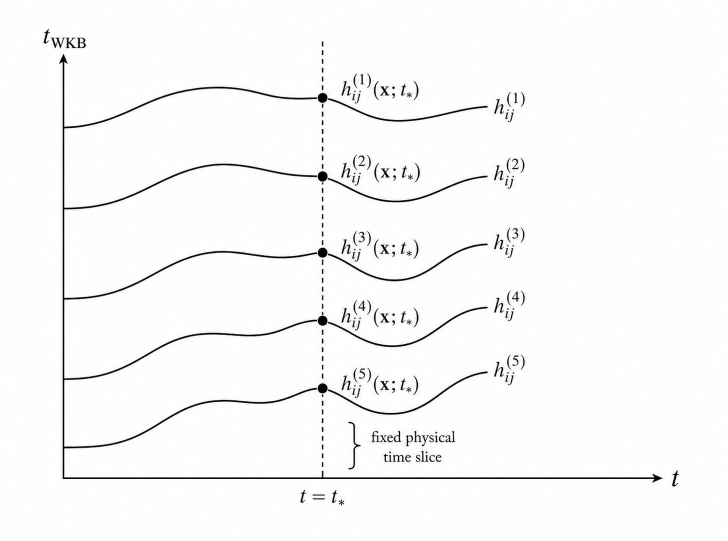}
\end{tabular}
\caption[]{The solid curves represent a family of candidate semiclassical branches
$\left\{
h_{ij}^{(\alpha)}
\right\}_{\alpha}$
in the $(t,t_{\rm WKB})$ plane. At fixed physical time $t=t_\ast$, the dashed vertical line intersects these different branches in different candidate on-shell configurations
$\left\{
h_{ij}^{(\alpha)}(\mathbf x;t_\ast)
\right\}_{\alpha}$.
They are not different values of $t_{\rm WKB}$ along one and the same FRW or de Sitter branch but represent different candidate semiclassical branches $h_{ij}^{(\alpha)}$. }
\label{trajectories}
\end{figure}

Once the boundary Hamiltonian depends on all degrees of freedom, the distinction between $t$ and $t_{\rm WKB}$ still remains, but the interpretation changes slightly. The exact boundary-time evolution now acts directly on the full state as in \eqref{eq:boundary_schrodinger_sector_split_revised}.
Thus matter, fermions, and ghosts now evolve directly with respect to the asymptotic time $t$. The emergent WKB time still arises from the bulk Wheeler--DeWitt equation as \eqref{eq:emergent_WKB_SE_precise},
but $t_{\rm WKB}$ now plays the role of an intrinsic semiclassical time whose evolution must be matched to the exact external time $t$ along a chosen branch. Accordingly, one requires:
\begin{equation}
t_{\rm WKB} = f(t)
\label{eq:tWKB_approx_t_revised}
\end{equation}
to the relevant order in the semiclassical expansion. We will soon discuss how to determine the function $f(t)$, but at least now we see that the WKB time does not replace all metric components by a single variable. Rather, it parametrizes the motion of the metric along a chosen semiclassical branch in superspace, while the remaining transverse metric directions survive as additional variables. The exact asymptotic time $t$ is generated by the boundary Hamiltonian. The issue of time is therefore resolved not by replacing the Wheeler--DeWitt equation with a Schr\"odinger equation, but by understanding how the intrinsic WKB time emerging from the bulk gravitational constraint becomes compatible, along a chosen semiclassical branch, with the external boundary time generated by the nonzero boundary Hamiltonian. This is, at least at face value, the content of \eqref{eq:tWKB_approx_t_revised}.

A more important point, which in fact provides the operational identification of the two time variables in \eqref{eq:tWKB_approx_t_revised}, arises when we ask the following question: what relation between the time parameter appearing in \eqref{eq:emergent_WKB_SE_precise} and the boundary time appearing in \eqref{eq:boundary_SE_metric_only_revised} ensures that the wavefunctional
$\Psi[h_{ij}({\bf x};t_{\rm WKB}),\widetilde\Xi({\bf x});t]$
evolves in exactly the same way according to both equations?
The answer is of course \eqref{eq:tWKB_approx_t_revised}, but we should keep in mind that,  
if one wants the analysis to be fully consistent, then the second equation \eqref{eq:boundary_SE_metric_only_revised} should also be understood along the same semiclassical branch whenever one is making the comparison with the first equation \eqref{eq:emergent_WKB_SE_precise}. The precise point is that the first equation is the \emph{reduced} matter-plus-ghost equation on a chosen WKB background, whereas the second equation is the \emph{exact} boundary-time equation for the full state before any reduction. So there are really two levels of description whose reconciliation happens in the way described above, namely,
when we bring the exact boundary Schr\"odinger equation to the same semiclassical branch for the metric degrees of freedom. Of course, 
as an exact equation, the boundary Schr\"odinger equation should remain written in terms of predominantly off-shell $h_{ij}(\mathbf x)$ not approximately on-shell $h_{ij}(\mathbf x;t_{\rm WKB})$.
However, once one restricts the exact equation to the same semiclassical branch used in the WKB analysis, then one may write its branch-restricted form as:
\begin{equation}
i\frac{\partial}{\partial t}
\Psi\Big[
h_{ij}(\mathbf x;t_{\rm WKB}),
\widetilde\Xi({\bf x});
t
\Big]
=
\widehat H_\partial\!\Big[\Xi; t_{\rm WKB})\Big]\,
\Psi\Big[
h_{ij}(\mathbf x;t_{\rm WKB}),
\widetilde\Xi({\bf x});
t
\Big] ,
\label{eq:boundary_equation_restricted_branch}
\end{equation}
which is no longer the fundamental exact equation; it is the exact equation \eqref{eq:boundary_schrodinger_sector_split_revised} evaluated on a particular semiclassical branch. In fact this is where \eqref{eq:tWKB_approx_t_revised} makes sense as a possible relation between the two times. Note however that one does \emph{not} insert
$t_{\rm WKB} = f(t)$
into the exact equations from the outset. Rather, the identification is made only after the full state has been restricted to a semiclassical branch and one wants to compare the WKB evolution with the branch-restricted boundary evolution.
So the logic is:
\begin{equation}
\text{\textcolor{blue}{exact boundary equation on full configuration space}} \nonumber
\label{eq:logic_step_one}
\end{equation}
\begin{equation}
\textcolor{red}{\downarrow} \nonumber
\label{eq:logic_arrow_one}
\end{equation}
\begin{equation}
\textcolor{blue}{\text{restrict to a semiclassical branch} ~h_{ij}({\bf x}; t_{\rm WKB})} \nonumber
\label{eq:logic_step_two}
\end{equation}
\begin{equation}
\textcolor{red}{\downarrow} \nonumber
\label{eq:logic_arrow_two}
\end{equation}
\begin{equation}
\text{\textcolor{blue}{compare branch-restricted boundary evolution with the WKB matter equation}} \nonumber
\label{eq:logic_step_three}
\end{equation}
\begin{equation}
\textcolor{red}{\downarrow} \nonumber
\label{eq:logic_arrow_three}
\end{equation}
\begin{equation}
\textcolor{blue}{t_{\rm WKB} = f(t)~~
\text{along that branch}} \nonumber
\label{eq:logic_step_four}
\end{equation}

\noindent This means the identification $\eqref{eq:tWKB_approx_t_revised}$, {\it i.e.} $t_{\rm WKB} = f(t)$
is made at the level of the \emph{comparison} between the two equations, not as a modification of the exact canonical formalism itself. In other words, 
if one wants to express both equations using a single time parameter after the matching, then the natural place is the reduced WKB equation. 
After this matching, the most useful form is to rewrite the WKB equation in terms of the boundary time $f(t)$ so that the matter-plus-ghost sector evolves on the semiclassical background $h_{ij}(\mathbf x; f(t))$. The generic evolution is shown in {\bf figures \ref{trajectories}}, {\bf \ref{trajectories2}} and {\bf \ref{trajectories3}}, and in the following let us elaborate on them.
In {\bf figure \ref{trajectories}}, one draws several colored curves, for example:
\begin{equation}
h_{ij}^{(1)},\qquad
h_{ij}^{(2)},\qquad
h_{ij}^{(3)},\qquad
h_{ij}^{(4)},\qquad
h_{ij}^{(5)},
\label{eq:family_of_candidate_branches_corrected_text}
\end{equation}
stacked in the $(t,t_{\rm WKB})$ plane. These should \emph{not} be interpreted as one single FRW branch evaluated at different values of $t_{\rm WKB}$. Rather, they represent different candidate semiclassical branches
$h_{ij}^{(\alpha)}(\mathbf x;t,t_{\rm WKB})$,
or, in a simpler notation when the explicit $t$-dependence is suppressed,
$h_{ij}^{(\alpha)}(\mathbf x;t_{\rm WKB})$.
By construction, each such branch is assumed to satisfy the Hamilton--Jacobi equation and the associated flow equation to the WKB order, and is therefore semiclassically on shell.
At a fixed physical time
$t=t_\ast$,
the various intersections:
\begin{equation}
h_{ij}^{(1)}(\mathbf x;t_\ast),\quad
h_{ij}^{(2)}(\mathbf x;t_\ast),\quad
h_{ij}^{(3)}(\mathbf x;t_\ast),\quad
h_{ij}^{(4)}(\mathbf x;t_\ast),\quad
h_{ij}^{(5)}(\mathbf x;t_\ast)
\label{eq:intersections_at_tstar_corrected}
\end{equation}
represent different candidate semiclassical configurations available at that same physical time. The vertical arrangement is therefore not a dynamical evolution in a second physical time. Rather, it is a bookkeeping device displaying a family of possible on-shell branches at fixed $t$.
Thus the first figure should be read as a picture of the set:
\begin{equation}
{\cal F}_{t_\ast}
=
\left\{
h_{ij}^{(\alpha)}(\mathbf x;t_\ast)
\right\}_{\alpha},
\label{eq:set_of_candidate_configs_corrected}
\end{equation}
that is, the family of candidate configurations at fixed $t=t_\ast$. The quantity $t_{\rm WKB}$ in that figure does not yet have the meaning of an independent physical time. It merely labels positions along each candidate branch. The important label distinguishing the different colored curves is really the branch label $\alpha$.

In {\bf figure \ref{trajectories2}} the viewpoint changes. Instead of displaying many candidate branches, one now draws a single blue curve in the $(t,t_{\rm WKB})$ plane. This figure is the correct picture \emph{after} a specific semiclassical branch has been chosen. For example, one may choose an FRW branch:
\begin{equation}
h_{ij}^{\rm FRW}(\mathbf x;t_{\rm WKB})
=
a^2(t_{\rm WKB})\,\gamma_{ij}(\mathbf x).
\label{eq:FRW_branch_for_figure_two}
\end{equation}
The blue curve then represents that one chosen semiclassical branch together with the matching relation
\begin{equation}
t_{\rm WKB}=t_{\rm WKB}(t)=f(t),
\label{eq:matching_relation_curve_corrected}
\end{equation}
which synchronizes the intrinsic WKB parameter with the exact physical time. The important conceptual change is that one is no longer discussing many competing branches at fixed $t$, but one selected physical history. The blue curve is therefore the graph of the matching relation for that chosen branch. Equivalently, the corresponding geometry may be represented as:
\begin{equation}
h_{ij}^{\rm phys}(\mathbf x;t)
=
h_{ij}^{\rm phys}\!\Big(
\mathbf x;\,
t_{\rm WKB}(t)
\Big) =
h_{ij}^{(\alpha_0)}\!\Big(
\mathbf x;\,
t_{\rm WKB}(t)
\Big) ,
\label{eq:physical_geometry_parametrized_corrected}
\end{equation}
or, as in the last equality, if one wishes to display the branch type explicitly.
Thus the blue curve shows how the exact time $t$ and the emergent parameter $t_{\rm WKB}$ are synchronized along one selected branch.

\begin{figure}[h]
\centering
\begin{tabular}{c}
\includegraphics[width=6in]{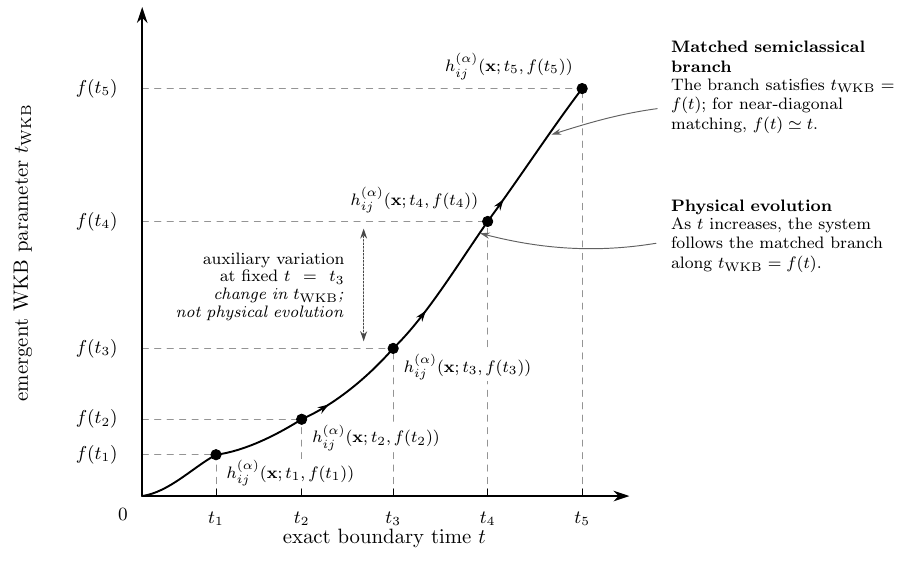}
\end{tabular}
\caption[]{The solid curve represents one selected semiclassical branch, together with the matching relation
$t_{\rm WKB}=f(t)$
between the intrinsic WKB parameter and the exact physical time. This is the appropriate picture after a definite on-shell branch, such as an FRW branch
$h_{ij}^{\rm FRW}(\mathbf x;t_{\rm WKB})=a^2(t_{\rm WKB})\gamma_{ij}(\mathbf x)$,
has been chosen.}
\label{trajectories2}
\end{figure}

\begin{figure}[h]
\centering
\begin{tabular}{c}
\includegraphics[width=6.2in]{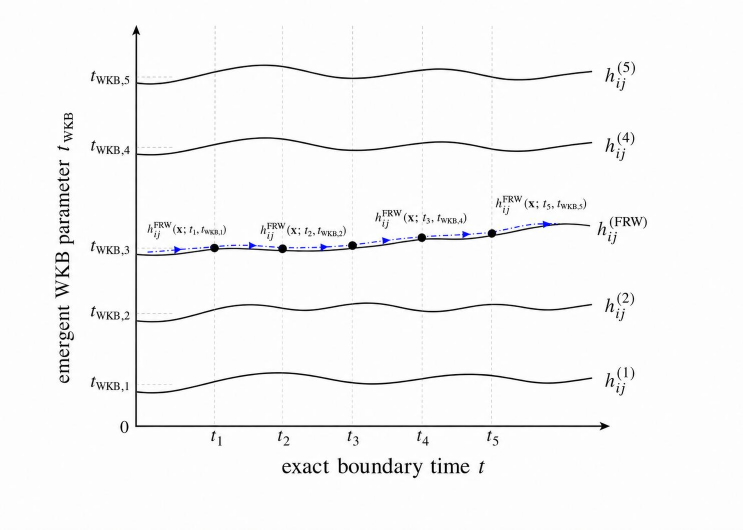}
\end{tabular}
\caption[]{The solid curves represent candidate semiclassical branches
$h_{ij}^{(\alpha)}$,
while the blue  dot-dashed curve represents the one selected branch matched to physical time:
$h_{ij}^{\rm phys}(\mathbf x;t)=h_{ij}^{(\alpha_0)}(\mathbf x;t_{\rm WKB}(t))$.
Thus the physical evolution proceeds along one chosen branch, while the remaining solid curves simply indicate the existence of other candidate semiclassical branches. The scale chosen to represent the evolution here differs from the scale chosen in {\bf figure \ref{trajectories2}}.}
\label{trajectories3}
\end{figure}

{\bf Figure \ref{trajectories3}} should then be understood as combining these two viewpoints. One again draws several colored curves representing candidate semiclassical branches
$h_{ij}^{(\alpha)}(\mathbf x;t_{\rm WKB})$,
but now one superposes a blue curve selecting one of them. The correct interpretation is that the blue curve does \emph{not} hop from one branch label $\alpha$ to another as time evolves. Rather, it identifies the one branch that has been selected by the matching to the exact physical evolution. So the corrected reading is that the blue curve should coincide with one particular candidate branch:
\begin{equation}
h_{ij}^{\rm phys}(\mathbf x;t)
=
h_{ij}^{(\alpha_0)}\!\Big(
\mathbf x;\,
t_{\rm WKB}(t)
\Big),
\label{eq:physical_path_single_selected_branch}
\end{equation}
for some selected $\alpha_0$. In other words, the blue curve is the matched representative of one chosen branch inside the larger family of possible branches. Therefore one should \emph{not} say that the physical evolution proceeds through the sequence
\begin{equation}
h_{ij}^{(1)}\!\big(\mathbf x;t_1,t_{{\rm WKB},1}\big)
\;\to\;
h_{ij}^{(2)}\!\big(\mathbf x;t_2,t_{{\rm WKB},2}\big)
\;\to\;
h_{ij}^{(3)}\!\big(\mathbf x;t_3,t_{{\rm WKB},3}\big)
\;\to\;\cdots,
\label{eq:wrong_sequence_of_branch_changes}
\end{equation}
because that would incorrectly suggest that the branch label itself changes along the physical history. Instead, once the branch is chosen, the physical evolution stays on that same branch:
\begin{equation}
h_{ij}^{(\alpha_0)}\!\big(\mathbf x;t_{{\rm WKB},1}\big)
\;\to\;
h_{ij}^{(\alpha_0)}\!\big(\mathbf x;t_{{\rm WKB},2}\big)
\;\to\;
h_{ij}^{(\alpha_0)}\!\big(\mathbf x;t_{{\rm WKB},3}\big)
\;\to\;\cdots.
\label{eq:correct_sequence_same_branch}
\end{equation}
Thus the role of the colored curves in the third figure is only to show that the chosen matched branch is selected from a larger family of candidates. The blue curve is not a path through different branches; it is the matched physical realization of one branch.

The three figures therefore illustrate three distinct but related viewpoints. {\bf Figure \ref{trajectories}} provides a family of candidate semiclassical branches at fixed physical time:
\begin{equation}
\left\{
h_{ij}^{(\alpha)}(\mathbf x;t_\ast)
\right\}_{\alpha}.
\label{eq:figure1_family_corrected}
\end{equation}
This is a kinematical picture of possible on-shell branches before one has made a definite choice.
{\bf Figure \ref{trajectories2}} provides a single matched semiclassical trajectory:
\begin{equation}
h_{ij}^{\rm phys}(\mathbf x;t)
=
h_{ij}^{(\alpha_0)}\!\big(\mathbf x;t_{\rm WKB}(t)\big),
\qquad
t_{\rm WKB}(t)=f(t).
\label{eq:figure2_single_curve_corrected}
\end{equation}
This is the correct dynamical picture once one definite branch has been selected. {\bf Figure \ref{trajectories3}} is a combined picture, showing how one chosen matched branch is selected from within a larger family of candidate branches:
\begin{equation}
h_{ij}^{\rm phys}(\mathbf x;t)
=
h_{ij}^{(\alpha_0)}\!\big(\mathbf x;t_{\rm WKB}(t)\big),
\qquad
\left\{
h_{ij}^{(\alpha)}
\right\}_{\alpha}
\text{ also exist as candidates.}
\label{eq:figure3_combined_corrected}
\end{equation}
This is the most refined picture, because it shows simultaneously the family of candidate branches and the one selected physical history.

\subsubsection{WdW equation, evolution and the meaning of time \label{sec8.1.2}}

Despite all the structure that we developed using {\bf figures \ref{trajectories}}, {\bf \ref{trajectories2}} and {\bf \ref{trajectories3}}, the story is far from complete. Additionally, we haven't been carefully keeping track of the background versus the fluctuations.  In this section we will try to rectify all the shortcomings of our earlier sections.

Let us assume that the full trans-series theory admits a time-independent minimum, for example a Minkowski vacuum, and that we expand around that background. The correct way to organize the problem is then to separate the fixed background data from the off-shell fluctuations around it.
Let the full set of fields be denoted by
$\Xi
=
\Big(
g_{\mu\nu},\,
\Phi,\,
\Psi_{\rm f},\,
{\rm ghosts}
\Big)$,
and let the chosen static minimum be
$\Xi_0
=
\Big(
\eta_{\mu\nu},\,
\Phi_0,\,
\Psi_{{\rm f},0},\,
{\rm ghosts}_0
\Big)$,
with
$\frac{\delta S_{\rm tot}[\Xi]}{\delta \Xi}\Big|_{\Xi=\Xi_0}=0$.
We then write the fluctuations as:
\begin{equation}
g_{\mu\nu}
=
\eta_{\mu\nu}
+
\delta g_{\mu\nu},
~~
\Phi
=
\Phi_0+\delta\Phi,
~~
\Psi_{\rm f}
=
\Psi_{{\rm f},0}+\delta\Psi_{\rm f},
~~
{\rm ghosts}
=
{\rm ghosts}_0+\delta{\rm ghosts}.
\label{eq:fluctuation_split_revised_final2}
\end{equation}
Equivalently, in canonical language,
$h_{ij}
=
\delta_{ij}
+
\delta h_{ij}$.
These fluctuations are, in general, off shell. Accordingly, once the background has been fixed, it is most natural to regard the wavefunctional as a functional of the fluctuation variables themselves:
\begin{equation}
\Psi
=
\Psi\Big[
\delta h_{ij}(\mathbf x),\,
\delta\Phi(\mathbf x),\,
\delta\Psi_{\rm f}(\mathbf x),\,
\delta{\rm ghosts}(\mathbf x);\,
t
\Big].
\label{eq:full_wavefunctional_fluctuations_revised_final2}
\end{equation}
Of course, this is simply a change of variables on configuration space, equivalent to writing the state as a functional of the full fields after the background values have been fixed.
The analysis then naturally splits into two parts. The first part
is the reduced matter-plus-ghost evolution from the Wheeler--DeWitt equation. The full Wheeler--DeWitt equation has the schematic form 
\eqref{eq:WdW_rearranged_boundary_time}
where $\widehat{H}_{\rm mat+gh}$ is an operator acting on the non-gravitational fluctuation variables, while depending parametrically on the gravitational configuration. Thus one may write schematically:
\begin{equation}
\widehat{H}_{\rm mat+gh}[\delta \mathbbmss{h}] \equiv
\int d^3x~ 
\widehat{\cal H}_{\rm matt+gh}
\Big[
\delta h_{ij}(\mathbf x);\,
\delta\Phi(\mathbf x),\,
\delta\Psi_{\rm f}(\mathbf x),\,
\delta{\rm ghosts}(\mathbf x)
\Big],
\label{eq:Hmattgh_operator_dependence_revised_final2}
\end{equation}
implying that $\widehat H_{\rm mat+gh}[\delta \mathbbmss{h}]$ is not merely a function of $\delta h_{ij}$ alone; rather, it is an operator on the matter-plus-ghost sector whose coefficients depend on the gravitational background configuration. In fact around the chosen static minimum, the full wavefunctional should be written with explicit dependence on the actual physical time $t$:
\begin{equation}
\Psi[\delta\Xi({\bf x}); t]
=
\Psi\Big[
\delta h_{ij}(\mathbf x),\,
\delta\Phi(\mathbf x),\,
\delta\Psi_{\rm f}(\mathbf x),\,
\delta{\rm ghosts}(\mathbf x);\,
t
\Big] ,
\label{eq:full_wavefunctional_exact_time_revised_final2}
\end{equation}
implying that, as before, the fluctuations are defined over a given Cauchy slice at any given time $t$. This is of course the boundary time $t$ that we had earlier.
A Born--Oppenheimer expansion of the full state then takes the form:
\begin{equation}
\Psi\big[\delta\Xi({\bf x});\,
t
\big]
=
A[\delta h;t]\,
\exp\!\big(
iM_{\rm P}^2 S_0[\delta h;t]
\big)\,
\psi\big[\delta\Xi({\bf x});\,
t
\big].
\label{eq:BO_ansatz_static_background_revised_final2}
\end{equation}
The appearance of the functional $S_0[\delta h;t]$ is not ad hoc. It is the leading Hamilton--Jacobi functional for the gravitational sector in the semiclassical WKB expansion of the Wheeler--DeWitt equation. Indeed, once the WKB ansatz \eqref{eq:BO_ansatz_static_background_revised_final2} is substituted into the bulk constraint, the highest-order terms in the large-$M_{p}$ expansion yield a functional Hamilton--Jacobi equation for $S_0$, while the next order yields the reduced evolution equation for $\psi$. Thus $S_0$ is the analogue, in gravitational configuration space, of Hamilton's principal functional in ordinary semiclassical mechanics.
Because the chosen minimum is time-independent, one may take the leading background data to be static at leading order, {\it i.e.}
$S_0[\delta h;t]
=
S_0[\delta h]$, and 
$A[\delta h;t]
=
A[\delta h]$. They eventually lead to, at leading order, the Hamilton-Jacobi equation and to next order, the Schr\"odinger type equation:
\bg\label{baptis2tag}
&& \frac{1}{2}\,
G_{ijkl}
\frac{\delta S_0}{\delta (\delta h_{ij})}
\frac{\delta S_0}{\delta (\delta h_{kl})}
+
U[\delta \mathbbmss{h}]
=
0 \nonumber\\
&-&\,i
\int d^{d-1}x\;
G_{ijkl}(x)\,
\frac{\delta S_0}{\delta (\delta h_{ij}(x))}
\frac{\delta}{\delta (\delta h_{kl}(x))}
\psi
+
\widehat H_{\rm mat+gh}[\delta \mathbbmss{h}]\,
\psi
=
0, \nd
where
one should emphasize that the arguments of $\psi[h_{ij}({\bf x}), \widetilde\Xi({\bf x}); t]$, the reduced wavefunctional,
are still generic configuration-space variables and therefore, in general, off shell. The reduced Schr\"odinger-type equation is obtained not by putting all of these variables on shell, but by restricting only the gravitational argument to the distinguished semiclassical Hamilton--Jacobi branch generated by $S_0[\delta h]$.
Thus one should understand that:
\begin{equation}
\delta h_{ij}(\mathbf x)
\quad\longrightarrow\quad
\delta h_{ij}(\mathbf x;t_{\rm WKB})
\label{eq:metric_argument_restricted_revised_final2}
\end{equation}
as the restriction of the generic off-shell gravitational fluctuation to the preferred semiclassical on-shell branch. By contrast, the non-gravitational fluctuation variables $\delta\widetilde\Xi({\bf x})$
remain as ordinary configuration-space arguments of the reduced wavefunctional and are not, at this stage, replaced by semiclassical trajectories.
This motivates the definition of the branchwise WKB time derivative
${\partial\over \partial t_{\rm WKB}}$
so that the reduced equation becomes:
\begin{equation}
i\left(
\frac{\partial \psi}{\partial t_{\rm WKB}}
\right)_{\!\!t,\delta\Phi,\delta\Psi_{\rm f},\delta{\rm ghosts}}
=
\widehat H_{\rm mat+gh}[\delta \mathbbmss{h}]\,
\psi ,
\label{eq:reduced_Schrodinger_matter_ghost_revised_final2}
\end{equation}
acting on $\psi[h_{ij}({\bf x}; t_{\rm WKB}), \widetilde\Xi({\bf x}); t]$.
The derivative in \eqref{eq:reduced_Schrodinger_matter_ghost_revised_final2} acts only on the implicit dependence of $\psi$ through the gravitational branch, while holding fixed the explicit time $t$ and the non-gravitational fluctuation arguments. This point is crucial: \eqref{eq:reduced_Schrodinger_matter_ghost_revised_final2} is not a second exact Schr\"odinger equation in an additional physical time. Rather, it is a reduced branchwise evolution equation extracted from the Wheeler--DeWitt constraint after the semiclassical separation of the gravitational sector.
One should therefore not simply replace:
\begin{equation}
\frac{\partial}{\partial t_{\rm WKB}}
\quad\longrightarrow\quad
\frac{\partial}{\partial t},
\label{eq:no_naive_replacement_revised_final2}
\end{equation}
because the left-hand side of \eqref{eq:reduced_Schrodinger_matter_ghost_revised_final2} does not act on the explicit $t$-dependence of $\psi$.
Instead, once a matching relation \eqref{eq:tWKB_approx_t_revised}
is imposed along a chosen branch, one must use the chain rule. This is a crucial point that point towards the novelty of our approach. We start by defining the wavefunctional over the semiclassical branch:
\begin{equation}
\widetilde\psi\Big[
\delta\Phi(\mathbf x),\,
\delta\Psi_{\rm f}(\mathbf x),\,
\delta{\rm ghosts}(\mathbf x);\, t
\Big]
\equiv
\psi\Big[
\delta h_{ij}(\mathbf x;f(t)),\,
\delta\Phi(\mathbf x),\,
\delta\Psi_{\rm f}(\mathbf x),\,
\delta{\rm ghosts}(\mathbf x);\,
t
\Big],
\label{eq:psi_tilde_definition_revised_final2}
\end{equation}
where the arguments in the LHS appear in this way 
because after the pullback only the non-gravitational variables remain as independent configuration-space arguments, while the metric argument has already been restricted to the chosen semiclassical branch. One may now propose the following chain-rule:
\begin{equation}
\frac{d}{dt}\widetilde\psi
=
\frac{df}{dt}
\left(
\frac{\partial \psi}{\partial t_{\rm WKB}}
\right)_{\!\!t,\delta\Phi,\delta\Psi_{\rm f},\delta{\rm ghosts}}
+
\left(
\frac{\partial \psi}{\partial t}
\right)_{\!\!t_{\rm WKB},\delta\Phi,\delta\Psi_{\rm f},\delta{\rm ghosts}} ,
\label{eq:chain_rule_revised_final2}
\end{equation}
where the first term is defined at constant $t$ and $\delta\widetilde\Xi({\bf x})$, whereas the second terms is defined at a constant $t_{\rm WKB}$ and $\delta\widetilde\Xi({\bf x})$. Note that there are 
no additional chain-rule term involving
$\delta\widetilde\Xi({\bf x}) = (\delta\Phi({\bf x}), 
\delta\Psi_{\rm f}({\bf x}), 
\delta{\rm ghosts}({\bf x}))$ 
appears in \eqref{eq:chain_rule_revised_final2}, because in the present construction these remain configuration-space arguments of the reduced wavefunctional, not externally prescribed classical trajectories depending on $t$.
Using \eqref{eq:reduced_Schrodinger_matter_ghost_revised_final2}, this gives:
\begin{equation}
i\frac{d}{dt}\widetilde\psi
=
\frac{df}{dt}\,
\widehat H_{\rm mat+gh}[\delta \mathbbmss{h}]\,
\widetilde\psi
+
i\left[
\left(
\frac{\partial \psi}{\partial t}
\right)_{\!\!t_{\rm WKB},\delta\Phi,\delta\Psi_{\rm f},\delta{\rm ghosts}}
\right]_{t_{\rm WKB}=f(t)}.
\label{eq:general_t_equation_revised_final2}
\end{equation}
Equation \eqref{eq:general_t_equation_revised_final2} is the correct consequence of the matching relation. In general, this equation is not yet a standard Schrödinger equation with respect to the exact time $t$, because it contains an additional term associated with the explicit, independent $t$-dependence of $\psi$. The natural question is then how this expression can be converted into a Schrödinger equation, or at least brought into a form closely analogous to one. This is precisely where the exact boundary-time evolution equation becomes relevant, namely 
$i\frac{\partial}{\partial t}\Psi
=
\widehat H_\partial\Psi$, and 
$\widehat{\cal H}_\perp\Psi=0$.
Substituting the Born--Oppenheimer decomposition into the first of these yields an exact relation for the explicit $t$-derivative of the reduced factor $\psi$:
\begin{align}
i
\left(
\frac{\partial \psi}{\partial t}
\right)_{\!\!t_{\rm WKB},\delta\Phi,\delta\Psi_{\rm f},\delta{\rm ghosts}}
&=
A^{-1}e^{-iM_{\rm P}^2S_0}\,
\widehat H_\partial[\delta\mathbbmss{h}]\,
A\,e^{iM_{\rm P}^2S_0}\psi
-
i\,A^{-1}\frac{\partial A}{\partial t}\,\psi
+
M_{\rm P}^2
\frac{\partial S_0}{\partial t}\,\psi ,
\label{eq:partial_t_psi_from_boundary_revised_final2}
\end{align}
for a fixed value of $t_{\rm WKB}$ and the off-shell fluctuations $\delta\widetilde\Xi({\bf x})$. We can now simplify \eqref{eq:partial_t_psi_from_boundary_revised_final2} by using the 
static-minimum approximation: 
$\frac{\partial S_0}{\partial t}=
\frac{\partial A}{\partial t}=0$. This approximation means that 
the Hamilton--Jacobi functional is not explicitly driven by the external time $t$. This is clear from the fact that 
the branch
$h_{ij}(\mathbf x;t_{\rm WKB})$
is obtained from the Hamilton--Jacobi flow equation \eqref{eq:metric_flow_equation_revised}
which involves the \emph{functional derivative} of $S_0[h]$ with respect to $h$, not the explicit partial derivative of $S_0$ with respect to the boundary time $t$. Thus there is no contradiction between
$\partial_t S_0=0$
and
$\partial_{t_{\rm WKB}} h_{ij}\neq 0$, and therefore implementing the stationary conditions simplifies \eqref{eq:partial_t_psi_from_boundary_revised_final2} to:
\begin{equation}
i
\left(
\frac{\partial \psi}{\partial t}
\right)_{\!\!t_{\rm WKB},\delta\Phi,\delta\Psi_{\rm f},\delta{\rm ghosts}}
=
A^{-1}e^{-iM_{\rm P}^2S_0}\,
\widehat H_\partial[\delta\mathbbmss{h}]\,
A\,e^{iM_{\rm P}^2S_0}\psi .
\label{eq:partial_t_psi_static_revised_final2}
\end{equation}
Accordingly, if the reduced sector is closed under the action of the conjugated boundary operator after restriction to the matched curve, then one may define an effective reduced Hamiltonian of the following form:
\bg\label{oppositetags}
\widehat H_{\rm eff}(t)
 &\equiv &
\frac{df}{dt}\,
\widehat H_{\rm matt+gh}[\delta\mathbbmss{h}]
+
\widehat H_{\partial}^{\rm red}[\delta\mathbbmss{h}] \nonumber\\
& \equiv &
\frac{df}{dt}\,
\widehat H_{\rm matt+gh}[\delta\mathbbmss{h}]
+
\left[
A^{-1}e^{-iM_{\rm P}^2S_0}\,
\widehat H_\partial[\delta\mathbbmss{h}]\,
A\,e^{iM_{\rm P}^2S_0}
\right]_{t_{\rm WKB}=f(t)},
\nd
which depends on the Hamilton-Jacobi function $S_0$ as well as the parameter $A$. The latter can be easily eliminated in favor of $S_0$ using the equation discussed in footnote \ref{lenapyar}. Also if $t_{\rm WKB}$ follows the trajectory very close to $t$, {\it i.e.} if $f(t) \approx t$ as shown in {\bf figure \ref{trajectories3}}, then the effective Hamiltonian may be further simplified. Generically however $t_{\rm WKB} = f(t)$, with $f(t)$ determined by restricted bulk and boundary equation, and $\widehat{H}_{\rm eff}(t)$ is a function of the boundary time $t$
so that the pulled-back reduced state obeys an effective Schr\"odinger equation of the form:
\bg\label{blazrog}
\fcolorbox{black}{white}{%
$\displaystyle
i\frac{d}{dt}\widetilde\psi\big[\delta\widetilde\Xi({\bf x}); t\big]
=
\widehat H_{\rm eff}(t)\,
\widetilde\psi\big[\delta\widetilde\Xi({\bf x}); t\big]
$%
}
\nd
which is different from the original bulk Wheeler--DeWitt equation rewritten in a trivial way (as in say \eqref{eq:naive_tprime_wrong_boundary_time}). Instead, it is an effective reduced evolution law obtained after semiclassical restriction to the gravitational branch and after incorporating the boundary-time dynamics.

This concludes our \textcolor{blue}{first step}, which is the study of the reduced matter-plus-ghost fluctuations on the chosen static semiclassical background, using the branchwise Schr\"odinger equation \eqref{eq:reduced_Schrodinger_matter_ghost_revised_final2}, the Hamilton--Jacobi reconstruction of the gravitational branch, and, when appropriate, the effective reduced Hamiltonian \eqref{oppositetags} after matching $t_{\rm WKB}$ to the exact time $t$. The \textcolor{blue}{second step} is different. Here one no longer freezes the gravitational background and studies only the reduced matter-plus-ghost sector. Instead one studies the full state:
\begin{equation}
\Psi\big[\delta\Xi({\bf x}); t]
=
\Psi\big[
\delta h_{ij}(\mathbf x),\,
\delta\Phi(\mathbf x),\,
\delta\Psi_{\rm f}(\mathbf x),\,
\delta{\rm ghosts}(\mathbf x);\,
t
\big]
\label{eq:full_state_again_revised_final2}
\end{equation}
in which we keep the argument of the wavefunctional to be completely off-shell on a given Cauchy-slice defined at the boundary time $t$. Since no semiclassical structures are needed, we won't have the Hamilton-Jacobi and the flow equations. Instead 
the exact evolution of \eqref{eq:full_state_again_revised_final2} is governed by the boundary Hamiltonian and the bulk constraints:
\begin{equation}
i\frac{\partial}{\partial t}
\Psi\Big[\delta\Xi({\bf x})
;\,
t
\Big]
=
\widehat H_\partial(t) \,
\Psi\Big[\delta\Xi({\bf x});\,
t
\Big], ~~
\widehat{\cal H}_\perp(t)\, \Psi\Big[\delta\Xi({\bf x});\,
t
\Big]=0,
~~
\widehat{\cal H}_i(t)\,\Big[\delta\Xi({\bf x});\,
t
\Big]\Psi=0 ,
\label{eq:bulk_constraints_again_full_revised_final2}
\end{equation}
where the Hamiltonians are functions of the  boundary time $t$ only since there is no $t_{\rm WKB}$ now.
This is the correct starting point for the full evolution of metric plus matter plus ghosts.
If one works perturbatively around the chosen Minkowski minimum, then one expands the full boundary Hamiltonian and the bulk constraints as:
\bg\label{rajarB}
&&\widehat H_\partial(t)
=
\widehat H_\partial^{(0)}(t)
+
\widehat H_\partial^{(2)}(t)
+
\widehat H_\partial^{(3)}(t)
+
\cdots \nonumber\\
&&\widehat{\cal H}_\perp(t)
=
\widehat{\cal H}_\perp^{(1)}(t)
+
\widehat{\cal H}_\perp^{(2)}(t)
+
\cdots,
\qquad
\widehat{\cal H}_i(t)
=
\widehat{\cal H}_i^{(1)}(t)
+
\widehat{\cal H}_i^{(2)}(t)
+
\cdots ,
\nd
where the superscripts denote the order in fluctuations.
Then the full evolution of all fluctuations is obtained by solving the exact boundary Schr\"odinger equation together with these linearized or nonlinear constraints.

In practice, step two is the study of the complete coupled fluctuation system $\delta\Xi({\bf x})$
subject simultaneously to
the exact boundary-time evolution and 
bulk gravitational constraints.
The two steps are therefore related hierarchically.
\textcolor{blue}{Step one} is a reduced description where 
we (a) treat the gravitational sector semiclassically, and (b)
derive a reduced branchwise Schr\"odinger equation for matter plus ghosts.
This is useful for studying propagation, spectra, vacuum structure, and matter/ghost dynamics on the chosen background branch exactly as one does for studying quantum field theories.  
\textcolor{blue}{Step two} is the full description where (a)
keep the metric fluctuations as dynamical, (b)
evolve the full wavefunctional with $\widehat H_\partial$,
and (c) impose the bulk constraints.
This is necessary if one wants to understand the coupled dynamics of geometry and matter, rather than matter alone on a prescribed background. Thus step one is contained within step two as the semiclassical reduction of the full problem. On the other hand, 
if the full action is trans-series, then both analyses receive corrections. For step one, the reduced matter-plus-ghost Hamiltonian becomes an effective trans-series Hamiltonian:
\begin{equation}
\widehat H_{\rm matt+gh}(t)
\quad\longrightarrow\quad
\widehat H_{\rm matt+gh}^{\rm eff}(t),
\label{eq:Hmatt_eff_transseries_revised_final2}
\end{equation}
and the WKB branch itself may become sector dependent.
For step two, the full boundary Hamiltonian and the full constraints are also corrected:
\begin{equation}
\widehat H_\partial(t)
\quad\longrightarrow\quad
\widehat H_\partial^{\rm eff}(t),
\quad
\widehat{\cal H}_\perp(t)
\quad\longrightarrow\quad
\widehat{\cal H}_\perp^{\rm eff}(t),
\quad
\widehat{\cal H}_i(t)
\quad\longrightarrow\quad
\widehat{\cal H}_i^{\rm eff}(t).
\label{eq:full_effective_operators_transseries_revised_final2}
\end{equation}
Thus the full trans-series theory modifies both the reduced semiclassical dynamics and the exact coupled dynamics. We will study this scenario a little later.

\subsubsection{Why do the boundary and the bulk evolution match up? \label{sec8.1.3}}

There remains, however, a couple of further puzzles. The first one, which we shall resolve in this section, may be formulated as follows. In the case of the bulk Hamiltonian, all fields can in principle contribute on an equal footing. The boundary Hamiltonian is different: the metric sector naturally gives rise to well-defined boundary contributions, whereas the same is not usually true for the other fields coupled to gravity. This raises the following question. If the boundary and bulk evolutions are compared after restricting only the metric sector to the same semiclassical branch, why should the resulting evolution of the remaining non-metric fields agree, or even take a similar form, in the two descriptions?

The correct resolution of this puzzle is that the two evolutions are \emph{not} similar in the na\"ive operator sense. Rather, they become compatible only after one distinguishes carefully between
(i) exact evolution of the full wavefunctional,
and (ii) reduced evolution of the non-gravitational sector on a chosen semiclassical gravitational branch.
The potential source of confusion is the following. In the bulk Wheeler--DeWitt analysis, the constraint operator contains all sectors:
\begin{equation}
\widehat{\cal H}_{\rm bulk}
=
\widehat{\cal H}_g
+
\widehat{\cal H}_{\rm mat+gh},
\label{eq:Hbulk_democratic}
\end{equation}
so the reduced branchwise equation naturally involves the matter-plus-ghost Hamiltonian
$\widehat H_{\rm mat+gh}[\delta\mathbbmss{h}, t]$.
By contrast, the exact boundary Hamiltonian is often dominated by the gravitational boundary term\footnote{The notation used to denote the Hamiltonian's dependence on the fluctuations and time $t$ may look inconsistent from our earlier sections, but it is not. We are following the enlarged notational scheme mentioned in footnote \ref{kanobile}, so $\widehat{H}(t)$ or $\widehat{H}({\rm fluc}, t)$  may look inconsistent but they both are derived from footnote \ref{kanobile} so are self-consistent.}:
\begin{equation}
\widehat H_\partial
\approx
\widehat H_\partial^{\rm grav}[\delta h, t],
\label{eq:Hboundary_metric_only_approx}
\end{equation}
with no separate matter boundary charge under the chosen asymptotic conditions. Then the puzzle is: how can a boundary evolution generated mainly by gravity reproduce the same non-gravitational dynamics as the bulk reduced equation?

The answer is that it does \emph{not} do so by acting democratically on all fields as an operator. It does so indirectly, through the fact that the full state is constrained and coupled. To see this, let us 
suppose that the exact full state satisfies:
\begin{equation}
i\frac{\partial}{\partial t}\Psi
=
\widehat H_\partial[\delta h, t]\,
\Psi,
\qquad
\widehat{\cal H}_\perp\Psi=0,
\qquad
\widehat{\cal H}_i\Psi=0 ,
\label{eq:exact_boundary_plus_constraints}
\end{equation}
where now we taking a restrictive picture in which the boundary Hamiltonian only has contributions from the gravitational sector and no contributions from the bulk matter sector. The crucial observation is that, 
even if $\widehat H_\partial$ depends only on the gravitational variables, it still evolves the \emph{full} wavefunctional
\begin{equation}
\Psi[\delta\Xi({\bf x}); t]
=
\Psi\Big[
\delta h_{ij}(\mathbf x),\,
\delta\Phi(\mathbf x),\,
\delta\Psi_{\rm f}(\mathbf x),\,
\delta{\rm ghosts}(\mathbf x);\,
t
\Big].
\label{eq:full_state_all_args}
\end{equation}
Therefore the matter, fermion, and ghost dependence of $\Psi$ can change with $t$ even if there is no independent matter boundary term.
The reason is simply that the state is not, in general, a tensor product of the form:
\begin{equation}
\Psi[\delta\Xi({\bf x}); t]
=
\Psi_g[\delta h({\bf x});t]\,
\Psi_{\rm mat+gh}[\delta\Phi({\bf x}),\delta\Psi_{\rm f}({\bf x}),\delta{\rm ghosts}({\bf x});t]
\label{eq:factorized_state_wrong}
\end{equation}
with completely independent evolution. Rather, because of the bulk constraints, the full state is a coupled functional on the joint configuration space. Hence a change in the gravitational sector can induce a nontrivial change in the conditional or reduced non-gravitational sector. Thus the key point is that a purely gravitational boundary Hamiltonian can still induce nontrivial time dependence in the non-gravitational sector because it evolves the full coupled wavefunctional, not an isolated gravitational factor.
In the semiclassical reduction, and as we saw before, one writes the following ans\"atze:

{\footnotesize
\begin{equation}
\Psi[\delta\Xi({\bf x}); t]
=
A[\delta h ({\bf x});t]\,
\exp\!\big(
iM_{\rm P}^2S_0[\delta h({\bf x});t]
\big)\,
\psi\Big[
\delta h ({\bf x}),\delta\Phi({\bf x}),\delta\Psi_{\rm f}({\bf x}),\delta{\rm ghosts}({\bf x});t
\Big],
\label{eq:BO_decomposition_puzzle}
\end{equation}}
and then restricts the metric fluctuation to the semiclassical branch
using the emergent time parameter $t_{\rm WKB}$ as $\delta h_{ij}(\mathbf x)
~~\longrightarrow ~~
\delta h_{ij}(\mathbf x;t_{\rm WKB})$ denoted by one of the colored curve in {\bf figure \ref{trajectories}}, or the one chosen in {\bf figure \ref{trajectories3}}.
The reduced branchwise equation then takes the form:
\begin{equation}
i\left(
\frac{\partial \psi}{\partial t_{\rm WKB}}
\right)_{\!\!t,\delta\Phi,\delta\Psi_{\rm f},\delta{\rm ghosts}}
=
\widehat H_{\rm matt+gh}[\delta\mathbbmss{h}]\,
\psi.
\label{eq:reduced_branchwise_equation_puzzle}
\end{equation}
This equation says that \emph{once the gravitational sector has been restricted to a semiclassical branch}, the remaining conditional evolution of the non-gravitational sector is governed by the bulk matter Hamiltonian evaluated on that evolving background.
So the reduced equation is a \emph{conditional} evolution law that answers the following question:
given the semiclassical branch $\delta h_{ij}(\mathbf x;t_{\rm WKB})$,
how does the non-gravitational sector evolve?
In this sense, the reduced bulk equation and the exact boundary equation are answering different questions:

{\footnotesize
\begin{equation}
\text{\textcolor{blue}{ exact boundary equation: evolution of the full state}}\nonumber
\label{eq:exact_question}
\end{equation}
\begin{equation}
\text{ \textcolor{blue}{ reduced bulk equation: conditional evolution on a chosen semiclassical gravitational branch}}\nonumber
\label{eq:reduced_question}
\end{equation}}
The reason the two descriptions can nevertheless agree is that the reduced matter state is extracted from the full state \emph{after} the latter has been evolved by the boundary Hamiltonian and \emph{after} one has restricted to the same semiclassical metric branch. More precisely, let:
\begin{equation}
\widetilde\psi[\delta\widetilde\Xi({\bf x}); t]
\equiv
\psi\Big[
\delta h_{ij}(\mathbf x;f(t)),\,
\delta\Phi(\mathbf x),\,
\delta\Psi_{\rm f}(\mathbf x),\,
\delta{\rm ghosts}(\mathbf x);\,
t
\Big]
\label{eq:psi_tilde_for_puzzle}
\end{equation}
be the reduced state pulled back to the matched curve
$t_{\rm WKB}=f(t)$.
Then the reduced effective equation may be written schematically as
\eqref{blazrog} with $\widehat H_{\rm eff}(t)$ is defined in terms of
$\widehat H_\partial^{\rm red}(t)$, the reduced boundary contribution, in \eqref{oppositetags}. 
Now, if $\widehat H_\partial$ is mainly gravitational, then $\widehat H_\partial^{\rm red}(t)$ is also mainly inherited from the gravitational sector. Yet this is \emph{not} inconsistent with the appearance of $\widehat H_{\rm matt+gh}[\delta\mathbbmss{h}; t]$, because the latter is the conditional bulk Hamiltonian describing how the non-gravitational sector responds to the evolving background branch. The exact boundary-time evolution and the reduced bulk evolution agree only after this restriction and only in the sense that they describe the same reduced physical branch.
So the correct comparison is not:
\begin{equation}
\widehat H_\partial
\stackrel{?}{=}
\widehat H_{\rm matt+gh}
\label{eq:not_equal_operators}
\end{equation}
but rather
the exact evolved full state, when restricted to the chosen semiclassical metric branch, induces the same reduced non-gravitational evolution that the bulk Wheeler--DeWitt reduction predicts via the constraints:
\begin{equation}
\widehat{\cal H}_\perp\Psi=0,
\qquad
\widehat{\cal H}_i\Psi=0.
\label{eq:constraints_essential}
\end{equation}
They correlate the gravitational and non-gravitational sectors. Thus the non-gravitational part of the exact state is not arbitrary even when $\widehat H_\partial$ contains no separate matter boundary term. In this sense the bulk constraints carry part of the dynamical burden that a separate matter boundary Hamiltonian would otherwise appear to supply. Therefore one may say that 
the similarity of the two evolutions does not come from identical operator content, but from the fact that the bulk constraints couple the gravitational and non-gravitational sectors, so that the exact boundary evolution of the full state induces the same reduced matter evolution after projection to the semiclassical branch.

So when is the match easier and when is it harder? 
There are two qualitatively different situations.
If the full theory produces genuine boundary contributions from all sectors, so that the boundary Hamiltonian takes the following form:
\begin{equation}
\widehat H_\partial[\delta\Xi, t]
=
\widehat H_\partial^{\rm grav}[\delta\Xi, t]
+
\widehat H_\partial^{\rm mat}[\delta\Xi, t]
+
\widehat H_\partial^{\rm ferm}[\delta\Xi, t]
+
\widehat H_\partial^{\rm gh}[\delta\Xi, t]
+\cdots,
\label{eq:full_boundary_all_sectors}
\end{equation}
then the match between exact boundary-time evolution and reduced non-gravitational evolution may appear more direct. On the other hand, the Schr\"odinger equation that appears from \eqref{eq:boundary_SE_exact} is now:
\bg
\label{blazrog2}
\fcolorbox{black}{white}{%
$\displaystyle
\begin{aligned}
i\frac{\partial}{\partial t}\Psi\big[\delta\Xi({\bf x});t\big]
&=
\widehat H_{\rm tot}(\delta\Xi,t)\,
\Psi\big[\delta\Xi({\bf x});t\big]
\\
&=
\Big(
\widehat{\mathcal H}^{\rm tot}_g(\delta\Xi,t)
+
\widehat{\mathcal H}^{\rm tot}_{\rm mat+gh}(\delta\Xi,t)
\Big)
\Psi\big[\delta\Xi({\bf x});t\big] 
\end{aligned}
$%
}
\nd
where $\widehat{\mathcal H}^{\rm tot}_g = \widehat{H}^{\rm grav}_\partial + \widehat{\mathcal H}_g$ and  
$\widehat{\mathcal H}^{\rm tot}_{\rm mat+gh} = \widehat H_\partial^{\rm mat}[\delta\Xi, t]
+
\widehat H_\partial^{\rm ferm}[\delta\Xi, t]
+
\widehat H_\partial^{\rm gh}[\delta\Xi, t]
+\cdots + \widehat{\mathcal H}_{\rm mat+gh}$
with $\widehat{\mathcal H}_g$ and $\widehat{\mathcal H}_{\rm mat+gh}$ from the bulk pieces in \eqref{eq:Hbulk_democratic} and $\widehat{H}^{\rm grav}_\partial$  and  $ \widehat H_\partial^{\rm mat}[\delta\Xi, t]
+
\widehat H_\partial^{\rm ferm}[\delta\Xi, t]
+
\widehat H_\partial^{\rm gh}[\delta\Xi, t]
+\cdots $ from the boundary pieces in \eqref{eq:full_boundary_all_sectors}. The difference from our earlier Schr\"odinger equation \eqref{blazrog} is threefold: (1) the wavefunctional appearing in \eqref{blazrog2} is the {\it total} wavefunctional \eqref{eq:full_state_all_args}, and not the WKB reduced one; (2) the LHS of the Schr\"odinger equation has a partial derivative with respect to the boundary time $t$ compared to the total derivative on the LHS of \eqref{blazrog}; and (3) both the gravitational and the matter + ghost Hamiltonians are modified from their bulk forms \eqref{eq:Hbulk_democratic} by their respective boundary contributions, compared to the unmodified matter + ghost Hamiltonian but no gravitational Hamiltonian in \eqref{blazrog}. 

By contrast, if only the gravitational sector contributes significantly at the boundary, then the match may appear more indirect and rests on the coupled nature of the full state together with the bulk constraints \eqref{eq:exact_boundary_plus_constraints}.
The Schr\"odinger equation now takes the following suggestive form:
\bg
\label{blazrog3}
\fcolorbox{black}{white}{%
$\displaystyle
\begin{aligned}
i\frac{\partial}{\partial t}\Psi\big[\delta\Xi({\bf x});t\big]
=
\Big(
\widehat{\mathcal H}^{\rm tot}_g(\delta\Xi,t)
+
\widehat{\mathcal H}_{\rm mat+gh}(\delta\Xi,t)
\Big)
\Psi\big[\delta\Xi({\bf x});t\big] 
\end{aligned}
$%
}
\nd
where $\widehat{\mathcal H}^{\rm tot}_g = \widehat{H}^{\rm grav}_\partial + \widehat{\mathcal H}_g$ is from the bulk and the boundary pieces \eqref{eq:Hbulk_democratic} and  \eqref{eq:Hboundary_metric_only_approx} respectively; and $\widehat{\mathcal H}_{\rm mat+gh}$ is from the bulk piece \eqref{eq:Hbulk_democratic} only. In fact \eqref{blazrog3} is the closest we can come to realizing the Schr\"odinger equation with no interference from the boundary term to the matter + ghost Hamiltonian and no semi-classical limits for the gravitational fluctuations. One may now compare this to \eqref{blazrog} and \eqref{blazrog2}: not only we now have the exact wavefunctional \eqref{eq:full_state_all_args}, as in \eqref{blazrog2}, and not the reduced wavefunctional as in \eqref{blazrog}, but also the Hamiltonian is purely the bulk matter + ghost Hamiltonian. Moreover we have ${\partial \Psi\over \partial t}$ and not ${d\psi\over dt}$ as in \eqref{blazrog}. However we should remember that \eqref{blazrog3} is derived assuming that the boundary contribution to the Hamiltonian comes predominantly from the gravitational sector. This is usually the case with a simple Einstein-Hilbert plus perturbative matter + ghost type action, but once we take the full trans-series form of the action, we can no longer claim that the matter + ghost sector will have no boundary contributions. All in all, if we take the most generic trans-series form of the action, the closest that we can come to realize the Schr\"odinger equation is \eqref{blazrog2}.

Thus the correct general statement may be divided into two parts. The first part is the following: if only the metric sector has a substantial boundary Hamiltonian, then the non-gravitational evolution seen in the reduced bulk description is reproduced not because the boundary Hamiltonian acts democratically on all fields, but because the exact full state is constrained and coupled, and its restriction to the same semiclassical metric branch induces the corresponding reduced non-gravitational dynamics. The second part is more concrete and precise: once the full trans-series form of the action is taken into consideration then the exact Schr\"odinger that appears from our analysis is \eqref{blazrog2}. This is the equation that leads (a) to the concept of the interacting vacuum $|\Omega\rangle$ that we have been using so far, and (b) the eigenstates from which we could construct the coherent and the GS states. 

However two questions still remain: how do we interpret the wavefunctional and where is the ordinary QFT hiding in this framework? For the second question, the answer has already been anticipated in the discussion surrounding \eqref{blazrog}. Regarding the first question, once it is stated explicitly that the theory is being organized around a Minkowski or warped-Minkowski minimum, the interpretation of the wavefunctional becomes considerably more transparent. But one must still distinguish carefully between: 
(i) the reference vacuum about which the expansion is performed, and
(ii) the full space of fluctuations being quantized.
With that distinction in place, the interpretation is indeed much clearer than in a fully background-free discussion. To see this, let us suppose one chooses a Minkowski or warped-Minkowski minimum:
\begin{equation}
\Xi_0
=
\Big(
\eta_{\mu\nu},\,
\Phi_0(\mathbf x),\,
\Psi_{{\rm f},0}(\mathbf x),\,
{\rm ghosts}_0(\mathbf x)
\Big),
\label{eq:minkowski_minimum_transparent}
\end{equation}
satisfying
$\frac{\delta S_{\rm tot}[\Xi]}{\delta \Xi}\big|_{\Xi=\Xi_0}=0$. As we discussed earlier, this is of course one of the many Minkowski minima in the system. The trans-series action $S_{\rm tot}[\Xi]$ is the consequence of all the neighboring minima  and therefore \eqref{eq:minkowski_minimum_transparent} continues to display the {\it same} minimum as we discussed in section \ref{sec5}. (Alternatively one can use a UV action that allows the Minkowski minima and then (a) integrate out the UV DOFs and (b) bring the low-energy action in the trans-series form over a chosen minimum. Both these ways allow the same physics.) One may then expand the fields as $\Xi = \Xi_0 + \delta\Xi$, whose components take the form: 
\begin{equation}
g_{\mu\nu}
=
\eta_{\mu\nu}
+
\delta h_{\mu\nu},
~~
\Phi
=
\Phi_0+\delta\Phi,
~~
\Psi_{\rm f}
=
\Psi_{{\rm f},0}+\delta\Psi_{\rm f},
~~
{\rm ghosts}
=
{\rm ghosts}_0+\delta{\rm ghosts} ,
\label{eq:fluctuation_split_transparent}
\end{equation}
with the fluctuations generically off-shell, although there could also be on-shell fluctuations. This is an important distinction that will be useful for interpreting the semiclassical backgrounds in the next section. 
In this setup, the Wheeler--DeWitt wavefunctional is most naturally interpreted as:
\begin{equation}
\Psi
=
\Psi\Big[
\delta h_{ij}(\mathbf x),\,
\delta\Phi(\mathbf x),\,
\delta\Psi_{\rm f}(\mathbf x),\,
\delta{\rm ghosts}(\mathbf x)
\Big],
\label{eq:wavefunctional_over_fluctuations_transparent}
\end{equation}
namely as the quantum amplitude for a given \emph{set of fluctuations} around the chosen Minkowski background.
So in this background-fixed language, the full Wheeler--DeWitt wavefunctional is the amplitude for the universe to contain a particular fluctuation configuration around the chosen Minkowski or warped-Minkowski vacuum. Although this is indeed much more transparent than speaking abstractly about arbitrary superspace, it is still not yet ordinary QFT. This is because, 
even in this Minkowski-based description, one must remember that the metric fluctuation
$\delta h_{ij}(\mathbf x)$
is still one of the quantum variables. Therefore the full wavefunctional is still not \emph{just} the ordinary QFT wavefunctional for matter on a fixed spacetime. It is the amplitude for metric fluctuations
together with matter, fermionic, and ghost fluctuations.
Therefore around a chosen Minkowski minimum, $\Psi$ is naturally interpreted as a wavefunctional of all fluctuations around that vacuum, including the gravitational ones. It is therefore more concrete than an abstract superspace wavefunction, but still more general than an ordinary QFT wavefunctional because the geometry itself is still fluctuating.

The familiar QFT wavefunctional appears only after one performs an additional step: namely, one separates the gravitational fluctuations into a semiclassical background part plus residual quantum fluctuations, and then freezes the background to a chosen branch. Schematically, therefore let us {\it split} the metric fluctuations into the following two parts:
\begin{equation}
\delta h_{\mu\nu}
=
\delta h_{\mu\nu}^{\rm sc}
+
\delta h_{\mu\nu}^{\rm off} ,
\label{eq:metric_split_transparent}
\end{equation}
where the first part is the semiclassical (``sc") one and the second part is the off-shell (``off") one.
If one then restricts to
$\delta h_{\mu\nu}^{\rm sc}$
and neglects or integrates out the remaining quantum gravitational fluctuations, the reduced wavefunctional becomes:
\begin{equation}
\psi\Big[
g_{\mu\nu}^{\rm sc}(t_{\rm WKB}),\,
\delta\Phi,\,
\delta\Psi_{\rm f},\,
\delta{\rm ghosts}
\Big],
\label{eq:reduced_wavefunctional_transparent}
\end{equation}
which is now a QFT-like wavefunctional on a chosen semiclassical background. Then, in the free limit, the vacuum and excited states indeed resemble oscillator-type wavefunctionals mode by mode.
So the hierarchy is:
\begin{equation}
\text{\textcolor{blue}{full WdW wavefunctional around Minkowski}} \nonumber
\label{eq:logic_step_one}
\end{equation}
\begin{equation}
\textcolor{red}{\Downarrow} \nonumber
\label{eq:logic_arrow_one}
\end{equation}
\begin{equation}
\textcolor{blue}{\text{wavefunctional of all fluctuations}} \nonumber
\label{eq:logic_step_two}
\end{equation}
\begin{equation}
\textcolor{red}{\Downarrow} \nonumber
\label{eq:logic_arrow_two}
\end{equation}
\begin{equation}
\text{\textcolor{blue}{semiclassical restriction of metric sector}} \nonumber
\label{eq:logic_step_three}
\end{equation}
\begin{equation}
\textcolor{red}{\Downarrow} \nonumber
\label{eq:logic_arrow_three}
\end{equation}
\begin{equation}
\textcolor{blue}{\text{ordinary QFT wavefunctional on that background}} \nonumber
\label{eq:logic_step_four}
\end{equation}
leading us basically to \eqref{blazrog}. However if one doesn't want to commit to the ``narrow" viewpoint, a much generic picture exists and is given by \eqref{blazrog2}. The interpretation of the wavefunctional becomes more complicated than before but in either case, a clear Schr\"odinger picture emerges.

\subsubsection{On/off-shell fluctuations around a Minkowski minimum \label{sec8.1.4}}

We are still left with one important conceptual issue. Once a Minkowski minimum has been chosen, it is not immediately clear how a nontrivial on-shell branch:
\begin{equation}
h_{ij}(\mathbf x,t_{\rm WKB})
\label{eq:nontrivial_branch_professional}
\end{equation}
can arise without simultaneously turning on nontrivial $t_{\rm WKB}$-dependent fluxes, matter expectation values, or nonperturbative contributions. Put differently, if one writes a reduced wavefunctional of the form:
\begin{equation}
\psi\Big[
h_{ij}(\mathbf x,t_{\rm WKB}),\,
\Phi(\mathbf x),\,
\Psi_{\rm f}({\bf x}),\,
{\rm ghosts}(\mathbf x);\,
t
\Big],
\label{eq:reduced_wavefunctional_professional}
\end{equation}
as in \eqref{blazrog}, with
$h_{ij}(\mathbf x,t_{\rm WKB})$
treated as on shell while the remaining fluctuations are kept off shell, one must explain how such a construction can consistently generate a broad class of semiclassical metric branches while the non-gravitational sectors are not simultaneously placed on shell around the same Minkowski minimum. To answer this, we will start with the full set of fields:
\begin{equation}
\Xi({\bf x}; t)
=
\Big(
g_{\mu\nu}({\bf x}),\,
\Phi({\bf x}),\,
\Psi_{\rm f}({\bf x}),\,
{\rm ghosts}({\bf x}); t
\Big),
\label{eq:full_field_content_clean2}
\end{equation}
defined on a given Cauchy slice at the boundary time $t$. This is of course the familiar set of fields that we have been taking so far. 
Suppose that the full trans-series theory admits a Minkowski or warped-Minkowski reference minimum:
\begin{equation}
\Xi_0({\bf x})
=
\Big(
\eta_{\mu\nu},\,
\Phi_0(\mathbf x),\,
\Psi_{{\rm f},0}(\mathbf x),\,
{\rm ghosts}_0(\mathbf x)
\Big),
\label{eq:minkowski_reference_minimum_clean2}
\end{equation}
which is a time-independent configuration that satisfies 
$\frac{\delta S_{\rm tot}[\Xi]}{\delta \Xi}\big|_{\Xi=\Xi_0}=0$.
In canonical language, the corresponding background spatial metric is
$h_{ij}^{(0)}(\mathbf x)=\delta_{ij}$, or its possible warped counter-part but we will avoid these complications and stick with the simplest choice.

The phrase ``static minimum'' refers only to this reference configuration. It does not mean that every semiclassical branch reconstructed later in the WKB analysis must itself be static. Thus one should distinguish carefully between (a) 
the static reference background $h_{ij}^{(0)}(\mathbf x)=\delta_{ij}$,
and (b) a semiclassical branch $h_{ij}^{\rm sc}(\mathbf x;t_{\rm WKB})$,
which is a one-parameter family of on-shell or semiclassical configurations associated with the Hamilton--Jacobi flow.
Expanding around the reference minimum, one writes $\Xi({\bf x}; t) = \Xi_0({\bf x}) + \delta\Xi({\bf x}; t)$ which, in components, take the following form:
\begin{equation}
g_{\mu\nu}
=
\eta_{\mu\nu}
+
\delta h_{\mu\nu},
~~
\Phi
=
\Phi_0+\delta\Phi,
~~
\Psi_{\rm f}
=
\Psi_{{\rm f},0}+\delta\Psi_{\rm f},
~~
{\rm ghosts}
=
{\rm ghosts}_0+\delta{\rm ghosts} ,
\label{eq:full_fluctuation_split_clean2}
\end{equation}
where we do not show the coordinates $({\bf x}, t)$ explicitly. 
At this stage all fluctuations are generic configuration-space variables and are therefore off shell. 

The correct semiclassical organization is not to keep only the metric fluctuation on shell while forcing all remaining sectors to stay purely off shell. Rather, one should allow every sector to split into a semiclassical expectation-value part and a residual off-shell fluctuation in the following way:
\begin{equation}
\delta h_{\mu\nu}
=
\delta h_{\mu\nu}^{\rm sc}
+
\delta h_{\mu\nu}^{\rm off},\nonumber
\label{eq:metric_split_sc_off_clean2}
\end{equation}
\begin{equation}
\delta\Phi
=
\delta\Phi_{\rm sc}
+
\delta\Phi_{\rm off},
~~
\delta\Psi_{\rm f}
=
\delta\Psi_{{\rm f},{\rm sc}}
+
\delta\Psi_{{\rm f},{\rm off}},
~~
\delta{\rm ghosts}
=
\delta{\rm ghosts}_{\rm sc}
+
\delta{\rm ghosts}_{\rm off} ,
\label{eq:matter_split_sc_off_clean2}
\end{equation}
which is somewhat similar to what we attempted in \eqref{eq:metric_split_transparent} for only the metric fluctuations, but now in \eqref{eq:matter_split_sc_off_clean2} we perform the split on {\it all} the fluctuations.
Equivalently, one may write the full semiclassical configuration as:
\begin{equation}
\Xi_{\rm sc} ({\bf x}; t)
=
\Big(
g_{\mu\nu}^{\rm sc}({\bf x}; t),\,
\Phi_{\rm sc}({\bf x}; t),\,
\Psi_{{\rm f},{\rm sc}}({\bf x}; t),\,
{\rm ghosts}_{\rm sc}({\bf x}; t)
\Big),
\label{eq:full_semiclassical_configuration_clean2}
\end{equation}
which is the advantage we gain from making the two way split in \eqref{eq:matter_split_sc_off_clean2}, {\it i.e.} we can combine the background values with the on-shell part of the split to construct a semiclassical background \eqref{eq:full_semiclassical_configuration_clean2} whose components are defined succinctly as:

{\footnotesize
\begin{equation}
g_{\mu\nu}^{\rm sc}
=
\eta_{\mu\nu}
+
\delta h_{\mu\nu}^{\rm sc},
~~~
\Phi_{\rm sc}
=
\Phi_0+\delta\Phi_{\rm sc},
~~~
\Psi_{{\rm f},{\rm sc}}
=
\Psi_{{\rm f},0}+\delta\Psi_{{\rm f},{\rm sc}},
~~~
{\rm ghosts}_{\rm sc}
=
{\rm ghosts}_0+\delta{\rm ghosts}_{\rm sc} ,
\label{eq:sc_configuration_components_clean2}
\end{equation}}
or more compactly as $\Xi^{\rm sc}({\bf x}; t) = \Xi_0({\bf x}) + \delta\Xi^{\rm sc}({\bf x}; t)$, with the semiclassical background developing dependence on ${\bf x}$ as well as the boundary time $t$. 
The remaining off-shell fluctuations are then:
\begin{equation}
\delta\Xi_{\rm off}({\bf x}; t)
=
\Big(
\delta h_{\mu\nu}^{\rm off}({\bf x}; t),\,
\delta\Phi_{\rm off}({\bf x}; t),\,
\delta\Psi_{{\rm f},{\rm off}}({\bf x}; t),\,
\delta{\rm ghosts}_{\rm off}({\bf x}; t)
\Big) ,
\label{eq:offshell_residuals_clean2}
\end{equation}
where, although the off-shell fluctuations \eqref{eq:offshell_residuals_clean2} could suggest fluctuations over the semiclassical background \eqref{eq:full_semiclassical_configuration_clean2}, in reality the fluctuations are simply over the Minkowski background \eqref{eq:minkowski_reference_minimum_clean2}. This way we can avoid issues like the non-existence of Wilsonian RG over temporally varying background and other nuances. 

The Wilsonian perspective suggests that the only consistent way for on-shell fluctuations to coexist around the Minkowski minimum is for them to be realized through coherent states or, more generally, through Glauber--Sudarshan states. It is important to emphasize, however, that the role of the coherent-state or GS-state construction is not merely to localize the quantum state on the metric branch. Rather, it localizes the state on the full semiclassical configuration, including all sectors of the theory:
\begin{equation}
\langle \delta\widehat h^{\rm sc}_{\mu\nu}(x)\rangle_\sigma
=
\delta h^{\rm sc}_{\mu\nu}(x),
~~~
\langle \delta\widehat \Phi_{\rm sc}(x)\rangle_\sigma
=
\delta\Phi_{\rm sc}(x),
~~~
\langle\delta\widehat \Psi_{\rm f, sc}(x)\rangle_\sigma
=
\delta\Psi_{{\rm f},{\rm sc}}(x) ,
\label{eq:GS_expectation_values_clean2}
\end{equation}
providing a proper identifications of the fluctuations in 
\eqref{eq:sc_configuration_components_clean2} with the ones constructed from the coherent or the GS states. In fact, once it comes to the on-shell configuration, there is no difference between the construction of the coherent or the GS states: both provide identical results for the semiclassical configurations. It should also be clear that:
\bg\label{comcontags}
\delta{\rm ghost}_{\rm sc} \equiv \langle \delta\widehat{\Upsilon}_{\rm sc}\rangle_\sigma = 0,~~~~ {\rm ghosts}_0 = 0, \nd
because there are no displacement operators associated with ghosts and no background ghost configurations.
Moreover, as discussed earlier, one advantage of the GS-state construction is that it can accommodate generic off-shell configurations, where the coherent-state construction would fail implying that:
\bg\label{ajkertagra}
\Xi({\bf x}; t) = \langle \widehat\Xi({\bf x}; t)\rangle_\sigma = 
\underbrace{\Xi_0({\bf x}) + \langle\delta\widehat\Xi_{\rm sc}({\bf x}; t)\rangle_\sigma}_{\langle\Xi_{\rm sc}({\bf x}; t)\rangle_\sigma} + \langle\delta\widehat\Xi_{\rm off}({\bf x}; t)\rangle_\sigma , \nd
thus accommodating all fluctuations around the solitonic vacuum including both on- and off-shell ones.
Moreover the off-shell spread of any GS state around a given on-shell or off-shell configuration prevents an uncontrolled spreading in configuration space. Such a consideration provides a precise meaning to the semiclassical configuration $\Xi_{\rm sc}({\bf x}; t)$ in 
\eqref{eq:full_semiclassical_configuration_clean2}, including the placement of the ghost sector, because the semiclassical pieces 
must solve the appropriate effective equations which in our case is given by:
\begin{equation}
\left\langle
\frac{\delta S_{\rm tot}[\Xi]}{\delta \Xi(x)}
\right\rangle_\sigma
=
\left\langle
\frac{\delta}{\delta \Xi(x)}
\log\Big(
D^\dagger(\sigma)D(\sigma)
\Big)
\right\rangle_\sigma,
\label{bootwa}
\end{equation}
which in the bootstrap language splits into two equations given by \eqref{bootsforwalking} for the on-shell DOFs. The ghost contribution appears more directly in the second equation that relates the off-diagonal states to the displacement operators, but indirectly in the first. The various sinks, sources, saddles and stationary points we saw in {\bf figures \ref{bootstrapbasins}}, {\bf \ref{bootsinksourcesaddle}} and {\bf \ref{sssrotate}} are all the semiclassical configurations that we could construct from the fluctuations over a Minkowski minimum using the GS states and not the coherent states because the bootstrap flows are best understood using the GS states. Thus a nontrivial metric branch: 
\begin{equation}
h_{ij}^{\rm sc}(\mathbf x;t_{\rm WKB}) \equiv 
\langle \widehat{h}_{ij}^{\rm sc}(\mathbf x;t_{\rm WKB})\rangle_\sigma =
\delta_{ij} + \langle \delta\widehat{h}_{ij}^{\rm sc}({\bf x}; t_{\rm WKB})\rangle_\sigma ,
\label{eq:nontrivial_metric_branch_clean2}
\end{equation}
forming say one of the basins in {\bf figures \ref{bootstrapbasins}}, {\bf \ref{bootsinksourcesaddle}} and {\bf \ref{sssrotate}}
cannot arise from the bare static Minkowski minimum alone. It requires nontrivial supporting expectation values in the other sectors, such as matter, fermionic, ghost, flux, or nonperturbative trans-series contributions. Once a semiclassical background has been chosen, one may study the residual off-shell fluctuations around it. The natural wavefunctional is then a functional of the remaining off-shell variables:
\begin{equation}
\Psi'
=
\Psi'\Big[
\delta h_{\mu\nu}^{\rm off},\,
\delta\Phi_{\rm off},\,
\delta\Psi_{{\rm f},{\rm off}}
\Big] \equiv \Psi'\Big[
\langle\delta\widehat h_{\mu\nu}^{\rm off}\rangle_\sigma,\,
\langle\delta\widehat\Phi_{\rm off}\rangle_\sigma,\,
\langle\delta\widehat\Psi_{{\rm f},{\rm off}}\rangle_\sigma
\Big] ,
\label{eq:wavefunctional_offshell_clean2}
\end{equation}
or more compactly as $\Psi'[\langle\delta\widehat\Xi^{\rm off}({\bf x}; t)\rangle_\sigma] = \Psi'[\delta\Xi^{\rm off}({\bf x}); t]$ where we showed explicitly the coordinates on the Cauchy slice,
but in general one could simply write the wavefunctional as $\Psi'[\langle\delta\widehat\Xi({\bf x}; t)\rangle_\sigma] = \Psi'[\delta\Xi({\bf x}); t]$ associated with generic on- and off-shell fluctuations around the Minkowski minimum. (Note the absence of ghost fluctuations, and also the difference from \eqref{eq:wavefunctional_over_fluctuations_transparent}.  We will discuss this in section \ref{sec8.1.6}.)
In a reduced description where the metric branch is kept explicitly semiclassical, one may write instead:
\begin{equation}
\psi'\Big[
g_{\mu\nu}^{\rm sc}(t_{\rm WKB}),\,
\delta\Phi_{\rm off},\,
\delta\Psi_{{\rm f},{\rm off}};\,
t
\Big] ,
\label{eq:reduced_wavefunctional_clean2}
\end{equation}
where the metric sector has already been restricted to the chosen semiclassical branch, while the remaining fields are still treated as off-shell quantum arguments with $\delta\Xi^{\rm off} = \langle \delta\widehat\Xi^{\rm off}\rangle_\sigma$. In other words, only the fluctuations around the matter sector are represented by the GS states. The parameter $t_{\rm WKB}$ is an intrinsic parameter along the chosen semiclassical branch. It is not, by itself, an additional exact physical time. The branch is obtained from the Hamilton--Jacobi functional $S_0$ through the equations \eqref{eq:gravitational_HJ_revised} and the flow equation \eqref{eq:metric_flow_equation_revised} which are now expressed as:
\bg\label{oshomtag}
&& \frac{1}{2}\,
G_{ijkl}
\frac{\delta S_0[\langle \widehat h\rangle_\sigma]}{\delta \langle\widehat h_{ij}\rangle_\sigma}
\frac{\delta S_0 [\langle \widehat h\rangle_\sigma] }{\delta \langle \widehat h_{kl}\rangle_\sigma}
+
U[\langle\widehat h\rangle_\sigma, ..]
=
0\nonumber\\
&& \frac{\partial \langle \widehat h_{ij}^{\rm sc}(\mathbf x;t_{\rm WKB})\rangle_\sigma}{\partial t_{\rm WKB}}
=
G_{ijkl}(\mathbf x)\,
\left.
\frac{\delta S_0 [\langle \widehat h\rangle_\sigma]  }{\delta \langle\widehat h_{kl}(\mathbf x)\rangle_\sigma}
\right|_{\langle \widehat h\rangle_\sigma=\langle\widehat h^{\rm sc}(\cdot;t_{\rm WKB})\rangle_\sigma} ,
\nd
where the dotted terms are the additional DOFs. Thus 
$\langle \widehat h_{ij}^{\rm sc}(\mathbf x;t_{\rm WKB})\rangle_\sigma$
is not the original Minkowski background itself, but a semiclassical on-shell branch generated {\it relative} to that background.

This also explains why the existence of such a branch does not contradict the statement that the theory has only Minkowski minima. A minimum is a stationary point of the action, whereas a semiclassical branch is a dynamical on-shell trajectory in configuration space. 
A theory may have only Minkowski minima and nevertheless admit nontrivial semiclassical branches around that minimum once the state carries nontrivial dressed expectation values. Hence the key observation is the following:
\begin{equation}
\text{\textcolor{blue}{vacuum minimum}}
~~~\neq ~~~
\text{\textcolor{blue}{semiclassical branch}.} \nonumber
\label{eq:minimum_not_branch_clean2}
\end{equation}
The same distinction appears in the discussion of coherent and Glauber--Sudarshan states. A generic off-shell fluctuation is simply an arbitrary configuration-space deformation. For example, expanding a fluctuation in modes:
\begin{equation}
\delta\Xi(x)
=
\sum_k f_k\,u_k(x),
\label{eq:mode_expansion_clean2}
\end{equation}
one finds that generic coefficients $f_k$ correspond to off-shell configurations, while only special coefficients satisfying the bootstrap equations \eqref{bootsforwalking} define on-shell fluctuations. Thus a generic GS state is a state on the full off-shell configuration space, whereas a semiclassical GS state is one that is sharply peaked on a full on-shell configuration:
\begin{equation}
\Xi_{\rm sc}({\bf x}; t) = \langle\widehat\Xi({x})\rangle_\sigma
=
\Big(
\langle\widehat g_{\mu\nu}^{\rm sc}(x)\rangle_\sigma,\,
\langle\widehat \Phi_{\rm sc}(x)\rangle_\sigma,\,
\langle\widehat \Psi_{{\rm f},{\rm sc}}(x)\rangle_\sigma,\,
\langle \widehat \Upsilon_{\rm sc}(x)\rangle_\sigma
\Big) ,
\label{eq:full_sc_config_GS_clean2}
\end{equation}
where $\Upsilon \equiv {\rm ghosts}$, and $\langle \widehat \Upsilon_{\rm sc}(x)\rangle_\sigma = 0$ would simply mean that no ghost configuration\footnote{This is in fact similar to \eqref{comcontags} where we interpret this as fluctuations.} is required to solve the bootstrap equations \eqref{bootsforwalking}, or alternatively, for finding the basins in  {\bf figures \ref{bootstrapbasins}}, {\bf \ref{bootsinksourcesaddle}} and {\bf \ref{sssrotate}}, although $\langle \lambda_3[\hat{\bf T}^{\rm nloc}_{\rm ghost}]_{AB}\rangle_\sigma$ in \eqref{bootsforwalking} could still be non-zero. Finally, once one has chosen a semiclassical branch
$\langle \widehat h_{ij}^{\rm sc}(\mathbf x;t_{\rm WKB})\rangle_\sigma$,
one may study off-shell fluctuations around it. But since we are actually defining theory over a static/Minkowski (or warped-Minkowski) minimum, as in \eqref{ajkertagra}, the fluctuations around such a ``time-dependent solution" is well defined and does not suffer from any renormalization group issue. This point has been demonstrated in \gspapers. Therefore
the clean overall picture that emerges from the above discussion is the following:
One organizes the theory around a Minkowski or warped-Minkowski minimum, splits all fields into semiclassical on-shell expectation-value parts and residual off-shell fluctuations, and then uses coherent or GS states to localize sharply on the full semiclassical configuration. The nontrivial metric branch $h_{ij}^{\rm sc}(\mathbf x;t_{\rm WKB})$ is supported not by the bare Minkowski minimum alone, but by the full dressed semiclassical expectation values of all sectors as $\langle\widehat h_{ij}^{\rm sc}(\mathbf x;t_{\rm WKB})\rangle_\sigma$, or more generically as in \eqref{ajkertagra}.

\subsubsection{Reference minima, semiclassical branches, and fluctuations \label{sec8.1.5}}

So far we saw that the clean way to organize the discussion is to separate three logically distinct levels:
(i) the reference minimum around which the theory is organized,
(ii) the exact equations satisfied by the full wavefunctional,
and (iii) the semiclassical branches realized as expectation values of suitable coherent or Glauber--Sudarshan states.
The key additional point is that, even when a semiclassical branch is time dependent, the microscopic theory may still be defined over a static Minkowski or warped-Minkowski minimum. In that case the time dependence appears only in the emergent semiclassical sector, while the ultraviolet definition, ghost structure, and renormalization prescription are inherited from the underlying static theory. This is the main technical and conceptual advantage of the Minkowski-based formulation. In the following we will compare three scenarios with reference minima (i) Minkowski or warped-Minkowski, (ii) de Sitter, and (iii) generic time-dependent background, and discuss the relative 
advantages over one another.

\subsubsection*{Minkowski (or warped-Minkowski) minimum}
Suppose one begins with a reference minimum as in \eqref{eq:minkowski_reference_minimum_clean2}, 
or more generally a warped-Minkowski minimum, which we will label as $\Xi_0^{\rm M}({\bf x})$. One then quantizes fluctuations \eqref{eq:full_fluctuation_split_clean2}  around it. 
The full exact state is therefore a wavefunctional of fluctuations:
\begin{equation}
\Psi
=
\Psi\Big[
\delta h_{ij}(\mathbf x),\,
\delta\Phi(\mathbf x),\,
\delta\Psi_{\rm f}(\mathbf x),\,
\delta{\rm ghosts}(\mathbf x);\,
t
\Big],
\label{eq:full_state_minkowski_case_final}
\end{equation}
and, when a genuine boundary Hamiltonian exists, the exact system is described by the Wheeler-DeWitt and the boundary Schr\"odinger equations \eqref{eq:boundary_SE_exact}.
In this setup, one may also try to realize semiclassical spacetime branches as expectation values in suitable coherent or GS states:
\begin{equation}
\langle \sigma|\widehat g_{\mu\nu}(x)|\sigma\rangle
=
g_{\mu\nu}^{\rm sc}(x),
\label{eq:GS_expectation_minkowski_case_final}
\end{equation}
with
$g_{\mu\nu}^{\rm sc}(x)$
possibly of FRW type, de Sitter-like type, or some other time-dependent form, provided the full dressed expectation values solve the effective equations.
The important refinement is the following. Once one has chosen a time-dependent semiclassical branch:
\begin{equation}
\langle \widehat\Xi_{\rm sc}(\mathbf x;t)\rangle_\sigma
=
\Xi_0^{\rm M}(\mathbf x)
+
\langle \delta\widehat\Xi_{\rm sc}(\mathbf x;t)\rangle_\sigma,
\label{eq:time_dep_sc_branch_minkowski_final}
\end{equation}
one may study off-shell fluctuations around it. This is the scenario that we studied in \eqref{ajkertagra}, and using the wavefunctional 
$\Psi'[\langle\delta\widehat\Xi(x)\rangle_\sigma]$ \eqref{eq:wavefunctional_offshell_clean2}, wherein the off-shell fluctuations are denoted by $\langle\delta\widehat\Xi_{\rm off}({\bf x}; t)\rangle_\sigma$. (At this stage we should note that $\Psi[\delta\Xi(x)]$ and $\Psi'[\langle\delta\widehat\Xi(x)\rangle_\sigma]$ are two different wavefunctional to realize the same system. This will be made clear in section \ref{sec8.1.6}.) The resulting fluctuation operator is generically time dependent and can be expressed schematically as:
\begin{equation}
{\cal D}_{\rm fluct}
=
{\cal D}_{\rm fluct}\!\Big[
\langle \widehat\Xi_{\rm sc}({\bf x}; t)\rangle_\sigma
\Big],
\label{eq:fluctuation_operator_time_dep_final}
\end{equation}
which happens because we are now measuring the fluctuations {\it relative} to the semiclassical background $\langle \widehat\Xi_{\rm sc}(\mathbf x;t)\rangle_\sigma$ from \eqref{eq:time_dep_sc_branch_minkowski_final} instead of the original Minkowski minimum $\Xi_0^{\rm M}({\bf x})$. Consequently,
the effective modes satisfy equations of the schematic form:
\begin{equation}
\ddot u_k^{\rm eff}(t)
+
\omega_k^2(t)\,
u_k^{\rm eff}(t)
=
0 ,
\label{eq:effective_mode_eq_time_dep_final}
\end{equation}
where $\omega_k(t)$ corresponds to the time-dependent frequencies. 
Therefore the usual time-dependent mode-mixing effects do still appear at the \emph{effective} level, so the natural question is whether this creates similar RG issues? The answer is of course no as we also discussed briefly in \gspapers. The reason is because these effective time-dependent modes may be expanded in the static basis of the underlying Minkowski or warped-Minkowski theory:
\begin{equation}
u_k^{\rm eff}(x;t)
=
\sum_n
\Big(
\alpha_{kn}(t)\,u_n^{(0)}(x)
+
\beta_{kn}(t)\,u_n^{(0)\ast}(x)
\Big),
\label{eq:bogoliubov_expansion_static_basis_final}
\end{equation}
where
$u_n^{(0)}(x)$
are the static vacuum modes. Thus the time dependence of the effective frequencies is not a fundamental pathology of the theory, but an emergent phenomenon inside a well-defined microscopic Hilbert space. So the precise conclusion is the following.
Around a Minkowski or warped-Minkowski minimum, the presence of a boundary Hamiltonian and the relative technical control of the vacuum sector make it the cleanest setting in which to organize fluctuations and attempt to realize a wide class of semiclassical time-dependent metric branches as suitably dressed GS states\footnote{The realization is restricted to a very {\it broad}, and quite possibly an exhaustive, class of physically admissible semiclassical branches. 
This non-uniqueness is not necessarily a problem. Rather, it reflects the
existence of different semiclassical branches, transient configurations, or
basins of attraction within the same underlying quantum theory. We can also provide an equivalent interpretation of the many semiclassical branches as arising from the backreaction of different
sets of quantum modes on the reference Minkowski, or warped-Minkowski,
background. More concretely, suppose that the fundamental theory admits a Minkowski
configuration $\Xi_{\rm M}$ satisfying
${\delta S_{\rm tot}\over \delta \Xi}
\big|_{\Xi=\Xi_{\rm M}}
=
0$.
A generic semiclassical configuration may then be written as
$\Xi_{\rm sc}
=
\Xi_{\rm M}
+
\delta\Xi_{\rm br}$,
where $\delta\Xi_{\rm br}$ denotes the collective backreaction of the excited
modes, fluxes, branes, or other sectors of the theory. The corresponding
semiclassical equation is not simply the linearized equation around
$\Xi_{\rm M}$, but rather
${\delta S_{\rm tot}\over \delta \Xi}
\big|_{\Xi=\Xi_{\rm M}+\delta\Xi_{\rm br}}
=
0$.
Expanding around the reference configuration gives:
\begin{equation}
\int d^d y\,
{\delta^2 S_{\rm tot}\over \delta\Xi(x)\delta\Xi(y)}
\bigg|_{\Xi=\Xi_{\rm M}}
\delta\Xi_{\rm br}(y)
+
{1\over 2}
\int d^d y\,d^d z\,
{\delta^3 S_{\rm tot}\over
\delta\Xi(x)\delta\Xi(y)\delta\Xi(z)}
\bigg|_{\Xi=\Xi_{\rm M}}
\delta\Xi_{\rm br}(y)\delta\Xi_{\rm br}(z)
+\cdots
=
0 . \nonumber
\end{equation}
Thus different solutions for $\delta\Xi_{\rm br}$ correspond to different
self-consistent backreacted branches of the same microscopic theory.
In the GS-state language, the corresponding statement is that the expectation
value of the full field configuration takes the form
$\langle \Xi(x)\rangle_\sigma
=
\Xi_{\rm M}(x)
+
\langle \delta\Xi(x)\rangle_\sigma$,
where the second term encodes the collective backreaction induced by the state
specified by $\sigma$. Different choices of $\sigma$ may therefore place the
system in different basins of attraction in the following quantitative way:
\begin{equation}
\sigma
\quad \longrightarrow \quad
\langle \Xi\rangle_\sigma
\quad \longrightarrow \quad
{\cal B}_\sigma \nonumber
\end{equation}
Here ${\cal B}_\sigma$ denotes the basin or semiclassical branch reached by the
state. These basins need not represent distinct microscopic theories; rather,
they may represent different backreacted semiclassical realizations of the same
underlying quantum theory. In this sense, the various fixed points or basins can indeed be alternatively interpreted as
the result of different patterns of backreaction of the modes on the reference
Minkowski background. The Minkowski configuration remains the organizing
vacuum of the microscopic theory, while the GS states generate the
semiclassical spread around it and select particular backreacted branches. \label{blazrog69}}. 
The effective fluctuations around such branches still exhibit time-dependent mode mixing, but these effects are embedded in a static microscopic theory and are therefore emergent rather than fundamental causing no issues with RG flows.

\subsubsection*{de Sitter minimum}
Now suppose instead that one takes de Sitter itself as the reference minimum, {\it i.e.}
$\Xi_0^{\rm dS}$.
Then, formally, one may again quantize fluctuations around that background and try to construct coherent or GS states representing other semiclassical branches.
However, here the conceptual difficulties are more severe as we discussed in some details in section \ref{sec2.0}. \textcolor{blue}{First}, the time dependence of the 
background $-$ encoded in the scale factor $a(t)$ for instance $-$ 
makes the mode decomposition and Wilsonian interpretation more subtle:
\begin{equation}
k_{\rm phys}(t)=\frac{k}{a(t)},
\label{eq:kphys_dS_case_final}
\end{equation}
so the UV/IR split is not fixed once and for all. \textcolor{blue}{Second}, one generally does not have the same clean and useful boundary-Hamiltonian formulation as in the Minkowski case, so the proper interpretation of time and of the bulk Schr\"odinger equation remain unclear. \textcolor{blue}{Third}, there is no underlying static vacuum in terms of which the time-dependent modes can be globally and unambiguously expanded. Thus the difficulty is not merely that the branch is time dependent. The deeper difficulty is that the \emph{reference minimum itself} is time dependent, so the microscopic definition of the theory already inherits the time-dependent ambiguities. Hence, if one takes de Sitter as the reference background, then one may formally study fluctuations around it and attempt to construct further semiclassical GS branches, but the Wilsonian RG interpretation becomes significantly more subtle, one generally loses the clean boundary-Hamiltonian structure available around Minkowski, and the time-dependent mode problem is now present already at the microscopic level rather than only in an emergent sector.

The above discussion raises a subtle but important issue. Suppose one starts from a Minkowski minimum and moves to an AdS$_4$ minimum. In the AdS case, one may still formulate a reasonably conventional Wilsonian effective field theory, since there is a clean notion of scale separation and no cosmological redshifting of physical momenta analogous to that in de Sitter 
space. 
By contrast, if one then attempts to uplift the AdS minimum to a de Sitter minimum, the physical momenta redshift according to \eqref{eq:kphys_dS_case_final}
so that modes continuously migrate from the ultraviolet to the 
infrared\footnote{If one writes AdS in a de Sitter slicing, then the metric coefficients do acquire explicit dependence on the de Sitter slicing time, and a mode decomposition performed with respect to that slicing will indeed exhibit behavior that looks very similar to cosmological redshifting. In that limited sense, one does see a ``de Sitter-like'' time dependence of the mode functions. However, this does \emph{not} mean that AdS has become physically equivalent to de Sitter, nor that the infrared issue is the same as in a genuine de Sitter vacuum. The crucial difference is that the apparent redshift in AdS written in dS slicing is largely a consequence of the \emph{choice of foliation}, whereas in genuine de Sitter space the cosmological expansion is an invariant feature of the spacetime itself.
More concretely, in a de Sitter slicing of AdS one has a metric of the following schematic form:
\begin{equation}
ds^2
=
d\rho^2
+
\sinh^2\!\Big(\frac{\rho}{L}\Big)\,
\Big(-\,dt^2
+
e^{2Ht}\,d\mathbf x^2\Big), \nonumber
\label{eq:AdS_dS_slicing_metric}
\end{equation}
where we have taken a flat slicing, for example, with $\rho$
being the AdS radial coordinate, labeling the different dS$_3$ slices; $L$ is the AdS$_4$ curvature radius; $H$ is the Hubble parameter of the de Sitter slices, {\it i.e.} the curvature scale of the dS$_3$ metric appearing inside the brackets; and
${\bf x} = (x_1, x_2)$ are the flat comoving spatial coordinates on each dS$_3$ slice. (This is a convenient slicing but by no means generic. One could always take other slicings, but the story remains unchanged.)  If one decomposes fields with respect to the time variable $t$ appearing above, then the effective physical momentum along the slice behaves like:
\begin{equation}
k_{\rm phys}(t)\sim k\,e^{-Ht}, \nonumber
\label{eq:apparent_redshift_ads_slicing}
\end{equation}
so one indeed finds de Sitter-like redshifting along the slice. But the important point is that this redshift is tied to the chosen slicing. The full AdS spacetime still has the global AdS structure, timelike boundary, and the usual better-controlled notion of asymptotic observables. In particular, the existence of a de Sitter slicing does not mean that AdS inherits the same intrinsic cosmological horizon structure or the same deep Wilsonian ambiguity as genuine de Sitter space.
Thus the appearance of redshifting in AdS written in de Sitter slicing is real at the level of that coordinate decomposition, but it is not the same as saying that AdS and de Sitter have the same physical RG problem. In AdS the effect is foliation-dependent, whereas in de Sitter the expansion is part of the spacetime itself.}. 
This means that the transition from Anti de-Sitter to de-Sitter is not merely a small quantitative shift in the vacuum energy, but a qualitative change in the infrared and Wilsonian structure of the theory.
For this reason, one should distinguish between two logically separate claims:
\vskip.1in
\noindent (i) \textcolor{blue}{A positive cosmological constant drastically changes the infrared structure of the theory.}

\vskip.1in

\noindent (ii) \textcolor{blue}{A controlled Wilsonian uplift from AdS to dS is highly nontrivial and perhaps obstructed.}
\vskip.1in
\noindent The first statement appears quite robust. The second is not a theorem, but it is a serious consistency concern. Indeed, if one begins with a controlled AdS effective theory and then adds a correction that is supposed to ``slightly'' uplift the minimum, one must also explain why the effective field theory remains meaningful once the infrared structure changes so drastically. In this sense, a successful uplift must do more than simply raise the value of the effective potential: it must also justify the continued validity of the EFT framework after the sign of the cosmological constant changes. Note that going to the static patch of de Sitter {\it does not} alleviate the problem. 

Thus the safest conclusion is not that quantum corrections can never change the sign of the cosmological constant, but rather that any such sign flip is conceptually much more delicate than a naive potential-based argument might suggest. Put differently, if the cosmological constant changes sign, then the EFT framework itself must be reinterpreted in an essential way. This makes controlled uplift highly nontrivial and casts serious doubt on naive uplifting arguments.

This line of reasoning also supports the alternative viewpoint that de Sitter-like behavior may be better understood not as a genuinely uplifted vacuum, but as a time-dependent excited state over a Minkowski or warped-Minkowski minimum. In that picture, the microscopic theory remains anchored in a vacuum with a cleaner EFT interpretation, while the positive effective energy density appears only at the level of a transient excited configuration. Then the cosmological redshift and the associated time dependence belong to the emergent state rather than to a newly fundamental vacuum. This suggests the following interpretation: the dramatic change in RG structure between AdS and dS suggests that a de Sitter-like phase may be more naturally interpreted as an excited or emergent state over a better-controlled vacuum, rather than as a genuinely uplifted de Sitter vacuum of the same Wilsonian EFT.

The resulting conclusion is therefore the following. The passage from AdS to dS is not simply a mild deformation of the vacuum energy; it is a structural change in the infrared behavior of the theory. This does not by itself prove a no-go theorem against uplift, but it does provide a strong conceptual and EFT-based argument against any overly naive uplift scenario. At the same time, it gives substantial support to the idea that de Sitter-like behavior may be more naturally viewed as a transient excited state rather than as a fundamental positive-cosmological-constant vacuum.

\subsubsection*{Any other time-dependent background as reference minimum}
The same logic extends to any other time-dependent reference background,
$\Xi_0^{\rm td}$,
such as FRW or any other cosmological/time-dependent solution. One may formally expand around that background, quantize fluctuations, and then try to construct other semiclassical branches as suitable GS states. But again the difficulties are of the same general kind as in de Sitter: time-dependent mode functions, subtler renormalization-group interpretation, absence of a preferred global vacuum notion, and a less clean exact Hamiltonian interpretation. In particular, there is no static microscopic vacuum whose basis modes can be used to reinterpret the time dependence as a derived effect. So the conclusion remains the same: 
for a generic time-dependent reference background, one may again quantize fluctuations and attempt to represent further semiclassical branches by suitable GS states, but the conceptual and technical subtleties of time-dependent backgrounds remain at the fundamental level.

\subsubsection*{The advantage of the Minkowski starting point}

So the real advantage of starting from Minkowski is not merely that it is one minimum among many. The deeper advantages are:
(i) the fluctuation problem is better controlled,
(ii) the interpretation of the full wavefunctional is clearer,
(iii) the boundary Hamiltonian structure is available in the cleanest form,
(iv) one may hope to generate more complicated semiclassical branches as dressed GS states without putting them in as exact minima from the start,
and, most importantly for the present discussion,
(v) the effective time-dependent fluctuations around an emergent semiclassical branch can still be expanded in the static mode basis of the underlying vacuum theory.
This last point means that, while time-dependent redshifting or frequency-mixing effects do appear in the effective description around
$\langle \Xi_{\rm sc}(\mathbf x;t)\rangle_\sigma$,
they {\it do not} signal a breakdown of the microscopic theory. Instead, they are Bogoliubov-type emergent effects inside a fixed, better-controlled Hilbert space built over the static vacuum. We can therefore summarize the situation in the following concise way:
the Minkowski-based formulation does not eliminate time-dependent mode mixing at the effective level, but it removes its status as a fundamental ambiguity by embedding the entire problem into a theory with a well-defined static vacuum, ultraviolet completion, ghost structure, and boundary-Hamiltonian interpretation. If instead one starts from de Sitter or any other time-dependent minimum, one may again study fluctuations and construct GS states, but one inherits the usual difficulties of time-dependent backgrounds, including the subtler Wilsonian RG interpretation and the loss of the clean Minkowski boundary-Hamiltonian structure. This is precisely what makes the Minkowski or warped-Minkowski starting point conceptually superior to taking a time-dependent spacetime itself as the fundamental minimum.

\subsubsection{Emergent wavefunctional and the two WdW equations \label{sec8.1.6}}

Our analysis above led to the notion of {\it two} wavefunctional \eqref{eq:wavefunctional_over_fluctuations_transparent} and \eqref{eq:wavefunctional_offshell_clean2}.
In this section we would like to carefully compare the two related but conceptually distinct wavefunctionals. The first, \eqref{eq:wavefunctional_over_fluctuations_transparent}, is the exact wavefunctional of the full gauge-fixed system, including the Faddeev--Popov ghosts and the gauge-fixing sector. The second, \eqref{eq:wavefunctional_offshell_clean2}, is an emergent wavefunctional written in terms of Glauber--Sudarshan expectation values, where the ghost sector no longer appears explicitly. Although these two objects describe the same underlying system, they do so at different levels of description. The purpose of this section is to explain precisely how they are related, why the absence of ghosts in the Glauber--Sudarshan description is not contradictory, and how the corresponding Wheeler--DeWitt equations should be understood. Also since our aim here is to isolate the ghosts, we will use a slightly different notation to denote the bosonic and fermionic DOFs. For example, let
${\Xi}
=
\big(
\check{\Xi},\,{\Upsilon}
\big)$,
where $\check{\Xi}$ denotes the full bosonic and fermionic physical degrees of freedom, while ${\Upsilon}$ denotes the ghost sector together with the variables required by gauge fixing. The full gauge-fixed action may be represented in the following way:
\begin{equation}
{S}_{\rm tot}(\check{\Xi},{\bf \Upsilon})
=
{S}_{\rm kin}(\check{\Xi})
+
{S}_{\rm NP}(\check{\Xi})
+
{S}_{\rm nloc}(\check{\Xi})
+
{S}^{({\rm g})}(\check{\Xi},{\Upsilon})
+
{S}^{({\rm g})}_{\rm nloc}(\check{\Xi},{\Upsilon}),
\label{eq:full_action_with_ghosts_section}
\end{equation}
where ${S}^{({\rm g})}$ and ${S}^{({\rm g})}_{\rm nloc}$ contain the ghost and gauge-fixing contributions. The exact wavefunctional of the full theory is therefore a functional on the full gauge-fixed configuration space expressed as:
\begin{equation}
{\Psi}
=
{\Psi}(\check{\Xi},{\Upsilon}) ,
\label{eq:exact_wavefunctional_full_section}
\end{equation}
which is the wavefunctional \eqref{eq:wavefunctional_over_fluctuations_transparent} that we studied before and which determines the interacting vacuum $|\Omega\rangle$ for the system (including all other excited states).  
At the formal level, the exact Schwinger--Dyson identity reads:
\begin{equation}
\left\langle
\frac{\delta {S}_{\rm tot}(\check{\Xi},{\Upsilon})}{\delta {\Xi}}
\right\rangle
=
0.
\label{eq:exact_SD_identity_section}
\end{equation}
This identity is always true inside the full path integral and does not, by itself, imply that one is sitting on a classical solution. Rather, it expresses the exact quantum equations of motion of the gauge-fixed theory. Similarly, the exact Hamiltonian constraint operator may be written schematically as:
\begin{equation}
\widehat{\mathcal H}_{\rm bulk}
=
\widehat{H}_{\rm grav}
+
\widehat{H}_{\rm matter}
+
\widehat{H}_{\rm mixed}
+
\widehat{H}_{\rm ghost}
+
\widehat{H}_{\rm gf},
\label{eq:exact_Hamiltonian_operator_section}
\end{equation}
where one may easily relate $\widehat{\mathcal H}_{\rm mat+gh}$ and $\widehat{\mathcal H}_g$ from \eqref{eq:Hbulk_democratic} with the relevant parts from \eqref{eq:exact_Hamiltonian_operator_section} after carefully distributing the ghost and the gauge-fixing terms.
After the dust settles, the exact Wheeler--DeWitt equation takes the form:
\begin{equation}
\widehat{\mathcal H}_{\rm bulk}(\widehat{\check\Xi},\widehat{\Upsilon})\,
\vert
{\Psi}(\check{\Xi},{\Upsilon})
\rangle
=
0.
\label{eq:exact_WdW_full_section}
\end{equation}
This is the exact microscopic equation of the full system. The ghost sector appears explicitly because the exact path integral and the exact state are defined on the gauge-fixed configuration space. Note that there are no expectation values over GS states included here, implying that we can study everything, including the semiclassical configurations discussed earlier, without introducing any GS states. The difference will be that the semiclassical configurations would immediately spread in the configuration space compared to the scenario where the GS states appear.

Now let us introduce a Glauber--Sudarshan state $|\sigma\rangle$. The corresponding expectation values define emergent semiclassical variables
$\langle\widehat{\check \Xi}\rangle_\sigma$.
The crucial point is that the ghost sector has already been taken into account in constructing these expectation values. Schematically:
\begin{equation}
\langle\widehat{\check \Xi}(x)\rangle_\sigma
=
\frac{
\int {\cal D}\check{\Xi}\,{\cal D}{\Upsilon}\;
e^{-i{S}_{\rm tot}[\check{\Xi},{\Upsilon}]}\,
{D}^\ast(\sigma;\check{\Xi})\,\check{\Xi}(x)\,{D}(\sigma;\check{\Xi})
}{
\int {\cal D}\check{\Xi}\,{\cal D}{\Upsilon}\;
e^{-i{S}_{\rm tot}[\check{\Xi},{\Upsilon}]}\,
{D}^\ast(\sigma;\check{\Xi})\,{D}(\sigma;\check{\Xi})
}.
\label{eq:GS_expectation_value_section}
\end{equation}
The integration over ${\Upsilon}$ has already removed the gauge redundancy and produced a reduced, gauge-fixed description in terms of the expectation-value variables $\langle\widehat{\check \Xi}\rangle_\sigma$. This motivates the definition of an emergent effective action:
\begin{equation}
\check{S}_{\rm tot}(\langle\widehat{\check\Xi}\rangle_\sigma)
=
{S}_{\rm kin}(\langle\widehat{\check\Xi}\rangle_\sigma)
+
\check{S}_{\rm NP}(\langle\widehat{\check\Xi}\rangle_\sigma)
+
\check{S}_{\rm nloc}(\langle\widehat{\check\Xi}\rangle_\sigma),
\label{eq:effective_action_GS_section}
\end{equation}
which is to be understood as the effective action governing the emergent semiclassical sector. It should also be compared to our earlier action \eqref{eq:full_action_with_ghosts_section}, where we see that the form of the non-perturbative and the non-local terms could in principle differ. There are several reasons for this, as discussed in \cite{wdwpaper}. However, we will not elaborate on them here, since they do not affect the main conclusion of the present analysis. 
The corresponding emergent wavefunctional is then the following:
\begin{equation}
\Psi'\big(\langle\widehat{\check\Xi}\rangle_\sigma\big) ,
\label{eq:emergent_wavefunctional_section}
\end{equation}
which appeared in \eqref{eq:wavefunctional_offshell_clean2}.
This object is \emph{not} the same as ${\Psi}(\check{\Xi},{\Upsilon})$. Rather, it is an effective wavefunctional obtained after projecting or reducing the exact theory to a Glauber--Sudarshan sector. Stated alternatively, we could only use $\Psi'\big(\langle\widehat{\check\Xi}\rangle_\sigma\big)$ to describe the system without worrying about $\Psi[\Xi]$. In fact one immediate advantage of using $\Psi'\big(\langle\widehat{\check\Xi}\rangle_\sigma\big)$ over $\Psi[\Xi]$ is that it can precisely tell us the probability amplitude for any semiclassical configuration appearing in say 
{\bf figures \ref{bootstrapbasins}}, {\bf \ref{bootsinksourcesaddle}} and {\bf \ref{sssrotate}} irrespective of whether they are sinks, sources, saddles or stationary points of the bootstrap map \eqref{bootsforwalking}.
Thus  
$\Psi'\big(\langle\widehat{\check\Xi}\rangle_\sigma\big)
\ne
{\Psi}(\check{\Xi},{\Upsilon})$,
but the relation between the two wavefunctionals is summarized by a weaker and more accurate statement of the following form\footnote{If the projection to the GS sector is sufficiently rich --- namely, if every physically admissible branch can be realized as $\langle\widehat{\check\Xi}\rangle_\sigma$ for some suitably dressed GS state --- then the emergent wavefunctional $\Psi'(\langle\widehat{\check\Xi}\rangle_\sigma)$ would indeed be exhaustive over the full physical branch space. The last step however is still an additional claim rather than an automatic consequence, thus extending the comment we made in footnote \ref{blazrog69}. We can also make this a bit more quantitative. 
Suppose the exact wavefunctional $\Psi[\Xi]$ is defined on the full microscopic space
${\cal C}_{\rm full}
=
\left\{
(\check\Xi,\Upsilon)
\right\}$,
and suppose it is exhaustive in the sense that every physically allowed microscopic configuration is represented in its support.
Now consider the GS projection:
\begin{equation}
{\cal P}_\sigma:\;
{\cal C}_{\rm full}
\longrightarrow
{\cal C}_{\rm GS}, \nonumber
\label{eq:projection_map_full_to_GS}
\end{equation}
where ${\cal C}_{\rm GS}$ is the space of configurations describable in terms of
$\langle\widehat{\check\Xi}\rangle_\sigma$.
Even if $\Psi[\Xi]$ is exhaustive on ${\cal C}_{\rm full}$, it does not automatically follow that ${\cal P}_\sigma({\cal C}_{\rm full})$ covers all conceivable semiclassical branches unless ${\cal P}_\sigma$ is known to have that property. It is quite possible that $\mathcal P_\sigma$ allows that property, but we haven't proved it yet, and a more detailed study is left for future work.}:
\bg\label{blazrog22}
\fcolorbox{black}{white}{%
$\displaystyle
\Psi'\big(\langle\widehat{\check\Xi}\rangle_\sigma\big)
\sim
{\cal P}_\sigma\,
{\Psi}(\check{\Xi},{\Upsilon})
$%
}
\nd
where ${\cal P}_\sigma$ denotes the effective projection, reduction, or coarse-graining map to the GS sector. The precise determination of $\mathcal P_\sigma$ will depend on the form of the emergent Wheeler-DeWitt equation that $\Psi'\big(\langle\widehat{\check\Xi}\rangle_\sigma\big)$ satisfies. We will determine the form of the emergent WdW equation below, but before that let us make one observation on the absence of ghosts in the emergent case.

The absence of explicit ghosts in the emergent wavefunctional is not a contradiction. The reason is that the ghosts have already done their job in the microscopic construction \eqref{eq:GS_expectation_value_section}. Once the path integral has been reduced to a section of the gauge bundle and rewritten in terms of the gauge-invariant or gauge-fixed expectation values $\langle\widehat{\check\Xi}\rangle_\sigma$, the resulting effective description no longer needs to display the ghost variables explicitly.
Thus the logic is the following: (a) 
existence of 
${S}_{\rm tot}(\check{\Xi},{\Upsilon})$
implies an 
exact gauge-fixed theory with explicit ghosts,
whereas (b) the existence of 
$\check{S}_{\rm tot}(\langle\widehat{\check\Xi}\rangle_\sigma)$ implies 
an emergent effective theory in which the ghost sector is implicit.
Therefore the ghosts have already been integrated out or absorbed into the definition of the emergent variables and the effective action.

In the same vein, 
the exact Wheeler--DeWitt equation is the constraint equation obeyed by the full gauge-fixed wavefunctional. Formally, this takes the form 
\eqref{eq:exact_WdW_full_section} which is the exact equation of the underlying theory. The ghosts and gauge-fixing contributions appear explicitly because they are part of the exact microscopic state space.

The emergent wavefunctional satisfies a different, effective Wheeler--DeWitt-type equation as described in \cite{wdwpaper}. To find such an equation, one first reduces the theory to the Glauber--Sudarshan sector and then constructs the effective Hamiltonian constraint operator appropriate to that reduced description following Weiss variation of the corresponding path-integral \cite{wdwpaper,weiss,feng}.
Thus the emergent equation should be written as
\begin{equation}
\widehat{\mathcal H}_{\rm eff}\big(\langle\widehat{\check\Xi}\rangle_\sigma\big)\,
\Psi'\big(\langle\widehat{\check\Xi}\rangle_\sigma\big)
=
0.
\label{eq:effective_WdW_section}
\end{equation}
This equation has the same physical role as the exact Wheeler--DeWitt equation, but it acts on a reduced state space and is governed by an effective Hamiltonian. The most one should say is that, in synchronous gauge with 
$N=1$ and 
$N^i=0$,
the effective Hamiltonian constraint operator may be represented schematically as being proportional to the operator form of the variation of the effective action with respect to the emergent $g^{00}$ variable:
\begin{equation}
\widehat{\mathcal H}_{\rm eff}
=
{\cal N}\,
\left[
\frac{\delta \check{S}_{\rm tot}(\langle\widehat{\check\Xi}\rangle_\sigma)}
{\delta \langle\widehat{g}^{00}\rangle_\sigma}
\right]_{\textcolor{blue}{{\rm oper}}},
\label{eq:effective_Hamiltonian_variation_section}
\end{equation}
where ${\cal N}$ is a convention-dependent normalization factor, and the subscript \textcolor{blue}{oper} represents the {\it operator} form of the bracket. The actual form of the bulk Hamiltonian\footnote{There is also an emergent boundary Hamiltonian possible but we will not worry about it here. Its presence serves the same purpose as before, namely, allows for a proper temporal evolution.} is as given in say \eqref{taylorchase} which, in the synchronous gauge makes $n^\mu = (1, 0, 0, 0)$ therefore giving credence to the form 
\eqref{eq:effective_Hamiltonian_variation_section}. Putting everything together, the emergent Wheeler--DeWitt equation may be written as:
\begin{equation}
\widehat{\mathcal H}_{\rm eff} (\langle\widehat{\check\Xi}\rangle_\sigma)
\Psi'\big(\langle\widehat{\check\Xi}\rangle_\sigma\big)
= {\cal N}\,
\left[
\frac{\delta \check{S}_{\rm tot}(\langle\widehat{\check\Xi}\rangle_\sigma)}
{\delta \langle\widehat{g}^{00}\rangle_\sigma}
\right]_{\textcolor{blue}{{\rm oper}}}
\Psi'\big(\langle\widehat{\check\Xi}\rangle_\sigma\big)
=
0.
\label{eq:emergent_WdW_variational_section}
\end{equation}
This is the correct way to formulate the emergent Wheeler--DeWitt equation: it is an effective projected version of the exact constraint equation, not merely the same equation with the fields replaced by their expectation values.

Since both wavefunctionals describe the same underlying system, one should indeed expect the two Wheeler--DeWitt equations to carry the same physical content. However, they do so at different levels: (a) at the exact level we have \eqref{eq:exact_WdW_full_section}, whereas  (b) at the emergent level we have \eqref{eq:emergent_WdW_variational_section}. 
The first is exact and microscopic. The second is reduced and effective. Therefore one should expect similarity of physical content, but not literal identity of variables, operators, or coefficients.

\subsubsection{GS States, WdW Constraints, and the physical  Hilbert space \label{hwlelo}}

The exact wavefunction discussed in \eqref{eq:exact_wavefunctional_full_section} should also tell us about the physical Hilbert space of our constrained system, an issue discussed briefly in footnote \ref{blazrog69}. Our aim here is to explain how the Glauber--Sudarshan states fit naturally into the physical Hilbert space of the constrained gravitational theory. The main point is that the physical Hilbert space is not the full kinematical Hilbert space. Rather, it is the subspace annihilated by the Hamiltonian and momentum constraints, with the remaining physical evolution generated by the boundary Hamiltonian. To this effect, 
let ${\cal H}_{\rm kin}$ denote the kinematical Hilbert space obtained before imposing the gravitational constraints. The physical Hilbert space ${\cal H}_{\rm phys}$ is defined by imposing the Wheeler--DeWitt and diffeomorphism constraints,
such that for the physical wavefunctional $|\Psi_{\rm phys}\rangle$:
\begin{equation}
|\Psi_{\rm phys}\rangle = |\Psi(\check\Xi, \Upsilon)\rangle   \in{\cal H}_{\rm phys}
~~
\Longleftrightarrow
~~
\widehat{\mathcal H}_{\perp}(x)|\Psi_{\rm phys}\rangle=0,
~~
\widehat{\mathcal H}_{i}(x)|\Psi_{\rm phys}\rangle=0.
\label{eq:Hphys_definition}
\end{equation}
States that do not satisfy these constraints are states in ${\cal H}_{\rm kin}$, but they do not by themselves represent physical states of the constrained gravitational system.

In a spacetime with an appropriate boundary or asymptotic structure, the total Hamiltonian takes the canonical ADM form \eqref{hotat1}, \eqref{hotat2}, and \eqref{hotat3} which may  now be expressed in the operator form as 
$\widehat H_{\rm tot}
=
\widehat H_{\rm bulk}
+
\widehat H_{\partial}$,
where 
$\widehat H_{\rm bulk}$ is in turn the operator form of \eqref{hotat2}. 
On physical states,
$\widehat H_{\rm bulk}|\Psi_{\rm phys}\rangle=0$,
so that the only nontrivial canonical generator is the boundary Hamiltonian:
\begin{equation}
\widehat H_{\rm tot}|\Psi_{\rm phys}\rangle
=
\widehat H_{\partial}|\Psi_{\rm phys}\rangle.
\label{eq:Htot_equals_Hpartial}
\end{equation}
Thus, although the Wheeler--DeWitt equation removes local bulk time evolution, the boundary Hamiltonian can still generate evolution with respect to the asymptotic or boundary time as we saw earlier. The eigenstates should satisfy the full constrained system \eqref{eq:boundary_SE_exact}. In other words, let $\{|\psi_n\rangle\}$ denote a complete set of normalizable physical states satisfying:
\begin{equation}
\widehat{\mathcal H}_{\perp}(x)|\psi_n\rangle=0,
\qquad
\widehat{\mathcal H}_{i}(x)|\psi_n\rangle=0,
\qquad
\widehat H_{\partial}|\psi_n\rangle=E_n|\psi_n\rangle.
\label{eq:physical_boundary_energy_basis}
\end{equation}
These states form a basis of ${\cal H}_{\rm phys}$, not of the full kinematical Hilbert space. Therefore any normalizable physical state admits the expansion
\begin{equation}
|\Psi_{\rm phys}\rangle
=
\sum_n c_n|\psi_n\rangle
+
\int dE\; c(E)|\psi_E\rangle , 
\label{eq:physical_basis_continuous}
\end{equation}
where we have generically assumed that 
the spectrum of $\widehat H_{\partial}$ also contain continuous components, implying that the usual discrete sum should be replaced by the corresponding sum plus integral. However if there are no continuous components, then $c(E) = 0$ and 
$c_n
=
\langle\psi_n|\Psi_{\rm phys}\rangle$.

Now consider a Glauber--Sudarshan state constructed by acting on a physical reference state $|\Omega\rangle$ with a displacement operator $D(\sigma)$ as in 
$|\sigma\rangle
\equiv
D(\sigma)|\Omega\rangle$.
where, as we saw before, $\sigma^{\mu\nu}$ denotes the source profile used to build the displacement operator with $D(\sigma)$ taking the form \eqref{sweenseyf}. 
The state $|\sigma\rangle$ admits an expansion in the physical basis $\{|\psi_n\rangle\}$ if and only if it is itself a physical state, {\it i.e.}
$|\sigma\rangle\in{\cal H}_{\rm phys}$.
Equivalently:

{\footnotesize
\begin{equation}
\widehat{\mathcal H}_{\perp}(x)D(\sigma)|\Omega\rangle=0 =
\widehat{\mathcal H}_{i}(x)D(\sigma)|\Omega\rangle ~~~ 
\Longleftrightarrow ~~
\left[
\widehat{\mathcal H}_{\perp}(x),D(\sigma)
\right]|\Omega\rangle=0 =
\left[
\widehat{\mathcal H}_{i}(x),D(\sigma)
\right]|\Omega\rangle ,
\label{eq:Dsigma_commutator_condition}
\end{equation}}
where the second identity works because 
$|\Omega\rangle$ is physical and therefore satisfies
$\widehat{\mathcal H}_{\perp}(x)|\Omega\rangle=0 =
\widehat{\mathcal H}_{i}(x)|\Omega\rangle$. 
Thus the relevant requirement is not that $D(\sigma)$ commutes with the constraints as an operator identity on all of ${\cal H}_{\rm kin}$. The weaker and physically sufficient condition is that $D(\sigma)$ preserve the physical subspace, or at least preserve physicality when acting on the chosen reference state $|\Omega\rangle$ as 
$D(\sigma){\cal H}_{\rm phys}
\subseteq
{\cal H}_{\rm phys}$,
or more weakly as
$D(\sigma)|\Omega\rangle
\in
{\cal H}_{\rm phys}$.
For every source $\sigma^{\mu\nu}$ satisfying this condition, the GS state can be expanded as\footnote{ One may also view the GS state using the products of the localized displacement operators that we studied earlier. To see this, we can discretize the interval as
$t_i=t_0<t_1<\cdots<t_N=t_f$.
The time-ordered exponential may be approximated by
$D_{[t_i,t_f]}(\sigma)
=
\lim\limits_{N\rightarrow\infty}\prod\limits_{j = N}^1 D_j(\sigma; t_j)$
where:
\begin{equation}
D_l(\sigma;t_l)
=
\exp\bigg[
\Delta t
\int_{\Sigma_{t_l}}d^3x\,
\sqrt{-\widehat{g}}\,
\sigma^{\mu\nu}(t_l,\mathbf x)
\widehat g_{\mu\nu}(t_l,\mathbf x)
\bigg] , \nonumber
\end{equation}
such that the order of the factors is essential, {\it i.e.}
$D_{l+1}D_l
\neq
D_lD_{l+1}$ in general. Now if we 
define recursively
$|\sigma;t_l\rangle
=
D_l(\sigma;t_l)|\sigma;t_{l-1}\rangle$
where 
$|\sigma;t_0\rangle
=
|\Omega\rangle$, then 
at every slice one may expand as
$|\sigma;t_l\rangle
=
\sum\limits_n c_n^{(l)}|\psi_n\rangle$, 
where the coefficients obey
$c_n^{(l)}
=
\sum\limits_m
\langle\psi_n|
D_l(\sigma;t_l)
|\psi_m\rangle
c_m^{(l-1)}$.
Introducing the transition matrix
$\left[D_l\right]_{nm}
\equiv
\langle\psi_n|
D_l(\sigma;t_l)
|\psi_m\rangle$,
one easily obtains
$c_n^{(l)}
=
\sum\limits_m
\left[D_l\right]_{nm}
c_m^{(l-1)}$. Putting everything together, the repeated kicks therefore results into:
\begin{equation}
c_n^{(N)}
=
\sum_{n_{N-1},\ldots,n_0}
\left[D_N\right]_{n n_{N-1}}
\left[D_{N-1}\right]_{n_{N-1}n_{N-2}}
\cdots
\left[D_1\right]_{n_1n_0}
c_{n_0}^{(0)}. \nonumber
\end{equation}
Thus, 
$|\sigma;t_f\rangle = \prod\limits_{j = N}^1 D_j |\Omega\rangle
= \sum\limits_n c_n^{(N)}|\psi_n\rangle$.
The first description, {\it i.e.} \eqref{eq:Dsigma_physical_expansion_continuous}, is therefore the final spectral decomposition of the
state, while the second description $-$ provided above $-$ gives the recursive construction of the
same coefficients. One should also note that there may be a very large or infinite family of coefficient vectors
${\bf c}^{(i)}(t) \equiv c_n^{(i, N)}(t)$, with $i = 1, ..., M$ and $M >> 1$ (or $M\to \infty$), each constructed from a different admissible
time-ordered sequence of kick operators. In other words:
\begin{equation}
\left\{
\mathbf c^{(i)}(t)
\right\}_{i=1}^{M}
\subset
\left\{
\mathbf c(t)
\;\middle|\;
i\dot{\mathbf c}(t)
=
H_{\partial}\mathbf c(t)
\right\}, \nonumber
\end{equation}
with $M = \infty$ unless further restrictions are imposed.
These vectors describe distinct GS
states and may all solve the same boundary Schr\"odinger equation, provided
the complete kick products preserve the physical Hilbert space and satisfy
the intertwining condition with the same boundary Hamiltonian. The
time-sliced construction changes how the coefficients are generated, but it
does not remove the large degeneracy of allowed Schr\"odinger solutions. \label{banfftagcross}}:
\begin{equation}
|\sigma\rangle = D(\sigma)|\Omega\rangle
=
\sum_n c_n(\sigma)|\psi_n\rangle
+
\int dE\;c(E;\sigma)|\psi_E\rangle ,
\label{eq:Dsigma_physical_expansion_continuous}
\end{equation}
assuming as before that the boundary spectrum has continuous sectors. If this is not the case then $c(E; \sigma) = 0$ and 
$c_n(\sigma)
=
\langle \psi_n|D(\sigma)|\Omega\rangle$. One may normalize this such that $0 < \langle \sigma |\sigma \rangle << \infty$, or keep it in the in-in path-integral format as we saw before.
This is the precise sense in which there can be a large class of sources $\sigma^{\mu\nu}$ compatible with the physical Hilbert space as alluded to in section \ref{sec3.3} and in footnote \ref{blazrog69}. The allowed class can be very large or even exhaustive. For example, it may include sources that are compatible with a chosen gauge fixing, sources that couple only to relational or Dirac observables, and sources for which the constraint equations are solved after the deformation. More generally, the admissible class is:
\begin{equation}
{\cal S}_{\rm phys}
=
\left\{
\sigma^{\mu\nu}
\;\middle|\;
\widehat{\mathcal H}_{\perp}(x)D(\sigma)|\Omega\rangle=0,
\quad
\widehat{\mathcal H}_{i}(x)D(\sigma)|\Omega\rangle=0
\right\} ,
\label{eq:allowed_sigma_constraint_definition}
\end{equation}
or alternatively as ${\cal S}_{\rm phys}
=
\left\{
\sigma^{\mu\nu}
\;\middle|\;
D(\sigma)|\Omega\rangle\in{\cal H}_{\rm phys}
\right\}$.
For $\sigma^{\mu\nu}\in{\cal S}_{\rm phys}$, the state $D(\sigma)|\Omega\rangle$ is a bona fide physical state, and the expansion \eqref{eq:Dsigma_physical_expansion_continuous} is guaranteed by completeness of the physical basis.
By contrast, for a generic source $\sigma^{\mu\nu}$ one may obtain a state in the kinematical Hilbert space,
$D(\sigma)|\Omega\rangle\in{\cal H}_{\rm kin}$,
which need not satisfy the gravitational constraints. In that case,
$D(\sigma)|\Omega\rangle
\notin
{\cal H}_{\rm phys}$,
and the state cannot be interpreted as a physical state of the constrained gravitational system. This puts a powerful constraint on the choice of $\sigma^{\mu\nu}$, implying that only its physical projection,
${\cal P}_{\rm phys}D(\sigma)|\Omega\rangle$,
where ${\cal P}_{\rm phys}$ denotes the projector or group-averaging map onto the constraint surface, is guaranteed to lie in ${\cal H}_{\rm phys}$. This projected state admits the expansion:
\begin{equation}
{\cal P}_{\rm phys}D(\sigma)|\Omega\rangle
=
\sum_n c_n^{\rm phys}(\sigma)|\psi_n\rangle,
\qquad
c_n^{\rm phys}(\sigma)
=
\langle\psi_n|D(\sigma)|\Omega\rangle ,
\label{eq:projected_Dsigma_expansion}
\end{equation}
assuming for simplicity no continuous spectrum. The choice of physical projection is easily implemented by restricting the functional form of $\sigma^{\mu\nu}$ to lie within an allowed domain. In fact within this domain all the sinks, sources, saddles and stationary points of the bootstrap map shown in {\bf figures \ref{bootstrapbasins}}, {\bf \ref{bootsinksourcesaddle}} and {\bf \ref{sssrotate}} can be exhaustively constructed!
The expectation value of a physical observable ${\cal O}_{\rm phys}$ in the GS state is therefore well defined for $\sigma^{\mu\nu}\in{\cal S}_{\rm phys}$:
\begin{equation}
\langle{\cal O}_{\rm phys}\rangle_\sigma
=
\frac{\langle\Omega|
D^\dagger(\sigma)
{\cal O}_{\rm phys}
D(\sigma)
|\Omega\rangle}{\langle\Omega | D^\dagger(\sigma) D(\sigma)|\Omega\rangle} ,
\label{eq:GS_expectation_value}
\end{equation}
which may then be brought in the in-in Schwinger-Keldysh form as shown in section \ref{sec6.0}.
This shows explicitly how the GS state behaves as a physical superposition of Wheeler--DeWitt states with definite boundary energy. Moreover, the boundary-time evolution is also transparent in this basis. Since
$\widehat H_{\partial}|\psi_n\rangle
=
E_n|\psi_n\rangle$,
the boundary-time evolved state is
\begin{equation}
|\sigma;t_\partial\rangle
=
e^{-i\widehat H_{\partial}t_\partial}
D(\sigma)|\Omega\rangle
=
\sum_n
c_n(\sigma)
e^{-iE_n t_\partial}
|\psi_n\rangle ,
\label{eq:GS_boundary_time_evolution}
\end{equation}
which is exactly what we had in section \ref{sec3.2}. Note that the computations in \eqref{eq:GS_expectation_value} and
\eqref{eq:GS_boundary_time_evolution} were formulated directly in the
Schwinger--Keldysh in--in framework precisely because an explicit
determination of the physical eigenstates $|\psi_n\rangle$ is, in
general, prohibitively difficult. Solving the full constrained system
in \eqref{eq:boundary_SE_exact}, including the Wheeler--DeWitt
constraints together with the boundary Schr\"odinger equation, is
substantially more involved than evaluating the corresponding
observables through the Schwinger--Keldysh path integral. Thus, the Schwinger--Keldysh in--in computation may be viewed as a
powerful alternative framework for extracting the physically relevant
information encoded in the coupled Schr\"odinger--Wheeler--DeWitt
system \eqref{eq:boundary_SE_exact}, without requiring an explicit
construction of its physical eigenstates.

Putting everything together, a remarkably clear picture emerges that may be elucidated in the following way. To start, recall that there are two logically distinct sets of conditions in the GS construction: (1) the exact Schwinger--Dyson relation \eqref{bootwa} satisfied by the state
$|\sigma\rangle
=
D(\sigma)|\Omega\rangle$, which expresses the balance between the microscopic equations of motion and the
state-dependent deformation generated by $D(\sigma)$; and (2) 
the semiclassical configuration
$\langle\Xi\rangle_\sigma$ lying on the diagonal mean-field shell given by the first equation in \eqref{bootsforwalking}
together with the off-diagonal balance condition given by the second equation in \eqref{bootsforwalking}.
These two sets of conditions should not be identified. Equation
\eqref{bootwa} is an exact state-dependent Schwinger--Dyson
relation, whereas
\eqref{bootsforwalking} are additional conditions imposed on
the mean-field configuration and on the off-diagonal sector.
Accordingly, one may classify GS states as follows: (a) 
off-shell GS states are those that satisfy \eqref{bootwa} but neither of the two equations from \eqref{bootsforwalking}; and (b) on-shell GS states are those that satisfy {\it both} \eqref{bootwa} and \eqref{bootsforwalking}. Now comes the important observation:
{\it whether the mean configuration is on-shell or off-shell, the underlying state
must still belong to the physical Hilbert space}\footnote{ There is one important qualification. If the operators $D_l$ are interpreted
as genuinely {\it external} kicks, then the evolution is not generated solely by
$\widehat H_\partial$. Instead, one has an effective generator
$\widehat H_{\rm eff}(t)
=
\widehat H_\partial(t)
+
\widehat H_{\sigma}(t)$, where $\widehat H_\sigma$ generates the additional kicks. However, for the construction to describe an excited state of one fixed microscopic
theory, rather than an externally driven theory, the appropriate interpretation is the one with $\widehat{H}_\partial$ and {\it not} the one with $\widehat{H}_{\rm eff}$. Even later, when we describe using trans-series, the trans-series content of the kicks must
be inherited from, and consistently matched to, the trans-series structure of
$\widehat H_\partial$. This is the precise criterion for a GS state to belong to the physical Hilbert space. See also the discussion after \eqref{eq:gs_continuous_drift}.}. In the canonical description
this requires the Wheeler-DeWitt and the Schr\"odinger constraints:
\begin{equation}
\widehat{\mathcal H}_{\perp}(x)
|\sigma;t\rangle
=
0,
\qquad
\widehat{\mathcal H}_{i}(x)
|\sigma;t\rangle
=
0,\qquad i\frac{\partial}{\partial t}
|\sigma;t\rangle
=
\widehat H_{\partial}
|\sigma;t\rangle ,
\label{eq:GS_local_constraints}
\end{equation}
or collectively
$\widehat H_{\rm bulk}
|\sigma;t\rangle
=
0$ along with solving the boundary Schr\"odinger equation.
The crucial distinction is then the following: all states have to necessarily solve \eqref{bootwa} as well as \eqref{eq:GS_local_constraints}. The special ``on-shell" GS states also need to solve the additional bootstrap equations \eqref{bootsforwalking}.
There is no contradiction here: the Wheeler--DeWitt and boundary equations
constrain the quantum state, whereas the mean-field shell condition constrains
a particular one-point function extracted from that state.

There is however an interesting extension to the physical boundary-energy basis once we take the trans-series form of the boundary Hamiltonian. (We will discuss this in more detail in section \ref{sec8.1.7}.)  Assuming that the physical basis satisfy the constraint equations \eqref{eq:physical_boundary_energy_basis} with the 
boundary Hamiltonian admitting a trans-series expansion of the form:
\begin{equation}
\widehat H_{\partial}
=
\sum_{\alpha}
\mathfrak s_\alpha\,
e^{-A_\alpha/g}
\sum_{k=0}^{\infty}
g^k\,
\widehat H_{\partial,k}^{(\alpha)},
\label{eq:Hboundary_transseries}
\end{equation}
where $\alpha$ labels perturbative and nonperturbative sectors,
$A_\alpha$ are the corresponding actions, and $\mathfrak s_\alpha$ denote
trans-series parameters or Stokes data, then:
\begin{equation}
i\dot c_n(t) =  \sum_m
H_{nm}^{\partial}(t)c_m(t)
=
\sum_m
\sum_{\alpha}
\mathfrak s_\alpha\,
e^{-A_\alpha/g}
\sum_{k=0}^{\infty}
g^k
H_{nm,k}^{(\alpha)}(t)
c_m(t) ,
\label{eq:coefficient_transseries_evolution}
\end{equation}
implying that each time-dependent coefficient receives contributions from the various
perturbative and nonperturbative sectors of the boundary Hamiltonian. In \eqref{eq:coefficient_transseries_evolution} we have defined $H_{nm,k}^{(\alpha)}
=
\langle\psi_n|
\widehat H_{\partial,k}^{(\alpha)}
|\psi_m\rangle$, and $H^\partial_{nm}$ is defined as:
\begin{equation}
H_{nm}^{\partial}
=
\langle\psi_n|
\widehat H_{\partial}
|\psi_m\rangle
=
\sum_{\alpha}
\mathfrak s_\alpha\,
e^{-A_\alpha/g}
\sum_{k=0}^{\infty}
g^k\,
H_{nm,k}^{(\alpha)}.
\label{eq:Hboundary_matrix_transseries}
\end{equation}
Note that \eqref{eq:coefficient_transseries_evolution} follows from the boundary Schr\"odinger
equation. It does not require the on-shell bootstrap conditions
\eqref{bootsforwalking}, although \eqref{bootwa} needs to be satisfied as mentioned earlier. The formal solution  of \eqref{eq:coefficient_transseries_evolution} may be expressed in the following way:
\begin{equation}
\vec c(t) =
\begin{pmatrix}
c_1(t)\\
c_2(t)\\
\vdots
\end{pmatrix}
=
\mathcal T
\exp\left[
-i\int_{t_i}^{t}dt'\,
H_{\partial}(t')
\right]
\vec c(t_i),
\label{eq:coefficient_formal_solution}
\end{equation}
implying that the trans-series structure of 
$\widehat H_{\partial}$ governs the temporal evolution of $c_n(t)$, irrespective of whether 
$\langle\Xi\rangle_\sigma$ is on-shell or off-shell.
Since $H_{\partial}$ itself contains all trans-series sectors, the coefficients
$c_n^{(l)}$ inherit the same trans-series structure:
\begin{equation}
c_n(t)
=
\sum_{\alpha}
\mathfrak s_\alpha\,
e^{-A_\alpha/g}
\sum_{k=0}^{\infty}
g^k
c_{n,k}^{(\alpha)}(t),
\label{eq:coefficient_transseries}
\end{equation}
possibly supplemented by logarithmic sectors, multi-instanton sectors, or
resonant terms when required. The on-shell equations \eqref{bootsforwalking} select a special subset of those solutions. In this precise sense, all physically admissible GS states evolve according to
the same boundary Hamiltonian, but only a restricted subset of them produces
on-shell mean configurations.

Therefore, after all the ingredients are taken into account, the
Glauber--Sudarshan state must satisfy several logically distinct but
mutually compatible consistency conditions. First, its expectation
values must obey the complete Schwinger--Dyson equations derived from
the resummed trans-series action. Second, the bootstrap split equations
must be satisfied, ensuring that the proposed expectation values define
a self-consistent on-shell configuration rather than an arbitrary
off-shell point in the space of fields. Third, the corresponding state
must satisfy the Wheeler--DeWitt and momentum constraints, so that it
lies in the physical Hilbert space of the constrained gravitational
theory. Finally, its remaining physical time dependence must obey the
boundary Schr\"odinger equation generated by
$\widehat H_{\partial}$.

Taken together, these conditions provide a unified characterization of
the physical GS state. The Schwinger--Dyson equations determine the
quantum dynamics of its expectation values, while the bootstrap split
equations impose the additional self-consistency conditions required
for those expectation values to describe an on-shell background of the
full trans-series theory. The Wheeler--DeWitt and momentum constraints
remove unphysical bulk degrees of freedom and eliminate any independent
notion of local bulk time evolution. The boundary Hamiltonian then
generates the genuine physical evolution with respect to the boundary,
or asymptotic, clock. Thus, at a bootstrap fixed point, the GS state is not merely a solution
of an effective equation of motion: it is an on-shell,
constraint-satisfying state whose boundary-time evolution remains
entirely within the physical Hilbert space. Away from the fixed points,
the state may no longer satisfy the bootstrap on-shell conditions,
satisfying instead only \eqref{bootwa}, but
its physicality within the Hilbert space continues to be preserved,
provided that the Wheeler--DeWitt and momentum constraints remain
satisfied throughout the corresponding boundary-time evolution.


\subsubsection{Trans-series structure of the WdW equation \label{sec8.1.7}}

This brings us to the last topic in the above sequence of ideas connecting to the Wheeler-DeWitt formulation in the presence of the GS states, namely, what happens if we take the full trans-series form of the action.
From the above discussion it is
clear that a fully self-consistent trans-series framework must ultimately
confront and resolve both the question of time and the Schr\"odinger evolution. Schematically, the relevant chain of logic that we have in mind is:
\begin{equation}
\textcolor{blue}{S_{\rm trans\text{-}series}}
\;\Longrightarrow\;
\text{\textcolor{blue}{Hamilton--Jacobi equation}}
\;\Longrightarrow\;
\textcolor{blue}{t_{\rm emergent}}
\;\Longrightarrow\;
\text{\textcolor{blue}{Schr\"odinger evolution}} \nonumber
\end{equation}
with the idea that a trans-series action should induce corresponding trans-series
corrections at every stage of the semiclassical construction. In fact all the elements in the above chain originally arose from the WdW equation \eqref{eq:rescaled_WdW_equation_simple} which should 
be understood as the simplest Einstein-plus-matter WKB structure. It is obtained under the assumption that one is working in a single dominant semiclassical sector, with a single Hamilton--Jacobi function $S_0[h]$, a single prefactor $A[h]$, and a single matter wavefunctional $\psi[h,\phi]$ propagating on that background. In other words, it corresponds to a one-branch truncation of the full problem. This simple structure is useful because it isolates the standard hierarchy: (a) Hamilton-Jacobi equation at 
${\cal O}(M_{\rm P}^4)$; (b) transport equation for $A[h]$
together with the emergent Schr\"odinger equation for $\psi$ at 
$\mathcal O(M_p^2)$; and (c) the first subleading quantum-gravitational correction at $\mathcal O(1)$.
However, once the full action is written in trans-series form, this entire hierarchy must be generalized.
The reason is that in the derivation of \eqref{eq:rescaled_WdW_equation_simple} one has implicitly assumed two simplifications: (1) 
the gravitational potential is exhausted by a single functional $U[h]$, and (2) 
the wavefunctional has the single-phase form $A[h]\,e^{iM_{\rm P}^2S_0[h]}\,\psi[h,\phi]$.
In a genuine trans-series setting, neither assumption is sufficient. Both the Hamiltonian constraint and the wavefunctional inherit contributions from infinitely many sectors. This may be easily inferred from the action itself. For example, 
suppose the full action takes the schematic trans-series form:
\begin{equation}
S_{\rm tot}[\Xi]
=
\sum_{\alpha}
{\cal C}_\alpha\,
\exp\!\Big(
-\,M_{\rm P}^{p_\alpha}\,{\cal A}_\alpha[\Xi]
\Big)\,
S_\alpha[\Xi],
\label{eq:transseries_action_general}
\end{equation}
where $\Xi$ denotes the full set of fields, including the spatial metric, matter, fermions, ghosts, and any other sectors. The label $\alpha$ runs over perturbative, instanton, multi-instanton, and possibly nonlocal or wormhole sectors. The coefficients ${\cal C}_\alpha$ encode the trans-series data, $\mathcal A_\alpha \equiv \mathcal A_\alpha[\Xi]$, and the exponential weights determine the nonperturbative hierarchy of the various contributions.

The point of \eqref{eq:transseries_action_general} is that the canonical Hamiltonian constraint must now be derived from the full action, not merely from a single perturbative piece. Once the total action takes the trans-series form \eqref{eq:transseries_action_general}, 
both the sectoral actions $S_\alpha[\Xi]$ and the exponential weights multiplying them depend on the full field configuration $\Xi$. This immediately implies that the canonical decomposition of $S_{\rm tot}$ is more complicated than in the purely perturbative case, because functional differentiation now acts not only on $S_\alpha[\Xi]$ but also on the field-dependent weights $\exp(-M_{\rm P}^{p_\alpha}{\cal A}_\alpha[\Xi])$.
Indeed, differentiating the full trans-series action gives:
\begin{align}
\frac{\delta S_{\rm tot}[\Xi]}{\delta \Xi}
&=
\sum_{\alpha}
{\cal C}_\alpha\,
\frac{\delta}{\delta \Xi}
\left[
\exp\!\Big(
-\,M_{\rm P}^{p_\alpha}\,{\cal A}_\alpha[\Xi]
\Big)\,
S_\alpha[\Xi]
\right]
\nonumber\\
&=
\sum_{\alpha}
{\cal C}_\alpha\,
\exp\!\Big(
-\,M_{\rm P}^{p_\alpha}\,{\cal A}_\alpha[\Xi]
\Big)
\left[
\frac{\delta S_\alpha[\Xi]}{\delta \Xi}
-
M_{\rm P}^{p_\alpha}
\frac{\delta {\cal A}_\alpha[\Xi]}{\delta \Xi}\,
S_\alpha[\Xi]
\right] ,
\label{eq:variation_transseries_action_revised}
\end{align}
where the second term inside the square brackets is new compared to the simpler case where ${\cal A}_\alpha$ is treated as a constant. Thus the non-perturbative weights are themselves dynamical functionals on configuration space and contribute directly to the equations of motion. It follows that the canonical Hamiltonian constraint, and therefore the Wheeler--DeWitt operator, must inherit a genuinely trans-series structure.
Schematically, one therefore expects the full bulk Wheeler--DeWitt operator to take the form:
\begin{equation}
\widehat{\cal H}_{\rm bulk}^{\rm tot}
=
\widehat{\cal H}_{\rm grav}^{(0)}
+
\widehat{\cal H}_{\rm matt}^{(0)}
+
\sum_{\alpha\neq 0}
{\cal C}_\alpha\,
\exp\!\Big(
-\,M_{\rm P}^{p_\alpha}\,{\cal A}_\alpha[\Xi]
\Big)\,
\Big(\widehat{\cal H}^{(\alpha)}
+
\Delta\widehat{{\cal H}}^{(\alpha)}\Big).
\label{eq:transseries_WdW_operator_revised}
\end{equation}
Here $\widehat{\cal H}_{\rm grav}^{(0)}+\widehat{\cal H}_{\rm matt}^{(0)}$ denotes the leading perturbative Wheeler--DeWitt operator, the operators $\widehat{\cal H}^{(\alpha)}$ encode the direct contribution of the individual trans-series sectors, and the terms $\Delta\widehat{{\cal H}}^{(\alpha)}$ schematically represent the additional pieces induced by the functional dependence of ${\cal A}_\alpha[\Xi]$. In other words, the field dependence of the trans-series weights generates extra contributions that would not be present if the non-perturbative exponents were mere constants.
Accordingly, the Wheeler--DeWitt equation is no longer the simple perturbative equation, but changes from:
\begin{equation}
\Big(
\widehat{\cal H}_{\rm grav}^{(0)}
+
\widehat{\cal H}_{\rm matt}^{(0)}
\Big)\Psi[\Xi]
=
0~~~ \longrightarrow ~~~
\widehat{\cal H}_{\rm bulk}^{\rm tot}\Psi[\Xi]
=
0 ,
\label{eq:full_transseries_WdW_revised}
\end{equation}
This means that the dynamics now contains explicit couplings between different sectors of the trans-series, and the resulting hierarchy is necessarily more involved than the single-sector WKB expansion. One can already see this from the effective potential and the effective supermetric.
In the simple equation \eqref{eq:rescaled_WdW_equation_simple}, the entire background information is encoded in the pair:
\begin{equation}
\Big(
U[h],\,
G_{ijkl}(h)
\Big).
\label{eq:simple_background_pair_revised}
\end{equation}
But once the action is trans-series, these objects are no longer purely perturbative. The potential must be replaced by an effective trans-series potential:
\begin{equation}
U[h]
\quad\longrightarrow\quad
U_{\rm eff}[\Xi]
=
U_0[h]
+
\sum_{\alpha\neq 0}
{\cal C}_\alpha\,
\exp\!\Big(
-\,M_{\rm P}^{p_\alpha}\,{\cal A}_\alpha[\Xi]
\Big)\,
U_\alpha[\Xi]
+
\cdots,
\label{eq:Ueff_transseries_revised}
\end{equation}
where $U_0[h]$ is the leading perturbative term and the $U_\alpha[h,\Xi]$ encode non-perturbative corrections. We have used the same label $\alpha$ for the sector appearing in the weight and in the induced potential term. This is natural in the present construction, since both originate from the same trans-series sector. One could certainly entertain a more general situation in which several different non-perturbative structures appear within the same trans-series family, but we shall not do so here and will keep this simpler sector-by-sector correspondence throughout.
In the same way, the supermetric entering the kinetic term may also receive trans-series corrections:
\begin{equation}
G_{ijkl}(h)
\quad\longrightarrow\quad
G^{\rm eff}_{ijkl}(\Xi)
=
G^{(0)}_{ijkl}(h)
+
\sum_{\alpha\neq 0}
{\cal C}_\alpha\,
\exp\!\Big(
-\,M_{\rm P}^{p_\alpha}\,{\cal A}_\alpha[\Xi]
\Big)\,
G^{(\alpha)}_{ijkl}(\Xi)
+
\cdots.
\label{eq:DeWitt_metric_transseries_revised}
\end{equation}
Again, we are using the same non-perturbative label $\alpha$ to characterize both the weight and the induced correction to the supermetric. More general possibilities could be considered, but we shall stick to this streamlined formulation unless stated otherwise.

The interpretation of \eqref{eq:Ueff_transseries_revised} and \eqref{eq:DeWitt_metric_transseries_revised} is that the geometry of superspace itself is corrected sector by sector. Thus even the notion of WKB flow becomes branch-dependent once the trans-series corrections are included.
For the same reason, the wavefunctional should not be assumed to have the single-branch form:
\begin{equation}
\Psi[\Xi({\bf x}); t]
=
A[h; t]\,
e^{\,iM_{\rm P}^2S_0[h; t]}\,
\psi[\Xi({\bf x}); t] ,
\label{eq:single_sector_ansatz_revised}
\end{equation}
which is appropriate only if one is ignoring all but one dominant branch. In a full trans-series treatment, the natural generalization is a sum over sectors, each with its own phase, its own prefactor, and its own non-perturbative weight:
\bg\label{lolashona1}
\Psi[\Xi]
 &= &
\sum_{\alpha}
{\cal N}_\alpha[\Xi]\,
\exp\!\Big(
iM_{\rm P}^2S_0^{(\alpha)}[\Xi]
+
iS_1^{(\alpha)}[\Xi]
+
\cdots
\Big)\,
\exp\!\Big(
-\,M_{\rm P}^{p_\alpha}\,{\cal A}_\alpha[\Xi]
\Big)\\
& = &
\sum_{\alpha}
A_\alpha[h,\widetilde\Xi]\,
\exp\!\Big(
iM_{\rm P}^2S_0^{(\alpha)}[h,\widetilde\Xi]
\Big)\,
\exp\!\Big(
-\,M_{\rm P}^{p_\alpha}\,{\cal A}_\alpha[\Xi]
\Big)\,
\psi_\alpha[h,\Phi,\Psi_{\rm f},{\rm ghosts}] ,\nonumber
\nd
where in the second line we have separated the matter factor more explicitly and $\widetilde\Xi = (\Phi, \Psi_{\rm f}, {\rm ghosts})$. (This wavefunction can be defined on a Cauchy slice so that $\Xi = \Xi({\bf x})$ and $\widetilde\Xi = \widetilde\Xi({\bf x})$ and there is an explicit dependence on the time $t$.)
Here each branch carries its own Hamilton--Jacobi function $S_0^{(\alpha)}$, its own prefactor $A_\alpha$, and its own reduced matter-plus-ghost wavefunctional $\psi_\alpha$. The dependence on $\Xi$ in the first two factors is kept explicit to emphasize that, in the full trans-series setting, the sectoral structures need not be purely gravitational.

In the simplest treatment, the leading order of \eqref{eq:rescaled_WdW_equation_simple} yields a single Hamilton--Jacobi equation \eqref{eq:gravitational_HJ_revised}.
This is the standard equation governing the dominant semiclassical background. However, once the full trans-series structure is restored, the action of the full operator \eqref{eq:transseries_WdW_operator_revised} on the full wavefunctional \eqref{lolashona1} generates a family of coupled equations. For each sector $\alpha$ one expects a Hamilton--Jacobi-type equation of the schematic form:
\begin{equation}
\frac{1}{2}\,
G^{\rm eff}_{ijkl}
\frac{\delta S_0^{(\alpha)}}{\delta h_{ij}}
\frac{\delta S_0^{(\alpha)}}{\delta h_{kl}}
+
U_{\rm eff}^{(\alpha)}[h,\Xi]
+
{\cal M}^{(\alpha)}[h,\Xi]
=
0,
\label{eq:sector_HJ_equation_revised}
\end{equation}
where ${\cal M}^{(\alpha)}[h,\Xi]$ denotes mixing terms induced by the other trans-series sectors as well as by the functional differentiation of the weights $\exp(-M_{\rm P}^{p_\alpha}{\cal A}_\alpha[\Xi])$.

The origin of ${\cal M}^{(\alpha)}[h,\Xi]$ is important. It arises because the Wheeler--DeWitt operator is acting on a sum of sectors rather than on a single branch, and because the exponential weights are themselves nontrivial functionals of the fields. Consequently, the equations for the various $S_0^{(\alpha)}$ are not independent: different sectors backreact on one another. Thus the full trans-series Wheeler--DeWitt problem is not described by one Hamilton--Jacobi equation, one transport equation, and one Schr\"odinger equation, but by a hierarchy of coupled equations.

Exactly the same logic applies to the reduced Schr\"odinger equation. In the simple one-branch approximation one obtains
\eqref{eq:emergent_WKB_SE_revised}.
But in the full trans-series setting, there is no reason to expect a unique $t_{\rm WKB}$ or a unique reduced matter wavefunction. Rather, each branch $\alpha$ has its own semiclassical flow and hence its own intrinsic WKB time:
\begin{equation}
\frac{\partial}{\partial t_{\rm WKB}^{(\alpha)}}
\equiv
\int d^{d-1}x\;
G_{ijkl}^{\rm eff}(x,\Xi)\,
\frac{\delta S_0^{(\alpha)}}{\delta h_{ij}(x)}
\frac{\delta}{\delta h_{kl}(x)} ,
\label{eq:sectorwise_tWKB_revised}
\end{equation}
where we have isolated $x = ({\bf x}, t)$ so that it becomes obvious that we are integrating over all the spatial directions. The net outcome is that the reduced matter-plus-ghost state in sector $\alpha$ obeys a sectorwise Schr\"odinger equation of the schematic form:
\begin{equation}
i\frac{\partial}{\partial t_{\rm WKB}^{(\alpha)}}\psi_\alpha
=
\widehat H_{{\rm matt+gh},\,\alpha}^{\rm eff}\,
\psi_\alpha
+
\sum_{\beta\neq\alpha}
{\cal V}_{\alpha\beta}\,
\psi_\beta.
\label{eq:sectorwise_SE_transseries_revised}
\end{equation}
The operator $\widehat H_{{\rm matt+gh},\,\alpha}^{\rm eff}$ is the effective matter-plus-ghost Hamiltonian appropriate to the $\alpha$-th branch, while the terms ${\cal V}_{\alpha\beta}$ encode the mixing between sectors.
Thus the full trans-series problem generally involves not one emergent time, but a family of branch-dependent intrinsic times, namely 
$t_{\rm WKB}^{(\alpha)}$,
each associated with a particular semiclassical background branch. In this sense, the standard WKB time is only the leading representative of a more complicated sectorwise structure.

Despite all these complications, the simpler equation \eqref{eq:rescaled_WdW_equation_simple} is still very useful. Its role is that of a controlled truncation of the full trans-series problem to a single dominant branch. In other words, it corresponds to keeping only one sector, say $\alpha=0$, and neglecting all inter-sector couplings. In that approximation the full wavefunctional reduces to:
\begin{equation}
\Psi
\approx
A[h]\,
e^{\,iM_{\rm P}^2S_0[h]}\,
\psi[h,\phi],
\label{eq:single_branch_truncation_revised}
\end{equation}
on a Cauchy slice labeled by $t$,
and the full Wheeler--DeWitt hierarchy collapses back to the familiar sequence: (a) 
leading Hamilton--Jacobi equation, (b)
transport equation for $A[h]$,
(c) emergent Schr\"odinger equation for $\psi$.
Thus the standard Born--Oppenheimer expansion is not wrong; it is simply the first truncation of the much larger trans-series structure.

In a similar vein, we could also work out the Wheeler-DeWitt equation for the emergent wavefunctional for the GS states exactly following \eqref{eq:emergent_WdW_variational_section}. The story is very similar
to the discussion developed in section \ref{sec8.1.6} with the emergent action replaced by the corresponding trans-series form, so we will not elaborate it further here. Instead we will study an interesting connection between the string vertex operators and the bulk GS states. This is what we turn to next.

\subsection{Analogy between stringy vertex operators and bulk GS states \label{sec8.2}}

So far, our analysis of GS states has not incorporated any direct string-theoretic input. The discussion has been carried out entirely from a four-dimensional perspective, although we have occasionally appealed to the trans-series structure of the effective action. In what follows, we will introduce a limited string-theoretic perspective, specifically from the world-sheet point of view, in order to highlight certain structural similarities between two seemingly different objects: two-dimensional world-sheet vertex operators and four-dimensional bulk GS states. This comparison is meant only at the structural level and is not intended to imply a dynamical equivalence between the two theories\footnote{A deeper relation between the two descriptions may nevertheless exist, but establishing such a relation is not the purpose of this section.}.

\subsubsection{Vertex operators and physical string states \label{sec8.2.1}}

In the context of string theory, the physical Hilbert space is equivalently described by BRST-invariant vertex operators. A closed-string physical state may be represented by a vertex operator $V$, and amplitudes are computed by inserting such operators in the worldsheet path integral. Schematically, in Lorentzian and Euclidean signatures we get respectively:
\bg\label{kataS}
{\cal A}_n
& = &
\int {\cal D}X\,{\cal D}h\,{\cal D}(\text{gh})\;
e^{\,iS_{\rm ws}[X,h,\text{gh}]}\,
V_1\cdots V_n \nonumber\\
{\cal A}_n^{(E)}
& = &
\int {\cal D}X\,{\cal D}h\,{\cal D}(\text{gh})\;
e^{-S_{\rm ws}^{(E)}[X,h,\text{gh}]}\,
V^{(E)}_1\cdots V^{(E)}_n ,
\nd
in terms of world-sheet fields $X^\mu$, world-sheet metric $h_{ab}$ and the world-sheet ghosts.
The important structural point is that the vertex operators are {\it multiplicative insertions}. They are not accompanied by an extra explicit factor of $i$ merely because the worldsheet theory is Lorentzian. The factor of $i$ belongs to the dynamical phase $e^{\,iS_{\rm ws}}$, not to the inserted operator itself. Thus $V_j^{(E)}$ may be interpreted as the vertex operator in the Euclidean space with the same structural representation as $V_j$. 
This is exactly parallel to the GS construction, where the operators
$D(\sigma)$ and 
$D^\dagger(\sigma)$
enter as contour insertions multiplying the Schwinger--Keldysh weight:
\begin{equation}
e^{\,iS[g_+]-iS[g_-]}\,
D(\sigma)\, D^\dagger(\sigma),
\label{eq:SK_weight_with_GS_detailed}
\end{equation}
rather than as elementary terms in the Lorentzian action carrying their own explicit $i$, as we saw from our earlier detailed discussion. 
Thus the direct operator-level analogy is:
\begin{equation}
e^{\,iS_{\rm ws}}\,
V_1\cdots V_n
\qquad
\longleftrightarrow
\qquad
e^{\,iS[g_+]-iS[g_-]}\,
D(\sigma)\,
 D^\dagger(\sigma) ,
\label{eq:direct_operator_analogy_detailed}
\end{equation}
which provides the first step in identifying these two theories. However the identification \eqref{eq:direct_operator_analogy_detailed} doesn't completely justify the equivalence because the displacement operators are integrated operators over the full $d+1$-dimensional space-time, whereas $V_j$ appears to be a local operator on the world-sheet. Therefore we need to consider 
an integrated worldsheet operator of the form:
\begin{equation}
{\cal V}
=
\int d^2z\;V(z,\bar z) ,
\label{eq:integrated_vertex_def_detailed}
\end{equation}
where we integrate over the full two-dimensional space-time. At this stage, 
it is essential to distinguish two logically different uses of ${\cal V}$. The \textcolor{blue}{first} use is the integrated operator insertion. This implies performing 
 the following operation:
\begin{equation}
\int {\cal D}X\,{\cal D}h\,{\cal D}(\text{gh})\;
e^{\,iS_{\rm ws}}
\exp\!\left(
\lambda {\cal V}
\right) ,
\label{eq:integrated_insertion_detailed}
\end{equation}
wherein we first exponentiate the operator \eqref{eq:integrated_vertex_def_detailed}, and then insert it directly to the Lorentzian path-integral.
In this form, the insertion behaves exactly like the GS contour factors
$W_\pm[g_\pm; \sigma]$ exponentiated as:
\begin{equation}
\exp\Big(W_+[g_+;\sigma]\Big)
\qquad
\text{or}
\qquad
\exp\Big(W_-[g_-;\sigma]\Big) ,
\label{eq:Wplus_Wminus_analogy_detailed}
\end{equation}
where, as like \eqref{eq:Wplus_Wminus_analogy_detailed}, 
there is no extra explicit factor of $i$ multiplying ${\cal V}$ at the level of the insertion itself either in the Lorentzian or the Euclidean case. This brings us to the \textcolor{blue}{second} use of $\mathcal V$, namely the absorbed action deformation. Here 
one may choose to regard ${\cal V}$ as a deformation of the world-sheet action in two possible ways:
\begin{equation}
S_{\rm ws}
\longrightarrow
S_{\rm ws}
+
\lambda \int d^2z\;V(z,\bar z), ~~~~~ S_{\rm ws}
\longrightarrow
S_{\rm ws}
-
i\lambda \int d^2z\;V(z,\bar z) ,
\label{eq:worldsheet_action_deformation_detailed}
\end{equation}
where in the first case we have an operator insertion to the action itself, whereas in the second case we have a {\it complex} deformation of the action. In fact it is the latter deformation that we are interested in. The reason is, taking this into account, the Lorentzian path integral takes the following form:
\begin{equation}
\exp\!\left[
i\left(
S_{\rm ws}
-
i\lambda\int d^2z\;V
\right)
\right] = e^{\,iS_{\rm ws}}
\exp\!\left(
\lambda \int d^2z\;V
\right) ,
\label{eq:absorbed_into_Lorentzian_action_detailed}
\end{equation}
which matches with \eqref{eq:integrated_insertion_detailed}. 
So an explicit $i$ appears only after one rewrites the insertion as part of a complexified effective action. It is not an intrinsic part of the vertex insertion itself.
The same distinction holds in the GS construction:
\begin{equation}
e^{\,iS[g_+]}\,e^{W_+[g_+;\sigma]}
=
\exp\!\left[
i\left(
S[g_+]-iW_+[g_+;\sigma]
\right)
\right],
\label{eq:GS_absorption_detailed}
\end{equation}
and similarly for the backward branch, as we saw earlier. The basic object is the insertion $e^{W_\pm}$, not an elementary action term $iW_\pm$. However in the Euclidean case, since neither the path integral weight nor the vertex operators or the GS displacement operators come with an $i$, one could in this case regard the operator insertions as genuine shifts of the Euclidean action. 

If one treats the integrated operator as a genuine deformation of the worldsheet theory, as in the second relation in \eqref{eq:worldsheet_action_deformation_detailed} then it modifies the Schwinger--Dyson equations in the standard way. Suppose:
\begin{equation}
S^{(E)}_{\rm ws}[X,h]
=
S^{(E)}_0[X,h]
-
\lambda \int d^2z\;V^{(E)}(z,\bar z).
\label{eq:deformed_ws_action_SD_detailed}
\end{equation}
where we are following the same strategic Euclidean construction as in 
\eqref{kataS}. This way we don't have to worry about the relative factor of $i$ in the subsequent analysis. For such a case, 
the variation with respect to the worldsheet metric gives:
\begin{equation}
\frac{\delta S^{(E)}_{\rm ws}}{\delta h_E^{ab}(z)}
=
\frac{\delta S^{(E)}_0}{\delta h_E^{ab}(z)}
-
\lambda\,
\frac{\delta}{\delta h_E^{ab}(z)}
\int d^2z'\;V^{(E)}(z',\bar z') ,
\label{eq:ws_variation_SD_detailed}
\end{equation}
whose vanishing would signal the worldsheet classical EOM. This is not what we need, so we can shift the worldsheet metric by a small amount to determine the quantum EOM. This is of course the worldsheet Schwinger-Dyson equation, and using the fact that the vertex operator depends on the worldsheet metric, we get the  
the exact SD identity to be:
\begin{equation}
\left\langle
\frac{\delta S^{(E)}_0}{\delta h_E^{ab}(z)}
\right\rangle_\lambda
=
\lambda\,
\left\langle
\frac{\delta}{\delta h_E^{ab}(z)}
\int d^2z'\;V^{(E)}(z',\bar z')
\right\rangle_\lambda ,
\label{eq:ws_SD_identity_detailed}
\end{equation}
where the subscript $\lambda$ signifies that the expectation value is taken over the $\lambda \ne 0$ case.
Thus the integrated operator shifts the worldsheet equations everywhere on the worldsheet, rather than merely changing boundary data. This is the logical worldsheet counterpart of the GS contour construction: an insertion extended over the whole dynamical domain modifies the effective equations governing the background, in contrast to a single-slice coherent-state preparation which changes only boundary data. However \eqref{eq:ws_SD_identity_detailed} doesn't appear to be of the same form as say \eqref{bootsforwalking} or, more appropriately, of the same form as \eqref{eq:EAB_def} and \eqref{eq:BAB_def}. 

We can also contrast it with a single-slice coherent state. 
A coherent state prepared on a single Cauchy slice should be written in terms of the canonical creation and annihilation operators, or equivalently in terms of the field and its conjugate momentum on that slice. Schematically, one may write:
\bg\label{vmojajuani}
|\alpha\rangle = D(\sigma)|\Omega\rangle
 &= &
\exp\!\left[
\int_{\Sigma_{t_0}} d^{d-1}x\;
\Big(
\alpha(\mathbf x)\,\hat a^\dagger(\mathbf x)
-
\alpha^\ast(\mathbf x)\,\hat a(\mathbf x)
\Big)
\right]
|\Omega\rangle \nonumber\\
& = &
\exp\!\left[
i\int_{\Sigma_{t_0}} d^{d-1}x\;
\Big(
\pi_\alpha(\mathbf x)\,\hat\phi(\mathbf x,t_0)
-
\phi_\alpha(\mathbf x)\,\hat\pi(\mathbf x,t_0)
\Big)
\right]
|\Omega\rangle ,
\nd
where the second equality is constructed from canonical field basis. The way we have defined $D(\sigma)$, it shifts the vacuum expectation values so that the coherent state is peaked around
$\langle \alpha|\hat\phi(\mathbf x,t_0)|\alpha\rangle
=
\phi_\alpha(\mathbf x)$,
and
$\langle \alpha|\hat\pi(\mathbf x,t_0)|\alpha\rangle
=
\pi_\alpha(\mathbf x)$ in the phase space. It is also interesting to compare this with the displacement operator for the coherent state in \eqref{sweenseyf}. In fact the one in \eqref{sweenseyf} is a \emph{specialized} version of the general canonical displacement operator in which one keeps only the momentum-generator piece. In that case the operator translates the metric operator while leaving the momentum expectation value unshifted at the level of the center of the packet. Indeed, using the canonical commutator, one finds schematically the following shifts of the canonical variables:

{\footnotesize
\begin{equation}
D^{-1}(\sigma)\,\hat g_{ij}(\mathbf x,t_0)\,D(\sigma)
=
\hat g_{ij}(\mathbf x,t_0)
+
\sigma_{ij}(\mathbf x,t_0),~~~~~~
D^{-1}(\sigma)\,\hat\pi^{ij}(\mathbf x,t_0)\,D(\sigma)
=
\hat\pi^{ij}(\mathbf x,t_0) ,
\label{eq:momentum_unshifted_from_pi_generator}
\end{equation}}
up to the precise density and the index conventions. Thus \eqref{sweenseyf} is the gravitational analogue of a \emph{configuration-space translation} coherent state: it shifts the metric profile but does not introduce an independent momentum shift. In the notation of \eqref{vmojajuani}, it corresponds to the restricted choice
$\phi_\alpha \neq 0$ and 
$\pi_\alpha = 0$, which 
after translating to the gravitational variables provide 
$\phi_\alpha \;\longrightarrow\; \sigma^{ij}$ and 
$\hat\pi \;\longrightarrow\; \hat\pi_{ij}$. Moreover
its path-integral representation modifies the initial or final boundary wavefunctional,
$\Psi_\alpha[\phi(t_0)]
=
\langle \phi(t_0)|\alpha\rangle$,
but it does not insert an operator throughout the full bulk history. Therefore it does not play the same structural role as either an integrated worldsheet vertex operator or the spacetime/contour insertion defining the Glauber--Sudarshan construction. The coherent state is fundamentally a boundary preparation, whereas the latter are deformations supported over the worldsheet or over the relevant spacetime/history contour. 

In particular, the GS construction is not captured by a single-slice coherent state. Rather, it is characterized by a displacement operator inserted over the relevant history, together with a nontrivial bootstrap/self-consistency condition for the backreacted configuration. Thus a coherent state is best viewed as a sharply peaked semiclassical packet on one branch, whereas the GS construction is broad enough to describe superpositions or distributions over such packets. (This story has been discussed in much details earlier, which the readers can go back for more information.)
Our worldsheet representation of the vertex operators suggests that they are structurally closer to the Glauber--Sudarshan construction than to a coherent state localized on a single Cauchy slice.

One may also see some self-consistency between the worldsheet beta functions and the bootstrap equations studied earlier. 
The true analogue of the GS bootstrap on the worldsheet is not merely the existence of an integrated vertex operator, but the requirement that the deformation be self-consistent. In a worldsheet conformal field theory, a deformation by couplings $\lambda^I$:
\begin{equation}
S_{\rm ws}
\;\longrightarrow\;
S_{\rm ws}
-
\sum_I \lambda^I \int d^2z\;V_I(z,\bar z),
\label{eq:ws_deformation_corrected}
\end{equation}
is consistent only if conformal invariance is preserved. This requires the vanishing of the beta functions,
$\beta^I(\lambda) \equiv \mu {d\lambda^I\over d\mu} =0$,
up to the usual scheme-dependent field redefinitions\footnote{More generically $\beta^I(\{\lambda^J\}) = 0$.}. Equivalently, the deformed worldsheet theory must remain a consistent conformal field theory, and this is precisely the condition that the corresponding target-space background satisfy the string equations of motion. However there is a subtlety here. The vanishing of the worldsheet beta functions and the worldsheet Schwinger-Dyson equation 
are connected but they are \emph{not} the same condition.
The worldsheet Schwinger--Dyson identity, now generalized from \eqref{eq:ws_SD_identity_detailed}, may be expressed as:
\begin{equation}
\left\langle
\frac{\delta S^{(E)}_0}{\delta h_E^{ab}(z)}
\right\rangle_\lambda
=
\sum_I\lambda^I\,
\left\langle
\frac{\delta}{\delta h_E^{ab}(z)}
\int d^2z'\;V_I^{(E)}(z',\bar z')
\right\rangle_\lambda ,
\label{eq:ws_SD_identity_answer}
\end{equation}
which is an exact quantum identity inside the deformed worldsheet path integral. It expresses the vanishing of the expectation value of the metric variation of the full deformed worldsheet action. It is true whether or not the deformation defines a conformal fixed point.
By contrast, the vanishing beta function condition
$\beta^I(\lambda)=0$
is the condition that the deformation by the integrated vertex operators remain conformal, {\it i.e.} that the deformed worldsheet theory defines a consistent background string vacuum. This is a much stronger requirement.
So the precise relation is the following.
The Schwinger--Dyson identity is an exact quantum identity of the deformed worldsheet theory, whereas the vanishing of the beta functions is the additional requirement that the deformation define a conformal fixed point.
The connection between them is that both concern the consistency of the deformation with respect to the worldsheet metric. Indeed, variation with respect to $h_E^{ab}$ inserts the worldsheet stress tensor, so \eqref{eq:ws_SD_identity_answer} is closely related to the condition that Weyl invariance be preserved. When the beta functions vanish, the trace of the stress tensor vanishes modulo contact terms and field redefinitions, and the deformation is exactly marginal. In that case the Schwinger--Dyson identity is compatible with conformal invariance in the strongest possible sense.
However, one should not identify the two statements. The Schwinger--Dyson identity can hold even when
$\beta^I(\lambda)\neq 0$,
in which case the deformed theory is still a perfectly good quantum field theory in two dimensions, but no longer a conformal field theory that yields a consistent background for string theory\footnote{In this case, the worldsheet theory must have an extra degree of freedom, like the Liouville mode, to be consistent.}.
Thus the proper logical relation is that the vanishing beta function implies that 
the worldsheet deformation is conformally consistent,
whereas
the Schwinger--Dyson identity holds more generally and does not by itself imply $\beta^I(\lambda)=0$.
 The former is an exact functional identity; the latter is the extra fixed-point condition selecting consistent string backgrounds.}

In the GS construction, the analogous requirement is that the displacement operator and the backreacted expectation values solve the coupled Schwinger--Dyson/bootstrap conditions. Schematically, one has from  \eqref{bootwa}:
\begin{equation}
\left\langle
\frac{\delta S_{\rm tot}[\Xi]}{\delta \Xi(x)}
\right\rangle_\sigma
=
\left\langle
\frac{\delta}{\delta \Xi(x)}
\log\Big(
D^\dagger(\sigma)D(\sigma)
\Big)
\right\rangle_\sigma,
\label{bootwa2}
\end{equation}
together with the requirement that the chosen GS data be self-consistent under the bootstrap map once we express, considering only the on-shell DOFs for the time being, the SD equation \eqref{bootwa2} as a set of two equations from \eqref{bootsforwalking}. In symbolic form, one may write the conditions as \eqref{eq:EAB_def} and \eqref{eq:BAB_def}
or equivalently, when such a formulation exists, as fixed points of the bootstrap map as in \eqref{eq:fixed_point_reordered}.
The precise form of aforementioned conditions depend on the detailed realization of the GS construction, but the conceptual point is clear: the insertion is admissible only if it is compatible with the backreacted equations of motion and with the full bootstrap consistency conditions.
Thus the proper analogy is not merely:
\begin{equation}
\int d^2z\;V
\qquad
\longleftrightarrow
\qquad
{D}(\sigma),
\label{eq:naive_analogy_corrected}
\end{equation}
since this only captures the structural similarity between the two constructions without delving into the actual dynamics of the two theories, 
but the deeper correspondence is instead the following:
\begin{equation}
\text{\textcolor{blue}{worldsheet marginality / conformal invariance of the deformation}} \nonumber
\label{eq:logic_step_one}
\end{equation}
\begin{equation}
\textcolor{red}{\Updownarrow} \nonumber
\label{eq:logic_arrow_one}
\end{equation}
\begin{equation}
\textcolor{blue}{\text{self-consistency of the GS insertion under the full bootstrap equations}} \nonumber
\label{eq:logic_step_two}
\end{equation}
Therefore, the real structural parallel is that both constructions involve two ingredients:
(i) a deformation operator,
and (ii) a nontrivial self-consistency condition selecting the admissible deformations. On the worldsheet, this condition is the vanishing of the beta functions. In the GS construction, it is the full Schwinger--Dyson/bootstrap consistency of the backreacted state.

\subsubsection{What the analogy does and does not mean \label{sec8.2.2}}

The analogy should therefore be understood in the following precise sense. An integrated worldsheet vertex operator and the GS contour insertion are similar because both are extended operator insertions that modify the full path-integral weight and thereby alter the corresponding effective equations. They are both much closer in structure to bulk deformations than to single-slice coherent-state preparations. However, one should not confuse the two settings:
\begin{equation}
\text{\textcolor{blue}{worldsheet CFT}}
~~
\neq
~~
\text{\textcolor{blue}{Schwinger--Keldysh spacetime path integral}}\nonumber
\label{eq:ws_not_SK_identical_detailed}
\end{equation}
The worldsheet theory is a two-dimensional constrained gauge system, while the GS construction is an in--in contour deformation in target spacetime. Their relation is therefore structural, not literal.
The worldsheet deformation by an integrated vertex operator and the bulk GS construction both involve extended insertions that modify the effective equations. However, in the GS case the insertion is tied to a {\it state-dependent bootstrap problem}, and this is what leads to the second equation involving
\begin{equation}
\left\langle
\frac{\delta}{\delta {\bf g}^{AB}}
\log\Big(D^\dagger(\sigma)D(\sigma)\Big)
\right\rangle_\sigma .
\label{eq:DD_log_term_main}
\end{equation}
That extra equation has no direct analogue in the simplest worldsheet Schwinger--Dyson identity for a fixed integrated deformation. The worldsheet Schwinger--Dyson equation \eqref{eq:ws_SD_identity_answer}
is a {\it single} shifted equation: the integrated operator contributes an extra term to the worldsheet equations of motion.
At this stage there is no automatic splitting into two coupled equations. One simply has a deformed theory and its corresponding Schwinger--Dyson identities. 

In the GS construction the situation is richer. The insertion is not just a fixed external deformation of the action. Rather, one has a contour insertion of $D(\sigma)$ and $D^\dagger(\sigma)$
which defines a deformed state, and one simultaneously asks for self-consistency of the backreacted mean fields generated by that state.
Because of this, the resulting problem is not exhausted by a single shifted Einstein or Schwinger--Dyson equation. Instead, the full bootstrap system separates into
(i) an effective Einstein-type equation for the mean fields,
and (ii) a second equation governing the GS insertion itself through 
$\left\langle
\frac{\delta}{\delta {\bf g}^{AB}}
\log\Big(D^\dagger D\Big)
\right\rangle_\sigma$.
So the pair:
\begin{equation}
\begin{cases}
{\bf G}_{AB}
-
{\bf T}^{\rm np}_{AB}
-
{\bf T}^{\rm nloc}_{AB}
=
0,
\\[4pt]
\left\langle
\frac{\delta}{\delta {\bf g}^{AB}}
\log\Big( D^\dagger(\sigma) D(\sigma)\Big)
\right\rangle_\sigma
=
\text{off-diagonal and ghost contributions},
\end{cases}
\label{eq:bulk_pair_compare}
\end{equation}
arises precisely because the GS insertion is not treated as a fixed background source. It is part of a self-consistent state construction.
The worldsheet equation \eqref{eq:ws_SD_identity_answer} is an exact Schwinger--Dyson identity in the deformed worldsheet theory. It says that the variation of the undeformed action with respect to the worldsheet metric is balanced by the metric variation of the deformation term. But here the right-hand side is just the response of the integrated vertex deformation. Unless one promotes the couplings $\lambda^I$ themselves to bootstrap variables, there is no extra GS-like self-consistency sector hiding inside the equation.
This is one of the key differences between the worldsheet realization of the vertex operators and the bulk GS case.

One may however ask as to 
why the worldsheet identity does not usually split this way.
In the simplest worldsheet treatment, $\lambda^I$ is just a coupling multiplying a chosen integrated operator. One asks whether the deformed theory is conformal, namely whether the corresponding beta function vanishes or the Schwinger--Dyson identity \eqref{eq:ws_SD_identity_answer} is satisfied. The latter
is certainly a self-consistency condition, but it is not usually written as a separate equation involving the derivative of
$\log\Big(
D^\dagger D
\Big)$
because there is no direct analogue of a GS displacement operator defining a new in--in state on a doubled contour.
Instead, the worldsheet self-consistency is encoded in the vanishing of beta functions or, equivalently, in BRST/conformal invariance of the deformed theory. So the worldsheet theory has
a deformed action plus conformal invariance conditions,
whereas the GS construction has a contour insertion plus a bootstrap split into a mean-field equation and a state equation.

If one nevertheless wants to imitate the bulk logic, the most sensible split is: (a) worldsheet stress-tensor or background equation; and (b) conformal consistency/marginality condition.
For case (a), the metric variation defines the worldsheet stress tensor, so the Schwinger--Dyson identity may be viewed as the condition:
\begin{equation}
\left\langle
T_{ab}^{(0)}(z)
\right\rangle_\lambda
+
\sum_I \lambda^I
\left\langle
\Delta T_{ab}^{(I)}(z)
\right\rangle_\lambda
=
0 \nonumber
\label{eq:ws_first_equation_split}
\end{equation}
\begin{equation}
\Delta T_{ab}^{(I)}(z)
\equiv
-\frac{2}{\sqrt{h_E(z)}}\,
\frac{\delta}{\delta h_E^{ab}(z)}
\int d^2z'\;V_I^{(E)}(z',\bar z').
\label{eq:deltaT_definition}
\end{equation}
This is the worldsheet analogue of the effective background equation: it expresses the balance of the stress tensor in the deformed theory.
For case (b), the true nontrivial self-consistency condition of the worldsheet deformation is not another SD equation, but the vanishing of the beta functions:
\begin{equation}
\beta^I(\{\lambda^J \})=0, ~~~
\forall ~~ (I, J)
\label{eq:beta_functions_second_equation}
\end{equation}
This is what ensures that the deformation defines a consistent conformal background rather than merely a deformed two-dimensional QFT. With this interpretation, the first equation means that 
the deformed worldsheet theory satisfies its local quantum metric-variation identity,
while the second means that 
the deformation is actually consistent as a string background, {\it i.e.} two-dimensional conformal invariance is preserved.

If one really wanted a bulk-GS-style two-equation split, one would have to reinterpret the worldsheet insertion itself as something bootstrap-like, for example by treating the deformation data $\lambda^I$ as variables selected by a self-consistency condition rather than as arbitrary couplings. Then one could imagine a split of the form (a) 
worldsheet background equation, 
together with
(b) bootstrap equation selecting admissible $\lambda^I$.
But in ordinary worldsheet CFT this second equation is nothing but
the vanishing beta function condition!
So the bootstrap content is already exhausted by the conformal fixed-point condition. Therefore, 
on the worldsheet, the natural ``second equation'' is not a separate GS-type insertion equation but simply the vanishing of the beta functions.

We may then summarize the above discussion in the following way.
The worldsheet deformation by an integrated vertex operator and the bulk GS construction are structurally analogous, but not identical. In the worldsheet theory, the integrated operator shifts the Schwinger--Dyson equations of the deformed CFT, while consistency is imposed separately through conformal invariance or vanishing beta functions. In the GS construction, the contour insertion defines a deformed state whose self-consistency leads to a genuine bootstrap split: one equation for the mean fields and a second equation for the insertion itself, encoded through 
$\left\langle
\frac{\delta}{\delta {\bf g}^{AB}}
\log\Big(D^\dagger(\sigma) D(\sigma)\Big)
\right\rangle_\sigma$.
The correct dictionary therefore is outlined in the following table:

\begin{table}[H]
\centering
\renewcommand{\arraystretch}{1.5}
\begin{tabular}{|p{0.47\textwidth}|p{0.47\textwidth}|}
\hline
\textbf{Worldsheet construction} & \textbf{GS bootstrap construction} \\
\hline

{BRST-invariant worldsheet vertex operator}
&
{GS contour operator insertion} 
\\
\hline
{integrated operator insertion over the worldsheet}
&
{operator insertion over the Schwinger--Keldysh contour}
\\
\hline
no extra explicit $i $ attached to the insertion itself 
&
no extra explicit $i $ attached to $D(\sigma), D^\dagger(\sigma)$
\\
\hline
SD identity + 
{beta-function equations } $\beta^I=0$
&
${\cal E}_{AB}=0,\ {\cal B}_{AB}=0 $
\\
\hline
{conformal invariance of the deformed worldsheet theory}
&
{bootstrap self-consistency of the GS-deformed spacetime branch} 
\\
\hline

\end{tabular}
\label{tab:WS_vs_GS_part0}
\end{table}

\noindent basically capturing a small aspect of what we said earlier. More detailed dictionary appears in {\bf Tables \ref{tab:WS_vs_GS_part1}}, {\bf \ref{tab:WS_vs_GS_part2}} and {\bf \ref{tab:WS_vs_GS_part3}}.

\subsubsection{Worldsheet WdW equation and the meaning of time \label{sec8.2.3}}

A worldsheet Wheeler--DeWitt equation is the statement that physical string states are annihilated by the worldsheet Hamiltonian and momentum constraints, {\it i.e.} by the Virasoro constraints. In canonical form one may write:
\begin{equation}
\widehat{\cal H}_\perp\,\Psi[X,h]=0,
\qquad
\widehat{\cal H}_\sigma\,\Psi[X,h]=0,
\label{eq:ws_WdW_constraints}
\end{equation}
where $X^\mu(\sigma)$ are the embedding fields and $h_{ab}$ is the worldsheet metric. Equivalently, after gauge fixing, these become the usual Virasoro conditions on physical states:
\begin{equation}
\hat T_{ab}\,\Psi=0
\qquad\Longleftrightarrow\qquad
L_n|\Psi\rangle=0,
\qquad
\widetilde L_n|\Psi\rangle=0,
\label{eq:ws_Virasoro_form}
\end{equation}
with the precise set depending on whether one is in the closed- or open-string case. Thus the worldsheet Wheeler--DeWitt equation is not an evolution equation in an external time, but rather the statement that the physical wavefunctional is annihilated by the worldsheet reparametrization constraints.

The above analysis now raises the following question: what is the meaning of time on the worldsheet? 
On the worldsheet there is no \emph{boundary time} analogous to the physical time $t$ that appeared in the bulk four-dimensional discussion with a boundary Hamiltonian. Instead, the worldsheet path integral is constrained by reparametrization and Weyl invariance, so the worldsheet Hamiltonian vanishes on physical states in the same general sense that gravitational constraints annihilate the bulk wavefunctional as we saw in \eqref{eq:ws_WdW_constraints}. Thus the worldsheet Schwinger--Dyson and Wheeler--DeWitt-type equations do not generate evolution in an external time variable. The correct interpretation is then the following.
Worldsheet ``time'' is not a physical external time. It is first an auxiliary worldsheet coordinate, and physical propagation emerges only after gauge fixing and solving the constraints.
In the covariant formulation one starts with worldsheet coordinates
$(\tau,\sigma)$,
but these are gauge redundancies rather than observables. The physical content comes from the embedding fields
$X^\mu(\tau,\sigma)$,
together with the Virasoro constraints. Because of these constraints, the worldsheet Hamiltonian is not an ordinary generator of physical target-space time evolution.

To obtain a more standard notion of propagation, one usually gauge-fixes. In conformal gauge, one keeps the residual Virasoro constraints and interprets propagation through worldsheet correlation functions and the string spectrum. In light-cone gauge, one chooses
\begin{equation}
X^+(\tau,\sigma) =
\frac{1}{\sqrt{2}}(X^0+X^{d-1})
=
x^+
+
2\alpha' p^+ \tau,
\label{eq:lightcone_gauge_clock}
\end{equation}
so that one of the target-space coordinates
plays the role of the evolution parameter. Then the remaining transverse fields evolve with respect to $X^+$, and one obtains an ordinary light-cone Hamiltonian description.

Thus the analogy with the bulk discussion is the following. In the bulk GS case one may have an exact boundary time and then an emergent WKB time. On the worldsheet there is no analogous exact boundary time. Instead, one gets physical propagation only after gauge fixing, where a target-space coordinate is chosen to parametrize evolution. The most standard choice is the light-cone gauge, in which
$X^+
\sim \tau$
and the worldsheet theory becomes an ordinary Hamiltonian system for the transverse modes.
The worldsheet Schwinger--Dyson and Wheeler--DeWitt-type equations imply that there is no physical boundary time intrinsic to the worldsheet theory. Physical propagation is instead recovered after gauge fixing. In particular, in light-cone gauge one uses a target-space embedding coordinate $X^+$ as the clock, and the remaining degrees of freedom propagate with respect to it.

One may however inquire what happens in the case of an open string {\it i.e.} a string attached to a D-brane, or a string attached to an orbifold fixed point. Does there exist a boundary Hamiltonian now? The answer is {\it no}, and the basic story is the same, but with important refinements.
For an open string, or a string ending on a D-brane, there is still no \emph{worldsheet boundary time} analogous to the bulk four-dimensional boundary time. The worldsheet coordinates
$(\tau,\sigma)$
remain gauge variables before gauge fixing, and the physical state conditions are again imposed by the worldsheet constraints. What changes is the boundary condition on the embedding fields:
\begin{equation}
\partial_\sigma X^\alpha\big|_{\partial\Sigma}=0, ~~~~
\delta X^i\big|_{\partial\Sigma}=0 ,
\label{eq:Dirichlet_bc}
\end{equation}
the first being the Neumann directions along the brane and the second being the Dirichlet directions orthogonal to the brane.
Thus the endpoints are constrained to lie on the D-brane worldvolume, but this does not by itself introduce a new physical boundary time on the worldsheet. Propagation is still interpreted only after gauge fixing, typically again by using conformal gauge or light-cone gauge.
So for open strings attached to D-branes, the worldsheet still has no physical boundary time. The difference from the closed-string case lies in the boundary conditions, not in the emergence of a new time variable. After gauge fixing, one may again use a target-space light-cone coordinate such as $X^+$ as the effective clock.

For a string attached to an orbifold fixed point, for example in a
$\mathbb{Z}_2$
orbifold, the situation is slightly different in language but not in principle. In such a case the string coordinates are subject to orbifold projection conditions. For a coordinate reflected by the orbifold,
$X^i \sim -X^i$,
one may have twisted-sector boundary conditions of the form:
\begin{equation}
X^i(\sigma+2\pi,\tau)
=
-\,X^i(\sigma,\tau).
\label{eq:twisted_boundary_condition}
\end{equation}
The fixed point is the locus left invariant by the orbifold action, and twisted strings are localized there in the orbifold directions. However, this localization does not create a physical boundary time either. It only changes the allowed sectors of the worldsheet theory and the corresponding mode expansions.
Thus in the orbifold case, strings localized at an orbifold fixed point are constrained in target space by the orbifold projection or twisted-sector conditions, but the worldsheet still does not acquire a physical boundary time. Propagation is again interpreted only after gauge fixing, typically by choosing a target-space coordinate such as $X^+$ as the effective clock.

There is, however, one useful distinction between the two cases.
For a D-brane, the endpoint condition is dynamical in the sense that the brane carries worldvolume degrees of freedom, and open-string excitations are naturally interpreted as fields living on the brane:
\begin{equation}
A_\alpha(X),\qquad \Phi^i(X),\qquad \Psi(X),\ldots
\label{eq:brane_worldvolume_fields}
\end{equation}
depending on the particular setup. So once gauge fixing is performed, the effective propagation may often be interpreted as target-space propagation along the brane worldvolume.
For an orbifold fixed point, by contrast, the localization is usually geometric and group-theoretic rather than a boundary condition on an actual dynamical brane. Twisted states are localized near the fixed point, but they do not arise from open-string endpoints satisfying Neumann/Dirichlet conditions. Instead they arise from nontrivial periodicity conditions in the closed-string sector or from projected open-string sectors, depending on the construction.

The above discussion then may be concluded in the folowing way. For open strings, D-branes, and orbifold fixed points alike, there is still no intrinsic worldsheet boundary time analogous to the bulk four-dimensional case. What changes are the boundary conditions or twisted-sector identifications. Physical propagation is recovered only after gauge fixing, most cleanly in light-cone gauge, where a target-space coordinate such as $X^+$ plays the role of the effective clock.

\section{Discussion and conclusion \label{sec9.0}}

In this paper we have developed a unified framework for understanding the role of coherent states \cite{dvali} and Glauber--Sudarshan states \cite{joydeep,wdwpaper,coherbeta,coherbeta2,borel2,hetborel,dileep,desitter2} in a diffeomorphism-invariant theory of gravity. The central tension addressed throughout the paper is the following. On the one hand, the bulk Hamiltonian of a gravitational theory is a sum of constraints and therefore vanishes on the physical constraint surface. On the other hand, semiclassical states in a fixed or asymptotically defined background do exhibit genuine time evolution. The resolution is that physical time evolution is not generated by an unconstrained bulk Hamiltonian, but by boundary data, asymptotic symmetries, or by an emergent WKB time on a chosen semiclassical branch.

The first part of the paper clarified this point in canonical language. Starting from the Einstein--Hilbert action supplemented by the Gibbons--Hawking--York term, and then extending the discussion to higher-curvature and matter-coupled systems, we reviewed how the total Hamiltonian takes the schematic form \cite{adm,teitelboim,ghy}:
\begin{equation}
H_{\rm tot}
=
\int_\Sigma d^{d}x\,
\left(
N{\cal H}_\perp
+
N^i{\cal H}_i
\right)
+
H_{\partial\Sigma}.
\end{equation}
The constraints ${\cal H}_\perp$ and ${\cal H}_i$ enforce refoliation and spatial diffeomorphism invariance. Consequently, on the physical constraint surface,
${\cal H}_\perp\approx 0,
{\cal H}_i\approx 0$,
and the nontrivial physical generator, when it exists, is the boundary Hamiltonian:
\begin{equation}
H_{\rm phys}=H_{\partial\Sigma}.
\end{equation}
This statement must be interpreted with the appropriate boundary conditions in mind. In asymptotically flat or warped-Minkowski sectors, the boundary Hamiltonian is the ADM, Brown--York, or Iyer--Wald charge associated with the chosen asymptotic time. It is this charge that gives a precise meaning to physical time evolution.

This canonical structure explains why ordinary coherent states can be consistently defined around fixed warped Minkowski backgrounds. Once a supersymmetric Minkowski branch is selected, the background supplies a preferred time, and small fluctuations admit a standard Hamiltonian description. In this regime, a coherent state of gravitational or matter fluctuations behaves much like a coherent state in ordinary QFT:
\begin{equation}
|\alpha\rangle
=
D_{{}_{\rm CS}}(\alpha)|0\rangle,
\end{equation}
with expectation values peaked around a classical profile, where $D_{{}_{\rm CS}}$ is defined in \eqref{sweenseyf}. However, such coherent states are naturally tied to perturbative expansions around a free or weakly interacting vacuum. In a gravitational theory with strong interactions, higher-curvature corrections, instantons, nonlocal terms and ghost sectors, this is not the most general or most natural state construction.

We need something that can capture not only the classical trajectories, {\it i.e.} the on-shell dynamics, but also the corresponding off-shell behavior.
This motivates the use of Glauber--Sudarshan (GS) states. In the framework developed here, a GS state is defined over an interacting gravitational vacuum or reference state:
\begin{equation}
|\sigma\rangle
=
D_{{}_{\rm GS}}(\sigma)|\Omega\rangle,
\end{equation}
where $|\Omega\rangle$ already incorporates the interactions of the underlying theory, and where $D_{{}_{\rm GS}}$ is defined in \eqref{sweenseyf}. This is a more elaborate and Lorentz invariant operator compared to the coherent state case. The expectation value of an operator ${\cal O}$ in such a state is schematically of the form:
\begin{equation}
\langle{\cal O}\rangle_\sigma
=
{
\langle\Omega|
D_{{}_{\rm GS}}^\dagger(\sigma)\,
{\cal O}\,
D_{{}_{\rm GS}}(\sigma)
|\Omega\rangle
\over
\langle\Omega|
D_{{}_{\rm GS}}^\dagger(\sigma)D_{{}_{\rm GS}}(\sigma)
|\Omega\rangle
}.
\end{equation}
This formula makes clear that the GS insertion is not an external source in the sense required for a Legendre transform. Rather, it is part of the definition of the state. Thus GS states are not constrained by the convexity logic of the exact 1PI effective action in the same way as candidate vacua. The relevant object is not a new minimum of an effective potential, but a normalized expectation value in a chosen excited state.

This distinction is especially important for the interpretation of de Sitter spacetime. The de Sitter-like configurations considered here are not eternal de Sitter vacua and are not claimed to be metastable minima of a Wilsonian or 1PI effective action. They are transient semiclassical profiles produced by GS states over supersymmetric Minkowski sectors. More precisely, the claim is that for a finite time interval
$t_i<t<t_f$,
one may obtain:
\begin{equation}
\langle \widehat g_{\mu\nu}(x,t)\rangle_\sigma
\simeq
g^{\rm qdS}_{\mu\nu}(x,t),
\qquad
t_i<t<t_f.
\end{equation}
Outside this interval the state need not remain close to a de Sitter configuration. This finite-time interpretation avoids the need to construct an eternal de Sitter vacuum, a de Sitter S-matrix, or a stable positive minimum of the exact effective action.

The Schwinger--Keldysh formalism provides the natural path-integral language for computing expectation values over these states. Since the relevant observables are expectation values in a time-dependent state, rather than transition amplitudes between in- and out-vacua, the in-in contour is the appropriate framework. The doubled fields on the forward and backward branches allow one to compute real-time observables while maintaining the normalization and reality properties of expectation values. In this language the GS state appears through branch-dependent insertions, and the expectation value takes the following schematic form:
\begin{equation}\label{blazrag100}
\langle{\cal O}\rangle_\sigma
=
{
\int {\cal D}\Xi_+{\cal D}\Xi_-\,
{\cal O}[\Xi_+]\,
\exp\left(iS_{\rm tot}[\Xi_+]-iS_{\rm tot}[\Xi_-]\right)
{D}_+[\sigma]{D}_-[\sigma]
\over
\int {\cal D}\Xi_+{\cal D}\Xi_-\,
\exp\left(iS_{\rm tot}[\Xi_+]-iS_{\rm tot}[\Xi_-]\right)
{D}_+[\sigma]{D}_-[\sigma]
}.
\end{equation}
Here $\Xi$ denotes collectively the physical fields, including FP ghosts, and ${D}_\pm[\sigma]$ denote the GS insertions on the two branches. This expression emphasizes that the construction is state-dependent and normalized.

The in-out formalism can still be useful, but only in a more limited sense. In-out amplitudes compute transition functions and need not be real expectation values. They can be derived from or compared with the SK formalism only under special assumptions, such as suitable matching of sources, boundary conditions, and contour reductions. For transient quasi--de Sitter configurations, the in-in formalism remains the primary description.

The trans-series {\transpapers} and Lefschetz-thimble \cite{picardlefschetz,wittenlefschetz} analyses give a further layer of structure. The full action entering the path integral \eqref{blazrag100} is not a simple finite local functional. It contains perturbative terms, higher-derivative corrections, instanton contributions, nonlocal terms, ghost contributions and gauge-fixing sectors. Schematically:
\begin{equation}\label{ellatagra}
S_{\rm tot}
=
S_{\rm pert}
+
\sum_I e^{-A_I/g}\,S_I
+
\sum_{I,J}e^{-(A_I+A_J)/g}\,S_{IJ}
+
S_{\rm nloc}
+
S_{\rm gh}
+
S_{\rm gf}
+
\cdots ,
\end{equation}
where $S_{\rm pert}$ denotes the perturbative fluctuations around the zero instanton sector; $S_{\rm nloc}$, $S_{\rm gh}$, and $S_{\rm gf}$ denote the non-local, ghost and gauge fixing contributions respectively (which may themselves be in the trans-series form); $S_{I}$ and $S_{IJ}$ denote the fluctuation determinants along the multi-instanton sectors (both real and complex ones); and the dotted terms are the higher instanton contributions.
The thimble decomposition organizes the corresponding path integral into saddle sectors:
\begin{equation}
{\cal C}
=
\sum_s n_s{\cal J}_s,
\end{equation}
where ${\cal J}_s$ are the relevant steepest-descent cycles and $n_s$ are intersection numbers. In this sense, the GS expectation value is naturally a trans-series object. Different saddles and thimbles contribute to the same state-dependent observable, and the semiclassical geometry arises only after this full organization is taken into account.

This brings us to one of the central result of the paper, namely, the emergence of a bootstrap equation for GS states. The basic Schwinger--Dyson identity in the presence of GS insertions has the schematic form:
\begin{equation}\label{ellatag22}
\left\langle
{\delta S_{\rm tot}\over \delta \Xi(x)}
\right\rangle_\sigma
=
\left\langle
{\delta\over \delta \Xi(x)}
\log
\left(
D^\dagger(\sigma)D(\sigma)
\right)
\right\rangle_\sigma .
\end{equation}
This equation expresses the condition that the chosen GS data and the resulting expectation values be mutually consistent. This equation is valid for any GS state $|\sigma\rangle$. It is not enough to choose an arbitrary function $\sigma$ and declare that it produces a desired geometry. Rather, the GS profile must lie in a basin of attraction of the bootstrap map which appears from above in a special {\it split} form \eqref{bootsforwalking} for the so-called ``on-shell" states. Fixed points of this map, appearing from \eqref{bootsforwalking} and {\it not} from generic form \eqref{ellatag22}, correspond to self-consistent semiclassical configurations, and quasi--de Sitter profiles may be interpreted as finite-time fixed points or approximate fixed points of this bootstrap dynamics.

The nonuniqueness of the bootstrap map is an added advantage. It reflects the fact that the theory may contain several semiclassical branches, each associated with different 
expectation values over the on-shell GS states that solve the bootstrap map emanating from \eqref{bootsforwalking} as shown in {\bf figures \ref{bootstrapbasins}}, {\bf \ref{bootsinksourcesaddle}} and {\bf \ref{sssrotate}}. These various semiclassical configurations have different form of the sources but they all appear from either directly solving \eqref{bootsforwalking}, or by following the bootstrap map. 
However it is worth emphasizing that, if the system of equations in
\eqref{bootsforwalking} could be solved exactly, the iterative
procedure would be unnecessary. In the present case, the
equations depend crucially on the trans-series structure of the action,
making it exceedingly difficult to anticipate, let alone construct,
their exact solutions. Under such circumstances, the bootstrap
procedure adopted here provides the most practical, and potentially the
only systematic framework for analyzing the system. The important requirement is not uniqueness, but consistency. A physically meaningful GS state must satisfy the split form of the Schwinger--Dyson equations \eqref{bootsforwalking}, respect the gauge-fixed or BRST structure of the theory, and produce expectation values whose time evolution is compatible with the boundary or WKB notion of time. The power of the bootstrap is that even if some semiclassical configuration is a source or a stationary point of the bootstrap map, one could still find a way to approach towards the relevant solution.

A more subtle issue is the problem of time that plagues any quantum gravity computation. For our case we have a slight advantage: the Wheeler--DeWitt equations are \eqref{eq:boundary_SE_exact}, which provides us with a clear notion of a boundary time, thus 
clarifying the meaning of time in this construction. At the microscopic level, the exact wavefunctional obeys a constraint equation of the form:
\begin{equation}
\widehat{\cal H}_{\rm bulk}\Psi[\Xi]=0 ,
\end{equation}
which appears from the first two equations in \eqref{eq:boundary_SE_exact}.
This is the frozen, background-independent description constraining the bulk dynamics in a way so as to tie up the gravitational and the matter DOFs together. One may now  choose a semiclassical branch by making the following WKB ans\"atze:
\begin{equation}
\Psi[h,\Phi]
=
\exp\left(
{i\over \hbar}S_0[h]
\right)
\psi[h,\Phi],
\end{equation}
where $S_0$ solves the corresponding Hamilton-Jacobi equation, and we have put the ghosts to zero. (In general $\Psi[\Xi] = \Psi[h, \Phi; \Upsilon]$, where $\Upsilon =$ ghosts.) This semiclassical branch is exactly one of the fixed point of the bootstrap map that appears in {\bf figures \ref{bootstrapbasins}}, {\bf \ref{bootsinksourcesaddle}} and {\bf \ref{sssrotate}}, thus further tying up the dynamics together. This is evident from the appearance of 
an {\it emergent} Schr\"odinger equation:
\begin{equation}
i{\partial \psi\over \partial t_{\rm WKB}}
=
\widehat H_{\rm eff}\psi ,
\end{equation}
expressed using the WKP time. 
The time $t_{\rm WKB}$ is branch-dependent and is determined by the classical gravitational Hamilton--Jacobi functional $S_0[h]$. At this point we take advantage of the third equation in the set \eqref{eq:boundary_SE_exact}, namely, the {\it boundary} Schr\"odinger equation. Combining the two Schr\"odinger equations together produces the exact bulk and boundary Schr\"odinger equation given by \eqref{blazrog}, \eqref{blazrog2} and \eqref{blazrog3} with increasing level of precision. 
Thus the boundary-time and WKB-time descriptions are not contradictory. They are two complementary ways of describing time after a semiclassical branch has been chosen.

There is yet another level of subtlety.
The GS construction also leads to an emergent, reduced wavefunctional. The exact microscopic wavefunctional $\Psi[\Xi]$ depends on the full gauge-fixed configuration space, including ghosts and gauge-fixing variables. By contrast, the GS-reduced wavefunctional depends on expectation-value variables
$\Psi'
=
\Psi'\left(\langle \widehat{\check\Xi}\rangle_\sigma\right)$, where $\Xi = (\check\Xi, \Upsilon)$ with $\Upsilon$ being the FP ghosts. 
It satisfies an effective Wheeler--DeWitt-type equation:
\begin{equation}
\widehat{\cal H}_{\rm eff}
\left(
\langle \widehat{\check\Xi}\rangle_\sigma
\right)
\Psi'
\left(
\langle \widehat{\check\Xi}\rangle_\sigma
\right)
=
0 ,
\end{equation}
where $\langle \widehat\Upsilon\rangle_\sigma = 0$ by definition.
This equation should not be obtained by naively replacing microscopic fields by their expectation values in the exact WdW equation. Rather, it is the constraint equation of the reduced GS sector, derived after integrating or projecting onto the appropriate {\it ghost free} effective variables, and after introducing non-trivial wavefunction renormalization following \cite{wdwpaper}.

Finally, we introduce a structural analogy between bulk GS insertions and world-sheet vertex operators. A physical string vertex operator is an insertion in the two-dimensional world-sheet path integral that creates or probes a target-space string state. Similarly, a GS insertion is an operator insertion in the bulk path integral that selects a semiclassical state. The analogy may be summarized schematically as
\begin{equation}
\text{\textcolor{blue}{world-sheet vertex insertion}}
\qquad
\Longleftrightarrow
\qquad
\text{\textcolor{blue}{bulk GS insertion}}
\end{equation}
However, this analogy is only structural. A world-sheet vertex operator must satisfy BRST and conformal-dimension constraints, while a bulk GS insertion must satisfy the gravitational Schwinger--Dyson and constraint/bootstrap equations. The two theories have different dynamics, different gauge symmetries, and different notions of physical state.

The world-sheet discussion also reinforces the interpretation of time. In the covariant world-sheet formulation, there is no external target-space time built into the two-dimensional path integral. Time emerges only after choosing a target-space interpretation, a semiclassical background, or a gauge such as light-cone gauge in which one embedding coordinate plays the role of a clock. This parallels the bulk discussion, where time emerges from boundary conditions, WKB branches, or relational structure rather than from a fundamental unconstrained Hamiltonian.

The overall conclusion is therefore as follows. Coherent states \cite{dvali} and GS states {\gspapers} play different roles in quantum gravity. Coherent states are well suited for describing perturbative excitations around a fixed background with a preferred time. GS states are more general: they are state-dependent constructions over an interacting gravitational vacuum, capable of producing transient semiclassical geometries that need not be minima of any potential. In particular, a quasi--de Sitter geometry may arise as:
\begin{equation}
\langle \widehat g_{\mu\nu}\rangle_\sigma
\simeq
g^{\rm qdS}_{\mu\nu}
\end{equation}
over a finite time interval without implying the existence of an eternal de Sitter vacuum, thus being perfectly consistent with the so-called trans-Planckian censorship \cite{tcc}. 

This perspective avoids several standard difficulties associated with de Sitter in quantum gravity. It does not require a positive minimum of the exact 1PI effective action, does not require a time-independent Wilsonian EFT valid over arbitrarily long de Sitter times, and does not require de Sitter asymptotics. Instead, it requires a well-defined reference Minkowski sector, a consistent GS state, an in-in path-integral formulation, and a bootstrap equation ensuring that the resulting expectation values are compatible with the full quantum dynamics. Although we will not discuss this point here, GS states can provide a natural explanation for dynamical dark energy \cite{desibao,jabbari,dieterobert}, as was recently shown in \cite{hetborel}.

Several open problems remain. First, the bootstrap equations should be solved in explicit examples, preferably in string or M-theory compactifications where the reference supersymmetric Minkowski sector is known. Second, the thimble decomposition should be made more explicit, including the role of Stokes phenomena and the choice of integration cycles. Third, the relation between the exact gauge-fixed Wheeler--DeWitt equation and the emergent GS-sector Wheeler--DeWitt equation should be developed more rigorously. Fourth, the analogy with world-sheet vertex operators should be sharpened by identifying the precise bulk analogues of BRST invariance, marginality, and physical-state conditions.

Despite these open questions, the framework developed here gives a coherent interpretation of transient quasi--de Sitter configurations in quantum gravity. The essential idea is not to replace the supersymmetric Minkowski vacuum by a de Sitter vacuum, but to construct a controlled excited state over the Minkowski sector whose expectation values temporarily mimic a quasi--de Sitter geometry. In this sense, GS states provide a natural language for describing finite-time cosmological backgrounds within a theory whose exact microscopic structure remains constrained by diffeomorphism invariance, trans-series dynamics, and the Wheeler--DeWitt equation.

\section*{Acknowledgements}

We would like to thank Robert Brandenberger, Joydeep Chakravarty, and Sumit Das for many helpful discussions. The work of TD, KD, and YKL
is supported in part by a grant from the Natural Sciences and
Engineering Research Council of Canada (NSERC). TD also acknowledges partial
support from a Trottier Space Institute Fellowship. The research at the University of Lethbridge is supported by Quantum Horizons Alberta and in part by the Natural Sciences and Engineering Research Council of Canada through Discovery Grant RGPIN-2026-05926.

\begin{table}[H]
\centering
\renewcommand{\arraystretch}{1.5}
\begin{tabular}{|p{0.47\textwidth}|p{0.47\textwidth}|}
\hline
\textbf{Worldsheet construction} & \textbf{GS bootstrap construction} \\
\hline

Two-dimensional generally covariant worldsheet theory with dynamical metric $h_{ab}$ and embedding fields $X^\mu$
&
Four-dimensional or higher-dimensional in--in bulk theory with Schwinger--Keldysh doubling of fields, including $g_+$, $g_-$, matter, and ghosts
\\
\hline

\begin{equation}
e^{\,iS_{\rm ws}[X,h,\text{gh}]} \nonumber
\end{equation}
in Lorentzian signature, or
\begin{equation}
e^{-S_{\rm ws}^{(E)}[X,h,\text{gh}]} \nonumber
\end{equation}
in Euclidean signature
&
\begin{equation}
e^{\,iS[g_+,\Phi_+,\text{gh}_+]-iS[g_-,\Phi_-,\text{gh}_-]} \nonumber
\end{equation}
with the GS insertions multiplying the contour weight
\\
\hline

Worldsheet Hamiltonian is a sum of constraints:

{\footnotesize
\begin{equation}
H_{\rm ws}
=
\int d\sigma\,
\left(
N{\cal H}_{\perp}
+
N^\sigma{\cal H}_{\sigma}
\right),
~~
H_{\rm ws}\approx 0 \nonumber
\end{equation}}
&
Boundary Hamiltonian exist in Minkowski space, but 
the bootstrap map is not a Hamiltonian flow in physical time. It is an iteration on the space of GS data:
\begin{equation}
\sigma_{n+1}={\cal F}(\sigma_n) \nonumber
\end{equation}
\\
\hline

Worldsheet time $\tau$ is gauge. Physical evolution is described after gauge fixing or relationally in terms of target-space time such as $X^0$ or $X^+$
&
The iteration label $n$ is not physical time. It counts steps of the self-consistency update in GS-state space
\\
\hline

Virasoro or BRST constraints:

{\footnotesize
\begin{equation}
L_n|\Psi\rangle=0,
~~
\widetilde L_n|\Psi\rangle=0,
~~
Q_{\rm BRST}|\Psi\rangle=0 \nonumber
\end{equation}}
&
Boundary Hamiltonian $\hat{\mathcal H}_{\rm phys}|\Psi\rangle = \hat{\mathcal H}_{\partial}|\Psi\rangle$, so doesn't annihilate $|\Psi\rangle$, leading to
coupled bootstrap conditions:

{\footnotesize
\begin{equation}
{\cal E}_{AB}[\Xi]=0,
\qquad
{\cal B}_{AB}[\sigma,\Xi]=0 \nonumber
\end{equation}}
or equivalently the fixed-point condition

{\footnotesize
\begin{equation}
{\cal F}(\sigma)=\sigma \nonumber
\end{equation}}
\\
\hline

\begin{equation}
{\cal H}_{\rm phys}
=
H^\bullet(Q_{\rm BRST}) \nonumber
\end{equation}
which is infinite-dimensional in ordinary string theory
&
Self-consistent GS branches correspond to fixed points of the bootstrap map. Different fixed points may describe static, de Sitter, FRW, or more exotic cosmological branches
\\
\hline

\end{tabular}
\caption{Comparison between the worldsheet construction and the bulk GS construction, part I.}
\label{tab:WS_vs_GS_part1}
\end{table}

\begin{table}[H]
\centering
\renewcommand{\arraystretch}{2.0}
\begin{tabular}{|p{0.47\textwidth}|p{0.47\textwidth}|}
\hline
\textbf{Worldsheet construction} & \textbf{GS bootstrap construction} \\
\hline

BRST-invariant vertex operators $V(z,\bar z)$ create physical string states
&
The GS operators
\begin{equation}
 D(\sigma),
\qquad
 D^\dagger(\sigma) \nonumber
\end{equation}
define the deformed in--in state
\\
\hline

\begin{equation}
e^{\,iS_{\rm ws}}\,V_1\cdots V_n \nonumber
\end{equation}
&
\begin{equation}
e^{\,iS[g_+]-iS[g_-]}\, D(\sigma)\,
 D^\dagger(\sigma) \nonumber
\end{equation}
\\
\hline

\begin{equation}
{\cal V}
=
\int d^2z\,V(z,\bar z) \nonumber
\end{equation}
may be inserted multiplicatively as
\begin{equation}
\exp\!\left(\lambda {\cal V}\right) \nonumber
\end{equation}
&
The contour insertion is encoded by
\begin{equation}
W_\pm[g_\pm;\sigma] \nonumber
\end{equation}
through
\begin{equation}
\exp\!\left(W_\pm[g_\pm;\sigma]\right) \nonumber
\end{equation}
\\
\hline

No extra explicit factor of $i$ multiplies the inserted vertex operator itself. The factor of $i$ belongs to the worldsheet action phase
&
No extra explicit factor of $i$ multiplies $D(\sigma)$ or $D^\dagger(\sigma)$ themselves. The factor of $i$ belongs to the SK action
\\
\hline

If one rewrites the insertion as part of the Lorentzian action, the action becomes effectively complex:

{\footnotesize
\begin{equation}
e^{\,iS_{\rm ws}}
e^{\,\lambda \int d^2z\,V}
=
\exp\!\left[
i\left(
S_{\rm ws}
-
i\lambda\int d^2z\,V
\right)
\right] \nonumber
\end{equation}}
&
Similarly,

{\footnotesize
\begin{equation}
e^{\,iS[g_+]}e^{\,W_+[g_+;\sigma]}
=
\exp\!\left[
i\left(
S[g_+]-iW_+[g_+;\sigma]
\right)
\right] \nonumber
\end{equation}}
and analogously on the backward branch
\\
\hline

An integrated vertex operator deforms the full worldsheet theory; it is not merely boundary data
&
The GS insertion acts over the full SK contour history; it is not merely an initial-slice preparation
\\
\hline

A single local or slice-based state preparation is not equivalent to a worldsheet-filling integrated deformation
&
An ordinary coherent state prepared on one Cauchy slice modifies only the boundary wavefunctional and is not equivalent to the GS contour insertion
\\
\hline

\end{tabular}
\caption{Comparison between the worldsheet construction and the bulk GS construction, part II.}
\label{tab:WS_vs_GS_part2}
\end{table}

\begin{table}[H]
\centering
\renewcommand{\arraystretch}{2.2}
\begin{tabular}{|p{0.47\textwidth}|p{0.47\textwidth}|}
\hline
\textbf{Worldsheet construction} & \textbf{GS bootstrap construction} \\
\hline

If treated as a genuine deformation of the worldsheet theory, the integrated operator shifts the worldsheet Schwinger--Dyson identities
&
The GS insertion shifts the effective in--in Einstein/Schwinger--Dyson equations entering the bootstrap system
\\
\hline

The deformation is consistent only if conformal invariance is preserved, namely
\begin{equation}
\beta^I(\lambda)=0 \nonumber
\end{equation}
&
The deformation is consistent only if the GS state and the backreacted mean fields satisfy
\begin{equation}
{\cal E}_{AB}=0,
\qquad
{\cal B}_{AB}=0 \nonumber
\end{equation}
\\
\hline

Vanishing beta functions select consistent string backgrounds
&
The coupled bootstrap equations select self-consistent GS-deformed branches
\\
\hline

The deformation parameter is not arbitrary if conformal symmetry is to be preserved
&
The GS label $\sigma$ is not arbitrary if self-consistency from WdW and Schr\"odinger equations is imposed
\\
\hline

A generic deformation away from marginality is off-shell with respect to conformal invariance
&
A generic perturbation of a GS fixed point is off-shell with respect to the coupled bootstrap system
\\
\hline

The issue is whether the deformation remains marginal under renormalization
&
The issue is whether the corresponding fixed point of ${\cal F}$ is a sink, source, or saddle under bootstrap iteration
\\
\hline

Different marginal branches correspond to different target-space backgrounds
&
Different fixed points of ${\cal F}$ correspond to different cosmological branches and may possess distinct basins of attraction
\\
\hline

Conformal invariance of the deformed worldsheet theory
&
Bootstrap self-consistency of the GS-deformed spacetime branch
\\
\hline

\end{tabular}
\caption{Comparison between the worldsheet construction and the bulk GS construction, part III.}
\label{tab:WS_vs_GS_part3}
\end{table}



\end{document}